\documentclass{osudissert96}

\usepackage[pdftitle=Searching for the Proton's Missing Spin,pdflang=en-US]{hyperref}

\usepackage{cite}

\usepackage[utf8]{inputenc}
\usepackage{float}
\usepackage{graphics,graphicx}
\usepackage{dcolumn}
\usepackage{bm}
\usepackage{amsmath}
\usepackage{tensor}
\usepackage{slashed}
\usepackage{amsmath,amssymb,amsfonts,mathrsfs,bbold}
\usepackage{yfonts}
\usepackage{xcolor}
\usepackage{mathtools}
\usepackage[framemethod=TikZ]{mdframed}
\usepackage{color}

\usepackage{multirow}
\usepackage{mathbbol}
\usepackage{mathtools}

\usepackage{slashed}
\usepackage{simplewick}

\usepackage[normalem]{ulem}

\usepackage{stackrel}
\usepackage[most]{tcolorbox}
\usepackage{braket}
\usepackage{caption}
\usepackage{subcaption}
\usepackage{simpler-wick}

\def\eq#1{{Eq.~(\ref{#1})}}

\def\fig#1{{Fig.~\ref{#1}}}
\newcommand{\thalf}{\tfrac{1}{2}}

\newcommand{\as}{\alpha_s}
\newcommand{\soz}{s_{10}}
\newcommand{\sto}{s_{21}}
\newcommand{\stt}{s_{32}}
\newcommand{\xoz}{x_{10}}
\newcommand{\xto}{x_{21}}
\newcommand{\xtt}{x_{32}}
\newcommand{\wint}{\int \frac{\mathrm{d}\omega}{2\pi i}}
\newcommand{\gint}{\int \frac{\mathrm{d}\gamma}{2\pi i}}

\newcommand{\dw}{\delta_{\omega}}
\newcommand{\gw}{\gamma_{\omega}}

\newcommand{\llangle}{\Big\langle \!\! \Big\langle}
\newcommand{\rrangle}{\Big\rangle \!\! \Big\rangle}

\newcommand{\abar}{\overline{\alpha}_s}

\newcommand{\gtwg}{G_{2\omega\gamma}}

\newcommand{\pd}{\partial}
\newcommand{\bas}{{\bar\alpha}_s}
\newcommand{\un}[1]{\underline{#1}}
\newcommand{\ul}[1]{\underline{#1}}
\newcommand{\tr}{\mbox{tr}}

\newcommand{\stf}{\frac{-\frac{3}{2}\omega\gamma + 4}{\gamma^2-\omega\gamma+1}}

\makeatletter
\DeclareRobustCommand{\cev}[1]{%
  {\mathpalette\do@cev{#1}}%
}
\newcommand{\do@cev}[2]{%
  \vbox{\offinterlineskip
    \sbox\z@{$\m@th#1 x$}%
    \ialign{##\cr
      \hidewidth\reflectbox{$\m@th#1\vec{}\mkern4mu$}\hidewidth\cr
      \noalign{\kern-\ht\z@}
      $\m@th#1#2$\cr
    }%
  }%
}
\makeatother

\def\BibTeX{{\rm B\kern-.05em{\sc i\kern-.025em b}\kern-.08em
    T\kern-.1667em\lower.7ex\hbox{E}\kern-.125emX}}

\renewcommand\typesetChapterTitle[1]{#1}

\begin{document}

\author{Jeremy Borden}

\title{Searching for the Proton's Missing Spin:\\Small-$x$ Helicity Evolution Equations and Their Analytic Solutions}

\authordegrees{B.A., M.S., Ph.D.}
\unit{Physics}

\advisorname{Yuri V. Kovchegov}
\member{Michael Annan Lisa}
\member{Samir D. Mathur}


\maketitle

\disscopyright


\begin{abstract}
  The proton spin puzzle denotes the challenge of describing the proton's spin in terms of the angular momenta of the quarks and gluons which comprise it. These quarks and gluons carry a fraction $x$ of the parent proton's momentum. A potentially important region of phase space is that of small $x$, where the quarks and gluons only possess a little of the proton's momentum. Contributions from these small-$x$ quarks and gluons are difficult to measure, though, since doing so requires very high energy experiments. Furthermore, the pioneering theoretical work of Bartels, Ermolaev, and Ryskin (BER) in the 1990s predicted substantial contributions to the proton spin from these small-$x$ particles. Clearly we need theoretical control over this corner of phase space in order to confidently resolve the spin puzzle.

In this dissertation, we build upon an existing framework for studying spin at small-$x$. Previously, several sets of small-$x$ evolution equations were derived in this formalism --- one in the large-$N_c$ limit and one in the large-$N_c\&N_f$ limit. Here $N_c$ and $N_f$ are the numbers of quark colors and flavors, respectively. These equations were numerically solved, resulting in several successful comparisons to existing work, but no analytic solutions had been found. In this dissertation we detail the construction of such analytic solutions. First we analytically solve the evolution equations in the large-$N_c$ limit. We subsequently derive an important correction to the existing large-$N_c\&N_f$ evolution equations coming from the contributions of quark-to-gluon transition operators. We then construct an analytic solution for the modified evolution in this limit as well.

From the solutions constructed here, we can predict the behavior of the quark and gluon helicity distributions, $\Delta \Sigma(x,Q^2)$ and $\Delta G(x,Q^2)$, and the related $g_1$ structure function at asymptotically small-$x$ (and large-$N_c$ or large-$N_c\&N_f$), finding
\begin{align*}
    \Delta \Sigma(x,Q^2) \sim \Delta G(x,Q^2) \sim g_1(x,Q^2) \sim \left(\frac{1}{x}\right)^{\alpha_h}\,,
\end{align*}
where we can obtain an exact analytic expression for the intercept $\alpha_h$ in the large-$N_c$ limit and the exact algebraic equation satisfied by $\alpha_h$ in the large-$N_c\&N_f$ limit. In addition to the general power law behavior governed by the intercept, we further obtain explicit analytic expressions for the helicity distributions in the asymptotic limit.

Our solutions also allow us to predict all four polarized DGLAP anomalous dimensions (at small-$x$ and in the large-$N_c$ or large-$N_c\&N_f$ limits), yielding expressions exact to all orders in the strong coupling $\as$. The expansions of our predictions in $\as$ agree completely with the full extent of existing finite-order calculations, to three loops.

The predictive power afforded by our analytic solutions also allows us to compare our predictions to the work of BER in more detail than ever before, despite the significant differences between their framework and our own. In this dissertation we will explore both the agreements and disagreements that result from these comparisons.

\end{abstract}


\begin{acknowledgements}

There is more gratitude to be shared than I have words to express here. The people (and cats) who have been with me these past years have done more than I will ever be able to repay, and I thank them all dearly.

I thank my advisor, Dr. Yuri Kovchegov. His knowledge and his expertise steered my PhD with a steady hand. The support, guidance, and education he offered me were the bedrock that bolstered my entire journey. I could not have done any of this without him. 

I thank all the friends that have made my life so special --- Mena, Ally, and Phoebe, my friends from college who became family forever; Justin and Nick, with whom I survived (barely) my physics major; Brandon, Brian, and Madhur, the physicists who have helped me grow and learn and laugh these past years in grad school. And to the many others I owe so much, invariably I thank you.

To my grandparents, whose wisdom and love shaped me from an early age. Your guidance has always been special. 

To my big brother, my first friend and my best. I have always looked up to you, and having your example to follow in life is an unwavering privilege.

To my parents, who have worked tirelessly to support me in every aspect of my life. I will never be able to thank you enough. You taught me to be a scientist, to be inquisitive, to think critically, to be caring and compassionate, to love and to share, to work hard and to laugh harder. You have both moved mountains to give me the world, and I am forever grateful.

To Emmy and to Pep, with all my love.

\end{acknowledgements}

\begin{vita}

\dateitem{December 26, 1997}{Born - Louisville, KY, USA}

\dateitem{2020}{B.A. Physics and Mathematics, Cornell University}

\dateitem{2022}{M.S. Physics, The Ohio State University}

\dateitem{2022 -- present}{Graduate Research and Teaching Associate, The Ohio State University.}

\begin{publist}

\pubitem{
J.~Borden and Y.~V. Kovchegov, \emph{{Analytic solution for the revised helicity evolution at small $x$ and large $N_c$: New resummed gluon-gluon polarized anomalous dimension and intercept}}, \href{https://doi.org/10.1103/PhysRevD.108.014001}{\emph{Phys. Rev. D} {\bfseries 108} (2023) 014001}, [\href{https://arxiv.org/abs/2304.06161}{{\ttfamily 2304.06161}}].}
\pubitem{
J. Borden, Y.V. Kovchegov, \& M. Li, \emph{{Helicity evolution at small $x$: quark to gluon and gluon to quark transition operators}}, \href{https://doi.org/10.1007/JHEP09(2024)037}{\emph{Journal of High Energy Physics} {\bfseries 09} (2024) 037}, [\href{https://arxiv.org/abs/2406.11647}{{\ttfamily 2406.11647}}].}
\pubitem{
J.~Borden and Y.~V. Kovchegov, \emph{{Analytic Solution for the Helicity Evolution Equations at Small $x$ and Large $N_c\&N_f$}}, 
\href{https://link.aps.org/doi/10.1103/ljl6-zvrq}{\emph{Phys. Rev. D} {\bfseries 113} (2026) 034002}, [\href{https://arxiv.org/abs/2508.00195}{{\ttfamily 2508.00195}}].}

\end{publist}

\begin{fieldsstudy}

\majorfield*
This material is based upon work supported by the U.S. Department of Energy, Office of Science, Office of Nuclear
Physics under Award Number DE-SC0004286 and within the framework of the Saturated Glue (SURGE) Topical
Theory Collaboration.

\end{fieldsstudy}

\end{vita}


\tableofcontents
\listoftables
\listoffigures


\chapter{Introduction}
\label{intro.ch}

Since the first measurements of the proton's form factors at the Stanford Linear Accelerator Center (SLAC) in the 1950s \cite{RevModPhys.28.214,PhysRev.92.978
}, it has been known that protons are not fundamental particles but are instead made of more fundamental ingredients. Early ideas about those ingredients came from Gell-Mann \cite{GELLMANN1964214} and Zweig \cite{Zweig:1964ruk}, who separately developed the so-called quark model. In fact, this model was developed to describe much more structure than just that of the proton. A whole host of composite particles called hadrons were being discovered and cataloged in the 1950s and 1960s, and physicists were searching for a way to describe all of their structure in terms of more basic building blocks. These more fundamental building blocks were the quarks.

\begin{figure}[!ht]
    \centering
    \includegraphics[width=0.4\linewidth]{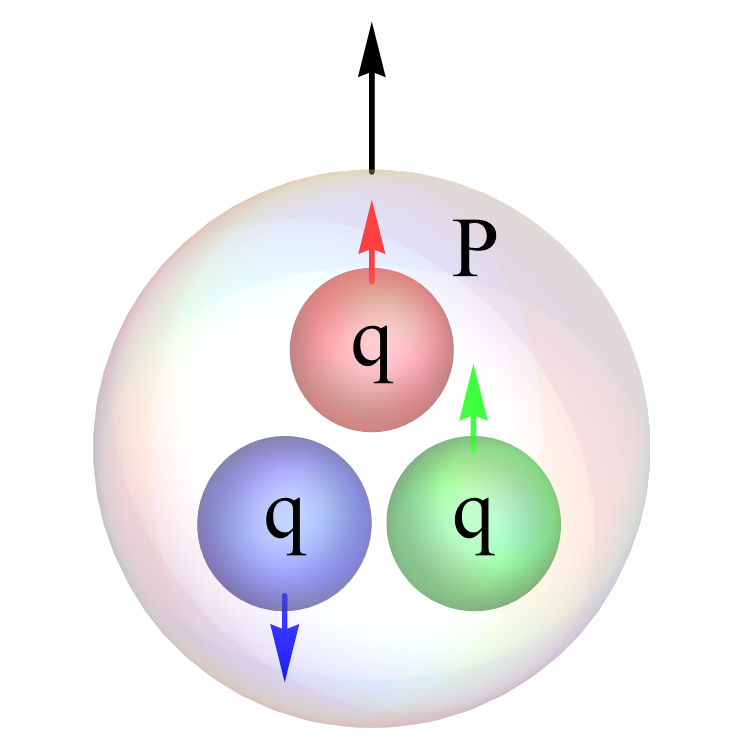}
    \caption{Naive quark model of the proton $P$ with two quark $q$ spins aligned and one anti-aligned relative to the proton spin.}
    \label{fig:constituentquarks}
\end{figure}

A relatively naive --- but in some ways still very successful --- model takes the proton to be made of three quarks (the general class of particles we call baryons are described in this way as bound states of three quarks). The quarks in this model are nonrelativistic and, like the proton, are spin-$1/2$ fermions. In such a model, it is easy to intuitively understand the proton's spin --- that is, its intrinsic angular momentum. Two of the constituent quarks have their spins pointed in the direction aligned with the proton's spin, while the third constituent quark's spin is in the opposite direction, as visualized in Fig.~\ref{fig:constituentquarks}.
In this model $100\%$ of the proton's spin is accounted for by the spin of the quarks. 

Perhaps unsurprisingly, more sophisticated models of the proton were also developed (see e.g. the bag model of \cite{PhysRevD.9.3471}). But even these sophisticated models which tried to accommodate more complicated phenomena like special relativity and confinement still predicted that a substantial quantity of the proton's spin must be carried by the quark spins, typically somewhere on the order of $60\%$.

Then in the late 1980s, the European Muon Collaboration (EMC) utilized polarized muon-proton scattering experiments to measure the net amount of the proton's spin carried by the quark spins. The shocking result was a measured value of around $6\%$ \cite{EuropeanMuon:1987isl,ASHMAN19891}. Even allowing for the maximal experimental uncertainties, this was an irreconcilable difference compared to theoretical predictions. Thus began the proton spin puzzle. The majority of the proton's spin could not be accounted for by the theoretical models of the time. 

The good news, however, is that today we have many powerful tools at our disposal to better understand the rich internal structure of the proton, chief among them Quantum Chromodynamics (QCD). Beginning in the 1970s, QCD began to emerge as the presumptive theoretical description of the strong force --- the force that binds protons and neutrons together in atomic nuclei, and as would come to be understood, the force that governs the complicated internal structures of the proton and neutron themselves, along with a host of other strongly-bound particles. 

QCD is a non-abelian $\text{SU}(N_c)$ gauge theory which describes the fundamental degrees of freedom of the strong force as quarks and gluons. The quarks of QCD are spin-$1/2$ fermions with fractional electric charges, although they are not exactly the same as the `constituent' quarks shown in Fig.~\ref{fig:constituentquarks}. There are six \textit{flavors} of quarks in the Standard Model, varying in their masses and electric charges. In addition to electric charge, the quarks are also charged under the strong force. This \textit{color} charge comes in $N_c = 3$ varieties called red, green, and blue (and the antiparticles of the quarks, the antiquarks, can carry anti-red, anti-green, or anti-blue color charge). The quarks form a color triplet and transform under the fundamental representation of $\text{SU}(N_c)$. Meanwhile gluons ---  the strong-force-carriers --- are spin-$1$ bosons that also carry a net color charge, a combination of color and anti-color (the color octet), and transform under the adjoint representation of $\text{SU}(N_c)$. 

Notably the fact that gluons are charged under the strong force means they can interact with other gluons. This is a crucial difference from abelian theories like quantum electrodynamics \cite{PhysRev.76.769} where the force-carrying particles (photons) do not self-interact. Among the consequences of the gluonic self-interactions in QCD is \textit{asymptotic freedom} \cite{PhysRevLett.30.1346,PhysRevLett.30.1343}, a remarkable property that tells us the particles of QCD interact very weakly at short distances (or large momentum transfer). This has critical implications for perturbative QCD calculations. The strong coupling --- the physical parameter that controls the strength of the force --- becomes relatively small at these short distance scales, and so we have a small dimensionless parameter in which we can make a reliable perturbative expansion. This perturbative regime of QCD is the backdrop for this entire dissertation and so the applicability of perturbation theory is critical here. Note however, that the running of the strong coupling --- that is, how the coupling changes with momentum scale --- also has important implications in the low-momentum/long distance regime. Whereas at high momentum scales the coupling is smaller and we can employ perturbation theory, at low momentum scales the coupling becomes very strong and perturbative methods break down. This also hints at the perplexing notion of confinement \cite{PhysRevD.10.2445}, whereby free color charges cannot be isolated. They are always \textit{confined} in color neutral combinations. A complete theoretical understanding of confinement is still lacking.

A particularly useful model of the the proton (and other hadrons) at high energy is Feynman's parton model \cite{PhysRevLett.23.1415}, where the proton is taken to be a system of point-like particles called \textit{partons}. Particularly effective in understanding the results of deep inelastic scattering (DIS) of electrons and protons\footnote{Much more on DIS in the next Chapter.} at SLAC\cite{PhysRevLett.23.930}, the model treats the proton (in a frame where the proton is moving ultrarelativistically) as a collection of these co-moving point-like partons which do not interact with each other. When colliding with the electron, the system of partons interacts with the electron probe incoherently. 

Feynman was agnostic about what particles these partons might be but the interpretation that emerged, and the one we still use today, is that they are the quarks and gluons of QCD. The parton framework serves as a powerful tool for understanding how the properties of the proton emerge from the properties of the intrinsic QCD degrees of freedom. But note that we are not limited to the three quark model like that in Fig.~\ref{fig:constituentquarks}. Instead we can have many partons and we will often label them with the Bjorken-$x$ variable, which corresponds to the longitudinal momentum fraction of a given parton relative to the parent proton. Intuitively, a parton could have as little as zero longitudinal momentum ($x=0$) and as much as the full momentum of the proton ($x=1$) and so $0<x<1$.

\begin{figure}[ht]
    \centering
    \includegraphics[width=0.5\linewidth]{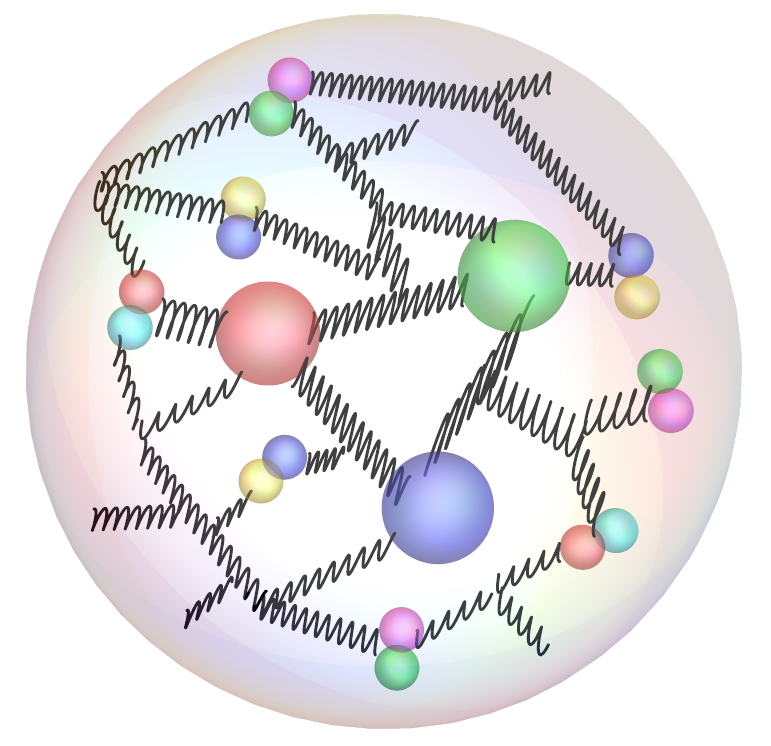}
    \caption{A more complicated but realistic illustration of the proton's structure. In addition to the three valence quarks (the large spheres), we now have a sea of quarks and antiquarks (the smaller colorful spheres) along with many gluons (the corkscrew lines). The spins of the particles are not represented here, but each quark and gluon can contribute its spin --- and also its orbital angular momentum --- to the proton's spin.}
    \label{fig:complicatedproton}
\end{figure}

The modern picture of the proton's structure that has emerged holds that there are indeed three quarks that live at relatively large $x$ (that is, close to $x=1$) --- these are the \textit{valence} quarks. But we also have a rich \textit{sea} of quarks and antiquarks at smaller values of $x$. These sea quarks can fluctuate in number, as particle-antiparticle pairs are created or annihilated, and the interactions among the sea quarks are mediated by gluons, which can themselves split into more gluons or recombine with each other. The interior structure of the proton is thus much less trivial than the naive diagram in Fig.~\ref{fig:constituentquarks}, but could instead look (for illustrative purposes only) more like the representation in Fig.~\ref{fig:complicatedproton}.

To describe the spin of the proton, we can make the following decomposition:
\begin{align}\label{intro_JMsumrule}
    S_q+L_q+S_G+L_G = \frac{1}{2}\,.
\end{align}
This is the Jaffe-Manohar sum rule \cite{Jaffe:1989jz}. \eq{intro_JMsumrule} says that we can break the spin of the proton, which is $1/2$ in units of $\hbar$, into the spins $S$ and orbital angular momenta (OAM) $L$ of the quarks $q$ and gluons $G$. There are other ways to decompose the proton spin, notably that due to Ji \cite{Ji:1996ek}, but in this dissertation we will use \eq{intro_JMsumrule}. Each of the terms on the left hand side of \eq{intro_JMsumrule} can be written as an integral over Bjorken-$x$ of an appropriate helicity or orbital angular momentum distribution\footnote{We will often use the terms \textit{helicity} and \textit{spin} interchangeably. Though strictly they are different (helicity is the projection of spin onto the direction of motion) for our purposes they are interchangeable.}.
\begin{subequations}\label{intro_spinsandorbital}\allowdisplaybreaks
\begin{align}\label{intro_spins}
    &S_q(Q^2) = \frac{1}{2}\int\limits_0^1\mathrm{d}x \Delta\Sigma(x,Q^2)\,, \quad\quad S_G(Q^2) = \int\limits_0^1 \mathrm{d}x \Delta G(x,Q^2) \,, \\
    \label{intro_oams}
    &L_q(Q^2) = \int\limits_0^1\mathrm{d}x L_q(x,Q^2)\,, \quad\quad L_G(Q^2) = \int\limits_0^1 \mathrm{d}x L_G(x,Q^2) \,.
\end{align}
\end{subequations}
A number of specifics regarding Eqs.~\eqref{intro_spins} will be explained in more detail in the next Chapter. For now, note the following. $\Delta\Sigma(x,Q^2)$ is the flavor-singlet quark helicity distribution,
\begin{align}\label{intro_DeltaSigmageneral}
    \Delta \Sigma(x,Q^2) = \sum_{f} \left[\Delta q_f(x,Q^2) + \Delta\overline{q}(x,Q^2) \right]\,,
\end{align}
where $\Delta q_f(x,Q^2)$ and $\Delta \overline{q}_f(x,Q^2)$ are the quark and antiquark helicity distributions of flavor $f$. The helicity distribution for each flavor of quark (antiquark) corresponds to the number of quarks (antiquarks) of a given flavor with spins aligned to the proton's spin minus the number anti-aligned. That is,
\begin{align}\label{quarkdisteachflavor}
    \Delta q_{f}(x,Q^2) = q_f^{(+)}(x,Q^2) - q_f^{(-)}(x,Q^2)
\end{align}
with a superscript $(+)$ denoting a quark spin aligned with the proton spin and a superscript $(-)$ denoting a quark spin anti-aligned with the proton spin. An equivalent expression can be written for the antiquark distribution $\Delta \overline{q}_f$. $\Delta G(x,Q^2)$ is the gluon helicity distribution. It similarly corresponds to the number of gluons with spins aligned with the proton's spin minus the number anti-aligned with the proton's spin. $L_q(x,Q^2)$ and $L_G(x,Q^2)$ are the quark and gluon orbital angular momentum distributions.

The variable $Q^2$ corresponds to the spatial resolving power of a deep inelastic scattering experiment (note that to study spin we are now in the context of \textit{polarized} DIS), with $1/Q$ the typical spatial resolution scale in the plane transverse to the collision axis. Larger values of $Q^2$ allow for finer-resolution measurements of hadronic structure. 

Next, note the bounds on the integrals in Eqs.~\eqref{intro_spinsandorbital}. In order to actually carry out the integrals, we would need to know the behavior of the helicity and OAM distributions at all values of $x$, from $x=0$ to $x=1$. Since $x$ is inversely proportional to the energy $E$ of a DIS experiment $x\sim\tfrac{1}{E}$,\footnote{This will be shown in more detail later} the small-$x$ regime corresponds to high collision energies and the large-$x$ regime to low collision energies. Then to meaningfully employ Eqs.~\eqref{intro_spinsandorbital} we need to understand the helicity and OAM distributions at all energies. In particular, the small-$x$ (high energy) behavior of the helicity distributions $\Delta\Sigma(x,Q^2)$ and $\Delta G(x,Q^2)$ is the main focus of this dissertation (though the large-$x$ region is not lacking for active research --- see \cite{ZEUS:2020ddd,Mukherjee:2023snp} for a taste of some recent experimental and theoretical developments related to large-$x$ physics).

The OAM distributions $L_q(x,Q^2)$ and $L_G(x,Q^2)$ will not be discussed in this dissertation, but they represent an indispensable part of the proton spin puzzle as formulated in \eq{intro_JMsumrule}. A rich ongoing research effort is being made to understand these distributions at small-$x$, much of which is closely connected to the study of the helicity distributions presented here. See \cite{Kovchegov:2019rrz,Kovchegov:2023yzd,Manley:2024pcl,Kovchegov:2024wjs,Bashinsky:1998if,Hagler:1998kg,Harindranath:1998ve,Hatta:2012cs,Ji:2012ba,Hoodbhoy:1998yb} for a selection of theoretical work on this front and \cite{Hatta:2016aoc,Bhattacharya:2022vvo,Bhattacharya:2023hbq,Bhattacharya:2024sck} for experimental proposals to extract measurements of parton orbital angular momentum.

The measurements of the total quark spin within the proton made by the European Muon Collaboration discussed earlier (which were made at a particular resolution $Q^2 \approx 10.7 \,\text{GeV}^2$) yielded
\begin{align}\label{intro_SqEMC}
    \frac{1}{2}\int\limits_{0.01}^1\mathrm{d}x \Delta\Sigma(x,Q^2\approx 10.7 \,\,\text{GeV}^2) \, \approx \, 0.030 \pm 0.024 \pm 0.035 \,.
\end{align}
Recall that if the very naive three-quark model was correct (three quarks carrying all the proton's spin, like in Fig.~\ref{fig:constituentquarks}), this number would be equal to $1/2$. Note also that the number cited above involved an extrapolation as $x\rightarrow 0$. The actual measurements went down to approximately $x = 0.01$. Since the EMC measurement, many more experiments have been done, resulting in more data and better constraints on both the quark and gluon spin contributions. The current figures sit at approximately (see \cite{Accardi:2012qut,Aschenauer:2013woa,Aschenauer:2015eha,Proceedings:2020eah,Ji:2020ena,Leader:2013jra})
\begin{subequations}\label{intro_currentdataSqSG}
\begin{align}
    & \frac{1}{2}\int\limits_{0.001}^{0.7}\mathrm{d}x \Delta\Sigma(x,Q^2\approx 10 \,\,\text{GeV}^2) \in [0.15,0.2]\,,\\
    & \int\limits_{0.05}^{0.7}\mathrm{d}x \Delta G(x,Q^2\approx 10 \,\,\text{GeV}^2) \in [0.13,0.26]\,.
\end{align}
\end{subequations}

There are several things to note regarding the values in Eqs.~\eqref{intro_currentdataSqSG}. First, even taking the upper bounds of both quantities, $S_q^{(\text{max})} + S_G^{(\text{max})} = 0.2 + 0.26 \neq \frac{1}{2}$. Thus there must be more spin hiding somewhere in the proton. The second thing to note is the possibility that some of that spin may live in the small-$x$ regime. Observe that the integrals in Eqs.~\eqref{intro_currentdataSqSG} do not go down all the way to $x = 0$. This is because, as discussed earlier, $x \sim \frac{1}{E}$ with $E$ the collision energy. That is to say, reaching $x=0$ would correspond to a collision at infinite energy. This is of course impossible and so any given experiment can only ever measure down to some minimum $x_{\text{min}}$. 

A priori this calls for theoretical input. Experiments can continue pushing to smaller and smaller values of $x$, and they should as these are valuable measurements crucial to our understanding of hadron structure. The future Electron-Ion Collider \cite{Accardi:2012qut,Boer:2011fh,Proceedings:2020eah,AbdulKhalek:2021gbh} will have a substantially improved coverage in both $x$ and $Q^2$ for polarized collisions along with very high luminosities and will be an indispensable machine for advancing our understanding of hadronic structure along with many other particle and nuclear phenomena. However, no experiment will ever be able to measure the parton distributions to $x=0$. Therefore if we are ever to confidently declare the proton spin puzzle resolved we need a solid theoretical understanding with which we can reliably extrapolate measured data down to $x=0$. This is where the notion of small-$x$ \textit{evolution} is critical --- given the helicity distributions at some input scale $x_0$, e.g. as measured in an experiment, we construct equations to \textit{evolve} these distributions from their starting scale to smaller values of $x$.

Beyond the intrinsic motivation for theoretical control of small-$x$ helicity, early work to study the helicity structure of the proton at small-$x$ yielded predictions for a substantial amount of spin in this regime. In the 1990s Bartels, Ermolaev, and Ryskin (BER) \cite{Bartels:1995iu,Bartels:1996wc} predicted a growth of the helicity distributions at asymptotically small-$x$ of
\begin{align}\label{intro_BERasympt}
        \Delta \Sigma (x, Q^2) \sim \Delta G (x, Q^2) \sim \left( \frac{1}{x} \right)^{3.66 \, \sqrt{\frac{\as N_c}{2\pi}}}\,,
\end{align}
where $\as$ is the strong coupling and $N_c$ is the number of quark colors. Note also that this result is obtained in the limit where the number of colors $N_c$ is taken to be a large parameter --- this essentially corresponds to a gluon-only treatment and will be discussed at greater length later. With the value of the strong coupling taken to be $0.18$ and with $N_c = 3$, the power law in \eq{intro_BERasympt} is $(1/x)^{1.07}$. Recalling that we obtain the total quark and gluon spin contributions by integrating over $x$ all the way down to $x=0$ (see \eq{intro_spins}), this looks like a potentially divergent amount of spin! That, of course, merits a detailed investigation. In this dissertation, we will discuss the results of exactly such an investigation.

This dissertation is structured as follows. In Ch.~\ref{background.ch} we discuss the physical background upon which the research in the following Chapters is based. We cover a more detailed treatment of the deep inelastic scattering process. We introduce the dipole picture --- a powerful paradigm for studying DIS at high energies. We discuss some of the other tools needed for a comprehensive treatment of high energy scattering and also introduce the notion of small-$x$ evolution, along the way touching on another important type of evolution --- evolution in the resolution scale $Q^2$. Then in Ch.~\ref{smallxhelicity.ch} we discuss the specific formalism used to perform calculations in the small-$x$ helicity framework. We present some of the main results of the small-$x$ helicity research program prior to the contributions of this dissertation. 

In Ch.~\ref{largeNcsoln.ch} we detail the construction of a fully analytic solution to the equations that describe the proton spin structure at small-$x$ \cite{Borden:2023ugd}. This solution is constructed in the gluons-only large-$N_c$ limit mentioned after \eqref{intro_BERasympt} and results in analytic descriptions of the helicity distributions at small-$x$ and large-$N_c$. The infinite precision associated with our (and any) analytic solution allows for more detailed comparisons with existing predictions than have been done before, and ultimately reveals small but meaningful discrepancies that had previously gone undetected. 

In Ch.~\ref{transitionops.ch} we work in the more general limit where both $N_f$ and $N_c$ are taken to be large parameters ($N_f$ is the number of quark flavors). This effectively amounts to restoring the quark contributions to our description of the proton spin. There we find an important correction to the equations as they existed prior to the work of this dissertation. Without this correction, the equations exhibit disagreements with the robust predictions of the finite-order framework. We explicitly construct the contributions from this correction, related to a class of quark-to-gluon and gluon-to-quark transition operators, and derive the new set of complete equations describing the proton spin at small-$x$ and large-$N_c\&N_f$, subsequently demonstrating that we have established full agreement with the existing finite-order predictions \cite{Borden:2024bxa}.

In Ch.~\ref{largeNcandNfsoln.ch} we use the analytic formalism developed in \cite{Borden:2023ugd} and employed in Ch.~\ref{largeNcsoln.ch} to construct a new analytic solution for the more complicated but more general equations derived in Ch.~\ref{transitionops.ch} in the large-$N_c\&N_f$ limit. This new solution represents a fully analytic description of the proton's spin in the most general framework of the small-$x$ formalism thus considered \cite{Borden:2025ehe}. We conclude in Ch.~\ref{conclusions.ch}.

\chapter{Background}
\label{background.ch}

In this Chapter we present of a number topics important to the small-$x$ helicity field, beginning first with a more detailed treatment of deep inelastic scattering.


\section{Deep Inelastic Scattering}\label{sec:dis}

\begin{figure}[ht]
    \centering
    \includegraphics[width=0.8\linewidth]
    {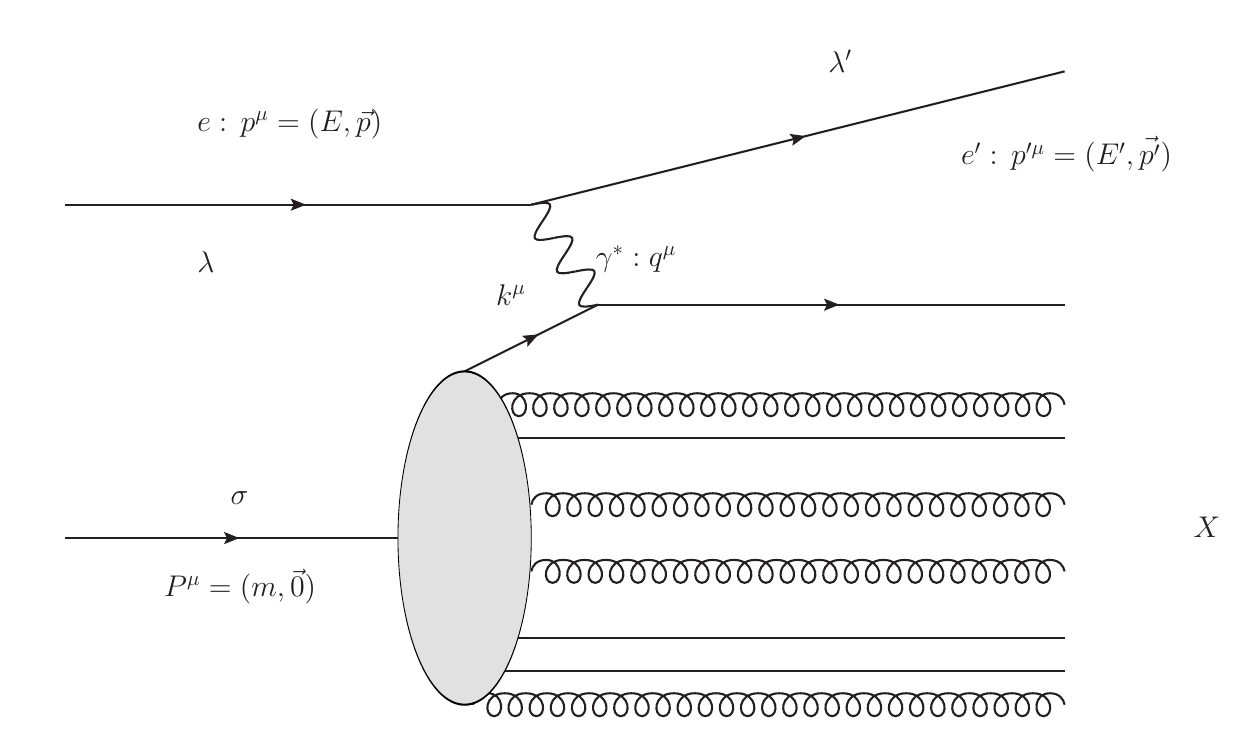}
    \caption{Diagrammatic representation of deep inelastic electron-proton scattering in the proton's rest frame. The proton (gray oval) is illustrated as a collection of quarks and gluons.}
    \label{fig:dis}
\end{figure}

Deep inelastic scattering (DIS) of an electron and proton is the name given to the process (at short distances)
\begin{align}\label{ch2_dis}
    e + p \rightarrow e' + X \,.
\end{align}
Here $e$ is an incoming electron which strikes the incoming proton $p$ and breaks it apart. $e'$ the outgoing electron and $X$ represents the particles produced from the fragmented proton. Of course, the electron could be replaced by a different lepton (the EMC experiments of \cite{EuropeanMuon:1987isl,ASHMAN19891} mentioned in the Ch.~\ref{intro.ch} used muons). In this Section, we largely follow the treatment of DIS presented in Ch.~2 of \cite{Kovchegov:2012mbw}. We can represent the process in the rest frame of the proton, as shown in Fig.~\ref{fig:dis}.
Here the incoming and outgoing electrons have momenta $p^{\mu} = (E,\vec{p})$ and $p^{\prime\mu} = (E',\vec{p'})$ and polarizations $\lambda$ and $\lambda'$, respectively. In the rest frame of the proton, its momentum is simply $(m,\vec{0})$, while its polarization is $\sigma$. Consistent with the parton model, we represent the proton as a co-moving system of quarks and gluons. The interaction between the incoming electron and the proton is mediated by a virtual photon $\gamma^*$ with momentum $q^{\mu} = (E-E', \vec{p}-\vec{p'})$. There are other possible interactions that can take place rather than photon exchange, principally those mediated by the weak interaction, but our focus here is on photon exchange which is the dominant contribution (see e.g. \cite{Anselmino:1994gn} for a review which also discusses the formalism associated with weakly-mediated DIS). In the context of the parton model, the electron scatters incoherently off the partons --- that is, the electron exchanges a virtual photon with one active quark (the uppermost quark in Fig.~\ref{fig:dis}) while the remaining partons are \textit{spectators}. These spectators do not interact with the virtual photon probe, nor do they interact with each other over the timescales of DIS. 

At high energies, the proton will be broken up by the energetic interaction and $X$ denotes the remnant state of the proton. We do not detect the individual partons, but rather the results of hadronization, the process whereby the remnant quarks and gluons recombine to form hadrons. A DIS experiment can be \textit{fully inclusive}, where we measure only the outgoing electron, or \textit{semi-inclusive}, where we measure the outgoing electron and tag on one of the resulting hadrons which can provide information on the flavor of the struck parton based on the composition of the detected hadron. 

To describe the DIS process we define the following Lorentz invariants:
\begin{align}\label{ch2_dis1}
    Q^2 = -q^2 \,,\qquad x_{Bj} = \frac{Q^2}{2P\cdot q} \,, \qquad y = \frac{P\cdot q}{P\cdot p}\,.
\end{align}
$Q^2$ is often called the virtuality, since it is a measure of how off-shell the virtual photon is. In what follows, we typically require $Q^2$ to be very large. Indeed for perturbation theory to be reliable, we need $Q^2$ much larger than any nonperturbative scale, like the square of the confinement scale $\Lambda_{QCD}^2$. $x_{Bj}$ is the Bjorken-$x$ variable (named after James Bjorken) which will be critical for us going forward. And $y$ can be shown to be the fractional energy transfer from the electron to the proton (in the proton rest frame). Now if we also define the center-of-mass energy squared of the interaction
\begin{align}\label{ch2_dis2}
    s= (P+p)^2\,,
\end{align}
one can show that 
\begin{align}\label{ch2_dis3}
    Q^2 \approx yx_{Bj}s\,,
\end{align}
where the approximation is valid at high energy $s \gg m^2 \gg m_e^2$ ($m_e$ is the mass of the electron).
\eq{ch2_dis3} tells us that $x_{Bj} \sim 1/s$, that is, Bjorken-$x$ scales inversely with energy. So indeed the small-$x$ regime is probed in high-energy collisions. This is the regime in which we work for most of this dissertation.

Following \cite{Kovchegov:2012mbw} we can write the differential cross section for electron-proton scattering (proceeding via photon exchange) as 
\begin{align}\label{ch2_dis4}
    \frac{d\sigma}{d^3p'} = \frac{\alpha_{EM}^2}{EE'Q^4} L_{\mu\nu}W^{\mu\nu}\,,
\end{align}
where the cross section has factored into a leptonic tensor $L_{\mu\nu}$ and a hadronic tensor $W^{\mu\nu}$. $\alpha_{EM}$ is the electromagnetic coupling. The leptonic tensor can be written as
\begin{align}\label{ch2_dis5}
    &L_{\mu\nu} = \sum_{\lambda' = \pm 1} \overline{u}_{\lambda'}(p')\gamma_{\mu} u_{\lambda}(p)\left[\overline{u}_{\lambda'}(p')\gamma_{\nu}u_{\lambda}(p) \right]^* \\
    &\hspace{0.7cm} = 2\left(p_\mu p'_\nu + p_\nu p'_\mu - p\cdot p' g_{\mu\nu} + m_e^2 g_{\mu\nu} - i\lambda\epsilon_{\mu\nu\alpha\beta}p^\alpha p^{\prime\beta} \right) \,, \notag
\end{align}
with $\epsilon_{\mu\nu\alpha\beta}$ the 4-dimensional Levi-Civita symbol and $\epsilon_{0123} = 1$. For our purposes the much more interesting piece is the hadronic tensor, which can be written as a sum of symmetric and antisymmetric parts,
\begin{align}\label{ch2_dis6}
    W_{\mu\nu} = W_{\mu\nu}^{(\text{sym})} + W_{\mu\nu}^{(\text{antisym})}\,,
\end{align}
which can themselves be written (for electron-proton scattering) as \cite{Kovchegov:2012mbw,Lampe:1998eu}
\begin{subequations}\label{ch2_dis7}
\begin{align}
    \label{ch2_dis7a}
    &W_{\mu\nu}^{(\text{sym})} = -\frac{1}{m}\left(g_{\mu\nu} - \frac{q_{\mu}q_{\nu}}{q^2}\right)F_1(x_{Bj},Q^2) \\
    &\hspace{1.75cm}+ \frac{2x_{Bj}}{mQ^2}\left(P_\mu - \frac{P\cdot q}{q^2}q_\mu \right)\left(P^\nu - \frac{P\cdot q}{q^2}q_\nu  \right)F_2(x_{Bj},Q^2) \,,\notag \\
    \label{ch2_dis7b}
    &W_{\mu\nu}^{(\text{antisym})} = i\epsilon_{\mu\nu\rho\sigma} \frac{q^\rho}{m\left(P\cdot q\right)} \left[S^\sigma g_1(x,Q^2) + \left( S^\sigma - \frac{S\cdot q}{P\cdot q}P^\sigma\right)g_2(x,Q^2)\right]\,.
\end{align}
\end{subequations}
Recall that $m$ is the mass of the proton, and the momenta are as labeled in Fig.~\ref{fig:dis}. We also have the proton's polarization vector $S^{\mu}$ which satisfies $P \cdot S = 0$ and $S^2 = - m^2$.

Eqs.~\eqref{ch2_dis7} are the most general tensor structures into which $W_{\mu\nu}$ can be decomposed (again for electron-proton scattering mediated by the exchange of a virtual photon), subject to the constraint of electromagnetic current conservation
\begin{align}\label{ch2_dis8}
    q_\mu W^{\mu\nu} = 0\,,\qquad q_\nu W^{\mu\nu} = 0\,.
\end{align}
One also has to employ several identities involving the metric $g_{\mu\nu}$ and Levi-Civita symbol $\epsilon_{\mu\nu\rho\sigma}$ in deriving the antisymmetric piece (see \cite{Lampe:1998eu} for details). 

The quantities attached to each tensor structure in Eqs.~\eqref{ch2_dis7} are called \textit{structure functions}. $F_1(x_{Bj},Q^2)$ and $F_2(x_{Bj},Q^2)$ are the unpolarized structure functions, while $g_1(x_{Bj},Q^2)$ and $g_2(x_{Bj},Q^2)$ are the polarized structure functions, distinguished as such because they are accessible only in polarized DIS experiments (since they live in $W_{\mu\nu}^{(\text{antisym})}$). As the name suggests, the structure functions contain all the information about the proton's structure that we can learn in (fully inclusive) DIS, without putting any constraints on the final state.

Thus far we have been discussing DIS in the proton's rest frame, but the structure functions (and the parton model itself) really take a clear meaning in what is often called the \textit{infinite momentum frame}. In this reference frame the proton is moving ultrarelativistically along, say, the $z$-axis. The proton's momentum in the vector notation $v = (v^0,v^1,v^2,v^3)$ is $P^{\mu} \approx \left(P + \frac{m^2}{2P}, 0, 0, P \right)$, where $P\gg m$ so that corrections of order $m^2/\vec{P}^2$ are negligible. Note also that in this frame, we can choose the virtual photon of momentum $q$ to have $q^3 = 0$ so that $q^{\mu} = (q^0,q^1,q^2,0)$. However, now is a convenient time to switch to light-cone coordinates, a coordinate system very useful for high energy scattering. We write a vector $v$ as $v = (v^+,v^-,\underline{v})$ with
\begin{align}\label{ch2_dis9}
    v^{\pm} = \frac{1}{\sqrt{2}}\left(v^0\pm
v^3\right)\,, \qquad \underline{v} = \left(v^1, v^2\right)\,,
\end{align}
and so the dot product between two vectors $u$ and $v$ is 
\begin{align}\label{ch2_dis10}
    u \cdot v = u^+v^- + u^-v^+ - \underline{u}\cdot\underline{v}\,.
\end{align}
If we need to refer to the magnitude of a transverse vector, $|\underline{v}|$, we will often write $v_\perp$.
These light-cone coordinates are useful at high energy because ultrarelativistic particles are moving approximately along their light cones, that is, predominantly in the $x^+$ or $x^-$ directions. 
In these coordinates, the proton momentum given above is $P^\mu \approx (P^+, 0, \underline{0})$ for $P^+ \approx 2 P$ very large. Meanwhile the virtual photon's momentum in these coordinates can be written $q^\mu = \left(q^+, \frac{\underline{q}^2 - Q^2}{q^+},\underline{q}\right)$ with $(q^+)^2 = \underline{q}^2 - Q^2$.

Now consider the quark that participates in the DIS process (the uppermost quark line in Fig.~\ref{fig:dis}). With momentum $k^\mu$  we can define the longitudinal momentum fraction of this quark as
\begin{align}\label{ch2_dis11}
    x \equiv \frac{k^+}{P^+}\,,
\end{align}
which is sometimes called the Feynman-$x$ variable. This variable tells us how much of the proton's momentum the quark carries along the light cone. With a DIS calculation in the framework of the parton model one can show that Feynman-$x$ is in fact equal to Bjorken-$x$, which was defined in Eqs.~\eqref{ch2_dis1} \cite{Kovchegov:2012mbw}. This is the prevailing interpretation of the Bjorken-$x$ variable in the parton model, and so we will often simply write $x$ instead of $x_{Bj}$.

We are in a position now to justify some of the assertions made thus far about DIS in the context of the parton model --- namely that there are no interactions between the partons during the DIS process and that the photon virtuality $Q^2$ provides an inverse spatial resolution scale for the process.

First we note that, based on general uncertainty principle arguments, the typical timescale for the DIS interaction will be inversely proportional to the $0^\text{th}$ component of the virtual photon's momentum. That is, in the infinite momentum frame,
\begin{align}\label{ch2_dis12}
    t_{\text{DIS}} \sim \frac{1}{q^{0}} \approx \frac{2xP}{Q^2}\,
\end{align}
where the approximation to obtain the final expression in \eq{ch2_dis12} follows from the fact that $Q^2 = 2 x\,P\cdot q $, which gives, using the momenta of the proton and virtual photon in the text before \eq{ch2_dis9}, $Q^2 \approx 2xPq^0$. Meanwhile the timescale for interactions between the partons within the proton is (first thinking about the proton's rest frame) governed by the nonperturbative, low-energy hadronic physics with which we can associate some nonperturbative timescale $\frac{1}{\mu}$. To return to the infinite momentum frame, we have to Lorentz boost this timescale by the boost factor $\frac{P}{m}$, so our inter-partonic interaction timescale in the infinite momentum frame is 
\begin{align}\label{ch2_dis13}
    t_{\text{partons}} \sim \frac{P}{\mu m}\,.
\end{align}
Then Eqs.~\eqref{ch2_dis12} and \eqref{ch2_dis13} give
\begin{align}\label{ch2_dis14}
\frac{t_{\text{DIS}}}{t_{\text{partons}}} \approx \frac{2x \, \mu m}{Q^2} \ll 1\,,
\end{align}
where the inequality follows from the fact that, for us, $x$ is small along with the fact that $\mu m\ll Q^2$ since we require $Q^2$ to be much greater than the scales of nonperturbative physics. In \eq{ch2_dis14} we have a justification for the claim that the partons do not interact with each other during the DIS process (in the infinite momentum frame), since the timescale for DIS is much shorter than that for partonic interactions.

Next, for the question of spatial resolution, note that the typical transverse scale $|\underline{x}| = x_\perp$ of the DIS process will be inversely proportional to the magnitude of the virtual photon's transverse momentum, $|\underline{q}| = q_\perp$:
\begin{align}\label{ch2_dis15}
    x_\perp \sim \frac{1}{q_\perp} \approx \frac{1}{Q}\,,
\end{align}
where to obtain the last approximation we simply note that, since $P$ is large, \eq{ch2_dis12} tells us that $q^0\ll Q$ and so $q^2 = - Q^2 = (q^0)^2 - \underline{q}^2 \approx -\underline{q}^2$. So indeed $1/Q$ is the typical transverse spatial scale we can resolve in DIS.

Returning to the structure functions that appear in Eqs.~\eqref{ch2_dis7}, it can be shown through a more involved calculation that $F_1$ and $F_2$ have the following simple forms (at leading order in $\as$) \cite{Kovchegov:2012mbw}:
\begin{subequations}\label{ch2_dis16}
\begin{align}
    \label{ch2_dis16a}
    &F_1(x) = \frac{1}{2}\sum_f Z_f^2 q_f(x) \,,\\
    \label{ch2_dis16b}
    &F_2(x) = \sum_f Z_f^2 xq_f(x)\,,
\end{align}
\end{subequations}
where $q_f(x)$ is a quark distribution function which ultimately counts the number of quarks of a given flavor with longitudinal momentum fraction $x$. $Z_f$ is the electric charge of a quark of flavor $f$, in units of $e$. In light of Eqs.~\eqref{ch2_dis16}, the structure functions have clear physical interpretations. $F_1$ counts the (charge-weighted) number of quarks in the proton with longitudinal momentum fraction $x$ and $F_2$ gives the (charge-weighted) average longitudinal momentum fraction of the quarks in the proton.

Similarly, the polarized structure function $g_1$ can be shown to be related (at leading order in $\as$) to the helicity distributions $\Delta q_f(x,Q^2)$, defined in \eq{quarkdisteachflavor}, which count the net number of quarks or antiquarks with spins aligned to the proton's spin:
\begin{align}\label{ch2_dis17}
    g_1(x,Q^2) = \frac{1}{2}\sum_f Z_f^2 \left[ \Delta q_f(x,Q^2) + \Delta \overline{q}_f(x,Q^2) \right]\,.
\end{align}
In particular, $g_1$ is experimentally accessible in the difference between the DIS cross section where the electron's ($\leftarrow$) and proton's ($\Rightarrow$) polarizations are anti-aligned and the cross section where they are aligned \cite{Lampe:1998eu}:
\begin{align}\label{ch2_measureg1}
    g_1(x,Q^2) \,\propto\, \sigma^{\leftarrow}_{\Rightarrow} - \sigma^{\leftarrow}_{\Leftarrow}\,.
\end{align}
The other polarized structure function $g_2$ that appears in $W_{\mu\nu}^{(\text{antisym})}$ in \eq{ch2_dis7b} is less amenable to an interpretation in the parton model \cite{ALTARELLI19821,PhysRevLett.67.552}. Though $g_2$ is important for transverse spin, its contributions are energy suppressed \cite{Lampe:1998eu,Cougoulic:2022gbk} and it will not be relevant for this dissertation, where we only consider longitudinal spin.

We omitted the details of the calculation that demonstrates it, but the structure functions $F_1$ and $F_2$ were written in Eqs.~\eqref{ch2_dis7} with arguments $x$ and $Q^2$, whereas in Eqs.~\eqref{ch2_dis16}, we have written them with only $x$ as the argument. This dependence of the structure functions only on one variable, $x$, and not on $Q^2$ is the phenomenon of \textit{Bjorken scaling}. This prediction from the framework of the parton model, originally made by J.D. Bjorken \cite{PhysRev.179.1547}, proved remarkably consistent with DIS data over quite a large range in $x$ and $Q^2$. As we will discuss in a later Section, though, Bjorken scaling is only approximately true and the structure functions are actually functions of both $x$ and $Q^2$. That $Q^2$-dependence is given by the Dokshitzer-Gribov-Lipatov-Altarelli-Parisi (DGLAP) evolution equations \cite{Gribov:1972ri,Altarelli:1977zs,Dokshitzer:1977sg}, which describe how the quark (and also gluon) distribution functions change or \textit{evolve} as one changes $Q^2$. The polarized structure function $g_1$ also exhibits its own approximate version of Bjorken scaling, which is ultimately also violated by a polarized version of the above-mentioned DGLAP evolution equations. However, in light of the much more detailed treatment $g_1$ will receive in this dissertation compared to $F_1$ and $F_2$, and the fact that its $Q^2$-dependence will be important for our purposes, we opt to write it with both arguments $x$ and $Q^2$ from the start, as in \eq{ch2_dis17}.

Throughout this Section, we have mainly been discussing quarks --- the framework here is sometimes referred to as the naive quark parton model. But we will soon need the full machinery of quantum chromodynamics, where gluons play a crucial role as well. Some of the results in this Chapter are ultimately modified by the transition to QCD. For example, the quark distribution functions introduced in Eqs.~\eqref{ch2_dis16} have a gluonic counterpart --- the gluon distribution function, sometimes called $G(x,Q^2)$. And of course there are polarized versions of both the quark and gluon distribution functions --- these are the quark and gluon helicity distributions we already introduced qualitatively in Eqs.~\eqref{intro_spins}. Furthermore, the polarized structure function $g_1(x,Q^2)$ has an additional contribution from the gluon helicity distribution that amends the quarks-only result in \eq{ch2_dis17}.


\section{Tools of High Energy Scattering}\label{sec:dipolepic}

In this Section, we review several of the tools most useful for exploring high energy scattering. The dipole picture of deep inelastic scattering is a framework for the scattering process at high energies that is not only convenient but also in some ways computationally simple. Wilson lines, sometimes also called \textit{gauge links}, offer another powerful tool, in this case for describing in a simple way the propagation of high energy particles through sources of color charge. Lastly, light-cone perturbation theory provides a framework for carrying out perturbative calculations. It is conceptually similar to traditional Feynman time-ordered perturbation theory, but offers a number of advantages at high energies.

\subsection{The Dipole Picture}
In Fig.~ \ref{fig:dis} we showed DIS as the interaction between an electron and proton where a virtual photon is exchanged between the electron and a quark within the proton. However, there is another way to view this process that further isolates the strong physics involved. Suppose that before interacting with the proton, the virtual photon fluctuates into a quark-antiquark pair. We call this a color dipole (note that since the photon is colorless, the net color charge of the dipole must also be zero). If we take the virtual photon's momentum (here considering the rest frame of the proton target) to be 
\begin{align}\label{ch2_dipolepicmomentum}
    q^\mu = \left(q^+, -\frac{Q^2}{2q^+}, \underline{0}\right) \,,
\end{align}
with the light-cone plus component $q^+$ very large, then its coherence length in the $+$ direction (the direction in which it is predominantly moving) is \cite{Kovchegov:2012mbw}
\begin{align}\label{ch2_cohlength}
    x^+ \approx \frac{1}{|q^-|} = \frac{2q^+}{Q^2} \sim \frac{1}{x\,m}
\end{align}
with $m$ the proton mass. The coherence length for a fluctuation of this virtual photon into a $q\overline{q}$ pair would then be on the same order. Compared to the size of the target proton (in the proton rest frame, $P^+ = P^- \sim m$, so $1/P^- \sim 1/m$), the coherence length of the $q\overline{q}$ fluctuation is thus quite long, given that we are at small $x$. That is to say, the Fock state of the virtual photon corresponding to a quark-antiquark pair remains coherent long enough to interact with the proton target.

\begin{figure}[ht]
    \centering
    \includegraphics[width=0.6\linewidth]{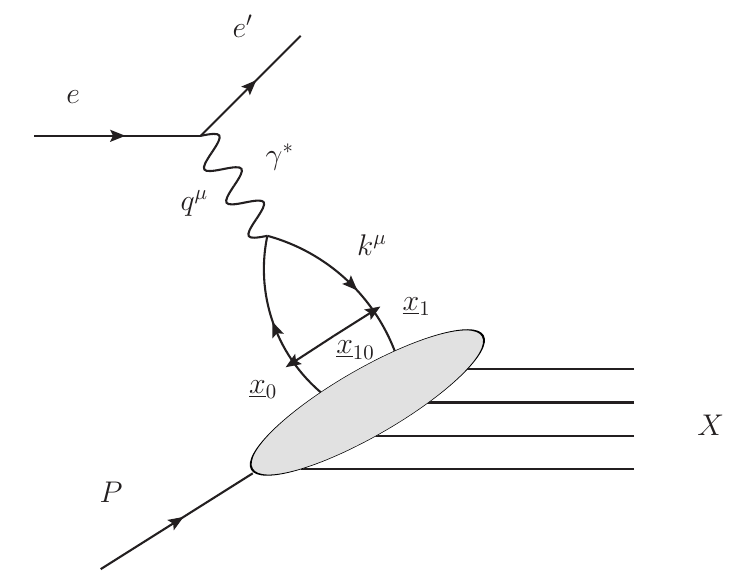}
    \caption{Deep inelastic scattering in the framework of dipole picture. The interaction between the $q\overline{q}$ dipole and the proton (P) is denoted with a gray oval. The antiquark is located at transverse coordinate $\underline{x}_0$, while the quark is located at transverse coordinate $\underline{x}_1$, with the transverse separation labeled $\underline{x}_{10} = \underline{x}_1 - \underline{x}_0$.}
    \label{fig:dipolepic}
\end{figure}

Then the DIS process can be thought of in two distinct steps. First, the virtual photon splits into a $q\overline{q}$ pair; and second, the $q\overline{q}$ pair interacts coherently with the proton. This is the dipole picture of deep inelastic scattering \cite{Kovchegov:2012mbw, PhysRevLett.47.297,Gribov:1969zz,Kopeliovich:1981pz,MUELLER1990115,Nikolaev:1990ja} and is represented in Fig.~\ref{fig:dipolepic}.
The dipole picture allows us to factorize the total virtual photon-target cross section as \cite{Kovchegov:2012mbw}
\begin{align}\label{ch2_dipolecrossection}
    \sigma_{\text{tot}}^{\gamma^* p}(x,Q^2) = \int \frac{\mathrm{d}^2 x_{10}}{4\pi}\int\limits_0^1 \frac{\mathrm{d}z}{z\left(1-z\right)} \left| \Psi^{\gamma^*\rightarrow q\overline{q}}(\underline{x}_{10},z) \right|^2 \sigma_{\text{tot}}^{q\overline{q} p}\left(\underline{x}_{10},\ln \tfrac{1}{x}\right)\,.
\end{align}
Here $z = k^+/q^+$ is the light-cone momentum fraction of the quark of momentum $k$ relative to the virtual photon of momentum $q$. $\underline{x}_{10} = \underline{x}_1 - \underline{x}_0$ is the transverse separation of the dipole, with magnitude $|\underline{x}_{10}| = x_{10}$. $\Psi^{\gamma^*\rightarrow q\overline{q}}(\underline{x}_{10},z)$ is the light-cone wavefunction for the virtual photon to split into a quark-antiquark pair.\footnote{More on light-cone wavefunctions shortly.} Importantly, $\Psi^{\gamma^*\rightarrow q\overline{q}}$ is entirely calculable within QED and so we have conveniently isolated the interesting strong interaction physics in $\sigma_{\text{tot}}^{q\overline{q} p}(\underline{x}_{10},\ln\tfrac{1}{x})$, which denotes the total cross section for the scattering of the $q\overline{q}$ dipole on the proton. This cross section is a function of both the transverse separation $\underline{x}_{10}$ and the log of inverse Bjorken-$x$. Note that sometimes the \textit{rapidity} variable $Y$ is used, where $Y = \ln(\hat{s}x_\perp^2) \approx \ln(1/x)$ for $\hat{s}$ the photon-target center of mass energy squared, where the approximation is achieved for $x_\perp^2 \sim 1/Q^2$. 

So in \eq{ch2_dipolecrossection} we have one very convenient consequence of the dipole picture. The physics of the strong interaction has been completely isolated into the total $q\overline{q}p$ cross section, with the remaining piece of $\sigma_{\text{tot}}^{\gamma^*p}$ --- that is, the $\gamma^*\to q\overline{q}$ light-cone wavefunction --- easy to calculate. As a reminder, this is a valid separation because the $q\overline{q}$ dipole remains coherent over scales considerably larger than the scale of the target, which as \eq{ch2_cohlength} told us, is only applicable at small-$x$.

Another simplification afforded by the dipole picture is that the transverse separation of the dipole $x_\perp$ ($x_{10}$ in the notation of Fig.~\ref{fig:dipolepic}) does not change during the interaction with the target \cite{Kovchegov:2012mbw,Kopeliovich:1981pz,MUELLER1990115,Brodsky:1994kf,Levin:1987xr} (this is implicitly why we wrote \eq{ch2_dipolecrossection} in a mixed representation of longitudinal momentum and transverse coordinates). To justify this we can construct the typical scale $\Delta x_{\perp}$ by which the transverse separation between the quark and antiquark will vary during the scattering. With $R$ the longitudinal size of our proton target, $k_\perp\sim Q \sim 1/x_\perp$ the relative transverse momentum kick given to the dipole during the scattering, and $E \sim q^0$ the dipole energy, we have, still in the rest frame of the target, \cite{Kovchegov:2012mbw}
\begin{align}\label{ch2_transversekick}
    \Delta x_\perp \approx R \frac{k_\perp}{E} \approx 2 m x R x_\perp \,.
\end{align}
Then recalling that in \eq{ch2_cohlength} we established the coherence length of the $q\overline{q}$ dipole as $l \sim 1/mx$, we have 
\begin{align}\label{ch2_transversekick2}
    \frac{\Delta x_\perp}{x_\perp} \sim \frac{R}{l} \ll 1\,
\end{align}
which is quite small since, as we already showed, the coherence length of the dipole is much longer than the longitudinal scale of our target (again, this requires us to be at small-$x$). So we conclude that the the transverse size of our dipole is approximately constant during high energy interactions. This property forms the basis of the \textit{eikonal} approximation. As we will later see, the eikonal approximation has to be relaxed in order to discuss helicity-dependent scattering. But the dipole picture remains a useful backdrop, with the constant transverse separation of the dipole being a leading order effect to which we can calculate sub-leading (\textit{sub-eikonal}) corrections. Such sub-eikonal corrections come with suppressing factors of energy, like that in \eq{ch2_transversekick}.

\subsection{Wilson Lines, Shock Waves, and Dipole Amplitudes}\label{sec:wilsonetc}

We have established that at the eikonal level (that is, to the leading power of energy) the scattering of the quark-antiquark dipole on the target causes negligible change in the transverse separation of the dipole. It turns out that we can use this fact to simplify our description of high energy quarks (and, as we will see, gluons) propagating through the color charge of the target. 

First, we must clarify in a bit more detail how we can model the scattering of the dipole on the proton target (that is, what happens in the gray oval in Fig.~\ref{fig:dipolepic}). For the purposes of this dissertation, the target of interest is the proton. However at high energies, we often think of modeling the proton target as a large and dilute nucleus, made out of $A \gg 1$ nucleons. The reason for this is that formally such a model gives a large parameter $A$ in which sub-leading effects may be suppressed, for example interactions between nucleons. This is often referred to as the Glauber-Gribov-Mueller model \cite{Gribov:1969zz,PhysRev.100.242,PhysRev.142.1195,GLAUBER1970135,Gribov:1968jf,Mueller:1989st}.  

\begin{figure}[h!]
\centering
\begin{subfigure}{\textwidth}
    \centering
    \includegraphics[width=0.8\textwidth]{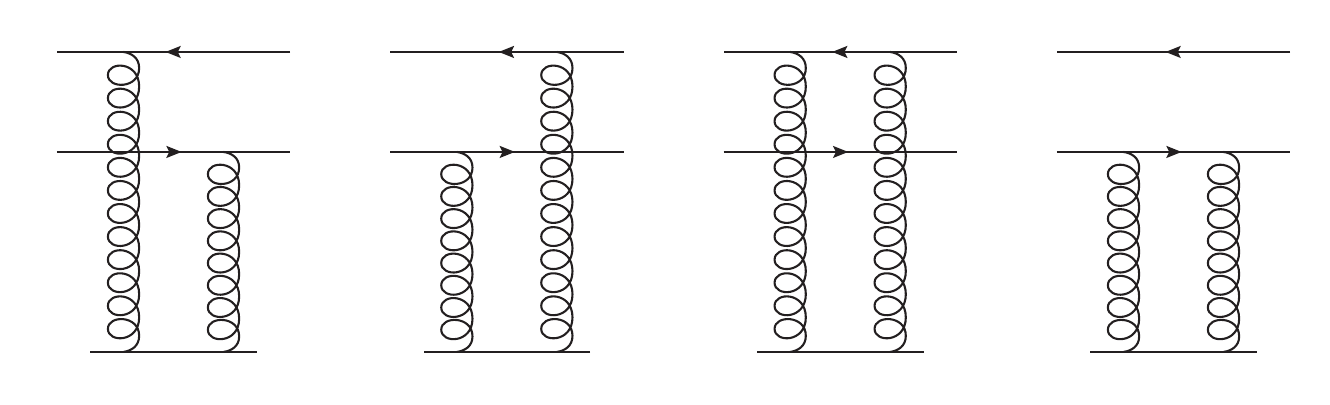}
    \caption{Scattering of a dipole on a single nucleon (the single line at the bottom of each sub-diagram) at the leading-order two-gluon exchange in the Glauber-Gribov-Mueller model. All four possible connections to the $q$ and $\overline{q}$ lines of the dipole are shown.}
    \label{fig:GGMsingle}
\end{subfigure}

\bigskip

\begin{subfigure}{\textwidth}
    \centering
    \includegraphics[width=0.8\textwidth]{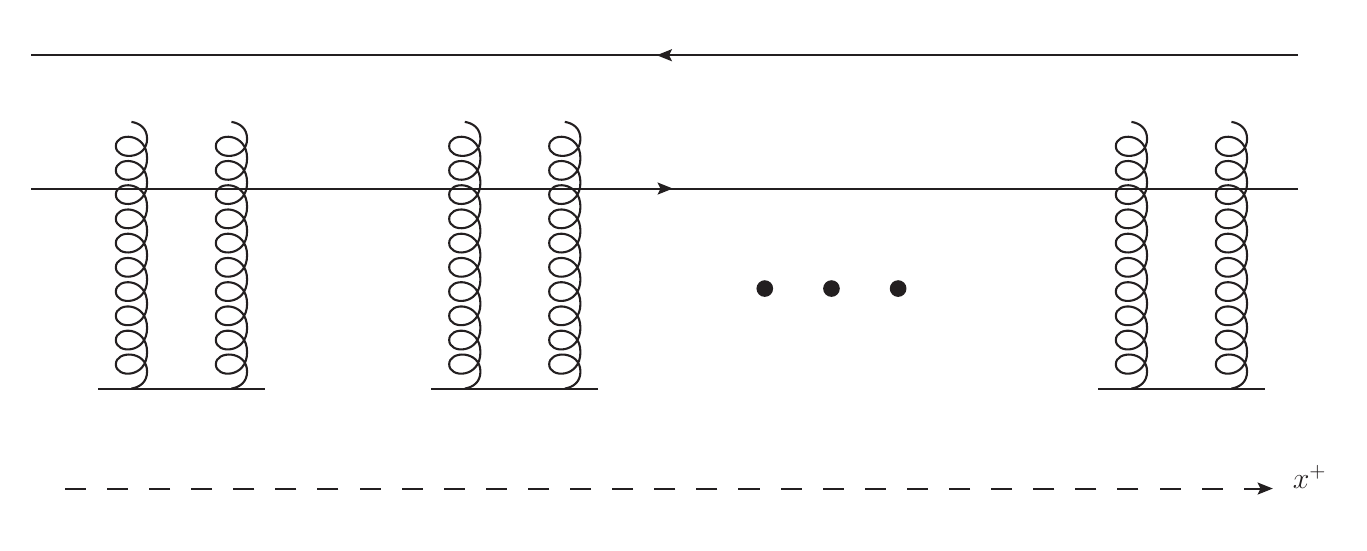}
    \caption{Generalizing the single dipole-nucleon scattering to the case of many nucleons, with the dipole encountering them as it moves along the $x^+$ direction. The unconnected gluons denote all four possible connections, as shown explicitly in Fig.~\ref{fig:GGMsingle}.}
    \label{fig:GGMmulti}
\end{subfigure}
\begin{subfigure}{\textwidth}
    \centering
    \includegraphics[width=0.8\textwidth]{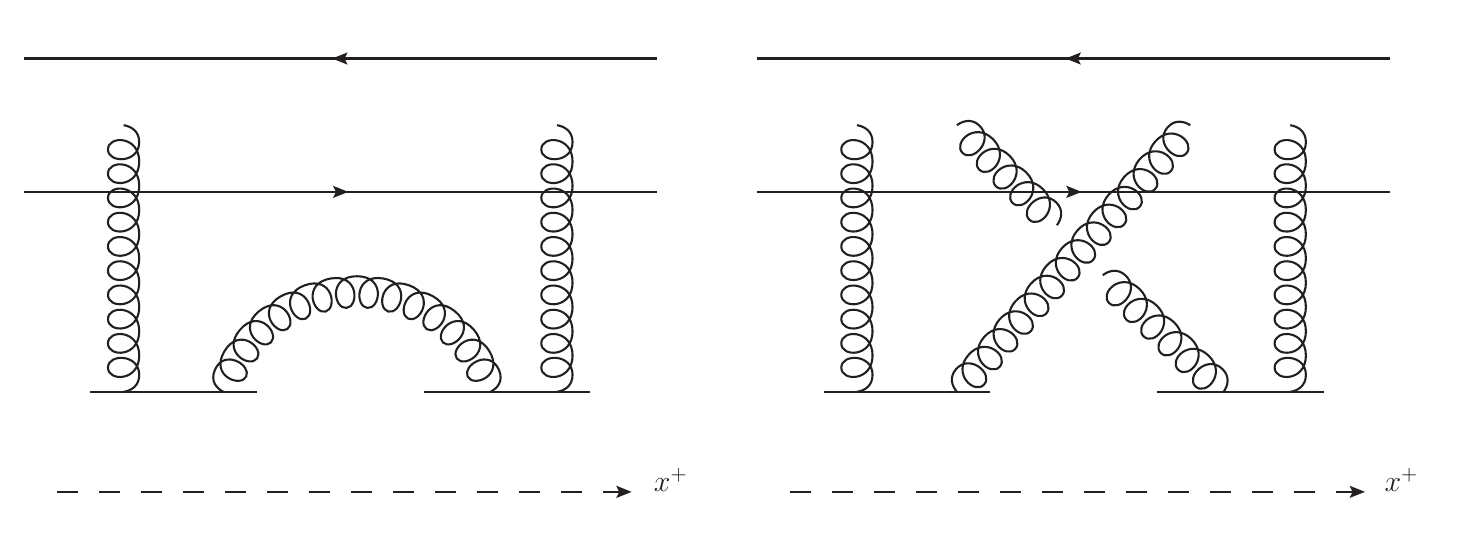}
    \caption{Examples of diagrams that are sub-leading relative to the ordered, separate interactions of Fig.~\ref{fig:GGMmulti}.}
    \label{fig:GGMcrosstalk}
\end{subfigure}

\label{GGM}
\caption{Dipole-nucleon scattering in the Glauber-Gribov-Mueller model. For more detailed versions of these figures, see Ch.~4 of \cite{Kovchegov:2012mbw}.}
\end{figure}

To start we can represent the scattering of a dipole on a single one of the nucleons in our target, like that shown in Fig.~\ref{fig:GGMsingle}. At leading order, the dominant contribution is a two-gluon exchange in the $t$-channel, as shown in the figure \cite{Kovchegov:2012mbw}. Generalizing to our large nucleus with many nucleons, we have Fig.~\ref{fig:GGMmulti}, where
for concreteness, we take the dipole to be moving with a large momentum in the $x^+$ direction and the nucleus with a large momentum in the $x^-$ direction. One can show that diagrams of the kind in Fig.~\ref{fig:GGMcrosstalk} are sub-leading relative to the exchanges shown in Fig.~\ref{fig:GGMmulti} --- that is, the leading contribution comes from separate gluon exchanges between the dipole and each separate nucleon, ordered along the $x^+$ direction as the dipole encounters them moving along its light cone \cite{Kovchegov:2012mbw}. Furthermore, one can show that the propagators of the lines making up the dipole are approximately on-shell between the gluon exchanges and that the exchanged gluons carry away a negligible amount of light-cone $+$ momentum (they are sometimes called Glauber gluons). 

\begin{figure}[h]
    \centering
    \includegraphics[width=\textwidth]{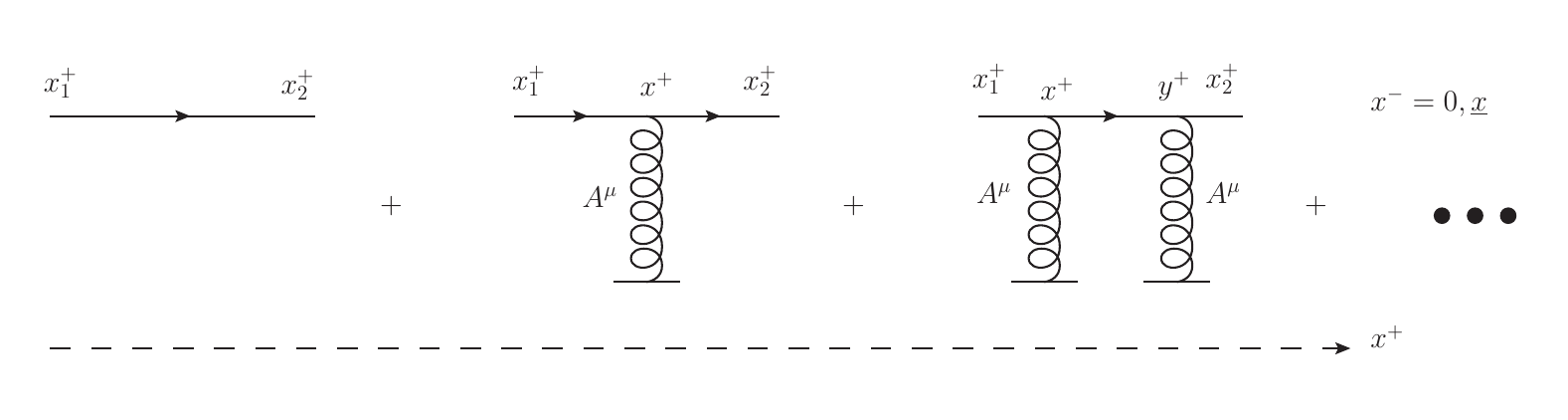}
    \caption{Summation of an arbitrary number of soft $t$-channel gluon exchanges ordered along the light-cone path of a high-energy quark.}
    \label{fig:wilsonlinetermbyterm}
\end{figure}

The ultimate generalization of this observation at high energies --- again, see \cite{Kovchegov:2012mbw} for details and a more rigorous derivation --- is that to describe the high-energy propagation of a quark or antiquark through a field of many color charges, we need to sum up a large number of $t$-channel gluon exchanges in an ordered way along the light cone propagation of the quark or antiquark. The first several terms of this sum, showing zero, one, and two gluon exchanges, are shown for the propagation of a quark line in Fig.~\ref{fig:wilsonlinetermbyterm}.
Note that we also employ a quasi-classical approximation where the gluon field of our nucleus $A^{\mu}$ is very strong and can be treated as approximately classical \cite{Mueller:1989st,McLerran:1993ni,McLerran:1993ka,McLerran:1994vd}. 
Denoting that gluon field $A^\mu = \sum_{a}A^{a\,\mu}t^a$ for $t^a$ the fundamental generators of $\text{SU}(N_c)$ (recall $N_c$ is the number of quark colors) and g the strong coupling, and noting that our quark line is propagating only in the $x^+$ direction so that $\mathrm{d}x^\mu A_{\mu} = \mathrm{d}x^+ A^-$, we can write the propagation shown in Fig.~\ref{fig:wilsonlinetermbyterm} as 
\begin{align}\label{ch2_wilsonlineterms}
    1 + ig\int\limits_{x_1^+}^{x_2^+}\mathrm{d}x^+ A^-(x^+, \underline{x}) + \left(ig\right)^2 \int\limits_{x_1^+}^{x_2^+}\mathrm{d}x^+ \int\limits_{x^+}^{x_2^+}\mathrm{d}y^+ A^-(y^+,\underline{x}) A^-(x^+,\underline{x}) +\, ...
\end{align}
We fixed our coordinates so that $x^- = 0$ and thus the gluon fields $A^\mu$ are functions only of the $+$ and $\perp$ coordinates. Note (i) the way we chain the integration limits in the term with two gluon exchanges. This ensures that the gluon at $y^+$ is exchanged further along the path than the gluon at $x^+$. And related to this, note that (ii) we write the gluon fields right-to-left so that the gluon farthest along the path is the leftmost one in our expression (this is important to define given the non-abelian nature of the gluon fields). (i) and (ii) define the notion of \textit{path ordering} \cite{Peskin:1995ev}. The generalization of \eq{ch2_wilsonlineterms} to an arbitrary number of gluon exchanges is the light-cone \textit{Wilson line}, an object of great importance to the rest of this dissertation. We denote it
\begin{align}\label{ch2_wilsonline}
    V_{\underline{x}}[x_2^+,x_1^+] = \mathcal{P}\, \text{exp} \left[ig\int\limits_{x_1^+}^{x_2^+}\mathrm{d}x^+ A^-\left(x^+,0^-,\underline{x}\right)   \right] \,.
\end{align}
Here $\mathcal{P}$ represents the path ordering operator. As can be checked explicitly, expanding the exponential in \eq{ch2_wilsonline} and effectuating the path ordering yields the terms in \eq{ch2_wilsonlineterms}. The important point is that we can describe the high-energy propagation of a quark through the color field of a target with the light-cone Wilson line in \eq{ch2_wilsonline}. Given that a high-energy quark's propagation is described by $V_{\underline{x}}$, it is straightforward to construct a high-energy antiquark's propagation as $V_{\underline{x}}^\dagger$. 

Since quarks transform under the fundamental representation of $\text{SU}(N_c)$ and correspondingly we wrote the Wilson line \eq{ch2_wilsonline} in terms of $A^\mu = \sum_a A^{a \mu}t^a $ for $t^a$ the fundamental generators of $\text{SU}(N_c)$, we often call $V_{\underline{x}}$ a fundamental Wilson line. In the same spirit as the above discussion, and since gluons transform under the adjoint representation of $\text{SU}(N_c)$, we can use an adjoint Wilson line to describe the high-energy propagation of a gluon through the color field of a target:
\begin{align}\label{ch2_wilsonlineadj}
    U_{\underline{x}}[x_f^+,x_i^+] = \mathcal{P}\, \text{exp} \left[ig\int\limits_{x_i^+}^{x_f^+}\mathrm{d}x^+ \mathcal{A}^-\left(x^+,0^-,\underline{x}\right)   \right] \,,
\end{align}
with $\mathcal{A}^{\mu} = \sum_a A^{a\,\mu}T^a$ and $T^a$ the adjoint $\text{SU}(N_c)$ generators, $(T^a)_{bc} = -if^{abc}$, for $f^{abc}$ the structure constants of $\text{SU}(N_c)$. 

\begin{figure}[h]
    \centering
    \includegraphics[width=.8\textwidth]{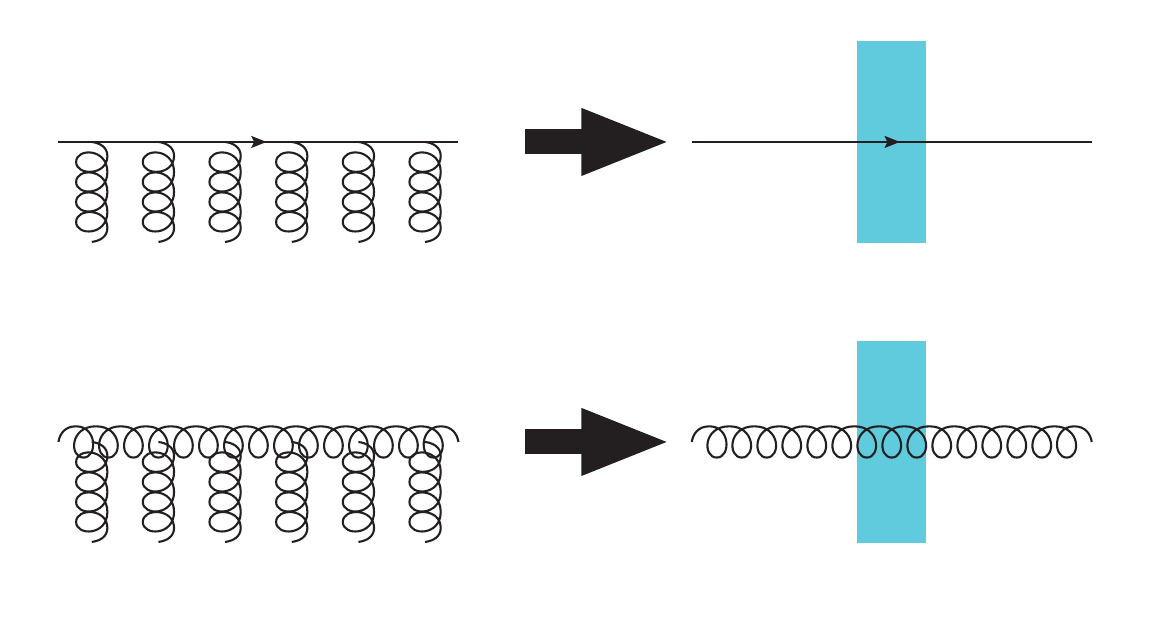}
    \caption{Wilson line and shockwave paradigms for high-energy scattering. We imagine the many soft gluon exchanges (the vertical corkscrews) of the Wilson lines to take place within the shock wave (the blue rectangles).}
    \label{fig:shockwaves}
\end{figure}

Moreover in the context of high energy scattering, what the scattering particle `sees' of the target is substantially flattened along the direction of propagation because of Lorentz contraction. In that sense, we typically call the interaction with the target a \textit{shock wave}. Fig.~\ref{fig:shockwaves} summarizes pictorially the framework we have established thus far --- namely that (1) the high energy propagation of quarks and gluons can be described by Wilson lines which sum up the path-ordered exchanges of many soft $t$-channel gluons and that (2) the interaction with the target can be thought of as a thin shock wave (the blue rectangle) in which all the gluon exchanges occur. 

Lastly, since we are still in the dipole picture of DIS, we can combine all the results of this Section to represent the propagation of the entire $q\overline{q}$ dipole through the color field of the target. The dipole $S$-matrix for this scattering can be written \cite{Kovchegov:2012mbw}
\begin{align}\label{ch2_unpolarizeddipole}
    S_{10}(zs) = \frac{1}{N_c} \bigg\langle \text{T tr}\left[V_{\underline{0}}V^{\dagger}_{\underline{1}} \right] \bigg\rangle (zs)\,,
\end{align}
where T denotes time ordering, the trace is over the fundamental indices, the subscripts denote transverse positions, and Wilson lines without specified bounds are infinite (that is, $V_{\underline{0}} = V_{\underline{x}_0}[\infty, -\infty]$). Also the angle brackets denote an averaging in the state of the target proton, $z$ denotes the smallest longitudinal momentum fraction of the lines making up the dipole, and $s$ is the center-of-mass energy squared of the interaction. Later we will also need to understand the contributions from gluon dipoles, and one can construct a similar object to that in \eq{ch2_unpolarizeddipole}, but in terms of adjoint Wilson lines rather than fundamental ones \cite{Kovchegov:2012mbw}:
\begin{align}\label{ch2_unpolarizeddipoleadj}
    S_{10}^{G}(zs) = \frac{1}{N_c^2 - 1}\bigg\langle \text{T Tr}\left[U_{\underline{0}}U^{\dagger}_{\underline{1}} \right] \bigg\rangle (zs)\,,
\end{align}
where $\text{Tr}$ denotes a trace over adjoint indices.
\begin{figure}[h]
    \centering
    \includegraphics[width=.8\textwidth]{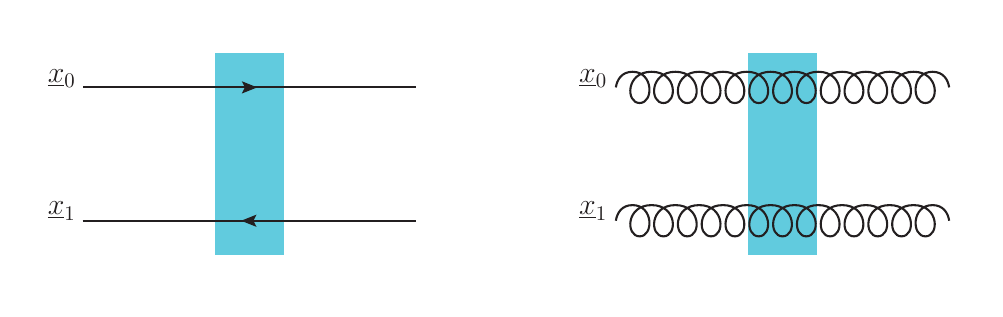}
    \caption{Diagrammatic representations of the fundamental and adjoint dipole amplitudes.}
    \label{fig:dipoleamplitudes}
\end{figure}
The objects in Eqs.~\eqref{ch2_unpolarizeddipole} and \eqref{ch2_unpolarizeddipoleadj} are represented diagrammatically in Fig.~\ref{fig:dipoleamplitudes} and are called \textit{dipole scattering amplitudes}. Though they are independent of helicity at the moment, these dipole amplitudes will form the basis of our entire discussion of helicity-dependent scattering in later Chapters.


\subsection{Light-Cone Perturbation Theory}\label{subsec_lcpt}

Light-cone perturbation theory (LCPT) \cite{Lepage:1980fj, Brodsky:1997de} is a method of performing quantum field theory calculations similar in some ways to the more traditional time-ordered perturbation theory, but with the important difference that time is replaced with light-cone time $x^+$. That is, instead of quantizing the theory at equal times, it is quantized at equal light-cone times. For the most part, the rules of LCPT are very similar to those of time-ordered perturbation theory. We do not restate the LCPT rules here, but a good summary can be found in Ch.~1 of \cite{Kovchegov:2012mbw}.

For the purposes of this dissertation, the main advantage of LCPT is that it lets us easily compute \textit{light-cone wavefunctions}. There are a number of reasons wavefunctions are more challenging objects to work with in an equal-time quantization. For one, equal-time wavefunctions are not boost invariant. So even if one can calculate a wavefunction in, say, the rest frame of a bound state of interest, it may be a considerable challenge to boost that wavefunction to a more relevant frame, like the infinite momentum frame in which DIS is frequently analyzed. Further, vacuum fluctuations mean that there are no equal-time wavefunctions of definite particle number \cite{Brodsky:1997de}. In contrast, light-cone wavefunctions can be defined independently of a reference frame and are boost invariant, plus the vacuum structure within LCPT is considerably simpler. Particles in LCPT are on-shell and have positive energies, which ultimately precludes vacuum fluctuations, leaving us with well-defined light-cone wavefunctions of definite particle number\footnote{This is a statement rife with subtlety. We typically understand the realistic QCD vacuum to be a very complicated picture with ties to topological structure, confinement, and a host of other phenomena. A truly trivial vacuum in LCPT would appear to contradict this. Some work has shown, however, that the complicated properties of the QCD vacuum can be captured in the zero modes of the LCPT language rather than the vacuum itself (though for this dissertation, these considerations are well beyond our scope). See Ch. 7 of \cite{Brodsky:1997de} for many more details which are far beyond the scope of this dissertation.} \cite{Brodsky:1997de}. 

To make our discussion of light-cone wavefunctions a bit more quantitative, and following \cite{Kovchegov:2012mbw,Brodsky:1997de}, consider a general hadronic bound state $\ket{\Psi}$ of momentum $P^\mu$ and mass $M$. In LCPT, the minus component of the four momentum $P^-$ plays the role of the light-cone Hamiltonian. Since particles in LCPT are on-shell we have
\begin{align}\label{ch2_wfsonshellcond}
    2P^+P^- - \underline{P}^2 = M^2\,,
\end{align}
and the Hamiltonian eigenvalue equation for our state $\ket{\Psi}$ can be written
\begin{align}\label{ch2_hamiltonian}
    H\ket{\Psi} = \frac{M^2 + \underline{P}^2}{2P^+}\ket{\Psi}\,.
\end{align}
To work with \eq{ch2_hamiltonian}, we decompose the hadronic state $\ket{\Psi}$ into a superposition of Fock states $\ket{n_G, n_q}$, corresponding to a particular number of gluons $n_G$ and quarks $n_q$, with relevant momentum, helicity, color, and flavor labels:
\begin{align}\label{ch2_fockstates}
    \ket{n_G,n_q} = \ket{n_G, \{k_i^+, \underline{k}_i, \lambda_i, a_i \}; n_q, \{p_j^+, \underline{p}_j, \sigma_j, \alpha_j, f_j\}}\,,
\end{align}
with $i = 1,...,n_G$ and $j = 1,...,n_q$; gluon momenta $k^\mu_i$, helicities $\lambda_i$, and colors $a_i$; and quark momenta $p^\mu_j$, helicities $\sigma_j$, colors $\alpha_j$, and flavors $f_j$. Any given Fock state represented by \eq{ch2_fockstates} can be written as a sequence of the appropriate creation operators acting on the vacuum $\ket{0}$.

For our hadronic state, we write
\begin{align}\label{ch2_fockstatedecomp}
    \ket{\Psi} = \sum_{n_G, n_q} \int \mathrm{d}\Omega_{n_G + n_q} \ket{n_G, n_q}\braket{n_G,n_q|\Psi}\,,
\end{align}
with $\int \mathrm{d}\Omega_{n_G + n_q}$ an integral over the phase space of the $n_G$ gluons and $n_q$ quarks and where the Fock states form a complete basis
\begin{align}\label{ch2_fockcompletebasis}
    \sum_{n_G, n_q} \int \mathrm{d}\Omega_{n_G + n_q} \ket{n_G, n_q}\bra{n_G, n_q} = 1\,.
\end{align}
The projection of the hadronic state onto a given Fock state is the light-cone wavefunction
\begin{align}\label{ch2_lightconewf}
    \Psi(n_G, n_q) = \braket{n_G,n_q|\Psi}\,.
\end{align}
Conveniently, calculating the light-cone wavefunctions of \eq{ch2_lightconewf} can be done in a diagrammatic way following the same rules (with a few minor modifications) as those used to calculate amplitudes in LCPT. Again we do not restate the rules here, but more comprehensive treatments can be found in, for example, \cite{Kovchegov:2012mbw,Brodsky:1997de}.


\section{Evolution}\label{sec:evol}

In Ch.~\ref{intro.ch} we introduced the notion of parton distribution functions (PDFs), as functions of both $x$ and $Q^2$. But as properties of QCD bound states, PDFs are nonperturbative objects. That is, we cannot fully calculate these distributions in a perturbative framework. However, if we have a PDF specified at some input scale (maybe from experimental data) and we are interested in its behavior at some other scale that lies in a perturbative domain, we can use techniques of perturbation theory to describe how the PDF changes as we move from the input scale to the new scale. This is the notion of \textit{evolution}. As the PDFs are functions of both $x$ and $Q^2$, the scale in question could be either of these variables. 

\subsection{\texorpdfstring{Evolution in $Q^2$}{Evolution in Q2}}

We often think of QCD observables in terms of collinear factorization, which allows us to separate short-distance, perturbative physics from the longer-distance, nonperturbative physics\footnote{The technical details of factorization are beyond the scope of this dissertation, but see, for example the book \cite{Sterman:1993hfp}.}. Very schematically, we may write \cite{Sterman:1993hfp}
\begin{align}\label{ch2_factorization}
    \sigma_{\text{tot}}(x,Q^2) \propto \int\limits_0^1 \mathrm{d}\xi \, f(\xi,\mu^2)\, C\left(\frac{x}{\xi}, \frac{Q^2}{\mu^2}\right)\,,
\end{align}
where $\sigma_{\text{tot}}$ is some cross section, $f(x,\mu^2)$ is some parton distribution function, $C$ is a coefficient function we can calculate perturbatively, and $\mu^2$ is the factorization scale --- the scale at which we imagine separating the short-- and long--distance physics. Of course, the cross section should not depend on the arbitrary choice of factorization scale:
\begin{align}\label{ch2_factorizationderiv}
    \frac{\mathrm{d}}{\mathrm{d}\ln \mu^2} \,\sigma_{\text{tot}}(x,Q^2) = 0\,.
\end{align}
In DIS we typically identify the factorization scale $\mu^2$ with $Q^2$. Doing so, one can show that Eqs.~\eqref{ch2_factorization} and \eqref{ch2_factorizationderiv} imply a $Q^2$ evolution for the distribution function of the form \cite{Sterman:1993hfp}
\begin{align}\label{ch2_evolutiongeneral}
    \frac{\partial}{\partial\ln Q^2} f(x,Q^2) = \int\limits_x^1 \frac{\mathrm{d}\xi}{\xi} P\left(\xi\right) f\left(\frac{x}{\xi},Q^2\right)\,,
\end{align}
where $P(\xi)$ is a perturbatively calculable evolution kernel. \eq{ch2_evolutiongeneral} illustrates the general structure of an evolution equation --- given the distribution $f(x, Q_0^2)$ at some input scale $Q_0^2$, \eq{ch2_evolutiongeneral} provides a differential equation in $\ln Q^2$ to evolve $f$ from its input scale to any $Q^2 > Q_0^2$. 

\begin{figure}[h!]
\centering
\begin{subfigure}{\textwidth}
    \centering
    \includegraphics[width=0.7\textwidth]{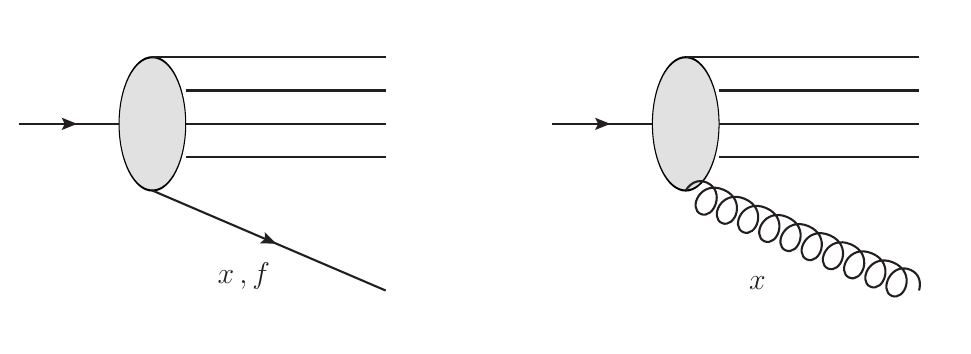}
    \caption{The quark (of flavor $f$) and gluon distributions within the proton. The straight horizontal lines represent the other partons.}
    \label{fig:qGwfs}
\end{subfigure}

\bigskip

\begin{subfigure}{\textwidth}
    \centering
    \includegraphics[width=0.7\textwidth]{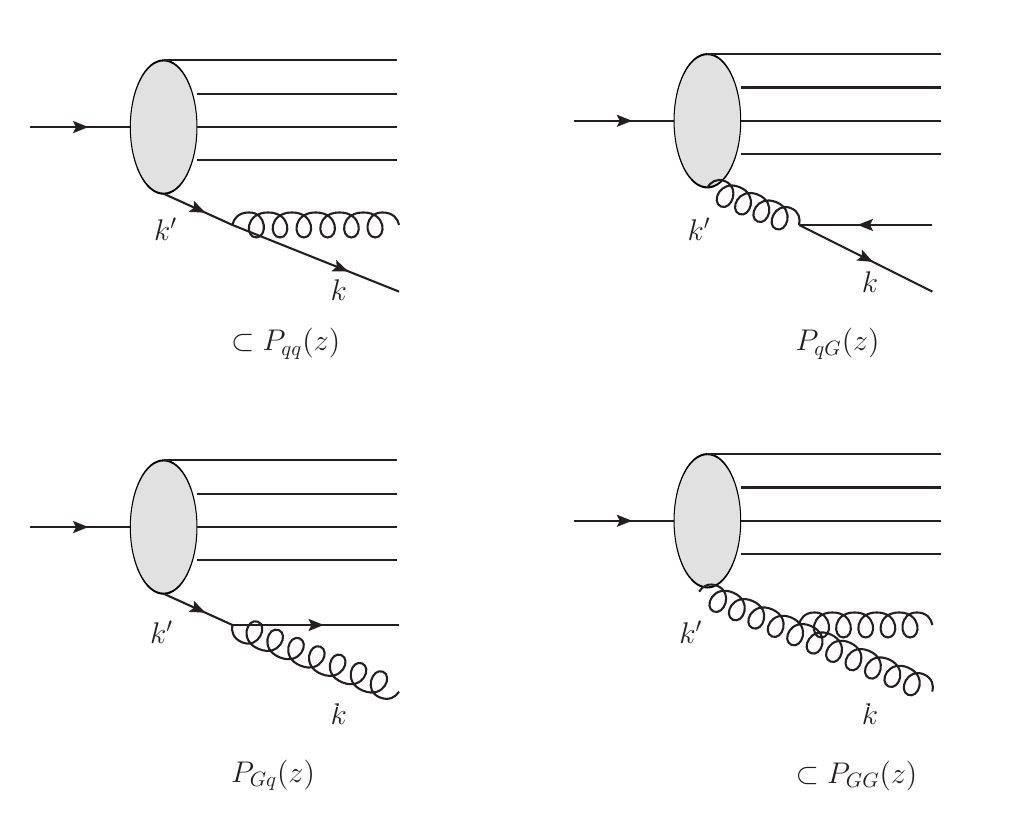}
    \caption{Leading-order contributions to the DGLAP splitting functions $P_{ij}$(z), i.e. finding a parton of type $i$ in the wavefunction of a parton of type $j$. Note that $z = k^+/k^{\prime\,+}$. The top left and bottom right are labeled as subsets of $P_{qq}$ and $P_{GG}$, respectively, as there are other leading-order contributions to these splitting functions.}
    \label{fig:dglapsplittingfuncs}
\end{subfigure}
\label{fig:dglapstuff}
\caption{Some diagrams necessary for constructing the DGLAP evolution equations. Note these are more accurately light-cone wavefunctions and only upon squaring them do we obtain contributions to the DGLAP evolution.}
\end{figure}

The full QCD generalization of the schematic evolution in \eq{ch2_evolutiongeneral} is a set of equations known as the Dokshitzer-Gribov-Lipatov-Altarelli-Parisi (DGLAP) evolution equations \cite{Gribov:1972ri,Altarelli:1977zs,Dokshitzer:1977sg} (see also \cite{Kovchegov:2012mbw} for a pedagogical treatment which we largely follow here). These equations describe the $Q^2$-evolution of the gluon distribution function $G(x,Q^2)$ and the flavor-singlet quark distribution function 
\begin{align}\label{ch2_quarkdist}
    \Sigma(x,Q^2) = \sum_f\left[q^f(x,Q^2) + q^{\overline{f}}(x,Q^2) \right]\,,
\end{align}
with $q^f$ and $q^{\overline{f}}$ the quark and antiquark distributions, respectively, of flavor $f$ (there is also a DGLAP evolution equation for the flavor non-singlet quark distribution function $q^{NS}(x,Q^2) \equiv q^f(x,Q^2) - q^{\overline{f}}(x,Q^2)$, but here we focus on the flavor-singlet distribution). The kernel of DGLAP evolution is controlled by a $2\times 2$ matrix of splitting functions, which multiplies the doublet of $(\Sigma(x,Q^2)\,,\,G(x,Q^2) )$. The four splitting functions are denoted $P_{ij}(z)$ for $i,j = q,G$. Each $P_{ij}(z)$ can be calculated perturbatively and corresponds to the probability of finding a parton of type $i$ in the light-cone wavefunction of a parton of type $j$. For example, $P_{Gq}$ represents the probability of finding a gluon in the wavefunction of a quark. The splitting functions are written as functions of $z$, the longitudinal momentum fraction of the daughter particle relative to the parent. We show the typical diagrammatic language for representing the quark and gluon distributions in Fig.~\ref{fig:qGwfs}, and we show the modifications of these distributions that contribute to the splitting functions in Fig.~\ref{fig:dglapsplittingfuncs}.
The full set of DGLAP evolution equations involve a convolution of the matrix of splitting functions with the parton distribution functions and is written as:
\begin{align}\label{ch2_dglap}
    \frac{\partial}{\partial \ln Q^2} \begin{pmatrix}
        \Sigma(x,Q^2) \\
        G(x,Q^2)
    \end{pmatrix} 
    = \frac{\alpha_s(Q^2)}{2\pi}\int_x^1 \frac{\mathrm{d}z}{z}
    \begin{pmatrix}
        P_{qq}(z) & P_{qG}(z) \\
        P_{Gq}(z) & P_{GG}(z)
    \end{pmatrix}
    \begin{pmatrix}
        \Sigma\left(\frac{x}{z},Q^2\right) \\
        G\left(\frac{x}{z},Q^2\right)
    \end{pmatrix}\,.
\end{align}
Note that this growth of the parton distribution functions in $Q^2$ is the violation of Bjorken scaling we mentioned very briefly in Sec.~\ref{sec:dis}. 

The leading order treatment of DGLAP is typically called the leading logarithmic approximation (LLA). In this LLA, each power of the strong coupling $\alpha_s$ brings one log of $Q^2$, the logs resulting from transverse momentum integrals cut off by $Q^2$ in the ultraviolet. Since we assume $Q^2 \gg \Lambda^2$ so that perturbation theory is valid (for $\Lambda$ an infrared cutoff), $\ln(Q^2/\Lambda^2)\gg 1$. Even though the coupling is small, we have the relevant parameter $\as \ln(Q^2/\Lambda^2) \sim \mathcal{O}(1)$. This is called the \textit{resummation} parameter since we re-sum a power series in it (to all orders) as opposed to the summation of a power series in just the small parameter $\as$ which we might truncate at some finite order.
\begin{figure}[ht!]
    \centering
    \includegraphics[width=0.5\linewidth]{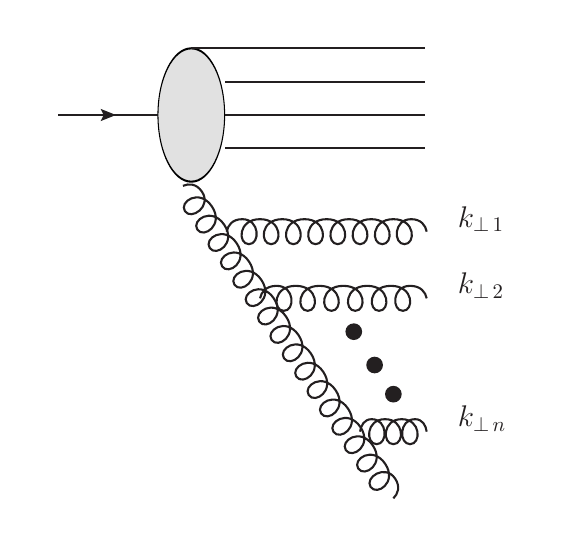}
    \caption{Subsequent emissions in DGLAP evolution.}
    \label{fig:dglapordering}
\end{figure}

We can view DGLAP evolution as a sequence of subsequent emissions. For example, imagine iterating more and more gluon emissions (the horizontal gluons) from the diagonal gluon line in the lower right corner of Fig.~\ref{fig:dglapsplittingfuncs} --- we show this explicitly in Fig.~\ref{fig:dglapordering}). In order to obtain the logs of $Q^2$, the transverse momenta of the emissions need to be strongly ordered:
\begin{align}\label{ch2_dglapordering}
    Q^2 \gg k_{\perp\,n}^2 \gg k_{\perp\,n-1}^2 \gg ... \gg k_{\perp\,1}^2\,, 
\end{align}
where the $k_{\perp\,n}$ is the transverse momentum of the last emission and $k_{\perp\,1}$ that of the first, as shown in Fig.~\ref{fig:dglapordering}.

We conclude here with a few closing notes regarding DGLAP. First, there is a polarized version of DGLAP evolution which describes the polarization-dependent evolution in $Q^2$, very much analogous to the unpolarized version presented in \eq{ch2_dglap}. In the polarized version we work with the helicity parton distribution functions $\Delta\Sigma(x,Q^2)$ and $\Delta G(x,Q^2)$ introduced previously, along with the polarized splitting functions $\Delta P_{ij}(z)$. This polarized DGLAP evolution is given by
\begin{align}\label{ch2_poldglap}
    \frac{\partial}{\partial \ln Q^2} \begin{pmatrix}
        \Delta\Sigma(x,Q^2) \\
        \Delta G(x,Q^2)
    \end{pmatrix} 
    = \int_x^1 \frac{\mathrm{d}z}{z}
    \begin{pmatrix}
        \Delta P_{qq}(z) & \Delta P_{qG}(z) \\
        \Delta P_{Gq}(z) & \Delta P_{GG}(z)
    \end{pmatrix}
    \begin{pmatrix}
        \Delta \Sigma\left(\frac{x}{z},Q^2\right) \\
        \Delta G\left(\frac{x}{z},Q^2\right)
    \end{pmatrix}\,.
\end{align}

Second, the small-$x$ regime brings an extra logarithm to the resummation parameter (from the longitudinal momentum integral), so that the small-$x$ DGLAP resummation parameter is $\as \ln(Q^2/\Lambda^2)\ln(1/x)$. It is often convenient, particularly in this double-logarithmic regime to work with DGLAP in moment space by introducing the Mellin transform
\begin{align}\label{ch2_Mellin}
    f_{\omega}(Q^2) = \int\limits_0^1 \mathrm{d}x \,x^{\omega-1}f(x,Q^2)\,,
\end{align}
and its inverse
\begin{align}\label{ch2_Mellininverse}
    f(x,Q^2) = \int \frac{\mathrm{d}\omega}{2\pi i}\,x^{-\omega}f_{\omega}(Q^2)\,,
\end{align}
where in \eq{ch2_Mellininverse} the integral is over a vertical contour in the complex-$\omega$ plane that lies to the right of all singularities of the function $f_\omega(Q^2)$. Employing the Mellin transform, one can write an analytic solution to the DGLAP evolution Eqs.~\eqref{ch2_dglap} in the form \cite{Kovchegov:2012mbw}
\begin{align}\label{ch2_dglapsoln}
    \begin{pmatrix}
        \Sigma(x,Q^2) \\
        G(x,Q^2)
    \end{pmatrix}
    = \wint e^{\omega\ln(1/x)}\text{exp}\left\{\begin{pmatrix}
        \gamma_{qq}(\omega) & \gamma_{qG}(\omega) \\
        \gamma_{Gq}(\omega) & \gamma_{GG}(\omega)
    \end{pmatrix} 
    \ln \frac{Q^2}{\Lambda^2}
    \right\}
    \begin{pmatrix}
        \Sigma_\omega(\Lambda^2) \\
        G_{\omega}(\Lambda^2)
    \end{pmatrix}\,,
\end{align}
where instead of the matrix of DGLAP splitting functions $P_{ij}(z)$, we have the matrix of DGLAP \textit{anomalous dimensions} $\gamma_{ij}(\omega)$ defined by the Mellin transform
\begin{align}\label{ch2_dglapanomalousdims}
    \gamma_{ij}(\omega) = \int\limits_0^1\mathrm{d}z\,z^{\omega-1} P_{ij}(z)\,. 
\end{align}
Note we have also written the initial scale of the evolution as $\Lambda^2$. The double logarithmic regime of the polarized DGLAP evolution, and in particular the expression of its solution in the form of \eq{ch2_dglapsoln}, will provide an important cross check of the small-$x$ helicity evolution in the coming Chapters. 


\subsection{\texorpdfstring{Evolution in $x$}{Evolution in x}}\label{subsec_evolinx}

Instead of asking how the parton distribution functions evolve as we increase $Q^2$, we can alternatively ask how they change as we decrease $x$. Note that instead of the collinear factorization sketched in \eq{ch2_factorization} which implied evolution in $Q^2$, small-$x$ evolution can be seen from an approach that factorizes in the rapidity variable $Y\sim \ln(1/x)$ \cite{Balitsky:1995ub,Balitsky:1998ya}. But the basic idea of small-$x$ evolution is the same as for $Q^2$ evolution --- given the distribution functions at some starting scale $x_0$ (likely inferred from experiment), evolution describes the behavior of these distributions as we decrease $x$. 

As mentioned in Sec.~\ref{sec:wilsonetc}, the leading contribution to high energy scattering of a dipole on a target is a two-gluon exchange in the $t$-channel of the forward amplitude. The Balitsky-Fadin-Kuraev-Lipatov (BFKL) evolution equation \cite{Kuraev:1977fs,Balitsky:1978ic} describes the QCD corrections to the leading two-gluon exchange for the unpolarized scattering of two such dipoles as one moves to higher energies (smaller $x$). Note that in this Section we largely follow the treatment of BFKL from the book \cite{Kovchegov:2012mbw}.

\begin{figure}[h!]
    \centering
    \includegraphics[width=0.6\linewidth]{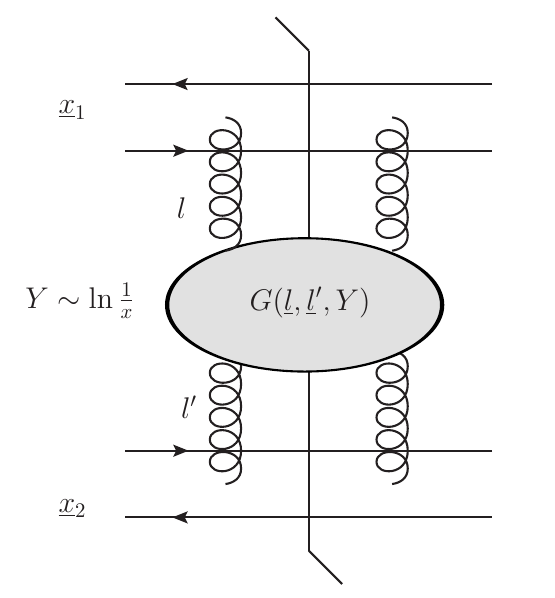}
    \caption{The BFKL Green function, as depicted in Ch.~3 of \cite{Kovchegov:2012mbw}.}
    \label{fig:bfklgreenfunction}
\end{figure}

BFKL evolution is often written in terms of a Green function $G(\underline{l},\underline{l}', Y)$ which propagates the two $t$-channel gluons. Its representation is shown, following \cite{Kovchegov:2012mbw}, in Fig.~\ref{fig:bfklgreenfunction}.
Here $l$ and $l'$ are the momenta of the gluons before and after the Green function and $Y = \ln(s x_{1\perp}x_{2\perp}) \sim \ln(1/x)$ for $\underline{x}_1$ and $\underline{x}_2$ the transverse sizes of the dipoles and $s$ the center-of-mass energy squared between them. As in Sec.~\ref{sec:wilsonetc} we use the unconnected gluon lines between the lines of the dipoles to denote all the possible connections. The vertical line denotes the final state cut, which separates amplitude from complex conjugate amplitude.

\begin{figure}[h!]
    \centering
    \includegraphics[width=0.8\linewidth]{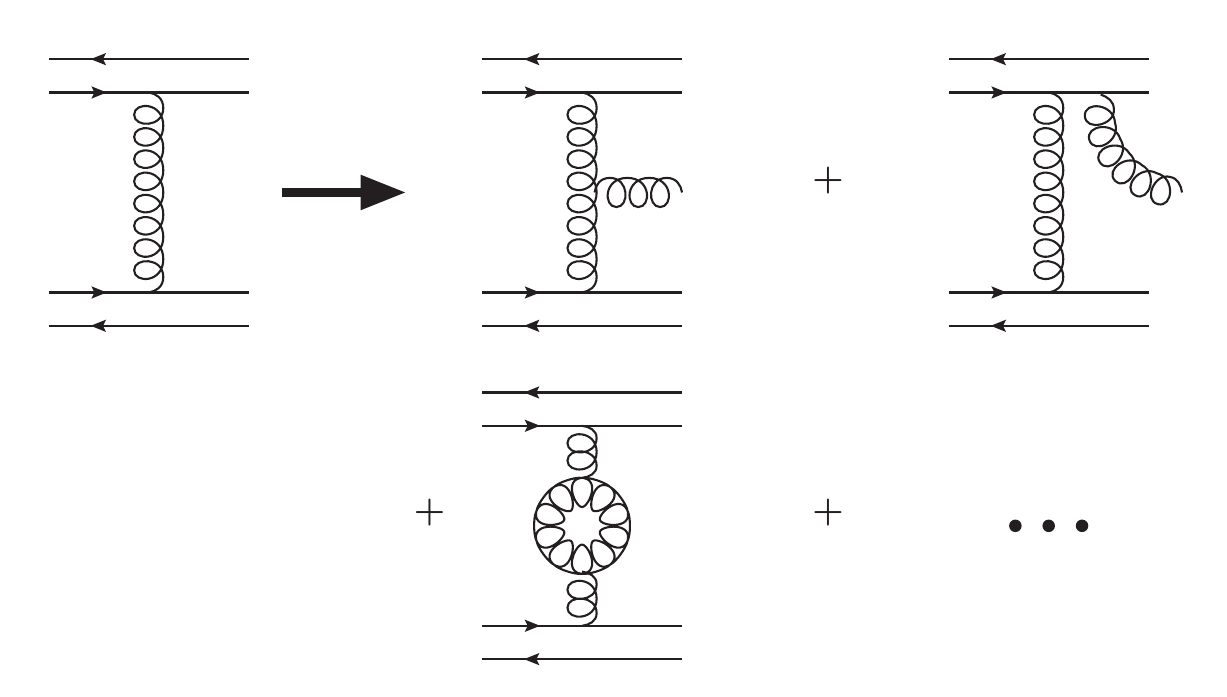}
    \caption{Examples of BFKL corrections to $t$-channel gluon exchange between two dipoles.}
    \label{fig:bfklcorrections}
\end{figure}

Some examples of the corrections that drive the evolution (only drawing one of the exchanged $t$-channel gluons) are shown in Fig.~\ref{fig:bfklcorrections}. 
Note that these corrections are contained within the Green function $G(\underline{l},\underline{l}', Y)$. The top row after the arrow shows two examples of real corrections --- where the extra emitted gluon is found in the final state --- while the bottom row shows an example of a virtual correction --- where the extra emitted gluon is not found in the final state. For concreteness, we show the contribution where the $t$-channel gluon connects to the quark line of each dipole, but of course it could connect in any of the four possible ways to the lines of the dipole.

\begin{figure}[t!]
\centering
\begin{subfigure}[t!]{0.4\textwidth}
    \centering
    \includegraphics[height=4cm]{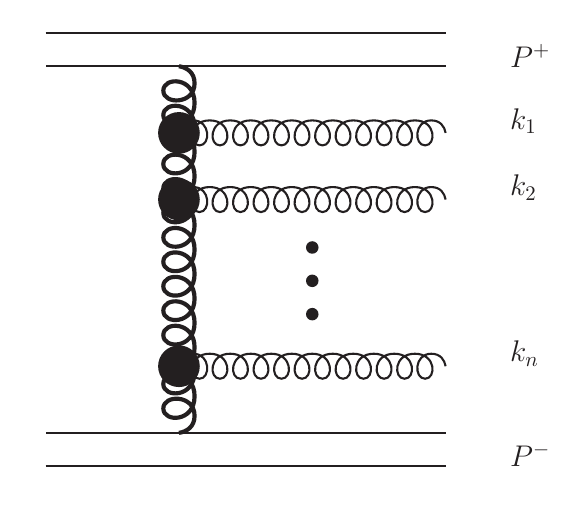}
    \caption{}
    \label{fig:bfklladder}
\end{subfigure}
~
\begin{subfigure}[t!]{0.5\textwidth}
    \centering
    \includegraphics[height=4cm]{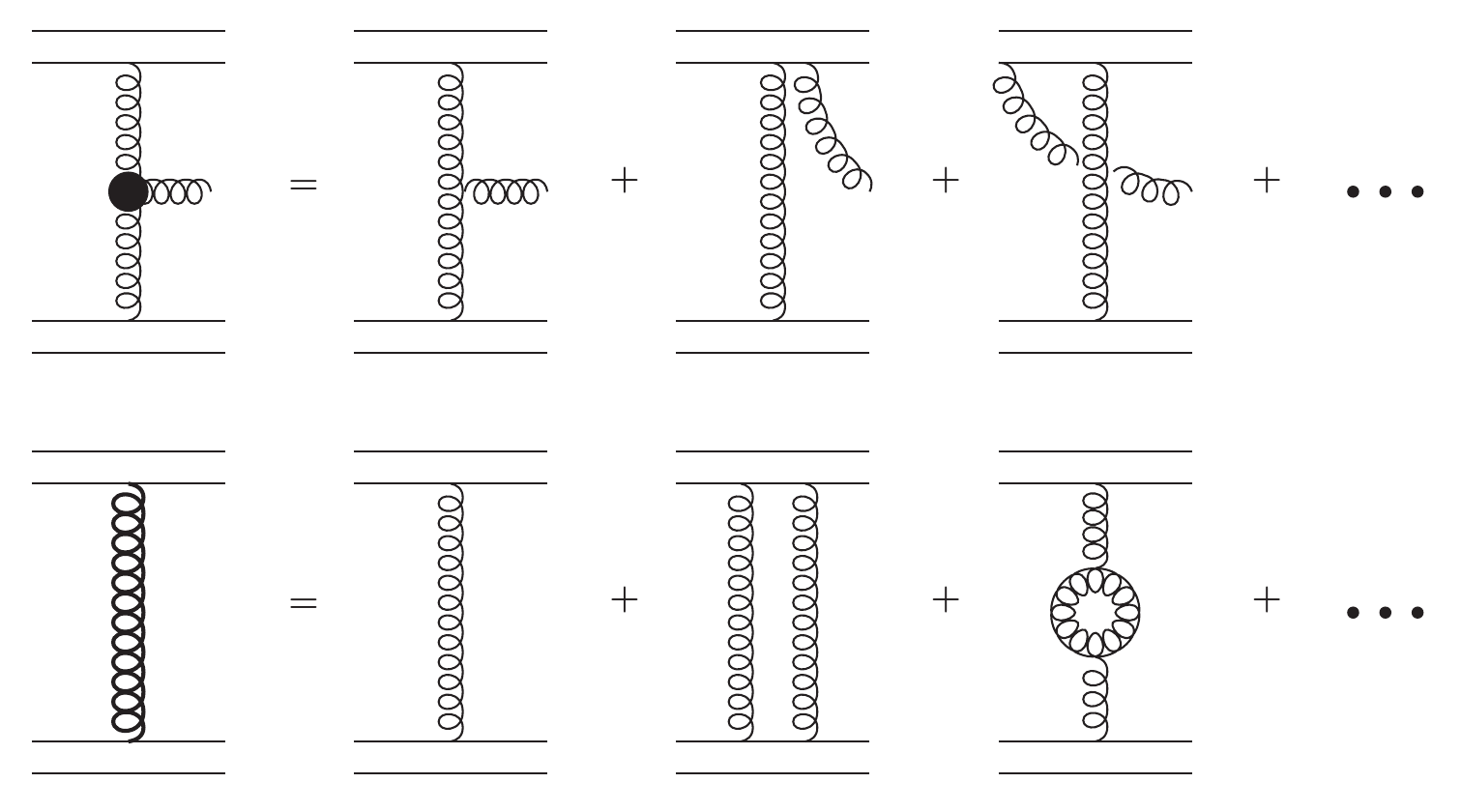}
    \caption{}
    \label{fig:lipatovandregge}
\end{subfigure}
\caption{The BFKL ladder (a) with momenta labeled and the ingredients out of which the ladder is built (b) --- the effective Lipatov vertex (b, top) and the reggeized gluon (b, bottom), illustrated in the style of Ch.~3 of \cite{Kovchegov:2012mbw}.}
\label{fig:bfklstuff}
\end{figure}

While the resummation parameter for DGLAP evolution presented in the previous Section was $\as \ln(Q^2/\Lambda^2)$, the resummation parameter for leading-log BFKL evolution is $\as\ln(1/x)$. By systematically including all the leading-log corrections to the $t$-channel gluon exchange --- that is, the corrections that bring in a log of energy $\ln s \sim \ln(1/x)$ for each power of the coupling --- BFKL evolution can be represented as a ladder or cascade, similar to the representation of DGLAP evolution in Fig.~\ref{fig:dglapordering}. We omit the derivation here, but see \cite{Kovchegov:2012mbw} for many more details. Fig.~\ref{fig:bfklladder} shows the ladder representation of the sequence of gluon emissions. 
The essential features which allow us to represent this evolution as a ladder are i) the thick black circle denoting the effective \textit{Lipatov vertex} and ii) the bold (vertical) gluon line denoting a \textit{reggeized} gluon. These two features are shown in more detail in Fig.~\ref{fig:lipatovandregge}. The Lipatov vertex denotes the sum of the leading-log real corrections where we find the extra emitted gluon in the final state. The reggeized gluon represents the single $t$-channel gluon exchange plus all the leading-log virtual corrections (in the color octet state), resulting in a single effective gluon propagator. Combining both these effects, BFKL evolution can be thought of as a ladder made up of reggeized gluon rails (the bold vertical gluon lines) and regular $s$-channel gluon rungs (the horizontal gluon lines) coming from successive emissions from the Lipatov vertices. Fig.~\ref{fig:bfklladder} represents an amplitude that makes up half the ladder (the complex conjugate amplitude would form the other half).

The full BFKL evolution equation can be written in terms of the Green function as
\begin{align}\label{ch2_bfkleq}
    \frac{\partial G\left(\underline{l},\underline{l}',Y\right)}{\partial Y} = \frac{\as N_c}{\pi^2}\int \frac{\mathrm{d}^2q}{\left(\underline{l}-\underline{q} \right)^2} \left[G\left(\underline{q},\underline{l}', Y\right) - \frac{l_{\perp}^2}{2q_{\perp}^2}G\left(\underline{l},\underline{l}' ,Y\right)   \right]
\end{align}
Recall that we obtained leading transverse logs in DGLAP evolution by ordering the transverse momenta, as in \eq{ch2_dglapordering}. However for BFKL we ultimately obtain longitudinal logs with a strong ordering of longitudinal momenta. With the momenta as labeled in Fig.~\ref{fig:bfklladder}, where we assume one of the dipoles moves with large $P^+$ and the other with large $P^-$, the momenta of the gluons forming the ladder are ordered
\begin{subequations}
\begin{align}
    &P^+ \gg k_1^+ \gg k_2^+ \gg ... \gg k_n^+ \,,\\
    & k_1^- \ll k_2^- \ll ... \ll k_n^- \ll P^-\,.
\end{align}
\end{subequations}

While the BFKL evolution of \eq{ch2_bfkleq} provides substantial insight into high energy scattering, it is not the full story. One can show that the unitarity of the $S$-matrix (along with QCD confinement) ultimately implies an upper bound on how quickly cross sections can grow with energy, called the Froissart-Martin bound \cite{PhysRev.123.1053,Martin:1969ina,Lukaszuk:1967zz}. It can be shown \cite{Kovchegov:2012mbw} that the total cross section resulting from BFKL evolution violates this bound and so fails to preserve unitarity. At very high energies, BFKL needs to be replaced with a new evolution, one that does not grow too quickly with energy.

\begin{figure}[h!]
    \centering
    \includegraphics[width=0.9\linewidth]{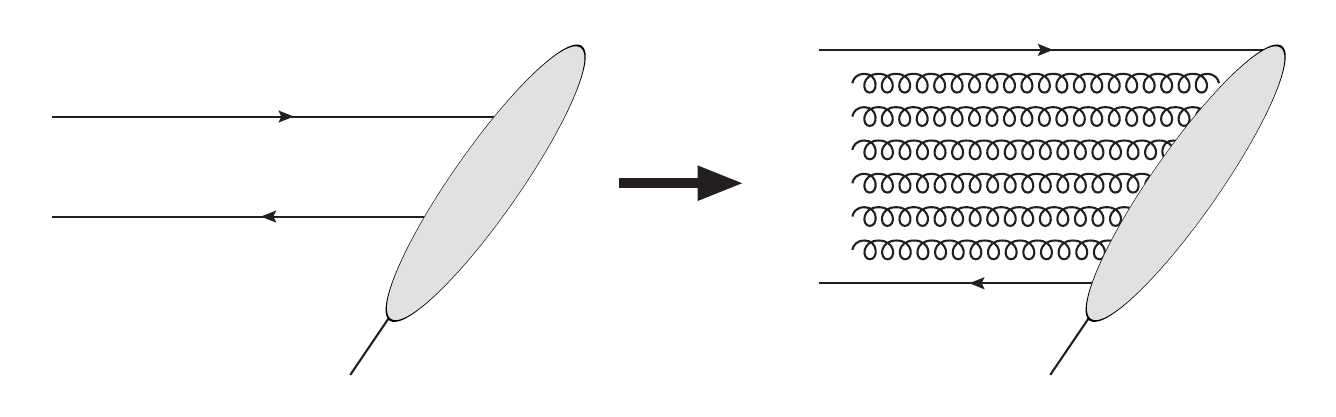}
    \caption{The effects of quantum evolution absorbed into the wavefunction of the incident dipole, which evolves into a system of many gluons that then scatters on the target (the gray oval).}
    \label{fig:dipolemodelevolution}
\end{figure}

\begin{figure}[h!]
    \centering
    \includegraphics[width=0.85\linewidth]{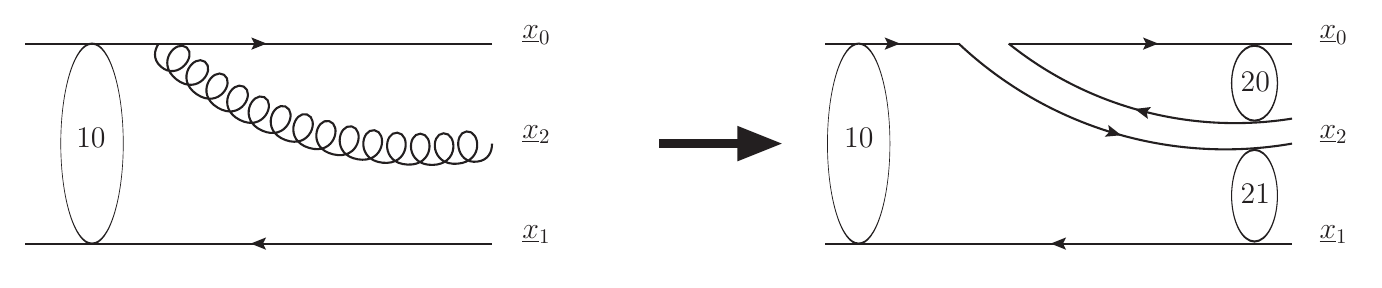}
    \caption{The decay of an initial dipole into several daughter dipoles.}
    \label{fig:daughterdipoles}
\end{figure}

The new evolution that restores unitarity at high energies is known as the Balitsky-Kovchegov (BK) equation \cite{Kovchegov:1999yj,Balitsky:1995ub}. This evolution is in some sense a generalization of the Glauber-Gribov-Mueller model discussed in Sec.~\ref{sec:wilsonetc}, with the BK equation explicitly accounting for the quantum corrections associated with evolution, which GGM does not do. BK evolution can be derived in the context of Mueller's dipole model \cite{Mueller:1993rr,Mueller:1995gb,Mueller:1994jq}. In this model we can start, as with BFKL evolution, from the wavefunction of a quark-antiquark pair, but now we absorb the effects of quantum evolution into the wavefunction of the $q\overline{q}$ pair. Just as with BFKL, evolution effects at high energy correspond to emissions of additional gluons. So the scattering of our dipole on a target has two components - the dipole evolves into a system of many many gluons and this system then scatters on the target. This framework is shown schematically in Fig.~\ref{fig:dipolemodelevolution}.
Further, in Mueller's dipole model, one takes the 't Hooft large-$N_c$ limit \cite{tHooft:1974pnl}, where the number of colors $N_c$ is taken to be a large parameter with the combination $\as N_c$ fixed and finite. In this limit a gluon can be represented as a double line, that is, a quark-antiquark pair in the color octet state. Then when our initial dipole emits a gluon, we can reinterpret the result as a system of two new dipoles. This is shown in Fig.~\ref{fig:daughterdipoles}, where each oval denotes a particular dipole. The initial dipole, labeled by the transverse positions of its quark and anti-quark lines as dipole 10, emits a gluon at transverse position $\underline{x}_2$. In the large-$N_c$ limit, this gluon is a quark-antiquark pair, whose $q$ and $\overline{q}$ lines then pair up with the original lines at $\underline{x}_1$ and $\underline{x}_0$ to form two new dipoles, dipoles 20 and 21. Another simplification afforded by the large-$N_c$ limit is the fact that non-planar contributions are $N_c$-suppressed, the consequence being that resulting dipoles evolve independently to leading order in $N_c$. For example, we would not have a contribution in the large-$N_c$ limit where dipole $20$ in the right panel of Fig.~\ref{fig:daughterdipoles} emits a gluon that connects to one of the lines making up dipole $21$.

Thus the right panel of Fig.~\ref{fig:dipolemodelevolution} is really a system of many dipoles, each of whose scattering on the target can be described with the GGM model.
In this framework, the BK equation describes the evolution of the imaginary part $N(\underline{x}_{1},\underline{x}_0, Y)$ of the forward dipole scattering amplitude. Here $\underline{x}_1$ and $\underline{x}_0$ are the transverse coordinates of lines of the original dipole, like those shown in Fig.~\ref{fig:daughterdipoles}. Their difference is $\underline{x}_{10} = \underline{x}_1 - \underline{x}_0$ with magnitude $|\underline{x}_{10}| = x_{10}$. $Y\sim\ln(1/x)$ is again the rapidity. The BK equation is 
\begin{align}\label{ch2_bk}
    &\frac{\partial}{\partial Y}N(\underline{x}_1,\underline{x}_0, Y) = \frac{\as N_c}{2\pi^2} \int \mathrm{d}^2 x_2 \frac{x_{10}^2}{x_{20}^2x_{21}^2} \bigg[N\left(\underline{x}_1, \underline{x}_2,Y\right) + N\left(\underline{x}_2, \underline{x}_0,Y\right) \\
    &\hspace{6.5cm} - N\left(\underline{x}_1, \underline{x}_0,Y\right) - N\left(\underline{x}_1, \underline{x}_2,Y\right)N\left(\underline{x}_2, \underline{x}_0,Y\right)   \bigg] \notag\,.
\end{align}
The first two terms in the square brackets of \eq{ch2_bk} correspond to the emission of a real gluon from dipole $10$ that generates two new daughter dipoles $21$ and $20$, exactly what is shown in Fig.~\ref{fig:daughterdipoles} (though for these terms evolution only continues in one of the daughter dipoles). The third term is a virtual correction within dipole $10$ so that the original dipole itself continues to evolve. These first three terms can be shown to exactly reproduce BFKL evolution \cite{Kovchegov:2012mbw}. The final term in the square brackets of \eq{ch2_bk} is nonlinear and new. It corresponds to the generation of two daughter dipoles which then both continue to evolve and, eventually, scatter on the target. Crucially, the minus sign in front of the nonlinear term slows down the growth of the cross section at high energies - this is the resolution of the unitarity violation of BFKL evolution.

There exists an all-$N_c$ generalization of BK evolution which can describe the evolution of an arbitrary operator built out of Wilson lines. This is the Jalilian-Marian-Iancu-McLerran-Weigert-Leonidov-Kovner (JIMWLK) evolution \cite{Jalilian-Marian:1997ubg,Jalilian-Marian:1997jhx,Weigert:2000gi,Iancu:2001ad,Iancu:2000hn,Ferreiro:2001qy}. Connecting back to our discussion in Sec.~\ref{sec:wilsonetc}, the relevant $S$-matrix operator for BK evolution --- recall the context of the dipole model --- is the fundamental dipole amplitude in \eq{ch2_unpolarizeddipole} (note that the amplitude $N$ in terms of which BK evolution is written in \eq{ch2_bk} is related to the $S$-matrix by $S= 1-N$). JIMWLK, which can evolve more complicated correlators of Wilson lines, realizes small-$x$ evolution in the context of an effective theory of high energy nuclear interactions called the \textit{color glass condensate} (CGC). A full presentation of the CGC effective theory is beyond the scope of this dissertation but for many more details see the early papers \cite{Iancu:2001ad,Iancu:2000hn}, or a number of review articles \cite{Gribov:1984tu,Iancu:2003xm,Weigert:2005us,JalilianMarian:2005jf,Gelis:2010nm,Albacete:2014fwa,Morreale:2021pnn}, or of course the book \cite{Kovchegov:2012mbw}.

Thus far we have mainly considered unpolarized scattering. But with the tools presented in this Section and with the basic principles of evolution we have established, we are ready to move to the polarized case and discuss the formalism behind small-$x$ helicity evolution. This is what we will do in the next Chapter.

 
\chapter{\texorpdfstring{Helicity at Small $x$}{Helicity at Small x}}
\label{smallxhelicity.ch}

We return now to the proton spin puzzle \cite{EuropeanMuon:1987isl, Jaffe:1989jz, Ji:1996ek, Boer:2011fh, Aidala:2012mv, Accardi:2012qut, Leader:2013jra, Aschenauer:2013woa, Aschenauer:2015eha,  Proceedings:2020eah, Ji:2020ena, AbdulKhalek:2021gbh}. In the interest of understanding the small-$x$ contributions to the proton's spin, we review in this Chapter the formalism with which we can study polarized small-$x$ evolution. This is a framework that allows us to describe the behavior of the helicity parton distribution functions (hPDFs) $\Delta \Sigma(x,Q^2)$ and $\Delta G(x,Q^2)$ along with the $g_1$ structure function at high energies (small-$x$), thereby probing the spin-content of the proton in that same regime. Early work on this subject in the 1990s by Bartels, Ermolaev, and Ryskin (BER) \cite{Bartels:1995iu,Bartels:1996wc} predicted a substantial contribution to the proton spin in the small-$x$ regime. More recently, the $s$-channel/shock wave approach \cite{Mueller:1994rr,Mueller:1994jq,Mueller:1995gb,Balitsky:1995ub,Balitsky:1998ya,Kovchegov:1999yj,Kovchegov:1999ua,Jalilian-Marian:1997dw,Jalilian-Marian:1997gr,Weigert:2000gi,Iancu:2001ad,Iancu:2000hn,Ferreiro:2001qy} has emerged as powerful tool to study the small-$x$ regime of hadronic structure \cite{Kovchegov:2015pbl, Hatta:2016aoc, Kovchegov:2016zex, Kovchegov:2016weo, Kovchegov:2017jxc, Kovchegov:2017lsr, Kovchegov:2018znm, Kovchegov:2019rrz, Boussarie:2019icw, Cougoulic:2019aja, Kovchegov:2020hgb, Cougoulic:2020tbc, Chirilli:2021lif, Kovchegov:2021lvz, Cougoulic:2022gbk,Adamiak:2023okq,Adamiak:2021ppq, Adamiak:2023yhz}. We explored some of the principles underlying this approach in Ch.~\ref{background.ch} and in this Chapter we will develop a more complete picture needed to study spin structure at small-$x$. In reviewing the formalism behind polarized small-$x$ evolution, we will use many results from a breadth of small-$x$ research that has developed over the past several years \cite{Kovchegov:2015pbl,Kovchegov:2018znm,Kovchegov:2017lsr,Kovchegov:2016weo,Kovchegov:2016zex,Kovchegov:2018zeq,Kovchegov:2021iyc,Cougoulic:2022gbk,Chirilli:2018kkw,PhysRevD.104.014019} and in particular we will often follow the comprehensive presentation in \cite{Tawabutr:2022gei}. Also note that, unless otherwise stated, we take our (proton) target to be moving in the light-cone plus direction with momentum $P^\mu = (P^+, 0^-, \underline{0})$ and $P^+$ large, while the projectile moves in the light-cone minus direction with momentum $k^\mu = (0^+, k^-, \underline{0})$ and $k^-$ large. We will work only in the light-cone gauge $A^- = 0$ and we will hold the strong coupling $\as$ fixed.


\section{Beyond the Eikonal Approximation}\label{sec:beyondeikonal}

\subsection{Polarized Wilson Lines}\label{subsec:polwilsonlines}

We discussed in Ch.~\ref{background.ch} the eikonal approximation \cite{Mueller:1994rr,Mueller:1994jq,Mueller:1995gb,Balitsky:1995ub,Balitsky:1998ya,Kovchegov:1999yj,Kovchegov:1999ua,Jalilian-Marian:1997dw,Jalilian-Marian:1997gr,Weigert:2000gi,Iancu:2001ad,Iancu:2000hn,Ferreiro:2001qy}. This is the leading order (in energy) treatment of high energy interactions where dipoles scatter on a target in a way that does not change their transverse separation. The relevant degrees of freedom are the infinite light-cone Wilson lines and the dipole amplitudes constructed from them. Notably, the unpolarized Wilson lines and dipole amplitudes (see Eqs.~\eqref{ch2_wilsonline}, \eqref{ch2_wilsonlineadj}, \eqref{ch2_unpolarizeddipole}, and \eqref{ch2_unpolarizeddipoleadj}) are independent of the target helicity. In order to probe the spin structure of the proton, we need polarization-dependent generalizations of these eikonal degrees of freedom. This leads to corrections at the \textit{sub-eikonal} level, corrections which do couple to helicity and are energy-suppressed relative to the leading eikonal contributions\footnote{The energy suppression of spin-dependent corrections is consistent with the well-known Low-Burnett-Kroll theorem. See \cite{Low:1958sn,Burnett:1967km}, along with \cite{Li:2024fdb} for a recent connection to the modern small-$x$ formalism.} \cite{Altinoluk:2014oxa,Balitsky:2015qba,Balitsky:2016dgz, Kovchegov:2017lsr, Kovchegov:2018znm, Chirilli:2018kkw, Jalilian-Marian:2018iui, Jalilian-Marian:2019kaf, Altinoluk:2020oyd, Boussarie:2020vzf, Boussarie:2020fpb, Kovchegov:2021iyc, Altinoluk:2021lvu, Kovchegov:2022kyy, Altinoluk:2022jkk, Altinoluk:2023qfr,Altinoluk:2023dww, Li:2023tlw}.

\begin{figure}[h!]
    \centering
    \includegraphics[width=0.9\linewidth]{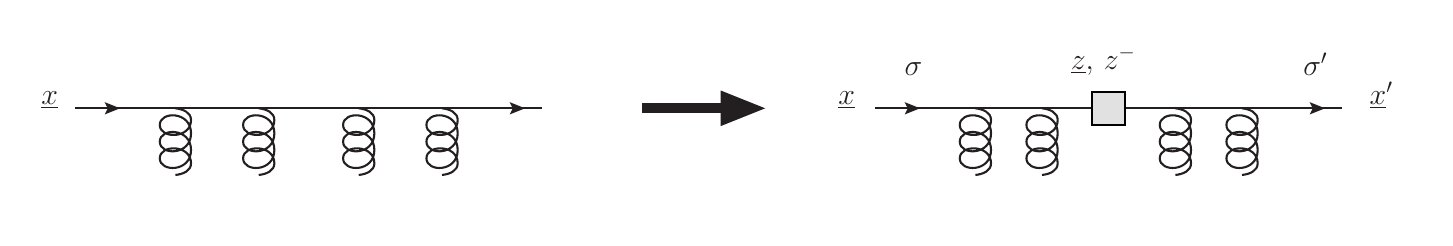}
    \caption{An ordinary Wilson line modified to allow for helicity dependence. The gray box denotes an additional interaction which is helicity-dependent and energy suppressed. This gray box can represent multiple different types of interactions, which we will see in an explicit way shortly.}
    \label{fig:polarizedwilsonline}
\end{figure}

Recall that the eikonal Wilson lines describe an arbitrary number of $t$-channel gluon exchanges with the target. We generalize these eikonal degrees of freedom to the spin-dependent case by allowing for one additional interaction with the target which is helicity-dependent and energy-suppressed. This generalization is shown schematically in Fig.~\ref{fig:polarizedwilsonline}. We note several important features of the helicity-dependent construction in Fig.~\ref{fig:polarizedwilsonline}. First, in the ordinary Wilson line, our quark remains at the same transverse coordinate $\underline{x}$ the whole time, whereas in the polarized case, the transverse coordinate can change from $\underline{x}$ to $\underline{x}'$ after the sub-eikonal interaction (denoted by the gray box). Furthermore, the sub-eikonal interaction may depend on the quark's helicity $\sigma$ (unlike the regular Wilson line where the gluon exchanges carry no polarization information) and we generally allow the quark's helicity to shift from $\sigma$ to $\sigma'$ after the sub-eikonal interaction.

\begin{figure}[h!]
\centering
\begin{subfigure}{\textwidth}
    \centering
    \includegraphics[width=0.7\textwidth]{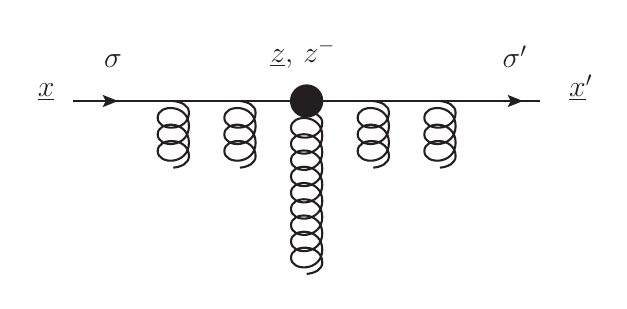}
    \caption{Sub-eikonal gluon exchange}
    \label{fig:subeikgluonexchange}
\end{subfigure}
~
\begin{subfigure}{\textwidth}
    \centering
    \includegraphics[width=0.8\textwidth]{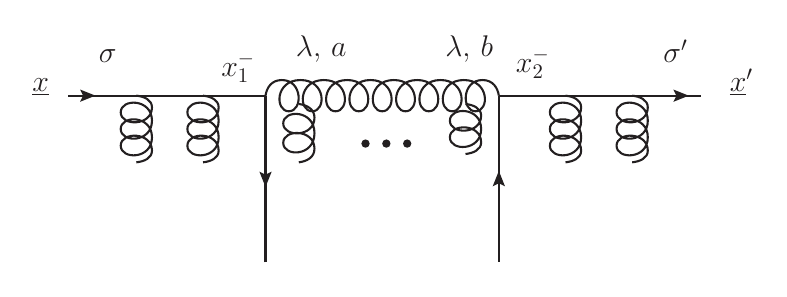}
    \caption{Sub-eikonal quark exchange}
    \label{fig:subeikquarkexchange}
\end{subfigure}
\caption{The additional sub-eikonal interactions needed to construct polarized Wilson lines.}
\label{fig:subeikonalexchanges}
\end{figure}

The eikonal (fundamental) Wilson line at transverse position $\underline{x}$ in the $A^- = 0$ gauge is (cf. \eq{ch2_wilsonline})
\begin{align}\label{ch3_fundamentalwilson}
    V_{\underline{x}}[x_f^-,x_i^-] = \mathcal{P}\, \text{exp}\left[ ig\int\limits_{x_i^-}^{x_f^-} \mathrm{d}x^- A^+\left(0^+,x^-,\underline{x}\right)\right]\,,
\end{align}
again with $A^\mu = \sum_a A^{a\,\mu}t^a$ and $t^a$ the fundamental generators of $\text{SU}(N_c)$.
The polarized generalization of this Wilson line is appropriately called a \textit{polarized Wilson line} and can take one of several forms, depending on whether the extra sub-eikonal interaction involves a gluon exchange or quark exchange \cite{Kovchegov:2018znm,Kovchegov:2021iyc,Kovchegov:2017lsr,Kovchegov:2018zeq,Chirilli:2018kkw,PhysRevD.104.014019,Tawabutr:2022gei}. These possible sub-eikonal exchanges are illustrated diagrammatically in Fig.~\ref{fig:subeikonalexchanges}. 
There the short gluon lines denote the regular eikonal exchanges associated with the ordinary Wilson line segments. We can exchange a sub-eikonal gluon as in Fig.~\ref{fig:subeikgluonexchange}, where the dark circle denotes a sub-eikonal gluon vertex whose expression we will present shortly. Alternatively, we can exchange two quarks in the $t$-channel as in Fig.~\ref{fig:subeikquarkexchange}. Note the intermediate horizontal gluon in Fig.~\ref{fig:subeikquarkexchange} is itself an adjoint Wilson line $U_{\underline{x}}$ containing eikonal gluon exchanges as well:
\begin{align}\label{ch3_adjwilsonline}
    U_{\underline{x}}[x_f^-,x_i^-] = \mathcal{P}\,\text{exp}\left[ig \int\limits_{x_i^-}^{x_f^-}\mathrm{d}x^- \mathcal{A}^+\left(0^+,x^-,\underline{x}\right)  \right]\,,
\end{align}
with $\mathcal{A}^{\mu} = \sum_a A^{a\,\mu}T^a$ and $(T^a)_{bc} = -if^{abc}$ the adjoint generators of $\text{SU}(N_c)$.

The general structures of these polarized Wilson lines are written for gluon and quark exchange, respectively, as \cite{Tawabutr:2022gei}
\begin{subequations}\label{ch3_polwilsonfund}
    \begin{align}
    \label{ch3_polwilsonfundg}
    &V^{G}_{\underline{x}',\underline{x};\sigma',\sigma} = \int\limits_{-\infty}^{\infty}\mathrm{d}z^- \mathrm{d}^2 z V_{\underline{x}'}[\infty,z^-]\delta^2(\underline{x}'-\underline{z})\hat{O}^G_{\sigma',\sigma}(z^-,\underline{z}) \delta^2(\underline{x}-\underline{z}) V_{\underline{x}}[z^-,-\infty]\,,\\
    \label{ch3_polwilsonfundq}
    & V^q_{\underline{x};\sigma',\sigma} = \int\limits_{-\infty}^{\infty}\mathrm{d}x_1^- \int\limits_{x_1^-}^{\infty}\mathrm{d}x_2^- \sum_\lambda \\
    &\hspace{3cm}\times V_{\underline{x}}[\infty,x_2^-]\hat{O}^{\prime\,q}_{\sigma',\lambda;b}(x_2^-,\underline{x})U_{\underline{x}}^{ba}[x_2^-,x_1^-]\hat{O}^q_{\lambda,\sigma;a}(x_1^-,\underline{x})V_{\underline{x}}[x_1^-,-\infty] \,,\notag
\end{align}
\end{subequations}
where in \eq{ch3_polwilsonfundq} $a,b$ are adjoint indices. Note also that $\sigma$ and $\sigma'$ are quark helicities while $\lambda$ is a gluon helicity, as labeled in Fig.~\ref{fig:subeikonalexchanges}. Both of Eqs.~\eqref{ch3_polwilsonfund} have the same general form. Reading right-to-left, we begin with an ordinary fundamental Wilson line starting at light-cone $-\infty$. At some point further along the path, we encounter a new operator structure mediating the sub-eikonal interaction, after which we pick back up with an ordinary fundamental Wilson line that goes off to light-cone $\infty$. In the gluon case of \eq{ch3_polwilsonfundg}, labeled with superscript $G$, the operator $\hat{O}^G_{\sigma',\sigma}(z^-,\underline{z})$ corresponds to a sub-eikonal gluon exchange (the dark circle in Fig.~\ref{fig:subeikgluonexchange}) which is distinct from the other eikonal gluon exchanges that make up the Wilson line since it can shift the propagating quark's transverse position and helicity. In the quark case of \eq{ch3_polwilsonfundq}, we have a sub-eikonal quark exchange mediated by the two quark-exchange operators $\hat{O}^q$ and $\hat{O}^{\prime q}$, with an adjoint Wilson line $U_{\underline{x}}^{ba}[x_2^-,x_1^-]$ running between them. We will see that it is the sub-eikonal gluon exchange that can shift the transverse position of the Wilson line and thus the transverse position of $V^{q}_{\underline{x};\sigma',\sigma}$ in \eq{ch3_polwilsonfundq} remains constant since all its gluon exchanges are eikonal. 

\begin{figure}[h!]
    \centering
    \includegraphics[width=0.9\linewidth]{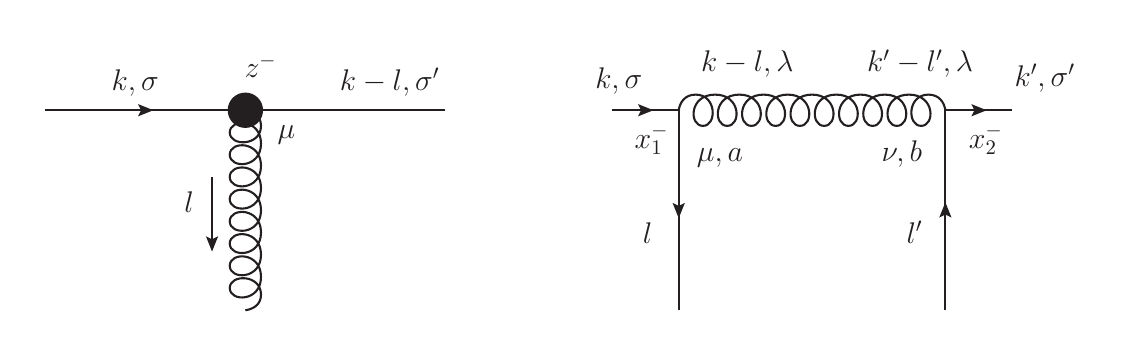}
    \caption{Sub-eikonal operators with momentum labels. Note the horizontal gluon in the right panel is an adjoint Wilson line, but the eikonal gluon exchanges are not shown.}
    \label{fig:subeikoperators}
\end{figure}

With the structure of the polarized Wilson lines established, all that remains is to write down expressions for the sub-eikonal operators $\hat{O}^G$, $\hat{O}^{q}$, and $\hat{O}^{\prime q}$. The operators are shown diagrammatically with momentum labels in Fig.~\ref{fig:subeikoperators}.
We have a quark propagator of the form (for the quark line with momentum~$k$) $i\slashed{k}/(k^2 + i\epsilon)$, where we take the quark's mass to be negligible. Omitting the factor $\slashed{k}$ and anticipating the Fourier transform we will eventually perform in order to write the operators in coordinate space, we can write \cite{Kovchegov:2021iyc,Tawabutr:2022gei}
\begin{align}\label{ch3_fourierpropagator}
    \int \frac{\mathrm{d}k^-}{2\pi} e^{-ik^+(x^--y^-)} \frac{i}{k^2 + i\epsilon} = \frac{1}{2k^-} e^{-i \frac{k_{\perp}^2}{2k^-}(x^- - y^-)} \approx \frac{1}{2k^-}\left[1 - i\frac{k_\perp^2}{2k^-}(x^- - y^-)  \right] 
\end{align}
where the last expression follows from expanding the exponential and dropping terms suppressed by further powers of the large $k^-$. Since we ultimately have two quark propagators (one before the sub-eikonal vertex and one after) we split the pre-factor $1/2k^-$ between them, assigning $1/\sqrt{2p^-}$ for each quark propagator with momentum $p$. With this rule and also assigning a gluon propagator $A^\mu(x^-,\underline{x})$ for the background gluon field exchanged with the target (and recalling we work in the $A^- = 0$ gauge), the sub-eikonal gluon exchange vertex in the left panel of Fig.~\ref{fig:subeikoperators} can be written (in momentum space) as \cite{Tawabutr:2022gei}
\begin{align}\label{ch3_gluonexchangeop}
    &\hat{O}^{G}_{\sigma',\sigma}(k,l) = \frac{1}{2\sqrt{k^-(k^--l^-)}} igA_{\mu}(l)\left[\overline{u}_{\sigma'}(k-l)\gamma^\mu u_\sigma(k)\right] + ...\\
    & \hspace{.5cm} \approx \delta_{\sigma\sigma'} ig A^+(l) + \sigma \delta_{\sigma\sigma'} \frac{g}{2k^-}\epsilon^{ij}\underline{A}^i(l)\underline{l}^j - \delta_{\sigma\sigma'}\frac{ig}{2k^-}\left[\underline{A}(l)\cdot\underline{k} + \underline{A}(l) \cdot \left(\underline{k}-\underline{l}\right) \right] + ...\notag\,,
\end{align}
where the second line is obtained in the approximation $l^- \ll k^-$ and the $...$ denotes the contribution from the second term in the square brackets of \eq{ch3_fourierpropagator} which we do not write out explicitly. Also note we use the $(\pm)$-interchanged Brodsky-Lepage spinors \cite{Lepage:1980fj}
\begin{subequations}\label{ch3_blspinors}
    \begin{align}
    \label{ch3_spinoru}
    &u_\sigma (p) = \frac{1}{\sqrt{\sqrt{2}p^-}}\left[\sqrt{2}p^- + m\gamma^0 + \gamma^0\underline{\gamma}\cdot \underline{p} \right] \rho(\sigma) \,\\
    \label{ch3_spinorv}
    &v_\sigma (p) = \frac{1}{\sqrt{\sqrt{2}p^-}}\left[\sqrt{2}p^- - m\gamma^0 + \gamma^0\underline{\gamma}\cdot \underline{p} \right] \rho(-\sigma)\,,
\end{align}
\end{subequations}
with
\begin{align}\label{ch3_rhospinors}
    \rho(+1) = \frac{1}{\sqrt{2}}
    \begin{pmatrix}
        1\\
        0\\
        -1\\
        0
    \end{pmatrix} \, \quad
    \rho(-1) = \frac{1}{\sqrt{2}}
    \begin{pmatrix}
        0\\
        1\\
        0\\
        1
    \end{pmatrix}\,.
\end{align}

Next we perform a Fourier transform. The first term of the last line of \eq{ch3_gluonexchangeop} yields exactly the eikonal contribution one would get by expanding the regular Wilson line to first order in the coupling. In the second term of the last line of \eq{ch3_gluonexchangeop}, we will obtain $\partial_1 A_2(z^-,\underline{z}) - \partial_2 A_1(z^-,\underline{z})$, which we can replace with $F_{12}(z^-,\underline{z})$ to sub-eikonal accuracy, neglecting the term nonlinear in the gluon field. In the third term of the second line of \eq{ch3_gluonexchangeop}, we will obtain $\underline{A}^i(z^-,\underline{z})\vec{\partial}^i_{\underline{z}} - \cev{\partial}^i_{\underline{z}}\underline{A}^i(z^-,\underline{z})$. One can show that we can instead use the combination $\underline{\cev{D}}^i \underline{\vec{D}}^i$, with the right- and left-acting covariant derivatives $\underline{\vec{D}}^i = \vec{\partial}^i -ig\underline{A}^i$ and $\underline{\cev{D}}^i = \cev{{\partial}}^i +ig\underline{A}^i$, to obtain this term from the gluon-exchange polarized Wilson line in \eq{ch3_polwilsonfundg}. In fact, employing the operator $\underline{\cev{D}}^i \underline{\vec{D}}^i$ in \eq{ch3_polwilsonfundg} yields three terms: the first has two derivatives and reproduces the second term in the square brackets of the propagator in \eq{ch3_fourierpropagator}; the second is the cross term and reproduces the $\underline{A}^i(z^-,\underline{z})\vec{\partial}^i_{\underline{z}} - \cev{\partial}^i_{\underline{z}}\underline{A}^i(z^-,\underline{z})$ above; the third is quadratic in the gluon field and can be dropped as sub-sub-eikonal. Furthermore, employing the $\underline{\cev{D}}^i \underline{\vec{D}}^i$ restores gauge covariance. Ultimately, we can write the sub-eikonal gluon exchange polarized Wilson line from \eq{ch3_polwilsonfundg} as \cite{Tawabutr:2022gei,Cougoulic:2022gbk}
\begin{align}\label{ch3_polwilsonfundg2}
    V^{G}_{\underline{x}',\underline{x};\sigma',\sigma} = V_{\underline{x}}\delta^2\left(\underline{x}'-\underline{x}\right)\delta_{\sigma\sigma'} + V^{G[1]}_{\underline{x}}\delta^2\left(\underline{x}' - \underline{x}\right) \sigma \delta_{\sigma\sigma'} + V^{G[2]}_{\underline{x}',\underline{x}}\delta_{\sigma\sigma'}\,,
\end{align}
where 
\begin{subequations}\label{ch3_pollineglue}
\begin{align}
    \label{ch3_vg1}
    &V^{G[1]}_{\underline{x}} = \frac{igP^+}{s}\int\limits_{-\infty}^\infty \mathrm{d}z^- V_{\underline{x}}\left[\infty,z^-\right]F^{12}\left(z^-,\underline{x}\right) V_{\underline{x}}\left[z^-,-\infty\right] \\
    &\text{and} \notag \\
    \label{ch3_vg2}
    &V^{G[2]}_{\underline{x}',\underline{x}} = -\frac{iP^+}{s} \int\limits_{-\infty}^\infty \mathrm{d}z^- \mathrm{d}^2 z V_{\underline{x}'}\left[\infty,z^-\right]\delta^2\left(\underline{x}'-\underline{z}\right)\underline{\cev{D}}^i\left(z^-,\underline{z}\right)\underline{\vec{D}}^i\left(z^-,\underline{z}\right) \\
    &\hspace{9cm}\times\delta^2\left(\underline{x}-\underline{z}\right) V_{\underline{x}}\left[z^-,-\infty\right] \notag
\end{align}
\end{subequations}
are known as the polarized gluon-exchange Wilson lines of type-1 and type-2. Note also that we employed the center of mass energy squared between the quark and target $s = 2P^+k^-$. 

Next, turning to the sub-eikonal quark exchange in the right panel of Fig.~\ref{fig:subeikoperators}, we can write the corresponding operators in momentum space as \cite{Tawabutr:2022gei}
\begin{subequations}\label{ch3_quarkops}
\begin{align}
    &\hat{O}^q_{\lambda,\sigma;a}(k,l) = \frac{1}{2\sqrt{k^-\left(k^--l^-\right)}} igt^a \epsilon^{\mu\,*}_\lambda(k-l)\bar{\psi}(l)\gamma_\mu u_\sigma(k) \,,\\
    &\hat{O}^{\prime,q}_{\sigma',\lambda;b}(k,l') = \frac{1}{2\sqrt{k^{\prime -}\left(k^{\prime -} - l^{\prime -}\right)}} igt^b \epsilon^\nu_\lambda(k'-l') \overline{u}_{\sigma'}(k')\gamma_\nu\psi(l')\,,
\end{align}
\end{subequations}
where the polarization vector for a gluon of polarization $\lambda$ and momentum $k$ is $\epsilon_\lambda^\mu(k) = ({\un \epsilon}_\lambda \cdot {\un k}/k^-, 0, {\un \epsilon}_\lambda)$ with ${\un \epsilon}_\lambda = -(1/\sqrt{2}) (\lambda, i)$ \cite{Lepage:1980fj}. Note also that the quark fields exchanged with the target (the vertical quark lines in the right panel of Fig.~\ref{fig:subeikoperators}) are the background fields $\psi$ and $\bar{\psi}$ rather than the projectile quark field to which we assign a $u$-spinor. 

With $k^- \approx k^{\prime -}$, $l^-,l^{\prime -} \ll k^-$, and with transverse momenta negligible, a direct calculation of the operators in Eqs.~\eqref{ch3_quarkops} Fourier transformed and substituted into the quark-exchange polarized Wilson line of \eq{ch3_polwilsonfundq} yields \cite{Tawabutr:2022gei,Cougoulic:2022gbk} 
\begin{align}\label{ch3_polwilsonfundq2}
    V^{q}_{\underline{x};\sigma',\sigma} = V^{q[1]}_{\underline{x}}\sigma \delta_{\sigma\sigma'} + V^{q[2]}_{\underline{x}}\delta_{\sigma\sigma'}\,,
\end{align}
where 
\begin{subequations}\label{ch3_pollinequark}
\begin{align}
    \label{ch3_vq1}
    &V^{q[1]}_{\underline{x}} = \frac{g^2 P^+}{2s}\int\limits_{-\infty}^\infty \mathrm{d}x_1^- \int\limits_{x_1^-}^\infty \mathrm{d}x_2^- V_{\underline{x}}\left[\infty,x_2^-\right]t^b \psi_\beta\left(x_2^-,\underline{x}\right) U^{ba}_{\underline{x}}\left[x_2^-,x_1^-\right]\left(\gamma^+\gamma_5\right)_{\alpha\beta}\\
    &\hspace{9cm}\times\bar{\psi}_\alpha\left(x_1^-,\underline{x}\right) t^a V_{\underline{x}}\left[x_1^-,-\infty\right] \notag\\
    \label{ch3_vq2}
    &V^{q[2]}_{\underline{x}} = -\frac{g^2P^+}{2s}\int\limits_{-\infty}^\infty \mathrm{d}x_1^- \int\limits_{x_1^-}^\infty \mathrm{d}x_2^- V_{\underline{x}}\left[\infty,x_2^-\right]t^b \psi_\beta\left(x_2^-,\underline{x}\right) U^{ba}_{\underline{x}}\left[x_2^-,x_1^-\right]\left(\gamma^+\right)_{\alpha\beta} \\
    &\hspace{9cm}\times \bar{\psi}_{\alpha}\left(x_1^-,\underline{x}\right) t^a V_{\underline{x}}\left[x_1^-,-\infty\right]\notag
\end{align}
\end{subequations}
are the polarized quark-exchange Wilson lines of type-1 and type-2. Note that $\alpha,\beta$ here are the spinor indices. Also note the difference in that \eq{ch3_vq1} contains the quark axial current $\overline{\psi}\gamma^+\gamma_5\psi$ whereas \eq{ch3_vq2} contains the vector current $\overline{\psi}\gamma^+\psi$ (with the latter ultimately not contributing to the small-$x$ helicity evolution).

At this point we have the full S-matrix for high-energy quark scattering, up to the sub-eikonal level. We write \cite{Cougoulic:2022gbk}
\begin{align}\label{fullquarksmatrix}
    V_{\underline{x},\underline{y};\sigma',\sigma} = V_{\underline{x}}\delta^2\left(\underline{x}-\underline{y}\right)\delta_{\sigma\sigma'} + \sigma \delta_{\sigma\sigma'}V^{\text{pol}[1]}_{\underline{x}}\delta^2\left(\underline{x}-\underline{y}\right) + \delta_{\sigma\sigma'}V^{\text{pol}[2]}_{\underline{x},\underline{y}}\,
\end{align}
where the second and third term contain the sub-eikonal contributions, grouped by their helicity structures ($\sigma\delta_{\sigma\sigma'}$ vs $\delta_{\sigma\sigma'}$):
\begin{subequations}\label{pollinestype12}
\begin{align}
    \label{pollinestype1}
    &V^{\text{pol}[1]}_{\underline{x}} = V^{G[1]}_{\underline{x}} + V^{q[1]}_{\underline{x}}\,,\\
    \label{pollinestype2}
    &V^{\text{pol}[2]}_{\underline{x},\underline{y}} = V^{G[2]}_{\underline{x},\underline{y}} + V^{q[2]}_{\underline{x}}\delta^2\left(\underline{x}-\underline{y}\right)\,
\end{align}
\end{subequations}
with $V^{G[1]}_{\underline{x}}$, $V^{G[2]}_{\underline{x},\underline{y}}$, $V^{q[1]}_{\underline{x}}$, $V^{q[2]}_{\underline{x}}$ as written in Eqs.~\eqref{ch3_pollineglue} and \eqref{ch3_pollinequark}.

\begin{figure}[h!]
\centering
\begin{subfigure}{\textwidth}
    \centering
    \includegraphics[width=0.6\textwidth]{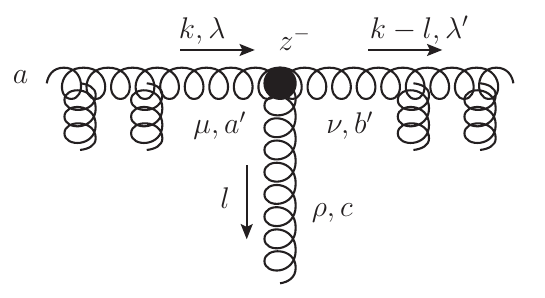}
    \caption{Sub-eikonal gluon exchange from an adjoint Wilson line.}
    \label{fig:polwilsonadjg}
\end{subfigure}
~
\begin{subfigure}{\textwidth}
    \centering
    \includegraphics[width=0.9\textwidth]{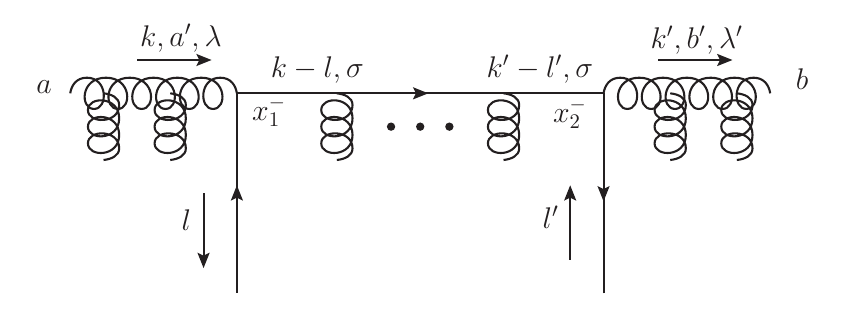}
    \caption{Sub-eikonal quark exchange from an adjoint Wilson line. Note there is another contribution where particle number on the quark lines flows in the opposite direction.}
    \label{fig:polwilsonadjq}
\end{subfigure}
\caption{The interactions making up the polarized adjoint Wilson lines.}
\label{fig:polarizedwilsonlineadj}
\end{figure}

Next we need to generalize the eikonal adjoint Wilson line to the polarized case. The generalization is very similar to the case of the polarized fundamental Wilson line. The polarized adjoint lines are shown in Fig.~\ref{fig:polarizedwilsonlineadj}, illustrated with the sub-eikonal gluon and quark exchanges from which they are built. 
Calculations similar to those in the fundamental case (again we take $l^-,l^{\prime -} \ll k^- \sim k^{\prime -}$, and also employ the target-projectile center of mass energy squared $s=2P^+k^-$) yield an $S$-matrix for high-energy gluon scattering, up to the sub-eikonal level of
\cite{Kovchegov:2018znm,Kovchegov:2015pbl,Kovchegov:2017lsr,Chirilli:2018kkw,Cougoulic:2022gbk,Tawabutr:2022gei}:
\begin{align}\label{ch3_gluonsmatrix}
    \left(U_{\underline{x},\underline{y};\lambda',\lambda}\right)^{ba} = \left(U_{\underline{x}}\right)^{ba}\delta^2\left(\underline{x}-\underline{y}\right)\delta_{\lambda\lambda'} + \lambda\delta_{\lambda\lambda'}U^{\text{pol}[1]}_{\underline{x}}\delta^2\left(\underline{x}-\underline{y}\right) + \delta_{\lambda\lambda'} U^{\text{pol}[2]}_{\underline{x},\underline{y}}\,,
\end{align}
where we again separate the sub-eikonal contributions by their helicity structures:
\begin{subequations}\label{ch3_poladjtype12}
\begin{align}
    \label{ch3_poladjtype1}
    &U^{\text{pol}[1]}_{\underline{x}} = U^{G[1]}_{\underline{x}} + U^{q[1]}_{\underline{x}}\,,\\
    \label{ch3_poladjtype2}
    &U^{\text{pol}[2]}_{\underline{x},\underline{y}} = U^{G[2]}_{\underline{x},\underline{y}} + U^{q[2]}_{\underline{x}}\delta^2\left(\underline{x}- \underline{y}\right)\,,
\end{align}
\end{subequations}
with
\begin{subequations}\label{ch3_pollinesadjall}\allowdisplaybreaks
\begin{align}
\label{ug1}
& (U_{\underline{x}}^{\textrm{G} [1]})^{ba} = \frac{2i g P^+}{s} \int\limits_{-\infty}^{\infty} \mathrm{d}{x}^- (U_{\underline{x}} [ \infty, x^-])^{bb'} \, ({\cal F}^{12})^{b'a'} (x^-, {\underline{x}}) \, (U_{\underline{x}} [ x^-, -\infty])^{a'a} \,, \\
\label{uq1}
& (U_{\underline{x}}^{\textrm{q} [1]})^{ba} = \frac{g^2 P^+}{2 \, s} \!\! \int\limits_{-\infty}^{\infty} \!\! \mathrm{d}{x}_1^- \! \int\limits_{x_1^-}^\infty \mathrm{d} x_2^- (U_{\underline{x}} [ \infty, x_2^-])^{bb'} \bar{\psi} (x_2^-,\underline{x}) \, t^{b'} V_{\underline{x}} [x_2^-,x_1^-] \, \gamma^+ \gamma^5 \, t^{a'} \\
&\hspace{5cm}\times\psi (x_1^-,\underline{x})  (U_{\underline{x}} [ x_1^-, -\infty])^{a'a} + \mbox{c.c.}  \,, \notag  \\
\label{ug2}
& (U_{{\underline{x}}, {\underline{y}}}^{\textrm{G} [2]})^{ba}  = - \frac{i \, P^+}{s} \int\limits_{-\infty}^{\infty} \mathrm{d}{z}^- \mathrm{d}^2 z \ (U_{\underline{x}} [ \infty, z^-])^{bb'} \, \delta^2 (\underline{x} - \underline{z}) \\
&\hspace{5cm}\times\cev{\underline{\mathscr{D}}}^{b'c} (z^-, {\underline{z}}) \, \cdot \, \vec{\underline{\mathscr{D}}}^{ca'}  (z^-, {\underline{z}}) (U_{\underline{y}} [ z^-, -\infty])^{a'a} \, \delta^2 (\underline{y} - \underline{z}) \,, \notag \\
\label{uq2}
& (U_{{\underline{x}}}^{\textrm{q} [2]} )^{ba} = - \frac{g^2 P^+}{2 \, s} \int\limits_{-\infty}^{\infty} \!\! \mathrm{d}{x}_1^- \! \int\limits_{x_1^-}^\infty \mathrm{d} x_2^- (U_{\underline{x}} [ \infty, x_2^-])^{bb'} \, \bar{\psi} (x_2^-,\underline{x}) \, t^{b'} \, V_{\un{x}} [x_2^-,x_1^-] \, \gamma^+ \, t^{a'} \\
&\hspace{5cm} \times\psi (x_1^-,\underline{x}) \,  (U_{\underline{x}} [ x_1^-, -\infty])^{a'a} - \mbox{c.c.} \,, \notag
\end{align}
\end{subequations}
and where ${\cal F}^{12} = \sum_a F^{a\,12}T^a$ is the adjoint field strength tensor and $\vec{\mathscr{D}}_i^{ab} = \vec{\partial}_i \delta^{ab} - ig(T^c)_{ab}A^c_i $ and $\cev{\mathscr{D}}_i^{ab} = \cev{\partial}_i \delta^{ab} + ig(T^c)_{ab}A^c_i $ are the right- and left-acting adjoint covariant derivatives.


\subsection{\texorpdfstring{Polarized Dipole Amplitudes, Helicity Distributions, and the $g_1$ Structure Function}{Polarized Dipole Amplitudes, Helicity Distributions, and the g1 Structure Function}}

\begin{figure}[h!]
    \centering
    \includegraphics[width=0.9\linewidth]{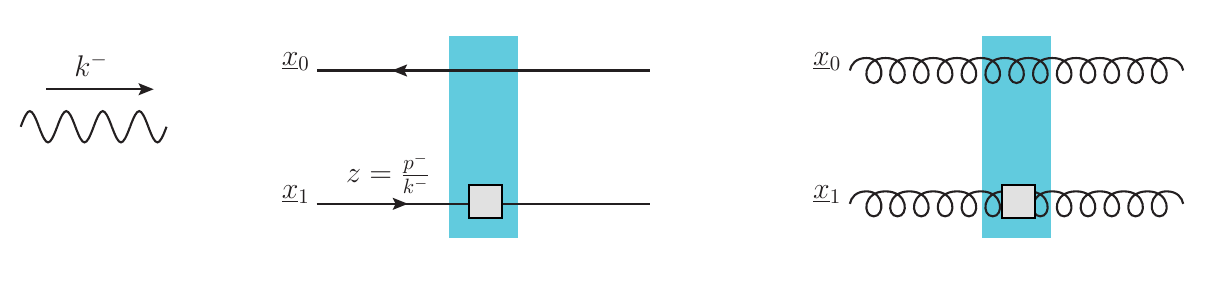}
    \caption{Schematic illustration of polarized dipole amplitudes. The gray boxes denote one of the sub-eikonal helicity-dependent interactions from Ch.~\ref{subsec:polwilsonlines}. We imagine the original projectile carries minus light-cone momentum $k^-$ and the quark line carries momentum $p^-$ so that its longitudinal momentum fraction is $z=p^-/k^-$. The other line in each diagram is an unpolarized (eikonal) Wilson line and the blue rectangle is the target shock-wave.}
    \label{fig:polarizeddipoles}
\end{figure}

In the shockwave and dipole pictures, we use the polarized Wilson lines of the previous Section to generalize the unpolarized dipole amplitudes from Ch.~\ref{sec:wilsonetc} (see Eqs.~\eqref{ch2_unpolarizeddipole} and \eqref{ch2_unpolarizeddipoleadj} along with Fig.~\ref{fig:dipoleamplitudes}). These new polarized generalizations of the dipole amplitudes are called, unsurprisingly, polarized dipole amplitudes. Instead of two eikonal Wilson lines propagating through the shock wave, we replace one of the eikonal Wilson lines with a polarized Wilson line of the type constructed in Ch.~\ref{subsec:polwilsonlines}. These new structures are illustrated in Fig.~\ref{fig:polarizeddipoles}. Note that the inclusion of one polarized Wilson line results in a (sub-eikonal) suppression by one power of energy. Including additional polarized Wilson lines would result in further suppression in additional powers of energy, which are beyond the sub-eikonal precision of our small-$x$ helicity calculations. 

Because we have multiple types of sub-eikonal interactions and therefore multiple types of polarized Wilson lines, we also end up with multiple types of polarized dipole amplitudes. For the fundamental dipoles we define the first type using the polarized Wilson line $V^{\text{pol}[1]}_{\underline{x}}$ from \eq{pollinestype1} \cite{Cougoulic:2022gbk,Kovchegov:2018znm,Kovchegov:2015pbl,Kovchegov:2016zex,Tawabutr:2022gei}
\begin{align}\label{ch3_Q10}
    Q_{10}(zs) &= \frac{zs}{2 N_c} \, \text{Re} \, \bigg\langle \text{T} \, \text{tr} \left[ V_{\underline{0}} \,  V_{\underline{1} }^{\textrm{pol} [1] \,\dagger} \right] + \text{T} \,  \text{tr} \left[ V_{\underline{1}}^{\textrm{pol} [1]} \, V_{\underline{0}}^\dagger \right]   \bigg\rangle (z) \\
    &=\frac{1}{2 N_c} \, \text{Re} \, \llangle \text{T} \, \text{tr} \left[ V_{\underline{0}} \,  V_{\underline{1} }^{\textrm{pol} [1] \,\dagger} \right] + \text{T} \,  \text{tr} \left[ V_{\underline{1}}^{\textrm{pol} [1]} \, V_{\underline{0}}^\dagger \right]   \rrangle (zs) \notag\,
\end{align}
where $z$ is the smallest longitudinal minus momentum fraction of the lines making up the dipole and $s$ is the center of mass energy squared between the original projectile and the target proton. The single angle brackets in the first line denote a helicity-dependent averaging in the target proton state and the double angle brackets constitute a rescaling to reflect the explicit energy suppression of the polarized dipole amplitudes relative to the unpolarized ones. Explicitly,
\begin{align}\label{ch3_doubleanglesdef}
    \llangle \,...\, \rrangle = s\Big\langle \,...\, \Big\rangle\,.
\end{align}
Again, our notation for the Wilson lines is that $V_{\underline{x}_0}[\infty,-\infty] \equiv V_{\underline{0}}$ is an infinite Wilson line at transverse position $\underline{x}_0$. 

For our purposes it is also convenient to integrate the dipole amplitudes over the impact parameter, defining \cite{Cougoulic:2022gbk,Kovchegov:2018znm,Kovchegov:2015pbl,Kovchegov:2016zex,Tawabutr:2022gei}
\begin{align}\label{ch3_Q}
    Q(x_{10}^2,zs) = \int \mathrm{d}^2 \left(\frac{x_1 + x_0}{2} \right)Q_{10}(zs)\,,
\end{align}
where as usual the transverse separation is denoted $\underline{x}_{10} \equiv \underline{x}_1 - \underline{x}_0$ with magnitude $x_{10}$.

For the other polarized fundamental Wilson line, $V_{\underline{x},\underline{y}}^{\text{pol}[2]}$ from \eq{pollinestype2}, we define an additional polarized line $V_{\underline{x}}^{iG[2]}$, which is related to the gluon exchange $V^{G[2]}_{\underline{x},\underline{y}}$ but omits the contribution of $V^{q[2]}_{\underline{x},\underline{y}}$ (the latter does not contribute to the helicity evolution in this work) \cite{Cougoulic:2022gbk}. The new polarized Wilson line is ($i$ is the transverse index)
\begin{align}\label{ch3_ViG2}
    V_{\underline{x}}^{i \, G [2]} =  \frac{P^+}{2 s} \, \int\limits_{-\infty}^{\infty} \mathrm{d} {z}^- \, V_{\underline{x}} [ \infty, z^-] \, \left[ \vec{D}^i (z^-, \underline{x}) - \cev{D}^i (z^-, \underline{x}) \right] \, V_{\underline{x}} [ z^-, -\infty] \,,
\end{align}
which we use to construct the polarized dipole amplitude
\begin{align}\label{ch3_Gi10}
    G^{i}_{10}(zs) = \frac{1}{2N_c}\,\text{Re}\,\llangle \text{tr}\left[V^{i\, G[2] \dagger}_{\underline{1}} V_{\underline{0}} \right] + \text{tr}\left[V_{\underline{0}}^{\dagger}V^{i\,G[2]}_{\underline{1}} \right]\rrangle(zs)\,.
\end{align}
Again we integrate over impact parameter, this time also decomposing the result into its possible tensor structures in the transverse separation $\underline{x}_{10}$:
\begin{align}\label{ch3_G10idecomp}
    \int \mathrm{d}^2\left(\frac{x_1 + x_0}{2} \right)\,G^i_{10}(zs) = \underline{x}_{10}^i G_1\left(\xoz^2,zs\right) + \epsilon^{ij}\underline{x}_{10}^j \,G_2\left(\xoz^2,zs\right)\,.
\end{align}
Ultimately the integrated dipole amplitude $G_1(\xoz^2,zs)$ will not contribute to any helicity distribution and we can simply use the term with $G_2(\xoz^2,zs)$ instead of the full impact-parameter-integrated decomposition in \eq{ch3_G10idecomp} \cite{Kovchegov:2017lsr,Cougoulic:2022gbk}. For concreteness we can also write \cite{Cougoulic:2022gbk}
\begin{align}\label{ch3_G2explicit}
    G_2\left(\xoz^2,zs\right) = \int \mathrm{d}^2\left(\frac{x_1 + x_0}{2}\right) \frac{\epsilon^{ij}\underline{x}_{10}^j}{x_{10}^2} G^i_{10}(zs)\,.
\end{align}

In the adjoint sector we can make similar definitions \cite{Kovchegov:2018znm,Kovchegov:2021lvz,Cougoulic:2022gbk}:
\begin{subequations}\label{ch3_adjamplitudes}
\begin{align}
    \label{ch3_G10adj}
    &G^{\text{adj}}_{10} (zs) \equiv \frac{1}{2 (N_c^2 -1)} \, \text{Re} \, \llangle \text{T} \, \text{Tr} \left[ U_{\underline{0}} \, U_{{\underline{1}}}^{\text{pol}\, [1] \, \dagger} \right] + \text{T} \, \text{Tr} \left[ U_{{\underline{1}}}^{\text{pol}\, [1]} \, U_{\underline{0}}^\dagger \right] \rrangle (zs)\,,\\
    \label{ch3_G10iadj}
    & G^{i\,\text{adj}}_{10}(zs) = \frac{1}{2\left(N_c^2-1\right)} \text{Re}\,\llangle \text{T}\,\text{Tr}\,\left[U_{\underline{0}}U^{i\,G[2]\dagger}_{\underline{1}} \right] + \text{T}\,\text{Tr}\,\left[U^{i\,G[2]}_{\underline{1}}U^{\dagger}_{\underline{0}} \right]\rrangle (zs)\,,
\end{align}
\end{subequations}
where here $\text{Tr}$ denotes a trace over adjoint indices and where $U_{\underline{x}}^{i\,G[2]}$ is the adjoint generalization of \eq{ch3_ViG2}:
\begin{align}\label{ch3_UiG2}
    U_{\underline{x}}^{i\,G[2]} = \frac{P^+}{2s}\int\limits_{-\infty}^{\infty}\mathrm{d}z^- U_{\underline{x}}\left[\infty,z^-\right]\left[\vec{\mathscr{D}}^i\left(z^-,\underline{x}\right) - \cev{\mathscr{D}}^i\left(z^-,\underline{x}\right)  \right]U_{\underline{x}}\left[z^-,-\infty\right]\,.
\end{align}
Ultimately we will rewrite the polarized adjoint amplitudes in Eqs.~\eqref{ch3_adjamplitudes} in terms of other amplitudes (to be defined shortly) and so we will not need impact-parameter-integrated versions like we had for the fundamental amplitudes in Eqs.~\eqref{ch3_Q} and \eqref{ch3_G2explicit}.

The real utility of the polarized amplitudes defined in this Section is the fact that they can be used to express the quark and gluon helicity PDFs, along with the related transverse-momentum-dependent PDFs (TMDs) and the $g_1$ structure function at small-$x$. For the helicity PDFs, we have \cite{Kovchegov:2017lsr,Cougoulic:2022gbk}
\begin{subequations}\label{ch3_helicitypdfs}
\begin{align}
    \label{ch3_DeltaG}
    & \Delta G(x,Q^2) = \frac{2N_c}{\as \pi^2} \left[\left(1+\xoz^2\frac{\partial}{\partial \xoz^2}\right)G_2\left(\xoz^2,zs=\frac{Q^2}{x}\right) \right]_{\xoz^2 = \frac{1}{Q^2}}\,,\\
    \label{ch3_DeltaSigma}
    &\Delta\Sigma(x,Q^2) = - \frac{N_cN_f}{2\pi^3}\int\limits_{\Lambda^2/s}^1\frac{\mathrm{d}z}{z}\int\limits_{\tfrac{1}{zs}}^{\text{min}\{\tfrac{1}{zQ^2},\tfrac{1}{\Lambda^2}\}} \frac{\mathrm{d}\xoz^2}{\xoz^2}\left[Q\left(\xoz^2,zs\right) + 2G_2\left(\xoz^2,zs\right) \right]\,,
\end{align}
\end{subequations}
where for our purposes $\Lambda$ is an explicit infrared cutoff. Note that $\Delta\Sigma$ is more correctly given by a sum over flavors of flavor-dependent dipole amplitudes, but here we assume all flavors contribute equally and thus we obtain the overall factor $N_f$. Also note that we will find a slight correction to \eq{ch3_DeltaSigma} in Ch.~\ref{transitionops.ch}. 

The TMDs are generalizations of the PDFs. Whereas the PDFs count a number of partons at a given momentum fraction $x$ (in the helicity case this is the number of spin-aligned partons minus the number of spin-anti-aligned partons relative to the parent hadron), the TMDs count a number of partons at a particular $x$ and at a particular transverse momentum $\underline{k}$. Naturally, then, the TMDs are functions of both $x$ and the transverse momentum $\underline{k}$.\footnote{The TMDs are not the main focus of this dissertation, but see \cite{Collins:2011zzd} for a more comprehensive introduction and \cite{Accardi:2012qut} for a survey of the role of TMDs in modern nuclear physics.} The dipole gluon helicity TMD $g_{1L}^{G\,dip}(x,k_T^2)$ and the flavor-singlet quark helicity TMD $g^S_{1L}(x,k_T^2)$ are given in terms of the polarized dipole amplitudes $Q(\xoz^2,zs)$ and $G_2(\xoz^2,zs)$ as \cite{Kovchegov:2018znm,Kovchegov:2018zeq,Kovchegov:2017lsr,Cougoulic:2022gbk}
\begin{subequations}\label{ch3_TMDs}
\begin{align}
    \label{ch3_gluon_TMD}
    & g^{G\,dip}_{1L}(x,k_T^2) = \frac{N_c}{\as 2\pi^4}\int \mathrm{d}^2 \xoz\, e^{-i\underline{k}\cdot\underline{x}_{10}} \left[1+\xoz^2\frac{\partial}{\partial\xoz^2}\right]G_2\left(\xoz^2,zs=\frac{Q^2}{x}\right)\,,\\
    \label{ch3_quark_TMD}
    & g^{S}_{1L}(x,k_T^2) = \frac{8iN_cN_f}{(2\pi)^5}\int\limits_{\Lambda^2/s}^{1}\frac{\mathrm{d}z}{z}\int \mathrm{d}^2 \xoz\, e^{i\underline{k}\cdot\underline{x}_{10}} \, \frac{\underline{x}_{10}}{\xoz^2}\cdot \frac{\underline{k}}{\underline{k}^2}\left[Q(\xoz^2,zs) + 2 \, G_2(\xoz^2,zs)\right]\,.
\end{align}
\end{subequations}
Finally the $g_1$ structure function is given by \cite{Cougoulic:2022gbk}
\begin{align}\label{ch3_g1}
    g_1(x,Q^2) = -\sum_{f}\frac{N_c Z_f^2}{4\pi^3}\int\limits_{\Lambda^2/s}^{1}\frac{\mathrm{d}z}{z} \int\limits_{\tfrac{1}{zs}}^{\min\left\{\tfrac{1}{zQ^2}, \tfrac{1}{\Lambda^2} \right\}} \frac{\mathrm{d}\xoz^2}{\xoz^2}\left[Q(\xoz^2,zs) + 2 \, G_2(\xoz^2,zs) \right].
\end{align}
As in \eq{ch3_DeltaSigma}, the integrand of \eq{ch3_g1} should in principle be distinct for each flavor $f$, but for simplicity we again assume each flavor comes with an equal contribution.

We conclude this Subsection with a brief note regarding the $g_1$ structure function. As discussed in Ch.~\ref{intro.ch}, the EMC measurements in the 1980s were initially thought to imply a tiny contribution to the proton spin from the spin of the quarks. However, what these experiments measured was not actually that quark spin content itself, but rather the first $x$-moment of the $g_1$ structure function, that is $\int_0^1 \mathrm{d}x\, g_1(x,Q^2)$. This is not a pure quark contribution as there is also a gluonic contribution to the $g_1$ structure function that was discovered after the EMC results \cite{ALTARELLI1988391,CARLITZ1988229}. Measuring the first moment of $g_1$ does not probe only the quark spin $S_q$ but rather the combination $S_q - \frac{\as N_f}{2\pi}S_G$, and so the EMC results did not really imply small net quark spin but rather a potentially significant cancellation between net quark and gluon spins \cite{Anselmino:1994gn}. And indeed we can see this gluonic contribution to the $g_1$ structure function here in our polarized dipole framework. Note the appearance of the polarized dipole amplitude $G_2(\xoz^2,zs)$ in the expression for the $g_1$ structure function in \eq{ch3_g1}. Per \eq{ch3_DeltaG}, we see that $G_2(\xoz^2,zs)$ is directly proportional to the gluon helicity distribution $\Delta G$, and so indeed we have a purely gluonic contribution to the $g_1$ structure function.


\section{Helicity Evolution}

In the previous Subsection we established the relevant degrees of freedom for the study of helicity at small-$x$ --- the polarized Wilson lines and polarized dipole amplitudes --- and showed how they related to the helicity PDFs and $g_1$ structure function. But in order to make use of these relations to actually calculate the helicity PDFs and $g_1$ at small-$x$, we need to obtain expressions for the polarized dipole amplitudes themselves. This we can do by constructing and solving their small-$x$ evolution equations. 

\begin{figure}[h!]
    \centering
    \includegraphics[width=0.9\linewidth]{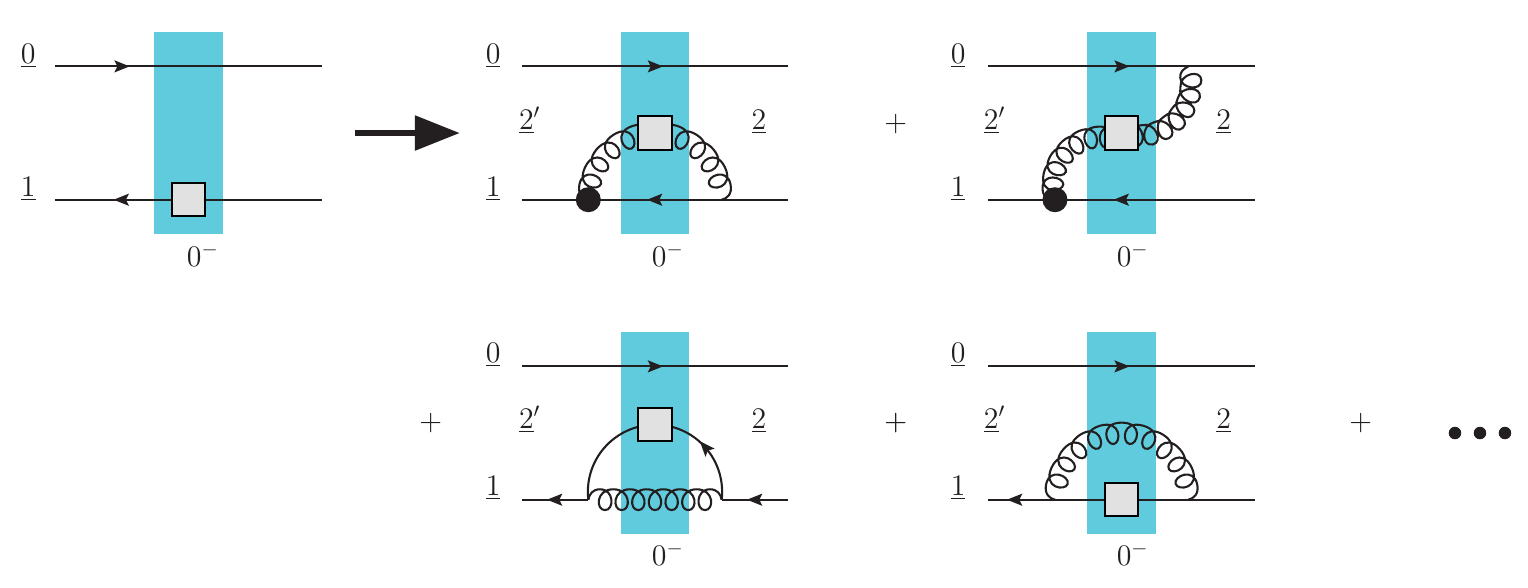}
    \caption{Some examples of diagrams contributing to the evolution of a polarized fundamental dipole.}
    \label{fig:someevolutiondiagrams}
\end{figure}

These evolution equations will sum up quantum corrections coming from additional quark and gluon exchanges. Diagrammatically this will correspond to modifications of Fig.~\ref{fig:polarizeddipoles} where one additional parton is emitted from the dipole. This parton could be emitted in a polarization-dependent way, via the sub-eikonal interactions presented in Ch.~\ref{subsec:polwilsonlines}, or could be an unpolarized eikonal emission \cite{Kovchegov:2015pbl}. Some examples of the types of diagrams we must calculate, for the case of the evolution of a polarized fundamental dipole, are shown in Fig.~\ref{fig:someevolutiondiagrams}. 
Again the gray box denotes a sub-eikonal interaction with the target shock-wave. The dark circle denotes the sub-eikonal gluon vertex (see Fig.~\ref{fig:polwilsonadjg}). Note also that the sub-eikonal vertex could appear on the right side of the shockwave as well, with a regular quark-gluon vertex on the left. The first three diagrams after the arrow denote sub-eikonal steps of evolution whereas the fourth includes a regular eikonal gluon emission. There are additional diagrams to compute beyond these few we showed. Furthermore we only showed the evolution of the dipole with a polarized antiquark line and unpolarized quark line --- one also has to add in the complex conjugate. In addition, one also has discriminate between the fundamental dipoles of type 1 and type 2, that is, between $Q_{10}(zs)$ and $G^i_{10}(zs)$. The latter involves gluon fields and so contains no quark emission like that in the second row of Fig.~\ref{fig:someevolutiondiagrams}. And of course one also has to construct the evolution of the adjoint polarized dipole amplitudes as well. The full details of all the diagrams contributing to the evolution and their calculations can be found in \cite{Kovchegov:2015pbl,Kovchegov:2016zex,Kovchegov:2017lsr,Kovchegov:2018znm,Cougoulic:2022gbk,Tawabutr:2022gei}.

The leading small-$x$ helicity evolution is double-logarithmic --- one can obtain logarithms of energy (equivalently $1/x$) from both the longitudinal and transverse integrals. The associated resummation parameters are 
\begin{align}\label{ch3_resummationparams}
    \as\ln^2(1/x) \quad \text{and} \quad \as\ln(1/x)\ln(Q^2/\Lambda^2)\,,
\end{align}
and resumming powers of these is known as the double-logarithmic approximation (DLA). While leading-order DGLAP requires strongly ordered transverse momenta to generate its logarithms, and while leading-order BFKL requires strongly ordered longitudinal momenta to generate its logarithms, one can show that the small-$x$ helicity evolution requires both the ordering of longitudinal momenta
\begin{align}\label{ch3_longitudinalordering}
    1 \gg z_1 \gg z_2 \gg ...
\end{align}
and also the ordering of the light-cone lifetimes
\begin{align}\label{ch3_lifetimeordering}
    z_1 x_1^2 \gg z_2 x_2^2 \gg z_3 x_3^2 \gg ...\,,
\end{align}
where $z_1$ is the smallest longitudinal momentum fraction in the first step of evolution, $z_2$ that in the second step, etc., and where $x_1^2$ is the squared magnitude of the transverse separation of the initial dipole, $x_2^2$ that of the next dipole in which evolution proceeds, etc. \cite{Kovchegov:2015pbl,Kovchegov:2018znm,Kovchegov:2017lsr,Cougoulic:2019aja,Kovchegov:2021lvz}.

\begin{figure}[h!]
    \centering
    \includegraphics[width=0.7\linewidth]{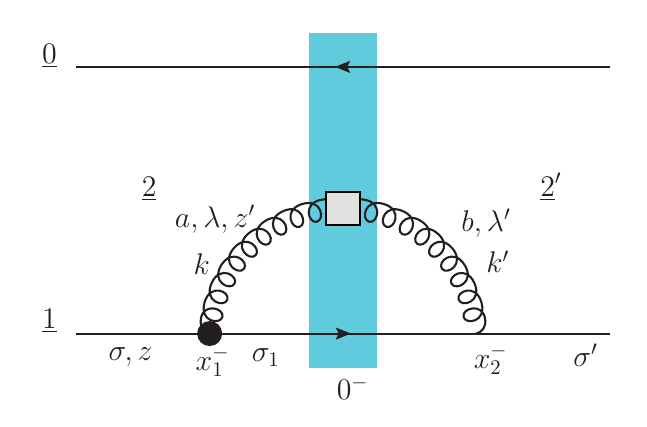}
    \caption{An example contribution to the evolution of the dipole amplitude $Q_{10}(zs)$ which we will calculate below.}
    \label{fig:examplecalc}
\end{figure}

We will present, but not derive here the full set of evolution equations. The full calculation is quite involved and its elements have been presented already several places in the literature, most recently in the paper \cite{Cougoulic:2022gbk} and the thesis \cite{Tawabutr:2022gei}. However in the interest of elucidating the formalism (some of which will also be used later in this dissertation) we will perform an example calculation of a particular diagram contributing to the small-$x$ evolution of the fundamental polarized dipole amplitude $Q_{10}(zs)$. We show this diagram in more detail in Fig.~\ref{fig:examplecalc}

There are several frameworks to calculate this diagram and the others involved in small-$x$ helicity evolution. Here we will show two of those frameworks. Both are blends of operator methods and light-cone perturbation theory, though to different extents. The first is a somewhat more elegant approach closer to a pure operator-based method --- it is known as the light-cone operator treatment (LCOT) \cite{Kovchegov:2017lsr,Kovchegov:2018znm,Cougoulic:2022gbk}. The second, while still employing operators to describe interactions with the target shock wave, relies on the traditional rules of light-cone perturbation theory (LCPT) \cite{Lepage:1980fj, Brodsky:1997de} to describe all splittings outside the shockwave \cite{Kovchegov:2015pbl}. We will see that both approaches yield the same results. There are additional ways to carry out these calculations which we will not discuss here, but see the second half of \cite{Cougoulic:2022gbk} for a framework employing the background field method and \cite{Li:2023tlw} for an effective Hamiltonian approach.


\subsection{Light-Cone Operator Treatment}\label{sec:ch3_lcot}
We will first calculate the diagram in Fig.~\ref{fig:examplecalc} --- which we will refer to as $(\delta Q)_I$ --- using the light-cone operator treatment. We begin with the definition of the polarized dipole $Q_{10}(zs)$, here only writing out the term with a polarized quark line and unpolarized anti-quark line. The complex conjugate would of course be included in the full evolution. Defined in \eq{ch3_Q10}, $Q_{10}(zs)$ is a function of the polarized Wilson line $V^{\text{pol}[1]}_{\underline{x}}$ (\eq{pollinestype1}) which contains both a gluon exchange and quark exchange term. Here we only need the gluon exchange term, so we explicitly write out the piece of $Q_{10}(zs)$ that depends on the polarized line $V^{G[1]}_{\underline{x}}$ (\eq{ch3_vg1}).
\begin{align}\label{ch3_Q10rewrite}
    Q_{10}(zs) \supset \frac{1}{2N_c} \frac{igP^+}{s} \int\limits_{-\infty}^{\infty} \mathrm{d}x_{1}^- \bigg\langle \text{T}\,\text{tr}\, \left[V_{\underline{1}}\left[\infty,x_1^-\right] F^{12}\left(x_1^-,\underline{x}_1\right)V_{\underline{1}}\left[x_1^-,-\infty\right]V_{\underline{0}}^\dagger \right] \bigg\rangle(zs)\,.
\end{align}
As can be seen in Fig.~\ref{fig:examplecalc}, we want to calculate a step of evolution where $Q_{10}(zs)$ emits an additional sub-eikonal gluon before the shock wave, at $x_1^- < 0 $ (the inequality follows from declaring the shock wave to be localized at $0^-$, which follows from the lifetime ordering constraint \eq{ch3_lifetimeordering} above). This sub-eikonal gluon propagates through the shock wave until it is absorbed at positive $x_2^-$. We can manipulate the expression in \eq{ch3_Q10rewrite} to generate exactly this contribution. Ultimately we need to obtain two gluon fields from $Q_{10}(zs)$ with which we can form a gluon propagator to describe the sub-eikonal gluon that traverses the shock wave. The $F^{12}(x_1^-,\underline{x}_1)$ operator in \eq{ch3_Q10rewrite} already provides one such gluon field, $F^{12} \approx \epsilon^{ij}\partial^i\underline{A}^j(x_1^-,\underline{x}_1) = \epsilon^{ij}\partial^i\underline{A}^{j\,a}(x_1^-,\underline{x}_1)t^a $ (to sub-eikonal accuracy in the $A^- = 0$ gauge). We can obtain another gluon field (this one at $x_2^- >0$) by expanding the fundamental Wilson $V_{\underline{1}}[\infty,x_1^-]$ in \eq{ch3_Q10rewrite}. Explicitly, we make the replacement
\begin{align}\label{ch3_lcotwilsonexpand}
    V_{\underline{1}}\left[\infty,x_1^-\right] \rightarrow \int\limits_0^{\infty}\mathrm{d}x_2^- V_{\underline{1}}\left[\infty,x_2^-\right]\,igA^+\left(x_2^-,\underline{x}_1\right) V_{\underline{1}}\left[x_2^-,x_1^-\right]\,.
\end{align}
Employing this replacement in \eq{ch3_Q10rewrite} we can write the contribution from the diagram in Fig.~\ref{fig:examplecalc} as \cite{Kovchegov:2017lsr,Tawabutr:2022gei}
\begin{align}\label{ch3_lcotexpanded}
    &\left(\delta Q\right)_I = \frac{1}{2N_c}\frac{-g^2P^+}{s}\int\limits_{-\infty}^{0}\mathrm{d}x_1^- \int\limits_0^\infty \mathrm{d}x_2^- \bigg\langle \text{T}\,\text{tr}\,\bigg[V_{\underline{1}}\left[\infty,x_2^-\right]\,A^{+\,b}\left(x_2^-,\underline{x}_1\right) t^b V_{\underline{1}}\left[x_2^-,x_1^-\right] \notag \\
    &\hspace{6cm}\times\epsilon^{ij}\left[\partial^i\underline{A}^{j\,a}\left(x_1^-,\underline{x}_1\right)\right] t^a V_{\underline{1}}\left[x_1^-,-\infty\right]V_{\underline{0}}^\dagger \bigg]\bigg\rangle(z') \\ \notag
    & = \frac{1}{2N_c}\frac{-g^2P^+}{s}\epsilon^{ij}\int\limits_{-\infty}^{0}\mathrm{d}x_1^- \int\limits_0^\infty \mathrm{d}x_2^- \bigg\langle
    \contraction[2ex]
    {}
    {A}
    {^{+\,b}\left(x_2^-,\underline{x}_1\right) [\partial^i }
    {\underline{A}}
    A^{+\,b}\left(x_2^-,\underline{x}_1\right) [\partial^i \underline{A}^{j\,a}\left(x_1^-,\underline{x}_1\right)]
    \, \text{T}\,\text{tr}\left[t^b V_{\underline{1}}t^a V_{\underline{0}}^\dagger \right] \bigg\rangle(z')\notag
\end{align}
where in the last line we explicitly denoted the contraction between the gluon fields in anticipation of the background-field propagator we will write down shortly and where we also set $V_{\underline{1}}[\infty,x_2^-] \rightarrow 1$, $V_{\underline{1}}[x_2^-,x_1^-] \rightarrow V_{\underline{1}}$, and $V_{\underline{1}}[x_1^-,-\infty] \rightarrow 1$. These last substitutions are consistent with the shock wave approximation whereby any Wilson lines not crossing the shock wave simply reduce to the identity and any Wilson lines crossing the shock wave can be treated as infinite due to the localized domain of the shock wave fields.

Next we need to construct the propagator denoted by the contraction in the last line of \eq{ch3_lcotexpanded}. The general structure for sub-eikonal propagators like this will be a sub-eikonal $S$-matrix element describing the polarized interaction with the target shock wave sandwiched in between free propagators, with the numerators of the propagators written as polarization sums. For the case considered here (with the momentum fractions, helicities, colors, and coordinates as labeled in Fig.~\ref{fig:examplecalc}) we write \cite{Kovchegov:2017lsr,Cougoulic:2022gbk,Tawabutr:2022gei}
\begin{align}\label{ch3_subeikpropagator}
    &\int\limits_{-\infty}^{0} \mathrm{d}x_1^- \int\limits_0^\infty \mathrm{d}x_2^- \,
    \contraction[2ex]
    {}
    {A}
    {^{+\,b}\left(x_2^-,\underline{x}_1\right) }
    {\underline{A}}
    A^{+\,b}\left(x_2^-,\underline{x}_1\right) \underline{A}^{j\,a}\left(x_1^-,\underline{x}_1\right) \\
    & = \sum_{\lambda,\lambda'}\int \mathrm{d}^2\underline{x}_2\mathrm{d}^2\underline{x}_{2'} \left[\int\limits_{-\infty}^{0}\mathrm{d}x_1^- \int \frac{\mathrm{d}^4k}{\left(2\pi\right)^4}\,e^{ik^+x_1^- + i\underline{k}\cdot\underline{x}_{21}} \frac{-i}{k^2 + i\epsilon} \underline{\epsilon}_{\lambda}^{j\,*}\left(k\right) \right] \notag \\
    &\hspace{3cm}\times\left[\left(U_{\underline{2}',\underline{2};\lambda',\lambda}^{ba}\right)_{\text{sub-eikonal}} \,2\pi\left(2k^-\right)\delta\left(k^--k^{\prime\,-}\right)\right] \notag \\
    &\hspace{3cm}\times\left[\int\limits_{0}^{\infty}\mathrm{d}x_2^- \int \frac{\mathrm{d}^4k'}{\left(2\pi\right)^4}\,e^{-ik^{\prime\,+}x_2^- - i\underline{k}'\cdot\underline{x}_{2'1}} \frac{-i}{\left(k^\prime\right)^2 + i\epsilon} \epsilon_{\lambda'}^{+}\left(k'\right) \right] \notag \,,
\end{align}
where $\epsilon_\lambda^\mu(k)$ is the gluon polarization vector defined after Eqs.~\eqref{ch3_quarkops}. Using the explicit gluon polarization vectors and the sub-eikonal $S$-matrix for gluon scattering in \eq{ch3_gluonsmatrix} (only $U_{\underline{x},\underline{y}}^{G[2]}$ contributes to $U_{\underline{x},\underline{y}}^{\text{pol}\,[2]}$) and carrying out all the integrals except those over $k^-$, $\underline{x}_2$, and $\underline{x}_{2'}$ we obtain \cite{Tawabutr:2022gei}
\begin{align}\label{ch3_propagatorsimplified}
     &\int\limits_{-\infty}^{0} \mathrm{d}x_1^- \int\limits_0^\infty \mathrm{d}x_2^- \,
    \contraction[2ex]
    {}
    {A}
    {^{+\,b}\left(x_2^-,\underline{x}_1\right) }
    {\underline{A}}
    A^{+\,b}\left(x_2^-,\underline{x}_1\right) \underline{A}^{j\,a}\left(x_1^-,\underline{x}_1\right) \\
    & = \frac{i}{4\pi^3}\int \mathrm{d}k^- \int\mathrm{d}^2x_2 \mathrm{d}^2x_{2'} \left[\frac{i\epsilon^{jl}\underline{x}^l_{21}}{x_{21}^2}U_{\underline{2}}^{\text{pol}[1]\,ba}\delta^2\left(\underline{x}_{2'2}\right) + \frac{\underline{x}_{2'1}^j}{x_{2'1}^2}U_{\underline{2'},\underline{2}}^{G[2]\,ba}\right] \ln \frac{1}{x_{21}\Lambda}\,, \notag
\end{align}
where $\Lambda$ is an infrared cutoff. Using this result in \eq{ch3_lcotexpanded}, we write the contribution $(\delta Q)_I$ of the diagram from Fig.~\ref{fig:examplecalc} as \cite{Tawabutr:2022gei}
\begin{align}\label{ch3_lcotresult}
    &\left(\delta Q\right)_I \to \frac{1}{2N_c}\frac{\as}{2\pi^2} \int \frac{\mathrm{d}z^{\prime}}{z'} \int \mathrm{d}^2 x_2\mathrm{d}^2x_{2'}\theta\left(\xoz^2z-\xto^2z'\right) \\
    &\hspace{2cm}\times\llangle\left[\frac{1}{x_{21}^2}U_{\underline{2}}^{\text{pol}[1]\,ba}\delta^2\left(\underline{x}_{2'2}\right) + \frac{i \epsilon^{ij}\underline{x}_{21}^i \underline{x}_{2'1}^j}{x_{2'1}^2x_{21}^2} U_{\underline{2'},\underline{2}}^{G[2]\,ba}\right] \text{T}\,\text{tr}\left[t^bV_{\underline{1}}t^a V_{\underline{0}}^\dagger\right]\rrangle(z's)\,, \notag
\end{align}
where we employed the center-of-mass energy squared $s = 2P^+(\tfrac{z}{z'}k^-)$ between the original quark line and the target and also employed the double-angle-brackets defined in \eq{ch3_Q10}. Note that we implicitly canceled the resulting factor of $1/zs$ since the same factor would appear upon applying the double angle brackets to the left hand side of the evolution (which would be $Q_{10}(zs)$ here, since that is the dipole we are currently evolving). Also note the insertion of the $\theta$-function $\theta(\xoz^2z-\xto^2z')$, consistent with the light-cone lifetime ordering requirements in \eq{ch3_lifetimeordering}. 

Any contribution to the small-$x$ evolution can be derived this way, namely by starting with a polarized dipole and forming the relevant contraction between fields to describe the emission of a new particle in a subsequent step of evolution. The propagator represented by that contraction can then be written as a sub-eikonal $S$-matrix element sandwiched between free propagators. Carrying out all the appropriate integrals will result in a contribution to the evolution, like that in \eq{ch3_lcotresult}. 

Next we will show how to obtain the contribution of the diagram in Fig.~\ref{fig:examplecalc} from the LCPT framework.


\subsection{Light-Cone Perturbation Theory}\label{sec:ch3_lcpt}
In the LCPT framework, we still treat the interaction with the target shock wave in terms of a sub-eikonal $S$-matrix element just as in the light-cone operator treatment (for the diagram  in Fig.~\ref{fig:examplecalc} this is still $(U^{ba}_{\underline{2},\underline{2'};\lambda',\lambda})_{\text{sub-eikonal}}$ from \eq{ch3_gluonsmatrix}). However, rather than forming a sub-eikonal propagator by contracting the relevant fields in the polarized dipole we are evolving, we instead build up the diagram using light-cone wavefunctions (see Ch.~\ref{subsec_lcpt}). This method was employed in the construction of Mueller's dipole model \cite{Mueller:1993rr,Mueller:1994gb}. It was also employed for helicity evolution in the early paper \cite{Kovchegov:2015pbl} and again received a detailed treatment in \cite{Tawabutr:2022gei}. 

Schematically, the diagram in Fig.~\ref{fig:examplecalc} can be built up out of the following ingredients: 1) the light-cone wavefunction for a quark to split into a quark and gluon, 2) the sub-eikonal operator interaction with the target, and 3) the complex conjugate of another light-cone wavefunction describing $q\rightarrow qG$ splitting.

\begin{figure}[h!]
    \centering
    \includegraphics[width=1\linewidth]{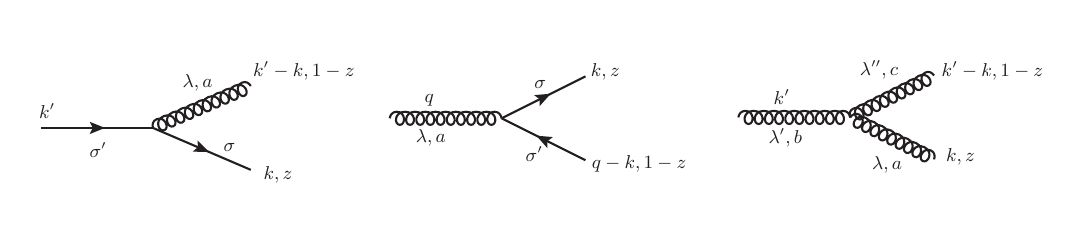}
    \caption{Leading order contributions to the $q\rightarrow qG$, $G\rightarrow q\overline{q}$, and $G\rightarrow GG$ light-cone wavefunctions.}
    \label{fig:ch3_lcwfs}
\end{figure}

To employ this LCPT technique we first need to write down the light-cone wavefunctions out of which we can build the diagrams. These are illustrated in Fig.~\ref{fig:ch3_lcwfs} (only the $q\to qG$ wavefunction in the figure is needed to construct the diagram from Fig.~\ref{fig:examplecalc}, but the other two are needed to construct additional contributions in the helicity evolution).
We begin by writing down the $q\rightarrow qG$ light-cone wavefunction from the leftmost panel of Fig.~\ref{fig:ch3_lcwfs}. Using the rules of LCPT \cite{Lepage:1980fj, Brodsky:1997de}, and with the momenta and helicities as labeled in the figure, we write the exact leading-order in $\as$ momentum-space wavefunction as \cite{Kovchegov:2015pbl,Kovchegov:2021lvz,Tawabutr:2022gei}
\begin{align}\label{ch3_qtoqGwf}
    \psi^{q\rightarrow qG}_{a,\lambda;\sigma,\sigma'}\left(k,k'\right) = 
    -g \overline{u}_{\sigma}(k) \slashed{\epsilon}_{\lambda}^*\left(k'-k\right) t^a u_{\sigma'}(k') \frac{1}{2k^{\prime\,-}} \frac{1}{k^{\prime\,+} - \left(k^{\prime}-k\right)^+ -k^+}\,,
\end{align}
where again our spinors are the $(\pm)$-interchanged Brodsky-Lepage spinors \cite{Lepage:1980fj} from Eqs.~\eqref{ch3_blspinors} and \eqref{ch3_rhospinors} and the gluon polarization vector $\epsilon_{\lambda}^\mu$ is defined in the text after Eqs.~\eqref{ch3_quarkops}. A direct calculation in \eq{ch3_qtoqGwf} yields
\begin{align}\label{ch3_qtoqGwf2}
    \psi^{q\rightarrow qG}_{a,\lambda;\sigma,\sigma'}\left(k,k'\right) = -gt^a \delta_{\sigma\sigma'} \sqrt{z} \frac{\underline{\epsilon}_{\lambda}^*\cdot \left(\underline{k} - z\underline{k}'\right)}{|\underline{k}-z\underline{k}'|^2}\left[1 + z + \sigma\lambda\left(1-z\right)\right]\,,
\end{align}
where $z = k^-/k^{\prime\,-}$. For the purposes of small-$x$ evolution, we need the limit of the above wavefunction where one of the outgoing particles is soft, i.e., has a small longitudinal momentum fraction. For the quark to be soft we would have $z\rightarrow 0$, while for the gluon to be soft we would have $1-z\rightarrow 0$, or equivalently $z\rightarrow 1$. These limits, Fourier-transformed to transverse coordinate space (which is more useful in our mixed-representation of longitudinal momenta/transverse coordinates) are
\begin{subequations}\label{ch3_qtoqGsoftlimits}
    \begin{align}\label{ch3_q_to_qG_wf_softq}
    & \psi^{q\rightarrow qG}_{a,\lambda;\sigma,\sigma'}\left(\underline{x}, z\right)|_{z\rightarrow 0} \approx -\frac{igt^a}{2\pi}\delta_{\sigma\sigma'}\sqrt{z}\, \frac{\underline{\epsilon}_\lambda^*\cdot \underline{x}}{\underline{x}^2}\left[1+\sigma\lambda\right] \quad \text{(soft quark)}, \\
\label{ch3_q_to_qG_wf_softG}
    & \psi^{q\rightarrow qG}_{a,\lambda;\sigma,\sigma'}\left(\underline{x}, z\right)|_{z\rightarrow 1} \approx \frac{igt^a}{2\pi}\delta_{\sigma\sigma'} \frac{\underline{\epsilon}_\lambda^*\cdot \underline{x}}{\underline{x}^2}\left[2+\sigma\lambda\left(1-z\right)\right] \quad \text{(soft gluon)}\,,
\end{align}
\end{subequations}
where we have defined $\underline{x}$ as the transverse separation between the soft outgoing particle and the incoming particle. Note also that \eq{ch3_q_to_qG_wf_softG} omits sub-eikonal polarization-independent terms that do not contribute to our helicity evolution. We can also use the expressions in Eqs.~\eqref{ch3_qtoqGsoftlimits} to describe the light-cone wavefunction of $\overline{q}\rightarrow \overline{q}G$ splitting as well, provided we include an additional overall minus sign to properly account for the Wick contractions associated with the antiquark fields. 

A similar calculation in the LCPT framework yields for the $G\rightarrow q\overline{q}$ light-cone wavefunction in the middle panel of Fig.~\ref{fig:ch3_lcwfs}
\begin{align}\label{ch3_G_to_qq_wf}
    \psi^{G\rightarrow q\bar{q}}_{a,\lambda;\sigma,\sigma'}\left(k,k'\right) = gt^a \sqrt{z\left(1-z\right)}\, \frac{\underline{\epsilon}_\lambda\cdot\left(\underline{k}-z\underline{q}\right)}{|\underline{k}-z\underline{q}|^2} \delta_{\sigma,-\sigma'}\left[1-2z-\sigma\lambda\right].
\end{align}
Again, we Fourier-transform to transverse coordinate space and take the limits where an outgoing particle is soft to obtain
\begin{subequations}\label{ch3_Gtoqqsoftlimits}
    \begin{align}\label{ch3_G_to_qq_wf_softq}
    & \psi^{G\rightarrow q\bar{q}}_{a,\lambda;\sigma,\sigma'}\left(\underline{x},z\right)|_{z\rightarrow 0} \approx \frac{igt^a}{2\pi} \sqrt{z}\, \frac{\underline{\epsilon}_\lambda\cdot\underline{x}}{\underline{x}^2} \delta_{\sigma,-\sigma'}\left[1-\sigma\lambda\right] \quad \text{(soft quark)}, \\
\label{ch3_G_to_qq_wf_softqbar}
    & \psi^{G\rightarrow q\bar{q}}_{a,\lambda;\sigma,\sigma'}\left(\underline{x},z\right)|_{z\rightarrow 1} \approx \frac{igt^a}{2\pi} \sqrt{1-z}\, \frac{\underline{\epsilon}_\lambda\cdot\underline{x}}{\underline{x}^2} \delta_{\sigma,-\sigma'}\left[1-\sigma'\lambda\right] \quad \text{(soft antiquark)}.
\end{align}
\end{subequations}

Finally for the $G\rightarrow GG$ wavefunction in the right panel of Fig.~\ref{fig:ch3_lcwfs} we write
\begin{align}\label{ch3_G_to_GG_wf}
    &\psi^{G\rightarrow GG}_{a,b,c;\lambda,\lambda',\lambda''}\left(k,k'\right) = 2igf^{abc}\, \frac{z\left(1-z\right)}{|\underline{k}-z\underline{k}'|^2} \bigg[\frac{1}{1-z} \underline{\epsilon}_{\lambda''}^*\cdot\left(\underline{k}-z\underline{k}' \right)\underline{\epsilon}_\lambda^*\cdot \underline{\epsilon}_{\lambda'} \\
    &\hspace{5.3cm}+ \frac{1}{z}\underline{\epsilon}_\lambda^*\cdot \left( \underline{k}-z\underline{k}'\right) \underline{\epsilon}_{\lambda''}^*\cdot \underline{\epsilon}_{\lambda'} - \underline{\epsilon}_{\lambda'}\cdot \left( \underline{k}-z\underline{k}'\right) \underline{\epsilon}_{\lambda''}^*\cdot \underline{\epsilon}_{\lambda'}^*  \bigg]\,. \notag
\end{align}
The transverse coordinate space expression in the limit of a soft outgoing gluon is, again keeping only the leading sub-eikonal, polarization-dependent terms,
\begin{align}\label{ch3_G_to_GG_wf_softG}
    &\psi^{G\rightarrow GG}_{a,b,c;\lambda,\lambda',\lambda''}\left(\underline{x},z\right)|_{z\rightarrow 0} \approx \frac{-g f^{abc}}{\pi} \frac{z}{\underline{x}^2}\bigg[\frac{1}{z}\underline{\epsilon}_{\lambda}^*\cdot\underline{x}\,\delta_{\lambda''\lambda'} + \underline{\epsilon}_{\lambda''}^*\cdot\underline{x}\,\delta_{\lambda\lambda'} + \underline{\epsilon}_{\lambda'}\cdot\underline{x}\,\delta_{\lambda'',-\lambda} \bigg]\,, \\
    &\hspace{12cm}\text{(soft gluon)} \notag.
\end{align}

Having established the light-cone wavefunctions, we return to the diagram in Fig.~\ref{fig:examplecalc}. We can write the contribution of this diagram to the evolution of $Q_{10}(zs)$ in terms of the relevant light-cone wavefunctions and Wilson lines, with the sub-eikonal interaction mediated by the sub-eikonal gluon $S$-matrix $(U^{ba}_{\underline{2}',\underline{2};\lambda',\lambda})_{\text{sub-eikonal}}$. With the momenta, helicities, coordinates and colors as labeled in that figure, we write this diagram as \cite{Kovchegov:2015pbl,Tawabutr:2022gei}
\begin{align}\label{ch3_lcptdiagram}
    &\left(\delta Q\right)_I = \frac{1}{2N_c}\left(\frac{1}{2}\sum_{\sigma,\sigma'}\sigma\right) \sum_{\lambda,\lambda',\sigma_1}\int\frac{\mathrm{d}z'}{z'}\int\frac{\mathrm{d}^2x_2\mathrm{d}^2x_{2'}}{4\pi}\theta\left(\xoz^2z-\xto^2z'\right) \\
    &\hspace{1.5cm}\times\bigg\langle \text{T}\,\text{tr}\, \bigg\{\left[\psi^{q\rightarrow qG}_{b,\lambda';\sigma',\sigma_1}\left(\underline{x}_{2'1},1-\frac{z'}{z}\right)\bigg|_{\tfrac{z'}{z}\rightarrow 0} \right]^* \left(U^{ba}_{\underline{2}',\underline{2};\lambda',\lambda}\right)_{\text{sub-eikonal}} \notag \\
    &\hspace{3cm}\times V_{\underline{1}} 
    \left[\psi^{q\rightarrow qG}_{a,\lambda;\sigma_1,\sigma}\left(\underline{x}_{21},1-\frac{z'}{z}\right)\bigg|_{\tfrac{z'}{z}\rightarrow 0} \right] \,V_{\underline{0}}^{\dagger} \bigg\}\bigg\rangle (z')\,. \notag \\
    &\hspace{1cm} = \frac{1}{2N_c}\left(\frac{1}{2}\sum_{\sigma,\sigma'}\sigma\right) \sum_{\lambda,\lambda',\sigma_1}\int\frac{\mathrm{d}z'}{z'}\int\frac{\mathrm{d}^2x_2\mathrm{d}^2x_{2'}}{4\pi}\theta\left(\xoz^2z-\xto^2z'\right) \notag \\
    & \hspace{1.5cm} \times \bigg\langle \text{T}\,\text{tr}\bigg\{\left[\frac{igt^b}{2\pi} \delta_{\sigma'\sigma_1}\frac{\epsilon_{\lambda'}^* \cdot \underline{x}_{2'1} }{x_{2'1}^2}\left(2+\sigma_1\lambda'\frac{z'}{z}\right) \right]^* \notag \\
    &\hspace{3cm}\times\left[\lambda\delta_{\lambda\lambda'}U_{\underline{2}}^{\text{pol}\,[1]}\delta^2\left(\underline{x}_{2'2}\right) + \delta_{\lambda\lambda'}U_{\underline{2}',\underline{2}}^{G[2]}  \right]^{ba} V_{\underline{1}}\notag \\
    &\hspace{3cm}\times\left[\frac{igt^a}{2\pi} \delta_{\sigma\sigma_1}\frac{\epsilon_{\lambda}^* \cdot \underline{x}_{21} }{x_{21}^2}\left(2+\sigma\lambda\frac{z'}{z}\right)   \right]   V_{\underline{0}}^\dagger \bigg\} \bigg\rangle (z') \,,\notag
\end{align}
where in the second equality we have substituted in the explicit light-cone wavefunctions along with the sub-eikonal gluon $S$-matrix, also replacing $U_{\underline{x},\underline{y}}^{\text{pol}\,[2]} \rightarrow U_{\underline{x},\underline{y}}^{G[2]}$ since $U_{\underline{x}}^{q[2]}$ does not contribute to helicity evolution. Note that the normalization $1/(2N_c)$ is inherited from the parent object $Q_{10}(zs)$ which we are evolving. We average over the incoming helicity $\sigma$ (a weighted average), sum over the outgoing helicity $\sigma'$, and sum over the internal helicities $\lambda,\lambda',\sigma_1$. We weight the average over incoming helicity by $\sigma$ --- this picks out the helicity-dependent term $V_{\underline{1}}^{\text{pol}[1]}$ from which $Q_{10}(zs)$ is built. Note also the summation over quark color indices is left implicit. Again we impose the lifetime-ordering $\theta$-function and treat the ordinary Wilson lines crossing the shockwave as infinite (here $V_{\underline{1}}$ and $V_{\underline{0}}^{\dagger}$). Carrying out all the helicity sums and employing the explicit gluon polarization vector we obtain
\begin{align}\label{ch3_lcptdiagram2}
    &\left(\delta Q\right)_I = \frac{1}{2N_c}\frac{\as}{2\pi^2} \int\frac{\mathrm{d}z'}{z} \int\mathrm{d}^2x_2\mathrm{d}^2x_{2'} \theta\left(\xoz^2z-\xto^2z'\right) \\
    &\hspace{1cm}\times \bigg\langle \left[\frac{1}{x_{21}^2}U_{\underline{2}}^{\text{pol}\,[1]\,ba}\delta^2\left(\underline{x}_{2'2}\right) + i\frac{\epsilon^{ij}\underline{x}_{21}^i \underline{x}_{2'1}^j}{x_{21}^2x_{2'1}^2} U_{\underline{2}',\underline{2}}^{G[2]\,ba}\right]\,\text{T}\,\text{tr}\left[t^bV_{\underline{1}}t^aV_{\underline{0}}^\dagger\right]\bigg\rangle(z') \notag \\
    &\hspace{1cm} \to \frac{1}{2N_c}\frac{\as}{2\pi^2} \int\frac{\mathrm{d}z'}{z'} \int\mathrm{d}^2x_2\mathrm{d}^2x_{2'} \theta\left(\xoz^2z-\xto^2z'\right)\notag \\
    &\hspace{1cm}\times \llangle \left[\frac{1}{x_{21}^2}U_{\underline{2}}^{\text{pol}\,[1]\,ba}\delta^2\left(\underline{x}_{2'2}\right) + i\frac{\epsilon^{ij}\underline{x}_{21}^i \underline{x}_{2'1}^j}{x_{21}^2x_{2'1}^2} U_{\underline{2}',\underline{2}}^{G[2]\,ba}\right]\,\text{T}\,\text{tr}\left[t^bV_{\underline{1}}t^aV_{\underline{0}}^\dagger\right]\rrangle(z's) \notag\,
\end{align}
where in the last equality we employed the double-angle brackets and, as before, canceled the resulting factor of $1/(zs)$ which we would also obtain by applying the double-angle brackets to $Q_{10}(zs)$ itself. As promised we have obtained the same expression as that derived with the LCOT formalism in \eq{ch3_lcotresult}. Also note that we have longitudinal and transverse integrals remaining, both of which can generate large logarithms, as is consistent with the framework of our double-logarithmic approximation.


\subsection{Full Evolution Equations}\label{sec:ch3_fullevoleqns}

The full set of evolution equations at the level of the polarized Wilson lines was derived in its current form in \cite{Cougoulic:2022gbk}. This is a set of four equations that describes the evolution of the polarized dipoles $Q_{10}(zs)$, $G^{\text{adj}}_{10}(zs)$, $G^i_{10}(zs)$ and $G^{i\,\text{adj}}_{10}(zs)$ defined in Ch.~\ref{sec:beyondeikonal} in terms of correlators of Wilson line operators. For completeness, we show the full set of evolution equations below: 
\begin{subequations}\label{ch3_fullevol}\allowdisplaybreaks
\begin{align}
\label{ch3_fullevolQ}
&\frac{1}{2N_c}\llangle \text{tr}\left[V_{\underline{0}}V_{\underline{1}}^{\text{pol}\,[1]\,\dagger}\right] + \text{c.c.}\rrangle(zs) = \frac{1}{2N_c}\llangle \text{tr}\left[V_{\underline{0}}V_{\underline{1}}^{\text{pol}\,[1]\,\dagger}\right] + \text{c.c.}\rrangle_0(zs) \\
&+ \frac{\as N_c}{2\pi^2}\int\limits_{\Lambda^2/s}^{z}\frac{\mathrm{d}z'}{z'}\int\mathrm{d}^2x_2 \notag \\
&\hspace{1cm}\times\bigg\{\left[\frac{1}{x_{21}^2} - \frac{\underline{x}_{21}}{x_{21}^2}\cdot \frac{\underline{x}_{20}}{x_{20}^2} \right] \frac{1}{N_c^2} \llangle \text{tr}\, \left[t^bV_{\underline{0}}t^aV_{\underline{1}}^\dagger \right] \left(U_{\underline{2}}^{\text{pol}\,[1]}\right)^{ba} + \text{c.c.}\rrangle(z's) \notag \\
&\hspace{1.4cm}+ \left[2\frac{\epsilon^{ij}x_{21}^j}{x_{21}^4} - \frac{\epsilon^{ij}\left(x_{20}^j + x_{21}^j \right)}{x_{20}^2x_{21}^2} - \frac{2\underline{x}_{20}\times\underline{x}_{21}}{x_{20}^2x_{21}^2}\left(\frac{x_{21}^i}{x_{21}^2} - \frac{x_{20}^i}{x_{20}^2} \right) \right] \notag \\
&\hspace{4cm}\times\frac{1}{N_c^2}\llangle \text{tr}\, \left[t^bV_{\underline{0}}t^aV_{\underline{1}}^\dagger\right]\left(U_{\underline{2}}^{i\,G[2]}\right)^{ba} + \text{c.c.}\rrangle(z's)    \bigg\}\notag \\
&+\frac{\as N_c}{4\pi^2}\int\limits_{\Lambda^2/s}^{z}\frac{\mathrm{d}z'}{z'}\int\frac{\mathrm{d}^2x_2}{x_{21}^2} \bigg\{\frac{1}{N_c^2}\llangle \text{tr}\,\left[t^bV_{\underline{0}}t^aV_{\underline{2}}^{\text{pol}\,[1]\dagger}\right]U_{\underline{1}}^{ba}\rrangle(z's) \notag \\
&\hspace{4.5cm}+ 2\frac{\epsilon^{ij}\underline{x}_{21}^j }{x_{21}^2}\frac{1}{N_c^2}\llangle \text{tr}\,\left[t^bV_{\underline{0}}t^a V_{\underline{2}}^{i\,G[2]\dagger}\right] U_{\underline{1}}^{ba}\rrangle(z's) + \text{c.c.} \bigg\}\notag \\
& + \frac{\as N_c}{2\pi^2}\int\limits_{\Lambda^2/s}^z \frac{\mathrm{d}z'}{z'} \int \mathrm{d}^2x_2 \frac{x_{10}^2}{x_{21}^2x_{20}^2} \bigg\{\frac{1}{N_c^2}\llangle \text{tr}\,\left[t^bV_{\underline{0}}t^a V_{\underline{1}}^{\text{pol}[1]\dagger} \right] U_{\underline{2}}^{ba}\rrangle (z's) \notag \\
&\hspace{6cm}- \frac{C_f}{N_c^2}\llangle \text{tr}\left[V_{\underline{0}}V_{\underline{1}}^{\text{pol}[1]\dagger}  \right]\rrangle(z's) + \text{c.c.}  
\bigg\} \,,\notag \\
\notag\\
\label{ch3_fullevolGadj}
&\frac{1}{2\left(N_c^2-1\right)}\text{Re}\,\llangle \text{Tr}\,\left[U_{\underline{0}}U_{\underline{1}}^{\text{pol}[1]\dagger} \right] + \text{c.c.}\rrangle(zs) \\
& = \frac{1}{2\left(N_c^2-1\right)}\text{Re}\,\llangle \text{Tr}\,\left[U_{\underline{0}}U_{\underline{1}}^{\text{pol}[1]\dagger} \right] + \text{c.c.}\rrangle_0(zs) \notag \\
& + \frac{\as}{\pi^2} \int\limits_{\Lambda^2/s}^{z}\frac{\mathrm{d}z'}{z'}\int\mathrm{d}^2x_2 \bigg\{\left[\frac{1}{x_{21}^2} - \frac{\underline{x}_{21}}{x_{21}^2}\cdot \frac{\underline{x}_{20}}{x_{20}^2} \right] \notag\\
&\hspace{4.8cm}\times\frac{1}{N_c^2 -1 } \llangle \text{Tr}\,\left[ T^bU_{\underline{0}}T^a U_{\underline{1}}^\dagger \right]\left(U_{\underline{2}}^{\text{pol}[1]}\right)^{ba} + \text{c.c.}\rrangle (z's) \notag\\
& \hspace{3.7cm}+ \left[2\frac{\epsilon^{ij}x_{21}^j}{x_{21}^4} - \frac{\epsilon^{ij}\left(x_{20}^j + x_{21}^j\right) }{x_{20}^2x_{21}^2} - \frac{2\underline{x}_{20}\times\underline{x}_{21}}{x_{20}^2x_{21}^2}\left(\frac{x_{21}^i}{x_{21}^2} - \frac{x_{20}^i}{x_{20}^2} \right)  \right] \notag \\
&\hspace{4.8cm}\times\frac{1}{N_c^2 -1 } \llangle \text{Tr}\,\left[T^bU_{\underline{0}}T^aU_{\underline{1}}^\dagger \right]\left(U_{\underline{2}}^{iG[2]}\right)^{ba} + \text{c.c.}\rrangle(z's)
\bigg\}\notag \\
& - \frac{\as N_f}{2\pi^2\left(N_c^2 -1\right)} \int\limits_{\Lambda^2/s}^{z}\frac{\mathrm{d}z'}{z'}\int\mathrm{d}^2x_2 \bigg\{ \frac{1}{x_{21}^2} \llangle \text{tr}\,\left[t^bV_{\underline{1}}t^a V_{\underline{2}}^{\text{pol}[1]\dagger} \right] U_{\underline{0}}^{ba}\rrangle (z's) \notag \\
&\hspace{5.5cm} + 2\frac{\epsilon^{ij}\underline{x}_{21}^j}{x_{21}^4} \llangle \text{tr}\,\left[t^bV_{\underline{1}}t^aV_{\underline{2}}^{iG[2]\dagger} \right]U_{\underline{0}}^{ba}\rrangle(z's) + \text{c.c.} \bigg\} \notag \\
& + \frac{\as}{2\pi^2} \int\limits_{\Lambda^2/s}^z \frac{\mathrm{d}z'}{z'}\int\mathrm{d}^2x_2 \frac{x_{10}^2}{x_{21}^2x_{20}^2}\frac{1}{N_c^2 - 1}\bigg\{\llangle \text{Tr}\,\left[T^bU_{\underline{0}}T^aU_{\underline{1}}^{\text{pol}[1]\dagger}\right] U_{\underline{2}}^{ba}\rrangle(z's) \notag \\
&\hspace{6.5cm}- N_c\llangle \text{Tr}\,\left[U_{\underline{0}}U_{\underline{1}}^{\text{pol}[1]\dagger} \right]\rrangle (z's) + \text{c.c.} \bigg\}\,, \notag \\
\notag \\
\label{ch3_fullevolGi}
&\frac{1}{2N_c}\llangle \text{tr}\left[V_{\underline{0}}V_{\underline{1}}^{iG[2]\dagger}\right] + \text{c.c.}\rrangle(zs) = \frac{1}{2N_c}\llangle \text{tr}\left[V_{\underline{0}}V_{\underline{1}}^{iG[2]\dagger}\right] + \text{c.c.}\rrangle_0(zs) \\
&+\frac{\as N_c}{4\pi^2}\int\limits_{\Lambda^2/s}^{z}\frac{\mathrm{d}z'}{z'}\int\mathrm{d}^2x_2 \bigg\{\left[\frac{\epsilon^{ij}x_{21}^j}{x_{21}^2} - \frac{\epsilon^{ij}x_{20}^j}{x_{20}^2} + 2x_{21}^i \frac{\underline{x}_{21}\times\underline{x}_{20}}{x_{21}^2x_{20}^2} \right] \notag \\
&\hspace{4cm} \times\frac{1}{N_c^2}  \llangle \text{tr}\,\left[t^bV_{\underline{0}}t^a V_{\underline{1}}^\dagger  \right]\left(U_{\underline{2}}^{\text{pol}[1]}\right)^{ba} + \text{c.c.}\rrangle(z's) \notag \\
& \hspace{2.3cm}+ \bigg[ \delta^{ij}\left(\frac{3}{x_{21}^2} - 2\frac{\underline{x}_{20}\cdot \underline{x}_{21} }{x_{20}^2x_{21}^2} - \frac{1}{x_{20}^2}\right) - 2\frac{x_{21}^ix_{20}^j }{x_{21}^2x_{20}^2}\left(2\frac{\underline{x}_{20}\cdot \underline{x}_{21}}{x_{20}^2} + 1 \right) \notag \\
&\hspace{3cm}+ 2\frac{x_{21}^ix_{21}^j}{x_{21}^2x_{20}^2} \left(2\frac{\underline{x}_{20}\cdot \underline{x}_{21}}{x_{21}^2} + 1  \right) + 2\frac{x_{20}^ix_{20}^j }{x_{20}^4} - 2\frac{x_{21}^ix_{21}^j}{x_{21}^4} \bigg] \notag\\
&\hspace{4cm}\times \frac{1}{N_c^2}\llangle \text{tr}\,\left[t^b V_{\underline{0}}t^aV_{\underline{1}}^\dagger \right]\left(U_{\underline{2}}^{jG[2]}\right)^{ba} + \text{c.c.}\rrangle(z's) \bigg\} \notag \\
& + \frac{\as N_c}{2\pi^2}\int\limits_{\Lambda^2/s}^{z} \frac{\mathrm{d}z'}{z'}\int\mathrm{d}^2x_2 \frac{x_{10}^2}{x_{21}^2x_{20}^2} \bigg\{ \frac{1}{N_c^2}\llangle \text{tr}\,\left[t^b V_{\underline{0}}t^aV_{\underline{1}}^{iG[2]\dagger} \right]\left(U_{\underline{2}}\right)^{ba}\rrangle(z's) \notag\\
&\hspace{5.5cm}- \frac{C_f}{N_c^2}\llangle \text{tr}\,\left[V_{\underline{0}}V_{\underline{1}}^{iG[2]\dagger}\right]\rrangle(z's) + \text{c.c.} \bigg\}\,, \notag \\
\notag \\
\label{ch3_fullevolGiadj}
&\frac{1}{2\left(N_c^2-1\right)}\llangle \text{Tr}\left[U_{\underline{0}}U_{\underline{1}}^{iG[2]\dagger}\right] + \text{c.c.}\rrangle(zs) = \frac{1}{2\left(N_c^2-1\right)}\llangle \text{Tr}\left[U_{\underline{0}}U_{\underline{1}}^{iG[2]\dagger}\right] + \text{c.c.}\rrangle_0(zs) \\
&+\frac{\as}{4\pi^2}\int\limits_{\Lambda^2/s}^{z}\frac{\mathrm{d}z'}{z'}\int\mathrm{d}^2x_2 \bigg\{\left[\frac{\epsilon^{ij}x_{21}^j}{x_{21}^2} - \frac{\epsilon^{ij}x_{20}^j}{x_{20}^2} + 2x_{21}^i \frac{\underline{x}_{21}\times\underline{x}_{20}}{x_{21}^2x_{20}^2} \right] \notag \\
&\hspace{4cm} \times\frac{1}{N_c^2-1}  \llangle \text{Tr}\,\left[T^bU_{\underline{0}}T^a U_{\underline{1}}^\dagger  \right]\left(U_{\underline{2}}^{\text{pol}[1]}\right)^{ba} + \text{c.c.}\rrangle(z's) \notag \\
& \hspace{2.3cm}+ \bigg[ \delta^{ij}\left(\frac{3}{x_{21}^2} - 2\frac{\underline{x}_{20}\cdot \underline{x}_{21} }{x_{20}^2x_{21}^2} - \frac{1}{x_{20}^2}\right) - 2\frac{x_{21}^ix_{20}^j }{x_{21}^2x_{20}^2}\left(2\frac{\underline{x}_{20}\cdot \underline{x}_{21}}{x_{20}^2} + 1 \right) \notag \\
&\hspace{3cm}+ 2\frac{x_{21}^ix_{21}^j}{x_{21}^2x_{20}^2} \left(2\frac{\underline{x}_{20}\cdot \underline{x}_{21}}{x_{21}^2} + 1  \right) + 2\frac{x_{20}^ix_{20}^j }{x_{20}^4} - 2\frac{x_{21}^ix_{21}^j}{x_{21}^4} \bigg] \notag\\
&\hspace{4cm}\times \frac{1}{N_c^2 - 1}\llangle \text{Tr}\,\left[T^b U_{\underline{0}}T^aU_{\underline{1}}^\dagger \right]\left(U_{\underline{2}}^{jG[2]}\right)^{ba} + \text{c.c.}\rrangle(z's) \bigg\} \notag \\
& + \frac{\as}{2\pi^2}\int\limits_{\Lambda^2/s}^{z} \frac{\mathrm{d}z'}{z'}\int\mathrm{d}^2x_2 \frac{x_{10}^2}{x_{21}^2x_{20}^2} \frac{1}{N_c^2-1}\bigg\{\llangle \text{Tr}\,\left[T^b U_{\underline{0}}T^aU_{\underline{1}}^{iG[2]\dagger} \right]\left(U_{\underline{2}}\right)^{ba}\rrangle(z's) \notag\\
&\hspace{6.5cm} - N_c\llangle \text{Tr}\,\left[U_{\underline{0}}U_{\underline{1}}^{iG[2]\dagger}\right]\rrangle(z's) + \text{c.c.} \bigg\}\,. \notag
\end{align}
\end{subequations}
In these equations, the correlators with subscript $0$ denote the initial conditions/inhomogeneous terms of the evolution \cite{Kovchegov:2015pbl,Kovchegov:2018znm,Kovchegov:2017lsr,Kovchegov:2021lvz}. These can be modeled explicitly and are often taken at Born level. Alternatively they can be fitted to data. However in this dissertation we will generally be agnostic about these initial conditions, allowing them to be arbitrary or, where necessary, making convenient and simple choices for them. Note also that in \eq{ch3_fullevol}, $\text{c.c.}$ denotes the complex conjugate, $C_F = (N_c^2-1)/2N_c$ is the fundamental Casimir operator, and we define the cross product between two vectors (in two transverse dimensions) as $\underline{x}\times\underline{y} = x^1y^2 - x^2y^1$.
The light-cone-lifetime-ordering $\theta$-functions are left implicit, but note that we require $z\xoz^2 \gg z'\xto^2 \gg 1/s$ and squared dipole sizes are cutoff by $1/\Lambda^2$ from above. Those two conditions together yield $z' \gg \Lambda^2/s$, hence the lower bound on the longitudinal integrals. We take the coupling $\as$ to be fixed. Eqs.~\eqref{ch3_fullevol} also contain some single-logarithmic terms. These will manifest shortly as unpolarized $S$-matrices (see \eq{ch2_unpolarizeddipole}), which contain important information regarding parton \textit{saturation} --- the phenomenon by which the growth of the parton distributions with energy slows down, allowing them to eventually \textit{saturate} at maximal occupation numbers \cite{Gribov:1984tu,Mueller:1985wy,McLerran:1993ni,McLerran:1993ka,McLerran:1994vd,Kovchegov:1996ty,Kovchegov:1997pc,JalilianMarian:1996xn}.

While quite general, Eqs.~\eqref{ch3_fullevol} are difficult to work with themselves. The major challenge is that they do not form a closed set of equations --- evolving our polarized dipoles resulted in new, more complicated operator expectation values. For example, \eq{ch3_fullevolQ} evolves the object $\llangle \text{tr}\, [V_{\underline{0}}V_{\underline{1}}^{\text{pol}1\dagger}]\rrangle$ (the left hand side of that equation). However on the right hand side we now have more complicated objects like $\llangle \text{tr}\, [t^bV_{\underline{0}}t^aV_{\underline{1}}^\dagger](U_{\underline{2}}^{\text{pol}})^{ba}\rrangle$, which is a correlator involving three Wilson lines rather than the two we started with. Thus we would have to construct an evolution equation for that three-Wilson-line correlator, but that equation would then introduce correlators with even more Wilson lines, which we would also need to evolve. In that sense, Eqs.~\eqref{ch3_fullevol} represent the first equations in an infinite hierarchy of evolution equations (the same thing happens in the unpolarized case and is known there as the Balitsky hierarchy \cite{Balitsky:1995ub,Balitsky:1998ya}).

Fortunately there are two limits where the evolution in Eqs.~\eqref{ch3_fullevol} reduces to a closed set of equations. The first is the 't Hooft large-$N_c$ limit \cite{tHooft:1974pnl} (discussed earlier after Fig.~\ref{fig:dipolemodelevolution}), where the number of colors $N_c$ is taken to be a large parameter with the combination $\as N_c$ fixed and finite. For us this will amount to ignoring all quark contributions to the evolution. We also have at our disposal Veneziano's large-$N_c\&N_f$ limit \cite{Veneziano:1976wm}. In this limit, both $N_c$ and $N_f$ are taken to be parametrically large while their ratio is held fixed ($\as N_c$ is also held fixed here). The small-$x$ helicity evolution in this limit is somewhat more complicated than that in the large-$N_c$ limit, though it is a more realistic picture since it takes into account the genuine quark contributions to the evolution.

For the large-$N_c$ evolution, we work with the following modified dipole amplitudes. Since we neglect quark contributions in this limit, we can make the replacement $U_{\underline{x}}^{\text{pol}[1]}\rightarrow U_{\underline{x}}^{G[1]}$, ignoring the contribution of $U_{\underline{x}}^{q[1]}$. We define the new dipole amplitude $G_{10}(zs)$ as the large-$N_c$ version of $Q_{10}(zs)$ from \eq{ch3_Q10}. This dipole amplitude is
\begin{align}\label{ch3_G10}
    G_{10}(zs) = \frac{1}{2N_c}\text{Re}\,\llangle \text{T}\,\text{tr}\,\left[V_{\underline{0}}V_{\underline{1}}^{G[1]\dagger}\right] + \text{T}\,\text{tr}\,\left[V_{\underline{1}}^{G[1]}V_{\underline{0}}^\dagger\right]\rrangle (zs)
\end{align}
along with its impact-parameter-integrated version
\begin{align}\label{ch3_G}
    G\left(\xoz^2,zs\right) = \int \mathrm{d}^2\left(\frac{x_0+x_1}{2}\right)G_{10}(zs)\,.
\end{align}
The large-$N_c$ evolution also involves the other fundamental polarized dipole amplitude $G_{10}^{i}$ and its impact-parameter-integrated version from \eq{ch3_G10idecomp}. As discussed after that equation, only $G_2(\xoz^2,zs)$ contributes to the helicity evolution. 

One can also show that the adjoint dipole amplitudes $G_{10}^{\text{adj}}$ and $G^{i\,\text{adj}}_{10}$ are very simply related to the fundamental ones --- $G_{10}(zs)$ and $G_{10}^i(zs)$, respectively --- in the large-$N_c$ limit \cite{Cougoulic:2022gbk}:
\begin{subequations}\label{ch3_fundandadjlargenc}
\begin{align}
    \label{ch3_fundandadjlargenc1}
    &G_{10}^{\text{adj}}(zs) = 4G_{10}(zs)S_{10}(zs)\,, \\
    \label{ch3_fundandadjlargenc2}
    &G_{10}^{i\,\text{adj}}(zs) = 2G_{10}^i(zs)S_{10}(zs)\,,
\end{align}
\end{subequations}
where as usual $S_{10}(zs)$ is the unpolarized dipole amplitude (\eq{ch2_unpolarizeddipole}). In the double-logarithmic approximation, we can simply set $S_{10}(zs) = 1$ since the 
evolution of the unpolarized dipole is single-logarithmic \cite{Balitsky:1995ub,Balitsky:1998ya,Kovchegov:1999yj,Kovchegov:1999ua,Jalilian-Marian:1997ubg,Jalilian-Marian:1997jhx,Weigert:2000gi,Iancu:2001ad,Iancu:2000hn,Ferreiro:2001qy}. Then the (impact-parameter-integrated) small-$x$ evolution in the large-$N_c$ limit only requires us to know the evolution of two objects: $G(\xoz^2,zs)$ and $G_2(\xoz^2,zs)$. 

There is a subtlety, however. A consistent enforcement of the DLA light-cone lifetime ordering actually requires us to introduce two additional objects. These are the so-called \textit{neighbor} dipole amplitudes \cite{Kovchegov:2015pbl,Kovchegov:2016zex,Kovchegov:2018znm,Kovchegov:2017lsr,Cougoulic:2022gbk}. Each polarized dipole amplitude has its corresponding neighbor: $\Gamma_{20,21}(zs)$ is the neighbor of $G_{10}(zs)$ and $\Gamma^i_{20,21}(zs)$ is the neighbor of $G_{10}^i(zs)$. Each neighbor's operator definition is the same as that of its associated non-neighbor dipole. The neighbors, however, depend on two transverse separations (the notation $\Gamma_{20,21}(zs)$ denotes the dependence on both $\underline{x}_{20}$ and $\underline{x}_{21}$), unlike the regular polarized dipoles which only depend on a single transverse separation. The physical consequence is that evolution of a neighbor dipole amplitude depends on the adjacent dipole, with a lifetime bound dependent on that adjacent dipole's size, as opposed to a non-neighbor dipole amplitude where a dipole evolves independently of the adjacent dipole \cite{Cougoulic:2022gbk}. We define impact-parameter-integrated neighbors analogously to the non-neighbors:
\begin{subequations}\label{ch3_impactparamneighbors}
\begin{align}
    \label{ch3_impactparamneighbors1}
    &\int \mathrm{d}^2\left(\frac{x_2+x_0}{2}\right) \Gamma_{20,21}(zs) = \Gamma(x_{20}^2,\xto^2,zs) \,,\\
    \label{ch3_impactparamneighbors2}
    &\int \mathrm{d}^2\left(\frac{x_2+x_0}{2}\right)\Gamma^i_{20,21}(zs) = \left(x_{20}\right)^i_\perp \Gamma_1(x_{20}^2,x_{21}^2,zs) + \epsilon^{ij}\left(x_{20}\right)^j_\perp \Gamma_2(x_{20}^2,x_{21}^2,zs)\,
\end{align}
\end{subequations}
where, as with $G_2(x_{10}^2,zs)$, only $\Gamma_2(x_{20}^2,x_{21}^2,zs)$ will contribute from \eq{ch3_impactparamneighbors2}. 

We now have the set of four objects which evolve under the large-$N_c$ version of the evolution from Eqs.~\eqref{ch3_fullevol}: $G(\xoz^2,zs)$, $G_2(\xoz^2,zs)$, $\Gamma(\xoz^2,\xto^2,zs)$, and $\Gamma_2(\xoz^2,\xto^2,zs)$. A number of identities help in reducing Eqs.~\eqref{ch3_fullevol} to a closed set of large-$N_c$ equations, chiefly the relation between fundamental and adjoint Wilson lines
\begin{align}\label{ch3_fundwilsontoadjwilson}
    \left(U_{\underline{x}}\right)^{ba} = 2\,\text{tr}\left[t^bV_{\underline{x}}t^aV_{\underline{x}}^\dagger \right]\,,
\end{align}
and the Fierz identities
\begin{subequations}\label{ch3_fierz}
\begin{align}
    \label{ch3_fierz1}
    &\text{tr}\,\left[t^aM_1t^aM_2\right] = \frac{1}{2}\text{tr}\,M_1\,\text{tr}\,M_2 - \frac{1}{2N_c}\text{tr}\,\left[M_1M_2\right]\,,\\
    \label{ch3_fierz2}
    &\text{tr}\,\left[t^a M_1\right]\text{tr}\,\left[t^aM_1\right] = \frac{1}{2}\text{tr}\left[M_1M_2\right] - \frac{1}{2N_c}\text{tr}\,M_1\,\text{tr}\,M_2\,,
\end{align}
\end{subequations}
where $M_1$ and $M_2$ are $N_c\times N_c$ matrices. Much of the utility of Eqs.~\eqref{ch3_fierz} is the ability to express a trace of multiple Wilson lines (e.g.~$\text{tr}\,[t^aV_{\underline{1}}t^aV_{\underline{0}}^\dagger]$) in terms of multiple traces of individual Wilson lines (e.g.~$\text{tr}\,[V_{\underline{1}}]\,\text{tr}\,[V_{\underline{0}}^\dagger]$) at leading order in $N_c$. This allows us to re-express more complicated Wilson line correlators in Eqs.~\eqref{ch3_fullevol} in terms of the original polarized dipole amplitudes and eventually obtain a closed set of linear equations in those dipole amplitudes, collapsing the infinite hierarchy discussed earlier.

The full simplification of Eqs.~\eqref{ch3_fullevol} into the closed set of large-$N_c$ equations is quite involved (see \cite{Kovchegov:2015pbl, Kovchegov:2018znm,Cougoulic:2022gbk,Tawabutr:2022gei} for details), but the result is the following set of four integral equations that resum the double-logarithmic contributions to the small-$x$ helicity evolution of
the four polarized dipoles $G(\xoz^2,zs)$, $G_2(\xoz^2,zs)$, $\Gamma(\xoz^2,\xto^2,zs)$, and $\Gamma_2(\xoz^2,\xto^2,zs)$:
\begin{subequations}\allowdisplaybreaks\label{ch3_largeNc_eqns_unscaled}
\begin{align}
    \label{ch3_evolG_unscaled}
    & G(\xoz^2,zs) = G^{(0)}(\xoz^2,zs) + \frac{\as N_c}{2\pi}\int_{\tfrac{1}{s\xoz^2}}^{z}\frac{\mathrm{d}z'}{z'}\int_{\tfrac{1}{z's}}^{\xoz^2}\frac{\mathrm{d}\xto^2}{\xto^2}\Bigg[\Gamma(\xoz^2,\xto^2,z's) + 3 \, G(\xto^2,z's) \\
    & \hspace*{8cm} + 2 \, G_2(\xto^2,z's) + 2 \, \Gamma_2(\xoz^2,\xto^2,z's)\Bigg]\,, \notag \\
    \label{ch3_evolGamma_unscaled}
    & \Gamma(\xoz^2,\xto^2,z's) = G^{(0)}(\xoz^2,z's) + \frac{\as N_c}{2\pi}\int_{\tfrac{1}{s\xoz^2}}^{z'}\frac{\mathrm{d}z''}{z''}\int_{\tfrac{1}{z''s}}^{\min\left[\xoz^2,\xto^2\tfrac{z'}{z''} \right] }\frac{\mathrm{d}\xtt^2}{\xtt^2}\\
    &\hspace*{2cm}\times\Bigg[\Gamma(\xoz^2,\xtt^2,z''s) + 3 \, G(\xtt^2,z''s) + 2 \, G_2(\xtt^2,z''s) + 2 \, \Gamma_2(\xoz^2,\xtt^2,z''s)\Bigg]\,,\notag  \\
    \label{ch3_evolG2_unscaled}
    & G_2(\xoz^2,zs) = G_2^{(0)}(\xoz^2,zs) + \frac{\as N_c}{\pi}\int_{\tfrac{\Lambda^2}{s}}^{z}\frac{\mathrm{d}z'}{z'}\int_{\max\left[\xoz^2,\tfrac{1}{z's}\right]}^{\min\left[\tfrac{z}{z'}\xoz^2, \tfrac{1}{\Lambda^2}\right]}\frac{\mathrm{d}\xto^2}{\xto^2}\\
    &\hspace*{9cm}\times[G(\xto^2,z's)+ 2 \, G_2(\xto^2,z's) ]\,,\notag \\
    \label{ch3_evolGamma2_unscaled}
    & \Gamma_2(\xoz^2,\xto^2,z's) =  G_2^{(0)}(\xoz^2,z's)  + \frac{\as N_c}{\pi}\int_{\tfrac{\Lambda^2}{s}}^{z'\tfrac{\xto^2}{\xoz^2}}\frac{\mathrm{d}z''}{z''}\int_{\max\left[\xoz^2,\tfrac{1}{z''s}\right]}^{\min\left[\tfrac{z'}{z''}\xto^2, \tfrac{1}{\Lambda^2}\right]}\frac{\mathrm{d}\xtt^2}{\xtt^2} \\
    &\hspace*{9cm}\times[G(\xtt^2,z''s)+ 2 \, G_2(\xtt^2,z''s) ] \notag\,, 
\end{align}
\end{subequations}
where $\Gamma(\xoz^2,\xto^2,z's)$ and $\Gamma_2(\xoz^2,\xto^2,z's)$ are only defined for $\xto \leq \xoz$ and again $\Lambda$ is an infrared (IR) cutoff such that we require all the dipole sizes $x_{ij}$ to satisfy $x_{ij} <1/\Lambda$. As mentioned earlier, a number of unpolarized $S$-matrices result from the simplification to the large-$N_c$ equations. The full set of equations obtained are actually nonlinear, with unpolarized dipole amplitudes multiplying polarized ones. But again, for our purposes and within the double-logarithmic approximation, the unpolarized dipoles can be set to $1$, resulting in the linearized set of equations above.

Now we move on to the more general large-$N_c\&N_f$ limit, where quark contributions are restored. As such we can no longer work with the large-$N_c$ polarized dipole amplitude $G_{10}(zs)$, but instead need the full type-1 fundamental dipole amplitude $Q_{10}(zs)$. We can still work with the other fundamental dipole $G^{i}_{10}(zs)$ and we still ultimately replace $G^{i}_{10}(zs)$ with the contribution of $G_2(\xoz^2,zs)$. In fact since $G_2(\xoz^2,zs)$, or equivalently $V_{\underline{x}}^{iG[2]}$ (see Eqs.~\eqref{ch3_ViG2}, \eqref{ch3_Gi10}, and \eqref{ch3_G2explicit}), has no sub-eikonal quark contribution, its evolution at large-$N_c\&N_f$ is almost the same as it was at large-$N_c$. We just have to replace the large-$N_c$ amplitude $G$ that enters the evolution kernel of $G_2$ in \eq{ch3_evolG2_unscaled} with its large-$N_c\&N_f$ generalization $\widetilde{G}$ (to be defined shortly) since quark exchanges can still occur at later steps of the evolution. 

As for the adjoint dipoles, the relationship between $G_{10}^{i\,\text{adj}}(zs)$ and $G_{10}^i(zs)$ from \eq{ch3_fundandadjlargenc2} still holds at large-$N_c\&N_f$ so we only need to consider the fundamental $G_{10}^i(zs)$. For the other adjoint dipole $G_{10}^{\text{adj}}(zs)$, it is convenient to define a new polarized Wilson line which is related to the adjoint line $U_{\underline{x}}^{\text{pol}[1]}$ \cite{Cougoulic:2022gbk},
\begin{align}\label{ch3_Wpol1}
    &W_{\underline{x}}^{\text{pol}[1]} = V_{\underline{x}}^{G[1]} - \frac{g^2P^+}{4s}\int\limits_{-\infty}^{\infty}\mathrm{d}x_1^-\int\limits_{x_1^-}^{\infty}\mathrm{d}x_2^- V_{\underline{x}}\left[\infty,x_2^-\right]\psi_{\alpha}\left(x_2^-,\underline{x}\right) \left(\frac{1}{2}\gamma^+\gamma_5\right)_{\beta\alpha}\\
    &\hspace*{8cm}\times\overline{\psi}_{\beta}\left(x_1^-,\underline{x}\right)V_{\underline{x}}\left[x_1^-,-\infty\right] \notag\,,
\end{align}
which we use to construct a new dipole amplitude 
\begin{align}\label{ch3_Gtilde}
    \widetilde{G}_{10}(zs) = \frac{1}{2N_c}\text{Re}\,\llangle \text{T}\,\text{tr}\,\left[V_{\underline{0}}W_{\underline{1}}^{\text{pol}[1]\dagger}\right] + \text{T}\,\text{tr}\,\left[W_{\underline{1}}^{\text{pol}[1]}V_{\underline{0}}^\dagger\right] \rrangle(zs)\,.
\end{align}
At large-$N_c\&N_f$ the new dipole amplitude $\widetilde{G}_{10}(zs)$ is related straightforwardly to the adjoint dipole $G_{10}^{\text{adj}}(zs)$ by \cite{Cougoulic:2022gbk}
\begin{align}\label{ch3_GtildeandGadj}
    G_{10}^{\text{adj}}(zs) = 4S_{10}(zs)\widetilde{G}_{10}(zs)\,.
\end{align}
As usual, we integrate each of the polarized dipoles over impact parameters. For $Q_{10}(zs)$ and $G^i_{10}(zs)$ we have already defined $Q(\xoz^2,zs)$ and $G_2(\xoz^2,zs)$, respectively, in Eqs.~\eqref{ch3_Q} and \eqref{ch3_G2explicit}. For $\widetilde{G}_{10}(zs)$ we similarly define
\begin{align}\label{ch3_Gtildeimpact}
    \widetilde{G}\left(\xoz^2,zs\right) = \int\mathrm{d}^2\left(\frac{x_0+x_1}{2}\right) \widetilde{G}_{10}(zs)\,.
\end{align}
Each of the three dipoles we consider --- $Q(\xoz^2,zs)$, $G_2(\xoz^2,zs)$, and $\widetilde{G}(\xoz^2,zs)$ --- has an associated neighbor dipole --- $\overline{\Gamma}_{20,21}(zs)$, $\Gamma^i_{20,21}(zs)$, and $\widetilde{\Gamma}_{20,21}(zs)$, respectively. Impact-parameter integrations are employed analogously:
\begin{subequations}\label{ch3_largencnfneighborsimpact}
\begin{align}
    &\int\mathrm{d}^2\left(\frac{x_0+x_2}{2}\right)\overline{\Gamma}_{20,21}(zs) = \overline{\Gamma}\left(x_{20}^2,x_{21}^2,zs\right) \,,\\
    &\int\mathrm{d}^2\left(\frac{x_0+x_2}{2}\right)\widetilde{\Gamma}_{20,21}(zs) = \widetilde{\Gamma}\left(x_{20}^2,x_{21}^2,zs\right) \,,
\end{align}
\end{subequations}
with $\Gamma^i_{20,21}(zs)$ integrated over impact parameter and decomposed exactly as it was in \eq{ch3_impactparamneighbors2}, ultimately to be replaced by just the contribution of $\Gamma_2$.

The resulting set of six objects --- $Q(\xoz^2,zs)$, $G_2(\xoz^2,zs)$, $\widetilde{G}(\xoz^2,zs)$ and their respective neighbors $\overline{\Gamma}(\xoz^2,\xto^2,zs)$, $\Gamma_2(\xoz^2,\xto^2,zs)$, and $\widetilde{\Gamma}(\xoz^2,\xto^2,zs)$ --- evolve under the large-$N_c\&N_f$ version of the small-$x$ helicity evolution. The full set of large-$N_c\&N_f$ evolution equations are \cite{Cougoulic:2022gbk}
\begin{subequations}\allowdisplaybreaks\label{ch3_LargeNcNfeqns}
\begin{align}
\label{ch3_lncnf1}
& Q(x^2_{10},zs) = Q^{(0)}(x^2_{10},zs) + \frac{\alpha_sN_c}{2\pi} \int_{1/s x^2_{10}}^{z} \frac{dz'}{z'}   \int_{1/z's}^{x^2_{10}}  \frac{dx^2_{21}}{x_{21}^2}    \bigg[ 2 \, {\widetilde G}(x^2_{21},z's)  \\
&\hspace*{6cm}+ 2 \, {\widetilde \Gamma}(x^2_{10},x^2_{21},z's) + \; Q(x^2_{21},z's) -  \overline{\Gamma}(x^2_{10},x^2_{21},z's) \notag\\
&\hspace*{6cm}+ 2 \, \Gamma_2(x^2_{10},x^2_{21},z's) + 2 \, G_2(x^2_{21},z's) \bigg] \notag \\
&\hspace*{2.5cm}+ \frac{\alpha_sN_c}{4\pi} \int_{\Lambda^2/s}^{z} \frac{dz'}{z'}   \int_{1/z's}^{\min \{ x^2_{10}z/z', 1/\Lambda^2 \}}  \frac{dx^2_{21}}{x_{21}^2} \left[Q(x^2_{21},z's) + 2 \, G_2(x^2_{21},z's) \right] ,  \notag  \\
\label{ch3_lncnf2}
&\overline{\Gamma}(x^2_{10},x^2_{21},z's) = Q^{(0)}(x^2_{10},z's) + \frac{\alpha_sN_c}{2\pi} \int_{1/s x^2_{10}}^{z'} \frac{dz''}{z''}   \int_{1/z''s}^{\min\{x^2_{10}, x^2_{21}z'/z''\}}  \frac{dx^2_{32}}{x_{32}^2}  \\
&\hspace*{4.8cm}\times\bigg[ 2\, {\widetilde G} (x^2_{32},z''s) + \; 2\, {\widetilde \Gamma} (x^2_{10},x^2_{32},z''s) +  Q(x^2_{32},z''s) \notag\\
&\hspace{5cm}-  \overline{\Gamma}(x^2_{10},x^2_{32},z''s) + 2 \, \Gamma_2(x^2_{10},x^2_{32},z''s) + 2 \, G_2(x^2_{32},z''s) \bigg] \notag \\
&\hspace*{2cm}+ \frac{\alpha_sN_c}{4\pi} \int_{\Lambda^2/s}^{z'} \frac{dz''}{z''}   \int_{1/z''s}^{\min \{ x^2_{21}z'/z'', 1/\Lambda^2 \}}  \frac{dx^2_{32}}{x_{32}^2} \left[Q(x^2_{32},z''s) + 2 \, G_2(x^2_{32},z''s) \right] , \notag \\
\label{ch3_lncnf3}
& {\widetilde G}(x^2_{10},zs) = {\widetilde G}^{(0)}(x^2_{10},zs) + \frac{\alpha_s N_c}{2\pi}\int_{1/s x^2_{10}}^z\frac{dz'}{z'}\int_{1/z's}^{x^2_{10}} \frac{dx^2_{21}}{x^2_{21}} \\
&\hspace*{4cm}\times \bigg[3 \, {\widetilde G}(x^2_{21},z's)
+ {\widetilde \Gamma}(x^2_{10},x^2_{21},z's) + \; 2\,G_2(x^2_{21},z's) \notag \\
&\hspace*{4.5cm}+  \left(2 - \frac{N_f}{2N_c}\right) \Gamma_2(x^2_{10},x^2_{21},z's) - \frac{N_f}{4N_c}\,\overline{\Gamma}(x^2_{10},x^2_{21},z's) \bigg] \notag \\
&\hspace*{2.5cm}- \frac{\alpha_sN_f}{8\pi}  \int_{\Lambda^2/s}^z \frac{dz'}{z'}\int_{\max\{x^2_{10},\,1/z's\}}^{\min \{ x^2_{10}z/z', 1/\Lambda^2 \}} \frac{dx^2_{21}}{x^2_{21}}  \left[   Q(x^2_{21},z's) +     2 \, G_2(x^2_{21},z's)  \right] , \notag \\
\label{ch3_lncnf4}
& {\widetilde \Gamma} (x^2_{10},x^2_{21},z's) = {\widetilde G}^{(0)}(x^2_{10},z's) + \frac{\alpha_s N_c}{2\pi}\int_{1/s x^2_{10}}^{z'}\frac{dz''}{z''}\int_{1/z''s}^{\min\{x^2_{10},x^2_{21}z'/z''\}} \frac{dx^2_{32}}{x^2_{32}} \\
&\hspace*{4cm}\times\bigg[3 \, {\widetilde G} (x^2_{32},z''s) + \; {\widetilde \Gamma}(x^2_{10},x^2_{32},z''s) + 2 \, G_2(x^2_{32},z''s) \notag\\
&\hspace*{4.5cm}+  \left(2 - \frac{N_f}{2N_c}\right) \Gamma_2(x^2_{10},x^2_{32},z''s) - \frac{N_f}{4N_c} \,\overline{\Gamma}(x^2_{10},x^2_{32},z''s) \bigg] \notag \\
&\hspace*{1.2cm}- \frac{\alpha_sN_f}{8\pi}  \int_{\Lambda^2/s}^{z'x^2_{21}/x^2_{10}} \frac{dz''}{z''}\int_{\max\{x^2_{10},\,1/z''s\}}^{\min \{ x^2_{21}z'/z'', 1/\Lambda^2 \} } \frac{dx^2_{32}}{x^2_{32}}  \left[   Q(x^2_{32},z''s) +  2  \,  G_2(x^2_{32},z''s)  \right] , \notag \\
\label{ch3_lncnf5}
& G_2(x_{10}^2, z s)  =  G_2^{(0)} (x_{10}^2, z s) + \frac{\as N_c}{\pi} \, \int_{\frac{\Lambda^2}{s}}^z \frac{d z'}{z'} \, \int_{\max \left[ x_{10}^2 , \frac{1}{z' s} \right]}^{\min \{\frac{z}{z'} x_{10}^2, 1/\Lambda^2 \}} \frac{d x^2_{21}}{x_{21}^2} \\
&\hspace*{9cm}\times\left[ {\widetilde G} (x^2_{21} , z' s) + 2 \, G_2 (x_{21}^2, z' s)  \right] ,\notag \\
\label{ch3_lncnf6}
& \Gamma_2 (x_{10}^2, x_{21}^2, z' s)  =  G_2^{(0)} (x_{10}^2, z' s) + \frac{\as N_c}{\pi}  \int_{\frac{\Lambda^2}{s}}^{z' \frac{x_{21}^2}{x_{10}^2}} \frac{d z''}{z''}  \int_{\max \left[ x_{10}^2 , \frac{1}{z'' s} \right]}^{\min \{ \frac{z'}{z''} x_{21}^2, 1/\Lambda^2 \}} \frac{d x^2_{32}}{x_{32}^2} \\
&\hspace*{9cm}\times\left[ {\widetilde G} (x^2_{32} , z'' s) + 2 \, G_2(x_{32}^2, z'' s)  \right] . \notag
\end{align}
\end{subequations}
Again, $\Lambda$ is an infrared cutoff here and we also require $\xto\leq\xoz$ for the neighbors $\overline{\Gamma}$, $\widetilde{\Gamma}$, and $\Gamma_2$.

If one solves the evolution equations for the polarized dipole amplitudes, one can then construct expressions (at small-$x$, large-$N_c$ or large-$N_c\&N_f$, and to DLA accuracy) for the helicity PDFs, helicity TMDs, and the $g_1$ structure function via Eqs.~\eqref{ch3_helicitypdfs}, \eqref{ch3_TMDs}, and \eqref{ch3_g1} (at large-$N_c$, $G(\xoz^2,zs) \approx Q(\xoz^2,zs)$ so one uses the $G(\xoz^2,zs)$ in place of $Q(\xoz^2,zs)$ in Eqs.~\eqref{ch3_DeltaSigma}, \eqref{ch3_quark_TMD}, \eqref{ch3_g1} in that limit). The double-logarithmic large-$N_c$ (Eqs.~\eqref{ch3_largeNc_eqns_unscaled}) and large-$N_c\&N_f$ (Eqs.~\eqref{ch3_LargeNcNfeqns}) evolution equations and their eventual solutions are the focus of the research presented in the following Chapters. 


\section{\texorpdfstring{Status of the Small-$x$ Helicity Evolution Before the Work Presented Below}{Status of the Small x Helicity Evolution Before the Work Presented Below}}

We conclude this Chapter by briefly summarizing the relevant status of the small-$x$ helicity research program as it stood prior to the research that will be presented in the following Chapters.\footnote{Of course there have been plenty of other developments in the small-$x$ helicity program that do not directly connect to the research presented in this dissertation. Unfortunately we cannot give a completely comprehensive overview of the program here.} The full evolution equations in their operator form (Eqs.~\eqref{ch3_fullevol}) along with the large-$N_c$ and large-$N_c\&N_f$ versions (Eqs.~\eqref{ch3_largeNc_eqns_unscaled} and \eqref{ch3_LargeNcNfeqns}) were derived in \cite{Cougoulic:2022gbk}. These equations were ultimately an update to the work of the previous papers \cite{Kovchegov:2015pbl,Kovchegov:2016zex,Kovchegov:2018znm}. The missing ingredient in those papers was the $\underline{\cev{D}}^i\underline{\vec{D}}^i$ operator that contributes to $V_{\underline{x}',\underline{x}}^{G[2]}$ in \eq{ch3_vg2} and ultimately to the polarized dipole amplitude $G_2(\xoz^2,zs)$ in Eqs.~\eqref{ch3_Gi10} and \eqref{ch3_G2explicit}\footnote{We often refer to the versions of the evolution by the initials of the authors affiliated with them. We denote the earlier evolution in \cite{Kovchegov:2015pbl,Kovchegov:2016zex,Kovchegov:2018znm} as the Kovchegov-Pitonyak-Sievert (KPS) evolution, while the updated version from \cite{Cougoulic:2022gbk} was amended by authors Cougoulic, Kovchegov, Tarasov, and Tawabutr and is thus known as the KPS-CTT evolution.}.

As mentioned earlier, pioneering studies of the parton helicity distributions at small-$x$ were carried out by Bartels, Ermolaev, and Ryskin (BER) in the 1990s \cite{Bartels:1995iu,Bartels:1996wc} using the very different framework of infrared evolution \cite{Gorshkov:1966ht,Kirschner:1983di,Kirschner:1994rq,Kirschner:1994vc,Griffiths:1999dj}. BER predicted small-$x$ asymptotic behavior for the helicity PDFs and the $g_1$ structure function in the large-$N_c$ limit of 
\begin{align}\label{ch3_BERasympt}
    \left(\text{BER}\right):\qquad \Delta \Sigma (x, Q^2) \sim \Delta G (x, Q^2) \sim g_1 (x, Q^2) \sim \left( \frac{1}{x} \right)^{3.66 \, \sqrt{\frac{\as N_c}{2\pi}}}\,.
\end{align}
Note that the power $3.66\sqrt{\frac{\as N_c}{2\pi}}$ on this power law is typically called the small-$x$ \textit{intercept}.
The original version of the small-$x$ helicity evolution derived in \cite{Kovchegov:2015pbl,Kovchegov:2016zex,Kovchegov:2018znm} (the KPS evolution derived prior to the inclusion of the new $\underline{\cev{D}}^i\underline{\vec{D}}^i$ operator) appeared to disagree significantly with the predictions of BER in \eq{ch3_BERasympt}, with predicted asymptotics for the quark helicity PDF and $g_1$ structure function of \cite{Kovchegov:2016weo,Kovchegov:2017jxc}
\begin{align}\label{ch3_earlykpsquarkasympt}
\Delta\Sigma(x,Q^2) \sim g_1(x,Q^2) \sim \left(\frac{1}{x}\right)^{2.31\sqrt{\frac{\as N_c}{2\pi}}}\,,
\end{align}
and predicted asymptotics for the gluon helicity PDF \cite{Kovchegov:2017lsr} of
\begin{align}\label{ch3_earlykpsglueasympt}
\Delta G(x,Q^2) \sim \left(\frac{1}{x}\right)^{1.88\sqrt{\frac{\as N_c}{2\pi}}}\,.
\end{align}
However the inclusion of the new operator led to the revised KPS-CTT helicity evolution in \cite{Cougoulic:2022gbk} (these revised evolution equations are the ones presented in the previous Section), and in that same paper the large-$N_c$ evolution from Eqs.~\eqref{ch3_largeNc_eqns_unscaled} was solved numerically. The result of that solution was a numerical prediction for the asymptotic behavior of the helicity distributions and $g_1$ structure function of 
\begin{align}\label{ch3_KPSCTTasympt}
    \left(\text{KPS-CTT}\right):\qquad \Delta \Sigma (x, Q^2) \sim \Delta G (x, Q^2) \sim g_1 (x, Q^2) \sim \left( \frac{1}{x} \right)^{3.66 \, \sqrt{\frac{\as N_c}{2\pi}}}\,,
\end{align}
which, up to the numerical precision of the calculation, appeared to match perfectly with the predictions of BER in \eq{ch3_BERasympt}. Also in \cite{Cougoulic:2022gbk}, the revised large-$N_c$ evolution equations were solved iteratively, order-by-order in $\as$, up to $\mathcal{O}(\as^3)$. This iterative solution contained a prediction, also order-by-order in $\as$, for the gluon-gluon polarized DGLAP splitting function $\Delta P_{GG}(z)$ at small-$x$ and large-$N_c$. Note that, in the gluon-only large-$N_c$ limit, the only remaining polarized DGLAP splitting function is $\Delta P_{GG}(z)$ so that the spin-dependent DGLAP equations in the large-$N_c$ limit become (cf.~\eq{ch2_poldglap})
\begin{align}\label{ch3_DGLAPlargeNc}
    \frac{\partial \Delta G\left(x,Q^2\right)}{\partial \ln Q^2} = \frac{\as\left(Q^2\right)}{2\pi}\int_z^1\frac{\mathrm{d}z}{z}\,\Delta P_{GG}\left(z\right)\,\Delta G\left(\frac{x}{z},Q^2\right)\,. 
\end{align}
The prediction for $\Delta P_{GG}(z)$ from \cite{Cougoulic:2022gbk} could be compared to the small-$x$, large-$N_c$ limit of the same quantity as calculated in the finite-order framework. Exact agreement was found between the prediction of \cite{Cougoulic:2022gbk} and the full three-loop expression of $\Delta P_{GG}(z)$ from the existing finite-order calculations \cite{Altarelli:1977zs,Dokshitzer:1977sg,Mertig:1995ny,Moch:2014sna,Zijlstra:1993sh,Moch:1999eb,vanNeerven:2000uj,Vermaseren:2005qc,Blumlein:2021ryt,Blumlein:2021lmf,Davies:2022ofz,Blumlein:2022gpp}\footnote{Note that one can also obtain a prediction for the small-$x$, large-$N_c$ limit of $\Delta P_{GG}(z)$ in the BER formalism of \cite{Bartels:1995iu,Bartels:1996wc}. That prediction is indeed also consistent with the full 3 loops of existing finite order calculations. In fact, moving to large-$N_c\&N_f$, one can obtain predictions in the BER framework for the other polarized DGLAP splitting functions as well. More on this in Ch.~\ref{transitionops.ch} and Ch.~\ref{largeNcandNfsoln.ch}.}. 

Thus, as published in \cite{Cougoulic:2022gbk}, the revised KPS-CTT small-$x$ helicity evolution had been numerically solved in the asymptotic limit and iteratively solved up to $\mathcal{O}(\as^3)$ in the coupling. The results appeared to indicate full agreement with both the BER predictions for the asymptotics and the finite order calculations of the $GG$ polarized DGLAP splitting function. The remainder of this dissertation details some of the recent developments of the research program dedicated to the double-logarithmic small-$x$ helicity evolution --- in particular, the construction of analytic solutions to the large-$N_c$ and large-$N_c\&N_f$ evolution equations. These analytic solutions will allow for a richer understanding of the small-$x$ helicity structure predicted by the evolution and will also yield more detailed comparisons to existing results. These comparisons will ultimately establish full consistency with finite-order results for all the polarized DGLAP splitting functions (at both large-$N_c$ and large-$N_c\&N_f$) but will also reveal some minor discrepancies with the predictions of BER --- both in polarized splitting functions and intercepts --- discrepancies that were unexpected in light of the apparent agreement of Eqs.~\eqref{ch3_BERasympt} and \eqref{ch3_KPSCTTasympt}. We begin in Ch.~\ref{largeNcsoln.ch} with the large-$N_c$ evolution.


\chapter{\texorpdfstring{Analytic Solution to the Small-$x$ Helicity Evolution at Large $N_c$}{Analytic Solution to the Small x Helicity Evolution at Large Nc}}
\label{largeNcsoln.ch}


In this Chapter, we construct an exact analytic solution of the revised double-logarithmic small-$x$ helicity evolution equations in the large-$N_c$ limit (Eqs.~\eqref{ch3_largeNc_eqns_unscaled}). This solution provides analytic small-$x$, large-$N_c$ expressions for the flavor-singlet quark and gluon helicity PDFs and for the $g_1$ structure function, with their leading small-$x$ asymptotics given by
\begin{align}
        \Delta \Sigma (x, Q^2) \sim \Delta G (x, Q^2) 
    \sim g_1 (x, Q^2) \sim \left( \frac{1}{x} \right)^{\alpha_h} .
\end{align}
Here we obtain an exact analytic expression for the intercept $\alpha_h$ (which had only been obtain numerically in \cite{Cougoulic:2022gbk}), with the numerical value of our analytic expression approximately given by $\alpha_h = 3.66074 \, \sqrt{\frac{\alpha_s \, N_c}{2 \pi}}$. Our solution also yields an all-order (in $\alpha_s$) resummed small-$x$ anomalous dimension $\Delta \gamma_{GG} (\omega)$\footnote{$\Delta \gamma_{GG}(\omega)$ is the Mellin transform of the polarized splitting function $\Delta P_{GG}(z)$, cf. \eq{ch2_dglapanomalousdims}.} which agrees with all the existing finite-order calculations (to three loops). Notably, our anomalous dimension is different from that obtained in the infrared evolution equation (IREE) framework developed earlier by BER \cite{Bartels:1996wc}, with the disagreement starting at four loops.  Despite the apparent numerical agreement observed in \cite{Cougoulic:2022gbk} between Eqs.~\eqref{ch3_BERasympt} and \eqref{ch3_KPSCTTasympt}, the intercept of our large-$N_c$ helicity evolution and that of BER actually disagree beyond that precision, with the BER intercept at large $N_c$ given by a different analytic expression from ours with the numerical value of $\alpha_h^{BER} = 3.66394 \, \sqrt{\frac{\alpha_s \, N_c}{2 \pi}}$. In Appendix~\ref{berdisagreement.app} we speculate on the origin of this disagreement. This Chapter is predominantly based on the paper \cite{Borden:2023ugd}. However we note that the calculation in Ch.~\ref{sec:largencasymptotics} was not originally derived by the author of this dissertation and did not appear in \cite{Borden:2023ugd}. It was derived and published in \cite{Kovchegov:2023yzd} but is included here with the permission of its authors.


\section{\texorpdfstring{Large-$N_c$ Equations}{Large Nc Equations}}

Eqs.~\eqref{ch3_largeNc_eqns_unscaled} contain the double-logarithmic, large-$N_c$, small-$x$ helicity evolution, which is written in terms of the polarized dipole amplitudes $G(\xoz^2,zs)$ and $G_2(\xoz^2,zs)$ along with their respective neighbor dipole amplitudes, denoted $\Gamma(\xoz^2,\xto^2,zs)$ and $\Gamma_2(\xoz^2,\xto^2,zs)$. These dipole amplitudes depend on the transverse separations $\xoz^2$ and $\xto^2$ along with a (small) longitudinal momentum fraction $z$ in the form of the center-of-mass energy squared $zs$ controlling the next step of evolution. For our purposes it is convenient to introduce the set of rescaled variables \cite{Kovchegov:2016weo}
\begin{gather}
\label{originalvars}
    \eta = \sqrt{\frac{\as N_c}{2\pi}}\ln\frac{zs}{\Lambda^2}\,, \quad \eta' = \sqrt{\frac{\as N_c}{2\pi}}\ln\frac{z's}{\Lambda^2}\,, \quad \eta'' = \sqrt{\frac{\as N_c}{2\pi}}\ln\frac{z''s}{\Lambda^2}\,, \\ 
    \soz = \sqrt{\frac{\as N_c}{2\pi}} \ln\frac{1}{x_{10}^2\Lambda^2}\,, \quad \sto = \sqrt{\frac{\as N_c}{2\pi}} \ln\frac{1}{x_{21}^2\Lambda^2}\,, \quad \stt = \sqrt{\frac{\as N_c}{2\pi}} \ln\frac{1}{x_{32}^2\Lambda^2}\,.\notag
\end{gather}
In terms of these, Eqs.~\eqref{ch3_largeNc_eqns_unscaled} can be written somewhat more compactly as
\begin{subequations}\label{largeNc_eqns}\allowdisplaybreaks
\begin{align}
\label{evolG}
& G(\soz,\eta) = G^{(0)}(\soz,\eta) + \int\limits_{\soz}^{\eta}\mathrm{d}\eta' \int\limits_{\soz}^{\eta'}\mathrm{d}\sto \, \bigg[ \Gamma(\soz,\sto,\eta') + 3G(\sto,\eta') \\
&\hspace*{8cm} + 2 \, G_2(s_{21},\eta') + 2 \, \Gamma_2(s_{10},s_{21},\eta')\bigg], \notag \\
\label{evolGamma}
& \Gamma(\soz,\sto,\eta') = G^{(0)}(\soz,\eta') + \Bigg[ \ \int\limits_{\soz}^{\sto}\mathrm{d} \stt \int\limits_{\stt}^{\eta'-\sto+\stt}\mathrm{d}\eta'' + \int\limits_{\sto}^{\eta'}\mathrm{d}\stt \int\limits_{\stt}^{\eta'}\mathrm{d}\eta''  \Bigg] \\ 
& \hspace*{3cm} \times\bigg[ \Gamma(\soz,\stt,\eta'') + 3 \, G(\stt,\eta'') + 2 \, G_2(\stt,\eta'') + 2 \, \Gamma_2(\soz,\stt,\eta'')\bigg], \notag \\
\label{evolG2}
& G_2(\soz,\eta) = G_2^{(0)}(\soz,\eta) + 2 \int\limits_0^{\soz}\mathrm{d}\sto \int\limits_{\sto}^{\eta-\soz+\sto}\mathrm{d}\eta' \bigg[G(\sto,\eta') + 2 \, G_2(\sto,\eta')\bigg], \\
\label{evolGamma2}
& \Gamma_2(\soz,\sto,\eta') = G_2^{(0)}(\soz,\eta') + 2 \int\limits_0^{\soz}\mathrm{d}\stt \int\limits_{\stt}^{\eta'-\sto+\stt}\mathrm{d}\eta'' \bigg[G(\stt,\eta'') + 2 \, G_2(\stt,\eta'')\bigg],
\end{align}
\end{subequations}
where we have changed the order of integration in the integral kernels of Eqs.~\eqref{ch3_evolGamma_unscaled}-\eqref{ch3_evolGamma2_unscaled} and where we require the ordering $0\leq\soz\leq\sto\leq\eta'$ in Eqs.~\eqref{evolGamma} and \eqref{evolGamma2}.

Once the dipole amplitudes $G$ and $G_2$ are determined by solving Eqs.~\eqref{largeNc_eqns}, they can be used to calculate 
the (dipole) gluon and (flavor-singlet) quark helicity TMDs $g^{G \, dip}_{1L}(x,k_T^2)$ and $g^{S}_{1L}(x,k_T^2)$, 
hPDFs $\Delta G(x,Q^2)$ and $\Delta \Sigma(x,Q^2)$, and the $g_1$ structure function at small-$x$ and large-$N_c$, by employing the relations in Eqs.~\eqref{ch3_helicitypdfs}, \eqref{ch3_TMDs}, and \eqref{ch3_g1} (recall at large-$N_c$ we replace $Q(\xoz^2,zs) \rightarrow G(\xoz^2,zs)$).


\section{Solution}

In this Section, we construct our analytic solution to the evolution in Eqs.~\eqref{largeNc_eqns}. Our method is based on a double-inverse-Laplace representation for the polarized dipole amplitudes. 


\subsection{\texorpdfstring{Double-Inverse-Laplace Representation for $G_2$, $\Gamma_2$, $G$}{Double Inverse Laplace Representation for G2, Gamma2, G}}

We begin by writing $G_2(\soz,\eta)$ as a double-inverse-Laplace transform over the variables $\eta-\soz$ and $\soz$:
\begin{align}
\label{G2}
    G_2(\soz,\eta) = \wint \gint \, e^{\omega(\eta-\soz)}e^{\gamma\soz}\gtwg.
\end{align}
The integrals here are taken over infinite straight-line contours in the complex $\omega$- and $\gamma$-planes, parallel to the imaginary axes and to the right of all the integrand's singularities.

We can also introduce corresponding double inverse Laplace transforms for the initial conditions/inhomogeneous terms $G^{(0)}(\soz,\eta)$ and $G_2^{(0)}(\soz,\eta)$,
\begin{subequations}\label{initial_conditions}
\begin{align}
    \label{G0}
    & G^{(0)}(\soz,\eta) = \wint \gint \, e^{\omega(\eta-\soz)}e^{\gamma\soz}G_{\omega\gamma}^{(0)}\,,\\
    \label{G20}
    & G_2^{(0)}(\soz,\eta) = \wint \gint \, e^{\omega(\eta-\soz)}e^{\gamma\soz}G_{2\omega\gamma}^{(0)}\,.
\end{align}
\end{subequations}

Next we observe that Eqs.~\eqref{evolG2} and \eqref{evolGamma2} admit the following scaling property:
\begin{align}
\label{G2Gamma2scaling}
    &\Gamma_2(\soz,\sto,\eta') - G_2^{(0)}(\soz,\eta') = G_2(\soz,\eta=\eta'+\soz-\sto) \\
    &\hspace{6cm}- G_2^{(0)}(\soz,\eta=\eta'+\soz-\sto). \notag
\end{align}
Using Eqs.~\eqref{G2} and \eqref{G20} in \eq{G2Gamma2scaling} we immediately have
\begin{align}
\label{Gamma2}
    \Gamma_2(\soz,\sto,\eta') = \wint\gint \left[e^{\omega(\eta'-\sto)}e^{\gamma\soz}\left(\gtwg - \gtwg^{(0)}\right) + e^{\omega(\eta'-\soz)}e^{\gamma\soz} \, \gtwg^{(0)} \right].
\end{align}

Now we write the amplitude $G(\soz,\eta)$ as a double inverse Laplace transform
\begin{align}
\label{Gprelim}
    G(\soz,\eta) = \wint \gint \, e^{\omega(\eta-\soz)}e^{\gamma\soz}G_{\omega\gamma}\,,
\end{align}
substitute Eqs.~\eqref{G2}, \eqref{G20}, and \eqref{Gprelim} into \eqref{evolG2} and perform the integrals over $\eta'$ and $\sto$. Next, applying the forward Laplace transforms over $\eta-\soz$ and $\soz$ (treating those as two independent variables) yields

\begin{align}\label{c3}
    G_{2 \omega \gamma} = G_{2 \omega \gamma}^{(0)} + \frac{2}{\omega \gamma} \left[G_{\omega\gamma} + 2 \, G_{2 \omega \gamma} \right].
\end{align}
Solving \eq{c3} for $G_{\omega\gamma}$ we arrive at
\begin{align}
\label{G_omegagamma_0}
    G_{\omega\gamma} = \tfrac{1}{2}\omega\gamma\left(\gtwg - \gtwg^{(0)} \right) - 2\gtwg\,,
\end{align}
so that \eq{Gprelim} gives
\begin{align}
\label{G}
    G(\soz,\eta) = \wint \gint e^{\omega(\eta-\soz)}e^{\gamma\soz}\left[\tfrac{1}{2}\omega\gamma\left(\gtwg-\gtwg^{(0)} \right) - 2\gtwg \right].
\end{align}
This way, we have obtained double-inverse-Laplace transform representations for the dipole amplitudes $G_2, \Gamma_2$, and $G$ given in Eqs.~\eqref{G2}, \eqref{Gamma2} and \eqref{G}, respectively.

From Eqs.~\eqref{evolG2} and \eqref{evolGamma2} we have several boundary conditions which our expressions for $G_2(\soz,\eta)$ and $\Gamma_2(\soz,\sto,\eta')$ must satisfy. We need
\begin{subequations}
\begin{align}
    \label{G2_bc0)}
    G_2(\soz=0,\eta) &= G_2^{(0)}(\soz=0,\eta)\,, \\
    \label{G2_bc1}
    G_2(\soz,\eta=\soz) &= G_2^{(0)}(\soz,\eta=\soz)\,, \\
    \label{Gamma2_bc0}
    \Gamma_2(\soz=0,\sto,\eta') &= G_2^{(0)}(\soz=0,\eta')\,, \\
    \label{Gamma2_bc1}
    \Gamma_2(\soz,\sto,\eta'=\sto) &= G_2^{(0)}(\soz,\eta'=\sto)\,.
\end{align}
\end{subequations}
Using Eqs.~\eqref{G2} and \eqref{G20}, we see that Eqs.~\eqref{G2_bc0)} and \eqref{G2_bc1} give respectively
\begin{subequations}\label{G2_bcs_laplace}
\begin{align}
    \label{G2_bcs_laplace_0}
    \wint \gint e^{\omega\eta}\gtwg &= \wint \gint e^{\omega\eta}\gtwg^{(0)}\,,\\
    \label{G2_bcs_laplace_1}
    \wint \gint e^{\gamma\soz}\gtwg &= \wint \gint e^{\gamma\soz}\gtwg^{(0)}. 
\end{align}
\end{subequations}
Note that the constraints resulting from Eqs.~\eqref{Gamma2_bc0} and \eqref{Gamma2_bc1} are equivalent to Eqs.~\eqref{G2_bcs_laplace}. 

Since $\gtwg, G_{\omega\gamma}$ must go to zero as $\omega\rightarrow\infty$ or $\gamma\rightarrow\infty$ in order for the Laplace transforms to exist, the second term on the right-hand side of \eq{c3} goes to zero faster than $1/\omega$ or $1/\gamma$ as $\omega\rightarrow\infty$ or $\gamma\rightarrow\infty$. This implies that
\begin{align}
    \int \frac{d \omega}{2\pi i} \, \frac{2}{\omega \gamma} \left[G_{\omega\gamma} + 2 \, G_{2 \omega \gamma} \right] = 0
\end{align}
and
\begin{align}
    \int \frac{d \gamma}{2\pi i} \, \frac{2}{\omega \gamma} \left[G_{\omega\gamma} + 2 \, G_{2 \omega \gamma} \right] = 0,
\end{align}
where the equalities follow from the $\omega$- and $\gamma$-contours being located to the right of all the singularities of the integrand, allowing us to close those contours to the right. We see that the conditions in Eqs.~\eqref{G2_bcs_laplace} are automatically satisfied by \eq{c3} and conclude that at this point Eqs.~\eqref{evolG2} and \eqref{evolGamma2} are completely solved.


\subsection{\texorpdfstring{Double Inverse Laplace Representation for $\Gamma$}{Double Inverse Laplace Representation for Gamma}}

Differentiating \eq{evolGamma} one can show that $\Gamma(\soz,\sto,\eta')$ satisfies the partial differential equation
\begin{align}
\label{Gamma_pde}
    &\frac{\partial^2\Gamma(\soz,\sto,\eta')}{\partial \sto^2} +  \frac{\partial^2\Gamma(\soz,\sto,\eta')}{\partial \sto \partial \eta'} + \Gamma(\soz,\sto,\eta') \\
    &\hspace{6cm} = -3 \, G(\sto,\eta') - 2 \, G_2(\sto,\eta') - 2\, \Gamma_2(\soz,\sto,\eta'). \notag
\end{align}
This second-order partial differential equation has two solutions, homogeneous and particular, which we label (h) and (p), respectively,
\begin{align}\label{Gamma_sol1}
    \Gamma(\soz,\sto,\eta') = \Gamma^{(h)}(\soz,\sto,\eta') + \Gamma^{(p)}(\soz,\sto,\eta') .
\end{align}
Looking for the homogeneous solution of the form 
\begin{align}
\Gamma^{(h)}(\soz,\sto,\eta')  = \wint\gint e^{\omega(\eta'-\sto)}e^{\gamma\sto}\Gamma_{\omega\gamma} (\soz)
\end{align}
and differentiating according to \eq{Gamma_pde} one arrives at the condition 
\begin{align}
\gamma^2 - \omega\gamma + 1 = 0 ,
\end{align}
which yields two solutions, $\gamma = \delta^+_\omega$ and $\gamma = \delta^-_\omega$, defined by
\begin{align}\label{delta_omega_pm}
   \dw^\pm \equiv \frac{\omega}{2}\left[1\pm\sqrt{1-\tfrac{4}{\omega^2}} \right].
\end{align}
Thus the homogeneous solution can be written as a linear combination of these two solutions,
\begin{align}
\label{HomogeneousGammaSimple2}
    \Gamma^{(h)}(\soz,\sto,\eta') = \wint \, e^{\omega(\eta'-\sto)}\big[ \Gamma_{\omega}^+(\soz) \, e^{\dw^+\sto} + \Gamma_{\omega}^-(\soz) \, e^{\dw^-\sto}   \big] ,
\end{align}
with some unknown functions $\Gamma_{\omega}^+(\soz)$ and $\Gamma_{\omega}^-(\soz)$. 

To construct a particular solution of \eq{Gamma_pde}, one can substitute Eqs.~\eqref{G}, \eqref{G2}, and \eqref{Gamma2} into the right hand side of \eq{Gamma_pde}, which motivates an ansatz for the particular solution of the form 
\begin{align}
\label{Gamma_particular_ansatz}
    &\Gamma^{(p)}(\soz,\sto,\eta') = \wint \gint \bigg[A_{\omega\gamma} \, e^{\omega(\eta'-\sto)}e^{\gamma\sto} + B_{\omega\gamma} \, e^{\omega(\eta'-\sto)}e^{\gamma\soz} \\
    &\hspace{6cm} + C_{\omega\gamma} \, e^{\omega(\eta'-\soz)}e^{\gamma\soz}\bigg].\notag
\end{align}
Substitution of \eq{Gamma_particular_ansatz} into \eq{Gamma_pde} allows one to determine the coefficients $A_{\omega\gamma}$, $B_{\omega\gamma}$, and $C_{\omega\gamma}$. The particular solution found this way is 
\begin{align}
\label{GammaParticularSimple}
    &\Gamma^{(p)}(\soz,\sto,\eta') = \wint \gint\\
    &\hspace{4cm}\times\bigg[e^{\omega(\eta'-\sto)}e^{\gamma\sto}\bigg(\stf \gtwg + 
    \frac{\frac{3}{2}\omega\gamma}{\gamma^2-\omega\gamma+1}\gtwg^{(0)} \bigg) \notag \\
    &\hspace{4.4cm}-2 \, e^{\omega(\eta'-\sto)}e^{\gamma\soz}\big[\gtwg - \gtwg^{(0)} \big]
    -2 \, e^{\omega(\eta'-\soz)}e^{\gamma\soz}\gtwg^{(0)} \bigg] . \notag
\end{align}

Combining the homogeneous \eqref{HomogeneousGammaSimple2} and particular \eqref{GammaParticularSimple} solutions we arrive at the general solution of \eq{Gamma_pde}, 
\begin{align}\label{Gamma}
    &\Gamma(\soz,\sto,\eta') = \wint e^{\omega(\eta'-\sto)}\left[\Gamma_{\omega}^+(\soz)e^{\dw^+\sto} + \Gamma_{\omega}^-(\soz)e^{\dw^-\sto} \right] \\
    &\hspace{2cm}+\wint\gint \Bigg[e^{\omega(\eta'-\sto)}e^{\gamma\sto}\left(\frac{- \tfrac{3}{2}\omega\gamma + 4}{\gamma^2 - \omega\gamma + 1}\gtwg + \frac{\tfrac{3}{2}\omega\gamma}{\gamma^2 - \omega\gamma + 1}\gtwg^{(0)}  \right)\notag \\
    &\hspace{4.8cm} -2 \, e^{\omega(\eta'-\sto)}e^{\gamma\soz}\left(\gtwg - \gtwg^{(0)}\right) - 2 \, e^{\omega(\eta'-\soz)}e^{\gamma\soz}\gtwg^{(0)}\Bigg]\notag.
\end{align}
The integral in \eq{Gamma} is not defined until we specify the location of the poles in the new denominator, $\gamma^2 - \omega\gamma + 1 = -\gamma[\omega-(\gamma+\tfrac{1}{\gamma}) ] = (\gamma-\dw^+)(\gamma-\dw^-)$, with respect to the $\omega$- and $\gamma$-contours. A simple analysis shows that one cannot have both the $\omega$- and $\gamma$-contours to the right of the new poles. Indeed, since
\begin{align}
    \gamma^2-\omega\gamma+1 = (\gamma - \delta_\omega^+) \, (\gamma - \delta_\omega^-) \stackrel{|\omega| \to \infty}{\longrightarrow} (\gamma - \omega) \, \left( \gamma - \frac{1}{\omega} \right),
\end{align}
we see that if Re~$\omega > $~Re~$\gamma$ then pole at $\gamma = \omega$ is to the right of the $\gamma$-contour and to the left of the $\omega$-contour; if Re~$\omega <$~Re~$\gamma$ then the $\gamma = \omega$ pole is to the left of the $\gamma$-contour and to the right of the $\omega$-contour.  
Thus if we choose the $\omega$-contour to be to the right of the singularity at $\omega = \gamma + 1/\gamma$ generated by the new denominator, the $\gamma$-contour must pass between the $\gamma = \dw^+$ and $\gamma = \dw^-$ poles. We stress that the locations of $\omega$- and $\gamma$-contours here are a choice, affecting both the homogeneous and particular solutions simultaneously: different choices for the contours' locations would result in different $\Gamma_{\omega}^+(\soz)$ and $\Gamma_{\omega}^-(\soz)$. As we will see below, the residue at $\gamma=\dw^+$ will be zero in the final answer. Therefore, this new pole to the right of the $\gamma$-contour will vanish, such that all the $\gamma$-singularities of the integrand will still be to the left of the $\gamma$-contour, as expected for an inverse Laplace transform. However, we will need to keep this pole in mind later when we invert an integral over $\gamma$.


\subsection{\texorpdfstring{Constraints on $\Gamma$}{Constraints on Gamma}}

Note that we might have lost some of the information of \eq{evolGamma} when we differentiated it to obtain \eq{Gamma_pde}. Hence, the expression \eqref{Gamma}, while a solution of the differential equation~\eqref{Gamma_pde}, may not yet be a solution of \eq{evolGamma}. To fully satisfy \eq{evolGamma} we take our expression \eqref{Gamma} for $\Gamma(\soz,\sto,\eta')$ along with the other three amplitudes given in Eqs.~\eqref{G2}, \eqref{Gamma2}, and \eqref{G} and substitute them all back into \eq{evolGamma}. Performing the integrals over $\eta''$ and $\stt$ and also making use of the facts that $\dw^+\dw^- = 1$ and $\dw^+ + \dw^- = \omega$ (as can be seen from \eq{delta_omega_pm}), we obtain
\begin{align}
\label{back_sub_for_Gamma_0}
    0 = &\wint \gint e^{\omega(\eta'-\sto)}e^{\gamma\soz} \\
    &\hspace{1.4cm}\times\Bigg\{ \frac{\gamma-\omega}{\omega}\left[\frac{- \tfrac{3}{2}\omega\gamma + 4}{\gamma^2-\omega\gamma+1}\gtwg + \frac{\tfrac{3}{2}\omega\gamma}{\gamma^2-\omega\gamma+1}\gtwg^{(0)} \right]+ 2\left(\gtwg-\gtwg^{(0)} \right) \Bigg\} \notag \\
    &+ \wint\gint e^{\omega(\eta'-\soz)}e^{\gamma\soz}\left(G_{\omega\gamma}^{(0)} + 2\gtwg^{(0)}\right)\notag\\
    &-\wint \bigg\{ \Gamma_{\omega}^+(\soz)\left[ \frac{e^{\omega(\eta'-\sto)}e^{\dw^+\soz}}{\omega \, \dw^+} + e^{\dw^+\eta'} - \frac{e^{\dw^+\soz}}{\omega \, \dw^+}\right] \notag \\
    &\hspace{1.5cm}+ \Gamma_{\omega}^- (\soz)\left[ \frac{e^{\omega(\eta'-\sto)}e^{\dw^- \soz}}{\omega \, \dw^-} + e^{\dw^-\eta'} - \frac{e^{\dw^-\soz}}{\omega \, \dw^-}\right]  \bigg\} \notag.
\end{align}
In arriving at \eq{back_sub_for_Gamma_0} we have dropped the following term:
\begin{align}
    \wint \gint \left(\frac{e^{\gamma\eta'}}{\gamma(\gamma-\omega)} + \frac{e^{\gamma\soz}}{\omega\gamma}\right)\gamma(\gamma-\omega)\left(\frac{(\tfrac{3}{2}\omega\gamma - 4)\gtwg}{\gamma^2-\omega\gamma+1} - \frac{\tfrac{3}{2}\omega\gamma\gtwg^{(0)}}{\gamma^2-\omega\gamma+1} \right).
\end{align}
With no $\omega$ in the exponent in this term, we can safely close the $\omega$-contour to the right. Then using the fact that $\gtwg$, $\gtwg^{(0)}$ must go to zero for $\omega\rightarrow\infty$ along with the property in \eq{G2_bcs_laplace_1}, one can show that this entire term is zero.

Now performing the forward Laplace transform over $\eta'$ in \eq{back_sub_for_Gamma_0} we obtain
\begin{align}
\label{back_sub_for_Gamma_1}
    0 &= e^{-\omega\sto} \, \gint \, e^{\gamma\soz}\\
    &\hspace{1.5cm}\times\Bigg\{ \frac{\gamma-\omega}{\omega}\left[\frac{\tfrac{-3}{2}\omega\gamma + 4}{\gamma^2-\omega\gamma+1}\gtwg + \frac{\tfrac{3}{2}\omega\gamma}{\gamma^2-\omega\gamma+1}\gtwg^{(0)} \right] + 2\left(\gtwg-\gtwg^{(0)} \right)\Bigg\}\notag \\
    &- \Gamma_\omega^+(\soz) \frac{e^{-\omega\sto}e^{\dw^+\soz}}{\omega \, \dw^+} - \Gamma_\omega^-(\soz) \frac{e^{-\omega\sto}e^{\dw^-\soz}}{\omega \, \dw^-}\notag \\
    &+\gint e^{-\omega\soz}e^{\gamma\soz}\bigg(G_{\omega\gamma}^{(0)} + 2\gtwg^{(0)}\bigg) + \int \frac{\mathrm{d}\omega'}{2\pi i}\Bigg(\frac{\Gamma_{\omega'}^+(\soz)}{\delta_{\omega'}^+ - \omega} + \frac{\Gamma_{\omega'}^-(\soz)}{\delta_{\omega'}^- - \omega} \Bigg) \notag \\
    & + \frac{1}{\omega}\int \frac{\mathrm{d}\omega'}{2\pi i}\left(\Gamma_{\omega'}^+(\soz)\frac{e^{\delta_{\omega'}^+\soz}}{\omega' \, \delta_{\omega'}^+} + \Gamma_{\omega'}^-(\soz)\frac{e^{\delta_{\omega'}^-\soz}}{\omega' \, \delta_{\omega'}^-}\right)\notag .
\end{align}
Note that the terms in the first two lines of \eq{back_sub_for_Gamma_1} have the same $\sto$-dependence, $\propto e^{-\omega\sto}$, whereas the last two lines are independent of $\sto$. Since \eq{back_sub_for_Gamma_1} must be valid for all $\sto >0$, we conclude that the sum of the first two lines in \eq{back_sub_for_Gamma_1} must be separately equal to zero, as must the sum of the last two lines. This gives two constraints
\begin{subequations}\label{b-cond}
\begin{align}
    \label{back_sub_for_Gamma_2}
    & \gint e^{\gamma\soz}\Bigg\{ \frac{\gamma-\omega}{\omega}\left[\frac{\tfrac{-3}{2}\omega\gamma + 4}{\gamma^2-\omega\gamma+1}\gtwg + \frac{\tfrac{3}{2}\omega\gamma}{\gamma^2-\omega\gamma+1}\gtwg^{(0)} \right] + 2\left(\gtwg-\gtwg^{(0)} \right)\Bigg\} \\
    &\hspace*{1cm} = \Gamma_\omega^+(\soz) \frac{e^{\dw^+\soz}}{\omega \, \dw^+} + \Gamma_\omega^-(\soz) \frac{e^{\dw^-\soz}}{\omega \, \dw^-}\,,\notag\\
    \label{back_sub_for_Gamma_3}
    & 0 = \gint e^{-\omega\soz}e^{\gamma\soz}\bigg(G_{\omega\gamma}^{(0)} + 2\gtwg^{(0)}\bigg) + \int \frac{\mathrm{d}\omega'}{2\pi i}\Bigg(\frac{\Gamma_{\omega'}^+(\soz)}{\delta_{\omega'}^+ - \omega} + \frac{\Gamma_{\omega'}^-(\soz)}{\delta_{\omega'}^- - \omega} \Bigg) \\
    & \hspace*{1cm} + \frac{1}{\omega}\int \frac{\mathrm{d}\omega'}{2\pi i}\left(\Gamma_{\omega'}^+(\soz)\frac{e^{\delta_{\omega'}^+\soz}}{\omega' \, \delta_{\omega'}^+} + \Gamma_{\omega'}^-(\soz)\frac{e^{\delta_{\omega'}^-\soz}}{\omega' \, \delta_{\omega'}^-}\right)\,.\notag
\end{align}    
\end{subequations}
If we satisfy the conditions \eqref{b-cond}, we will have solved \eq{evolGamma}.

Before we do that, we observe that the evolution equation~\eqref{evolG} for $G(\soz,\eta)$ is just a special case of \eq{evolGamma} for $\Gamma(\soz,\sto,\eta)$. Then to ensure that \eq{evolG} is satisfied as well, we impose the condition
\begin{align}\label{G=Gamma}
\Gamma(\soz,\sto=\soz,\eta) = G(\soz,\eta), 
\end{align}
which follows from Eqs.~\eqref{evolG} and \eqref{evolGamma}. We then substitute the dipole amplitudes from Eqs.~\eqref{G} and \eqref{Gamma} into \eq{G=Gamma} and perform the forward Laplace transform over $\eta-\soz$, obtaining
\begin{align}
\label{G_Gamma_scaling}
    &\gint e^{\gamma\soz}\left[ \frac{\tfrac{3}{2}\omega\gamma-4}{\gamma^2-\omega\gamma+1}\gtwg -  \frac{\tfrac{3}{2}\omega\gamma}{\gamma^2-\omega\gamma+1}\gtwg^{(0)} + \tfrac{1}{2}\omega\gamma\left(\gtwg - \gtwg^{(0)} \right)  \right] 
    \\
    &\hspace*{1cm}= \Gamma_{\omega}^+(\soz) \, e^{\dw^+\soz} + \Gamma_{\omega}^-(\soz) \, e^{\dw^-\soz}\,. \notag
\end{align}

Equations~\eqref{back_sub_for_Gamma_2} and \eqref{G_Gamma_scaling} can be solved to give individual expressions for $\Gamma_{\omega}^{+}(\soz)$ and $\Gamma_{\omega}^{-}(\soz)$. Making use of the properties $(\dw^\pm)^2 - \omega\dw^\pm + 1 = 0$, $\dw^+\dw^- = 1$, $\dw^+ + \dw^- = \omega$, which follow from \eq{delta_omega_pm}, the results can be written as
\begin{subequations}\label{Gamma_omega_pm}
\begin{align}
    \label{Gamma_omega_plus}
    &\Gamma_{\omega}^+(\soz) = \frac{e^{-\dw^+\soz}}{\dw^+-\dw^-}\gint e^{\gamma\soz}\frac{\omega \, \dw^+}{2 \, (\gamma-\dw^+)}\\
    &\hspace*{4cm}\times\left[\gtwg\left(\gamma^2-\omega\gamma+4-\tfrac{8}{\omega} \, \dw^- \right) - \gtwg^{(0)}\left(\gamma^2-\omega\gamma+4\right)  \right], \notag\\
    \label{Gamma_omega_minus}
    &\Gamma_{\omega}^-(\soz) = \frac{e^{-\dw^-\soz}}{\dw^--\dw^+}\gint e^{\gamma\soz}\frac{\omega \, \dw^-}{2 \, (\gamma-\dw^-)}\\
    &\hspace*{4cm}\times\left[\gtwg\left(\gamma^2-\omega\gamma+4-\tfrac{8}{\omega} \, \dw^+ \right) - \gtwg^{(0)}\left(\gamma^2-\omega\gamma+4\right)  \right]. \notag
\end{align}
\end{subequations}
We are only left with \eq{back_sub_for_Gamma_3} to satisfy. Employing \eq{c3} in Eqs.~\eqref{Gamma_omega_pm} along with the fact that $G_{2\omega \gamma}, G_{2\omega \gamma}^{(0)} \rightarrow 0$ as $\omega \rightarrow \infty$ we conclude that 
\begin{align}
 e^{\dw^+\soz} \, \Gamma_{\omega}^+(\soz) \to 0, \ \ \  e^{\dw^- \soz} \, \omega \, \Gamma_{\omega}^- (\soz) \to 0 \ \ \ \textrm{when} \ \ \  \omega \to \infty.
\end{align}
We have also employed the fact that $\dw^+ \to \omega$ and $\dw^- \to 1/\omega$ as $\omega \to \infty$, which again follows from \eq{delta_omega_pm}. This allows us to close the $\omega'$ contour to the right in the last term of \eq{back_sub_for_Gamma_3}, obtaining zero.  
\eq{back_sub_for_Gamma_3} then becomes
\begin{align}\label{back_sub_for_Gamma_4}
    &\gint e^{-\omega\soz}e^{\gamma\soz}\bigg(G_{\omega\gamma}^{(0)} + 2 \, \gtwg^{(0)}\bigg) = \int \frac{\mathrm{d}\omega'}{2\pi i}\Bigg(\frac{\Gamma_{\omega'}^+(\soz)}{\omega - \delta_{\omega'}^+} + \frac{\Gamma_{\omega'}^-(\soz)}{\omega - \delta_{\omega'}^-} \Bigg) \\
    &\hspace{3.8cm}= -\int \frac{\mathrm{d}\omega'}{2\pi i}\Bigg(\frac{\omega-\delta_{\omega'}^-}{\omega}\frac{\Gamma_{\omega'}^+(\soz)}{\omega'-\left(\omega+\tfrac{1}{\omega} \right)} + \frac{\omega-\delta_{\omega'}^+}{\omega}\frac{\Gamma_{\omega'}^-(\soz)}{\omega'-\left(\omega+\tfrac{1}{\omega}\right)} \Bigg)\notag\,,
\end{align}
where we have again used the properties $\dw^+\dw^- = 1$, $\dw^+ + \dw^- = \omega$ to obtain the second line of \eq{back_sub_for_Gamma_4}. Now we can close the $\omega'$-contour to the right, picking up the pole at $\omega' = \omega + \tfrac{1}{\omega}$ in each term. Using $\delta^-_{\omega+\frac{1}{\omega}} = \frac{1}{\omega}$ and $\delta^+_{\omega+\frac{1}{\omega}} = \omega$ (as can be seen from \eq{delta_omega_pm}) we see that the $\Gamma^-_{\omega'}(\soz)$ term vanishes. Then \eq{back_sub_for_Gamma_4} has become
\begin{align}\label{back_sub_for_Gamma_5}
    \gint e^{\gamma\soz}\bigg(G_{\omega\gamma}^{(0)} + 2\gtwg^{(0)}\bigg) = \left(1-\frac{1}{\omega^2}\right) \,\Gamma^+_{\omega+\frac{1}{\omega}}\,,
\end{align}
where we have defined $\Gamma_\omega^+$ without the argument $\soz$
by $\Gamma_\omega^+(\soz) \equiv \Gamma_\omega^+ \, e^{-\dw^+\soz}$. Again using $\delta^+_{\omega + \frac{1}{\omega}} = \omega$, $\delta^-_{\omega + \frac{1}{\omega}} = 1/\omega$, we rewrite \eq{back_sub_for_Gamma_5} as
\begin{align}\label{back_sub_for_Gamma_5.3}
    \gint e^{\gamma\soz}\bigg(G_{\delta^+_{\omega + \frac{1}{\omega}} \gamma}^{(0)} + 2 \, G^{(0)}_{2 \, \delta^+_{\omega + \frac{1}{\omega}} \gamma} \bigg) = \left(1-\frac{1}{\left[ \delta^+_{\omega + \frac{1}{\omega}} \right]^2}\right) \, \Gamma^+_{\omega+\frac{1}{\omega}}\,,
\end{align}
or, replacing $\omega + \tfrac{1}{\omega} \to \omega$, as
\begin{align}\label{back_sub_for_Gamma_5.7}
    \gint e^{\gamma\soz}\bigg(G_{\delta^+_{\omega} \gamma}^{(0)} + 2 \, G^{(0)}_{2 \, \delta^+_{\omega} \gamma} \bigg) = \left(1-\frac{1}{\left[ \delta^+_{\omega} \right]^2}\right) \, \Gamma^+_{\omega}\, .
\end{align}
Next we need to invert the remaining transform. Beginning this process and employing \eq{Gamma_omega_plus} we write
\begin{align}\label{back_sub_for_Gamma_fix}
    &G_{\delta^+_{\omega} \gamma}^{(0)} + 2 \, G^{(0)}_{2 \, \delta^+_{\omega} \gamma} = e^{-\dw^+\soz}\int\frac{\mathrm{d}\gamma'}{2\pi i} \frac{1}{\gamma-\gamma'}\left(1-\frac{1}{\left[\dw^+\right]^2}\right) \frac{1}{\dw^+-\dw^-}\,\frac{\omega\dw^+}{2\left(\gamma'-\dw^+\right)} \\
    &\hspace*{3cm}\times\left[G_{2\omega\gamma'}\left(\left(\gamma'\right)^2 - \omega\gamma' + 4 -\tfrac{8}{\omega}\dw^- \right) - G^{(0)}_{2\omega\gamma'}\left(\left(\gamma'\right)^2 - \omega\gamma' + 4\right) \right]\,. \notag 
\end{align}
Now we need to carry out the remaining integral on the right hand side of \eq{back_sub_for_Gamma_fix}. We do this by closing the $\gamma'$-contour to the right and picking up the pole at $\gamma'=\gamma$. However there is another pole we must pick up in this procedure. Recalling the discussion after \eq{Gamma}, the pole at $\gamma'=\delta^+_\omega$ lies to the right of the $\gamma'$ contour. Thus we close the $\gamma'$-contour to the right, pick up both poles, and write the result as the sum of the two residues:
\begin{align}\label{G2_omega_gamma_constraint0}
    G^{(0)}_{\dw^+\gamma} + 2 \, G^{(0)}_{2 \, \dw^+\gamma} = \frac{\omega}{2 \, (\gamma-\dw^+)}\bigg[&\gtwg\left(\gamma-\gamma^-_\omega\right)\left(\gamma-\gamma^+_\omega\right) - \gtwg^{(0)}\left(\gamma^2-\omega\gamma+4\right) \\
    &- G_{2\omega\dw^+}\left(\dw^+ - \gamma^-_\omega \right)\left(\dw^+ - \gamma^+_\omega \right) + 3 \, G^{(0)}_{2\omega\dw^+}\bigg]\notag.
\end{align}
In arriving at \eq{G2_omega_gamma_constraint0} we have again used $(\dw^+)^2 - \omega\dw^+ + 1 =0$ and have also defined
\begin{align}
\gamma^2 -\omega\gamma + 4 - \frac{8}{\omega} \, \dw^- \equiv \left(\gamma-\gamma^-_\omega\right)\left(\gamma-\gamma^+_\omega\right)
\end{align}
with
\begin{align}\label{gamma_omega_pm}
    \gamma^{\pm}_\omega = \frac{\omega}{2}\left[1 \pm \sqrt{1 - \frac{16}{\omega^2}\sqrt{1-\frac{4}{\omega^2}}}  \right].
\end{align}
Note also that the pole at $\gamma = \dw^+$ is not present on the right-hand side of \eq{G2_omega_gamma_constraint0} (and therefore is also not present on the equation's left-hand side), ensuring that our inverse-Laplace transforms remain well-defined with no poles to the right of their integration contours.

Satisfying the constraint in \eq{G2_omega_gamma_constraint0} would complete the solution of Eqs.~\eqref{largeNc_eqns} by expressing $\gtwg$ in terms of $\gtwg^{(0)}$ and $G^{(0)}_{\omega \gamma}$. Equation~\eqref{G_omegagamma_0} would then allow us to find $G_{\omega\gamma}$, after which all the dipole amplitudes can be constructed using Eqs.~\eqref{G2}, \eqref{Gamma2}, \eqref{G}, \eqref{Gamma}, and \eqref{Gamma_omega_pm}. The quantities $\gtwg^{(0)}$ and $G^{(0)}_{\omega \gamma}$ are specified by the initial conditions/inhomogeneous terms.

The only remaining problem is that \eq{G2_omega_gamma_constraint0} contains $\gtwg$ with two different arguments --- $\gtwg$ itself along with $G_{2\omega\dw^+}$. This makes the equation harder to solve for $\gtwg$. However, we can solve \eq{G2_omega_gamma_constraint0} for $G_{2\omega\dw^+}$ by setting $\gamma = \gamma^+_\omega$ in it.  The term $\gtwg\left(\gamma-\gamma^-_\omega\right)\left(\gamma-\gamma^+_\omega\right)$ will vanish as long as $\gtwg$ does not have a pole at $\gamma=\gamma^+_\omega$. However, we assumed this to be true from the beginning: in writing the inverse Laplace transform \eqref{G2}, we assumed that all singularities of $G_{2 \omega \gamma}$ are to the left of the $\gamma$ and $\omega$ integration contours. Since $\gamma^+_\omega \to \omega$ as $\omega \to \infty$, the pole at $\gamma=\gamma^+_\omega$ becomes a pole at $\gamma = \omega$ for large $\omega$. If the $\omega$ and $\gamma$ integration contours were chosen in \eq{G2} such that Re~$\omega >$~Re~$\gamma$, then a pole at $\gamma = \omega$ would violate the assumption of the $\gamma$-contour being to the right of all the singularities of the integrand. If the $\omega$ and $\gamma$ integration contours were chosen such that Re~$\omega <$~Re~$\gamma$, the same argument would apply to the $\omega$ contour. Hence, by writing \eq{G2} we assumed that the pole at $\gamma=\gamma^+_\omega$ in $\gtwg$ does not exist. 

Putting $\gamma = \gamma^+_\omega$ in \eq{G2_omega_gamma_constraint0} and dropping the $\gtwg\left(\gamma-\gamma^-_\omega\right)\left(\gamma-\gamma^+_\omega\right)$  term allows one to solve for $G_{2 \omega \delta_\omega^+}$, yielding
\begin{align}\label{G2_eq_5}
    &G_{2 \omega \delta_\omega^+} = \frac{1}{\omega \, \left(\dw^+ - \gamma^-_\omega \right) \, \left(\dw^+ - \gamma^+_\omega \right)} \bigg\{ 2 \, \left(\dw^+ - \gamma^+_\omega \right) \left[ G_{\dw^+\gamma_\omega^+}^{(0)} + 2G_{2 \, \dw^+\gamma_\omega^+}^{(0)} \right] \\
    &\hspace{6cm}- 8 \, \delta_\omega^- \, G_{2 \omega \gamma_\omega^+}^{(0)} + 3 \, \omega \, G_{2 \omega \delta_\omega^+}^{(0)} \bigg\}.\notag
\end{align}
Substituting this result back into \eq{G2_omega_gamma_constraint0} and solving for $\gtwg$, we obtain
\begin{align}\label{G2_omega_gamma}
    \gtwg = \gtwg^{(0)} + \frac{1}{\omega\left(\gamma-\gamma^-_\omega\right)\left(\gamma-\gamma^+_\omega\right)} &\Bigg[2\left(\gamma-\dw^+\right)\left(G^{(0)}_{\dw^+\gamma} + 2 \, G^{(0)}_{2\dw^+\gamma} \right) \\
    &- 2\left(\gamma^+_\omega-\dw^+\right)\left(G^{(0)}_{\dw^+\gamma^+_\omega} + 2 \, G^{(0)}_{2\dw^+\gamma^+_\omega} \right)
    \notag \\
    &+ 8 \, \dw^-\left(\gtwg^{(0)} - G^{(0)}_{2\omega\gamma^+_\omega} \right) \Bigg]\notag .
\end{align}
Note that indeed, by construction, there is no $\gamma = \gamma^+_\omega$ pole on the right of \eq{G2_omega_gamma}.

We have now completely solved Eqs.~\eqref{largeNc_eqns}. The polarized dipole amplitudes in our solution are given by Eqs.~\eqref{G2}, \eqref{Gamma2}, \eqref{G}, and \eqref{Gamma}, with the ingredients of these expressions constructed in Eqs.~\eqref{delta_omega_pm}, \eqref{Gamma_omega_pm}, \eqref{gamma_omega_pm}, and \eqref{G2_omega_gamma}, for the initial conditions specifying $\gtwg^{(0)}$ and $G^{(0)}_{\omega \gamma}$.


\section{\texorpdfstring{Summary of Results and Small-$x$ Asymptotics}{Summary of Results and Small x Asymptotics}}

Let us now summarize our solution and construct its small-$x$ asymptotics. For brevity, we define
\begin{align}\label{bas_def}
    \bas= \frac{\as N_c}{2\pi}
\end{align}
Rescaling
\begin{align}\label{rescalingbas}
\omega \to \frac{\omega}{\sqrt{\bas}}, \ \ \ \gamma \to \frac{\gamma}{\sqrt{\bas}}, \ \ \ G_{2\omega\gamma} \to \bas \, G_{2\omega\gamma}, \ \ \  G_{2\omega\gamma}^{(0)} \to \bas \, G_{2\omega\gamma}^{(0)}, \ \ \ G_{\omega\gamma}^{(0)} \to \bas \, G_{\omega\gamma}^{(0)},
\end{align}
with the lowercase $\omega, \gamma$ indices not reflecting the rescaling of those variables, we write our solution as follows:
\begin{subequations}\label{alleqs}\allowdisplaybreaks
\begin{align}
    \label{alleqs_G2}
    &G_2(\xoz^2,zs) = \wint \gint \, e^{\omega\ln(zs\xoz^2) +\gamma \ln \left(\tfrac{1}{\xoz^2\Lambda^2} \right)} \, G_{2\omega\gamma}, \\
    \label{alleqs_Gamma2}
    &\Gamma_2(\xoz^2,\xto^2,z's) = \wint\gint \\
    &\hspace{2cm} \times \left[e^{\omega \ln(z's\xto^2) + \gamma \ln\left(\tfrac{1}{\xoz^2\Lambda^2}\right)}\left(G_{2\omega\gamma} - G^{(0)}_{2\omega\gamma}\right) + e^{\omega\ln(z's\xoz^2) + \gamma \ln\left(\tfrac{1}{\xoz^2\Lambda^2}\right)} \, G^{(0)}_{2\omega\gamma} \right], \notag \\
    \label{alleqs_G}
    &G(\xoz^2,zs) = \wint \gint \, e^{\omega\ln(zs\xoz^2) + \gamma \ln \left(\tfrac{1}{\xoz^2\Lambda^2}\right) } \\
    &\hspace{6cm}\times\left[\frac{\omega\gamma}{2\, \bas}\left(G_{2\omega\gamma} -G^{(0)}_{2\omega\gamma} \right) - 2 \, G_{2\omega\gamma} \right], \notag \\
    \label{alleqs_Gamma}
    &\Gamma(\xoz^2,\xto^2,z's) \\
    &\quad = \wint \, e^{\omega\ln(z's\xto^2)} \left[\Gamma_{\omega}^+(\xoz^2) \, e^{\dw^+ \ln \left(\tfrac{1}{\xto^2\Lambda^2}\right)} + \Gamma_{\omega}^-(\xoz^2) \, e^{\dw^- \ln \left(\tfrac{1}{\xto^2\Lambda^2}\right)} \right] \notag \\ 
    & \quad + \wint\gint \notag \, e^{\omega\ln(z's\xto^2) + \gamma \ln \left(\tfrac{1}{\xto^2\Lambda^2}\right)} \left[\frac{\left(-\tfrac{3}{2}\omega\gamma + 4 \, \bas\right)G_{2\omega\gamma} + \tfrac{3}{2}\omega\gamma \, G^{(0)}_{2\omega\gamma}  }{\gamma^2 - \omega\gamma + \bas}  \right]\notag \\
    & \quad -\wint\gint \bigg[2 \, e^{\omega \ln(z's\xto^2) + \gamma \ln \left( \tfrac{1}{\xoz^2\Lambda^2} \right)}\left(G_{2\omega\gamma} - G^{(0)}_{2\omega\gamma}\right) \notag \\
    &\hspace{6cm}+ 2 \, e^{\omega\ln(z's\xoz^2) + \gamma \ln \left(\tfrac{1}{\xoz^2\Lambda^2} \right)} \, G^{(0)}_{2\omega\gamma}\bigg],\notag
\end{align}
\end{subequations}
with
\begin{subequations}\label{alleqs2}\allowdisplaybreaks
\begin{align}
    \label{alleqs_G2_omega_gamma}
    &\gtwg = \gtwg^{(0)} + \frac{\bas}{\omega\left(\gamma-\gamma^-_\omega\right)\left(\gamma-\gamma^+_\omega\right)} \Bigg[2 \left(\gamma-\dw^+\right)\left(G^{(0)}_{\dw^+\gamma} + 2 \, G^{(0)}_{2 \, \dw^+\gamma} \right) \\
    &\hspace{4cm}- 2\left(\gamma^+_\omega-\dw^+\right)\left(G^{(0)}_{\dw^+\gamma^+_\omega} + 2 \, G^{(0)}_{2 \, \dw^+\gamma^+_\omega} \right) + 8 \, \dw^-\left(\gtwg^{(0)} - G^{(0)}_{2\omega\gamma^+_\omega} \right) \Bigg], \notag \\
    \label{alleqs_G0}
    &G^{(0)}(\xoz^2,zs) = \wint \gint \, e^{\omega\ln\left(zs\xoz^2\right) + \gamma \ln\left(\tfrac{1}{\xoz^2\Lambda^2}\right)}G^{(0)}_{\omega\gamma}\,, \\
     \label{alleqs_G20}
    &G_2^{(0)}(\xoz^2,zs) = \wint \gint \, e^{\omega\ln\left(zs\xoz^2\right) + \gamma\ln\left(\tfrac{1}{\xoz^2\Lambda^2}\right)}G^{(0)}_{2\omega\gamma}\,, \\
    \label{alleqs_Gamma_plus}
    &\Gamma_{\omega}^+(\xoz^2) = \frac{e^{-\dw^+\ln\left(\tfrac{1}{\xoz^2\Lambda^2}\right)}}{\bas \, (\dw^+-\dw^-)}\gint \, e^{\gamma\ln\left(\tfrac{1}{\xoz^2\Lambda^2}\right)}\frac{\omega \, \dw^+}{2 \, (\gamma-\dw^+)}\\
    &\hspace{4cm}\times\bigg[\gtwg\left(\gamma^2-\omega\gamma + 4 \, \bas- \tfrac{8 \, \bas}{\omega} \, \dw^- \right) - \gtwg^{(0)}\left(\gamma^2-\omega\gamma+4 \, \bas\right)  \bigg], \notag \\
    \label{alleqs_Gamma_minus}
    &\Gamma_{\omega}^-(\xoz^2) = \frac{e^{-\dw^-\ln\left(\tfrac{1}{\xoz^2\Lambda^2}\right)}}{\bas \, (\dw^- -\dw^+)}\gint \, e^{\gamma\ln\left(\tfrac{1}{\xoz^2\Lambda^2}\right)}\frac{\omega \, \dw^-}{2 \, (\gamma-\dw^-)}\\
    &\hspace{4cm}\times\bigg[\gtwg\left(\gamma^2-\omega\gamma + 4 \, \bas-\tfrac{8 \, \bas}{\omega} \, \dw^+  \right) - \gtwg^{(0)}\left(\gamma^2-\omega\gamma+4 \, \bas\right)  \bigg], \notag \\
    \label{alleqs_delta_pm}
    &\dw^\pm = \frac{\omega}{2}\left[1\pm\sqrt{1-\frac{4\,\bas}{\omega^2}} \right], \\
    \label{alleqs_gamma_pm}
    &\gamma^{\pm}_\omega = \frac{\omega}{2}\left[1 \pm \sqrt{1 - \frac{16\,\bas}{\omega^2} \, \sqrt{1-\frac{4\,\bas}{\omega^2}}}  \right].
\end{align}
\end{subequations}

With the four polarized dipole amplitudes known, Eqs.~\eqref{ch3_helicitypdfs}, \eqref{ch3_TMDs}, and \eqref{ch3_g1} give us the gluon and (flavor-singlet) quark helicity PDFs and TMDs along with the $g_1$ structure function. We begin by substituting \eq{alleqs_G2} into \eq{ch3_gluon_TMD} for the gluon dipole TMD, while neglecting the derivative term at DLA. Integrating out ${\un x}_{10}$ we arrive at
\begin{align}\label{gluon_TMD_3}
     g^{G\,dip}_{1L}(x,k_T^2) = \frac{2 \, N_c}{\as \, \pi^3} \, \frac{1}{k_T^2} \wint \gint \, e^{\omega\ln\left(\tfrac{Q^2}{xk_T^2}\right) + \gamma\ln\left(\frac{k_T^2}{\Lambda^2}\right)} \, 2^{2\omega-2\gamma} \, \frac{\Gamma\left(\omega-\gamma+1\right)}{\Gamma\left(\gamma-\omega\right)} \, \gtwg.
\end{align}
Note that the $\Gamma$ in \eq{gluon_TMD_3} is the $\Gamma$-function and not the neighbor dipole.
The gluon helicity PDF at DLA follows immediately from substituting \eq{alleqs_G2} into \eq{ch3_DeltaG},
\begin{align}\label{gluon_PDF_1}
    \Delta G(x,Q^2) = \frac{2 \, N_c}{\as \pi^2} \, \wint \gint \, e^{\omega\ln\left(\tfrac{1}{x}\right) + \gamma\ln\left(\tfrac{Q^2}{\Lambda^2}\right)}\gtwg \,.
\end{align}
Next we substitute Eqs.~\eqref{alleqs_G} and \eqref{alleqs_G2} into \eq{ch3_quark_TMD} for the flavor-singlet quark helicity TMD (while remembering that $Q=G$ at large $N_c$). Integrating out ${\un x}_{10}$ and $z$ yields
\begin{align}\label{quark_TMD_4}
    &g^{S}_{1L}(x,k_T^2) = - \frac{N_f}{\as \, 2 \pi^3} \, \frac{1}{k_T^2}\wint\gint \, \left[  e^{\omega\ln\left(\tfrac{Q^2}{x \, k_T^2}\right) + \gamma\ln\left(\tfrac{k_T^2}{\Lambda^2}\right)} - e^{(\gamma - \omega) \, \ln \left( \tfrac{k_T^2}{\Lambda^2} \right)} \right]\\
    &\hspace*{6cm}\times2^{2\omega-2\gamma}\,\frac{\Gamma\left(1+\omega-\gamma\right)}{\Gamma\left(1-\omega+\gamma\right)} \,\gamma\left(\gtwg - \gtwg^{(0)}\right)\,. \notag
\end{align}
To obtain the quark helicity PDF in the DLA we substitute Eqs.~\eqref{alleqs_G} and \eqref{alleqs_G2} into \eq{ch3_DeltaSigma}, again remembering that $Q=G$ at large $N_c$. 
Carrying out the $x_{10}^2$ and $z$-integrals and employing \eq{c3} we arrive at 
\begin{align}\label{quark_PDF_2}
    &\Delta \Sigma(x,Q^2) = - \frac{N_f}{\as \, 2 \pi^2} \, \wint\gint \, \frac{\omega}{\omega-\gamma}\left(\gtwg - \gtwg^{(0)} \right) \, e^{\omega\ln\left(\tfrac{1}{x}\right)} \, e^{\gamma\ln\left(\tfrac{Q^2}{\Lambda^2}\right)} \, ,  
\end{align}
where Re~$\omega >$~Re~$\gamma$ along their contours. 
To obtain the $g_1$ structure function, we replace $N_f \to \tfrac{1}{2} \sum_f Z_f^2$ in \eq{quark_PDF_2}, which gives
\begin{align}\label{g1_3}
    & g_1(x,Q^2)  = - \frac{1}{2} \sum_f Z_f^2 \, \frac{1}{\as \, 2 \pi^2} \, \wint\gint \, \frac{\omega}{\omega-\gamma}\left(\gtwg - \gtwg^{(0)} \right) \, e^{\omega\ln\left(\tfrac{1}{x}\right)} \, e^{\gamma\ln\left(\tfrac{Q^2}{\Lambda^2}\right)} ,
\end{align}
again with Re~$\omega >$~Re~$\gamma$ on the integration contours.

Thus in Eqs.~\eqref{gluon_TMD_3}, \eqref{quark_TMD_4}, \eqref{gluon_PDF_1}, \eqref{quark_PDF_2}, and \eqref{g1_3} we have analytic small-$x$ large-$N_c$ expressions for the quark and gluon helicity TMDs, PDFs, and the $g_1$ structure function.

Importantly, the small-$x$ asymptotics of the dipole amplitudes in \eq{alleqs} are governed by the rightmost singularity in the complex $\omega$-plane. One can show that this rightmost singularity is a branch point of the large square root in $\gamma^-_{\omega}$. Setting the expression under that large square root in $\gamma^-_{\omega}$ from \eq{alleqs_gamma_pm} to zero gives the equation
\begin{align}\label{intercept_eqn}
    1 - \frac{16\,\bas}{\omega^2} \, \sqrt{1-\frac{4\,\bas}{\omega^2}} = 0\,,
\end{align}
whose rightmost solution in the complex $\omega$-plane is
\begin{align}\label{intercept}
    \omega = \alpha_h \equiv \frac{4}{3^{1/3}} \, \sqrt{\textrm{Re} \left[ \left( - 9 + i \, \sqrt{111} \right)^{1/3} \right] } \,\sqrt{\frac{\as \, N_c}{2 \pi}} \approx 3.66074 \, \sqrt{\frac{\as \, N_c}{2 \pi}}\,.
\end{align}
We thus arrive at the small-$x$ asymptotics of all the helicity-dependent quantities discussed above, driven by the following leading power of $1/x$:
\begin{align}\label{asympt_all}
    \Delta \Sigma (x, Q^2) \sim \Delta G (x, Q^2) 
    \sim g_1 (x, Q^2) \sim g^{G\,dip}_{1L}(x,k_T^2) \sim g^{S}_{1L}(x,k_T^2) \sim \left( \frac{1}{x} \right)^{\alpha_h}\,.
\end{align}
Together with the general solution of the large-$N_c$ small-$x$ helicity evolution given in Eqs.~\eqref{alleqs} and \eqref{alleqs2}, the asymptotics \eqref{asympt_all} are the main result of this Chapter.


\section{Resummed Anomalous Dimension and Cross-Checks}

Now let us perform several cross-checks of our solution.  As a first cross-check, we consider \cite{Cougoulic:2022gbk} where our Eqs.~\eqref{ch3_largeNc_eqns_unscaled} were solved iteratively with the initial conditions $G^{(0)}_2(\xoz^2,zs) = 1 $ and $G^{(0)}(\xoz^2,zs) = 0$. With these initial conditions, Eqs.~\eqref{alleqs_G0}, \eqref{alleqs_G20}, and \eqref{alleqs_G2_omega_gamma} give
\begin{align}\label{specific_ICs}
    G^{(0)}_{\omega\gamma} = 0\,,\hspace{1cm} G^{(0)}_{2\omega\gamma} = \frac{1}{\omega\gamma}\,,\hspace{1cm} \gtwg = \frac{1}{\omega(\gamma-\gamma^{-}_{\omega})}.
\end{align} 
One can then expand Eqs.~\eqref{alleqs} in powers of $\as$ and integrate over $\gamma$ and $\omega$ (it is easier to carry out the $\gamma$-integrals first, then expand in powers of $\as$, then carry out the $\omega$-integrals). We have confirmed up to $\mathcal{O}(\as^2)$ that such an expansion of our analytic solution is in complete agreement with the iterative solution from \cite{Cougoulic:2022gbk}.

As another cross-check we can use the gluon helicity PDF $\Delta G(x,Q^2)$ given in \eq{gluon_PDF_1}. Employing $G^{(0)}_{\omega\gamma} $, $G^{(0)}_{2\omega\gamma}$ and $\gtwg$ from \eq{specific_ICs} in \eq{gluon_PDF_1} we obtain
\begin{align}\label{gluon_PDF_soln2}
    \Delta G(x,Q^2) = \frac{2N_c}{\as \pi^2}\wint e^{\omega\ln\left(\tfrac{1}{x}\right) + \gamma^-_{\omega}\ln\left(\tfrac{Q^2}{\Lambda^2}\right)}\frac{1}{\omega}\,.
\end{align}
We see that (cf. \eq{ch2_dglapsoln}) $\Delta \gamma_{GG}(\omega) \equiv \gamma^-_{\omega}$ is our prediction for the resummed $GG$ polarized DGLAP anomalous dimension (at small $x$ and in the large-$N_c$ limit), exact to all orders in $\as$: 
\begin{align}\label{anomalous_dim}
\Delta \gamma_{GG}(\omega) = \gamma^-_{\omega} = \frac{\omega}{2}\left[1 - \sqrt{1 - \frac{16\,\bas}{\omega^2}\sqrt{1-\frac{4\,\bas}{\omega^2}} } \ \right] .
\end{align}
Expanding this in powers of $\as$ we obtain
\begin{align}\label{anomalous_dim_exp}
\Delta \gamma_{GG}(\omega)  = \frac{4\,\bas}{\omega} + \frac{8\,\bas^2}{\omega^3} + \frac{56\,\bas^3}{\omega^5} + \frac{496\,\bas^4}{\omega^7} + \mathcal{O}(\as^5). 
\end{align}
Thus our all-order resummed small-$x$ anomalous dimension $\Delta \gamma_{GG}(\omega)$ agrees with the fixed-order calculations to the full existing three-loop order \cite{Altarelli:1977zs,Dokshitzer:1977sg,Mertig:1995ny,Moch:2014sna}, with novel predictions at $\mathcal{O}(\as^4)$ and beyond. This accomplishes another cross-check of our solution.


\section{Comparison to BER}

Here we compare our results to the earlier resummation for helicity distributions at small $x$ done by Bartels, Ermolaev, and Ryskin (BER) \cite{Bartels:1995iu,Bartels:1996wc} in the infrared evolution equations framework. In order to do so, we will need to simplify the expression for the anomalous dimension obtained by BER in \cite{Bartels:1996wc} for the pure-glue case. Following \cite{Bartels:1996wc} we write the $g_1$ structure function as
\begin{align}
    g_1 (x, Q^2) = - \frac{1}{2 \pi} \, \mbox{Im} \, T_3 (x, Q^2) 
\end{align}
with the signature-odd scattering amplitude $T_3$ given by
\begin{align}\label{T3}
    T_3^S (x, Q^2) = \int \frac{d \omega}{2 \pi i} \, \xi (\omega) \, \left( \frac{1}{x} \right)^\omega \, \left( \frac{Q^2}{\Lambda^2} \right)^{F_0 (\omega) / 8 \pi^2} \, \frac{1}{\omega - F_0 (\omega) / 8 \pi^2} \, R_B
\end{align}
for the flavor-singlet case (denoted by the superscript $S$ on the amplitude). Here
\begin{align}
    \xi (\omega) = \frac{e^{- i \pi \omega} -1}{2} \approx \frac{- i \pi \omega}{2}
\end{align}
is the signature factor, $R_B$ is given by the Mellin transform of the Born initial conditions, while $\Lambda$ is an IR cutoff, denoted by $\mu$ in \cite{Bartels:1996wc}. 

The anomalous dimension $F_0 (\omega) / 8 \pi^2$ was found in \cite{Bartels:1996wc} to be (see Eq.~(4.8) in \cite{Bartels:1996wc})
\begin{align}\label{F0}
    \frac{F_0 (\omega)}{8 \pi^2} = \frac{\omega}{2} \, \left[ 1 - \sqrt{1 - \frac{2 \, \as}{\pi \, \omega^2} \, M_0 + \frac{\as}{\pi^3 \, \omega^3} \, G_0 \, F_8 (\omega)  } \right]. 
\end{align}
Here $M_0$ and $G_0$ are $2\times2$ matrices in the quark-gluon distributions space. Their gluon--gluon components are $(M_0)_{GG} = 4 N_c$ and $(G_0)_{GG} = N_c$. The adjoint (octet) amplitude $F_8 (\omega)$ has to be found by solving the following non-linear differential equation,
\begin{align}\label{F8_equation}
    F_8 (\omega) = \frac{4 \pi \as}{\omega} \, M_8 + \frac{\as \, N_c}{2 \pi \, \omega} \, \frac{d F_8 (\omega)}{d \omega} + \frac{1}{8 \pi^2 \, \omega} \, [F_8 (\omega)]^2 .
\end{align}
Again, $M_8$ is a $2\times2$ matrix in the quark and gluon distributions space: its gluon-gluon component is $(M_8 )_{GG} = 2 N_c$. 

To obtain the pure-glue anomalous dimension we discard quarks, and replace the matrices $M_0$, $G_0$, and $M_8$ by their $GG$ components. The matrix functions $F_0 (\omega)$ and $F_8 (\omega)$ also become single-component objects, which we will label $F_{0 \, GG} (\omega)$ and $F_{8 \, GG} (\omega)$, respectively. The solution of \eq{F8_equation} can then be found by using the substitution \cite{Kirschner:1983di}
\begin{align}
    F_{8 \, GG} (\omega) = 4 \pi \as N_c  \, \frac{\pd}{\pd \omega} \ln u (z)
\end{align}
where 
\begin{align}
    z = \frac{\omega}{\omega_0} \ \ \ \mbox{with} \ \ \ \omega_0 = \sqrt{\frac{\as \, N_c}{2 \pi}}.
\end{align}
This reduces \eq{F8_equation} to
\begin{align}\label{u_eq}
    u''(z) - z \, u'(z) + 2 \, u(z) =0.
\end{align}
The solution of \eq{u_eq} giving the correct perturbative expansion of $F_{8 \, GG} (\omega)$ in powers of $\as$ (that is, giving $F_{8 \, GG} (\omega) = 8 \pi \as \, N_c/\omega$ at order-$\as$) is quite simple, 
\begin{align}
    u(z) = z^2 - 1, 
\end{align}
leading to
\begin{align}
    F_{8 \, GG} (\omega) = \frac{8 \pi \as \, N_c}{\omega} \, \frac{1}{1- \frac{\as \, N_c}{2 \pi} \, \frac{1}{\omega^2}}.
\end{align}
Using this result in \eq{F0} along with $(M_0)_{GG} = 4 N_c$ and $(G_0)_{GG} = N_c$ yields the re-summed $GG$ polarized small-$x$ large-$N_c$ anomalous dimension\footnote{While we did not take the large-$N_c$ limit in our calculation, taking it now would not modify anything in \eq{BER_anom_dim}: it appears that pure glue and large-$N_c$ approximations are identical for BER evolution.}
\begin{align}\label{BER_anom_dim}
    \Delta \gamma^{BER}_{GG} (\omega) \equiv \frac{F_{0 \, GG} (\omega)}{8 \pi^2} = \frac{\omega}{2} \, \left[ 1 - \sqrt{1 - \frac{16 \, \bas}{\omega^2} \, \frac{1 - \frac{3 \, \bas}{\omega^2}}{1 - \frac{\bas}{\omega^2} } } \ \right].
\end{align}
Comparing this with \eq{anomalous_dim}, we conclude that our re-summed $GG$ polarized small-$x$ anomalous dimension is different from the one which follows from the evolution obtained by BER. Curiously, the perturbative expansion of $\Delta \gamma^{BER}_{GG} (\omega)$ in powers of $\as$ yields
\begin{align}\label{BER_exp}
    \Delta \gamma^{BER}_{GG} (\omega) = \frac{4 \, \bas}{\omega} + \frac{8 \, \bas^2}{\omega^3} + \frac{56 \, \bas^3}{\omega^5} + \frac{504 \, \bas^4}{\omega^7} + {\cal O} (\as^5) .
\end{align}
Comparing this with \eq{anomalous_dim_exp}, we see that the two anomalous dimensions, ours and BER, agree at the one-, two-, and three-loop levels, which have also been verified by the perturbative calculations \cite{Altarelli:1977zs,Dokshitzer:1977sg,Mertig:1995ny,Moch:2014sna}. However, at the four-loop level, our and BER anomalous dimensions disagree by a small amount. Disagreement persists at higher orders, reflecting the fact that Eqs.~\eqref{anomalous_dim} and \eqref{BER_anom_dim} indeed contain different functions. 

One may wonder about the agreement between the BER intercept and the one found numerically in \cite{Cougoulic:2022gbk}: both intercepts were reported to be $\alpha_h = 3.66 \, \sqrt{\bas}$ \cite{Bartels:1996wc,Cougoulic:2022gbk}. To find an analytic expression for the intercept in the BER calculation, we need to find the right-most singularity of $\Delta \gamma^{BER}_{GG} (\omega)$. Equating the expression under the square root of \eq{BER_anom_dim} to zero, we see that the rightmost branch point is given by
\begin{align}\label{BER_intercept2}
    \omega = \alpha_h^{BER} \equiv \sqrt{\frac{17 + \sqrt{97}}{2}} \, \sqrt{\frac{\as \, N_c}{2 \pi}} \approx 3.66394 \, \sqrt{\frac{\as \, N_c}{2 \pi}},
\end{align}
as first reported in \cite{Kovchegov:2016zex}.
Comparing this to our \eq{intercept}, we see that the two intercepts are indeed also different, though in both cases the numerical pre-factor rounds up to $3.66$. 

We thus observe a difference between the solution of our large-$N_c$ helicity evolution equations and the corresponding results of the BER IREE-based resummation. The difference of the intercept appears to be numerically insignificant. The disagreement between the perturbative expansion of our anomalous dimension in \eq{anomalous_dim_exp} and the expansion in \eq{BER_exp} at the four-loop level implies that potential future perturbative calculations of the $GG$ polarized anomalous dimension at four loops can determine which approach is correct.  

The origin of this apparent disagreement between our calculation and that of BER is not entirely clear. We note here that some questions about the validity of one of the approximations made in \cite{Bartels:1996wc} were raised earlier in Appendix~B of \cite{Kovchegov:2016zex}. The questions addressed the role of non-ladder hard (large transverse momentum) gluons in the IREE for helicity: it appears that in \cite{Bartels:1996wc} BER had stated that such hard gluons cannot contribute in the DLA, while in \cite{Kovchegov:2016zex} a counter-example was constructed for the quark--quark scattering amplitude at order $\as^3$. Since then, a suggestion has been put forward that such hard-gluon non-ladder contributions can be accounted for in the IREE obtained by BER at order $\as^3$ by re-defining the ladder to include diagrams with uncut rungs. Below, in Appendix~\ref{berdisagreement.app}, we describe how the order-$\as^3$ non-ladder diagrams from \cite{Kovchegov:2016zex} may yet be included into BER IREE, potentially explaining the agreement between the anomalous dimensions \eqref{anomalous_dim_exp} and \eqref{BER_exp} at order $\as^3$. However, when trying to apply the same line of reasoning to a diagram at order $\as^4$ containing hard non-ladder gluons, we run into potential problems and cannot unambiguously incorporate it into BER IREE. This issue at order $\as^4$ may be a possible explanation of the discrepancy between our \eqref{anomalous_dim_exp} and BER \eqref{BER_exp} anomalous dimensions at four loops. We note once again that a full four-loop calculation of polarized DGLAP anomalous dimensions would unambiguously resolve this discrepancy.


\section{Behavior of the Helicity Distributions in the Asymptotic Limit}\label{sec:largencasymptotics}

In addition to the intercept derived in \eq{intercept} and the power law growth in \eq{asympt_all}, we can also exploit the structure of our double-Laplace solution to obtain explicit functional forms for the helicity distributions in the asymptotic limit. As stated in the introduction to this Chapter, we again emphasize that the calculation in this Section was not the work of this author and was originally done in \cite{Kovchegov:2023yzd}. The results are reproduced here with the permission of the authors of that paper.

To avoid extraneous factors of $\as$, we opt to work in the notation prior to the rescaling done in \eq{rescalingbas}. First we fix the simple initial conditions $G_2^{(0)}(\xoz,zs) = G^{(0)}(\xoz^2,zs)$, with which one can show that the double-Laplace images $G^{(0)}_{\omega\gamma}$ and $G^{(0)}_{2\omega\gamma}$ are given by (see Eqs.~\eqref{alleqs_G0} and \eqref{alleqs_G20})
\begin{align}\label{ch4_asy_ics}
    G^{(0)}_{\omega\gamma} = G^{(0)}_{2\omega\gamma} = \frac{1}{\omega\gamma},
\end{align}
with $G_{2\omega\gamma}$ from \eq{G2_omega_gamma} subsequently given by
\begin{align}\label{ch4_asy_icsg2omegagamma}
    G_{2\omega\gamma} = \frac{1}{\omega\left(\gamma-\gamma_\omega^-\right)}\left( 1 + \frac{2}{\gamma\gamma_\omega^+}\right)\,
\end{align}
where $\gamma_\omega^{\pm}$ are given by the unscaled expression in \eq{gamma_omega_pm}. In this notation, and carrying out the integrals over $\gamma$ by closing the contour to the left and picking up the appropriate poles, the helicity PDFs become (see Eqs.~\eqref{ch3_helicitypdfs})
\begin{subequations}\label{ch4_asy_pdfs}
\begin{align}
    &\Delta\Sigma(y,t) = -\frac{N_f}{2\as\pi^2}\wint \,\frac{e^{\omega y}}{\omega} \left[e^{\gamma_\omega^- t} - 1 \right]\left( 1+ \frac{2}{\gamma_\omega^-\gamma_\omega^+}\right),\\
    &\Delta G(y,t) = \frac{2 N_c}{\as\pi^2}\wint \,\frac{e^{\omega y}}{\omega}\left[e^{\gamma_\omega^- t}\left(1+\frac{2}{\gamma_\omega^+\gamma_\omega^-}\right) - \frac{2}{\gamma_\omega^-\gamma_\omega^+}\right],
\end{align}
\end{subequations}
where we have defined $y=\sqrt{\abar}\ln(1/x)$ and $t = \sqrt{\abar}\ln(Q^2/\Lambda^2)$. 

\begin{figure}[ht]
    \centering
    \includegraphics[width=0.7\linewidth]{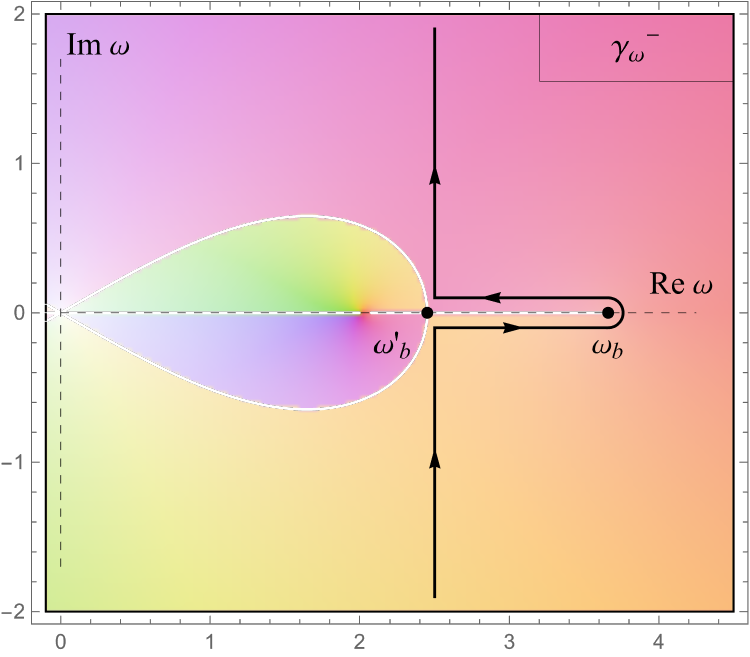}
    \caption{Plot of the large-$N_c$ $GG$ anomalous dimension $\Delta\gamma_{GG}(\omega) = \gamma_\omega^-$ with the distorted inverse-Laplace integration contour overlaid. Note that color corresponds to the Arg of the function while color intensity corresponds to magnitude (paler color $\leftrightarrow$ larger magnitude). $\omega_b$ denotes the right-most branch point while $\omega'_b$ denotes a sub-leading branch point.}
    \label{fig:largencanomdimplot}
\end{figure}

In the interest of obtaining explicit functional forms for $\Delta\Sigma$ and $\Delta G$ in the asymptotic limit, we can approximate the $\omega$-integrals in Eqs.~\eqref{ch4_asy_pdfs} in the following way. Recall that the intercept corresponds to the right-most singularity in the complex-$\omega$ plane and that this singularity comes from the polarized anomalous dimension $\Delta \gamma_{GG}(\omega) = \gamma_\omega^-$. The structure of $\gamma_\omega^-$ is represented in Fig.~\ref{fig:largencanomdimplot} where we plot the anomalous dimension itself along with the overlaid inverse-Laplace integration contour.
 By distorting our integration around the leading branch cut, which runs from the rightmost branch point $\omega_b \approx 3.66$ to a sub-leading branch point $\omega_b'\approx 2.4$, and by neglecting the vertical segments of the integration since they have smaller real parts, we can obtain approximate expressions for the helicity PDFs in Eqs.~\eqref{ch4_asy_pdfs} in the asymptotic limit. 

 Then defining $\xi = \omega_b - \omega$ we approximate the helicity PDFs in terms of the discontinuities of their integrands across the leading branch cut \cite{Kovchegov:2023yzd}:
 \begin{subequations}\label{ch4_asy_disc}
\begin{align}
    &\Delta\Sigma(y,t) \approx -\lim_{\epsilon\rightarrow 0^+}\int\limits_0^\infty \frac{\mathrm{d}\xi}{2\pi i} \left(\Delta\Sigma_{\omega_b-\xi + i\epsilon} - \Delta\Sigma_{\omega_b-\xi-i\epsilon}  \right) ,\\
    &\Delta G(y,t) \approx -\lim_{\epsilon\rightarrow 0^+}\int\limits_0^\infty \frac{\mathrm{d}\xi}{2\pi i} \left(\Delta G_{\omega_b-\xi + i\epsilon} - \Delta G_{\omega_b-\xi-i\epsilon}  \right) ,
 \end{align}
\end{subequations}
where we denote the integrands of $\Delta \Sigma$ and $\Delta G$ in Eqs.~\eqref{ch4_asy_pdfs} as $\Delta\Sigma_\omega$ and $\Delta G_\omega$ and
where for simplicity we integrate all the way to $\infty$ since the factor $e^{\omega y}$ in Eqs.~\eqref{ch4_asy_pdfs} will ensure the dominant contributions to the integrals come from values of $\omega$ with the largest real parts (or equivalently, the smallest values of $\xi$)\footnote{Note that the leading minus signs in front of each of Eqs.~\eqref{ch4_asy_disc} are corrections to the results of \cite{Kovchegov:2023yzd}.}. Note that in addition to $\gamma_\omega^-$, the function $\gamma_\omega^+$ also appears in the the integrands of Eqs.~\eqref{ch4_asy_pdfs} in the factor $\gamma_\omega^-\gamma_\omega^+$. However, as can be easily shown,
\begin{align}\label{ch4_asy_gpgm}
    \gamma_\omega^-\gamma_\omega^+ = 4\sqrt{1-\frac{4}{\omega^2}},
\end{align}
so that no new branch cuts emerge in the vicinity of the one between $\omega_b$ and $\omega_b'$. Thus indeed the only branch cut we need to worry about for our integration is the rightmost one shown in Fig.~\ref{fig:largencanomdimplot}. 

Evaluating the discontinuities in $\Delta\Sigma_{\omega_b-\xi}$ and $\Delta G_{\omega_b-\xi}$, Eqs.~\eqref{ch4_asy_pdfs} and \eqref{ch4_asy_disc} yield \cite{Kovchegov:2023yzd}
\begin{subequations}\label{ch4_asy_disc1}
\begin{align}
    &\Delta\Sigma(y,t) \approx \frac{N_f}{2\as\pi^2} I\left(\omega_b;y,t\right),\\
    &\Delta G(y,t) \approx -\frac{2N_c}{\as\pi^2} I(\omega_b;y,t),
\end{align}
\end{subequations}
where we have defined
\begin{align}\label{ch4_asy_discintegral}
    I\left(\omega_b;y,t\right) = \int\limits_0^\infty \frac{\mathrm{d}\xi}{2\pi i}\,\frac{e^{\left(\omega_b - \xi\right)y}}{\omega_b-\xi}\left[e^{\gamma^-_{\omega_b-\xi+i\epsilon}t} - e^{\gamma^-_{\omega_b-\xi-i\epsilon}t} \right]\left(1+\frac{2}{\gamma^-_{\omega_b-\xi}{\gamma^+_{\omega_b-\xi}}}\right)\,
\end{align}
and where in \eq{ch4_asy_discintegral} we drop the $\pm i\epsilon$ in the last factor since, according to \eq{ch4_asy_gpgm}, the product $\gamma_\omega^-\gamma_\omega^+$ is continuous in the region we are considering.

To evaluate $I(\omega_b;y,t)$ in \eq{ch4_asy_discintegral}, recalling that the integral is dominated by the smallest values of $\xi$, we can expand the integrand in powers of $\xi$ and then carry out the integration term-by-term. To expand the integrand, we first need the expansion of $\gamma^-_{\omega_b-\xi\pm i\epsilon}$, which can be written \cite{Kovchegov:2023yzd}
\begin{align}\label{ch4_asy_gammaexpansion}
    \gamma^-_{\omega_b-\xi \pm i\epsilon} = \frac{\omega_b}{2} \mp i a_1\xi^{1/2} - \frac{\xi}{2} \pm i a_3\xi^{3/2} + \mathcal{O}\left(\xi^{7/2}\right)
\end{align}
where we define
\begin{align}\label{ch4_asy_as}
    &a_1 = \frac{8\sqrt{2}\sqrt{\omega_b^2-6}}{\omega_b^{5/2}}\,,\hspace{2cm} a_3 = \frac{512\sqrt{2}\left(\omega_b^2-2\right)}{\omega_b^{15/2}\sqrt{\omega_b^2 - 6}}\,.
\end{align}
Then we employ \eq{ch4_asy_gammaexpansion} along with the identity in \eq{ch4_asy_gpgm} to expand the integrand in \eq{ch4_asy_discintegral}. We also employ the fact that
\begin{align}\label{ch4_asy_bpid}
    1-\frac{16}{\omega_b^2}\sqrt{1-\frac{4}{\omega_b^2}} = 0
\end{align}
since the branch point $\omega_b$ is a zero of the overall square root in $\gamma_\omega^-$ in \eq{gamma_omega_pm}. We subsequently carry out the integral over $\xi$ term by term and arrive at the following explicit forms of the helicity PDFs in the asymptotic limit \cite{Kovchegov:2023yzd}:
\begin{subequations}\label{ch4_asy_pdfsfinal}
\begin{align}
    &\Delta \Sigma(y,t) \approx \left[\frac{c_{1,q}(t)}{y^{3/2}} + \frac{c_{2,q}(t)}{y^{5/2}} + \mathcal{O}\left(\frac{1}{y^{7/2}}\right) \right]e^{\omega_b y} \\
    &\Delta G(y,t) \approx \left[\frac{c_{1,G}(t)}{y^{3/2}} + \frac{c_{2,G}(t)}{y^{5/2}} + \mathcal{O}\left(\frac{1}{y^{7/2}}\right) \right]e^{\omega_b y}\,
\end{align}
\end{subequations}
where 
\begin{subequations}\label{ch4_asy_cs}
\begin{align}
    \label{ch4_asy_c1q}
    &c_{1,q}(t) = -\frac{N_f}{4\pi^{5/2}\as}t\,e^{\omega_b\tfrac{t}{2}}\,\frac{a_1\left(\omega_b^2+8\right)}{\omega_b^3},\\
    \label{ch4_asy_c2q}
    &c_{2,q}(t) = \frac{3N_f}{16\pi^{5/2}\as} \, t\, e^{\omega_b\tfrac{t}{2}}\frac{1}{3\omega_b^{5}}\bigg[6a_3\omega_b^2\left(\omega_b^2+8\right) + 3a_1\omega_b\left(t\omega_b\left(\omega_b^2+8\right)-48-2\omega_b^2 \right) \\
    &\hspace{6cm}+ a_1^3\left(192 + 8t^2\omega_b^2 + t^2\omega_b^4\right)   \bigg], \notag\\
    \label{ch4_asy_c1G}
    &c_{1,G}(t) = \frac{N_c}{\pi^{5/2}\as}t\,e^{\omega_b\tfrac{t}{2}}\,\frac{a_1\left(\omega_b^2+8\right)}{\omega_b^3},\\
    \label{ch4_asy_c2G}
    &c_{2,G}(t) = -\frac{3N_c}{4\pi^{5/2}\as} \, t\, e^{\omega_b\tfrac{t}{2}}\frac{1}{3\omega_b^{5}}\bigg[6a_3\omega_b^2\left(\omega_b^2+8\right) + 3a_1\omega_b\left(t\omega_b\left(\omega_b^2+8\right)-48-2\omega_b^2 \right) \\
    &\hspace{6cm}+ a_1^3\left(192 + 8t^2\omega_b^2 + t^2\omega_b^4\right)   \bigg] .\notag
\end{align}
\end{subequations}
Notably, the results in Eqs.~\eqref{ch4_asy_pdfsfinal} and \eqref{ch4_asy_cs} predict a ratio (for $N_f=4,N_c=3$) of $\Delta G(y,t)/\Delta \Sigma(y,t) = -3$ at large-$N_c$ in the asymptotic limit\footnote{Again note an overall sign difference for both $\Delta \Sigma$ and $\Delta G$ relative to \cite{Kovchegov:2023yzd}. However the predicted ratio of $\Delta G $ to $\Delta \Sigma$ from that reference is unaffected.}, in reasonable qualitative agreement with the asymptotic ratio of $\Delta G(y,t)/\Delta \Sigma(y,t) \approx -2.29$ predicted in \cite{Boussarie:2019icw} based on work done in the framework of the BER IREE approach. The latter prediction was also made for $N_f=4, N_c=3$ but was made without invoking the large-$N_c$ (or large-$N_c\&N_f$) limit, hence some discrepancy with the large-$N_c$ prediction in this Chapter is to be expected.


\section{Chapter Summary}

In this Chapter we have successfully derived a completely analytic solution to the revised  large-$N_c$ helicity evolution of Eqs.~\eqref{ch3_largeNc_eqns_unscaled}. Using our solution, we have obtained corresponding analytic predictions for the helicity PDFs and TMDs along with the $g_1$ structure function at small-$x$ and in the large-$N_c$ limit. We exploited the inverse-Laplace structure of our solution to obtain an analytic prediction for the intercept governing the power-law growth of the helicity distributions at asymptotically small-$x$\footnote{As Ch.~\ref{sec:largencasymptotics} also showed, we can extract the more detailed functional forms of the helicity distributions in the asymptotic limit as well \cite{Kovchegov:2023yzd}.} and we also extracted an analytic, resummed prediction for the polarized $GG$ DGLAP anomalous dimension, exact to all orders in $\as$ (also at large-$N_c$).

Notably, in comparison to existing work, we find that the expansion of our predicted polarized anomalous dimension agrees completely with the full extent of the finite order calculations that exist, to $\mathcal{O}(\as^3)$. In comparison to the predictions of BER, however, we find some very small discrepancies. Our resummed anomalous dimension in \eq{anomalous_dim} is a different function than that predicted by BER in \eq{BER_anom_dim}. The expansions of both functions remarkably agree to the first three orders in $\as$ (both are in agreement with finite order calculations) but disagree with each other at the four-loop level and beyond. We find a similarly small disagreement in our predicted intercept when comparing to BER. The two analytic expressions (ours in \eq{intercept}, BER in \eq{BER_intercept2}) are indeed different. When evaluated numerically, both round to $3.66\sqrt{\abar}$, but disagree beyond that precision. 

The extension of the finite-order calculation of $\Delta \gamma_{GG}(\omega)$ to the four loop level would hopefully resolve the discrepancy. Nevertheless, the very close agreement of our predictions with those of BER, especially given that the two sets of predictions were made in quite different approaches, ought to inspire some confidence in the accuracy and the predictive power of the small-$x$ helicity formalism used here.

The large-$N_c$ limit has been a valuable place to begin. We established a formalism with which to solve the complicated small-$x$ evolution equations and we had a useful formal limit that could also be applied to the calculations of both finite-order and BER in order to yield meaningful comparisons. However, the large-$N_c$ treatment ignores quark contributions to the evolution, and as Ch.~\ref{subsec:polwilsonlines} established, quark exchange is an important element of the polarized Wilson lines and of our description of small-$x$ helicity. Thus our next step is to move to the more general large-$N_c\&N_f$ limit, where with the number of quark flavors $N_f$ now taken to be parametrically large, quark contributions are restored to the evolution. The result is the more general, more complicated, but still closed set of evolution equations in Eqs.~\eqref{ch3_LargeNcNfeqns}. These equations will form the basis for the coming two Chapters.

\chapter{Quark-to-Gluon and Gluon-to-Quark Transition Operators}
\label{transitionops.ch}

With the large-$N_c$ evolution equations completely solved in Ch.~\ref{largeNcsoln.ch} we would like, if possible, to solve the more general set of large-$N_c\&N_f$ evolution equations~\eqref{ch3_LargeNcNfeqns}. However, as we will soon see, there is actually a problem with those equations --- they disagree with robust predictions from the finite-order framework. In this Chapter, we present that disagreement and ultimately identify a contribution to the large-$N_c\&N_f$ evolution that was previously excluded, coming from a class of quark-to-gluon and gluon-to-quark transition operators. We explicitly construct the contributions of these transition operators to the large-$N_c\&N_f$ evolution and demonstrate that their inclusion restores full consistency with finite-order predictions. This Chapter is based on the paper \cite{Borden:2024bxa}.


\section{\texorpdfstring{Problems with the Large-$N_c\&N_f$ Evolution}{Problems with the Large Nc\&Nf Evolution}}

In \cite{Adamiak:2023okq} the large-$N_c\&N_f$ small-$x$ helicity evolution equations~\eqref{ch3_LargeNcNfeqns} were solved iteratively, similarly to that done in \cite{Cougoulic:2022gbk} for the large-$N_c$ evolution equations.\footnote{The paper \cite{Adamiak:2023okq} was not the work of this author. It was written by other researchers working in the small-$x$ helicity program.} From that iterative solution the authors of \cite{Adamiak:2023okq} extracted expressions, order-by-order in $\as$, for the small-$x$ large-$N_c\&N_f$ polarized DGLAP splitting functions. Unlike the large-$N_c$ evolution where $\Delta P_{GG}(z)$ is the only polarized splitting function, the large-$N_c\&N_f$ evolution contains all four of the polarized DGLAP splitting functions: $\Delta P_{qq}(z)$, $\Delta P_{qG}(z)$, $\Delta P_{Gq}(z)$, and $\Delta P_{GG}(z)$. The following results were extracted, to two loops, from the large-$N_c\&N_f$ helicity evolution in \cite{Adamiak:2023okq}:
\begin{subequations}\label{ch5_splfuncsbad}
\begin{align}
    &\Delta P_{qq}(x) = \left(\frac{\as N_c}{4\pi}\right) + \left(\frac{\as N_c}{4\pi}\right)^2 \left(\frac{1}{2}-\textcolor{red}{4}\frac{N_f}{N_c}\right)\ln^2\frac{1}{x} + \mathcal{O}\left(\as^3\right) ,\\
    &\Delta P_{qG}(x) = -\left(\frac{\as N_c}{4\pi}\right)\frac{2N_f}{N_c} - \left(\frac{\as N_c}{4\pi}\right)^2 \textcolor{red}{13}\frac{N_f}{N_c}\ln^2\frac{1}{x} + \mathcal{O}\left(\as^3\right),\\
    &\Delta P_{Gq}(x) = 2\left(\frac{\as N_c}{4\pi}\right) + \textcolor{red}{8}\left(\frac{\as N_c}{4\pi}\right)^2 \ln^2\frac{1}{x} + \mathcal{O}\left(\as^3\right) ,\\
    &\Delta P_{GG}(x) = 8\left(\frac{\as N_c}{4\pi}\right) + \left(\frac{\as N_c}{4\pi}\right)^2\left(16 - \textcolor{red}{0}\frac{N_f}{N_c} \right) \ln^2\frac{1}{x} + \mathcal{O}\left(\as^3\right).
\end{align}
\end{subequations}
These are to be compared with the small-$x$, large-$N_c\&N_f$ limit of calculations done at finite-order. In the $\overline{\text{MS}}$ scheme, the first two loops of the polarized splitting functions $\Delta\overline{P}_{ij}(z)$ at small-$x$ and large-$N_c\&N_f$ are \cite{Altarelli:1977zs,Dokshitzer:1977sg,Mertig:1995ny,Moch:2014sna}
\begin{subequations}\label{ch5_MSbar2loops}
\begin{align}
\label{ch5_Pqq2loops}
&\Delta \overline{P}_{qq}(x) = \left(\frac{\alpha_sN_c}{4\pi}\right) + \left(\frac{\alpha_s N_c}{4\pi}\right)^2 \left( \frac{1}{2}-\textcolor{blue}{2}\frac{N_f}{N_c} \right)\ln^2\frac{1}{x} + \mathcal{O}\left(\alpha_s^3\right) \,,  \\
\label{ch5_PqG2loops}
&\Delta \overline{P}_{qG}(x) =  - \left(\frac{\alpha_sN_c}{4\pi}\right)\frac{2N_f}{N_c} - \left(\frac{\alpha_sN_c}{4\pi}\right)^2 \textcolor{blue}{5}\frac{N_f}{N_c}\ln^2\frac{1}{x} + \mathcal{O}\left(\alpha_s^3\right) \,, \\
\label{ch5_PGq2loops}
&\Delta \overline{P}_{Gq}(x) =   2\left(\frac{\alpha_sN_c}{4\pi}\right) + \textcolor{blue}{5}\left(\frac{\alpha_s N_c}{4\pi}\right)^2 \ln^2\frac{1}{x} + \mathcal{O}\left(\alpha_s^3\right)  \,, \\
\label{ch5_PGG2loops}
&\Delta \overline{P}_{GG}(x) =   8 \left(\frac{\alpha_sN_c}{4\pi}\right) + \left(\frac{\alpha_s N_c}{4\pi}\right)^2 \left( 16-\textcolor{blue}{2}\frac{N_f}{N_c} \right) \ln^2\frac{1}{x} + \mathcal{O}\left(\alpha_s^3\right)\,.
\end{align}
\end{subequations}
The factors in red in Eqs.~\eqref{ch5_splfuncsbad} show where the predictions of the small-$x$ evolution disagreed with the finite order calculations, whose corresponding factors are shown in blue in Eqs.~\eqref{ch5_MSbar2loops}. In that same paper \cite{Adamiak:2023okq} it was shown that the differences between Eqs.~\eqref{ch5_splfuncsbad} and \eqref{ch5_MSbar2loops} could not be attributed to any scheme dependence.

We note also that the author of this dissertation observed a similar discrepancy by constructing an analytic solution to the large-$N_c\&N_f$ equations~\eqref{ch3_LargeNcNfeqns}. In that solution, the discrepancy with finite-order calculations appears in the eigenvalues of the matrix of polarized DGLAP anomalous dimensions\footnote{Details of this (unpublished) solution will not be presented here, but we will return to the challenge of analytically solving the large-$N_c\&N_f$ equations in Ch.~\ref{largeNcandNfsoln.ch}. In particular we will have much more to say about how the eigenvalues of the anomalous dimension matrix emerge from the solution.} 
\begin{align}\label{ch5_anomalousdimmatrix}
    \begin{pmatrix}
        \Delta\gamma_{qq}(\omega) & \Delta\gamma_{qG}(\omega) \\
        \Delta\gamma_{Gq}(\omega) & \Delta\gamma_{GG}(\omega)
    \end{pmatrix}\,,
\end{align}
(recall \eq{ch2_dglapanomalousdims} for the definition of the anomalous dimension in terms of its corresponding splitting function). The analytic expressions for these eigenvalues obtained from that solution could be expanded in powers of $\as$ to give 
\begin{align}\label{ch5_eigexpansion}
    &\bigg(\text{Eigenvalues from small-}x \text{ evolution}\bigg) \\
    &=\left(\frac{\as}{4\pi} \right)\frac{1}{2}\left[9N_c \pm \sqrt{N_c\left(49N_c - 16N_f\right)}  \right] \frac{1}{\omega} \notag\\
    & + \left(\frac{\as}{4\pi} \right)^2 \frac{1}{2}\frac{N_c}{\left(49N_c - 16N_f\right)} \bigg[\left(49N_c-16N_f\right)\left(33N_c-8N_f\right) \notag\\
    &\hspace*{5cm}\pm \sqrt{N_c\left(49N_c - 16N_f \right) }\left(217N_c - \textcolor{red}{112}N_f \right) \bigg] \frac{1}{\omega^3} \notag \\ 
    &+ \left(\frac{\as}{4\pi} \right)^3 \frac{N_c}{\left(49N_c - 16N_f\right)^2} \bigg[N_c\left(49N_c - 16N_f\right)^2 \left(225N_c - \textcolor{red}{84}N_f \right) \notag \\
    &\hspace*{1cm} \pm \sqrt{N_c\left(49N_c - 16N_f \right)} \left(76489N_c^3 - \textcolor{red}{68412}N_c^2N_f + \textcolor{red}{16896}N_cN_f^2 - 1024 N_f^3 \right) \bigg] \frac{1}{\omega^5} \notag \\
    &+ \mathcal{O}\left(\as^4\right). \notag
\end{align}
We compare this to the eigenvalues constructed from the $\overline{\text{MS}}$ anomalous dimensions (here we include the full 3 loop result):
\begin{align}\label{ch5_eigexpansionmsbar}\allowdisplaybreaks
    &\bigg(\text{Eigenvalues from } \overline{\text{MS}} \bigg) \\
    &=\left(\frac{\as}{4\pi} \right)\frac{1}{2}\left[9N_c \pm \sqrt{N_c\left(49N_c - 16N_f\right)}  \right] \frac{1}{\omega} \notag\\
    & + \left(\frac{\as}{4\pi} \right)^2 \frac{1}{2}\frac{N_c}{\left(49N_c - 16N_f\right)} \bigg[\left(49N_c-16N_f\right)\left(33N_c-8N_f\right) \notag\\
    &\hspace*{5cm}\pm \sqrt{N_c\left(49N_c - 16N_f \right) }\left(217N_c - \textcolor{blue}{80}N_f \right) \bigg] \frac{1}{\omega^3} \notag \\ 
    &+ \left(\frac{\as}{4\pi} \right)^3 \frac{N_c}{\left(49N_c - 16N_f\right)^2} \bigg[N_c\left(49N_c - 16N_f\right)^2 \left(225N_c - \textcolor{blue}{64}N_f \right) \notag \\
    &\hspace*{1cm} \pm \sqrt{N_c\left(49N_c - 16N_f \right)} \left(76489N_c^3 - \textcolor{blue}{60712}N_c^2N_f + \textcolor{blue}{14784}N_cN_f^2 - 1024N_f^3 \right) \bigg] \frac{1}{\omega^5} \notag \\
    &+ \mathcal{O}\left(\as^4\right). \notag
\end{align}
As before we show the disagreeing terms in red (small-$x$ evolution) and blue (finite-order). This comparison also makes it clear that the discrepancy cannot be explained with any scheme dependence --- even though individual anomalous dimensions (or splitting functions) can vary between schemes, the eigenvalues of the matrix of anomalous dimensions are in principle scheme-independent (for fixed coupling).\footnote{This is another issue we will discuss in more detail in Ch.~\ref{largeNcandNfsoln.ch}.} Then Eqs.~\eqref{ch5_splfuncsbad}, \eqref{ch5_MSbar2loops}, \eqref{ch5_eigexpansion}, and \eqref{ch5_eigexpansionmsbar} tell us that the large-$N_c\&N_f$ helicity evolution in its current form from Eqs.~\eqref{ch3_LargeNcNfeqns} cannot be fully correct. Ultimately there is a missing contribution that needs to be included in the small-$x$ evolution.

The helicity evolution derived in \cite{Kovchegov:2015pbl, Kovchegov:2016zex, Kovchegov:2018znm, Cougoulic:2022gbk} only involved polarized Wilson lines where the incoming quark becomes an outgoing quark after interacting with the shock wave, or an incoming gluon becomes an outgoing gluon. The \textit{transition} operators, where an incoming (anti)quark interacts with the shock wave and becomes an outgoing gluon, and vice versa, an incoming gluon becomes an outgoing (anti)quark were considered in \cite{Kovchegov:2015pbl, Kovchegov:2016zex}. However they were not included in the calculation since they were perceived to cancel for flavor-singlet quantities \cite{Kovchegov:2015pbl, Kovchegov:2016zex}, thus not contributing to the flavor-singlet helicity evolution equations in the DLA investigation performed there.  The contribution of such transition operators was extensively studied by Chirilli in \cite{Chirilli:2021lif}: they do enter the DLA helicity evolution equations derived in that reference. However the flavor-singlet versus non-singlet decomposition is done differently in \cite{Chirilli:2021lif} from \cite{Kovchegov:2015pbl, Kovchegov:2016zex, Cougoulic:2022gbk} and comparison of the two frameworks has been challenging. The importance of the transition operators for the flavor non-singlet helicity evolution was also recognized in \cite{Kovchegov:2016zex}, but the contribution of these operators was (correctly) neglected in that reference as sub-leading in the large-$N_c$ limit employed there, leading to the same small-$x$ asymptotics as that obtained by BER in \cite{Bartels:1995iu} for the flavor non-singlet case (if one takes the large-$N_c$ limit of the latter).

In this Chapter we will show that the $q/{\bar q} \to G$ and $G \to q/{\bar q}$ shock-wave transition operators do indeed generate DLA terms and contribute to helicity evolution in the flavor-singlet channel. We explicitly derive the corrections due to the transition operators and incorporate them into the large-$N_c \& N_f$ helicity evolution equations constructed in \cite{Cougoulic:2022gbk} and presented earlier in Eqs.~\eqref{ch3_LargeNcNfeqns}, resulting in an updated set of large-$N_c\&N_f$ evolution equations. We perform our calculations using both the light-cone operator treatment (LCOT, see Ch.~\ref{sec:ch3_lcot}) and light-cone perturbation theory (LCPT, see Ch.~\ref{sec:ch3_lcpt}).The large-$N_c$ helicity evolution equations ~\eqref{ch3_largeNc_eqns_unscaled} from \cite{Cougoulic:2022gbk} remain unchanged, since they include the gluon contributions only. And thus the analytic solution to the large-$N_c$ evolution constructed in Ch.~\ref{largeNcsoln.ch} also remains unchanged. We verify the corrected large-$N_c \& N_f$ helicity evolution equations we obtain below by solving them iteratively: this results in predictions for the polarized DGLAP splitting functions that agree with BER \cite{Bartels:1996wc} to three loops. Our splitting functions also agree with the (small-$x$ and large-$N_c \& N_f$ limit of the) finite-order $\overline{\text{MS}}$ polarized splitting functions \cite{Altarelli:1977zs,Dokshitzer:1977sg,Mertig:1995ny,Moch:2014sna} at one and two loops. The agreement at three loops is achieved after a minor scheme transformation: the same is true for BER splitting functions \cite{Moch:2014sna}. The observed agreement applies to all $N_c$- and $N_f$-dependent terms in the splitting functions. 
We also predict the 4-loop polarized DGLAP splitting functions at small-$x$ and large-$N_c \& N_f$.


\section{Derivation of the Transition Operators}\label{sec:ch5_transitionops}

\begin{figure}[h]
    \centering
    \includegraphics[width=\textwidth]{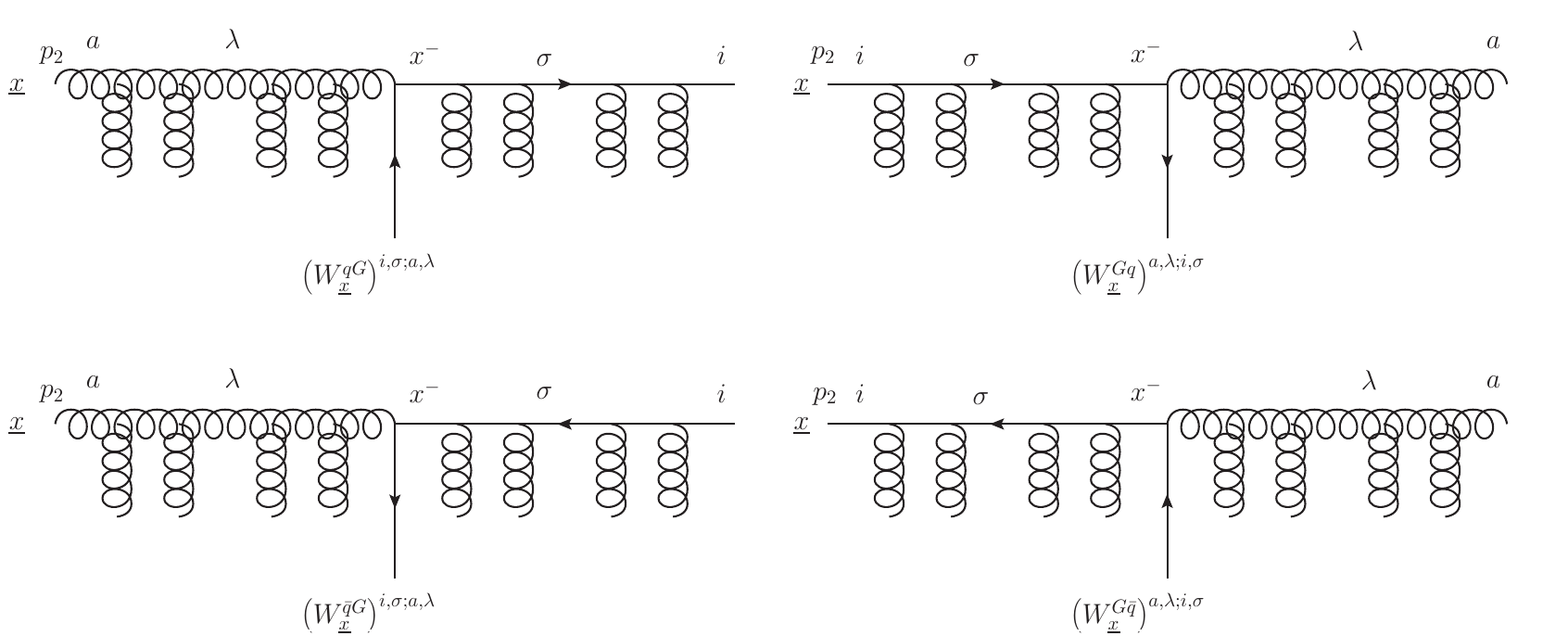}
    \caption{A diagrammatic representation of the $q/{\bar q} \to G$ and $G \to q/{\bar q}$ transition operators. The incoming particle's momentum $p_2$ has a large light-cone minus component. The polarizations and colors of the incoming and outgoing partons are labeled explicitly.}
    \label{fig:newoperators}
\end{figure}

We begin by constructing the (anti)quark to gluon and gluon to (anti)quark transition operators. They are shown diagrammatically in \fig{fig:newoperators}. As usual the background gluon and quark fields in \fig{fig:newoperators} are represented by vertical corkscrew and straight lines, respectively. The non-eikonal interaction is due to the quark background field: the interactions with the gluon background field in \fig{fig:newoperators} are, therefore, assumed to be eikonal, and will be described by light-cone Wilson lines in the appropriate (adjoint or fundamental) color representations.  

A straightforward calculation along the lines of those done in \cite{Kovchegov:2017lsr, Kovchegov:2018znm, Kovchegov:2021iyc} yields for the transition operators defined in \fig{fig:newoperators}
\begin{subequations}\label{Ws}
\begin{align}
& W_{\un x}^{Gq} [\infty, - \infty] = \frac{- i g}{2 \, \sqrt{\sqrt{2} \, p_2^-}} \, \delta_{\lambda, \sigma} \, \int\limits_{-\infty}^\infty d x^- \, U_{\un x}^{ab} [\infty, x^-] \, {\bar \psi} (x^-, {\un x}) \, t^b \, \\
&\hspace*{6cm}\times\left[ \rho(+) + \rho(-) - \sigma \, \left( \rho(+) - \rho(-) \right) \right] \, V_{\un x} [x^-, - \infty], \notag \\
& W_{\un x}^{G{\bar q}} [\infty, - \infty] = \frac{i g}{2 \, \sqrt{\sqrt{2} \, p_2^-}} \, \delta_{\lambda, \sigma} \, \int\limits_{-\infty}^\infty d x^- \, U_{\un x}^{ab} [\infty, x^-] \, V_{\un x} [-\infty, x^-]  \, \\
&\hspace*{6cm}\times\left[ {\bar \rho}(+) + {\bar \rho}(-) + \sigma \, \left( {\bar \rho}(+) - {\bar \rho}(-) \right) \right] \, t^b \, {\psi} (x^-, {\un x}), \notag\\
& W_{\un x}^{qG} [\infty, - \infty] =  \frac{- i g}{2 \, \sqrt{\sqrt{2} \, p_2^-}} \, \delta_{\lambda, \sigma} \, \int\limits_{-\infty}^\infty d x^- \, V_{\un x} [\infty, x^-]  \, \left[ {\bar \rho}(+) + {\bar \rho}(-) - \sigma \, \left( {\bar \rho}(+) - {\bar \rho}(-) \right) \right] \\
&\hspace*{6cm}\times t^b \, {\psi} (x^-, {\un x}) \, U_{\un x}^{ba} [x^-, -\infty] , \notag \\
& W_{\un x}^{{\bar q}G} [\infty, - \infty] = \frac{i g}{2 \, \sqrt{\sqrt{2} \, p_2^-}} \, \delta_{\lambda, \sigma} \, \int\limits_{-\infty}^\infty d x^- \, {\bar \psi} (x^-, {\un x}) t^b \, \left[ \rho(+) + \rho(-) + \sigma \, \left( \rho(+) - \rho(-) \right) \right] \\
&\hspace{6cm}\times V_{\un x} [x^-, \infty]  \, U_{\un x}^{ba} [x^-, -\infty] .\notag
\end{align}
\end{subequations}
We are suppressing the polarization and color indices, which are shown in \fig{fig:newoperators}. The superscripts of the $W$-operators in Eq.~\eqref{Ws} (read right-to-left) denote the incoming and outgoing particles: for instance, $W_{\un x}^{Gq}$ denotes the transition operator for an incoming quark and outgoing gluon, as shown in the upper right panel of \fig{fig:newoperators}. As usual we are working in $A^- =0$ gauge for the gluons; an incoming or outgoing gluon with momentum $k$ is described by the usual polarization four-vector (listed in the text after Eqs.~\eqref{ch3_quarkops}), while the external quark and anti-quark lines are described by the ``anti"-Brodsky--Lepage spinors in Eqs.~\eqref{ch3_blspinors} and \eqref{ch3_rhospinors}.

Relations between the transition operators are
\begin{align}
\left( W_{\un x}^{Gq} [\infty, - \infty] \right)^\dagger = W_{\un x}^{qG}  [-\infty, \infty], \ \ \ \left( W_{\un x}^{G{\bar q}} [\infty, - \infty] \right)^\dagger = W_{\un x}^{{\bar q}G}  [-\infty, \infty] .
\end{align}

Defining
\begin{subequations}\label{W12}
\begin{align}
& W_{\un x}^{a; i \, [1]} [b^-, a^-] \\
&\hspace*{1cm}= \frac{i g}{2 \, \sqrt{\sqrt{2} \, p^-}} \, \int\limits_{a^-}^{b^-} d x^- \, U_{\un x}^{ab} [b^-, x^-] \, {\bar \psi}^{i'} (x^-, {\un x}) \, t^b \, \left[ \rho(+) - \rho(-) \right] \, \left( V_{\un x} [x^-, a^-] \right)^{i'i}, \notag\\
& W_{\un x}^{a; i \, [2]} [b^-, a^-] \\
&\hspace*{1cm}= - \frac{i g}{2 \, \sqrt{\sqrt{2} \, p^-}} \, \int\limits_{a^-}^{b^-} d x^- \, U_{\un x}^{ab} [b^- , x^-] \, {\bar \psi}^{i'} (x^-, {\un x}) \, t^b \, \left[ \rho(+) + \rho(-) \right] \, \left( V_{\un x} [x^-, a^-] \right)^{i'i}, \notag
\end{align}
\end{subequations}
which are rows in quark color space,
we rewrite Eqs.~\eqref{Ws} as
\begin{subequations}\label{Ws2}
\begin{align}
& \left( W_{\un x}^{Gq}[\infty, - \infty] \right)^{a,\lambda ; i, \sigma} = \sigma \, \delta_{\lambda, \sigma} \, W_{\un x}^{a; i \, [1]} [\infty, - \infty] + \delta_{\lambda, \sigma} \, W_{\un x}^{a; i \, [2]} [\infty, - \infty] , \\
& \left( W_{\un x}^{G{\bar q}} [\infty, - \infty] \right)^{a,\lambda ; i, \sigma} = - \sigma \, \delta_{\lambda, \sigma} \, \left( W_{\un x}^{a; i \, [1]} [\infty, - \infty] \right)^\dagger + \delta_{\lambda, \sigma} \, \left( W_{\un x}^{a; i \, [2]} [\infty, - \infty] \right)^\dagger , \\
& \left( W_{\un x}^{qG} [\infty, - \infty] \right)^{i, \sigma ; a,\lambda} = \sigma \, \delta_{\lambda, \sigma} \, \left( W_{\un x}^{a; i \, [1]} [-\infty, \infty] \right)^\dagger + \delta_{\lambda, \sigma} \, \left( W_{\un x}^{a; i \, [2]} [-\infty, \infty] \right)^\dagger , \\
& \left( W_{\un x}^{{\bar q}G} [\infty, - \infty] \right)^{i, \sigma ; a,\lambda} =  - \sigma \, \delta_{\lambda, \sigma} \, W_{\un x}^{a; i \, [1]} [-\infty, \infty] + \delta_{\lambda, \sigma} \, W_{\un x}^{a; i \, [2]} [-\infty, \infty] .
\end{align}
\end{subequations}

Having constructed the shock-wave transition operators, we are now ready to calculate their contributions to the flavor-singlet helicity evolution.


\section{Evolution of the Adjoint Dipole of the First Type}\label{sec:ch5_evoladj}

\begin{figure}[h!]
\centering
\includegraphics[width= 0.95 \textwidth]{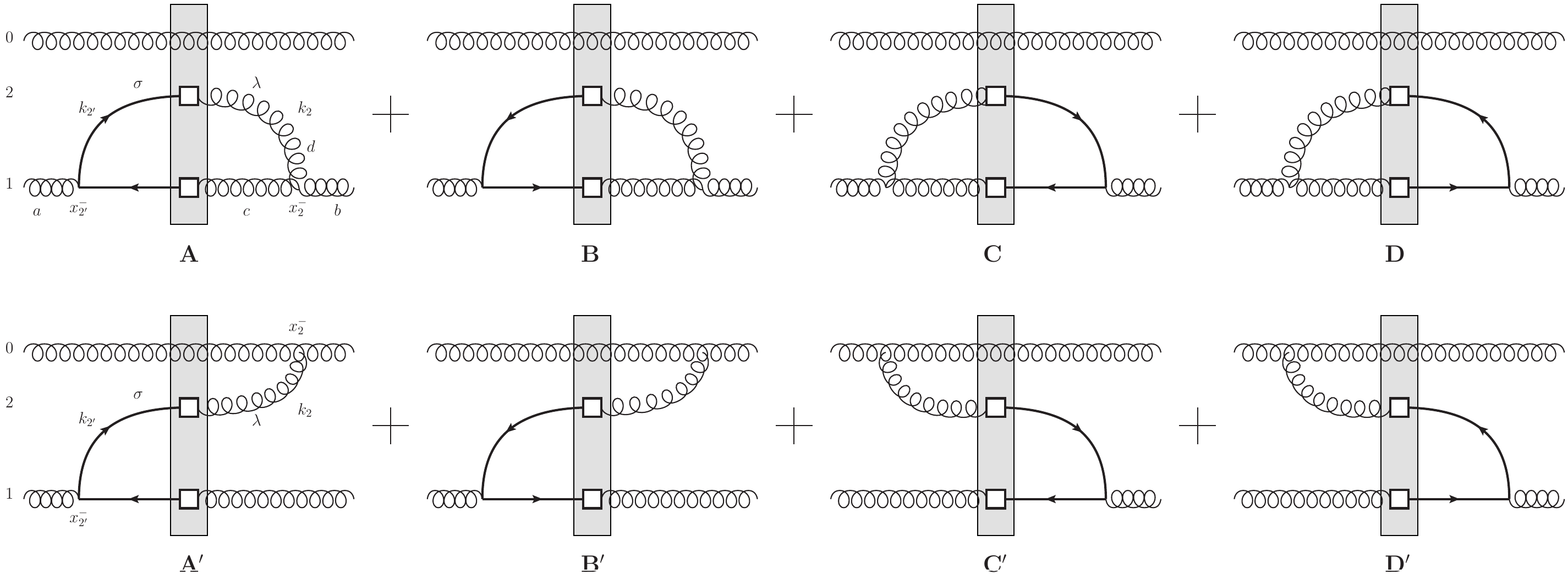}
\caption{The diagrams for the evolution of $G^\textrm{adj}_{10}$ due to the $q/{\bar q} \to G$ and $G \to q/{\bar q}$ shock-wave transition operators calculated here. The shaded rectangle denotes the shock wave, while the white square denotes the non-eikonal interaction with the shock wave mediated by the transition operators from \fig{fig:newoperators}.}
\label{FIG:Gadj_evol}
\end{figure}

The shock-wave transition operators \eqref{Ws2} may contribute to the flavor-singlet helicity evolution of the adjoint type-1 polarized dipole amplitude $G^\textrm{adj}_{10}$ defined in \eq{ch3_G10adj}. The diagrams in the shock wave formalism are shown in \fig{FIG:Gadj_evol}, with the shock wave represented by a vertical shaded rectangle. These diagrams arise due to the contribution of the polarized Wilson line operator $\left( U_{\un x}^{\textrm{q} [1]} \right)^{ba}$ to $G^\textrm{adj}_{10}$ (see \eq{uq1}). Note that the transition operators come in pairs --- the diagrams in \fig{FIG:Gadj_evol} each include two transition operators from Eqs.~\eqref{Ws2}. These diagrams were not included in the analysis of \cite{Kovchegov:2015pbl, Kovchegov:2018znm, Cougoulic:2022gbk}.

\subsection{LCOT}\label{sec:ch5_adjevol_lcot}

To calculate diagrams A and A' from \fig{FIG:Gadj_evol} using LCOT we will need the following background-field propagator, which can be constructed along the lines of \cite{Kovchegov:2017lsr, Kovchegov:2018znm, Kovchegov:2021iyc, Cougoulic:2022gbk} (see also the discussion in Ch.~\ref{sec:ch3_lcot}). Quarks are assumed to be massless.
\begin{align}\label{psibar_a+}
&\int\limits_{-\infty}^0 dx_{2'}^- \, 
\int\limits_0^\infty dx_2^- \, 
\contraction[2ex]
{}
{{\bar \psi}}{_\alpha^i
(x_{2'}^- , \ul{x}_1) \:}
{a}
\: 
{\bar \psi}_\alpha^i (x_{2'}^- , \ul{x}_1) \:
a^{+ \, d} (x_2^- , \ul{x}_0) \\
& \hspace*{1cm}=
\frac{1}{4 \pi^3} \, \int\limits_0^{p_2^-} d k^- \,   \int d^2 x_2 \,  \frac{x_{20}^m}{x_{20}^2} \, \bigg[ \left( \epsilon_+^m \, {\bar \rho} (+)_\beta - \epsilon_-^m \, {\bar \rho} (-)_\beta \right) \, W_{\un 2}^{d;i \, [1]} [\infty, - \infty] \notag \\
&\hspace*{5.6cm}+ \left( \epsilon_+^m \, {\bar \rho} (+)_\beta + \epsilon_-^m \, {\bar \rho} (-)_\beta \right) \, W_{\un 2}^{d;i \, [2]} [\infty, - \infty] \bigg]  \notag \\
&\hspace*{3cm} \times \, \left[ i \, \sqrt{\sqrt{2} k^-} \, \ln \left( \frac{1}{x_{21} \Lambda} \right) + \frac{1}{\sqrt{\sqrt{2} k^-}} \,  \frac{x_{21}^l}{x_{21}^2} \, \gamma^0 \, \gamma^l \right]_{\beta\alpha} . \notag
\end{align}
The notations used in \eq{psibar_a+} are defined in \fig{FIG:Gadj_evol}, with the typical (large) minus momentum of the original (parent) dipole being $p_2^-$. The first term in the last square brackets of \eq{psibar_a+} does not contribute to the evolution at hand. The second term gives, after some algebra,
\begin{align}\label{AA'}
A + A' & = \frac{1}{N_c^2 -1} \, \frac{\as}{2 \pi^2} \, \sum_f \, \int\limits_0^{p_2^-} \frac{d k^-}{k^-} \, \int d^2 x_2 \\
&\times\llangle \left( U^\dagger_{\un 0} \right)^{ab} \, f^{dbc} \, \left[ i \, W_{\un 2}^{d \, [1]} [\infty, - \infty] \, t^a \, \left( W_{\un 1}^{c \, [2]} [\infty, - \infty] \right)^\dagger \frac{{\un x}_{20}}{x_{20}^2}  \cdot \frac{{\un x}_{21}}{x_{21}^2}  \right. \notag\\ 
&\hspace*{.4cm}+ W_{\un 2}^{d \, [1]} [\infty, - \infty] \, t^a \, \left( W_{\un 1}^{c \, [1]} [\infty, - \infty] \right)^\dagger \frac{{\un x}_{20}}{x_{20}^2}  \times \frac{{\un x}_{21}}{x_{21}^2}  \notag \\
&\hspace*{.4cm}+ i \, W_{\un 2}^{d \, [2]} [\infty, - \infty] \, t^a \, \left( W_{\un 1}^{c \, [1]} [\infty, - \infty] \right)^\dagger \frac{{\un x}_{20}}{x_{20}^2}  \cdot \frac{{\un x}_{21}}{x_{21}^2} \notag \\ 
&\hspace*{.4cm}  \left.  + W_{\un 2}^{d \, [2]} [\infty, - \infty] \, t^a \, \left( W_{\un 1}^{c \, [2]} [\infty, - \infty] \right)^\dagger \frac{{\un x}_{20}}{x_{20}^2}  \times \frac{{\un x}_{21}}{x_{21}^2}  \right] \rrangle - ({\un x}_{20} \to {\un x}_{21}) , \notag
\end{align} 
where we have suppressed the quark color indices on the $W$'s.

Here, the double angle brackets are defined as 
\begin{align}\label{double_def}
\llangle W_{\un 2}^{d \, [k]} \, \ldots \, \left( W_{\un 1}^{c \, [l]} \right)^\dagger \rrangle \equiv \sqrt{2 k^- p_1^+} \, \sqrt{2 p_2^- p_1^+} \, \Big\langle W_{\un 2}^{d \, [k]} \, \ldots \, \left( W_{\un 1}^{c \, [l]} \right)^\dagger \Big\rangle 
\end{align} 
with the superscripts $k, l = 1,2$. This is a generalization of the definition introduced in \eq{ch3_Q10} to the case of two non-eikonal interactions considered in \fig{FIG:Gadj_evol}.

Next we move on to diagrams B and B'. We now need another propagator, which one can also construct in the usual way,
\begin{align}\label{psi_a+}
\int\limits_{-\infty}^0 dx_{2'}^- \, & 
\int\limits_0^\infty dx_2^- \, 
\contraction[2ex]
{}
{\psi}{_\alpha^i
(x_{2'}^- , \ul{x}_1) \:}
{a}
\: 
\psi_\alpha^i (x_{2'}^- , \ul{x}_1) \:
a^{+ \, d} (x_2^- , \ul{x}_0) \\
&=
\frac{1}{4 \pi^3} \, \int\limits_0^{p_2^-} d k^- \,   \int d^2 x_2 \,  \frac{x_{20}^m}{x_{20}^2} \, \left[ i \, \sqrt{\sqrt{2} k^-} \, \ln \left( \frac{1}{x_{21} \Lambda} \right) + \frac{1}{\sqrt{\sqrt{2} k^-}} \,  \frac{x_{21}^l}{x_{21}^2} \, \gamma^l \, \gamma^0 \right]_{\alpha\beta}  \notag \\
&\hspace*{3.6cm}\times \, \bigg[ \left( \epsilon_+^m \, {\rho} (-)_\beta - \epsilon_-^m \, {\rho} (+)_\beta \right) \, \left( W_{\un 2}^{d; i \, [1]} [\infty, - \infty] \right)^\dagger \notag\\
&\hspace*{4cm}- \left( \epsilon_+^m \, {\rho} (-)_\beta + \epsilon_-^m \, {\rho} (+)_\beta \right) \, \left( W_{\un 2}^{d; i \, [2]} [\infty, - \infty] \right)^\dagger \bigg].  \notag 
\end{align}
Again the first term in the (now) first square brackets does not contribute to evolution. The second term yields
\begin{align}\label{BB'}
B + B' & = \frac{1}{N_c^2 -1} \, \frac{\as}{2 \pi^2} \, \sum_f \, \int\limits_0^{p_2^-} \frac{d k^-}{k^-} \, \int d^2 x_2 \\
&\times\llangle \left( U^\dagger_{\un 0} \right)^{ab} \, f^{dbc} \, \left[ - i \, W_{\un 1}^{c \, [2]} [\infty, - \infty] \, t^a \, \left( W_{\un 2}^{d \, [1]} [\infty, - \infty] \right)^\dagger \frac{{\un x}_{20}}{x_{20}^2}  \cdot \frac{{\un x}_{21}}{x_{21}^2}  \right. \notag \\ 
&\hspace*{.5cm} + W_{\un 1}^{c \, [1]} [\infty, - \infty] \, t^a \, \left( W_{\un 2}^{d \, [1]} [\infty, - \infty] \right)^\dagger \frac{{\un x}_{20}}{x_{20}^2}  \times \frac{{\un x}_{21}}{x_{21}^2}  \notag\\
&\hspace*{.5cm}- i \, W_{\un 1}^{c \, [1]} [\infty, - \infty] \, t^a \, \left( W_{\un 2}^{d \, [2]} [\infty, - \infty] \right)^\dagger \frac{{\un x}_{20}}{x_{20}^2}  \cdot \frac{{\un x}_{21}}{x_{21}^2} \notag \\
& \hspace*{.5cm}\left. + W_{\un 1}^{c \, [2]} [\infty, - \infty] \, t^a \, \left( W_{\un 2}^{d \, [2]} [\infty, - \infty] \right)^\dagger \frac{{\un x}_{20}}{x_{20}^2}  \times \frac{{\un x}_{21}}{x_{21}^2}  \right] \rrangle - ({\un x}_{20} \to {\un x}_{21}) . \notag
\end{align} 
We observe that
\begin{align}
B + B' = (A + A')^* .
\end{align}

Considering diagrams C and C' next, we need the following propagator:
\begin{align}\label{psi_a+_2}
\int\limits_{-\infty}^0 dx_{2'}^- \, & 
\int\limits_0^\infty dx_2^- \, 
\contraction[2ex]
{}
{\psi}{_\alpha^i
(x_{2}^- , \ul{x}_1) \:}
{a}
\: 
\psi_\alpha^i (x_{2}^- , \ul{x}_1) \:
a^{+ \, d} (x_{2'}^- , \ul{x}_0) \\
&=
- \frac{1}{4 \pi^3} \, \int\limits_0^{p_2^-} d k^- \,   \int d^2 x_2 \,  \frac{x_{20}^m}{x_{20}^2} \, \left[ i \, \sqrt{\sqrt{2} k^-} \, \ln \left( \frac{1}{x_{21} \Lambda} \right) + \frac{1}{\sqrt{\sqrt{2} k^-}} \,  \frac{x_{21}^l}{x_{21}^2} \, \gamma^0 \, \gamma^l \right]_{\alpha\beta}  \notag \\
& \hspace*{3.6cm}\times \, \bigg[ \left( \epsilon_+^{m \, *} \, {\rho} (+)_\beta - \epsilon_-^{m \, *} \, {\rho} (-)_\beta \right) \, \left( W_{\un 2}^{d; i \, [1]} [- \infty, \infty] \right)^\dagger \notag\\
&\hspace*{4cm}+ \left( \epsilon_+^{m \, *} \, {\rho} (+)_\beta + \epsilon_-^{m \, *} \, {\rho} (-)_\beta \right) \, \left( W_{\un 2}^{d; i \, [2]} [- \infty, \infty] \right)^\dagger \bigg].  \notag 
\end{align}
Only the second term in the first square bracket contributes. We obtain
\begin{align}\label{CC'}
C + C' & = \frac{1}{N_c^2 -1} \, \frac{\as}{2 \pi^2} \, \sum_f \, \int\limits_0^{p_2^-} \frac{d k^-}{k^-} \, \int d^2 x_2 \\
&\times\llangle \left( U^\dagger_{\un 0} \right)^{ab} \, f^{dac} \, \left[ - i \, W_{\un 1}^{c \, [2]} [-\infty, \infty] \, t^b \, \left( W_{\un 2}^{d \, [1]} [- \infty, \infty] \right)^\dagger \frac{{\un x}_{20}}{x_{20}^2}  \cdot \frac{{\un x}_{21}}{x_{21}^2}  \right. \notag \\ 
& \hspace*{.5cm}+ W_{\un 1}^{c \, [1]} [- \infty, \infty] \, t^b \, \left( W_{\un 2}^{d \, [1]} [- \infty, \infty] \right)^\dagger \frac{{\un x}_{20}}{x_{20}^2}  \times \frac{{\un x}_{21}}{x_{21}^2}  \notag\\
&\hspace*{.5cm}- i \, W_{\un 1}^{c \, [1]} [- \infty, \infty] \, t^b \, \left( W_{\un 2}^{d \, [2]} [- \infty, \infty] \right)^\dagger \frac{{\un x}_{20}}{x_{20}^2}  \cdot \frac{{\un x}_{21}}{x_{21}^2} \notag \\
&\hspace*{.5cm} \left. + W_{\un 1}^{c \, [2]} [- \infty, \infty] \, t^b \, \left( W_{\un 2}^{d \, [2]} [- \infty, \infty] \right)^\dagger \frac{{\un x}_{20}}{x_{20}^2}  \times \frac{{\un x}_{21}}{x_{21}^2}  \right] \rrangle - ({\un x}_{20} \to {\un x}_{21}) . \notag
\end{align}

One may now guess that 
\begin{align}
D + D' = (C + C')^* ,
\end{align}
such that
\begin{align}\label{DD'}
D + D' & = \frac{1}{N_c^2 -1} \, \frac{\as}{2 \pi^2} \, \sum_f \, \int\limits_0^{p_2^-} \frac{d k^-}{k^-} \, \int d^2 x_2 \\
&\times\llangle \left( U^\dagger_{\un 0} \right)^{ab} \, f^{dac} \, \left[ i \, W_{\un 2}^{d \, [1]} [- \infty, \infty] \, t^b \, \left( W_{\un 1}^{c \, [2]} [-\infty, \infty] \right)^\dagger \frac{{\un x}_{20}}{x_{20}^2}  \cdot \frac{{\un x}_{21}}{x_{21}^2}  \right. \notag \\ 
& \hspace*{.5cm} + W_{\un 2}^{d \, [1]} [- \infty, \infty] \, t^b \, \left( W_{\un 1}^{c \, [1]} [- \infty, \infty] \right)^\dagger \frac{{\un x}_{20}}{x_{20}^2}  \times \frac{{\un x}_{21}}{x_{21}^2}  \notag\\
&\hspace*{.5cm}+  i \,  W_{\un 2}^{d \, [2]} [- \infty, \infty] \, t^b \, \left( W_{\un 1}^{c \, [1]} [- \infty, \infty] \right)^\dagger \frac{{\un x}_{20}}{x_{20}^2}  \cdot \frac{{\un x}_{21}}{x_{21}^2} \notag \\
&\hspace*{.5cm} \left. + W_{\un 2}^{d \, [2]} [- \infty, \infty] \, t^b \, \left( W_{\un 1}^{c \, [2]} [- \infty, \infty]  \right)^\dagger \frac{{\un x}_{20}}{x_{20}^2}  \times \frac{{\un x}_{21}}{x_{21}^2}  \right] \rrangle - ({\un x}_{20} \to {\un x}_{21}) . \notag
\end{align} 
This is indeed what one obtains by a direct calculation using the propagator
\begin{align}\label{psibar_a+_2}
\int\limits_{-\infty}^0 dx_{2'}^- \, & 
\int\limits_0^\infty dx_2^- \, 
\contraction[2ex]
{}
{{\bar \psi}}{_\alpha^i
(x_{2}^- , \ul{x}_1) \:}
{a}
\: 
{\bar \psi}_\alpha^i (x_{2}^- , \ul{x}_1) \:
a^{+ \, d} (x_{2'}^- , \ul{x}_0) \\
&=
\frac{1}{4 \pi^3} \, \int\limits_0^{p_2^-} d k^- \,   \int d^2 x_2 \,  \frac{x_{20}^m}{x_{20}^2} \, \bigg[ \left( - \epsilon_+^{m \, *} \, {\bar \rho} (-)_\beta + \epsilon_-^{m \, *} \, {\bar \rho} (+)_\beta \right) \, W_{\un 2}^{d; i \, [1]} [- \infty, \infty] \notag\\
&\hspace*{4.5cm}+ \left( \epsilon_+^{m \, *} \, {\bar \rho} (-)_\beta + \epsilon_-^{m \, *} \, {\bar \rho} (+)_\beta \right) \, W_{\un 2}^{d; i \, [2]} [- \infty, \infty] \bigg]  \notag \\
&\hspace*{3.6cm} \times \, \left[ i \, \sqrt{\sqrt{2} k^-} \, \ln \left( \frac{1}{x_{21} \Lambda} \right) + \frac{1}{\sqrt{\sqrt{2} k^-}} \,  \frac{x_{21}^l}{x_{21}^2} \, \gamma^l \, \gamma^0 \right]_{\beta\alpha} . \notag
\end{align}

The sum of the diagrams from \fig{FIG:Gadj_evol} is obtained by adding the contributions found in Eqs.~\eqref{AA'}, \eqref{BB'}, \eqref{CC'}, and \eqref{DD'}.

Simplifying Eqs.~\eqref{AA'}, \eqref{BB'}, \eqref{CC'}, and \eqref{DD'} in the DLA, we neglect the terms with the cross products as not double-logarithmic, and write
\begin{align}\label{total}
& A+A'+B+B'+C+C'+D+D' \bigg|_{DLA} \\
&= - \frac{1}{N_c^2 -1} \, \frac{\as}{2 \pi} \, \sum_f \, \int\limits_0^{p_2^-} \frac{d k^-}{k^-} \, \int\limits^{x_{10}^2} \frac{d x_{21}^2}{x_{21}^2} \notag\\
&\hspace{2cm}\times\llangle \left( U^\dagger_{\un 0} \right)^{ab} \, f^{dbc} \, \bigg[ i \, W_{\un 2}^{d \, [1]} [\infty, - \infty] \, t^a \, \left( W_{\un 1}^{c \, [2]} [\infty, - \infty] \right)^\dagger \notag\\
&\hspace*{5cm}+ i \, W_{\un 2}^{d \, [2]} [\infty, - \infty] \, t^a \, \left( W_{\un 1}^{c \, [1]} [\infty, - \infty] \right)^\dagger  + \mbox{c.c.} \bigg] \notag \\ 
&\hspace{2.3cm} + \left( U^\dagger_{\un 0} \right)^{ab} \, f^{dac} \, \bigg[ i \, W_{\un 2}^{d \, [1]} [- \infty, \infty] \, t^b \, \left( W_{\un 1}^{c \, [2]} [-\infty, \infty] \right)^\dagger \notag\\
&\hspace*{5.3cm}+  i \,  W_{\un 2}^{d \, [2]} [- \infty, \infty] \, t^b \, \left( W_{\un 1}^{c \, [1]} [- \infty, \infty] \right)^\dagger + \mbox{c.c.}  \bigg] \rrangle . \notag 
\end{align}
Here c.c. stands for complex conjugate.

Using the definitions \eqref{W12} one can readily show that the expression in the first square brackets of \eq{total} is
\begin{align}\label{sq1}
&  i \, W_{\un 2}^{d \, [1]} [\infty, - \infty] \, t^a \, \left( W_{\un 1}^{c \, [2]} [\infty, - \infty] \right)^\dagger + i \, W_{\un 2}^{d \, [2]} [\infty, - \infty] \, t^a \, \left( W_{\un 1}^{c \, [1]} [\infty, - \infty] \right)^\dagger  + \mbox{c.c.}  \\ 
& = - i \frac{g^2}{4 \sqrt{k^- \, p_2^-}} \, \int\limits_{-\infty}^\infty dy^- \, \int\limits_{-\infty}^\infty dz^- \, U_{\un 2}^{dd'} [\infty, y^-] \, U_{\un 1}^{cc'} [\infty, z^-] \, {\bar \psi} (y^-, {\un x}_2 ) \, t^{d'} \, \gamma^+ \, \gamma^5 \notag\\
& \hspace*{5cm}\times V_{\un 2} [y^-, - \infty] t^a \, V_{\un 1} [-\infty, z^-] \, t^{c'} \, \psi (z^-, {\un x}_1) + \mbox{c.c.} \,. \notag 
\end{align}
The expression in the second square brackets of \eq{total} is obtained from the above by simply interchanging the $\infty \leftrightarrow - \infty$ limits: 
\begin{align}\label{sq2}
&  i \, W_{\un 2}^{d \, [1]} [- \infty, \infty] \, t^b \, \left( W_{\un 1}^{c \, [2]} [-\infty, \infty] \right)^\dagger +  i \,  W_{\un 2}^{d \, [2]} [- \infty, \infty] \, t^b \, \left( W_{\un 1}^{c \, [1]} [- \infty, \infty] \right)^\dagger  + \mbox{c.c.}  \\ 
&= - i \frac{g^2}{4 \sqrt{k^- \, p_2^-}} \, \int\limits_{-\infty}^\infty dy^- \, \int\limits_{-\infty}^\infty dz^- \, U_{\un 2}^{dd'} [-\infty, y^-] \, U_{\un 1}^{cc'} [-\infty, z^-] \, {\bar \psi} (y^-, {\un x}_2 ) \, t^{d'} \, \gamma^+ \, \gamma^5 \notag \\
&\hspace*{5cm}\times V_{\un 2} [y^-, \infty]
t^b \, V_{\un 1} [\infty, z^-] \, t^{c'} \, \psi (z^-, {\un x}_1) + \mbox{c.c.} . \notag 
\end{align}

Employing \eq{sq1} along with
\begin{align}\label{Uab}
U^{ab} = 2 \, \tr [t^a \, V \, t^b \, V^\dagger],
\end{align}
applying the Fierz identity several times (see Eqs.~\eqref{ch3_fierz}) while neglecting the $N_c$-suppressed terms (either explicitly, or by neglecting the terms with fewer color traces, since each color trace brings in a factor of $N_c$), we arrive at
\begin{align}\label{sq12}
& \llangle \left( U^\dagger_{\un 0} \right)^{ab} \, f^{dbc} \, \bigg[ i \, W_{\un 2}^{d \, [1]} [\infty, - \infty] \, t^a \, \left( W_{\un 1}^{c \, [2]} [\infty, - \infty] \right)^\dagger \\
&\hspace*{4cm} + i \, W_{\un 2}^{d \, [2]} [\infty, - \infty] \, t^a \, \left( W_{\un 1}^{c \, [1]} [\infty, - \infty] \right)^\dagger  + \mbox{c.c.} \bigg] \rrangle \notag \\
& =  \int\limits_{-\infty}^\infty dy^- \int\limits_{-\infty}^\infty dz^- \,  \left\langle \tr \left[ V_{\un 2} \, V^\dagger_{\un 0} \right] \right\rangle \, \left\langle \tr \left[V_{\un 0} \, V^\dagger_{\un 1} \right] \right\rangle \notag\\
&\hspace*{2cm} \times \llangle \frac{g^2}{16 \sqrt{k^- \, p_2^-}} \, {\bar \psi} (y^-, {\un x}_2 ) \, \gamma^+ \, \gamma^5  V_{\un 2} [y^-, \infty] V_{\un 1} [\infty, z^-] \, \psi (z^-, {\un x}_1) \rrangle + \mbox{c.c.} . \notag
\end{align}
Similarly, for the expression \eqref{sq2}, we get
\begin{align}\label{sq22}
& \llangle \left( U^\dagger_{\un 0} \right)^{ab} \, f^{dac} \, \bigg[ i \, W_{\un 2}^{d \, [1]} [- \infty, \infty] \, t^b \, \left( W_{\un 1}^{c \, [2]} [-\infty, \infty] \right)^\dagger \notag\\
&\hspace*{4cm} +  i \,  W_{\un 2}^{d \, [2]} [- \infty, \infty] \, t^b \, \left( W_{\un 1}^{c \, [1]} [- \infty, \infty] \right)^\dagger + \mbox{c.c.}  \bigg] \rrangle \\
& =  \int\limits_{-\infty}^\infty dy^- \int\limits_{-\infty}^\infty dz^- \,  \left\langle \tr \left[ V^\dagger_{\un 2} \, V_{\un 0} \right] \right\rangle \, \left\langle \tr \left[V^\dagger_{\un 0} \, V_{\un 1} \right] \right\rangle \notag\\
&\hspace*{1.5cm} \times \llangle \frac{g^2}{16 \sqrt{k^- \, p_2^-}} \, {\bar \psi} (y^-, {\un x}_2 ) \, \gamma^+ \, \gamma^5 \, V_{\un 2} [y^-, - \infty]  V_{\un 1} [- \infty, z^-] \, \psi (z^-, {\un x}_1) \rrangle + \mbox{c.c.} . \notag
\end{align}

Employing Eqs.~\eqref{sq12} and \eqref{sq22} in \eq{total} we obtain (at large $N_c \& N_f$)
\begin{align}\label{total2}
& A+A'+B+B'+C+C'+D+D' \bigg|_{DLA} \\
&= - \frac{\as}{2 \pi} \, \sum_f \, \int\limits_0^{p_2^-} \frac{d k^-}{k^-} \, \int\limits^{x_{10}^2} \frac{d x_{21}^2}{x_{21}^2} \, S_{20} (2 k^- p_1^+)  \, S_{10} (2 k^- p_1^+) \notag\\
&\hspace*{1cm}\times \llangle \frac{g^2}{16 \sqrt{k^- \, p_2^-}} \, \int\limits_{-\infty}^\infty dy^- \int\limits_{-\infty}^\infty dz^- \, \bigg[ {\bar \psi} (y^-, {\un x}_2 ) \, \gamma^+ \, \gamma^5 \, V_{\un 2} [y^-, \infty] \,  V_{\un 1} [\infty, z^-] \, \psi (z^-, {\un x}_1) \notag \\ 
&\hspace*{3.8cm} + {\bar \psi} (y^-, {\un x}_2 ) \, \gamma^+ \, \gamma^5 \, V_{\un 2} [y^-, - \infty] \,  V_{\un 1} [- \infty, z^-] \, \psi (z^-, {\un x}_1) + \mbox{c.c.} \bigg] \rrangle  . \notag 
\end{align}

\begin{figure}[h!]
\centering
\begin{subfigure}[t!]{0.47\textwidth}
    \centering
    \includegraphics[height=5.25cm]{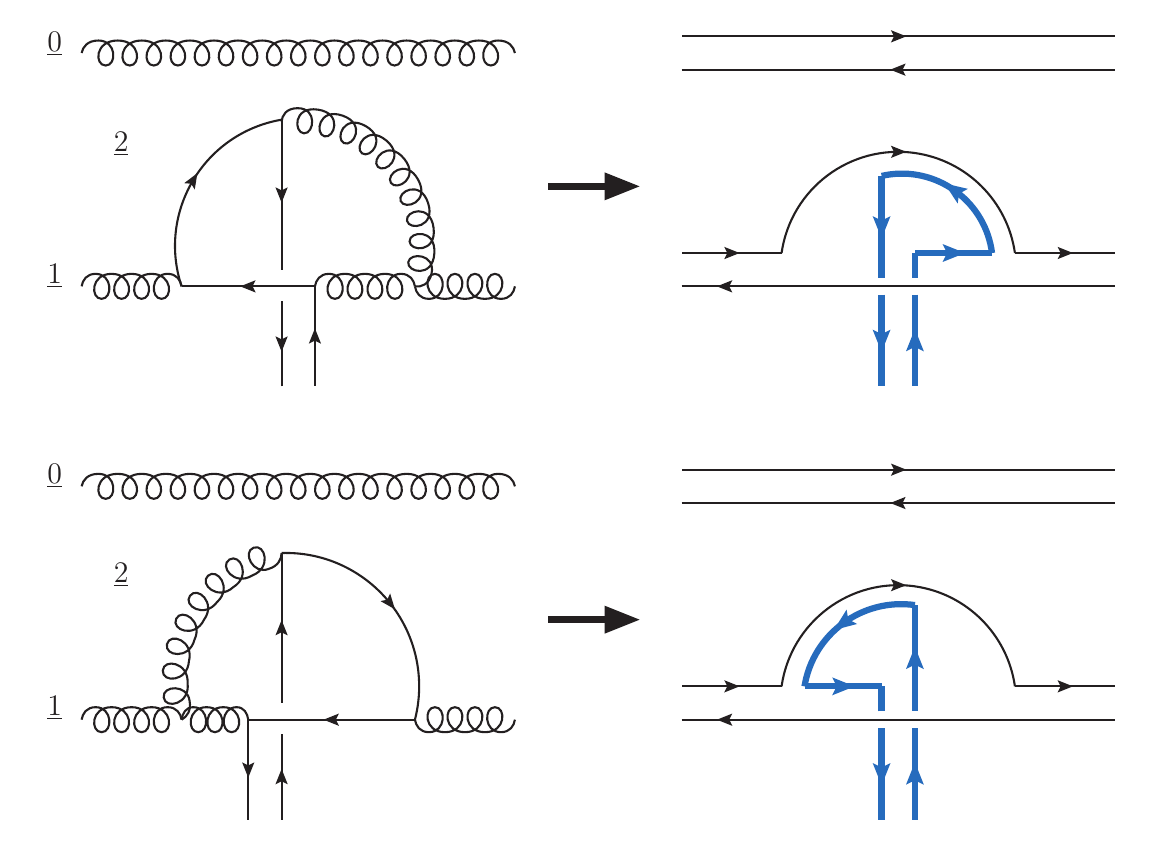}
    \label{fig:ch5_qtildedef}
\end{subfigure}
~
\begin{subfigure}[t!]{0.5\textwidth}
    \centering
    \includegraphics[height=2.5cm]{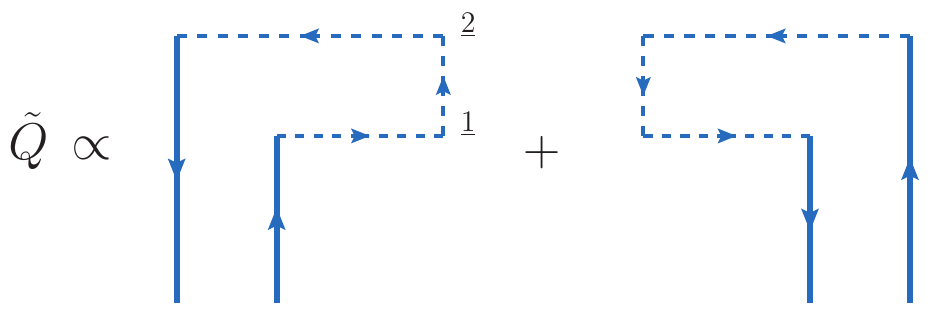}
    \label{fig:ch5_qtildestaples}
\end{subfigure}
\caption{(Left) Illustration of the Wilson line staple structure emerging from the large-$N_c$ ($\&N_f$) limit of some transition diagrams. (Right) The future- and past-pointing Wilson line staples that combine to give $\widetilde{Q}$. The solid vertical blue lines in both panels are the background quark fields. In the right panel, we used dashed blue lines to denote the Wilson line staples.}
\label{fig:ch5_qtildestuff}
\end{figure}

By analogy to Eqs.~\eqref{ch3_Wpol1} and \eqref{ch3_Gtilde} we define a new object called $\widetilde{Q}_{12}(s)$.
\begin{align}\label{Qtilde}
&{\widetilde Q}_{12} (s) \\
&\equiv \llangle \frac{g^2}{16 \sqrt{k^- \, p_2^-}} \, \int\limits_{-\infty}^\infty dy^- \int\limits_{-\infty}^\infty dz^- \, \bigg[ {\bar \psi} (y^-, {\un x}_2 ) \, \left( \frac{1}{2} \,  \gamma^+ \, \gamma^5 \right) \, V_{\un 2} [y^-, \infty] \,  V_{\un 1} [\infty, z^-] \, \psi (z^-, {\un x}_1)  \notag\\ 
& \hspace*{2.4cm} + {\bar \psi} (y^-, {\un x}_2 ) \, \left( \frac{1}{2} \,  \gamma^+ \, \gamma^5 \right) \, V_{\un 2} [y^-, - \infty] \,  V_{\un 1} [- \infty, z^-] \, \psi (z^-, {\un x}_1) + \mbox{c.c.} \bigg] \rrangle (s) . \notag
\end{align}
One cannot properly call $\widetilde{Q}$ a dipole scattering amplitude, since it is more like a quark helicity TMD --- one term contains a future-pointing (SIDIS) Wilson line staple and another term contains a past-pointing (Drell-Yan (DY)) Wilson line staple. We illustrate $\widetilde{Q}$ in more detail in Fig.~\ref{fig:ch5_qtildestuff}, showing in the left panel how a `post-and-staple' structure emerges from some transition diagrams in the large-$N_c$ ($\&N_f$) limit. We do not show the shock wave, but the quark-to-gluon and gluon-to-quark vertices in the middle of the diagrams are the shock wave transition operators, denoted previously as white squares in Fig.~\ref{FIG:Gadj_evol}. The composition of $\widetilde{Q}$ itself is shown in more detail in the right panel of the figure, where the vertical blue solid line `posts' are the background quark/antiquark fields exchanged with the target while the dashed blue lines are the Wilson line `staples.' Note that the transverse Wilson lines (the vertical blue dashed lines in the right panel connecting transverse positions $\underline{2}$ and $\underline{1}$) are implicitly set to $1$, consistent with the definition in \eq{Qtilde}.

Employing the definition in \eq{Qtilde}, we rewrite \eq{total2}
\begin{align}\label{total3}
&A+A'+B+B'+C+C'+D+D' \bigg|_{DLA} \\
&\hspace*{3cm} = - \frac{\as}{\pi} \, \sum_f \, \int\limits_0^{p_2^-} \frac{d k^-}{k^-} \, \int\limits^{x_{10}^2} \frac{d x_{21}^2}{x_{21}^2} \, S_{20} (2 k^- p_1^+)  \, S_{10} (2 k^- p_1^+) \, {\widetilde Q}_{12}  (2 k^- p_1^+) . \notag
\end{align}
In DLA we put $S=1$, since the unpolarized evolution is single-logarithmic. In addition, for simplicity, we replace $\sum_f \to N_f$, assuming that all flavors contribute equally (which, indeed, is not the case in phenomenology \cite{Adamiak:2023yhz}). Writing $k^- = z' p_2^-$ and generalizing the calculation to the case of the parent dipole not being the original projectile (but rather being produced by previous steps of the evolution and described by the minus momentum fraction $z > z'$) we obtain, after imposing the light-cone lifetime constraints \cite{Kovchegov:2015pbl, Kovchegov:2018znm, Cougoulic:2019aja},
\begin{align}\label{total4}
A+A'+B+B'+C+C'+D+D' \bigg|_{DLA} = - \frac{\as \, N_f}{\pi} \, \int\limits_\frac{\Lambda^2}{s}^{z} \frac{d z'}{z'} \, \int\limits^{x_{10}^2}_{1/(z' s)} \frac{d x_{21}^2}{x_{21}^2} \, {\widetilde Q}_{12}  (z' \, s) . 
\end{align}

Note that \eq{total4} gives the contribution of the transition operators to the evolution of $G^\textrm{adj}_{10} (zs)$. However, the large-$N_c \& N_f$ helicity evolution equations from \cite{Cougoulic:2022gbk} are written in terms of ${\widetilde G}_{10} (zs)$ instead of $G^\textrm{adj}_{10} (zs)$. Employing \eq{ch3_GtildeandGadj} we write
\begin{align}\label{Gadj_Gtilde2}
G^\textrm{adj}_{10} (zs) = 4 \, S_{10} (zs)  \, {\widetilde G}_{10} (zs) \approx 4 \, {\widetilde G}_{10} (zs),
\end{align}
with the last step valid in DLA, where $S_{10} =1$. We see that while \eq{total4} is the contribution to the evolution of $G^\textrm{adj}_{10} (zs)$, the contribution to the evolution of ${\widetilde G}_{10} (zs)$ in DLA is 4 times smaller.
Combining Eqs.~\eqref{Gadj_Gtilde2} and \eqref{total4} we arrive at the following contribution of the transition operators to the evolution of the dipole amplitude ${\widetilde G}_{10} (zs)$,
\begin{align}\label{total5}
{\widetilde G}_{10} (zs) \supset - \frac{\as \, N_f}{4 \pi} \, \int\limits_\frac{\Lambda^2}{s}^{z} \frac{d z'}{z'} \, \int\limits^{x_{10}^2}_{1/(z' s)} \frac{d x_{21}^2}{x_{21}^2} \ {\widetilde Q}_{12}  (z' \, s) .
\end{align}
We conclude that the transition operators do indeed contribute to the evolution of ${\widetilde G}_{10} (zs)$. The evolution equation for ${\widetilde G}_{10} (zs)$ found in \cite{Cougoulic:2022gbk} and written earlier in \eq{ch3_lncnf3} needs to be augmented by including the new additive term from \eq{total5} into its right-hand side. The evolution equation \eqref{ch3_lncnf4} for the corresponding neighbor dipole amplitude $\widetilde \Gamma$ needs to be modified in a similar way, though with a slightly different upper integration limit on the dipole sizes. 

To complete the emerging new set of large-$N_c \& N_f$ helicity evolution equations we now need to derive an evolution equation for $\widetilde Q$. Since $\widetilde Q$ enters the evolution of $\widetilde G$ in \eq{total5} in the $x_{21} \ll x_{10}$ regime, it always describes the evolution of the smaller of the two daughter ``dipoles" ($x_{21} \ll x_{20} \approx x_{10}$), and the lifetime of the subsequent evolution in ${\widetilde Q}_{12}  (z' \, s)$ is bounded by $\sim z' \, x_{21}^2$ from above, a limit dependent on the transverse size of the $x_{21}$ ``dipole" and independent of $x_{10}$. We thus conclude that, unlike for $Q, G_2$ and ${\widetilde G}$, no neighbor ``dipole amplitude" is needed for $\widetilde Q$. We will only need to construct the evolution for $\widetilde Q$ proper: this is what we will do in Ch.~\ref{sec:ch5_evolQtilde}. However first, we will re-derive the contribution of the transition operators to the evolution of $G_{10}^{\text{adj}}$ in the LCPT approach introduced in Ch.~\ref{sec:ch3_lcpt}.


\subsection{LCPT}\label{sec:ch5_adjevol_lcpt}

\begin{figure}[h!]
    \centering
    \includegraphics[width=\textwidth]{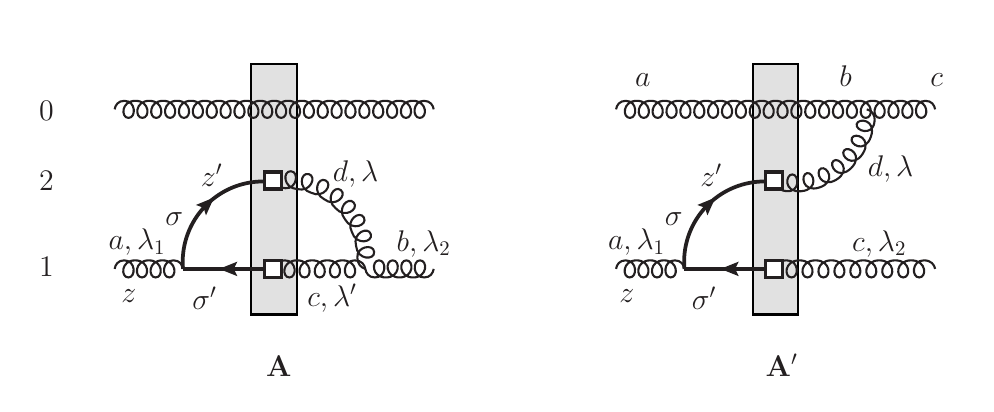}
    \caption{Diagrams A and A', originally considered in \fig{FIG:Gadj_evol}.}
    \label{fig:appA_diagramsAA'}
\end{figure}

We begin with diagrams A and A' from \fig{FIG:Gadj_evol}, reproduced in Fig.~\ref{fig:appA_diagramsAA'} with a few modifications to the labels.
As a reminder, in this framework we still treat the interactions with the shock wave in terms of operators, here the transition operators in Eqs.~\eqref{Ws}. For everything outside the shock wave, we employ the light cone wavefunctions from Ch.~\ref{sec:ch3_lcpt}. We write for diagram A:
\begin{align}\label{appA_diagramA1}
    &A = \frac{1}{2\left(N_c^2 - 1\right)}\left(\frac{1}{2}\sum_{\lambda_1,\lambda_2}\lambda_1\right)\sum_f\sum_{\sigma,\sigma',\lambda,\lambda'} \int\frac{\mathrm{d}z'}{z'}\int \frac{\mathrm{d}^2\underline{x}_2}{4\pi}\theta\left(x_{10}^2z - x_{21}^2z'\right)\\
    &\hspace{1cm}\times\bigg\langle U_{\underline{0}}^{ba}\left[\psi^{G\rightarrow GG}_{dbc;\lambda,\lambda_2,\lambda'}\left(\underline{x}_{21}, \frac{z'}{z}\right)\bigg|_{\tfrac{z'}{z}\rightarrow 0}\right]^*\left(W_{\underline{2}}^{Gq}\left[\infty,-\infty\right]\right)^{d,\lambda;i,\sigma} \notag\\
    &\hspace{2cm}\times\left[\left(\psi^{G\rightarrow q\bar{q}}_{a,\lambda_1;\sigma,\sigma'}\right)^{ij}\left(\underline{x}_{21},\frac{z'}{z}\right)\bigg|_{\tfrac{z'}{z}\rightarrow 0}  \right] \left(W_{\underline{1}}^{G\bar{q}}\left[\infty,-\infty\right]\right)^{c,\lambda'; j,\sigma'} \bigg\rangle \notag.
\end{align}
The factor of $1/2$ in front of the helicity sum averages over incoming helicities. We also weight our sum by the incoming helicity $\lambda_1$ in order to pick out the helicity-dependent term of the sub-eikonal gluon S-matrix, since this is the one that contributes to $G^{\text{adj}}_{10}$
(see Eqs.~\eqref{ch3_gluonsmatrix}, \eqref{ch3_poladjtype1}, and \eqref{ch3_G10adj}). 
Here we have included both adjoint and fundamental color indices as well as polarization labels for clarity. Note also that we treat both the shock-wave transition operators as being part of the (forward) amplitude rather than one being in the amplitude and another one in the complex conjugate amplitude. Hence we use $W^{Gq}$ and $W^{G\bar{q}}$, both with endpoints running from $-\infty$ to $\infty$. In addition, we have a theta function in the first line of \eq{appA_diagramA1} to enforce light-cone lifetime ordering \cite{Kovchegov:2015pbl,Kovchegov:2018znm,Cougoulic:2022gbk}.

Now we substitute our splitting wavefunctions and shock-wave transition operators into \eq{appA_diagramA1}. Here we opt to employ the explicit expressions in Eqs.~\eqref{Ws} for the transition operators from the outset, obtaining
\begin{align}\label{appA_diagramA2}
    &A = \frac{1}{2\left(N_c^2 - 1\right)}\left(\frac{1}{2}\sum_{\lambda_1,\lambda_2}\lambda_1\right)\sum_f\sum_{\sigma,\sigma',\lambda,\lambda'} \int\frac{\mathrm{d}z'}{z'}\int \frac{\mathrm{d}^2\underline{x}_2}{4\pi}\theta\left(x_{10}^2z - x_{21}^2z'\right)\\
    &\hspace{0.25cm}\times\Bigg\langle U_{\underline{0}}^{ba}\left[\frac{-g}{\pi} f^{dbc} \delta_{\lambda_2,\lambda'} \frac{\underline{\epsilon}_{\lambda}^*\cdot\underline{x}_{21}}{x_{21}^2}\right]^* \notag \\
    &\hspace*{0.75cm}\times\bigg(\int\limits_{-\infty}^{\infty}\mathrm{d}w_2^- \,U_{\underline{2}}^{de}\left[\infty,w_2^-\right] \frac{-ig}{2\sqrt{\sqrt{2}k^-}}\bar{\psi}_\alpha\left(w_2^-,\underline{x}_2\right) t^{e} \delta_{\sigma,\lambda} \notag\\
    &\hspace*{2cm}\times \left[\sigma\left(\rho_\alpha(-)-\rho_\alpha(+) \right) + \left(\rho_\alpha(-)+\rho_\alpha(+) \right)   \right] V_{\underline{2}}\left[w_2^-,-\infty \right]\bigg) \notag \\
    &\hspace{0.75cm}\times \left[\frac{ig}{2\pi}t^a\delta_{\sigma,-\sigma'}\sqrt{\frac{z'}{z}}\left(1-\sigma\lambda_1 \right)\frac{\underline{\epsilon}_{\lambda_1}\cdot\underline{x}_{21}}{x_{21}^2}   \right] \notag \\
    &\hspace{0.75cm}\times\bigg(\int\limits_{-\infty}^{\infty}\mathrm{d}w_1^- \,U_{\underline{1}}^{cf}\left[\infty,w_1^-\right] \frac{-ig}{2\sqrt{\sqrt{2}p_2^-}}    V_{\underline{1}}^\dagger\left[w_1^-,-\infty \right] \delta_{\sigma',\lambda'}\notag\\
    &\hspace*{2cm}\times\left[ \sigma'\left(\bar{\rho}_\beta(-)-\bar{\rho}_\beta(+) \right) - \left(\bar{\rho}_\beta(-)+\bar{\rho}_\beta(+) \right)  \right]  t^f \psi_\beta\left(w_1^-,\underline{x}_1 \right)\bigg)\bigg\rangle \notag.
\end{align}
We have now suppressed the fundamental color indices but explicitly included the spinor indices $\alpha,\beta$. We also suppose that the soft quark entering the shock wave at transverse position 2 comes in with momentum $k$, while the momentum of the parent dipole is $p_2$. Both of these momenta are assumed to have large minus components. With these explicit factors of momentum, and with the factor $\tfrac{1}{\sqrt{z}}$ from the $G\rightarrow q\bar{q}$ light-cone wave function, we can write
\begin{align}\label{appA_diagramA3}
    \frac{1}{\sqrt{k^-p_2^-}}\frac{1}{\sqrt{z}} \approx \frac{1}{\sqrt{\left(z'P^- \right)\left(zP^-\right) } } \frac{1}{\sqrt{z}} = \frac{1}{\sqrt{z'}}\frac{2p_1^+}{zs}
\end{align}
where we have taken the projectile to have a large minus component of momentum $P^-$ and the target proton to have a large plus component of momentum $p_1^+$ with $s=2p_1^+P^-$ the center of mass energy squared between them.The adjoint parent dipole in \fig{fig:appA_diagramsAA'} need not be the original projectile and could have instead been produced by previous steps of evolution. Employing this fact and carrying out some algebra we arrive at 
\begin{align}\label{appA_diagramA4}
    &A = \frac{1}{z}\frac{1}{2\left(N_c^2 - 1\right)}4i \frac{\alpha_s N_f}{\pi^2}\int\frac{\mathrm{d}z'}{z'}\sqrt{z'}\int \frac{\mathrm{d}^2\underline{x}_2}{x_{21}^2}\theta\left(x_{10}^2z - x_{21}^2z'\right) \\
    &\hspace{1cm}\times\bigg\langle \frac{g^2p_1^+}{8\sqrt{z'}s}\int\limits_{-\infty}^{\infty}\mathrm{d}w_1^-\int\limits_{-\infty}^{\infty}\mathrm{d}w_2^- \,U_{\underline{0}}^{ba} f^{dbc} U_{\underline{2}}^{de}\left[\infty, w_2^-\right] \bar{\psi}_\alpha\left(w_2^-,\underline{x}_2\right) t^e \notag \\
    &\hspace{1.5cm} \times V_{\underline{2}}\left[w_2^-,-\infty\right] t^a U_{\underline{1}}^{cf}\left[\infty,w_1^-\right] V_{\underline{1}}^\dagger\left[w_1^-,-\infty\right] t^f \left(\frac{1}{2}\gamma^+\gamma^5 \right)_{\alpha\beta}\psi_{\beta}\left(w_1^-,\underline{x}_1\right)\bigg\rangle . \notag
\end{align}
We have replaced $\sum_f$ with $N_f$ for simplicity. Employing the double angle brackets defined in \eq{double_def}, we obtain
\begin{align}\label{appA_diagramA5}
    &A = \frac{1}{zs}\frac{1}{2\left(N_c^2 - 1\right)}4i \frac{\alpha_s N_f}{\pi^2}\int\frac{\mathrm{d}z'}{z'}\int \frac{\mathrm{d}^2\underline{x}_2}{x_{21}^2}\theta\left(x_{10}^2z - x_{21}^2z'\right) \\
    &\hspace{1cm}\times\llangle \frac{g^2p_1^+}{8\sqrt{z'}s}\int\limits_{-\infty}^{\infty}\mathrm{d}w_1^-\int\limits_{-\infty}^{\infty}\mathrm{d}w_2^- \,U_{\underline{0}}^{ba} f^{dbc} U_{\underline{2}}^{de}\left[\infty, w_2^-\right] \bar{\psi}_\alpha\left(w_2^-,\underline{x}_2\right) t^e \notag \\
    &\hspace*{1.5cm}\times V_{\underline{2}}\left[w_2^-,-\infty\right] t^a U_{\underline{1}}^{cf}\left[\infty,w_1^-\right] V_{\underline{1}}^\dagger\left[w_1^-,-\infty\right] t^f \left(\frac{1}{2}\gamma^+\gamma^5 \right)_{\alpha\beta}\psi_{\beta}\left(w_1^-,\underline{x}_1\right)\rrangle \notag.
\end{align}
Diagram A' is similar. We write
\begin{align}\label{appA_diagramA'1}
    &A' = \frac{1}{2\left(N_c^2 - 1\right)}\left(\frac{1}{2}\sum_{\lambda_1,\lambda_2}\lambda_1\right)\sum_f\sum_{\sigma,\sigma',\lambda} \int\frac{\mathrm{d}z'}{z'}\int \frac{\mathrm{d}^2\underline{x}_2}{4\pi}\theta\left(x_{10}^2z - \text{max}\{x_{20}^2,x_{21}^2\}z'\right)\\
    &\hspace{1cm}\times\bigg\langle U_{\underline{0}}^{ba}\left[\psi^{G\rightarrow GG}_{dcb;\lambda}\left(\underline{x}_{20}, \frac{z'}{z}\right)\bigg|_{\tfrac{z'}{z}\rightarrow 0}\right]^*\left(W_{\underline{2}}^{Gq}\left[\infty,-\infty\right]\right)^{d,\lambda;i,\sigma} \notag \\
    &\hspace*{2cm} \times\left[\left(\psi^{G\rightarrow q\bar{q}}_{a,\lambda_1;\sigma,\sigma'}\right)^{ij}\left(\underline{x}_{21},\frac{z'}{z}\right)\bigg|_{\tfrac{z'}{z}\rightarrow 0}  \right]\left(W_{\underline{1}}^{G\bar{q}}\left[\infty,-\infty\right]\right)^{c,\lambda_2; j,\sigma'} \bigg\rangle \notag.
\end{align}
Note that the light-cone lifetime-ordering theta function has been modified to account for both resulting (daughter) dipole sizes \cite{Kovchegov:2015pbl,Kovchegov:2018znm,Cougoulic:2022gbk}. Following the same steps as with diagram A, we obtain the following result for diagram A',
\begin{align}\label{appA_diagramA'3}
    &A' = \frac{1}{zs}\frac{1}{2\left(N_c^2-1\right)}4i \frac{\alpha_s N_f}{\pi^2}\int\frac{\mathrm{d}z'}{z'}\int \mathrm{d}^2\underline{x}_2 \theta\left(x_{10}^2z - \text{max}\{x_{20}^2,x_{21}^2\}z'\right)\frac{\underline{x}_{20}\cdot\underline{x}_{21}}{x_{20}^2x_{21}^2} \\ &\hspace{1cm}\times\llangle \frac{g^2p_1^+}{8\sqrt{z'}s} \int\limits_{-\infty}^{\infty}\mathrm{d}w_1^-\int\limits_{-\infty}^{\infty}\mathrm{d}w_2^-\, U_{\underline{0}}^{ba}f^{dcb}U_{\underline{2}}^{de}\left[\infty,w_2^-\right] \bar{\psi}_{\alpha}\left(w_2^-,\underline{x}_2\right)  t^e V_{\underline{2}}\left[w_2^-,-\infty\right] \notag\\
    &\hspace{3.5cm} \times t^a U_{\underline{1}}^{cf}\left[\infty,w_1^-\right] V_{\underline{1}}^\dagger\left[w_1^-,-\infty \right]  t^f \left(\frac{1}{2}\gamma^+\gamma^5\right)_{\alpha\beta}\psi_{\beta}\left(w_1^-,\underline{x}_1 \right)\rrangle \notag.
\end{align}
We have neglected a term proportional to $\underline{x}_{20} \times \underline{x}_{21}$ which does not contribute in DLA. 

For diagrams B and B' from \fig{FIG:Gadj_evol} it remains true as in Sec. \ref{sec:ch5_adjevol_lcot} that 
\begin{align}\label{appA_diagramBB'1}
    \left(B+B'\right) = \left(A+A'\right)^*.
\end{align}
A direct calculation within the framework of this Section bears this out, but for brevity we do not show it here. 

Next we consider diagrams C and C' from \fig{FIG:Gadj_evol}. For diagram C we can write
\begin{align}\label{appA_diagramC1}
    &C = \frac{1}{2\left(N_c^2 - 1\right)}\left(\frac{1}{2}\sum_{\lambda_1,\lambda_2}\lambda_1\right)\sum_f\sum_{\sigma,\sigma',\lambda,\lambda'} \int\frac{\mathrm{d}z'}{z'}\int \frac{\mathrm{d}^2\underline{x}_2}{4\pi}\theta\left(x_{10}^2z - x_{21}^2z'\right)\\
    &\hspace{1cm}\times\bigg\langle U_{\underline{0}}^{ba}\left[\psi^{G\rightarrow GG}_{dac;\lambda,\lambda_1,\lambda'}\left(\underline{x}_{21}, \frac{z'}{z}\right)\bigg|_{\tfrac{z'}{z}\rightarrow 0}\right]\left(W_{\underline{1}}^{\bar{q}G}\left[\infty,-\infty\right]\right)^{j,\sigma';c,\lambda'} \notag \\
    &\hspace*{2cm}\times\left[\left(\psi^{G\rightarrow q\bar{q}}_{b,\lambda_2;\sigma,\sigma'}\right)^{ji}\left(\underline{x}_{21},\frac{z'}{z}\right)\bigg|_{\tfrac{z'}{z}\rightarrow 0}  \right]^* \left(W_{\underline{2}}^{qG}\left[\infty,-\infty\right]\right)^{i,\sigma;d,\lambda} \bigg\rangle \notag.
\end{align}
Just as we did for diagram A, we substitute in the light-cone wave functions and the explicit expressions for the shock-wave transition operators. Some simplification yields
\begin{align}\label{appA_diagramC2}
    &C = -\frac{1}{zs}\frac{1}{2\left(N_c^2-1\right)} 4i\frac{\alpha_s N_f}{\pi^2}\int\frac{\mathrm{d}z'}{z'}\int\frac{\mathrm{d}^2\underline{x}_2}{x_{21}^2}\theta\left(x_{10}^2z-x_{21}^2z'\right) \\
    &\hspace{1.5cm}\times\llangle \frac{g^2 p_1^+}{8\sqrt{z'}s} \int\limits_{-\infty}^{\infty}\mathrm{d}w_1^-\int\limits_{-\infty}^{\infty}\mathrm{d}w_2^- \,U_{\underline{0}}^{ba}f^{dac}\bar{\psi}_{\alpha}\left(w_1^-,\underline{x}_{1}\right) t^{e} V_{\underline{1}}^\dagger\left[\infty,w_1^-\right] U_{\underline{1}}^{ec}\left[w_1^-,-\infty\right] \notag\\
    &\hspace*{4cm}\times t^{b}V_{\underline{2}}\left[\infty,w_2^-\right]t^{f}\left(\frac{1}{2}\gamma^+\gamma^5 \right)_{\alpha\beta} \psi_\beta\left(w_2^-,\underline{x}_2 \right) U_{\underline{2}}^{fd}\left[w_2^-,-\infty\right]\rrangle. \notag
\end{align}

Diagram C' from \fig{FIG:Gadj_evol} is calculated similarly. The result is 
\begin{align}\label{appA_diagramC'1}
    &C' = -\frac{1}{zs}\frac{1}{2\left(N_c^2-1\right)} 4i \frac{\alpha_s N_f}{\pi^2} \int\frac{\mathrm{d}z'}{z'}\int\mathrm{d}^2\underline{x}_2\, \theta\left(x_{10}^2z - \text{max}\{x_{20}^2,x_{21}^2\}z' \right) \frac{\underline{x}_{20}\cdot \underline{x}_{21}}{x_{20}^2x_{21}^2} \\
    &\hspace{1.2cm}\times \llangle \frac{g^2p_1^+}{8\sqrt{z'}s} \int\limits_{-\infty}^{\infty}\mathrm{d}w_1^-\int\limits_{-\infty}^{\infty}\mathrm{d}w_2^-\, U_{\underline{0}}^{ba} f^{dca} \bar{\psi}_{\alpha}\left(w_1^-,\underline{x}_1\right) t^{e} V_{\underline{1}}^{\dagger}\left[\infty,w_1^-\right] U_{\underline{1}}^{ec}\left[w_1^-,-\infty\right] \notag\\
    &\hspace{3.8cm} \times t^{d} V_{\underline{2}}\left[\infty,w_2^-\right] t^{f} \left(\frac{1}{2}\gamma^+\gamma^5 \right)_{\alpha\beta} \psi_{\beta}\left(w_2^-,\underline{x}_2 \right) U_{\underline{2}}^{fd}\left[w_2^-,-\infty\right]\rrangle \notag.
\end{align}
Again a straightforward calculation in the LCPT framework shows that
\begin{align}\label{appA_diagramDD'}
    (D+D') = (C+C')^*
\end{align}
Adding together all the contributions to the evolution of $G^{\text{adj}}_{10}$ we obtain
\begin{align}\label{appA_alladjdiagrams1}
    &A+A'+B+B'+C+C'+D+D' \bigg|_{\text{DLA}} \\
    &= 
    \frac{4i}{2\left(N_c^2 - 1\right)} \frac{\alpha_s N_f}{\pi}\int\limits_{\Lambda^2/s}^{z}\frac{\mathrm{d}z'}{z'}\int\limits_{1/z's}^{x_{10}^2}\frac{\mathrm{d}x_{21}^2}{x_{21}^2} \llangle \frac{g^2p_1^+}{8\sqrt{z'}s}\int\limits_{-\infty}^{\infty}\mathrm{d}w_1^-\int\limits_{-\infty}^{\infty}\mathrm{d}w_2^- \notag\\
    &\hspace*{2cm}\times\bigg(U_{\underline{0}}^{ba} f^{dbc} U_{\underline{2}}^{de}\left[\infty, w_2^-\right] \bar{\psi}_\alpha\left(w_2^-,\underline{x}_2\right) t^e V_{\underline{2}}\left[w_2^-,-\infty\right] t^a U_{\underline{1}}^{cf}\left[\infty,w_1^-\right] \notag \\
    &\hspace*{4.5cm} \times V_{\underline{1}}^\dagger\left[w_1^-,-\infty\right] t^f \left(\frac{1}{2}\gamma^+\gamma^5 \right)_{\alpha\beta}\psi_{\beta}\left(w_1^-,\underline{x}_1\right) \notag \\
    &\hspace*{2cm} + U_{\underline{0}}^{ba}f^{dca}\bar{\psi}_{\alpha}\left(w_1^-,\underline{x}_{1}\right) t^{e} V_{\underline{1}}^\dagger\left[\infty,w_1^-\right] U_{\underline{1}}^{ec}\left[w_1^-,-\infty\right] t^{b}V_{\underline{2}}\left[\infty,w_2^-\right]t^{f} \notag \\
    &\hspace*{4.5cm}\times\left(\frac{1}{2}\gamma^+\gamma^5 \right)_{\alpha\beta} \psi_\beta\left(w_2^-,\underline{x}_2 \right) U_{\underline{2}}^{fd}\left[w_2^-,-\infty\right] \quad +\quad \text{c.c.} \bigg) \rrangle \notag
\end{align}
where we canceled the factor of $1/zs$ since the same factor appears on the left hand side of the evolution equation (upon applying the double angle brackets) and we also simplified our kernel in DLA, where the transverse integral is only logarithmic for $x_{21}\ll x_{10} \sim x_{20}$. Next we take the large-$N_c$ limit, using \eq{Uab}
along with the Fierz identity. We neglect all terms sub-leading in $N_c$ and set the unpolarized S-matrices we obtain to 1 (again, this is valid at DLA). At large-$N_c$ the structure in double angle brackets in \eq{appA_alladjdiagrams1} becomes $\widetilde{Q}_{21}$ from Sec.~\ref{sec:ch5_adjevol_lcot} (see \eq{Qtilde} and simply use the relations $k^- = z'p_2^-$, $s = 2p_1^+p_2^-$ from that Section to reconcile the pre-factors). The result at large $N_c\&N_f$ is
\begin{align}
    A+A'+B+B'+C+C'+D+D' \bigg|_{\text{DLA}} = - \frac{\alpha_s N_f}{\pi}\int\limits_{\Lambda^2/s}^{z}\frac{\mathrm{d}z'}{z'}\int\limits_{1/z's}^{x_{10}^2}\frac{\mathrm{d}x_{21}^2}{x_{21}^2} \, \widetilde{Q}_{12}\left(z's\right).
\end{align}
Lastly, to write the evolution equations in terms of the dipole amplitude $\widetilde{G}_{10}$ instead of $G^{\text{adj}}_{10}$, we divide the contribution obtained here by $4$ (see \eq{Gadj_Gtilde2}). The result is 
\begin{align}\label{appA_alldiagrams4}
    \widetilde{G}_{10}\left(zs\right) \supset -\frac{\alpha_s N_f}{4\pi}\int\limits_{\Lambda^2/s}^{z}\frac{\mathrm{d}z'}{z'}\int\limits_{1/z's}^{x_{10}^2}\frac{\mathrm{d}x_{21}^2}{x_{21}^2} \widetilde{Q}_{12}\left(z's\right)
\end{align}
in full agreement with \eq{total5}.


\section{\texorpdfstring{Evolution of $\widetilde{Q}$}{Evolution of Qtilde}}\label{sec:ch5_evolQtilde}

Now we will construct the small-$x$ evolution of the new object $\widetilde{Q}$. Again we will do this two ways --- first employing LCOT and then again employing LCPT.


\subsection{LCOT}\label{sec:ch5_Qtildeevollcot}

\begin{figure}[h!]
\centering
\includegraphics[width= 0.95 \textwidth]{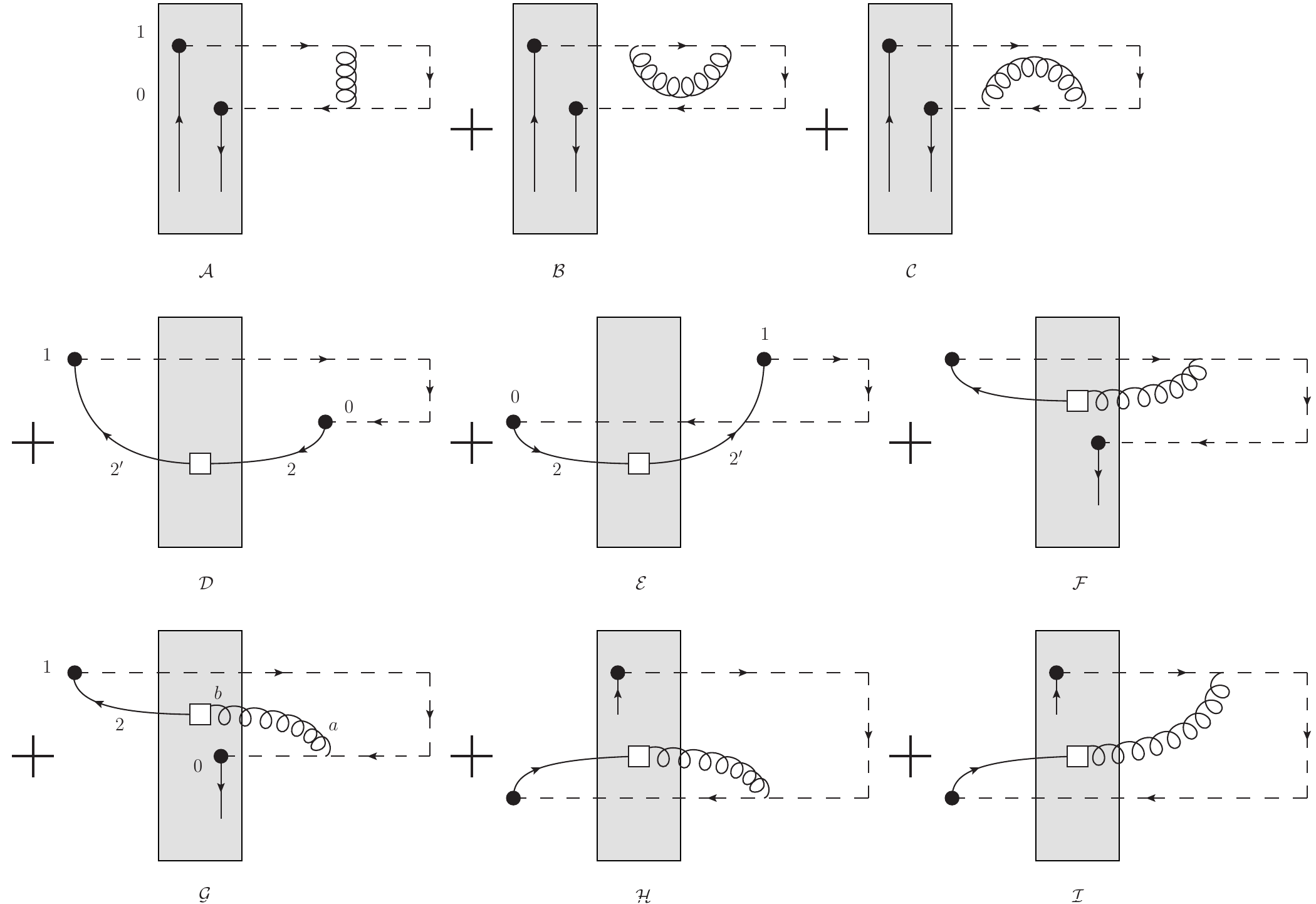}
\caption{Diagrams contributing to the DLA evolution of ${\widetilde Q}_{10} (zs)$. The shaded rectangle is, again, the shock wave. The white square denotes the non-eikonal interactions, either due to the transition operators \eqref{Ws} or due to the previous sub-eikonal operators introduced in Ch.~\ref{subsec:polwilsonlines}. Vertical straight lines denote the quark and anti-quark fields of the shock wave. The dashed line denotes the (future-pointing) Wilson-line staple.}
\label{FIG:Qtilde_evol}
\end{figure}

Diagrams contributing to the evolution of ${\widetilde Q}_{10} (zs)$ are shown in \fig{FIG:Qtilde_evol}, with only the future-pointing Wilson line staple shown: the diagrams for the past-pointing staple are constructed and calculated by analogy. Since ${\widetilde Q}_{10} (zs)$ is similar to the quark helicity TMD (SIDIS plus DY), the diagrams in \fig{FIG:Qtilde_evol} closely parallel those in Fig.~2 of \cite{Kovchegov:2018znm}. The analysis carried out in \cite{Kovchegov:2018znm} suggests that only diagrams $\cal D$ and $\cal E$ in our \fig{FIG:Qtilde_evol} should contribute in the DLA: this is what we will indeed confirm below. 

Diagrams $\cal A$, $\cal B$ and $\cal C$ from \fig{FIG:Qtilde_evol} are just the eikonal corrections for a ``half-dipole" amplitude \cite{Kovchegov:2015zha}. Their contribution is known from the unpolarized small-$x$ evolution \cite{Mueller:1994rr,Mueller:1994jq,Mueller:1995gb,Balitsky:1995ub,Balitsky:1998ya,Kovchegov:1999yj,Kovchegov:1999ua,Jalilian-Marian:1997dw,Jalilian-Marian:1997gr,Weigert:2000gi,Iancu:2001ad,Iancu:2000hn,Ferreiro:2001qy}:
\begin{align}\label{ABC}
&{\cal A} + {\cal B} + {\cal C} = - \frac{\as \, N_c}{4 \pi^2} \, \int\limits_\frac{\Lambda^2}{s}^{z} \frac{d z'}{z'} \, \int d^2 x_2 \, \frac{x_{10}^2}{x_{21}^2 \, x_{20}^2} \, {\widetilde Q}_{10} (z' s) \\
&\hspace{1.9cm} \approx - \frac{\as \, N_c}{4 \pi} \, \int\limits_\frac{\Lambda^2}{s}^{z} \frac{d z'}{z'} \, \left[ \int\limits^{x_{10}^2}_{1/(z' s)} \frac{d x_{21}^2}{x_{21}^2} + \int\limits^{x_{10}^2}_{1/(z' s)} \frac{d x_{20}^2}{x_{20}^2} \right] \, {\widetilde Q}_{10} (z' s) . \notag
\end{align}
Here we include the past-pointing (DY) Wilson line staple's contribution as well to assemble the entire $\widetilde Q$ object. The last step in \eq{ABC} is the simplification in the DLA limit. 

Next, let us calculate diagrams $\cal F$ and $\cal G$ in \fig{FIG:Qtilde_evol}. We will need the propagator \eqref{psi_a+} again. After some algebra we end up with
\begin{align}\label{FG}
& {\cal F} + {\cal G} = - 2 P^+ \, \frac{g^4}{256 \, \pi^3} \int\limits_0^{p_2^-} \frac{d k^-}{k^-} \, \int\limits_{-\infty}^\infty dy^- \int\limits_{-\infty}^\infty dz^- \, \int d^2 x_2 \\
&\hspace*{1cm}\times\bigg\langle {\bar \psi} (y^-, {\un x}_0) \, V_{\un 0} [y^-, \infty] \, t^a \, V_{\un 1} \, U_{\un 2}^{ab} [\infty, z^-] \, V_{\un 2} [-\infty , z^-]  \notag\\
&\hspace*{1.5cm} \times \, \left[ \frac{{\un x}_{20}}{x_{20}^2}  \cdot \frac{{\un x}_{21}}{x_{21}^2}  \, \gamma^+ \, \gamma^5 - i \, \frac{{\un x}_{20}}{x_{20}^2}  \times \frac{{\un x}_{21}}{x_{21}^2} \, \gamma^+  \right] \, t^b \, \psi (z^-, {\un x}_2) \bigg\rangle - ({\un x}_{20} \to {\un x}_{21}) + \mbox{c.c.} . \notag
\end{align}
Dropping the cross-product term as non-DLA and simplifying the dot-product term in DLA we obtain
\begin{align}\label{FG2}
& {\cal F} + {\cal G} = 2 P^+ \, \frac{g^4}{256 \, \pi^2} \int\limits_0^{p_2^-} \frac{d k^-}{k^-} \, \int\limits_{-\infty}^\infty dy^- \int\limits_{-\infty}^\infty dz^- \, \int\limits^{x_{10}^2} \frac{d x^2_{21}}{x^2_{21}} \, \bigg\langle {\bar \psi} (y^-, {\un x}_0) \, V_{\un 0} [y^-, \infty] \, t^a \, V_{\un 1} \\
&\hspace*{5cm}\times U_{\un 2}^{ab} [\infty, z^-] \, V_{\un 2} [-\infty , z^-] \gamma^+ \, \gamma^5 \, t^b \, \psi (z^-, {\un x}_2) \bigg\rangle + \mbox{c.c.}  . \notag
\end{align}
Taking the large-$N_c$ limit with the help of \eq{Uab} and the Fierz identity yields
\begin{align}\label{FG3}
& {\cal F} + {\cal G} = \frac{\as \, N_c}{4 \, \pi} \int\limits_\frac{\Lambda^2}{s}^{z} \frac{d z'}{z'} \,  \int\limits^{x_{10}^2}_{1/(z' s)} \frac{d x^2_{21}}{x^2_{21}} \, S_{21} (z' s)  \, {\widetilde Q}_{20} (z' s) \\
&\hspace*{1.2cm} \approx \frac{\as \, N_c}{4 \, \pi} \int\limits_\frac{\Lambda^2}{s}^{z} \frac{d z'}{z'} \,  \int\limits^{x_{10}^2}_{1/(z' s)} \frac{d x^2_{21}}{x^2_{21}} \, {\widetilde Q}_{20} (z' s) , \notag
\end{align}
where in the last step we have put $S = 1$, which is valid in DLA. Again we have included the past-pointing Wilson line staple's contribution by simply interchanging $\infty \leftrightarrow -\infty$ in the above.
Adding \eq{FG3} to the first term on the right of \eq{ABC} we get 
\begin{align}\label{Q-Q}
\frac{\as \, N_c}{4 \, \pi} \int\limits_\frac{\Lambda^2}{s}^{z} \frac{d z'}{z'} \,  \int\limits^{x_{10}^2}_{1/(z' s)} \frac{d x^2_{21}}{x^2_{21}} \, \left[ {\widetilde Q}_{20} (z' s) - {\widetilde Q}_{10} (z' s) \right] \approx 0,
\end{align}
where the combination is approximately zero in the DLA. It is now straightforward to conclude that diagrams $\cal H$ and $\cal I$ from \fig{FIG:Qtilde_evol} cancel the second term on the right of \eq{ABC} with the same DLA accuracy. 

Therefore, only the diagrams $\cal D$ and $\cal E$ from \fig{FIG:Qtilde_evol}, along with their past-pointing Wilson line staple counterparts, contribute to the evolution of the dipole amplitude ${\widetilde Q}$, corresponding to the diagram B and its complex conjugate in Fig.~2 of \cite{Kovchegov:2018znm}. We have confirmed the analysis carried out in \cite{Kovchegov:2018znm}. Now, we need to calculate diagrams $\cal D$ and $\cal E$. For diagram $\cal E$ we will employ the propagator
\begin{align}\label{q_propagator1}
& \int\limits_{-\infty}^0 dx_{2}^- \, 
\int\limits_0^\infty dx_{2'}^- \, 
\contraction
{}
{\bar\psi^i_\alpha}
{(x_2^- , {\un x}_0) \:}
{psi^j_\beta}
\bar\psi^i_\alpha (x_2^- , {\un x}_0) \: 
\psi^j_\beta (x_{2'}^- , {\un x}_1) 
\\
&= \frac{1}{\pi} \sum_{\sigma}  \int\limits_0^{p_2^-} d k^- \, k^-  \int d^2 x_2 \, d^2 x_{2'} \, \left[ \int
 \frac{d^2 k_{2'}}{(2\pi)^2} \, 
e^{-i \ul{k}_{2'} \cdot \ul{x}_{2'1}} \,  \frac{1}{{\un k}_{2'}^2 }  \, \left( u_{\sigma} (k_{2'}) \right)_\beta \right] 
\notag\\
& \hspace*{1cm}\times
\left( \sigma \, V_{\un 2}^{\textrm{pol} [1]} \, \delta^2 ({\un x}_{22'}) + V_{{\ul 2}', {\un 2}}^{\textrm{pol} [2]} \right)^{ji} \, \left[ \int \frac{d^2 k_2}{(2\pi)^2} \, e^{i \ul{k}_2 \cdot \ul{x}_{20} }  \, \frac{1}{{\un k}_2^2} \, \left( {\bar u}_\sigma (k_2) \right)_\alpha \right] , \notag
\end{align}
which includes a minus sign arising due to Wick contractions. For diagram $\cal D$ we need the following propagator,
\begin{align}\label{q_propagator2}
& \int\limits_{-\infty}^0 dx_{2'}^- \, 
\int\limits_0^\infty dx_2^- \, 
\contraction
{}
{\bar\psi^i_\alpha}
{(x_2^- , {\un x}_0) \:}
{psi^j_\beta}
\bar\psi^i_\alpha (x_2^- , {\un x}_0) \: 
\psi^j_\beta (x_{2'}^- , {\un x}_1) 
\\
&= \frac{1}{\pi} \sum_{\sigma}  \int\limits_0^{p_2^-} d k^- \, k^-  \int d^2 x_2 \, d^2 x_{2'} \, \left[ \int
 \frac{d^2 k_{2'}}{(2\pi)^2} \, 
e^{i \ul{k}_{2'} \cdot \ul{x}_{2'1}} \,  \frac{1}{{\un k}_{2'}^2 }  \, \left( v_{\sigma} (k_{2'}) \right)_\beta \right] 
\notag \\ 
&\hspace*{1cm} \times
\left( - \sigma \, V_{\un 2}^{\textrm{pol} [1] \, \dagger} \, \delta^2 ({\un x}_{22'}) + V_{{\ul 2}, {\un 2}'}^{\textrm{pol} [2] \, \dagger} \right)^{ji} \, \left[ \int \frac{d^2 k_2}{(2\pi)^2} \, e^{- i \ul{k}_2 \cdot \ul{x}_{20} }  \, \frac{1}{{\un k}_2^2} \, \left( {\bar v}_\sigma (k_2) \right)_\alpha \right] . \notag
\end{align}
Note the overall sign difference when comparing our \eq{q_propagator2} to Eq.~(92) in \cite{Cougoulic:2022gbk}: this is due to the Wick contraction included in our \eq{q_propagator2}. There is also a sign difference in front of the $\sigma$ term in \eq{q_propagator2} compared to Eq.~(92) in \cite{Cougoulic:2022gbk}. The typo does not propagate in Sec.~IV of \cite{Cougoulic:2022gbk}.

Using these propagators, we arrive at
\begin{align}\label{DE_app1}
&{\cal D} + {\cal E} = \frac{\as}{4 \, \pi^2} \, \int\limits_\frac{\Lambda^2}{s}^{z} \frac{d z'}{z'} \, \bigg\{ - \int d^2 x_2 \, \frac{{\un x}_{20}}{x_{20}^2}  \cdot \frac{{\un x}_{21}}{x_{21}^2} \llangle \tr \left[ V_{\un 1} \, V_{\un 2}^{\textrm{pol} [1] \, \dagger} \right] \rrangle \\
&\hspace*{4cm} + i \int d^2 x_2 \, d^2 x_{2'} \, \frac{{\un x}_{2'1}}{x_{2'1}^2} \times \frac{{\un x}_{20}}{x_{20}^2}  \llangle \tr \left[ V_{\un 1} \, V_{{\ul 2}, {\un 2}'}^{\textrm{pol} [2] \, \dagger} \right] \rrangle  + \mbox{c.c.} \bigg\} ,\notag
\end{align}
where we have inserted a factor of $2$ to account for the past-pointing Wilson lines. Anticipating the impact-parameter integration, we perform a calculation parallel to that on page 12 of \cite{Cougoulic:2022gbk}, obtaining
\begin{align}
{\cal D} + {\cal E} = & \  - \frac{\as}{4 \, \pi^2} \, \int\limits_\frac{\Lambda^2}{s}^{z} \frac{d z'}{z'} \, \int d^2 x_2 \,  \left\{ \frac{{\un x}_{20}}{x_{20}^2}  \cdot \frac{{\un x}_{21}}{x_{21}^2} \llangle \tr \left[ V_{\un 1} \, V_{\un 2}^{\textrm{pol} [1] \, \dagger} \right] \rrangle \right. \\
& - \left.  \left[ - \epsilon^{ik} \, \frac{x_{20}^k + x_{21}^k}{x_{21}^2 \, x_{20}^2} + 2 \frac{{\un x}_{21} \times {\un x}_{20}}{x_{21}^2 \, x_{20}^2}  \left( \frac{x_{21}^i}{x_{21}^2} -  \frac{x_{20}^i}{x_{20}^2} \right) \right]  \llangle \tr \left[ V_{\un 1} \, V_{{\ul 2}}^{i \, \textrm{G} [2] \, \dagger} \right] \rrangle  + \mbox{c.c.} \right\} , \notag
\end{align}
or, equivalently, 
\begin{align}\label{DE}
{\cal D} + {\cal E} = & \  - \frac{\as \, N_c}{2 \, \pi^2} \, \int\limits_\frac{\Lambda^2}{s}^{z} \frac{d z'}{z'} \, \int d^2 x_2 \,  \left\{ \frac{{\un x}_{20}}{x_{20}^2}  \cdot \frac{{\un x}_{21}}{x_{21}^2} \, Q_{21} (z' s) \right. \\
& \hspace*{2cm}- \left.  \left[ - \epsilon^{ik} \, \frac{x_{20}^k + x_{21}^k}{x_{21}^2 \, x_{20}^2} + 2 \frac{{\un x}_{21} \times {\un x}_{20}}{x_{21}^2 \, x_{20}^2}  \left( \frac{x_{21}^i}{x_{21}^2} -  \frac{x_{20}^i}{x_{20}^2} \right) \right]  \, G_{21}^i (z' s) \right\} . \notag
\end{align}
Integrating \eq{DE} over impact parameter and extracting the DLA contribution we get
\begin{align}\label{DE_app2}
&\int d^2 \left( \frac{x_0 + x_1}{2} \right) \, [{\cal D} + {\cal E} ] \\
&\hspace*{2.6cm} = - \frac{\as \, N_c}{2 \, \pi} \, \int\limits_\frac{\Lambda^2}{s}^{z} \frac{d z'}{z'} \, \int\limits_{\max \left[ x_{10}^2, \frac{1}{z' s} \right]}^{\min \left[ \frac{z}{z'} x_{10}^2, \frac{1}{\Lambda^2} \right]} \frac{d x^2_{21}}{x^2_{21}} \, \left[ Q (x_{21}^2, z' s) + 2 \, G_2 (x_{21}^2, z' s) \right] , \notag
\end{align}
with the $x_{21}^2$-integration limits resulting from the structure of the kernel in \eq{DE}, from the light-cone lifetime limits, and from $1/\Lambda$ being the upper cutoff on the dipole sizes.

The evolution equation for the impact-parameter integrated $\widetilde Q$, 
\begin{align}
{\widetilde Q} (x^2_{10}, z s) \equiv \int d^2 \left( \frac{x_0 + x_1}{2} \right) \, {\widetilde Q}_{10} (z s) ,
\end{align}
is thus
\begin{align}\label{eq:Qevol}
&{\widetilde Q} (x^2_{10}, z s) =  {\widetilde Q}^{(0)} (x^2_{10}, z s) \\
&\hspace*{3cm}- \frac{\as \, N_c}{2 \, \pi} \, \int\limits_\frac{\Lambda^2}{s}^{z} \frac{d z'}{z'} \, \int\limits_{\max \left[ x_{10}^2, \frac{1}{z' s} \right]}^{\min \left[ \frac{z}{z'} x_{10}^2, \frac{1}{\Lambda^2} \right]} \frac{d x^2_{21}}{x^2_{21}} \, \left[ Q (x_{21}^2, z' s) + 2 \, G_2 (x_{21}^2, z' s) \right] . \notag
\end{align}
Here ${\widetilde Q}^{(0)}$ denotes the inhomogeneous term in the integral equation, resulting from the initial condition for the evolution of $\widetilde Q$. In the next Section, we re-derive this result using LCPT.


\subsection{LCPT}\label{sec:ch5_Qtildeevollcpt}

\begin{figure}[h!]
    \centering
    \includegraphics[width=0.9\textwidth]{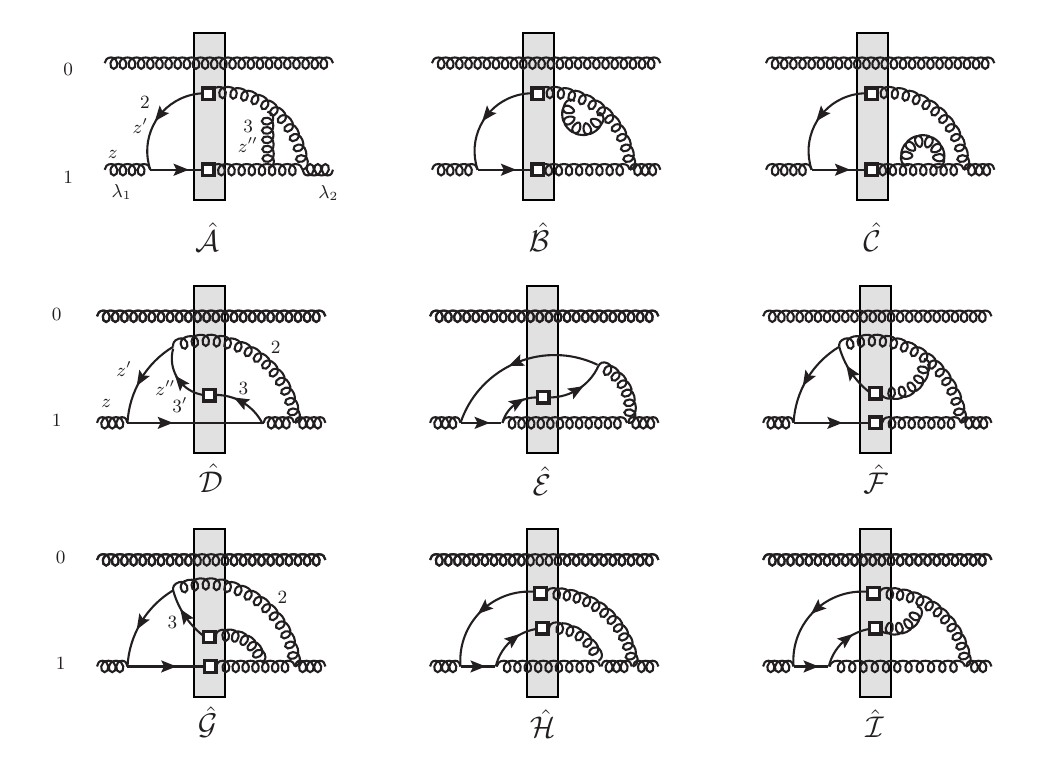}
    \caption{The diagrams containing two steps of evolution: a step of evolution for $G_{10}^\textrm{adj}$ calculated in Ch.~\ref{sec:ch5_adjevol_lcot} or Ch.~\ref{sec:ch5_adjevol_lcpt} followed by a step of evolution for $\widetilde{Q}_{21}$. Upon taking the large-$N_c$ limit and factoring out the evolution kernel for $G_{10}^\textrm{adj}$ found in the previous Sections, the contributions of these diagrams will reduce to those in \fig{FIG:Qtilde_evol}.}
    \label{fig:appA_Qtilde_alldiagrams}
\end{figure}

To reproduce the results of Sec. \ref{sec:ch5_Qtildeevollcot}, we again consider the diagrams in \fig{FIG:Qtilde_evol}. However, in order to use the LCPT light-cone wavefunctions we begin without yet having taken the large-$N_c$ limit. That is, we start from the same diagrams for the evolution of the adjoint dipole we calculated in Ch.~ \ref{sec:ch5_adjevol_lcot} and Ch.~\ref{sec:ch5_adjevol_lcpt} and add an additional step of evolution to those diagrams. In this procedure we will obtain two evolution kernels. We can factor out and remove the first kernel, corresponding to the first step of our evolution calculated above, and what remains will be the second step of the evolution that we wish to calculate here, the evolution of $\widetilde{Q}$. The diagrams we consider are shown in \fig{fig:appA_Qtilde_alldiagrams}. As these diagrams are similar to but distinct from those in \fig{FIG:Qtilde_evol}, we denote them by the same calligraphic letters but with hats. Upon ultimately taking the large-$N_c$ limit we will obtain the contributions of the diagrams in \fig{FIG:Qtilde_evol}. 

We begin with the virtual corrections labeled $\hat{\cal A}$, $\hat{\cal B}$, and $\hat{\cal C}$ shown in the first row of Fig.~\ref{fig:appA_Qtilde_alldiagrams}. To calculate these diagrams we treat the extra emitted gluon (the one at transverse position $\underline{x}_3$) as real (going through the shock wave, but not interacting with it). Subsequently we multiply the result of that calculation by an overall factor of $-\frac{1}{2}$, in accordance with the unitarity arguments relating the sum of real and virtual corrections \cite{Kovchegov:2012mbw} which we use to extract the virtual contribution we are after here.
For diagram $\hat{\cal A}$, which represents the sum of two time-orderings, we write
\begin{align}\label{appA_eikA}
    &\hat{\cal A} = \left(-\frac{1}{2}\right)\frac{1}{2\left(N_c^2 - 1\right)} \left(\frac{1}{2}\sum_{\lambda_1,\lambda_2}\lambda_1\right)\sum_f\sum_{\text{internal}}\int\frac{\mathrm{d}z'}{z'}\int\frac{\mathrm{d}^2\underline{x}_2}{4\pi}\theta\left(x_{10}^2z - x_{21}^2z'\right) \\
    &\times \int\frac{\mathrm{d}z''}{z''}\int\frac{\mathrm{d}^2\underline{x}_3}{4\pi}\theta\left(x_{21}^2z' - \text{max}\{x_{32}^2,x_{31}^2\}z''\right) \bigg\langle \left[\psi^{G\rightarrow GG}_{ebf;\lambda_3,\lambda_2,\lambda_4}\left(\underline{x}_{21}, \frac{z'}{z}\right)\bigg|_{\tfrac{z'}{z}\rightarrow 0}\right]^*  \notag \\
    &\times \left(W^{Gq}_{\underline{1}}\left[\infty,-\infty\right] \right)^{c,\lambda';j,\sigma'}\left[\left(\psi^{G\rightarrow q\bar{q}}_{a\lambda_1;\sigma',\sigma}\right)^{ji}\left(\underline{x}_{21},1-\frac{z'}{z}\right)\bigg|_{1-\tfrac{z'}{z}\rightarrow 1}\right] \notag \\
    &\times\left(W^{G\bar{q}}_{\underline{2}}\left[\infty,-\infty\right]\right)^{d,\lambda; i, \sigma} U_{\underline{0}}^{ba} \notag \\
    &\times \bigg(\left[\psi^{G\rightarrow GG}_{gde; \lambda_5,\lambda,\lambda_3}\left(\underline{x}_{32},\frac{z''}{z'} \right)\bigg|_{\tfrac{z''}{z'}\rightarrow 0}\right] \left[\psi^{G\rightarrow GG}_{gfc; \lambda_5,\lambda_4,\lambda'}\left(\underline{x}_{31},\frac{z''}{z'} \right)\bigg|_{\tfrac{z''}{z'}\rightarrow 0}\right]^* \notag \\
    &\hspace*{2cm} + \left[\psi^{G\rightarrow GG}_{gcf; \lambda_5,\lambda',\lambda_4}\left(\underline{x}_{31},\frac{z''}{z'} \right)\bigg|_{\tfrac{z''}{z'}\rightarrow 0}\right] \left[\psi^{G\rightarrow GG}_{ged; \lambda_5,\lambda_3,\lambda}\left(\underline{x}_{32},\frac{z''}{z'} \right)\bigg|_{\tfrac{z''}{z'}\rightarrow 0}\right]^*    \bigg) \bigg\rangle. \notag
\end{align}

Meanwhile, for diagrams $\hat{\cal B}$ and $\hat{\cal C}$ we write
\begin{align}\label{appA_eikBC}
    &\hat{\cal B} + \hat{\cal C} = \frac{1}{2}\left(-\frac{1}{2}\right)\frac{1}{2\left(N_c^2 - 1\right)} \left(\frac{1}{2}\sum_{\lambda_1,\lambda_2}\lambda_1\right)\sum_f\sum_{\text{internal}}\int\frac{\mathrm{d}z'}{z'}\int\frac{\mathrm{d}^2\underline{x}_2}{4\pi}\theta\left(x_{10}^2z - x_{21}^2z'\right) \\
    &\times \int\frac{\mathrm{d}z''}{z''}\int\frac{\mathrm{d}^2\underline{x}_3}{4\pi}\theta\left(x_{21}^2z' - x_{32}^2 z''\right) \bigg\langle \left[\psi^{G\rightarrow GG}_{ebf;\lambda_3,\lambda_2,\lambda_4}\left(\underline{x}_{21}, \frac{z'}{z}\right)\bigg|_{\tfrac{z'}{z}\rightarrow 0}\right]^*\notag\\
    &\times\left(W^{Gq}_{\underline{1}\left[\infty,-\infty\right]} \right)^{c,\lambda';j,\sigma'} \left[\left(\psi^{G\rightarrow q\bar{q}}_{a\lambda_1;\sigma',\sigma}\right)^{ji}\left(\underline{x}_{21},1-\frac{z'}{z}\right)\bigg|_{1-\tfrac{z'}{z}\rightarrow 1}\right]\left(W^{G\bar{q}}_{\underline{2}}\left[\infty,-\infty\right]\right)^{d,\lambda; i, \sigma} U_{\underline{0}}^{ba} \notag \\
    &\times \left[\psi^{G\rightarrow GG}_{gde; \lambda_5,\lambda,\lambda_4}\left(\underline{x}_{32},\frac{z''}{z'} \right)\bigg|_{\tfrac{z''}{z'}\rightarrow 0}\right] \left[\psi^{G\rightarrow GG}_{gfe; \lambda_5,\lambda_3,\lambda_4}\left(\underline{x}_{32},\frac{z''}{z'} \right)\bigg|_{\tfrac{z''}{z'}\rightarrow 0}\right]^* \bigg \rangle \,+\, \left(\underline{x}_{32}\rightarrow \underline{x}_{31} \right). \notag
\end{align}
We have multiplied diagrams $\hat{\cal B}$ and $\hat{\cal C}$ by an additional factor of $\tfrac{1}{2}$. When we eventually take the large-$N_c$ limit, the gluon that connects transverse positions $\underline{x}_2$ and $\underline{x}_1$ becomes a quark-antiquark pair and the virtual gluon at transverse position $\underline{x}_3$ could couple to either of these lines. But only the quark line (in this diagram) becomes a part of the large-$N_c$ structure $\widetilde{Q}_{21}$, and so we only want the contribution where the virtual gluon couples to this line. The factor of $1/2$ gives us this contribution only.

The diagrams we consider here represent an additional step of evolution done to diagram B in \fig{FIG:Gadj_evol}. In addition we need to include this same second step of evolution done to all the other diagrams in \fig{FIG:Gadj_evol}. This is straightforward as the contribution from diagram A is simply the conjugate of what we have just written, while that from diagrams C and D adds in the contribution to the past-pointing staple and its conjugate (see the second term in parentheses in \eq{Qtilde}). Meanwhile diagrams A', B', C', and D' simply modify the DLA kernel for the first step of the evolution (the same modification occurred in arriving at \eq{appA_alladjdiagrams1}). Adding in all these contributions, carrying out the algebra, taking the large-$N_c$ limit, and multiplying by an additional factor of $\tfrac{1}{4}$ to switch from $G^{\text{adj}}_{21}$ to $\widetilde{G}_{21}$ we obtain
\begin{align}\label{appA_alleik1}
    &\hat{\cal A} + \hat{\cal B} + \hat{\cal C} = \left(-\frac{\alpha_s N_f}{4\pi} \int\limits_{\Lambda^2/s}^{z}\frac{\mathrm{d}z'}{z'}\int\limits_{1/z's}^{x_{10}^2}\frac{\mathrm{d}x_{21}^2}{x_{21}^2} \right) \left(-\frac{\alpha_s N_c}{4\pi^2} \right)\int\frac{\mathrm{d}z''}{z''}\int\mathrm{d}^2\underline{x}_3 \\
    &\hspace{2cm}\times\bigg[\frac{1}{x_{32}^2}\theta\left(x_{21}^2z'-x_{32}^2z''\right) + \frac{1}{x_{31}^2}\theta\left(x_{21}^2z'-x_{31}^2z''\right) \notag \\
    & \hspace*{4cm}- 2 \frac{\underline{x}_{32}\cdot \underline{x}_{31}}{x_{32}^2x_{31}^2}\theta\left(x_{21}^2z' - \text{max}\{x_{32}^2,x_{31}^2\}z'' \right)\bigg] \widetilde{Q}_{21}(z''s).\notag
\end{align}
We have canceled a factor of $\tfrac{1}{zs}$ for the same reason stated following \eq{appA_alladjdiagrams1}. In our usual DLA simplification we have also set unpolarized S-matrices to $1$. Now the first kernel in \eq{appA_alleik1} (the large parentheses in the first line) is simply the kernel for the first step of evolution. So we remove it, leaving just the second step of evolution which, now that we have taken the large-$N_c$ limit, corresponds to diagrams $\cal A$, $\cal B$, and $\cal C$ in \fig{FIG:Qtilde_evol}. Simplifying the kernel at DLA we obtain
\begin{align}\label{appA_alleik2}
    {\cal A}+{\cal B}+{\cal C} = -\frac{\alpha_s N_c}{4 \pi} \int\limits_{\Lambda^2/s}^{z'}\frac{\mathrm{d}z''}{z''} \left[\int\limits_{1/(z''s)}^{x_{21}^2} \frac{\mathrm{d}x_{32}^2}{x_{32}^2} +  \int\limits_{1/(z''s)}^{x_{21}^2} \frac{\mathrm{d}x_{31}^2}{x_{31}^2} \right] \widetilde{Q}_{21}(z''s),
\end{align}
which agrees with \eq{ABC} in the main text upon substituting $\underline{x}_1 \rightarrow \underline{x}_{2}$, $\underline{x}_0 \rightarrow \underline{x}_{1}$, $\underline{x}_2 \rightarrow \underline{x}_{3}$, $z' \rightarrow z''$, and $z\rightarrow z''$ in the latter. 

The remaining diagrams we need are shown in the last two rows of \fig{fig:appA_Qtilde_alldiagrams}. 
We first consider diagrams $\hat{\cal F}$ and $\hat{\cal G}$. For diagram $\hat{\cal F}$ we write 
\begin{align}\label{appA_Qtilde_2steps1}
    &\hat{\cal F} = \frac{1}{2\left(N_c^2 - 1\right)}\left(\frac{1}{2}\sum_{\lambda_1,\lambda_2}\lambda_1 \right)\sum_{f} \sum_{\text{internal}} \int\frac{\mathrm{d}z'}{z'}\int\frac{\mathrm{d}^2\underline{x}_2}{4\pi} \int\frac{\mathrm{d}z''}{z''}\int\frac{\mathrm{d}^2\underline{x}_3}{4\pi} \\
    &\times \bigg\langle \left[\psi^{G\rightarrow GG}_{fbc;\lambda_4,\lambda_2,\lambda'}\left(\underline{x}_{21}, \tfrac{z'}{z} \right)\bigg|_{\tfrac{z'}{z}\rightarrow 0} \right]^* \left[\psi^{G\rightarrow GG}_{gfd;\lambda_3,\lambda_4,\lambda}\left(\underline{x}_{32},\tfrac{z''}{z'} \right)\bigg|_{\tfrac{z''}{z'}\rightarrow 0} \right]^* U_{\underline{2}}^{de}U_{\underline{0}}^{ba} \notag \\
    &\times\left(W_{\underline{1}}^{Gq} \left[\infty,-\infty\right]\right)^{c,\lambda';j,\sigma_0'}  \left[\left(\psi^{G\rightarrow q\bar{q}}\right)^{ji}_{a\lambda_1;\sigma_0',\sigma_0}\left(\underline{x}_{21},1-\tfrac{z'}{z} \right)\bigg|_{1-\tfrac{z'}{z}\rightarrow 1} \right] \notag\\
    &\times\left[-\left(\psi^{q\rightarrow qG}\right)^{ii'}_{e,\lambda;\sigma_0,\sigma}\left(\underline{x}_{32},\tfrac{z''}{z'} \right)\bigg|_{\tfrac{z''}{z}\rightarrow 0}  \right] \left(W_{\underline{3}}^{G\bar{q}}\left[\infty,-\infty \right]\right)^{g,\lambda_3;i'\sigma}\bigg\rangle \notag .
\end{align}
Here we have used used $-\psi^{q\rightarrow qG}$ for the $\bar{q}\rightarrow \bar{q}G$ splitting, as mentioned in the paragraph just after \eq{ch3_qtoqGsoftlimits}. We have also suppressed the light-cone lifetime-ordering theta functions for brevity. Carrying out the algebra we obtain 
\begin{align}\label{appA_Qtilde_2steps3}
    &\hat{\cal F} = \frac{1}{2\left(N_c^2-1\right)}\left(-\frac{\alpha_s N_f}{\pi^2}\right) \int\frac{\mathrm{d}z'}{z'}\int\frac{\mathrm{d}^2\underline{x}_2}{x_{21}^2} \left(\frac{2\alpha_s}{\pi^2}\right)\int\frac{\mathrm{d}z''}{z''}\int\frac{\mathrm{d}^2\underline{x}_3}{x_{32}^2} \\
    &\hspace{0.5cm}\times\llangle \frac{g^2P^+}{8\sqrt{z''}s}\int\limits_{-\infty}^{\infty}\mathrm{d}w_1^-\int\limits_{-\infty}^{\infty}\mathrm{d}w_3^- f^{fbc}f^{gfd}U_{\underline{1}}^{ch}\left[\infty,w_1^-\right]\bar{\psi}_{\alpha}\left(w_1^-,\underline{x}_1\right) t^h \left(\frac{1}{2}\gamma^+\gamma^5\right)_{\alpha\beta} \notag \\
    &\hspace*{1.5cm}\times V_{\underline{1}}\left[w_1^-,-\infty\right] t^a t^e U_{\underline{3}}^{gh'}\left[\infty,w_3^-\right] V_{\underline{3}}^\dagger\left[w_3^-,-\infty\right] t^{h'} \psi_\beta\left(w_3^-,\underline{x}_3\right) U_{\underline{2}}^{de}U_{\underline{0}}^{ba}\rrangle \notag,
\end{align}
where we have again canceled a factor of $\tfrac{1}{zs}$ for the same reason as above. Again, diagram $\hat{\cal F}$ represents a second step of evolution done to diagram B in \fig{FIG:Gadj_evol}, so we need to add in this same step of evolution done to the rest of the diagrams in \fig{FIG:Gadj_evol}, just as discussed before \eq{appA_alleik1}. We again obtain the original kernel from the first step of evolution, which we remove. Doing this, then taking the large-$N_c$ limit, setting the unpolarized S-matrices that result to $1$, and including the factor of $\tfrac{1}{4}$ for the normalization of $\widetilde{G}$, we obtain the contribution of diagram ${\cal F}$. Diagram $\hat{\cal G}$ is calculated similarly. Simplifying the kernel at DLA, we have for the contributions of diagrams ${\cal F}$ and ${\cal G}$
\begin{align}\label{appA_Qtilde_2steps4}
    {\cal F} + {\cal G} = \frac{\alpha_s N_c}{4\pi}\int\limits_{\Lambda^2/s}^{z'}\frac{\mathrm{d}z''}{z''} \int\limits_{1/(z''s)}^{x_{21}^2}\frac{\mathrm{d}x_{32}^2}{x_{32}^2} \widetilde{Q}_{31}(z''s).
\end{align} 
\eq{appA_Qtilde_2steps4} agrees with \eq{FG3} upon substituting $\underline{x}_1 \rightarrow \underline{x}_{2}$, $\underline{x}_0 \rightarrow \underline{x}_{1}$, $\underline{x}_2 \rightarrow \underline{x}_{3}$, $z' \rightarrow z''$, and $z\rightarrow z'$ in the latter. Further, \eq{appA_Qtilde_2steps4} cancels the first term of \eq{appA_alleik2} with DLA accuracy, just as in \eq{Q-Q}. Subsequently diagrams $\hat{\cal H}$ and $\hat{\cal I}$ from Fig.~ \ref{fig:appA_Qtilde_alldiagrams} yield contributions that cancel the second term of \eq{appA_alleik2} with DLA accuracy, just as in Ch.~\ref{sec:ch5_Qtildeevollcot}.

What remains is to calculate the contributions from diagrams $\hat{\cal D}$ and $\hat{\cal E}$. We write
\begin{align}\label{appA_Qtilde_2steps5}
    &\hat{\cal D} + \hat{\cal E} = (-1) \frac{1}{2\left(N_c^2 - 1\right)}\left(\frac{1}{2}\sum_{\lambda_1,\lambda_2}\lambda_1 \right) \sum_{f} \sum_{\text{internal}} \int\frac{\mathrm{d}z'}{z'}\int\frac{\mathrm{d}^2\underline{x}_2}{4\pi} \int\frac{\mathrm{d}z''}{z''}\int\frac{\mathrm{d}^2\underline{x}_3\mathrm{d}^2\underline{x}_3'}{4\pi} \\
    &\hspace{0.5cm} \times \bigg\langle \left[\psi^{G\rightarrow GG}_{dbc;\lambda,\lambda_2,\lambda'}\left(\underline{x}_{21},\frac{z'}{z} \right)\bigg|_{\tfrac{z'}{z}\rightarrow 0} \right]^* \left[\left(\psi^{G\rightarrow q\bar{q}}\right)^{j'j''}_{c\lambda'; \sigma',\sigma_0'}\left(\underline{x}_{31},\frac{z''}{z}  \right)\bigg|_{\tfrac{z''}{z}\rightarrow 0}\right]^* V_{\underline{1}}^{j''j} \notag \\
    &\hspace{0.85cm}\times\left[\left(\psi^{G\rightarrow q\bar{q}}\right)^{ji}_{a\lambda_1;\sigma_0'\sigma_0}\left(\underline{x}_{21}, 1-\frac{z'}{z}\right)\bigg|_{1-\tfrac{z'}{z}\rightarrow 1 }  \right] \left[-\left(\psi^{q\rightarrow qG}\right)^{ii'}_{e\lambda; \sigma_0,\sigma}\left(\underline{x}_{3'2}, \frac{z''}{z'} \right)\bigg|_{\tfrac{z''}{z'}\rightarrow 0} \right] \notag \\
    &\hspace{1cm}\times\left(-\sigma \delta_{\sigma,\sigma'}V_{\underline{3}}^{\text{pol}[1]\,\dagger} \delta^2\left(\underline{x}_3-\underline{x}_{3'} \right) + \delta_{\sigma,\sigma'}V_{\underline{3},\underline{3'}}^{\text{pol}[2]\,\dagger}  \right)^{i'j'} U_{\underline{2}}^{de}U_{\underline{0}}^{ba}  \quad + \quad \text{c.c.}  \bigg\rangle \notag.
\end{align}
As we did for diagrams $\hat{\cal F}$ and $\hat{\cal G}$, we again use $-\psi^{q\rightarrow qG}$ for the $\bar{q}\rightarrow \bar{q}G$ splitting. In addition we have an overall minus sign for the fermion loop. This is still required even with an operator insertion mediating the interaction with the shock-wave. We have also once again omitted the lifetime-ordering theta functions. Following the same steps as in the other diagrams and taking the large-$N_c$ limit, we obtain for the contribution of diagrams ${\cal D} + {\cal E}$:
\begin{align}\label{appA_Qtilde_2steps7}
    &{\cal D}+{\cal E} = \frac{\alpha_s}{4\pi^2}\int\limits_{\Lambda^2/s}^{z'}\frac{\mathrm{d}z''}{z''}\bigg\{-\int \mathrm{d}^2\underline{x}_3 \frac{\underline{x}_{32}\cdot \underline{x}_{31}}{x_{32}^2x_{31}^2} \llangle \text{tr}\left[V_{\underline{2}}V_{\underline{3}}^{\text{pol}[1]\dagger} \right] \rrangle \\
    &\hspace*{4cm} + i\int\mathrm{d}^2\underline{x}_{3}\mathrm{d}^2\underline{x}_{3'} \frac{\underline{x}_{3'2}}{x_{3'2}^2}\times\frac{\underline{x}_{31}}{x_{31}^2} \llangle \text{tr}\left[V_{\underline{2}}V_{\underline{3},\underline{3'}}^{\text{pol}[2]\dagger} \right]\rrangle  + \text{c.c} \bigg\}. \notag
\end{align}
\eq{appA_Qtilde_2steps7} agrees with \eq{DE_app1} upon substituting $\underline{x}_0 \rightarrow \underline{x}_{1}$, $\underline{x}_1 \rightarrow \underline{x}_{2}$, $\underline{x}_{2} \rightarrow \underline{x}_{3}$, $\underline{x}_{2'} \rightarrow \underline{x}_{3'}$, $z\rightarrow z'$, and $z' \rightarrow z''$ in the latter equation. The rest of the calculation then proceeds exactly as that following \eq{DE_app1} in the main text, resulting in 
\begin{align}\label{appA_Qtilde_2steps8}
    &\int \mathrm{d}^2\left(\frac{x_1+x_2}{2}\right)\left[{\cal D} + {\cal E} \right] \\
    &\hspace*{2.6cm}= -\frac{\alpha_s N_c}{2\pi}\int\limits_{\Lambda^2/s}^{z'}\frac{\mathrm{d}z''}{z''}\int\limits_{\text{max}\left[x_{21}^2,\tfrac{1}{z''s}\right]}^{\text{min}\left[\tfrac{z'}{z''}x_{21}^2, \tfrac{1}{\Lambda^2} \right]} \frac{\mathrm{d}x_{32}^2}{x_{32}^2}\left[Q(x_{32}^2,z''s) + 2G_2(x_{32}^2,z''s)  \right] ,\notag
\end{align}
in agreement with \eq{DE_app2} upon making the same substitutions. Thus we have re-derived the evolution of the new structure $\widetilde{Q}$ using LCPT and obtained full agreement with the results of Ch.~\ref{sec:ch5_Qtildeevollcot}.

At this point we have completed the modification of the large-$N_c \& N_f$ evolution equations~\eqref{ch3_LargeNcNfeqns} from \cite{Cougoulic:2022gbk} by calculating the contributions of the transition operators in Eqs.~\eqref{Ws}. In addition to the correction to the evolution of $\widetilde G$ and $\widetilde \Gamma$ resulting from the additional term in \eq{total5}, we also have the evolution equation~\eqref{eq:Qevol} for the new operator $\widetilde Q$. Before we summarize the new corrected large-$N_c \& N_f$ evolution equations, we pause to make a few observations. 


\subsection{\texorpdfstring{A Relation Between $\widetilde{Q}$ and the Quark Helicity TMD and PDF}{A Relation Between Qtilde and the Quark Helicity TMD and PDF}}\label{sec:Qtilde_relations}

As we have mentioned above, the operator definition of $\widetilde Q$ in \eq{Qtilde} resembles that for the quark helicity TMD (SIDIS plus DY). Let us try to better quantify this similarity. We start with Eq.~(8) of \cite{Kovchegov:2018znm} for the SIDIS quark helicity TMD, in which we un-insert the complete set of states $X$:
\begin{align}\label{TMD11}
&g_{1L}^{S, \, \textrm{SIDIS}} (x, k_T^2) = 2 \, \frac{2 p^+}{(2\pi)^3} \:  \int d^{2} \zeta \, d \zeta^- \, d^{2} \xi \, d \xi^-
\, e^{- i {\un k} \cdot ({\un \zeta} - {\un \xi})} \\
&\hspace*{5cm}\times
\left\langle \bar\psi (\xi) \, \left( \thalf \gamma^+ \gamma^5 \right) \, V_{\ul \xi} [\xi^-, \infty] \,
 V_{\ul \zeta} [\infty , \zeta^-] \, \psi (\zeta) \right\rangle .\notag
\end{align}
We have also inserted an overall factor of 2 to account for the anti-quark contribution and approximated $k \cdot (\zeta - \xi) \approx - {\un k} \cdot ({\un \zeta} - {\un \xi})$ in the exponent, which is valid at small $x$. Fourier-transforming \eq{TMD11} we arrive at
\begin{align}\label{TMD12}
&\int d^2 k_\perp \, e^{i {\un k} \cdot {\un x}_{10}} \, g_{1L}^{S, \, \textrm{SIDIS}} (x, k_T^2) \\
&\hspace*{1cm}= \frac{2 p^+}{\pi} \,  \int d \zeta^- \, d \xi^- \, d^{2} \xi \,  \left\langle \bar\psi (\xi) \, \left( \thalf \gamma^+ \gamma^5 \right) \, V_{\ul \xi} [\xi^-, \infty] \, V_{{\ul \xi} + {\un x}_{10}} [\infty , \zeta^-] \, \psi (\xi + x_{10}) \right\rangle . \notag
\end{align}
Similarly, for the DY quark helicity TMD, we write
\begin{align}\label{TMD_DY}
&\int d^2 k_\perp \, e^{i {\un k} \cdot {\un x}_{10}} \, g_{1L}^{S, \, \textrm{DY}} (x, k_T^2) \\
&\hspace*{.8cm}= \frac{2 p^+}{\pi} \,  \int d \zeta^- \, d \xi^- \, d^{2} \xi \,  \left\langle \bar\psi (\xi) \, \left( \thalf \gamma^+ \gamma^5 \right) \, V_{\ul \xi} [\xi^-, -\infty] \, V_{{\ul \xi} + {\un x}_{10}} [-\infty , \zeta^-] \, \psi (\xi + x_{10}) \right\rangle .\notag
\end{align}
Comparing Eqs.~\eqref{TMD12} and \eqref{TMD_DY} to the definition of ${\widetilde Q}$ in \eq{Qtilde}, we readily see that 
\begin{align}\label{gQ_relation}
&\int d^2 k_\perp \, e^{i {\un k} \cdot {\un x}_{10}} \, \left[ g_{1L}^{S, \, \textrm{SIDIS}} (x, k_T^2) + g_{1L}^{S, \, \textrm{DY}} (x, k_T^2) \right] \\
&\hspace{5cm}= \frac{1}{\pi} \, \frac{16}{g^2} \, \frac{1}{2} \, \int d^2 \left( \frac{x_0 + x_1}{2} \right) \, {\widetilde Q}_{10} (s = Q^2/x) .\notag
\end{align}
The factor of $1/2$ on the right of \eq{gQ_relation} removes the double-counting due to the complex conjugate terms: indeed, since TMDs are real, one can readily show that the expression in \eq{gQ_relation} is also real. The factor of $2 p^+$ from Eqs.~\eqref{TMD12} and \eqref{TMD_DY} is present in \eq{Qtilde} owing to the definition of the double angle brackets in \eq{double_def}. We thus obtain
\begin{align}\label{Qg_relation}
  {\widetilde Q} (x^2_{10}, Q^2/x) = \frac{\as \, \pi^2}{2} \, \int d^2 k_\perp \, e^{i {\un k} \cdot {\un x}_{10}} \, \left[ g_{1L}^{S, \, \textrm{SIDIS}} (x, k_T^2) + g_{1L}^{S, \, \textrm{DY}} (x, k_T^2) \right]. 
\end{align}
Since helicity TMDs are PT-even, the DY and SIDIS helicity TMDs are equal, $g_{1L}^{S, \, \textrm{SIDIS}} (x, k_T^2) = g_{1L}^{S, \, \textrm{DY}} (x, k_T^2) \equiv g_{1L}^S (x, k_T^2)$, such that we can rewrite \eq{Qg_relation} as
\begin{align}\label{Qg_relation2}
 {\widetilde Q} (x^2_{10}, Q^2/x) = \as \, \pi^2 \, \int d^2 k_\perp \, e^{i {\un k} \cdot {\un x}_{10}} \, g_{1L}^S (x, k_T^2), 
\end{align}
or equivalently, 
\begin{align}\label{Qg_relation3}
g_{1L}^S (x, k_T^2) = \frac{1}{4 \pi^4 \as} \, \int d^2 x_{10} \, e^{- i {\un k} \cdot {\un x}_{10}} \ {\widetilde Q} (x^2_{10}, Q^2/x). 
\end{align}
Equations~\eqref{Qg_relation2} and \eqref{Qg_relation3} provide a relation between the quark helicity TMDs and the new object $\widetilde Q$ we have introduced in this Chapter. 

We can also connect $\widetilde Q$ to the quark helicity PDFs by writing the flavor-singlet quark helicity PDF as an integral of the corresponding TMD over $k_T$,
\begin{align}\label{DSigma_Qtilde}
\Delta \Sigma (x, Q^2) = \sum_f \int\limits^{Q^2} d^2 k_\perp \, g_{1L}^S (x, k_T^2) = \frac{N_f}{\as \, \pi^2} \ {\widetilde Q} \left( x^2_{10} = \frac{1}{Q^2} , s = \frac{Q^2}{x} \right).
\end{align}
Here, again, for simplicity we assume that all quark flavors contribute equally. Using \eq{eq:Qevol} we write
\begin{align}\label{DSigma}
&\Delta \Sigma (x, Q^2)  = \frac{N_f}{\as \, \pi^2} \, {\widetilde Q}^{(0)} \left( x^2_{10} = \frac{1}{Q^2} , s = \frac{Q^2}{x} \right) \\
&\hspace*{2.5cm} - \frac{N_c \, N_f}{2 \, \pi^3} \, \int\limits_\frac{\Lambda^2}{s}^{1} \frac{d z}{z} \, \int\limits_{\max \left[ \frac{1}{Q^2}, \frac{x}{z \, Q^2} \right]}^{\min \left[ \frac{1}{z \, Q^2}, \frac{1}{\Lambda^2} \right]} \frac{d x^2_{10}}{x^2_{10}} \, \left[ Q (x_{10}^2, z s) + 2 \, G_2 (x_{10}^2, z s) \right] \notag\\
&\hspace*{1.8cm} =  \frac{N_f}{\as \, \pi^2} \, {\widetilde Q}^{(0)} \left( x^2_{10} = \frac{1}{Q^2} , s = \frac{Q^2}{x} \right) \notag\\
&\hspace*{2.5cm}- \frac{N_c \, N_f}{2 \, \pi^3} \, \int\limits_{\frac{1}{Q^2}}^{\frac{1}{\Lambda^2}} \frac{d x^2_{10}}{x^2_{10}} \, \int\limits_\frac{1}{s \, x_{10}^2}^{\frac{1}{Q^2 \, x_{10}^2}} \frac{d z}{z} \,  \left[ Q (x_{10}^2, z s) + 2 \, G_2 (x_{10}^2, z s) \right] . \notag
\end{align}
Comparing \eq{ch3_DeltaSigma} to \eq{DSigma}, we see that the latter has the extra inhomogeneous term ${\widetilde Q}^{(0)}$ which was neglected in \eq{ch3_DeltaSigma} (and in \cite{Kovchegov:2015pbl, Kovchegov:2016zex, Kovchegov:2016weo, Kovchegov:2018znm, Cougoulic:2022gbk, Adamiak:2023yhz}) as not evolving at small $x$. In addition, the lower limit of the $x_{10}$ integral is now different in the first line of \eq{DSigma} as compared to \eq{ch3_DeltaSigma}.

We conclude that equations~\eqref{DSigma_Qtilde} and \eqref{DSigma} result in an expression for $\Delta \Sigma$ at small $x$ with a slight modification as compared to \eq{ch3_DeltaSigma} used earlier in the literature. As we will see below, the present expression for $\Delta\Sigma$ appears to be close to the $\overline{\text{MS}}$ scheme for calculating the polarized DGLAP splitting functions. (The previous definition \eqref{ch3_DeltaSigma} can be thought of corresponding to the ``polarized DIS scheme" \cite{Adamiak:2023okq}.)


\section{\texorpdfstring{Contribution to the Evolution of the Dipole Amplitude $Q$}{Contribution to the Evolution of the Dipole Amplitude Q}}\label{sesc:ch5_fundevol}

In this Section we now determine the contribution of the transition operators in Eqs.~\eqref{Ws2} to the DLA evolution of the dipole amplitude $Q$ defined in \eq{ch3_Q}. Note that the other fundamental dipole $G_2$ is defined by a purely gluonic operator, such that its evolution does not contain quarks and receives no contribution from the transition operators. As usual, we will derive the contributions to the evolution of $Q$ using both LCOT and LCPT. Ultimately we will see that the contribution is $N_c$-suppressed and so can be neglected in the large-$N_c\&N_f$ limit we consider in this Chapter.


\subsection{LCOT}\label{sec:Qevol_lcot}

\begin{figure}[ht]
\centering
\includegraphics[width= 0.95 \textwidth]{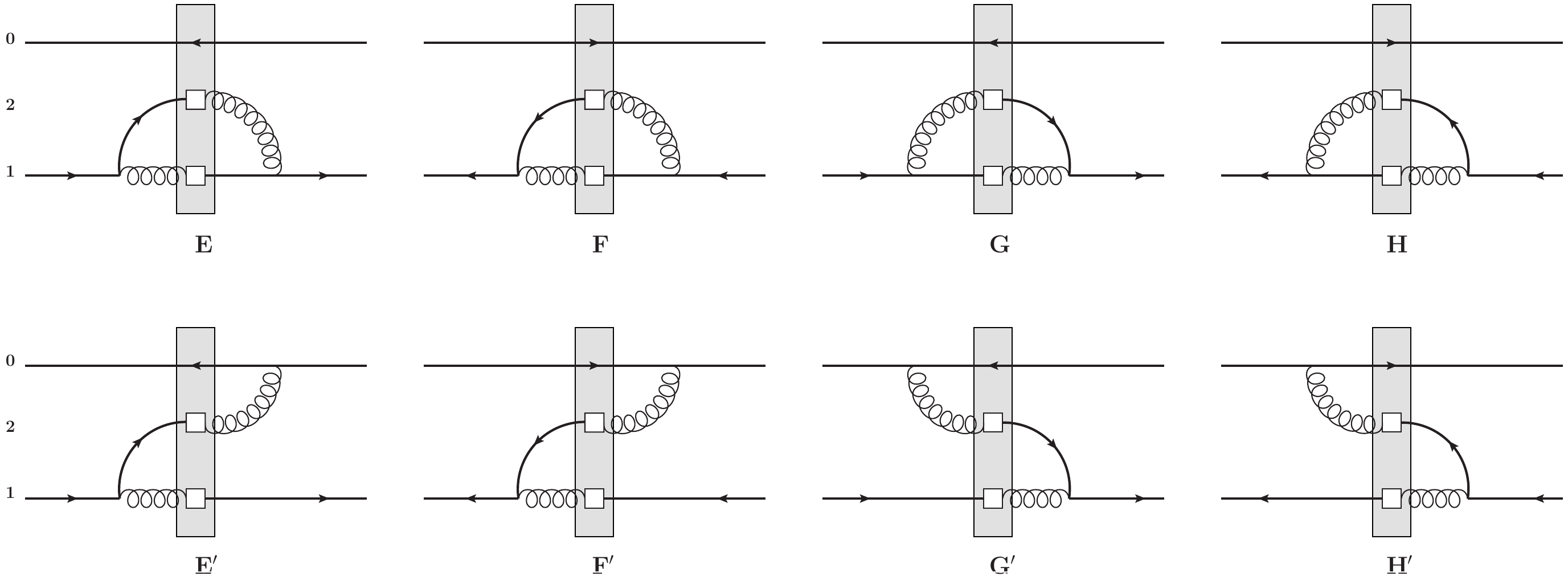}
\caption{The diagrams for the evolution of $Q_{10}$ due to the $q/{\bar q} \to G$ and $G \to q/{\bar q}$ shock-wave transition operators. The shaded rectangle denotes the shock wave, while the white square denotes the non-eikonal interaction with the shock wave mediated by the transition operators from \fig{fig:newoperators}.}
\label{FIG:Q_evol}
\end{figure}

The relevant diagrams containing contributions of the transition operators to the evolution of the dipole amplitude $Q$ are shown in \fig{FIG:Q_evol}. 
Their calculation in the LCOT framework parallels that in Ch.~\ref{sec:ch5_adjevol_lcot} and employs the background field propagators \eqref{psibar_a+}, \eqref{psi_a+}, \eqref{psi_a+_2}, and \eqref{psibar_a+_2}. One obtains
\begin{subequations}
\begin{align}
    & E + E' = - \frac{\as}{4 \pi^2 N_c} \, \int\limits_0^{p_2^-} \frac{d k^-}{k^-} \, \int d^2 x_2 \\
    &\llangle \tr \Bigg[ t^a \, V_{\un 0}^\dagger \, t^b \, \bigg\{ - \frac{{\un x}_{20}}{x_{20}^2}  \cdot \frac{{\un x}_{21}}{x_{21}^2} \, \bigg[  \left( W_{\un 1}^{a \, [2]} [- \infty, \infty] \right)^\dagger \, W_{\un 2}^{b \, [1]} [\infty, -\infty] \notag\\ 
    &\hspace*{4.5cm} + \left( W_{\un 1}^{a \, [1]} [- \infty, \infty] \right)^\dagger \, W_{\un 2}^{b \, [2]} [\infty, -\infty] \bigg] \notag\\
    &\hspace*{2.2cm}+ i \, \frac{{\un x}_{20}}{x_{20}^2}  \times \frac{{\un x}_{21}}{x_{21}^2} \, \, \bigg[  \left( W_{\un 1}^{a \, [1]} [- \infty, \infty] \right)^\dagger \, W_{\un 2}^{b \, [1]} [\infty, -\infty] \notag \\ 
    &\hspace*{4.5cm} + \left( W_{\un 1}^{a \, [2]} [- \infty, \infty] \right)^\dagger \, W_{\un 2}^{b \, [2]} [\infty, -\infty] \bigg] \bigg\} \Bigg] \rrangle - ({\un x}_{20} \to {\un x}_{21}) , \notag \\
    & F + F' = (E + E')^* , \\
    & G + G' = - \frac{\as}{4 \pi^2 N_c} \, \int\limits_0^{p_2^-} \frac{d k^-}{k^-} \, \int d^2 x_2 \\
    &\llangle \tr \Bigg[ t^a \, V_{\un 0}^\dagger \, t^b \, \bigg\{ - \frac{{\un x}_{20}}{x_{20}^2}  \cdot \frac{{\un x}_{21}}{x_{21}^2} \, \bigg[  \left( W_{\un 2}^{a \, [1]} [- \infty, \infty] \right)^\dagger \, W_{\un 1}^{b \, [2]} [\infty, -\infty] \notag\\ 
    &\hspace*{4.5cm} + \left( W_{\un 2}^{a \, [2]} [- \infty, \infty] \right)^\dagger \, W_{\un 1}^{b \, [1]} [\infty, -\infty] \bigg] \notag\\
    &\hspace*{2.2cm}- i \, \frac{{\un x}_{20}}{x_{20}^2}  \times \frac{{\un x}_{21}}{x_{21}^2} \, \, \bigg[  \left( W_{\un 2}^{a \, [1]} [- \infty, \infty] \right)^\dagger \, W_{\un 1}^{b \, [1]} [\infty, -\infty] \notag \\ 
    &\hspace*{4.5cm} + \left( W_{\un 2}^{d \, [2]} [- \infty, \infty] \right)^\dagger \, W_{\un 1}^{b \, [2]} [\infty, -\infty] \bigg] \bigg\} \Bigg] \rrangle - ({\un x}_{20} \to {\un x}_{21}) , \notag \\
    & H + H' = (G + G')^* .
\end{align}    
\end{subequations}
It is worth noting that the diagrams in Fig.~\ref{FIG:Q_evol} were calculated in \cite{Chirilli:2021lif}. (See Eq.(4.28), Eq.(6.81) and Eq.(6.98) in \cite{Chirilli:2021lif}.) 

We are particularly interested in the DLA limit of these diagrams. Adding up the contributions and neglecting the cross-product terms in the DLA yields
\begin{align}\label{EFGH1}
    & E + E' + F + F' + G + G' + H + H' = - \frac{\as}{4 \pi N_c} \int\limits_0^{p_2^-} \frac{d k^-}{k^-} \int\limits^{x_{10}^2} \frac{d x^2_{21}}{x^2_{21}} \, \llangle \tr \Bigg[ t^a \, V_{\un 0}^\dagger \, t^b \\ 
    & \times \, \bigg[  \left( W_{\un 1}^{a \, [2]} [- \infty, \infty] \right)^\dagger \, W_{\un 2}^{b \, [1]} [\infty, -\infty] + \left( W_{\un 1}^{a \, [1]} [- \infty, \infty] \right)^\dagger \, W_{\un 2}^{b \, [2]} [\infty, -\infty] \notag \\ 
    & + \left( W_{\un 2}^{a \, [1]} [- \infty, \infty] \right)^\dagger \, W_{\un 1}^{b \, [2]} [\infty, -\infty] + \left( W_{\un 2}^{a \, [2]} [- \infty, \infty] \right)^\dagger \, W_{\un 1}^{b \, [1]} [\infty, -\infty]\bigg] \Bigg] \rrangle + \mbox{c.c.} . \notag 
\end{align}
To simplify \eq{EFGH1} we first note that
\begin{align}\label{W4}
    & \tr \Bigg[ t^a \, V_{\un 0}^\dagger \, t^b \, \, \bigg[  \left( W_{\un 1}^{a \, [2]} [- \infty, \infty] \right)^\dagger \, W_{\un 2}^{b \, [1]} [\infty, -\infty] 
    + \left( W_{\un 1}^{a \, [1]} [- \infty, \infty] \right)^\dagger \, W_{\un 2}^{b \, [2]} [\infty, -\infty] \bigg] \Bigg] \\
    & = \frac{g^2}{4 \sqrt{p_2^- \, k^-}} \, \int\limits_{-\infty}^\infty d y^- \, \int\limits_{-\infty}^\infty d z^- \, U_{\un 1}^{ca} [z^-, - \infty] \, U_{\un 2}^{bd} [\infty, y^-] \, {\bar \psi} (y^-, {\un x}_2) \, \gamma^+ \, \gamma^5 \, t^d \, V_{\un 2} [y^-, - \infty] \notag\\ 
    &\hspace*{5cm}\times t^a \, V_{\un 0}^\dagger \, t^b \, V_{\un 1} [\infty, z^-] \, t^c \, \psi (z^-, {\un x}_1) , \notag
\end{align}
which allows us to rewrite \eq{EFGH1} as
\begin{align}\label{EFGH2}
    & E + E' + F + F' + G + G' + H + H' \\
    &= - \frac{\as}{4 \pi N_c} \int\limits_0^{p_2^-} \frac{d k^-}{k^-} \int\limits^{x_{10}^2} \frac{d x^2_{21}}{x^2_{21}} \, \int\limits_{-\infty}^\infty d y^- \, \int\limits_{-\infty}^\infty d z^- \, \llangle  \frac{g^2}{4 \sqrt{p_2^- \, k^-}} \, U_{\un 1}^{ca} [z^-, - \infty] \, U_{\un 2}^{bd} [\infty, y^-] \notag \\ 
    & \times \, {\bar \psi} (y^-, {\un x}_2) \, \gamma^+ \, \gamma^5 \, t^d \, V_{\un 2} [y^-, - \infty] \, t^a \, V_{\un 0}^\dagger \, t^b \, V_{\un 1} [\infty, z^-] \, t^c \, \psi (z^-, {\un x}_1) + ({\un 1} \leftrightarrow {\un 2})  \rrangle + \mbox{c.c.} .  \notag
\end{align}
Employing \eq{Uab} and the Fierz identity we can recast \eq{W4} as
\begin{align}\label{W42}
    & \llangle \tr \Bigg[ t^a \, V_{\un 0}^\dagger \, t^b \, \, \bigg[  \left( W_{\un 1}^{a \, [2]} [- \infty, \infty] \right)^\dagger \, W_{\un 2}^{b \, [1]} [\infty, -\infty] 
    \\
    &\hspace*{3cm}+ \left( W_{\un 1}^{a \, [1]} [- \infty, \infty] \right)^\dagger \, W_{\un 2}^{b \, [2]} [\infty, -\infty] \bigg] \Bigg] \rrangle \notag \\
    & = \frac{g^2}{16 \sqrt{p_2^- \, k^-}} \, \int\limits_{-\infty}^\infty d y^- \, \int\limits_{-\infty}^\infty d z^- \notag \\
    &\hspace*{1cm}\times\Bigg\{ \llangle {\bar \psi} (y^-, {\un x}_2) \, \gamma^+ \, \gamma^5 \, V_{\un 2} [y^-,\infty] \, V_{\un 1} \, V_{\un 0}^\dagger \, V_{\un 2} \, V_{\un 1} [-\infty, z^-] \, \psi (z^-, {\un x}_1) \rrangle \notag \\
    &\hspace*{1.4cm} - S_{10} \, \llangle {\bar \psi} (y^-, {\un x}_2) \, \gamma^+ \, \gamma^5 \, V_{\un 2} [y^-,-\infty] \, V_{\un 1} [-\infty, z^-] \, \psi (z^-, {\un x}_1) \rrangle \notag \\
    &\hspace*{1.4cm}- S_{20} \, \llangle {\bar \psi} (y^-, {\un x}_2) \, \gamma^+ \, \gamma^5 \, V_{\un 2} [y^-,\infty] \, V_{\un 1} [\infty, z^-] \, \psi (z^-, {\un x}_1) \rrangle \notag \\
    &\hspace*{1.4cm} + \frac{1}{N_c^2} \, \llangle {\bar \psi} (y^-, {\un x}_2) \, \gamma^+ \, \gamma^5 \, V_{\un 2} [y^-,-\infty] \, V_{\un 0}^\dagger \, V_{\un 1} [\infty, z^-] \, \psi (z^-, {\un x}_1) \rrangle  \Bigg\} , \notag 
\end{align}
where we have employed the large-$N_c$ limit to factor out the unpolarized $S$-matrices $S_{10}$ and $S_{20}$. 

Using \eq{W42} in \eq{EFGH2} we conclude that the contribution of the diagrams in \fig{FIG:Q_evol} is sub-leading at large $N_c$ (and at large-$N_c \& N_f$): see the explicit overall factor of $1/N_c$ in \eq{EFGH2}. We conclude that these diagrams do not contribute to the large-$N_c \& N_f$ (or large-$N_c$) evolution of the dipole amplitude $Q$ in the DLA. Therefore, we can safely neglect the contribution of the diagrams in \fig{FIG:Q_evol}.


\subsection{LCPT}

To re-derive the contribution of the transition operators to the evolution of the fundamental dipole $Q$, we begin with diagrams $E$ from Fig.~\ref{FIG:Q_evol}. Using the light-cone wavefunctions from Ch.~\ref{sec:ch3_lcpt} we write
\begin{align}\label{ch5_Elcpt}
    &E = \frac{1}{2N_c}\frac{1}{2}\sum_{\sigma,\sigma_2}\sigma \sum_{\text{internal}} \int\frac{\mathrm{d}z'}{z'}\frac{\mathrm{d}^2\underline{x}_2}{4\pi}\bigg\langle \text{tr}\,\bigg\{\left[\psi^{q\rightarrow qG}_{b\lambda';\sigma_1\sigma_2}\left(\underline{x}_{21},1-\frac{z'}{z}\right)\bigg|_{1-\tfrac{z'}{z}\rightarrow 1} \right]^* \\
    &\hspace{1cm}\times\left(W^{qG}\left[\infty,-\infty\right]\right)^{a\lambda;\sigma_1} \left(W^{Gq}\left[\infty,-\infty\right]\right)^{b\lambda';\sigma'} \left[\psi^{q\rightarrow qG}_{a\lambda;\sigma\sigma'}\left(\underline{x}_{21},\frac{z'}{z}\right)\bigg|_{\tfrac{z'}{z}\rightarrow0}  \right] V_{\underline{0}}^\dagger \bigg\}\bigg\rangle \notag \\
    & = \frac{1}{2N_c} \frac{1}{2}\sum_{\sigma,\sigma_2}\sigma \sum_{\text{internal}} \int\frac{\mathrm{d}z'}{z'}\frac{\mathrm{d}^2\underline{x}_2}{4\pi} \notag \\
    &\bigg\langle \text{tr}\,\bigg\{\left[\frac{ig}{2\pi}t^b \delta_{\sigma_1\sigma_2}\frac{\underline{\epsilon}_{\lambda'}^*\cdot \underline{x}_{21}}{x_{21}^2}\left(2+\sigma_1\lambda'\frac{z'}{z'}\right)  \right]^* \notag\\
    &\times\bigg(\int\limits_{-\infty}^{\infty}\mathrm{d}y^- V_{\underline{1}}\left[\infty,y^-\right]\bigg(-\frac{igt^c}{2\sqrt{\sqrt{2}p_2^-}}\bigg)\delta_{\sigma_1\lambda}\left[\sigma_1\left(\bar{\rho}_\beta(-)-\bar{\rho}_\beta(+)\right) + \left(\bar{\rho}_\beta(-)+\bar{\rho}_\beta(+)\right)\right] \notag\\
    &\hspace{4.4cm}\times \psi_\beta\left(y^-,\underline{x}_1\right)   U_{\underline{1}}^{ca}\left[y^-,-\infty\right] \bigg) \notag\\
    &\times\bigg(\int\limits_{-\infty}^{\infty}\mathrm{d}z^- U_{\underline{2}}^{bd}\left[\infty,z^-\right]\overline{\psi}_\alpha\left(z^-,\underline{x}_2\right) \left(-\frac{igt^d}{2\sqrt{\sqrt{2}k^-}}\right)\delta_{\sigma'\lambda'}\notag \\
    &\hspace*{4.4cm}\times\left[\sigma'\left(\rho_\alpha(-)-\rho_\alpha(+) \right) + \left(\rho_\alpha(-)+\rho_\alpha(+) \right)\right]   V_{\underline{2}}\left[z^-,-\infty\right] \bigg)  \notag\\
    &\times \left[-\frac{ig}{2\pi}t^a \delta_{\sigma\sigma'}\sqrt{\frac{z'}{z}}\left(1+\sigma'\lambda)\right)\frac{\underline{\epsilon}_{\lambda}^*\cdot\underline{x}_{21}}{x_{21}^2} \right] V_{\underline{0}}^\dagger
    \bigg\}\bigg\rangle \notag
\end{align}
where we have left implicit the lifetime-ordering $\theta$-function and suppressed the fundamental indices on the transition operators in order to write them in the trace. The rest of the color and helicity labels can be inferred from Fig.~\ref{FIG:Q_evol}. In the second equality we have explicitly substituted the expressions for the light-cone wavefunctions and the transition operators. We have also supposed, just as in Ch.~\ref{sec:ch5_adjevol_lcpt}, that the soft quark at transverse position $2$ has momentum $k$ and the parent dipole has momentum $p_2$ such that \eq{appA_diagramA3} still holds. Following similar steps to those employed in Ch.~\ref{sec:ch5_adjevol_lcpt} gives
\begin{align}\label{ch5_Elcpt1}
    &E = -\frac{\as}{4\pi N_c}\int\limits_0^{p_2^-} \frac{\mathrm{d}k^-}{k^-}\int\frac{\mathrm{d}\xto^2}{\xto^2}\int\limits_{-\infty}^{\infty}\mathrm{d}y^-\int\limits_{-\infty}^{\infty}\mathrm{d}z^- \llangle \frac{g^2}{4\sqrt{p_2^-k^-}}U_{\un 1}^{ca} [z^-, - \infty] \, U_{\un 2}^{bd} [\infty, y^-]  \\ 
    &\hspace*{2cm} \times \, {\bar \psi} (y^-, {\un x}_2) \, \gamma^+ \, \gamma^5 \, t^d \, V_{\un 2} [y^-, - \infty] \, t^a \, V_{\un 0}^\dagger \, t^b \, V_{\un 1} [\infty, z^-] \, t^c \, \psi (z^-, {\un x}_1) \rrangle  ,\notag
\end{align}
where we also switched the longitudinal integration $z'\rightarrow k^-$ with $z' = k^-/p_2^-$. 

Diagram $E'$ is obtained analogously and in the DLA just modifies the upper limit of the transverse integral in the contribution of diagram $E$ (just as we saw in Ch.~\ref{sec:ch5_evoladj}). For the other diagrams from Fig.~\ref{FIG:Q_evol}, it is of course still true that $F+F' = (E+E')^*$. The same relationship we have between diagrams $E$ and $E'$ also holds between $G$ and $G'$, and we also have $H+H' = (G+G')^*$. So all that remains is to calculate diagram $G$. We write
\begin{align}\label{ch5_Glcpt}
    &G = \frac{1}{2N_c}\frac{1}{2}\sum_{\sigma,\sigma_2}\sigma \sum_{\text{internal}} \int\frac{\mathrm{d}z'}{z'}\frac{\mathrm{d}^2\underline{x}_2}{4\pi}\bigg\langle \text{tr}\,\bigg\{\left[\psi^{q\rightarrow qG}_{a\lambda;\sigma'\sigma_2}\left(\underline{x}_{21},\frac{z'}{z}\right)\bigg|_{\tfrac{z'}{z}\rightarrow 0} \right]^* \\
    &\hspace{.2cm}\times\left(W^{qG}\left[\infty,-\infty\right]\right)^{b\lambda';\sigma'} \left(W^{Gq}\left[\infty,-\infty\right]\right)^{a\lambda;\sigma_1} \left[\psi^{q\rightarrow qG}_{b\lambda';\sigma\sigma_1}\left(\underline{x}_{21},1-\frac{z'}{z}\right)\bigg|_{1-\tfrac{z'}{z}\rightarrow 1}  \right] V_{\underline{0}}^\dagger \bigg\}\bigg\rangle .\notag 
\end{align}
Following the usual steps we obtain
\begin{align}\label{ch5_Glcpt1}
    &G = -\frac{\as}{4\pi N_c}\int\limits_0^{p_2^-} \frac{\mathrm{d}k^-}{k^-}\int\limits^{\xoz^2}\frac{\mathrm{d}\xto^2}{\xto^2}\int\limits_{-\infty}^{\infty}\mathrm{d}y^-\int\limits_{-\infty}^{\infty}\mathrm{d}z^- \llangle \frac{g^2}{4\sqrt{p_2^-k^-}}U_{\un 2}^{ca} [z^-, - \infty] \, U_{\un 1}^{bd} [\infty, y^-] \\ 
    &\hspace*{2cm} \times \, {\bar \psi} (y^-, {\un x}_1) \, \gamma^+ \, \gamma^5 \, t^d \, V_{\un 1} [y^-, - \infty] \, t^a \, V_{\un 0}^\dagger \, t^b \, V_{\un 2} [\infty, z^-] \, t^c \, \psi (z^-, {\un x}_2) \rrangle  .\notag
\end{align}
Then adding together all the diagrams from Fig.~\ref{FIG:Q_evol} we have, in the DLA,
\begin{align}\label{ch5_fundevolinlcpt}
    & E + E' + F + F' + G + G' + H + H' \\
    &= - \frac{\as}{4 \pi N_c} \int\limits_0^{p_2^-} \frac{d k^-}{k^-} \int\limits^{x_{10}^2} \frac{d x^2_{21}}{x^2_{21}} \, \int\limits_{-\infty}^\infty d y^- \, \int\limits_{-\infty}^\infty d z^- \, \llangle  \frac{g^2}{4 \sqrt{p_2^- \, k^-}} \, U_{\un 1}^{ca} [z^-, - \infty] \, U_{\un 2}^{bd} [\infty, y^-] \notag \\ 
    & \times \, {\bar \psi} (y^-, {\un x}_2) \, \gamma^+ \, \gamma^5 \, t^d \, V_{\un 2} [y^-, - \infty] \, t^a \, V_{\un 0}^\dagger \, t^b \, V_{\un 1} [\infty, z^-] \, t^c \, \psi (z^-, {\un x}_1) + ({\un 1} \leftrightarrow {\un 2})  \rrangle + \mbox{c.c.} , \notag
\end{align}
in full agreement with \eq{EFGH2}. The $N_c$ suppression follows exactly as it did in the end of Ch.~\ref{sec:Qevol_lcot} and thus we can indeed neglect the contribution of the transition operators to the evolution of $Q$ in the large-$N_c\&N_f$ limit.


\section{\texorpdfstring{A New Version of the Large-$N_c\&N_f$ Evolution Equations}{A New Version of the Large Nc and Nf Evolution Equations}}\label{sec:ch5_updatedeqs}

Let us now summarize the large-$N_c \& N_f$ evolution, which has in this Chapter been modified from its previous form in Eqs.~\eqref{ch3_LargeNcNfeqns}. Combining our results in \eq{total5} and \eq{eq:Qevol} and also incorporating \eq{total5} into the evolution for the neighbor dipole amplitude $\widetilde \Gamma$, we arrive at a new set of large-$N_c\&N_f$ evolution equations:

\begin{subequations}\label{eq_LargeNcNf}\allowdisplaybreaks
\begin{align}
& Q(x^2_{10},zs) = Q^{(0)}(x^2_{10},zs) + \frac{\alpha_sN_c}{2\pi} \int_{1/s x^2_{10}}^{z} \frac{dz'}{z'}   \int_{1/z's}^{x^2_{10}}  \frac{dx^2_{21}}{x_{21}^2}    \bigg[ 2 \, {\widetilde G}(x^2_{21},z's) \\
&\hspace*{5cm} + 2 \, {\widetilde \Gamma}(x^2_{10},x^2_{21},z's) + \; Q(x^2_{21},z's) -  \overline{\Gamma}(x^2_{10},x^2_{21},z's) \notag\\
&\hspace*{5cm}+ 2 \, \Gamma_2(x^2_{10},x^2_{21},z's) + 2 \, G_2(x^2_{21},z's)   \bigg] \notag \\
&\hspace*{2cm}+ \frac{\alpha_sN_c}{4\pi} \int_{\Lambda^2/s}^{z} \frac{dz'}{z'}   \int_{1/z's}^{\min \{ x^2_{10}z/z', 1/\Lambda^2 \}}  \frac{dx^2_{21}}{x_{21}^2} \left[Q(x^2_{21},z's) + 2 \, G_2(x^2_{21},z's) \right] ,  \notag  \\
&\overline{\Gamma}(x^2_{10},x^2_{21},z's) = Q^{(0)}(x^2_{10},z's) + \frac{\alpha_sN_c}{2\pi} \int_{1/s x^2_{10}}^{z'} \frac{dz''}{z''}   \int_{1/z''s}^{\min\{x^2_{10}, x^2_{21}z'/z''\}}  \frac{dx^2_{32}}{x_{32}^2}\\
&\hspace*{3cm}\times\bigg[ 2\, {\widetilde G} (x^2_{32},z''s) + \; 2\, {\widetilde \Gamma} (x^2_{10},x^2_{32},z''s) +  Q(x^2_{32},z''s)\notag\\
&\hspace*{3cm}-  \overline{\Gamma}(x^2_{10},x^2_{32},z''s) + 2 \, \Gamma_2(x^2_{10},x^2_{32},z''s) + 2 \, G_2(x^2_{32},z''s) \bigg] \notag \\
&\hspace*{2cm}+ \frac{\alpha_sN_c}{4\pi} \int_{\Lambda^2/s}^{z'} \frac{dz''}{z''}   \int_{1/z''s}^{\min \{ x^2_{21}z'/z'', 1/\Lambda^2 \}}  \frac{dx^2_{32}}{x_{32}^2} \left[Q(x^2_{32},z''s) + 2 \, G_2(x^2_{32},z''s) \right] , \notag \\
& {\widetilde G}(x^2_{10},zs) = {\widetilde G}^{(0)}(x^2_{10},zs) + \frac{\alpha_s N_c}{2\pi}\int_{1/s x^2_{10}}^z\frac{dz'}{z'}\int_{1/z's}^{x^2_{10}} \frac{dx^2_{21}}{x^2_{21}} \\
&\hspace*{2.4cm}\times\bigg[3 \, {\widetilde G}(x^2_{21},z's) + {\widetilde \Gamma}(x^2_{10},x^2_{21},z's)  + \; 2\,G_2(x^2_{21},z's)  \notag \\
&\hspace*{2.4cm}+  \left(2 - \frac{N_f}{2N_c}\right) \Gamma_2(x^2_{10},x^2_{21},z's) - \frac{N_f}{4N_c}\,\overline{\Gamma}(x^2_{10},x^2_{21},z's) \color{blue}{ - \frac{N_f}{2 N_c} \, {\widetilde Q}(x^2_{21},z's) }  \bigg] \notag \\
&\hspace*{2cm}- \frac{\alpha_sN_f}{8\pi}  \int_{\Lambda^2/s}^z \frac{dz'}{z'}\int_{\max\{x^2_{10},\,1/z's\}}^{\min \{ x^2_{10}z/z', 1/\Lambda^2 \}} \frac{dx^2_{21}}{x^2_{21}}  \left[   Q(x^2_{21},z's) +     2 \, G_2(x^2_{21},z's)  \right] , \notag \\
& {\widetilde \Gamma} (x^2_{10},x^2_{21},z's) = {\widetilde G}^{(0)}(x^2_{10},z's) + \frac{\alpha_s N_c}{2\pi}\int_{1/s x^2_{10}}^{z'}\frac{dz''}{z''}\int_{1/z''s}^{\min\{x^2_{10},x^2_{21}z'/z''\}} \frac{dx^2_{32}}{x^2_{32}} \\
&\hspace*{2.2cm}\times\bigg[3 \, {\widetilde G} (x^2_{32},z''s)  + \; {\widetilde \Gamma}(x^2_{10},x^2_{32},z''s) + 2 \, G_2(x^2_{32},z''s)  \notag \\
&\hspace*{2.2cm}+  \left(2 - \frac{N_f}{2N_c}\right) \Gamma_2(x^2_{10},x^2_{32},z''s) - \frac{N_f}{4N_c} \,\overline{\Gamma}(x^2_{10},x^2_{32},z''s)  \color{blue}{ - \frac{N_f}{2 N_c} \, {\widetilde Q}(x^2_{32},z''s) } \bigg] \notag \\
&\hspace*{1.5cm}- \frac{\alpha_sN_f}{8\pi}  \int_{\Lambda^2/s}^{z'x^2_{21}/x^2_{10}} \frac{dz''}{z''}\int_{\max\{x^2_{10},\,1/z''s\}}^{\min \{ x^2_{21}z'/z'', 1/\Lambda^2 \} } \frac{dx^2_{32}}{x^2_{32}}  \left[   Q(x^2_{32},z''s) +  2  \,  G_2(x^2_{32},z''s)  \right] , \notag \\
& G_2(x_{10}^2, z s)  =  G_2^{(0)} (x_{10}^2, z s) + \frac{\as N_c}{\pi} \, \int\limits_{\frac{\Lambda^2}{s}}^z \frac{d z'}{z'} \, \int\limits_{\max \left[ x_{10}^2 , \frac{1}{z' s} \right]}^{\min \{\frac{z}{z'} x_{10}^2, 1/\Lambda^2 \}} \frac{d x^2_{21}}{x_{21}^2} \\
&\hspace*{9cm}\times\left[ {\widetilde G} (x^2_{21} , z' s) + 2 \, G_2 (x_{21}^2, z' s)  \right] ,\notag \\
& \Gamma_2 (x_{10}^2, x_{21}^2, z' s)  =  G_2^{(0)} (x_{10}^2, z' s) + \frac{\as N_c}{\pi}  \int\limits_{\frac{\Lambda^2}{s}}^{z' \frac{x_{21}^2}{x_{10}^2}} \frac{d z''}{z''}  \int\limits_{\max \left[ x_{10}^2 , \frac{1}{z'' s} \right]}^{\min \{ \frac{z'}{z''} x_{21}^2, 1/\Lambda^2 \}} \frac{d x^2_{32}}{x_{32}^2} \\
&\hspace*{9cm}\times\left[ {\widetilde G} (x^2_{32} , z'' s) + 2 \, G_2(x_{32}^2, z'' s)  \right] , \notag \\
& \color{blue}{ {\widetilde Q} (x^2_{10}, z s) =  {\widetilde Q}^{(0)} (x^2_{10}, z s) - \frac{\as \, N_c}{2 \, \pi} \, \int\limits_\frac{\Lambda^2}{s}^{z} \frac{d z'}{z'} \, \int\limits_{\max \left[ x_{10}^2, \frac{1}{z' s} \right]}^{\min \{\frac{z}{z'} x_{10}^2, 1/\Lambda^2 \}} \frac{d x^2_{21}}{x^2_{21}}} \\
&\hspace*{9.3cm}\textcolor{blue}{\times\left[ Q (x_{21}^2, z' s) + 2 \, G_2 (x_{21}^2, z' s) \right] }.  \notag
\end{align}
\end{subequations}

The new terms --- which are the contributions of the transition operators to the evolution of the adjoint dipole along with the new equation for the evolution of the transition operators themselves (in the object $\widetilde{Q}$) --- are shown in blue. Just as Eqs.~\eqref{ch3_LargeNcNfeqns}, Eqs.~\eqref{eq_LargeNcNf} are written treating $\Lambda$ as an infrared cutoff and we impose the usual ordering $\xto \leq \xoz$ for the neighbors.

For completeness, let we also restate the expressions for helicity PDFs at small $x$: 
\begin{subequations}\label{eq:DeltaGSigma_G2_Qtilde}
\begin{align}
& \Delta G (x, Q^2) = \frac{2 N_c}{\as \, \pi^2} \, G_2 \left(  x_{10}^2 = \frac{1}{Q^2} ,  s = \frac{Q^2}{x} \right) , \\
& \Delta \Sigma (x, Q^2) = \frac{N_f}{\as \, \pi^2} \, {\widetilde Q} \left( x^2_{10} = \frac{1}{Q^2} , s = \frac{Q^2}{x} \right) .
\end{align}
\end{subequations}
While the expression for $\Delta G$ was obtained earlier \cite{Kovchegov:2017lsr}, the result for $\Delta \Sigma$ is new, slightly correcting the older result \cite{Kovchegov:2015pbl, Kovchegov:2016zex, Kovchegov:2018znm, Cougoulic:2022gbk}.


\section{Iterative Solution of the New Evolution}\label{sec:ch5_iterativesoln}

In this Section, we will solve Eqs.~\eqref{eq_LargeNcNf} iteratively --- that is, one step of evolution at a time --- similar to what was done in \cite{Cougoulic:2022gbk, Adamiak:2023okq}. We assume non-vanishing constant (independent of $x$ and $Q^2$) initial conditions $\Delta \Sigma^{(0)}$ and $\Delta G^{(0)}$ for the flavor-singlet quark and gluon helicity PDFs. The (also constant) initial conditions for the polarized dipole amplitudes are 
\begin{equation}\label{init_cond}
G_2^{(0)} = \frac{\alpha_s \pi^2}{2N_c} \Delta G^{(0)},\quad
\widetilde{Q}^{(0)} = \frac{\alpha_s \pi^2}{N_f} \Delta \Sigma^{(0)}, \quad Q^{(0)} = -\frac{1}{2} \widetilde{Q}^{(0)}, \quad \widetilde{G}^{(0)} = \frac{N_f}{4N_c} \widetilde{Q}^{(0)}. 
\end{equation}
The first two conditions follow from Eqs.~\eqref{eq:DeltaGSigma_G2_Qtilde}. The last two conditions can be derived from the original definitions of $Q(x_{10}^2, s)$ in Eq.~\eqref{ch3_Q10}, $\widetilde{G}(x_{10}^2, s)$ in Eq.~\eqref{ch3_Gtilde}, and $\widetilde{Q}(x_{10}^2, s)$ in Eq.~\eqref{Qtilde} by taking the zero dipole size limit, $x_{10}^2 \rightarrow 0$. Indeed, we first notice that only $V_{\un x}^{\textrm{q} [1]}$ in $V_{\un x}^{\textrm{pol} [1]}$ contributes to $Q(x_{10}^2, s)$ in this limit, and only the quark part of $W_{{\un x}}^{\textrm{pol} [1] }$ in \eq{ch3_Wpol1} contributes to $\widetilde{G}(x_{10}^2, s)$. Performing a little algebra and clarifying that the zero dipole size limit corresponds to $x_{10}^2 \to 1/s$, with the center-of-mass energy squared $s$ being the highest momentum scale in the problem, we arrive at the following relations between the (impact-parameter integrated) polarized dipole amplitudes:
\begin{subequations}\label{relationsGQQ}
    \begin{align}
    & \widetilde{G} \left(x_{10}^2 = \frac{1}{s}, s \right) = - \frac{N_f}{4 C_F} \, Q \left(x_{10}^2 = \frac{1}{s}, s \right) \approx - \frac{N_f}{2 N_c} \, Q \left(x_{10}^2 = \frac{1}{s}, s \right), \label{GtildeQ} \\
    & \widetilde{G} \left(x_{10}^2 = \frac{1}{s}, s \right) = \frac{N_f}{4 N_c} \, {\widetilde Q} \left(x_{10}^2 = \frac{1}{s}, s \right). \label{GtildeQtilde}
\end{align}
\end{subequations}
Here we have summed over flavors in $\widetilde G$, assuming that all flavors contribute equally (otherwise, one would have to replace $N_f \, Q \to \sum_f Q_f$ in \eq{GtildeQ} and $N_f \, {\widetilde Q} \to \sum_f {\widetilde Q}_f$ in \eq{GtildeQtilde}, with the flavor-dependent dipole amplitudes $Q_f$ and ${\widetilde Q}_f$). The last transition in \eq{GtildeQ} is valid in the large-$N_c$ and large-$N_c \& N_f$ limits. Equations~\eqref{relationsGQQ} are otherwise exact. Since all the integral kernels in Eqs.~\eqref{eq_LargeNcNf} vanish in the $x_{10}^2 \to 1/s$ limit (which in those equations, corresponds to the $x_{10}^2 \to 1/(z s)$ limit) we see that the conditions \eqref{relationsGQQ} need to be satisfied by the inhomogeneous terms (the initial conditions) in Eqs.~\eqref{eq_LargeNcNf}, after which they will be always satisfied by the dipole amplitudes solving those evolution equations. Equations~\eqref{relationsGQQ} allowed us to fix the initial amplitudes $Q^{(0)}$ and $\widetilde{G}^{(0)}$ in \eq{init_cond}.

In what follows, we will solve Eqs.~\eqref{eq_LargeNcNf} iteratively, using Eqs.~\eqref{eq:DeltaGSigma_G2_Qtilde} to obtain order-by-order in $\as$ expansions for the hPDFs,
\begin{subequations}\label{pert_exp}
    \begin{align}
        & \Delta \Sigma (x, Q^2) = \Delta\Sigma^{(0)} + \Delta\Sigma^{(1)}(x,Q^2) + \Delta\Sigma^{(2)}(x,Q^2) + \ldots , \\
        & \Delta G (x, Q^2) = \Delta G^{(0)} + \Delta G^{(1)}(x,Q^2) + \Delta G^{(2)}(x,Q^2) + \ldots , 
    \end{align}
\end{subequations}
where the index in the superscript parentheses matches the power of $\as$ correction to $\Delta\Sigma^{(0)}$ and $\Delta G^{(0)}$. The iterative expansion of the dipole amplitudes and the DGLAP splitting functions will be labeled in the same way.


\subsection{Step 1}

After one step of evolution --- that is, acting the kernels of the evolution in Eqs.~\eqref{eq_LargeNcNf} on the inhomogeneous terms from \eq{init_cond} --- one obtains
\begin{subequations}\label{eq:G_2_(1)}
    \begin{align}
        G_2^{(1)}(x_{10}^2, zs) =&\frac{\alpha_s N_c}{\pi}\ln (zs x_{10}^2)\ln \left(\frac{1}{\Lambda^2 x_{10}^2}\right) \left[\frac{N_f}{4N_c} \widetilde{Q}^{(0)} + 2G_2^{(0)}\right], \\
\widetilde{Q}^{(1)}(x_{10}^2, zs) =&- \frac{\alpha_s N_c}{2\pi}\ln (zs x_{10}^2)\ln \left(\frac{1}{\Lambda^2 x_{10}^2}\right) \left[-\frac{1}{2}\widetilde{Q}^{(0)} + 2G_2^{(0)}\right]. 
    \end{align}
\end{subequations}

The hPDFs at this step follow from Eq.~\eqref{eq:DeltaGSigma_G2_Qtilde}:
\begin{subequations}
    \begin{align}
        &\Delta G^{(1)}(x, Q^2) =   \frac{\alpha_s N_c}{4\pi}\ln\frac{1}{x}\ln \frac{Q^2}{\Lambda^2}  \left[2 \Delta \Sigma^{(0)} + 8 \Delta G^{(0)}\right],\\
&\Delta \Sigma^{(1)}(x, Q^2) =  \frac{\alpha_s N_c}{4\pi }\ln\frac{1}{x}\ln \frac{Q^2}{\Lambda^2}  \left[ \Delta \Sigma^{(0)} - 2\frac{N_f}{N_c} \Delta G^{(0)}\right]. 
    \end{align}
\end{subequations}
Having these two expressions for the hPDFs, we can employ the spin-dependent DGLAP evolution from Eqs.~\eqref{ch2_poldglap}, restated below,
\begin{align}\label{DGLAP_diff}
&\frac{\pd}{\pd \ln Q^2}
\begin{pmatrix}
 \Delta {\Sigma} (x, Q^2) \\
 \Delta {G} (x, Q^2)
\end{pmatrix}
= 
\int\limits_x^1 \frac{dz}{z} \, 
\begin{pmatrix}
\Delta P_{qq} (z) & \Delta P_{qG} (z) \\
\Delta P_{Gq} (z)  & \Delta P_{GG} (z) 
\end{pmatrix}
\,
\begin{pmatrix}
 \Delta {\Sigma} \left( \frac{x}{z} , Q^2 \right) \\
 \Delta {G} \left( \frac{x}{z} , Q^2 \right)
\end{pmatrix} \\
&\hspace*{3.9cm}
\equiv 
\left[\Delta\mathbf{P} \otimes
\begin{pmatrix} 
\Delta {\Sigma}   \\
\Delta {G} 
\end{pmatrix} \right] (x, Q^2 )
\end{align}
to readily read out the polarized small-$x$ large-$N_c \& N_f$ splitting functions at the one-loop order 
\begin{equation}\label{eq:DeltaP_(0)}
\Delta \mathbf{P}^{(0)} (x) = \frac{\alpha_s N_c}{4\pi} \begin{pmatrix}
    1 & -2\frac{N_f}{N_c}\\[10pt]
    2 & 8 \\
\end{pmatrix}. \notag
\end{equation}

The other polarized dipole amplitudes and the neighbor dipole amplitudes after one step of the evolution are given by
\begin{subequations}\label{eq:Q(1)}
\begin{align} 
& Q^{(1)}(x_{10}^2, zs) 
= \frac{\alpha_s N_c}{4\pi}\ln^2(zsx_{10}^2) \left[\left(\frac{N_f}{N_c} - \frac{1}{4}\right)\widetilde{Q}^{(0)} + 5 G_2^{(0)}\right] \\
&\hspace*{2.3cm}+  \frac{\alpha_s N_c}{4\pi} \ln(zsx_{10}^2) \ln\frac{1}{\Lambda^2x_{10}^2}\left[-\frac{1}{2} \widetilde{Q}^{(0)} + 2G_2^{(0)}\right], \notag\\   & \widetilde{G}^{(1)}(x_{10}^2, zs) = \frac{\alpha_s N_c}{4\pi} \ln^2(zsx_{10}^2) \left[\frac{5N_f}{8N_c} \widetilde{Q}^{(0)} + 4 \left( 1-\frac{N_f}{8N_c} \right) G_2^{(0)}\right] \label{eq:Gtilde(1)}  \\ 
& \hspace*{2.3cm} - \frac{\alpha_s N_c}{4\pi} \ln(zsx_{10}^2)\ln\left(\frac{1}{\Lambda^2x_{10}^2}\right) \frac{N_f}{2N_c}\left[-\frac{1}{2} \widetilde{Q}^{(0)} + 2G_2^{(0)}\right] , \notag \\
& \bar{\Gamma}^{(1)}(x_{10}^2, x_{21}^2, z's) 
=\frac{\alpha_s N_c}{4\pi}  \ln^2(z'sx_{21}^2) \left[\left(\frac{N_f}{N_c} - \frac{1}{4}\right) \widetilde{Q}^{(0)}  + 5 G_2^{(0)}\right] \\ 
& \hspace*{3.1cm} + \frac{\alpha_s N_c}{4\pi}  \ln(z'sx_{21}^2)\ln\frac{1}{\Lambda^2 x_{21}^2} \left[\left( \frac{2N_f}{N_c}-\frac{1}{2}\right) \widetilde{Q}^{(0)} + 10 G_2^{(0)}\right] \notag\\
&\hspace*{3.1cm}-\frac{\alpha_s N_c}{4\pi}  \ln(z'sx_{21}^2)\ln\frac{1}{\Lambda^2 x_{10}^2} \left[\frac{2N_f}{N_c}\widetilde{Q}^{(0)} + 8 G_2^{(0)}\right] , \notag \\
& \widetilde{\Gamma}^{(1)}(x_{10}^2, x_{21}^2, z's) 
= \frac{\alpha_s N_c}{4\pi}  \ln^2(z'sx_{21}^2)  \left[\frac{5N_f}{8N_c} \widetilde{Q}^{(0)} + \left( 4-\frac{N_f}{2N_c} \right) G_2^{(0)}\right] \\
& \hspace*{3.1cm} + \frac{\alpha_s N_c}{4\pi}  \ln(z'sx_{21}^2) \ln\frac{1}{\Lambda^2 x_{21}^2}   \left[\frac{5N_f}{4N_c} \widetilde{Q}^{(0)} + \left( 8-\frac{N_f}{N_c} \right) G_2^{(0)}\right]\notag\\
&\hspace*{3.1cm}-\frac{\alpha_s N_c}{4\pi} \ln(z's x_{21}^2) \ln\frac{1}{\Lambda^2x_{10}^2}\left[\frac{N_f}{N_c} \, \widetilde{Q}^{(0)} + 8 G_2^{(0)}\right] , \notag \\ 
& \Gamma_2^{(1)}(x_{10}^2, x_{21}^2, z's) =4\frac{\alpha_s N_c}{4\pi}  \ln(z's x_{21}^2) \ln \left(\frac{1}{\Lambda^2x_{10}^2}\right)\left[\frac{N_f}{4N_c} \, \widetilde{Q}^{(0)} + 2G_2^{(0)}\right].
    \end{align}
\end{subequations}


\subsection{Step 2}

After two steps of the evolution (that is, after acting the evolution equations on the dipole amplitudes from our first iteration), we obtain
\begin{subequations}
    \begin{align}
        G_2^{(2)}(x_{10}^2, zs) 
=&\left(\frac{\alpha_s N_c}{4\pi} \right)^2 \frac{1}{3}\ln^3(zsx_{10}^2)\ln\left(\frac{1}{\Lambda^2x_{10}^2}\right)  \left[\frac{5N_f}{2N_c}  \widetilde{Q}^{(0)} + 2 \left( 8-\frac{N_f}{N_c} \right) G_2^{(0)}\right]\\
&+\left(\frac{\alpha_s N_c}{4\pi} \right)^2 \ln^2(zsx_{10}^2) \ln^2\left(\frac{1}{\Lambda^2x_{10}^2}\right)\left[\frac{9N_f}{4N_c} \widetilde{Q}^{(0)} + \left( 16- \frac{N_f}{N_c} \right) G_2^{(0)})\right], \notag \\
\widetilde{Q}^{(2)}(x_{10}^2, zs)
=&-\left(\frac{\alpha_s N_c}{4\pi} \right)^2 \frac{2}{3} \ln^3(zsx_{10}^2) \ln \frac{1}{\Lambda^2x_{10}^2}\left[ \left( \frac{N_f}{N_c}-\frac{1}{4} \right) \widetilde{Q}^{(0)} + 5 G_2^{(0)}\right]\\
&-\left(\frac{\alpha_s N_c}{4\pi} \right)^2  \ln^2(zsx_{10}^2)\ln^2\left(\frac{1}{\Lambda^2x_{10}^2}\right)\left[\left(-\frac{1}{4} + \frac{N_f}{N_c}\right)\widetilde{Q}^{(0)} + 9 G_2^{(0)}\right]. \notag
    \end{align}
\end{subequations}
The hPDFs from the second iteration are
\begin{subequations}
    \begin{align}
        \Delta G^{(2)} (x,Q^2) 
=&\left(\frac{\alpha_s N_c}{4\pi} \right)^2 \frac{1}{3}\ln^3\frac{1}{x}\ln\left(\frac{Q^2}{\Lambda^2}\right)  \left[5 \Delta \Sigma^{(0)} + 2 \left( 8-\frac{N_f}{N_c} \right) \Delta G^{(0)}\right]\\
&+\left(\frac{\alpha_s N_c}{4\pi} \right)^2 \ln^2\frac{1}{x} \ln^2\left(\frac{Q^2}{\Lambda^2}\right)\left[\frac{9}{2} \Delta\Sigma^{(0)} + \left(16- \frac{N_f}{N_c} \right)\Delta G^{(0)}\right], \notag \\
\Delta \Sigma^{(2)}(x, Q^2) 
=&\left(\frac{\alpha_s N_c}{4\pi} \right)^2 \frac{1}{3}\ln^3\frac{1}{x} \ln\frac{Q^2}{\Lambda^2}\left[ \left( \frac{1}{2} - \frac{2N_f}{N_c} \right) \Delta \Sigma^{(0)} - \frac{5N_f}{N_c} \Delta G^{(0)}\right]\\
&+\left(\frac{\alpha_s N_c}{4\pi} \right)^2\ln^2\frac{1}{x} \ln^2\frac{Q^2}{\Lambda^2}  \left[ \left( \frac{1}{4}-\frac{N_f}{N_c} \right) \Delta \Sigma^{(0)} -\frac{9N_f}{2N_c} \Delta G^{(0)}\right]. \notag
    \end{align}
\end{subequations}
From the terms containing $\ln^3\frac{1}{x} \ln\frac{Q^2}{\Lambda^2}$, we can read out the polarized small-$x$ large-$N_c \& N_f$ splitting functions at the two-loop order:
\begin{equation}\label{eq:DeltaP_(1)}
    \Delta \mathbf{P}^{(1)} (x) = \left(\frac{\alpha_s N_c}{4\pi}\right)^2\ln^2\frac{1}{x}\begin{pmatrix}
        \frac{1}{2} - \frac{2N_f}{N_c} & -\frac{5N_f}{N_c}\\[10pt]
        5 & 16-\frac{2N_f}{N_c} \\
    \end{pmatrix}.
\end{equation}
The other polarized dipole amplitudes after two steps of the evolution that we will need for the next iteration of the solution are
\begin{subequations}
\begin{align}
&Q^{(2)}(x_{10}^2, zs) \\
&\hspace*{0.4cm}= \ \frac{1}{6}\left(\frac{\alpha_s N_c}{4\pi} \right)^2\ln^4(zsx_{10}^2)\left[\left(\frac{5N_f}{N_c} -\frac{1}{4}\right)\widetilde{Q}^{(0)} + \left( 35-\frac{4N_f}{N_c} \right) G_2^{(0)}\right] \notag \\
&\hspace*{0.6cm}+\frac{1}{3} \left(\frac{\alpha_s N_c}{4\pi} \right)^2\ln^3(zsx_{10}^2) \ln \frac{1}{\Lambda^2x_{10}^2} \left[\left(\frac{7N_f}{N_c}-\frac{1}{2}\right)\widetilde{Q}^{(0)}+ \left(46-\frac{4N_f}{N_c}\right) G_2^{(0)}\right] \notag \\
&\hspace*{0.6cm}+\frac{1}{4}\left(\frac{\alpha_s N_c}{4\pi} \right)^2\ln^2(zsx_{10}^2)\ln^2\left(\frac{1}{\Lambda^2x_{10}^2}\right)\left[\left(\frac{2N_f}{N_c}-\frac{1}{2}\right)\widetilde{Q}^{(0)} + 18G_2^{(0)} \right], \notag \\
\label{eq:Gtilde_(2)_final}
&\widetilde{G}^{(2)}(x_{10}^2, zs) \\
&\hspace*{0.4cm}= \ \frac{1}{6}\left(\frac{\alpha_s N_c}{4\pi} \right)^2 \ln^4(zsx_{10}^2) \left[\frac{N_f}{N_c} \left( -\frac{N_f}{2N_c} + \frac{35}{8} \right) \, \widetilde{Q}^{(0)} + \left( 28 - \frac{11N_f}{2N_c} \right) \, G_2^{(0)}\right] \notag \\
&\hspace*{0.6cm}+\frac{1}{6}\left(\frac{\alpha_s N_c}{4\pi} \right)^2 \ln^3(zsx_{10}^2)\ln\frac{1}{\Lambda^2x_{10}^2} \left[\frac{N_f}{N_c} \left( \frac{19}{2} -\frac{2N_f}{N_c} \right) \widetilde{Q}^{(0)} + \left( 64-\frac{18N_f}{N_c} \right) \, G_2^{(0)}\right] \notag \\
&\hspace*{0.6cm}-\frac{1}{8}\left(\frac{\alpha_s N_c}{4\pi} \right)^2 \ln^2(zsx_{10}^2)\ln^2\frac{1}{\Lambda^2x_{10}^2}\left[\frac{N_f}{N_c} \left( \frac{2N_f}{N_c} - \frac{1}{2} \right) \, \widetilde{Q}^{(0)} + \frac{18N_f}{N_c}\,  G_2^{(0)}\right]. \notag
    \end{align}
\end{subequations}


\subsection{Step 3}

After three steps of the evolution, the polarized dipole amplitudes that we need are
\begin{subequations}
\begin{align}
&G^{(3)}_2(x_{10}^2, zs)
\\
&=\frac{1}{30} \left(\frac{\alpha_s N_c}{4\pi} \right)^3 \ln^5(zsx_{10}^2)\ln\frac{1}{\Lambda^2x_{10}^2} \left[\frac{N_f}{N_c} \left(-\frac{2N_f}{N_c} + \frac{35}{2} \right) \widetilde{Q}^{(0)} + \left( 112 - \frac{22N_f}{N_c} \right) G_2^{(0)}\right] \notag\\
&+\frac{1}{24}\left(\frac{\alpha_s N_c}{4\pi} \right)^3 \ln^4(zsx_{10}^2) \ln^2\frac{1}{\Lambda^2 x_{10}^2} \left[\frac{N_f}{N_c} \left( 39 -\frac{4N_f}{N_c} \right) \widetilde{Q}^{(0)} + 2 \left( 128-\frac{26N_f}{N_c} \right) G_2^{(0)}\right] \notag \\
&+\frac{1}{36} \left(\frac{\alpha_s N_c}{4\pi} \right)^3\ln^3(zsx_{10}^2) \ln^3\frac{1}{\Lambda^2 x_{10}^2}\left[\frac{N_f}{N_c} 
\left(73-\frac{4N_f}{N_c} \right) \widetilde{Q}^{(0)} + 2 \left( 256-\frac{34N_f}{N_c} \right) G_2^{(0)}\right] \notag \\
&\widetilde{Q}^{(3)}(x_{10}^2, zs) 
\\
&= -\frac{1}{30}\left(\frac{\alpha_s N_c}{4\pi} \right)^3\ln^5(zsx_{10}^2)\ln\frac{1}{\Lambda^2x_{10}^2}  \left[ \left( \frac{10N_f}{N_c}-\frac{1}{2} \right)\widetilde{Q}^{(0)} + 2 \left( 35-\frac{4N_f}{N_c} \right) G_2^{(0)}\right] \notag \\
&\hspace*{.4cm}-\frac{1}{24}\left(\frac{\alpha_s N_c}{4\pi} \right)^3\ln^4(zsx_{10}^2)\ln^2\frac{1}{\Lambda^2x_{10}^2} \left[ \left( \frac{24N_f}{N_c} -1 \right ) \widetilde{Q}^{(0)} + 2 \left( 78 - \frac{8N_f}{N_c} \right) G_2^{(0)}\right] \notag \\
&\hspace*{.4cm}-\frac{1}{36}\left(\frac{\alpha_s N_c}{4\pi} \right)^3 \ln^3(zsx_{10}^2)\ln^3\frac{1}{\Lambda^2x_{10}^2} \left[ \left( \frac{40N_f}{N_c}-1 \right) \widetilde{Q}^{(0)} + 2 \left( 146-\frac{8N_f}{N_c} \right) G_2^{(0)}\right]. \notag 
\end{align}
\end{subequations}
The corresponding hPDFs are 
\begin{subequations}\allowdisplaybreaks
\begin{align}
&\Delta G^{(3)}(x, Q^2) 
\\
&=\frac{1}{30}\left(\frac{\alpha_s N_c}{4\pi} \right)^3 \ln^5\frac{1}{x}\ln\frac{Q^2}{\Lambda^2}  \left[ \left( 35-\frac{4N_f}{N_c} \right) \Delta \Sigma^{(0)} + 2 \left( 56 - \frac{11N_f}{N_c} \right ) \Delta G^{(0)}\right] \notag \\
&\hspace*{.1cm}+\frac{1}{24}\left(\frac{\alpha_s N_c}{4\pi} \right)^3 \ln^4\frac{1}{x} \ln^2\frac{Q^2}{\Lambda^2} \left[2 \left( 39 -\frac{4N_f}{N_c} \right) \Delta\Sigma^{(0)} + 4 \left( 64-\frac{13N_f}{N_c} \right) \Delta G^{(0)}\right] \notag \\
&\hspace*{.1cm}+ \frac{1}{36}\left(\frac{\alpha_s N_c}{4\pi} \right)^3\ln^3\frac{1}{x} \ln^3\frac{Q^2}{\Lambda^2 }\left[2 \left( 73-\frac{4N_f}{N_c} \right) \Delta \Sigma ^{(0)} + 4 \left(128-\frac{17N_f}{N_c} \right) \Delta G^{(0)}\right], \notag \\ 
&\Delta \Sigma^{(3)}(x, Q^2) 
\\
&= -\frac{1}{30}\left(\frac{\alpha_s N_c}{4\pi} \right)^3\ln^5\frac{1}{x}\ln\frac{Q^2}{\Lambda^2} \left[\left( \frac{10N_f}{N_c}-\frac{1}{2} \right) \Delta \Sigma^{(0)} + \frac{N_f}{N_c}\left( 35-\frac{4N_f}{N_c} \right) \Delta G^{(0)}\right] \notag \\
&\hspace*{.4cm}-\frac{1}{24}\left(\frac{\alpha_s N_c}{4\pi} \right)^3\ln^4\frac{1}{x}\ln^2\frac{Q^2}{\Lambda^2} \left[\left( \frac{24N_f}{N_c} -1 \right) \Delta\Sigma^{(0)} + \frac{N_f}{N_c} \left(78 - \frac{8N_f}{N_c} \right) \Delta G^{(0)}\right] \notag \\
&\hspace*{.4cm}-\frac{1}{36}\left(\frac{\alpha_s N_c}{4\pi} \right)^3 \ln^3\frac{1}{x}\ln^3\frac{Q^2}{\Lambda^2} \left[\left( \frac{40N_f}{N_c}-1 \right) \Delta \Sigma^{(0)} + \frac{N_f}{N_c} \left( 146-\frac{8N_f}{N_c} \right) \Delta G^{(0)}\right]. \notag 
    \end{align}
\end{subequations}
Again, from the terms containing $\ln^5\frac{1}{x} \ln \frac{Q^2}{\Lambda^2}$, one reads out the polarized small-$x$ large-$N_c \& N_f$ splitting functions at the three-loop order:
\begin{equation}\label{eq:DeltaP_(2)}
    \Delta \mathbf{P}^{(2)} (x) = \frac{1}{6}\left(\frac{\alpha_s N_c}{4\pi}\right)^3 \ln^4\frac{1}{x}\begin{pmatrix}
    (\frac{1}{2} - \frac{10N_f}{N_c})\, & \, - \frac{N_f}{N_c}(35-\frac{4N_f}{N_c})\\[10pt]
    35-\frac{4N_f}{N_c}\, &\, 2(56-\frac{11N_f}{N_c}) \\
    \end{pmatrix}.
\end{equation}


\subsection{Summary and Comparisons up to Three Loops}

To summarize the above results, we have obtained the polarized splitting functions up to three loops by iteratively solving the small-$x$ helicity evolution equations. Collecting Eqs.~\eqref{eq:DeltaP_(0)},~\eqref{eq:DeltaP_(1)} and~\eqref{eq:DeltaP_(2)}, we have the following expressions
\begin{subequations}\label{eq:DeltaP_BER}
\begin{align}
\label{eq:P'_qq}
& \Delta P_{qq} (x) = \left(\frac{\as N_c}{4 \pi}\right) +\left( \frac{\as N_c}{4 \pi} \right)^2 \left(\frac{1}{2} - 2 \, \frac{N_f}{N_c} \right)\ln^2 \frac{1}{x} \\
&\hspace*{3.1cm}+\left( \frac{\as N_c}{4 \pi} \right)^3 \frac{1}{12} \left(1-20\frac{N_f}{N_c}\right)\ln^4 \frac{1}{x}\,+ {\cal O} (\as^4) , \notag\\
\label{eq:P'_qG}
& \Delta P_{qG} (x) = - \left(\frac{\as N_c}{4 \pi}\right) \frac{2N_f}{N_c}-\left( \frac{\as N_c}{4 \pi} \right)^2 \, \frac{5N_f}{N_c}\ln^2 \frac{1}{x} \\
&\hspace*{5.1cm}- \left( \frac{\as N_c}{4 \pi} \right)^3\frac{1}{6}\frac{N_f}{N_c}\left( 35 - 4  \, \frac{N_f}{N_c} \right)\, \ln^4 \frac{1}{x} \, + {\cal O} (\as^4)  , \notag\\
\label{eq:P'_Gq}
& \Delta P_{Gq} (x) = 2\left(\frac{\as N_c}{4 \pi}\right) + 5 \, \left( \frac{\as N_c}{4\pi} \right)^2 \, \ln^2 \frac{1}{x} \\
&\hspace*{4.4cm}+\left( \frac{\as N_c}{4 \pi} \right)^3 \, \frac{1}{6}\left( 35 - 4  \, \frac{N_f}{N_c} \right)\, \ln^4 \frac{1}{x} + {\cal O} (\as^4)  , \notag\\
\label{eq:P'_GG}
& \Delta P_{GG} (x) = 8 \, \left(\frac{\as N_c}{4\pi}\right) + \left( \frac{\as N_c}{4\pi} \right)^2 \, \left( 16 - 2\frac{N_f}{ \, N_c} \right) \, \ln^2 \frac{1}{x} \\
&\hspace{4.2cm}+ \left( \frac{\as N_c}{4\pi} \right)^3 \, \frac{1}{3} \,  \left( 56 - 11 \, \frac{N_f}{N_c} \right) \, \ln^4 \frac{1}{x} + {\cal O} (\as^4) .\notag
\end{align}
\end{subequations}
These splitting functions completely agree with the results of BER IREE expansion (obtained by applying the large-$N_c \& N_f$ limit to Eq.~(14) in \cite{Blumlein:1996hb}).

For comparison, in the $\overline{\text{MS}}$ scheme, the small-$x$ limits of the polarized splitting functions at large-$N_c \& N_f$ are \cite{Altarelli:1977zs,Dokshitzer:1977sg,Mertig:1995ny,Moch:2014sna},
\begin{subequations}\label{eq:DeltaP_MSbar}
\begin{align}
\label{Pqq} 
&\Delta \overline{P}_{qq}(x) = \left(\frac{\alpha_sN_c}{4\pi}\right) + \left(\frac{\alpha_s N_c}{4\pi}\right)^2 \left( \frac{1}{2}-2\frac{N_f}{N_c} \right)\ln^2\frac{1}{x} \\
&\hspace*{3.7cm}+ \left(\frac{\alpha_sN_c}{4\pi}\right)^3\frac{1}{12} \left( 1-20\frac{N_f}{N_c} \right) \ln^4\frac{1}{x} + {\cal O} (\alpha_s^4) \,,  \notag  \\
\label{eq:PqG}
&\Delta \overline{P}_{qG}(x) =  - \left(\frac{\alpha_sN_c}{4\pi}\right)\frac{2N_f}{N_c} - \left(\frac{\alpha_sN_c}{4\pi}\right)^2 5\frac{N_f}{N_c}\ln^2\frac{1}{x} \\
&\hspace*{5.1cm}- \left(\frac{\alpha_sN_c}{4\pi}\right)^3\frac{1}{6}\frac{N_f}{N_c} \left( 34-4\frac{N_f}{N_c} \right) \ln^4\frac{1}{x} + {\cal O} (\alpha_s^4) \,,  \notag    \\
\label{eq:PGq}
&\Delta \overline{P}_{Gq}(x) =   2\left(\frac{\alpha_sN_c}{4\pi}\right) + 5\left(\frac{\alpha_s N_c}{4\pi}\right)^2 \ln^2\frac{1}{x} \\
&\hspace*{4.4cm}+ \left(\frac{\alpha_s N_c}{4\pi}\right)^3\frac{1}{6} \left( 36-4\frac{N_f}{N_c} \right) \ln^4\frac{1}{x} + {\cal O} (\alpha_s^4)  \,,   \notag \\
\label{PGG} 
&\Delta \overline{P}_{GG}(x) =   8 \left(\frac{\alpha_sN_c}{4\pi}\right) + \left(\frac{\alpha_s N_c}{4\pi}\right)^2 \left( 16-2\frac{N_f}{N_c} \right) \ln^2\frac{1}{x} \\
&\hspace*{4.2cm}+ \left(\frac{\alpha_s N_c}{4\pi}\right)^3\frac{1}{3} \left( 56-11\frac{N_f}{N_c} \right) \ln^4\frac{1}{x} + {\cal O} (\alpha_s^4) \,.  \notag
\end{align}
\end{subequations}
We see that our (and BER) polarized splitting functions $\Delta P_{qq}$ and $\Delta P_{GG}$ completely agree with the $\overline{\text{MS}}$ ones, while $\Delta P_{qG}$ and $\Delta P_{Gq}$ appear to disagree slightly at order-$\as^3$ with their $\overline{\text{MS}}$ counterparts. The same disagreement between the BER 3-loop results and $\overline{\text{MS}}$ was attributed to scheme dependence of the splitting functions in \cite{Moch:2014sna}: in Appendix~\ref{scheme.app} we construct the scheme transformation relating Eqs.~\eqref{eq:DeltaP_BER} and \eqref{eq:DeltaP_MSbar} explicitly. Here we note that the three-loop calculation \cite{Moch:2014sna} was carried out in the Larin scheme \cite{Larin:1991tj, Larin:1993tq}, and the results were then transformed into the $\overline{\text{MS}}$ scheme (see \cite{Blumlein:2024euz} for a recent comparison of the two schemes).

Let us also note the observation made in \cite{Blumlein:1996hb} that the splitting functions obtained from solving the BER IREE evolution equations have the property that 
\begin{equation}\label{PqGGq}
    \Delta P_{qG}(x) = - \frac{N_f}{N_c} \Delta P_{Gq}(x)
\end{equation}
to all orders in $\alpha_s$. The same property seems to be true for our splitting functions \eqref{eq:DeltaP_BER} to three loops and, as we will see below, at four loops as well. However, the relation \eqref{PqGGq} does not hold for the polarized splitting functions \eqref{eq:DeltaP_MSbar} in the $\overline{\text{MS}}$ scheme. It is violated at order-$\alpha_s^3$ as can be seen from Eqs.~\eqref{eq:PqG} and \eqref{eq:PGq}.


\subsection{Step 4}

We can continue iteratively solving the evolution equations beyond step 3 (for simplicity further iteration was done while setting $\Delta\Sigma^{(0)} = 0$). The resulting four-loop splitting functions we obtain this way are
\begin{subequations}\label{splittingfunctions_4loops}
\begin{align}
    \label{Pqq_4loops}
    &\Delta P_{qq}^{(3)}(x) = \left(\frac{\alpha_s N_c}{4\pi}\right)^4 \frac{1}{720}\left(5 - 748\frac{N_f}{N_c} + 80\frac{N_f^2}{N_c^2} \right)\ln^6\frac{1}{x}, \\
    \label{PqG_4loops}
    &\Delta P_{qG}^{(3)}(x) = -\left(\frac{\alpha_s N_c}{4\pi}\right)^4 \frac{1}{360}\frac{N_f}{N_c}\left(1213 - 224\frac{N_f}{N_c}\right)\ln^6\frac{1}{x}, \\
    \label{PGq_4loops}
    &\Delta P_{Gq}^{(3)}(x) = \left(\frac{\alpha_s N_c}{4\pi}\right)^4 \frac{1}{360}\left(1213 - 224\frac{N_f}{N_c}\right)\ln^6\frac{1}{x}, \\
     \label{PGG_4loops}
    &\Delta P_{GG}^{(3)}(x) = \left(\frac{\alpha_s N_c}{4\pi}\right)^4 \frac{1}{180}\left(1984 - 549\frac{N_f}{N_c} + 20\frac{N_f^2}{N_c^2} \right)\ln^6\frac{1}{x}.
\end{align}
\end{subequations}
These are our predictions for the small-$x$ large-$N_c \& N_f$ polarized splitting functions at four loops. They indeed obey the property \eqref{PqGGq} stated above. We hope to be able to compare them to the appropriate limit of the finite-order four-loop result, when the latter is calculated. 

In the meantime, the four-loop splitting functions in Eqs.~\eqref{splittingfunctions_4loops} are to be compared with the four loop predictions from the BER IREE expansion. Taking Eqs.~(15) in \cite{Blumlein:1996hb} and applying the large-$N_c\&N_f$ limit, these are 
\begin{subequations}\label{splittingfunctions_4loops_BER}
\begin{align}
    \label{Pqq_4loops_BER}
    &\Delta P_{qq}^{(3)\,\text{(BER)}}(x) = \left(\frac{\alpha_s N_c}{4\pi}\right)^4 \frac{1}{720}\left(5 - 764\frac{N_f}{N_c} + 80\frac{N_f^2}{N_c^2} \right)\ln^6\frac{1}{x}, \\
    \label{PqG_4loops_BER}
    &\Delta P_{qG}^{(3)\,\text{(BER)}}(x) = -\left(\frac{\alpha_s N_c}{4\pi}\right)^4 \frac{1}{360}\frac{N_f}{N_c}\left(1229 - 224\frac{N_f}{N_c}\right)\ln^6\frac{1}{x}, \\
    \label{PGq_4loops_BER}
    &\Delta P_{Gq}^{(3)\,\text{(BER)}}(x) = \left(\frac{\alpha_s N_c}{4\pi}\right)^4 \frac{1}{360}\left(1229 - 224\frac{N_f}{N_c}\right)\ln^6\frac{1}{x}, \\
     \label{PGG_4loops_BER}
    &\Delta P_{GG}^{(3)\,\text{(BER)}}(x) = \left(\frac{\alpha_s N_c}{4\pi}\right)^4 \frac{1}{180}\left(2016 - 557\frac{N_f}{N_c} + 20\frac{N_f^2}{N_c^2} \right)\ln^6\frac{1}{x}.
\end{align}
\end{subequations}
By comparing Eqs.~\eqref{splittingfunctions_4loops} to Eqs.~\eqref{splittingfunctions_4loops_BER} we see that evidently the polarized splitting functions we obtain exhibit a very minor disagreement with those from the BER IREE expansion beginning at four loops. This is consistent with the results of Ch.~\ref{largeNcsoln.ch}, where our analytic all-orders prediction for the polarized small-$x$ $GG$ anomalous dimension at large-$N_c$ disagreed with that derived from the BER IREE. There the disagreement also began at four loops. The difference between Eqs.~\eqref{splittingfunctions_4loops} and \eqref{splittingfunctions_4loops_BER} cannot be accounted for by a scheme transformation since it persists even in the gluons-only large-$N_c$ limit where there is only one splitting function, $\Delta P_{GG}$ \cite{Borden:2023ugd}.


\section{Chapter Summary}

In this Chapter we introduced the notion that the existing large-$N_c\&N_f$ evolution equations from \cite{Cougoulic:2022gbk} (and presented in Ch.~\ref{sec:ch3_fullevoleqns}) were in some way incomplete. We showed the disagreements between the predictions of the evolution and the results of finite-order calculations for the small-$x$, large-$N_c\&N_f$ polarized DGLAP splitting functions (and additionally for the eigenvalues of the anomalous dimension matrices in the two frameworks). We identified the class of operators --- the $q/{\bar q} \to G$ and $G \to q/{\bar q}$ shock wave transition operators --- which needed to be incorporated into the small-$x$ evolution. Using the frameworks of both LCOT and LCPT, we calculated the contributions of these operators to the small-$x$ evolution of the flavor-singlet quark and gluon helicity PDFs.\footnote{Again, these operators were previously considered in \cite{Chirilli:2021lif} in the framework of helicity evolution. However, a direct comparison with \cite{Chirilli:2021lif} appears to be difficult.} While the transition operators do not contribute to helicity evolution in the large-$N_c$ limit which contains no quarks (thus leaving unaffected the results of Ch.~\ref{largeNcsoln.ch}), they do contribute in the large-$N_c \& N_f$ limit. The main result of this Chapter is the set of modified large-$N_c \& N_f$ helicity evolution equations~\eqref{eq_LargeNcNf}. We showed that this modified evolution agrees with the small-$x$ and large-$N_c \& N_f$ limit of the existing finite-order calculations of the polarized DGLAP splitting functions up to the three known loops, indicating that the modified equations~\eqref{eq_LargeNcNf} are likely to include all the contributions needed for small-$x$ helicity evolution. In the next Chapter, we will return to the challenge of analytically solving the helicity evolution equations, like we did in Ch.~\ref{largeNcsoln.ch}. However, this time we focus on the newly revised set of large-$N_c\&N_f$ equations derived here. Though this is a more complicated set of equations than the comparatively simpler large-$N_c$ evolution, we will see that the same tools developed in Ch.~\ref{largeNcsoln.ch} can be applied once more.

\chapter{\texorpdfstring{Analytic Solution to the Modified Small-$x$ Helicity Evolution at Large $N_c\&N_f$}{Analytic Solution to the Modified Small x Helicity Evolution at Large Nc \& Nf}}
\label{largeNcandNfsoln.ch}

We have modified the large-$N_c\&N_f$ evolution equations with the inclusion of the quark-to-gluon and gluon-to-quark transition operators of Ch.~\ref{transitionops.ch}. With the cross-checks performed in that Chapter, we can be confident that we now have the full set of equations needed to describe the double-logarithmic small-$x$ helicity evolution in the large-$N_c\&N_f$ limit. We are now in a position to re-purpose the techniques we developed in Ch.~\ref{largeNcsoln.ch} in order to analytically solve the new set of large-$N_c\&N_f$ evolution equations~\eqref{eq_LargeNcNf}. In this Chapter we will develop exactly this analytic solution. As with our solution to the large-$N_c$ evolution, our solution here will allow us to make a number of critical analytic predictions, now in the more general and more realistic limit of large $N_c\&N_f$. 

Using our solution we will obtain analytic, small-$x$ large-$N_c\&N_f$ expressions for the hPDFs and the $g_1$ structure function. We will obtain analytic expressions for the eigenvalues of the matrix of polarized DGLAP anomalous dimensions and subsequently, analytic expressions for each of the four individual anomalous dimensions themselves, exact to all orders in $\as$. We can extract the leading small-$x$ asymptotic power law growth of the helicity distributions, given by the familiar
\begin{align}
    \Delta \Sigma(x,Q^2) \sim \Delta G(x,Q^2) \sim g_1(x,Q^2) \sim \left(\frac{1}{x}\right)^{\alpha_h}\,,
\end{align}
where we obtain the algebraic equation satisfied by the intercept $\alpha_h$. Although the algebraic complexity of the equation prevents us from obtaining a general analytic expression for the intercept, we determine $\alpha_h$ numerically (and in a special case, analytically) for various values of $N_c$ and $N_f$. We further obtain explicit expressions for the helicity distributions in the asymptotic limit, which also yield a prediction for the asymptotic ratio of $\Delta G/ \Delta \Sigma$. This Chapter is based on the paper \cite{Borden:2025ehe
}.


\section{Large-\texorpdfstring{$N_c\&N_f$}{Nc\&Nf} Equations}\label{ch6sec:equations}

We begin with the revised set of large-$N_c\&N_f$ evolution equations~\eqref{eq_LargeNcNf}. As in Ch.~\ref{largeNcsoln.ch}, we again introduce the rescaled variables in Eqs.~\eqref{originalvars}, in terms of which the large-$N_c\&N_f$ equations can be written 
\begin{subequations}\label{ch6evoleqs_IR_1}\allowdisplaybreaks
\begin{align}
    \label{ch6Q_IR2}
    &Q(\soz,\eta) = Q^{(0)}(\soz,\eta) + \int\limits_{\soz}^{\eta}\mathrm{d}\eta' \int\limits_{\soz}^{\eta'} \mathrm{d}\sto \bigg[2\widetilde{G}(\sto,\eta') + 2\widetilde{\Gamma}(\soz,\sto,\eta') + Q(\sto,\eta') \\
    &\hspace{6cm} - \overline{\Gamma}(\soz,\sto,\eta') + 2\Gamma_2(\soz,\sto,\eta') + 2G_2(\sto,\eta') \bigg] \notag \\
    &\hspace{2.4cm}+ \frac{1}{2}\left[ \int\limits_{0}^{\soz}\mathrm{d}\sto\int\limits_{\sto}^{\eta+\sto-\soz} \mathrm{d}\eta' + \int\limits_{\soz}^{\eta}\mathrm{d}\sto\int\limits_{\sto}^{\eta}\mathrm{d}\eta'\right] \bigg[Q(\sto,\eta') + 2G_2(\sto,\eta') \bigg] \notag\,,\\
    \label{ch6Gammabar_IR2}
    &\overline{\Gamma}(\soz,\sto,\eta') = Q^{(0)}(\soz,\eta') + \Bigg[\ \int\limits_{\soz}^{\sto}\mathrm{d}\stt \int\limits_{\stt}^{\eta'-\sto+\stt}\mathrm{d}\eta'' + \int\limits_{\sto}^{\eta'}\mathrm{d}\stt \int\limits_{\stt}^{\eta'}\mathrm{d}\eta'' \Bigg]\\
    &\hspace*{3.5cm}\times
    \bigg[2\widetilde{G}(\stt,\eta'') + 2\widetilde{\Gamma}(\soz,\stt,\eta'') + Q(\stt,\eta'') - \overline{\Gamma}(\soz,\stt,\eta'') \notag \\
    &\hspace*{8.5cm}+ 2\Gamma_2(\soz,\stt,\eta'') + 2G_2(\stt,\eta'') \bigg] \notag \\
    &\hspace{2cm}+ \frac{1}{2} \Bigg[\int\limits_{0}^{\sto}\mathrm{d}\stt  \int\limits_{\stt}^{\eta'-\sto+\stt}\mathrm{d}\eta'' + \int\limits_{\sto}^{\eta'}\mathrm{d}\stt   \int\limits_{\stt}^{\eta'}\mathrm{d}\eta''
    \Bigg]\bigg[Q(\stt,\eta'') + 2G_2(\stt,\eta'') \bigg] \notag \,,\\
    \label{ch6Gtilde_IR2}
    &\widetilde{G}(\soz,\eta) = \widetilde{G}^{(0)}(\soz,\eta) + \int\limits_{\soz}^{\eta}\mathrm{d}\eta' \int\limits_{\soz}^{\eta'} \mathrm{d}\sto \bigg[3\widetilde{G}(\sto,\eta') + \widetilde{\Gamma}(\soz,\sto,\eta') + 2G_2(\sto,\eta') \\
    &\hspace*{2.7cm} + \left(2-\frac{N_f}{2N_c} \right)\Gamma_2(\soz,\sto,\eta') - \frac{N_f}{4N_c}\overline{\Gamma}(\soz,\sto,\eta') -\frac{N_f}{2N_c}\widetilde{Q}(\sto,\eta')  \bigg] \notag \\
    &\hspace{2cm} - \frac{N_f}{4N_c} \int\limits_{0}^{\soz}\mathrm{d}\sto \int\limits_{\sto}^{\eta+\sto-\soz}\mathrm{d}\eta' \bigg[Q(\sto,\eta') + 2G_2(\sto,\eta') \bigg]\notag \,,\\
    \label{ch6Gammatilde_IR2}
    &\widetilde{\Gamma}(\soz,\sto,\eta') = \widetilde{G}^{(0)}(\soz,\eta') + \Bigg[\ \int\limits_{\soz}^{\sto}\mathrm{d}\stt \int\limits_{\stt}^{\eta'-\sto+\stt}\mathrm{d}\eta'' + \int\limits_{\sto}^{\eta'}\mathrm{d}\stt \int\limits_{\stt}^{\eta'}\mathrm{d}\eta''  \Bigg] \\
    &\hspace*{2cm}\times
    \bigg[3\widetilde{G}(\stt,\eta'') + \widetilde{\Gamma}(\soz,\stt,\eta'') + 2G_2(\stt,\eta'') \notag \\
    &\hspace*{2.7cm}+ \left(2-\frac{N_f}{2N_c} \right)\Gamma_2(\soz,\stt,\eta'') - \frac{N_f}{4N_c}\overline{\Gamma}(\soz,\stt,\eta'') -\frac{N_f}{2N_c}\widetilde{Q}(\stt,\eta'')  \bigg]  \notag \\
    &\hspace{2cm} - \frac{N_f}{4N_c} \int\limits_{0}^{\soz}\mathrm{d}\stt \int\limits_{\stt}^{\eta'-\sto+\stt}\mathrm{d}\eta'' \bigg[Q(\stt,\eta'') + 2G_2(\stt,\eta'') \bigg]\notag \,,\\
    \label{ch6G2_IR2}
    &G_2(\soz,\eta) = G_2^{(0)}(\soz,\eta) + 2 \int\limits_{0}^{\soz}\mathrm{d}\sto \int\limits_{\sto}^{\eta+\sto-\soz}\mathrm{d}\eta' \bigg[\widetilde{G}(\sto,\eta') + 2G_2(\sto,\eta') \bigg]\,,\\
    \label{ch6Gamma2_IR2}
    &\Gamma_2(\soz,\sto,\eta') = G_2^{(0)}(\soz,\eta') + 2 \int\limits_{0}^{\soz}\mathrm{d}\stt \int\limits_{\stt}^{\eta'-\sto+\stt}\mathrm{d}\eta'' \bigg[\widetilde{G}(\stt,\eta'') + 2G_2(\stt,\eta'') \bigg]\,,\\
    \label{ch6Qtilde_IR2}
    &\widetilde{Q}(\soz,\eta) = \widetilde{Q}^{(0)}(\soz,\eta) - \int\limits_0^{\soz}\mathrm{d}\sto \int\limits_{\sto}^{\eta+\sto-\soz}\mathrm{d}\eta' \bigg[Q(\sto,\eta')+2G_2(\sto,\eta')\bigg]\,,
    \end{align}
\end{subequations}
where we assume the ordering $0\leq\soz\leq\sto\leq\eta'$ in Eqs.~\eqref{ch6Gammabar_IR2}, \eqref{ch6Gammatilde_IR2}, and \eqref{ch6Gamma2_IR2}.

Once analytic expressions for the dipole amplitudes\footnote{As discussed in Ch.~\ref{transitionops.ch}, the structure of $\widetilde{Q}$ is like that of a sum of TMDs with the forward- and past pointing Wilson-line staples, so $\widetilde{Q}(\xoz^2,zs)$ cannot be properly called a dipole amplitude. However, for simplicity, we will often just refer to the collection of seven objects that evolve under Eqs.~\eqref{ch6evoleqs_IR_1} as `dipole amplitudes,' though strictly speaking one should remember this caveat.} are known, one can obtain the flavor-singlet quark and gluon hPDFs using Eqs.~\eqref{eq:DeltaGSigma_G2_Qtilde} from Ch.~\ref{transitionops.ch}, restated below:
\begin{subequations}\label{ch6pdfsfromdipoles}
\begin{align}
    \label{ch6DeltaGfromdipoles}
    &\Delta G(x,Q^2) = \frac{2N_c}{\as \pi^2} \, G_2\left( \xoz^2 = \frac{1}{Q^2}, s=\frac{Q^2}{x}\right), \\
    \label{ch6DeltaSigmafromdipoles}
    &\Delta \Sigma(x,Q^2) = \frac{N_f}{\as \pi^2} \, \widetilde{Q} \left( \xoz^2 = \frac{1}{Q^2}, s=\frac{Q^2}{x}\right).
\end{align}
\end{subequations}
Recall that \eq{ch6DeltaGfromdipoles} is consistent with previous versions of the small-$x$ helicity evolution \cite{Kovchegov:2015pbl, Kovchegov:2017lsr, Kovchegov:2018znm,Cougoulic:2022gbk,Kovchegov:2016zex}, while \eq{ch6DeltaSigmafromdipoles} was derived in Ch.~\ref{transitionops.ch} and is a slight modification from the previous small-$x$ helicity results, ultimately representing a scheme transformation relative to the previous result. In addition, the $g_1$ structure function is given in terms of the dipole amplitudes by \eq{ch3_g1}, which we repeat here:
\begin{align}\label{ch6g1_DLA}
g_1 (x, Q^2)  = - \sum_f \frac{N_c \, Z^2_f}{4 \pi^3} \int\limits_{\Lambda^2/s}^1 \frac{dz}{z} \,  \int\limits^{\min \left\{ \frac{1}{z Q^2} , \frac{1}{\Lambda^2} \right\}}_\frac{1}{zs} \frac{d x^2_{10}}{x_{10}^2} \, \left[ Q (x_{10}^2, zs) + 2 \, G_2 (x_{10}^2, zs) \right],
\end{align}
with $Z_f$ the fractional electric charge of the quark. For simplicity, we assume here that the objects $\widetilde Q$ and $Q$, whose operator definitions include quark fields of a fixed flavor \cite{Borden:2024bxa}, are independent of the quark flavors: to bring back the flavor dependence, one needs to replace $\widetilde Q \to \widetilde Q_f$, $Q \to Q_f$, and $N_f \to \sum_f$ in Eqs.~\eqref{ch6evoleqs_IR_1}, \eqref{ch6DeltaSigmafromdipoles}, and \eqref{ch6g1_DLA} (cf. \cite{Adamiak:2021ppq, Adamiak:2023okq, Adamiak:2023yhz}).


\section{Solution}\label{ch6sec:solution}

\subsection{\texorpdfstring{Double Inverse Laplace Representations for \\ $G_2, \Gamma_2, \widetilde{G}, Q,\widetilde{Q}$}{Double Inverse Laplace Representations for G2, Gamma2, Gtilde, Q, Qtilde}}

The solution we construct here follows quite closely to that constructed in Ch.~\ref{largeNcsoln.ch} for the large-$N_c$ evolution equations. We begin by introducing the following double-inverse Laplace transforms for $G_2(\soz^2,\eta)$, $\widetilde{G}(\soz^2,\eta)$, $Q(\soz^2,\eta)$, $\widetilde{Q}(\soz^2,\eta)$ and their initial conditions/inhomogeneous terms:
\begin{subequations}\label{ch6doubleLaplacestart}
    \begin{align} \label{ch6doubleLaplaceG2}
    &G_2(\soz,\eta) = \wint \gint e^{\omega(\eta-\soz)}e^{\gamma\soz}G_{2\omega\gamma}\,,
    \\ 
    \label{ch6doubleLaplaceG20}
    &G_2^{(0)}(\soz,\eta) = \wint \gint e^{\omega(\eta-\soz)}e^{\gamma\soz}G^{(0)}_{2\omega\gamma}\,,
    \\ 
    \label{ch6doubleLaplaceGtilde}
    &\widetilde{G}(\soz,\eta) = \wint \gint e^{\omega(\eta-\soz)}e^{\gamma\soz}\widetilde{G}_{\omega\gamma}\,,
    \\ 
    \label{ch6doubleLaplaceGtilde0}
    &\widetilde{G}^{(0)}(\soz,\eta) = \wint \gint e^{\omega(\eta-\soz)}e^{\gamma\soz}\widetilde{G}^{(0)}_{\omega\gamma}\,,
    \\ 
    \label{ch6doubleLaplaceQ}
    &Q(\soz,\eta) = \wint \gint e^{\omega(\eta-\soz)}e^{\gamma\soz}Q_{\omega\gamma}\,,
    \\ 
    \label{ch6doubleLaplaceQ0}
    &Q^{(0)}(\soz,\eta) = \wint \gint e^{\omega(\eta-\soz)}e^{\gamma\soz}Q^{(0)}_{\omega\gamma}\,,
    \\ 
    \label{ch6doubleLaplaceQtilde}
    &\widetilde{Q}(\soz,\eta) = \wint \gint e^{\omega(\eta-\soz)}e^{\gamma\soz}\widetilde{Q}_{\omega\gamma}\,,
    \\ 
    \label{ch6doubleLaplaceQtilde0}
    &\widetilde{Q}^{(0)}(\soz,\eta) = \wint \gint e^{\omega(\eta-\soz)}e^{\gamma\soz}\widetilde{Q}^{(0)}_{\omega\gamma}.
    \end{align}
\end{subequations}
As usual, these integrals are taken along vertical contours parallel to the imaginary axes in the $\omega$ and $\gamma$ planes, with all singularities of the integrands located to the left of the contours. 

As can be seen from Eqs.~\eqref{ch6G2_IR2} and \eqref{ch6Gamma2_IR2}, the dipole amplitudes $G_2$ and $\Gamma_2$ obey the following property:
\begin{align}\label{ch6G2Gamma2scaling}
    &\Gamma_2(\soz,\sto,\eta') - G_2^{(0)}(\soz,\eta') = G_2(\soz,\eta=\eta'+\soz-\sto) \\
    &\hspace*{6cm}- G_{2}^{(0)}(\soz,\eta=\eta'+\soz-\sto).\notag
\end{align}
Then using Eqs.~\eqref{ch6doubleLaplaceG2}, \eqref{ch6doubleLaplaceG20}, and \eqref{ch6G2Gamma2scaling} we straightforwardly have
\begin{align}\label{ch6doubleLaplaceGamma2}
    &\Gamma_2(\soz,\sto,\eta') = \wint \gint \left[e^{\omega(\eta'-\sto)}e^{\gamma\soz}\left(G_{2\omega\gamma} - G^{(0)}_{2\omega\gamma}  \right)  + e^{\omega(\eta'-\soz)}e^{\gamma\soz} \, G^{(0)}_{2\omega\gamma}  \right].
\end{align}
Next, we substitute our double Laplace transforms from Eqs.~\eqref{ch6doubleLaplaceG2} and \eqref{ch6doubleLaplaceGtilde} into the evolution equation \eqref{ch6G2_IR2}. Carrying out the integrals over $\sto$ and $\eta'$ and then inverting the transforms, we find
\begin{align}\label{ch6G2minusG20}
    G_{2\omega\gamma} - G^{(0)}_{2\omega\gamma} = \frac{2}{\omega\gamma}\left(\widetilde{G}_{\omega\gamma} + 2 \, G_{2\omega\gamma}\right),
\end{align}
or equivalently
\begin{align}\label{ch6Gtildeomegagamma}
    \widetilde{G}_{\omega\gamma} = \frac{\omega\gamma}{2}\left(G_{2\omega\gamma} - G^{(0)}_{2\omega\gamma} \right) - 2 \, G_{2\omega\gamma}.
\end{align}
Since all double-Laplace images ($G_{2\omega\gamma}, G^{(0)}_{2\omega\gamma}, \widetilde{G}_{\omega\gamma}, \widetilde{G}^{(0)}_{\omega\gamma}, Q_{\omega\gamma}, Q^{(0)}_{\omega\gamma},\widetilde{Q}_{\omega\gamma},\widetilde{Q}^{(0)}_{\omega\gamma}$) must go to zero as $\omega$ or $\gamma$ go to infinity, Eq.~\eqref{ch6G2minusG20} implies that the difference between $G_{2\omega\gamma}$ and $G^{(0)}_{2\omega\gamma}$ goes to zero faster than $1/\omega$ or $1/\gamma$ as $\omega$ or $\gamma$, respectively, go to infinity. We can write
\begin{subequations}\label{ch6G2minusG20atinfty}
    \begin{align}
    &\wint \left(G_{2\omega\gamma} - G^{(0)}_{2\omega\gamma}\right) = \wint \frac{2}{\omega\gamma}\left(\widetilde{G}_{\omega\gamma} + 2 \, G_{2\omega\gamma}\right) = 0 
    \end{align}
    and
    \begin{align}
    &\gint \left(G_{2\omega\gamma} - G^{(0)}_{2\omega\gamma}\right) = \gint \frac{2}{\omega\gamma}\left(\widetilde{G}_{\omega\gamma} + 2G_{2\omega\gamma}\right) = 0,
    \end{align}
\end{subequations}
where the last equality in each line follows from closing the $\omega$- or $\gamma$-contour to the right. This fact can be used along with the double Laplace representations in Eqs.~\eqref{ch6doubleLaplacestart} to straightforwardly show that the boundary conditions implied by Eqs.~\eqref{ch6G2_IR2} and \eqref{ch6Gamma2_IR2},
\begin{align}
    &G_2(\soz=0,\eta) = G^{(0)}_2(\soz=0,\eta)\,,\\
    &G_2(\soz,\eta=\soz) = G^{(0)}_2(\soz,\eta=\soz)\,, \\
    &\Gamma_2(\soz=0,\sto,\eta') = G^{(0)}_2(\soz=0,\eta')\,, \\
    &\Gamma_2(\soz,\sto,\eta'=\sto) = G^{(0)}_2(\soz,\eta'=\sto)\,,
\end{align}
are automatically satisfied. All the above steps have closely followed those in Ch.~\ref{largeNcsoln.ch}.

Next, we can substitute the double Laplace expressions Eqs.~\eqref{ch6doubleLaplaceG2}, \eqref{ch6doubleLaplaceQ}, and \eqref{ch6doubleLaplaceQtilde} into the evolution equation \eqref{ch6Qtilde_IR2}. Doing this, carrying out the integrals over $\sto$ and $\eta'$, and then inverting the Laplace transforms, we find
\begin{align}\label{ch6QtildeminusQtilde0}
    \widetilde{Q}_{\omega\gamma}- \widetilde{Q}^{(0)}_{\omega\gamma} = -\frac{1}{\omega\gamma}\left(Q_{\omega\gamma} + 2G_{2\omega\gamma}\right),
\end{align}
or equivalently
\begin{align}\label{ch6Qtildeomegagamma}
    Q_{\omega\gamma} = -\omega\gamma\left(\widetilde{Q}_{\omega\gamma} - \widetilde{Q}^{(0)}_{\omega\gamma}\right) - 2G_{2\omega\gamma}.
\end{align}
Again we see that \eq{ch6QtildeminusQtilde0} implies that the difference between $\widetilde{Q}_{\omega\gamma}$ and $\widetilde{Q}^{(0)}_{\omega\gamma}$ goes to zero faster than $1/\omega$ or $1/\gamma$ as $\omega\rightarrow \infty$ or $\gamma\rightarrow\infty$, which allows us to write
\begin{subequations}\label{ch6QtildeminusQtilde0atinfty}
    \begin{align}
    &\wint \left(\widetilde{Q}_{\omega\gamma} - \widetilde{Q}^{(0)}_{\omega\gamma}\right) = -\wint \frac{1}{\omega\gamma} \left(Q_{\omega\gamma} + 2G_{2\omega\gamma}\right) = 0
    \end{align}
    and
    \begin{align}
    &\gint \left(\widetilde{Q}_{\omega\gamma} - \widetilde{Q}^{(0)}_{\omega\gamma}\right) = -\gint \frac{1}{\omega\gamma}\left(Q_{\omega\gamma} + 2G_{2\omega\gamma}\right) = 0\,,
\end{align}
\end{subequations}
where again the last equality in each line follows from closing the contour to the right. Eqs.~\eqref{ch6QtildeminusQtilde0atinfty} can be used along with the double-Laplace representations in \eq{ch6doubleLaplacestart} to show that the two boundary conditions for $\widetilde{Q}$ implied by \eq{ch6Qtilde_IR2},
\begin{align}\label{ch6QtildeBCs}
    &\widetilde{Q}(\soz=0,\eta) = \widetilde{Q}^{(0)}(\soz=0,\eta)\,,\\
    &\widetilde{Q}(\soz,\eta=\soz) = \widetilde{Q}^{(0)}(\soz,\eta=\soz)\,,
\end{align}
are automatically satisfied.

At this point, the evolution equations \eqref{ch6G2_IR2}, \eqref{ch6Gamma2_IR2}, and \eqref{ch6Qtilde_IR2} are completely satisfied, and we have obtained expressions for the double-inverse Laplace transforms of  $\Gamma_2$, $\widetilde{G}$, and $Q$ in terms of the yet unknown double-Laplace images $G_{2\omega\gamma}$ and $\widetilde{Q}_{\omega\gamma}$. It remains to satisfy Eqs.~\eqref{ch6Q_IR2}, \eqref{ch6Gammabar_IR2}, \eqref{ch6Gtilde_IR2}, and \eqref{ch6Gammatilde_IR2}, obtain double-Laplace expressions for the remaining dipole amplitudes $\overline{\Gamma}$ and $\widetilde{\Gamma}$, and ultimately solve for the double-Laplace images $G_{2\omega\gamma}$ and $\widetilde{Q}_{\omega\gamma}$.


\subsection{Double Inverse Laplace Representations for \texorpdfstring{$\overline{\Gamma}$, $\widetilde{\Gamma}$}{Gammabar, Gammatilde}}

Upon differentiating Eqs.~\eqref{ch6Gammabar_IR2} and \eqref{ch6Gammatilde_IR2}, one can show that $\overline{\Gamma}$ and $\widetilde{\Gamma}$ satisfy the following second-order partial differential equations:
\begin{subequations}\label{ch6secondorderpdes}
\begin{align}
    \label{ch6GammabarPDE}
    &\frac{\partial^2 \overline{\Gamma}(\soz,\sto,\eta')}{\partial\sto\partial\eta'} + \frac{\partial^2 \overline{\Gamma}(\soz,\sto,\eta')}{\partial\sto^2} = -2\widetilde{G}(\sto,\eta') - 2\widetilde{\Gamma}(\soz,\sto,\eta') - \frac{3}{2}Q(\sto,\eta') \\
    &\hspace{6cm}+ \overline{\Gamma}(\soz,\sto,\eta') - 2\Gamma_2(\soz,\sto,\eta') - 3G_2(\sto,\eta') \,,\notag \\
    \label{ch6GammatildePDE}
    &\frac{\partial^2 \widetilde{\Gamma}(\soz,\sto,\eta')}{\partial\sto\partial\eta'} + \frac{\partial^2 \widetilde{\Gamma}(\soz,\sto,\eta')}{\partial\sto^2} = -3\widetilde{G}(\sto,\eta') - \widetilde{\Gamma}(\soz,\sto,\eta') - 2G_2(\sto,\eta') \\
    &\hspace{3cm}- \left(2-\frac{N_f}{2N_c} \right)\Gamma_2(\soz,\sto,\eta') + \frac{N_f}{4N_c}\overline{\Gamma}(\soz,\sto,\eta') +\frac{N_f}{2N_c}\widetilde{Q}(\sto,\eta'). \notag
\end{align}
\end{subequations}
Similar to \cite{Borden:2023ugd}, we proceed to solve these two partial differential equations by constructing their homogeneous and particular solutions. 
We begin with the homogeneous solutions, employing the following ansatz:
\begin{subequations}
\begin{align}\label{ch6Gammatildehansatz}
    \widetilde{\Gamma}^{(h)}(\soz,\sto,\eta') = \wint\gint e^{\omega(\eta'-\sto)}e^{\gamma\sto}\widetilde{\Gamma}_{\omega\gamma}(\soz), \\
    \overline{\Gamma}^{(h)}(\soz,\sto,\eta') = \wint\gint e^{\omega(\eta'-\sto)}e^{\gamma\sto}\overline{\Gamma}_{\omega\gamma}(\soz) .
\end{align}
\end{subequations}
Plugging these into the homogeneous part of Eqs.~\eqref{ch6secondorderpdes} we obtain
\begin{subequations}\label{ch6Gamma_eqs2}
\begin{align}
& [\gamma (\gamma - \omega) - 1] \, \overline{\Gamma}_{\omega\gamma}(\soz) = - 2 \, \widetilde{\Gamma}_{\omega\gamma}(\soz), \\
& [\gamma (\gamma - \omega) + 1] \,  \widetilde{\Gamma}_{\omega\gamma}(\soz) = \frac{N_f}{4 N_c} \, \overline{\Gamma}_{\omega\gamma}(\soz) . 
\end{align}
\end{subequations}
Solving these gives
\begin{align}\label{ch6gammafourthorder}
\gamma^2 \, (\gamma - \omega)^2 = 1 - \frac{N_f}{2 \, N_c} \,,
\end{align}
which can readily be solved for $\gamma$, giving
\begin{align}\label{ch6deltapmpm}
    \gamma = \dw^{\pm\pm} = \tfrac{1}{2}\left(\omega \pm \sqrt{\omega^2 \pm 4\sqrt{1-\tfrac{N_f}{2N_c}}}\,\right).
\end{align}
The notation here is such that the $\pm$ indices on $\dw^{\pm\pm}$ should be read left to right as they are encountered on the right hand side of \eq{ch6deltapmpm}.
In addition, Eqs.~\eqref{ch6Gamma_eqs2} give
\begin{align}\label{ch6relationbtwgammatildeandbar}
  \overline{\Gamma}_{\omega\gamma}(\soz) = \frac{1 \pm \sqrt{ 1 -  \frac{N_f}{2 \, N_c} }}{N_f / (4 N_c) }  \,  \widetilde{\Gamma}_{\omega\gamma}(\soz) ,  
\end{align}
where the $\pm$ here is the same as the second index in $\dw^{\pm\pm}$ . 
We thus have the homogeneous solutions, written as linear combinations of the solutions corresponding to each of the four solutions $\gamma = \dw^{\pm\pm}$ from \eq{ch6deltapmpm},
\begin{subequations}\label{ch6homogeneous_sols}
\begin{align}\label{ch6Gammatildeh}
    & \widetilde{\Gamma}^{(h)}(\soz,\sto,\eta') = \wint e^{\omega(\eta'-\sto)} \sum_{\alpha,\beta = +,-} e^{\dw^{\alpha\beta}\sto} \, \widetilde{\Gamma}_\omega^{(\alpha\beta)}(\soz) , \\
    \label{ch6Gammabarh}
     & \overline{\Gamma}^{(h)}(\soz,\sto,\eta') = \wint e^{\omega(\eta'-\sto)} \sum_{\alpha,\beta = +,-} e^{\dw^{\alpha\beta}\sto}  \frac{1 + \beta \, \sqrt{ 1 - \frac{N_f}{2 \, N_c} }}{N_f / (4 N_c) }  \,  \widetilde{\Gamma}_\omega^{(\alpha\beta)}(\soz) ,
\end{align}
\end{subequations}
where we have also employed \eq{ch6relationbtwgammatildeandbar} in writing \eq{ch6Gammabarh}.

Moving on to the particular solutions of Eqs.~\eqref{ch6secondorderpdes}, we look for them in the following form: 
\begin{subequations}
\begin{align}\label{ch6Gammatildepansatz}
   &  \widetilde{\Gamma}^{(p)}(\soz,\sto,\eta') = \wint\gint \bigg[e^{\omega(\eta'-\sto)}e^{\gamma\sto}A_{\omega\gamma} + e^{\omega(\eta'-\sto)}e^{\gamma\soz}B_{\omega\gamma} \\
   &\hspace*{9cm}+ e^{\omega(\eta'-\soz)}e^{\gamma\soz}C_{\omega\gamma}  \bigg] , \notag \\
   &  \overline{\Gamma}^{(p)}(\soz,\sto,\eta') = \wint\gint \bigg[e^{\omega(\eta'-\sto)}e^{\gamma\sto}  \overline{A}_{\omega\gamma} + e^{\omega(\eta'-\sto)}e^{\gamma\soz}  {\overline B}_{\omega\gamma} \\
   &\hspace*{9cm}+ e^{\omega(\eta'-\soz)}e^{\gamma\soz}  {\overline C}_{\omega\gamma} \bigg] . \notag
\end{align}
\end{subequations}
Plugging those into the full Eqs.~\eqref{ch6secondorderpdes}, and employing Eqs.~\eqref{ch6doubleLaplacestart},  \eqref{ch6doubleLaplaceGamma2}, \eqref{ch6Gtildeomegagamma}, and \eqref{ch6Qtildeomegagamma}, yields
\begin{subequations}\label{ch6ABC3}\allowdisplaybreaks
\begin{align}
& A_{\omega\gamma} =  \frac{1}{\gamma^2 \, (\gamma - \omega)^2 - 1 + \tfrac{N_f}{2 N_c}} \, \Bigg\{  \frac{\omega\gamma}{2} \, \left[ 3 - \tfrac{N_f}{2 N_c} - 3 \,  \gamma \, (\gamma - \omega) \right] \, \left(G_{2\omega\gamma} - G^{(0)}_{2\omega\gamma} \right) \\
&\hspace*{4.5cm}+ \tfrac{3 N_f}{8 N_c} \, \omega \, \gamma \, \left( \widetilde{Q}_{\omega\gamma} - \widetilde{Q}^{(0)}_{\omega\gamma} \right) + \left[ 4 \, \gamma \, (\gamma - \omega) - 4 + \tfrac{N_f}{N_c} \right] \, G_{2\omega\gamma} \notag \\
&\hspace*{4.5cm}+ \tfrac{N_f}{2 N_c} \, \left[ \gamma \, (\gamma - \omega) - 1 \right] \, \widetilde{Q}_{\omega\gamma} \Bigg\}, \notag \\
& \overline{A}_{\omega\gamma} =  \frac{1}{\gamma^2 \, (\gamma - \omega)^2 - 1 + \tfrac{N_f}{2 N_c}} \, \Bigg\{ \omega\gamma \, [2 - \gamma \, (\gamma - \omega)] \, \left(G_{2\omega\gamma} - G^{(0)}_{2\omega\gamma} \right)- \tfrac{N_f}{N_c} \, \widetilde{Q}_{\omega\gamma} \\
&\hspace*{3.5cm}+ 4 \, [\gamma \, (\gamma - \omega) -1] \, G_{2\omega\gamma} + \tfrac{3}{2} \, \omega\gamma \, [\gamma \, (\gamma - \omega) + 1] \, \left( \widetilde{Q}_{\omega\gamma} - \widetilde{Q}^{(0)}_{\omega\gamma} \right)  \Bigg\},\notag \\
& B_{\omega\gamma} = {\overline B}_{\omega\gamma} = - 2 \, \left(G_{2\omega\gamma} - G^{(0)}_{2\omega\gamma} \right) ,\\
& C_{\omega\gamma} = {\overline C}_{\omega\gamma} = - 2 \, G^{(0)}_{2\omega\gamma} .
\end{align}
\end{subequations}
Since
\begin{subequations}
\begin{align}
    & \widetilde{\Gamma} (\soz,\sto,\eta') = \widetilde{\Gamma}^{(h)}(\soz,\sto,\eta') +\widetilde{\Gamma}^{(p)}(\soz,\sto,\eta') , \\
    &  \overline{\Gamma} (\soz,\sto,\eta') =   \overline{\Gamma}^{(h)}(\soz,\sto,\eta') +  \overline{\Gamma}^{(p)}(\soz,\sto,\eta'), 
\end{align}
\end{subequations}
at this point we have double-Laplace expressions for all the dipole amplitudes, which we collect together here:
\begin{subequations}\label{ch6all}\allowdisplaybreaks
\begin{align}
    \label{ch6all_G2}
    &G_2(\soz,\eta) = \wint \gint e^{\omega(\eta-\soz)}e^{\gamma \soz}\gtwg , \\
    \label{ch6all_Gamma2}
    &\Gamma_2(\soz,\sto,\eta') = \wint \gint \left[e^{\omega(\eta'-\sto)}e^{\gamma\soz}\left(\gtwg - \gtwg^{(0)}\right) + e^{\omega(\eta'-\soz)}e^{\gamma\soz}\gtwg^{(0)} \right] , \\
    \label{ch6all_Gtilde}
    &\widetilde{G}(\soz,\eta) = \wint\gint e^{\omega(\eta-\soz)}e^{\gamma\soz}\left[ \frac{\omega\gamma}{2}\left(G_{2\omega\gamma} - G_{2\omega\gamma}^{(0)}\right) - 2G_{2\omega\gamma}\right] , \\
    \label{ch6all_Q}
    &Q(\soz,\eta) = \wint \gint e^{\omega(\eta-\soz)}e^{\gamma\soz} \, \left[ - \omega\gamma \, \left( \widetilde{Q}_{\omega\gamma} - \widetilde{Q}^{(0)}_{\omega\gamma} \right) - 2G_{2\omega\gamma} \right] , \\
    \label{ch6all_Qtilde}
    &\widetilde{Q}(\soz,\eta) = \wint \gint e^{\omega(\eta-\soz)}e^{\gamma\soz} \, \widetilde{Q}_{\omega\gamma} , \\
    \label{ch6all_Gammatilde}
    &\widetilde{\Gamma}(\soz,\sto,\eta') =  \wint e^{\omega(\eta'-\sto)}\sum_{\alpha,\beta = +,-} e^{\dw^{\alpha\beta}\sto} \widetilde{\Gamma}_\omega^{(\alpha\beta)}(\soz) \\
    &\hspace{2cm} + \wint\gint \bigg[e^{\omega(\eta'-\sto)}e^{\gamma\sto}A_{\omega\gamma} -2e^{\omega(\eta'-\sto)}e^{\gamma\soz}\left(G_{2\omega\gamma} - G^{(0)}_{2\omega\gamma}\right) \notag\\
    &\hspace*{5cm}-2e^{\omega(\eta'-\soz)}e^{\gamma\soz}G^{(0)}_{2\omega\gamma} \bigg] , \notag \\
    \label{ch6all_Gammabar}
    &\overline{\Gamma}(\soz,\sto,\eta') = \wint \, e^{\omega(\eta'-\sto)}\sum_{\alpha,\beta = +,-} e^{\dw^{\alpha\beta}\sto} \frac{1 + \beta \, \sqrt{ 1 - \frac{N_f}{2 \, N_c} }}{N_f / (4 N_c) }  \,  \widetilde{\Gamma}_\omega^{(\alpha\beta)}(\soz) \\
    &\hspace{2cm} + \wint\gint \bigg[ e^{\omega(\eta'-\sto)}e^{\gamma\sto} \, {\overline A}_{\omega\gamma} - 2e^{\omega(\eta'-\sto)}e^{\gamma\soz}\left(G_{2\omega\gamma} - G^{(0)}_{2\omega\gamma}\right) \notag \\
    &\hspace*{5cm}- 2e^{\omega(\eta'-\soz)}e^{\gamma\soz}G^{(0)}_{2\omega\gamma} \bigg] , \notag \\
    &\text{with} \notag \\
    \label{ch6all_deltapms}
    & \delta_{\omega}^{\pm\pm} \equiv \frac{1}{2}\left[\omega \pm \sqrt{\omega^2 \pm 4\sqrt{1-\frac{N_f}{2N_c} } } \right] , \\
    \label{ch6Aeq}
    & A_{\omega\gamma} =  \frac{1}{\left(\gamma-\dw^{++} \right)\left(\gamma-\dw^{+-} \right)\left(\gamma-\dw^{-+} \right)\left(\gamma-\dw^{--} \right)} \\
    &\hspace*{1cm}\times\Bigg\{  \frac{\omega\gamma}{2} \, \left[ 3 - \tfrac{N_f}{2 N_c} - 3 \,  \gamma \, (\gamma - \omega) \right] \, \left(G_{2\omega\gamma} - G^{(0)}_{2\omega\gamma} \right) + \tfrac{3 N_f}{8 N_c} \, \omega \, \gamma \, \left( \widetilde{Q}_{\omega\gamma} - \widetilde{Q}^{(0)}_{\omega\gamma} \right) \notag \\ 
&\hspace*{2cm} + \left[ 4 \, \gamma \, (\gamma - \omega) - 4 + \tfrac{N_f}{N_c} \right] \, G_{2\omega\gamma} + \tfrac{N_f}{2 N_c} \, \left[ \gamma \, (\gamma - \omega) - 1 \right] \, \widetilde{Q}_{\omega\gamma} \Bigg\}, \notag \\
\label{ch6Abar_eq}
& \overline{A}_{\omega\gamma} =  \frac{1}{\left(\gamma-\dw^{++} \right)\left(\gamma-\dw^{+-} \right)\left(\gamma-\dw^{-+} \right)\left(\gamma-\dw^{--} \right)} \\
&\hspace*{1cm}\times\Bigg\{ \omega\gamma \, [2 - \gamma \, (\gamma - \omega)] \, \left(G_{2\omega\gamma} - G^{(0)}_{2\omega\gamma} \right) + 4 \, [\gamma \, (\gamma - \omega) -1] \, G_{2\omega\gamma}   \notag \\
& \hspace*{2cm}+ \tfrac{3}{2} \, \omega\gamma \, [\gamma \, (\gamma - \omega) + 1] \, \left( \widetilde{Q}_{\omega\gamma} - \widetilde{Q}^{(0)}_{\omega\gamma} \right) - \tfrac{N_f}{N_c} \, \widetilde{Q}_{\omega\gamma}  \Bigg\} \notag. 
\end{align}
\end{subequations}
Note that we have used Eqs.~\eqref{ch6gammafourthorder} and \eqref{ch6deltapmpm} to rewrite the denominators of $A_{\omega\gamma}$ and $\overline{A}_{\omega\gamma}$ in Eqs.~\eqref{ch6Aeq} and \eqref{ch6Abar_eq}. This makes it clear that our procedure to solve the partial differential equations in Eqs.~\eqref{ch6secondorderpdes} has introduced additional poles in the integrand. For large $\omega$, we note the following behavior of the functions $\dw^{\alpha\beta}$:
\begin{subequations}\label{ch6polescalingslargew}
    \begin{align}
        &\dw^{++} \sim \omega \, , \\
        &\dw^{+-} \sim \omega \, , \\
        &\dw^{-+} \sim -\sqrt{1-\tfrac{N_f}{2N_c}}\frac{1}{\omega} \, , \\
        &\dw^{--} \sim \sqrt{1-\tfrac{N_f}{2N_c}}\frac{1}{\omega}.
    \end{align}
\end{subequations}
Since $\dw^{++},\dw^{+-} \sim \omega$ for large $\omega$, these poles cannot lie to the left of both the $\omega$ and $\gamma$ contours. This is the same situation encountered in the analytic solution of the large-$N_c$ evolution equations constructed in Ch.~\ref{largeNcsoln.ch}, although there was only one such pole in that solution, while here we have two. Nevertheless we can follow the procedure of Ch.~\ref{largeNcsoln.ch} and declare that the poles at $\gamma = \dw^{++}$ and $\gamma = \dw^{+-}$ lie to the left of the $\omega$-contour but to the right of the $\gamma$-contour. This implies that we are choosing $\text{Re}\,\omega > \text{Re}\,\gamma$ along the integration contours.

Although we have solved the partial differential equations \eqref{ch6secondorderpdes}, these solutions are not yet solutions of the full integral evolution equations~\eqref{ch6Gammabar_IR2} and \eqref{ch6Gammatilde_IR2} from which we obtained those PDEs. We have also yet to satisfy the non-neighbor partners of these equations, Eqs.~\eqref{ch6Q_IR2} and \eqref{ch6Gtilde_IR2}. The next step is to substitute the double-Laplace results from Eqs.~\eqref{ch6all} back into the evolution equations for $\widetilde{\Gamma}$ and $\overline{\Gamma}$ (Eqs.~\eqref{ch6Gammatilde_IR2} and \eqref{ch6Gammabar_IR2}, respectively) in order to obtain the remaining constraints necessary to ensure the full evolution equations are satisfied. Then, since the evolution equations for $\widetilde{G}$ and $Q$ (Eqs.~\eqref{ch6Gtilde_IR2} and \eqref{ch6Q_IR2}) are special cases of those for $\widetilde{\Gamma}$ and $\overline{\Gamma}$, respectively, we can ensure the evolution equations for $\widetilde{G}$  and $Q$ are also satisfied by setting
\begin{subequations}
\begin{align}
    &\widetilde{\Gamma}(\soz,\sto=\soz,\eta') = \widetilde{G}(\soz,\eta') , 
    \\
    &\overline{\Gamma}(\soz,\sto=\soz,\eta') = Q(\soz,\eta').
\end{align}
\end{subequations}
This will ultimately allow us to solve for the unknown functions $\widetilde{\Gamma}_\omega^{(++)}(\soz)$, $\widetilde{\Gamma}_\omega^{(+-)}(\soz)$, $\widetilde{\Gamma}_\omega^{(-+)}(\soz)$, $\widetilde{\Gamma}_\omega^{(--)}(\soz)$, $G_{2\omega\gamma}$, and $\widetilde{Q}_{\omega\gamma}$. We will do this in the next Sections.


\subsection{Obtaining the Remaining Constraints}

We begin with the evolution equation for $\widetilde{\Gamma}(\soz,\sto,\eta')$,  \eq{ch6Gammatilde_IR2}. Substituting all the relevant double-Laplace expressions from Eqs.~\eqref{ch6all}, carrying out all the integrals over $\stt$ and $\eta''$, and performing the forward Laplace transform over $\eta'$ yields    
\begin{align}\label{ch6simplifying2}
    &0 =  e^{- \omega \sto} \sum_{\alpha,\beta = +,-} \widetilde{\Gamma}_\omega^{(\alpha\beta)}(\soz) \, \frac{\dw^{\alpha\beta}-\omega}{\omega}e^{\dw^{\alpha\beta}\soz} \\
    &\hspace{.3cm} + e^{- \omega \, \sto} \gint \, e^{\gamma\soz} \, \left[ A_{\omega\gamma} \frac{\gamma-\omega}{\omega} + \frac{N_f}{4N_c} \,  \left( \widetilde{Q}_{\omega\gamma} - \widetilde{Q}^{(0)}_{\omega\gamma} \right) + 2 \, \left(G_{2\omega\gamma} - G^{(0)}_{2\omega\gamma}\right) \right] \notag \\
    &\hspace{.3cm} + \int \frac{d \omega'}{2\pi i } \sum_{\alpha,\beta = +,-}\widetilde{\Gamma}_{\omega'}^{(\alpha\beta)}(\soz) \left[ \frac{1}{\delta_{\omega'}^{\alpha\beta} - \omega} \,  +  
   \frac{1}{\omega} \frac{\omega'-\delta_{\omega'}^{\alpha\beta}}{\omega'}\, e^{\delta_{\omega'}^{\alpha\beta}\soz} \right]\notag\\
   &\hspace{.3cm}
   + e^{- \omega \soz} \gint  e^{\gamma\soz} \, \left[ \widetilde{G}^{(0)}_{\omega\gamma} + 2 \, G^{(0)}_{2\omega\gamma} \right]. \notag  
\end{align}
Along the way, we have dropped several terms which are zero, as can be shown by closing either the $\omega$ or the $\gamma$ integration contour to the right. Now we observe that two of the terms in \eq{ch6simplifying2} have the same $\sto$ dependence, $e^{-\omega\sto}$, whereas the other two terms are independent of $\sto$. Since the equation is valid for any $\sto$, we conclude that both sets of terms must separately equal zero and arrive at the following two constraints:
\begin{subequations}\label{ch6conditions1}
\begin{align}\label{ch6condition1a}
& \gint \, e^{\gamma\soz} \, \left[ A_{\omega\gamma} \frac{\gamma-\omega}{\omega} + \frac{N_f}{4N_c} \,  \left( \widetilde{Q}_{\omega\gamma} - \widetilde{Q}^{(0)}_{\omega\gamma} \right) + 2 \, \left(G_{2\omega\gamma} - G^{(0)}_{2\omega\gamma}\right) \right] \\
&\hspace*{1cm}= \sum_{\alpha,\beta = +,-} \widetilde{\Gamma}_\omega^{(\alpha\beta)}(\soz) \, \frac{\omega-\dw^{\alpha\beta}}{\omega} e^{\dw^{\alpha\beta}\soz}  \notag\\
&\text{and} \notag \\
\label{ch6condition1b}
& 0 = \int \frac{d \omega'}{2\pi i } \sum_{\alpha,\beta = +,-}\widetilde{\Gamma}_{\omega'}^{(\alpha\beta)}(\soz) \left[ \frac{1}{\delta_{\omega'}^{\alpha\beta} - \omega} \,  +  
   \frac{1}{\omega} \frac{\omega'-\delta_{\omega'}^{\alpha\beta}}{\omega'}\, e^{\delta_{\omega'}^{\alpha\beta}\soz} \right] \\
   &\hspace*{1cm}
   + e^{- \omega \soz} \gint  e^{\gamma\soz} \, \left[ \widetilde{G}^{(0)}_{\omega\gamma} + 2 \, G^{(0)}_{2\omega\gamma} \right]. \notag
\end{align}
\end{subequations}
We can obtain a third constraint and satisfy the evolution equation \eq{ch6Gtilde_IR2} by requiring that 
\begin{align}\label{ch6thirdconstraint}
    \widetilde{\Gamma}(\soz,\sto=\soz,\eta') = \widetilde{G}(\soz,\eta').
\end{align}
Using our double-Laplace expressions \eqref{ch6all_Gammatilde} and \eqref{ch6all_Gtilde} and applying the inverse transform over $\eta'-\soz$ (here treating $\eta'-\soz$ and $\soz$ as independent variables), \eq{ch6thirdconstraint} gives
\begin{align}\label{ch6simplifying3}
    \sum_{\alpha,\beta=+,-} e^{\dw^{\alpha\beta}\soz} \widetilde{\Gamma}_{\omega}^{(\alpha\beta)}(\soz) = \gint e^{\gamma\soz}\left[\frac{\omega\gamma}{2}\left(G_{2\omega\gamma} - G^{(0)}_{2\omega\gamma} \right) - A_{\omega\gamma} \right].
\end{align}
In Eqs.~\eqref{ch6conditions1} and \eqref{ch6simplifying3} we have thus obtained the three constraints necessary to fully satisfy the evolution equations for $\widetilde{\Gamma}$ and $\widetilde{G}$. Next we can apply this same procedure to the evolution equations for $\overline{\Gamma}$ and $Q$. 

Beginning with \eq{ch6Gammabar_IR2} we substitute in all the relevant double-Laplace expressions from Eqs.~\eqref{ch6all}, carry out the integrals over $\stt$ and $\eta'$, perform the forward Laplace transform over $\eta'$, and again drop several terms which can be shown to be zero. The result is
\begin{align}\label{ch6simplifying5}
    &0 = -e^{-\omega\sto}\sum_{\alpha,\beta = +,-}e^{\dw^{\alpha\beta}\soz}\widetilde{\Gamma}^{(\alpha\beta)}_\omega(\soz) \frac{\omega-\dw^{\alpha\beta}}{\omega}\frac{4N_c}{N_f}\left(1+\beta\sqrt{1-\tfrac{N_f}{2N_c}}\right) \\
    &\hspace{0.4cm}  + e^{-\omega\sto}\gint e^{\gamma\soz}\left[ {\overline A}_{\omega\gamma} \frac{\gamma-\omega}{\omega} - \frac{1}{2} \,  \left( \widetilde{Q}_{\omega\gamma} - \widetilde{Q}^{(0)}_{\omega\gamma} \right) + 2 \, \left(G_{2\omega\gamma} - G^{(0)}_{2\omega\gamma}\right) \right] \notag \\
    &\hspace{0.4cm} + e^{- \omega \soz} \gint  e^{\gamma\soz} \, \left[ Q^{(0)}_{\omega\gamma} + 2 \, G^{(0)}_{2\omega\gamma} \right] \notag\\
    &\hspace{0.4cm}- \int \frac{d \omega'}{2\pi i } \sum_{\alpha,\beta = +,-} \frac{1}{\omega - \delta_{\omega'}^{\alpha\beta}} \, \widetilde{\Gamma}_{\omega'}^{(\alpha\beta)}(\soz) \, \frac{4N_c}{N_f}\left(1 + \beta \, \sqrt{ 1 - \frac{N_f}{2 \, N_c} }\right)  \notag \\
    &\hspace{0.4cm}  +  
   \frac{1}{\omega} \int \frac{d \omega'}{2\pi i } \sum_{\alpha,\beta = +,-} \widetilde{\Gamma}_{\omega'}^{(\alpha\beta)}(\soz) \frac{\omega'-\delta_{\omega'}^{\alpha\beta}}{\omega'}\, \frac{4N_c}{N_f}\left(1 + \beta \, \sqrt{ 1 - \frac{N_f}{2 \, N_c} }\right) \, e^{\delta_{\omega'}^{\alpha\beta}\soz} \notag
\end{align}
Just as with \eq{ch6simplifying2}, two of the terms here share the same $\sto$ dependence, $e^{-\omega\sto}$, whereas the other terms are independent of $\sto$, giving us two separate constraints,
\begin{subequations}\label{ch6conditions2}
    \begin{align}\label{ch6conditions2a}
    &\sum_{\alpha,\beta = +,-}e^{\dw^{\alpha\beta}\soz}\widetilde{\Gamma}^{(\alpha\beta)}_\omega(\soz) \frac{\omega-\dw^{\alpha\beta}}{\omega}\frac{4N_c}{N_f}\left(1+\beta\sqrt{1-\tfrac{N_f}{2N_c}}\right) \\
    &\hspace{3cm} = \gint e^{\gamma\soz}\left[ {\overline A}_{\omega\gamma} \frac{\gamma-\omega}{\omega} - \frac{1}{2} \,  \left( \widetilde{Q}_{\omega\gamma} - \widetilde{Q}^{(0)}_{\omega\gamma} \right) + 2 \, \left(G_{2\omega\gamma} - G^{(0)}_{2\omega\gamma}\right) \right]\notag \\
    &\text{and} \notag \\
    \label{ch6conditions2b}
    &0 = e^{- \omega \soz} \gint  e^{\gamma\soz} \, \left[ Q^{(0)}_{\omega\gamma} + 2 \, G^{(0)}_{2\omega\gamma} \right] \\
    &\hspace{0.4cm}- \int \frac{d \omega'}{2\pi i } \sum_{\alpha,\beta = +,-} \frac{1}{\omega - \delta_{\omega'}^{\alpha\beta}} \, \widetilde{\Gamma}_{\omega'}^{(\alpha\beta)}(\soz) \, \frac{4N_c}{N_f}\left(1 + \beta \, \sqrt{ 1 - \frac{N_f}{2 \, N_c} }\right) \notag \\
    &\hspace{0.4cm}  +  
   \frac{1}{\omega} \int \frac{d \omega'}{2\pi i } \sum_{\alpha,\beta = +,-} \widetilde{\Gamma}_{\omega'}^{(\alpha\beta)}(\soz) \frac{\omega'-\delta_{\omega'}^{\alpha\beta}}{\omega'}\, \frac{4N_c}{N_f}\left(1 + \beta \, \sqrt{ 1 - \frac{N_f}{2 \, N_c} }\right) \, e^{\delta_{\omega'}^{\alpha\beta}\soz} .\notag
    \end{align}
\end{subequations}
For one final constraint, and to satisfy the evolution equation for $Q$ (\eq{ch6Q_IR2}), we require
\begin{align}\label{ch6GammabartoQconstraint}
    \overline{\Gamma}(\soz,\sto=\soz,\eta') = Q(\soz,\eta').
\end{align}
Using the double-Laplace expressions from Eqs.~\eqref{ch6all} and performing the forward transform over $\eta'-\soz$ (again treating $\eta'-\soz$ and $\soz$ as independent variables) gives
\begin{align}\label{ch6condition2c}
\gint e^{\gamma \soz} \, \left[ - {\overline A}_{\omega\gamma} - \omega\gamma \, \left(\widetilde{Q}_{\omega\gamma} - \widetilde{Q}^{(0)}_{\omega\gamma}  \right) \right] = \sum_{\alpha,\beta = +,-} \widetilde{\Gamma}_\omega^{(\alpha\beta)}(\soz) \, \frac{1 + \beta \, \sqrt{ 1 - \frac{N_f}{2 \, N_c} }}{N_f / (4 N_c) } \, e^{\dw^{\alpha\beta}\soz}.
\end{align}

To summarize we note that in Eqs.~\eqref{ch6conditions1}, \eqref{ch6simplifying3}, \eqref{ch6conditions2}, and \eqref{ch6condition2c} we have a total of six constraints that must be satisfied by our double-Laplace constructions in order to ensure that all the evolution equations in \eqref{ch6evoleqs_IR_1} are satisfied. What remains is to solve those constraints for the currently unknown functions $\widetilde{\Gamma}^{(\alpha\beta)}_{\omega}(\soz)$, $G_{2\omega\gamma}$, and $\widetilde{Q}_{\omega\gamma}$.


\subsection{Solving the Constraints}

\subsubsection{Four Straightforward Constraints}
The four equations~\eqref{ch6condition1a}, \eqref{ch6simplifying3}, \eqref{ch6conditions2a}, \eqref{ch6condition2c} constitute a system of equations we can straightforwardly solve for all four of the $\widetilde{\Gamma}^{(\alpha\beta)}_{\omega}(\soz)$ functions as integrals over $\gamma$. First we rewrite the common structure shared by each as
\begin{align}\label{ch6Gammatildestructure}
    \widetilde{\Gamma}_{\omega}^{(\alpha\beta)}(\soz) = e^{-\dw^{\alpha\beta}\soz}\gint e^{\gamma\soz} \widetilde{\Gamma}_{\omega\gamma}^{(\alpha\beta)}
\end{align}
(note the distinction of $\widetilde{\Gamma}^{(\alpha\beta)}_{\omega\gamma}$ on the right-hand side written now as a function of $\omega$ and $\gamma$ and not of $\soz$).
Then we undo the inverse Laplace transform over $\gamma$ in all terms of Eqs.~\eqref{ch6condition1a}, \eqref{ch6simplifying3}, \eqref{ch6conditions2a}, \eqref{ch6condition2c}, solving the resulting linear system of equations algebraically for the functions $\widetilde{\Gamma}^{(\alpha\beta)}_{\omega\gamma}$. The results can be written compactly as 
\begin{align}\label{ch6Gammatildescompact}
& \widetilde{\Gamma}_{\omega\gamma}^{(\alpha\beta)} = \frac{\beta}{8 (\dw^{\alpha\beta} - \dw^{-\alpha, \beta}) \, \sqrt{4 - 2 n}} \, \Big\{ \left[ \, {\overline A}_{\omega\gamma} 2 n - 4 A_{\omega\gamma} (2- \beta \sqrt{4 - 2 n}) \right] (\dw^{-\alpha, \beta} - \gamma) \\ 
&\hspace*{4.5cm} + n \omega  \left( \widetilde{Q}_{\omega\gamma} - \widetilde{Q}^{(0)}_{\omega\gamma} \right) (3 - \beta \sqrt{4 - 2 n} - 2 \gamma \dw^{\alpha, \beta}  ) \notag \\
&\hspace*{4.5cm}+ 2 \omega \left(G_{2\omega\gamma} - G^{(0)}_{2\omega\gamma}\right) \left[ (2- \beta \sqrt{4 - 2 n}) (4 - \gamma \dw^{\alpha, \beta}  ) - 2 n  \right] \Big\}. \notag 
\end{align}
In writing \eq{ch6Gammatildescompact} we have defined 
\begin{align}\label{ch6littlen}
n \equiv \frac{N_f}{N_c}
\end{align}
and have also used the fact that 
\begin{align}\label{ch6deltasumrule}
\dw^{\alpha\beta} + \dw^{-\alpha, \beta} = \omega.
\end{align}
Combining Eqs.~\eqref{ch6Gammatildestructure} and \eqref{ch6Gammatildescompact} we have
\begin{align}\label{ch6Gamma_final}
  &  \widetilde{\Gamma}_{\omega}^{(\alpha\beta)}(\soz) = e^{-\dw^{\alpha\beta}\soz}\gint e^{\gamma\soz} \frac{\beta}{8 (\dw^{\alpha\beta} - \dw^{-\alpha, \beta}) \, \sqrt{4 - 2 n}} \\
  &\hspace{1cm}\times\Big\{ \left[ \, {\overline A}_{\omega\gamma} 2 n - 4 A_{\omega\gamma} (2- \beta \sqrt{4 - 2 n}) \right] (\dw^{-\alpha, \beta} - \gamma) \notag \\ 
&\hspace{2cm} + n \omega  \left( \widetilde{Q}_{\omega\gamma} - \widetilde{Q}^{(0)}_{\omega\gamma} \right) (3 - \beta \sqrt{4 - 2 n} - 2 \gamma \dw^{\alpha, \beta}  ) \notag\\
&\hspace{2cm}+ 2 \omega \left(G_{2\omega\gamma} - G^{(0)}_{2\omega\gamma}\right) \left[ (2- \beta \sqrt{4 - 2 n}) (4 - \gamma \dw^{\alpha, \beta}  ) - 2 n   \right] \Big\} \notag .
\end{align}


\subsubsection{Two Remaining Constraints}

Two constraints remain to be satisfied in Eqs.~\eqref{ch6condition1b} and \eqref{ch6conditions2b}. Solving these will allow us to obtain expressions for our final two remaining unknowns, the double-Laplace images $G_{2\omega\gamma}$ and $\widetilde{Q}_{\omega\gamma}$ and will complete our solution of the evolution equations \eqref{ch6evoleqs_IR_1}. We rewrite the two constraints below:
\begin{subequations}\label{ch6conditions_b}
\begin{align}
& 0 = e^{- \omega \soz} \gint  e^{\gamma\soz} \, \left[ \widetilde{G}^{(0)}_{\omega\gamma} + 2 \, G^{(0)}_{2\omega\gamma} \right] - \int \frac{d \omega'}{2\pi i } \sum_{\alpha,\beta = +,-} \frac{1}{\omega - \delta_{\omega'}^{\alpha\beta}} \,  \widetilde{\Gamma}_{\omega'}^{(\alpha\beta)}(\soz)  \\
    &\hspace{1cm} +  
   \frac{1}{\omega} \int \frac{d \omega'}{2\pi i } \sum_{\alpha,\beta = +,-} \widetilde{\Gamma}_{\omega'}^{(\alpha\beta)}(\soz) \frac{\omega'-\delta_{\omega'}^{\alpha\beta}}{\omega'}\, e^{\delta_{\omega'}^{\alpha\beta}\soz} , \notag \\
   & 0 = e^{- \omega \soz} \gint  e^{\gamma\soz} \, \left[ Q^{(0)}_{\omega\gamma} + 2 \, G^{(0)}_{2\omega\gamma} \right] \\
   &\hspace{1cm}- \int \frac{d \omega'}{2\pi i } \sum_{\alpha,\beta = +,-} \frac{1}{\omega - \delta_{\omega'}^{\alpha\beta}} \,  \widetilde{\Gamma}_{\omega'}^{(\alpha\beta)}(\soz) \, \frac{1 + \beta \, \sqrt{ 1 - \frac{n}{2}}}{n / 4 }  \notag \\
    &\hspace{1cm} +  
   \frac{1}{\omega} \int \frac{d \omega'}{2\pi i } \sum_{\alpha,\beta = +,-} \widetilde{\Gamma}_{\omega'}^{(\alpha\beta)}(\soz) \frac{\omega'-\delta_{\omega'}^{\alpha\beta}}{\omega'}\, \frac{1 + \beta \, \sqrt{ 1 - \frac{n}{2} }}{n/ 4  }  \, e^{\delta_{\omega'}^{\alpha\beta}\soz} . \notag
\end{align}
\end{subequations}
It is straightforward to show using \eq{ch6Gamma_final} that
\begin{align}\label{ch6termscaling}
e^{\dw^{+\beta}\soz} \,  \widetilde{\Gamma}_{\omega}^{(+ \beta)}(\soz)  \to 0, \ \ \  e^{\dw^{-\beta}\soz} \, \omega \,  \widetilde{\Gamma}_{\omega}^{(- \beta)}(\soz)  \to 0, \ \ \ \textrm{when} \ \ \ \omega \to \infty.  
\end{align}
Then we can close the $\omega'$ contour to the right in the last term of each of the equations \eqref{ch6conditions_b} and obtain zero. What remains of Eqs.~\eqref{ch6conditions_b} can be rewritten using the identity in \eq{ch6deltasumrule} along with another identity relating the $\dw^{\alpha\beta}$,
\begin{align}\label{ch6deltaproductrule}
    \dw^{\alpha\beta}\dw^{-\alpha,\beta} = -\beta\sqrt{1-\frac{n}{2}} ,
\end{align}
which can be straightforwardly shown from the definition in \eq{ch6deltapmpm}. We obtain
\begin{subequations}\label{ch6conditions_b3}
\begin{align}
&  e^{- \omega \soz} \gint  e^{\gamma\soz} \, \left[ \widetilde{G}^{(0)}_{\omega\gamma} + 2 \, G^{(0)}_{2\omega\gamma} \right] \\
&\hspace{2.5cm}= -  \frac{1}{\omega} \int \frac{d \omega'}{2\pi i } \sum_{\alpha,\beta = +,-}  \frac{\omega - \delta_{\omega'}^{-\alpha , \beta}}{\omega' - \left( \omega - \frac{\beta}{\omega} \, \sqrt{ 1 - \frac{n}{2}} \right) } \, \widetilde{\Gamma}_{\omega'}^{(\alpha\beta)}(\soz) , \notag \\
   & e^{- \omega \soz} \gint  e^{\gamma\soz} \, \left[ Q^{(0)}_{\omega\gamma} + 2 \, G^{(0)}_{2\omega\gamma} \right] \\
   &\hspace{2.5cm}= -  \frac{1}{\omega} \int \frac{d \omega'}{2\pi i } \sum_{\alpha,\beta = +,-} \frac{\omega - \delta_{\omega'}^{-\alpha , \beta}}{\omega' - \left( \omega - \frac{\beta}{\omega} \, \sqrt{ 1 - \frac{n}{2}} \right) }  \,  \widetilde{\Gamma}_{\omega'}^{(\alpha\beta)}(\soz) \, \frac{1 + \beta \, \sqrt{ 1 - \frac{n}{2} }}{n/ 4 } .\notag
\end{align}
\end{subequations}
On the right hand side of each of Eqs.~\eqref{ch6conditions_b3}, we can close the contour to the right, picking up the poles at $\omega' = \omega - \tfrac{\beta}{\omega}\sqrt{1-\tfrac{n}{2}}$. One can show using \eq{ch6deltapmpm} that
\begin{align}\label{ch6deltasatpole}
\delta_{\omega - \frac{\beta}{\omega} \, \sqrt{ 1 - \frac{n}{2}} }^{\alpha , \beta} = \thalf \, \left[ \omega \, (1+\alpha) - (1-\alpha) \, \frac{\beta}{\omega} \, \sqrt{ 1 - \frac{n}{2}}  \right],
\end{align}
which subsequently gives
\begin{align}\label{ch6deltasatpole2}
\delta_{\omega - \frac{\beta}{\omega} \, \sqrt{ 1 - \frac{n}{2}} }^{+ , \beta} = \omega, \ \ \ \delta_{\omega - \frac{\beta}{\omega} \, \sqrt{ 1 - \frac{n}{2 }} }^{- , \beta} = - \frac{\beta}{\omega} \, \sqrt{ 1 - \frac{n}{2}} . 
\end{align}
Then in view of the factor $\omega - \delta_{\omega'}^{-\alpha,\beta}$ in both of Eqs.~\eqref{ch6conditions_b3}, we conclude that the residues of the $\omega' = \omega - \tfrac{\beta}{\omega}\sqrt{1-\tfrac{n}{2}}$ poles for $\alpha = -$ are zero, leaving only the residues from the $\alpha = +$ contribution. Eqs.~\eqref{ch6conditions_b3} thus become
\begin{subequations}\label{ch6conditions_b4}
\begin{align}
&  e^{- \omega \soz} \gint  e^{\gamma\soz} \, \left[ \widetilde{G}^{(0)}_{\omega\gamma} + 2 \, G^{(0)}_{2\omega\gamma} \right] = \sum_{\beta = +,-} \left(1 +  \frac{\beta}{\omega^2} \, \sqrt{ 1 - \frac{n}{2}}  \right) \,   \widetilde{\Gamma}_{\omega - \frac{\beta}{\omega} \, \sqrt{ 1 - \frac{n}{2}}}^{(+\beta)}(\soz) ,  \\
   & e^{- \omega \soz} \gint  e^{\gamma\soz} \, \left[ Q^{(0)}_{\omega\gamma} + 2 \, G^{(0)}_{2\omega\gamma} \right] \\
   &\hspace{4cm}= \sum_{\beta = +,-} \left(1 +  \frac{\beta}{\omega^2} \, \sqrt{ 1 - \frac{n}{2}}  \right) \,   \widetilde{\Gamma}_{\omega - \frac{\beta}{\omega} \, \sqrt{ 1 - \frac{n}{2}}}^{(+\beta)}(\soz)\frac{1 + \beta \, \sqrt{ 1 - \frac{n}{2} }}{n / 4 } . \notag
\end{align}
\end{subequations}

Next we recall \eq{ch6Gammatildestructure}, which, along with the first equality in \eq{ch6deltasatpole2} tells us that
\begin{align}
    \widetilde{\Gamma}_{\omega-\tfrac{\beta}{\omega}\sqrt{1-\tfrac{n}{2}}}^{(+\beta)}(\soz) = e^{-\omega\soz}\gint e^{\gamma\soz} \widetilde{\Gamma}_{\omega-\tfrac{\beta}{\omega}\sqrt{1-\tfrac{n}{2}},\gamma}^{(+\beta)}.
\end{align}
Using this in Eqs.~\eqref{ch6conditions_b4} and writing out the sum over $\beta$ explicitly, we have
\begin{subequations}\label{ch6conditions_b6}
\begin{align}
&  \gint  e^{\gamma\soz} \, \left[ \widetilde{G}^{(0)}_{\omega\gamma} + 2 \, G^{(0)}_{2\omega\gamma} \right] = \left(1 +  \frac{1}{\omega^2} \, \sqrt{ 1 - \frac{n}{2}}  \right) \,   \gint e^{\gamma\soz} \widetilde{\Gamma}_{\omega - \frac{1}{\omega} \, \sqrt{ 1 - \frac{n}{2}},\gamma}^{(++)} \\
&\hspace{5.1cm} + \left(1 -  \frac{1}{\omega^2} \, \sqrt{ 1 - \frac{n}{2}}  \right) \,   \gint e^{\gamma\soz}\widetilde{\Gamma}_{\omega + \frac{1}{\omega} \, \sqrt{ 1 - \frac{n}{2}},\gamma}^{(+-)} \notag\,,  \\
   &\gint  e^{\gamma\soz} \, \left[ Q^{(0)}_{\omega\gamma} + 2 \, G^{(0)}_{2\omega\gamma} \right] \\
   &\hspace{4cm}=  \left(1 +  \frac{1}{\omega^2} \, \sqrt{ 1 - \frac{n}{2}}  \right) \,   \gint e^{\gamma\soz}\widetilde{\Gamma}_{\omega - \frac{1}{\omega} \, \sqrt{ 1 - \frac{n}{2}},\gamma}^{(++)} \, \frac{1 + \sqrt{ 1 - \frac{n}{2} }}{n / 4 } \notag \\
   & \hspace{4cm} +  \left(1 -  \frac{1}{\omega^2} \, \sqrt{ 1 - \frac{n}{2}}  \right) \,   \gint e^{\gamma\soz}\widetilde{\Gamma}_{\omega + \frac{1}{\omega} \, \sqrt{ 1 - \frac{n}{2}},\gamma}^{(+-)} \, \frac{1 - \sqrt{ 1 - \frac{n}{2} }}{n / 4 }.\notag
\end{align}
\end{subequations}
Equations~\eqref{ch6conditions_b6} can be written more compactly as a single equation,
\begin{align}\label{ch6conditions_b8}
& \gint  e^{\gamma\soz} \, \left\{\frac{n}{4 } \left[ Q^{(0)}_{\omega\gamma} + 2 \, G^{(0)}_{2\omega\gamma} \right] - \left[1 - \beta \, \sqrt{ 1 - \frac{n}{2}}  \right]  \left[ \widetilde{G}^{(0)}_{\omega\gamma} + 2 \, G^{(0)}_{2\omega\gamma} \right] \right\} \\ 
& \hspace{4cm} \notag = \beta \, \left(1 +  \frac{\beta}{\omega^2} \, \sqrt{ 1 - \frac{n}{2}}  \right) \,   \gint e^{\gamma\soz}\widetilde{\Gamma}_{\omega - \frac{\beta}{\omega} \, \sqrt{ 1 - \frac{n}{2}},\gamma}^{(+\beta)} \, 2 \, \sqrt{ 1 - \frac{n}{2}} .
\end{align}

Recalling \eq{ch6deltasatpole2}, we can write this is as
\begin{align}\label{ch6conditions_b9}
& \gint  e^{\gamma\soz} \, \Bigg\{\frac{n}{4} \left[ Q^{(0)}_{\delta_{\omega - \frac{\beta}{\omega} \, \sqrt{ 1 - \frac{n}{2}} }^{+ , \beta} \ \gamma} + 2 \, G^{(0)}_{2 \ \delta_{\omega - \frac{\beta}{\omega} \, \sqrt{ 1 - \frac{n}{2}} }^{+ , \beta} \ \gamma} \right] \\
&\hspace{2.5cm} - \left[1 - \beta \, \sqrt{ 1 - \frac{n}{2}}  \right] \times  \left[ \widetilde{G}^{(0)}_{\delta_{\omega - \frac{\beta}{\omega} \, \sqrt{ 1 - \frac{n}{2}} }^{+ , \beta} \ \gamma} + 2 \, G^{(0)}_{2 \ \delta_{\omega - \frac{\beta}{\omega} \, \sqrt{ 1 - \frac{n}{2}} }^{+ , \beta} \ \gamma} \right] \Bigg\} \notag \\
&\hspace{0.5cm}= \beta \, \left(1 +  \frac{\beta}{\left[ \delta_{\omega - \frac{\beta}{\omega} \, \sqrt{ 1 - \frac{n}{2}} }^{+ , \beta} \right]^2} \, \sqrt{ 1 - \frac{n}{2}}  \right) \,   \gint e^{\gamma\soz}\widetilde{\Gamma}_{\omega - \frac{\beta}{\omega} \, \sqrt{ 1 - \frac{n}{2}},\gamma}^{(+\beta)} \, 2 \, \sqrt{ 1 - \frac{n}{2 }} , \notag
\end{align}
which, upon the substitution 
\begin{align}
\omega - \frac{\beta}{\omega} \, \sqrt{ 1 - \frac{n}{2}} \to \omega \, ,
\end{align}
yields
\begin{align}\label{ch6conditions_b11}
& \gint  e^{\gamma\soz} \, \left\{\frac{n}{4} \left[ Q^{(0)}_{\delta_{\omega }^{+ , \beta} \ \gamma} + 2 \, G^{(0)}_{2 \ \delta_{\omega  }^{+ , \beta} \ \gamma} \right] - \left[1 - \beta \, \sqrt{ 1 - \frac{n}{2}}  \right] \,  \left[ \widetilde{G}^{(0)}_{\delta_{\omega  }^{+ , \beta} \ \gamma} + 2 \, G^{(0)}_{2 \ \delta_{\omega }^{+ , \beta} \ \gamma} \right] \right\} \\
& \hspace{4cm}  = \beta \, \left(1 +  \frac{\beta}{\left[ \delta_{\omega }^{+ , \beta} \right]^2} \, \sqrt{ 1 - \frac{n}{2}}  \right) \, 2 \, \sqrt{ 1 - \frac{n}{2}} \,  \gint  e^{\gamma\soz} \, \widetilde{\Gamma}^{+ , \beta}_{\omega \gamma}  . \notag
\end{align}

We begin to invert the remaining inverse Laplace transform by writing
\begin{align}\label{ch6conditions_b12}
& \frac{n}{4} \left[ Q^{(0)}_{\delta_{\omega }^{+ , \beta} \ \gamma} + 2 \, G^{(0)}_{2 \ \delta_{\omega  }^{+ , \beta} \ \gamma} \right] - \left[1 - \beta \, \sqrt{ 1 - \frac{n}{2}}  \right] \,  \left[ \widetilde{G}^{(0)}_{\delta_{\omega  }^{+ , \beta} \ \gamma} + 2 \, G^{(0)}_{2 \ \delta_{\omega }^{+ , \beta} \ \gamma} \right]  \\
& \hspace{4cm}  = \beta \, \left(1 +  \frac{\beta}{\left[ \delta_{\omega }^{+ , \beta} \right]^2} \, \sqrt{ 1 - \frac{n}{2}}  \right) \, 2 \, \sqrt{ 1 - \frac{n}{2}} \,  \int \frac{d \gamma'}{2 \pi i} \, \frac{1}{\gamma - \gamma'} \, \widetilde{\Gamma}^{+ , \beta}_{\omega \gamma'}  . \notag
\end{align}
Next, we would like to complete the inversion of the  Laplace transform on the right hand side of \eq{ch6conditions_b12}. To do so, we must carefully recall the structure of $\widetilde{\Gamma}^{+,\beta}_{\omega\gamma}$. \eq{ch6Gammatildescompact} tells us that $\widetilde{\Gamma}^{+,\beta}_{\omega\gamma}$ depends on the functions $A_{\omega\gamma}$ and $\overline{A}_{\omega\gamma}$. Defined in Eqs.~\eqref{ch6Aeq} and \eqref{ch6Abar_eq}, these functions have poles at $\gamma=\dw^{++}$ and $\gamma=\dw^{+-}$. As discussed in the text following Eqs.~\eqref{ch6all}, although these poles lie to the left of the $\omega$-contour, they lie to the right of the $\gamma'$-contour. Then closing the $\gamma'$-contour to the right in \eq{ch6conditions_b12} requires us to pick up three poles: $\gamma'=\gamma$, $\gamma' = \dw^{++}$, and $\gamma' = \dw^{+-}$. Doing so, we obtain 
\begin{align}\label{ch6conditions_b13}
& \frac{n}{4} \left[ Q^{(0)}_{\delta_{\omega }^{+ , \beta} \ \gamma} + 2 \, G^{(0)}_{2 \ \delta_{\omega  }^{+ , \beta} \ \gamma} \right] - \left[1 - \beta \, \sqrt{ 1 - \frac{n}{2}}  \right] \,  \left[ \widetilde{G}^{(0)}_{\delta_{\omega  }^{+ , \beta} \ \gamma} + 2 \, G^{(0)}_{2 \ \delta_{\omega }^{+ , \beta} \ \gamma} \right]  \\
&  =  \frac{1}{8 (\dw^{+\beta} - \dw^{-, \beta}) } \,  \left(1 +  \frac{\beta}{\left[ \delta_{\omega }^{+ , \beta} \right]^2} \, \sqrt{ 1 - \frac{n}{2}}  \right) \notag \\
&\times\Bigg\{ \left[ \, {\overline A}_{\omega\gamma} 2 n - 4 A_{\omega\gamma} (2- \beta \sqrt{4 - 2 n}) \right] (\dw^{-, \beta} - \gamma) \notag  \\ 
&\hspace{1cm} + n \omega  \left( \widetilde{Q}_{\omega\gamma} - \widetilde{Q}^{(0)}_{\omega\gamma} \right) (3 - \beta \sqrt{4 - 2 n} - 2 \gamma \dw^{+, \beta}  ) \notag\\
&\hspace{1cm}+ 2 \omega \left(G_{2\omega\gamma} - G^{(0)}_{2\omega\gamma}\right) \left[ (2- \beta \sqrt{4 - 2 n}) (4 - \gamma \dw^{+, \beta}  ) - 2 n   \right] \notag \\ 
&\hspace{1cm} - \frac{\dw^{-, \beta} - \dw^{++}}{\gamma - \dw^{++}} \lim_{\gamma' \to \dw^{++}} \Big[ (\gamma' -  \dw^{++}) \left[ \, {\overline A}_{\omega\gamma'} 2 n - 4 A_{\omega\gamma'} (2- \beta \sqrt{4 - 2 n}) \right]  \Big] \notag \\ 
&\hspace{1cm} - \frac{\dw^{-, \beta} - \dw^{+-}}{\gamma - \dw^{+-}} \lim_{\gamma' \to \dw^{+-}} \Big[ (\gamma' -  \dw^{+-}) \left[ \, {\overline A}_{\omega\gamma'} 2 n - 4 A_{\omega\gamma'} (2- \beta \sqrt{4 - 2 n}) \right]  \Big]  \Bigg\} . \notag
\end{align}
Explicitly substituting $A_{\omega\gamma}$ and $\overline{A}_{\omega\gamma}$ from Eqs.~\eqref{ch6Aeq} and \eqref{ch6Abar_eq}, respectively, into \eq{ch6conditions_b13}, after some lengthy but straightforward algebra, we can evaluate the residues (limits) in \eq{ch6conditions_b13} and recast these two constraints (the first for $\beta = +$ and the second for $\beta = -$) as
\begin{subequations}\label{ch6ppandpmcombo6}
\begin{align}
    \label{ch6ppcombo6}
    &Q^{(0)}_{\dw^{++}\gamma} + 2G^{(0)}_{2\dw^{++}\gamma} - \frac{4}{n}\left(1-\sqrt{1-\frac{n}{2}}\right) \left( \widetilde{G}^{(0)}_{\dw^{++}\gamma} + 2G^{(0)}_{2\dw^{++}\gamma}\right) \\
    & = \frac{4}{n}\frac{1}{\gamma-\dw^{++}} \Bigg\{\frac{\omega}{4}\left(-2+\sqrt{4-2n}\right) \Bigg[G_{2\omega\gamma}\left(\gamma-r_1^{++}\right)\left(\gamma-r_1^{-+}\right) \notag \\
    &\hspace{6cm}- G^{(0)}_{2\omega\gamma}\left(\left(\gamma-\dw^{++}\right)\left(\gamma-\dw^{-+}\right) + 2 - \frac{1}{2}\sqrt{4-2n}\right)\Bigg] \notag\\
    &\hspace{2.4cm} -\frac{n}{4} \, \omega \, \Bigg[ \widetilde{Q}_{\omega\gamma}\left(\gamma-r_2^{++}\right)\left(\gamma-r_2^{-+}\right) - \widetilde{Q}^{(0)}_{\omega\gamma}\left(\left(\gamma-\dw^{++}\right)\left(\gamma-\dw^{-+}\right) + \frac{3}{2}\right)\Bigg] \notag \\
    &\hspace{2.4cm}- \left(\gamma\rightarrow\dw^{++}\right) \Bigg\} ,\notag \\  
    \label{ch6pmcombo6}
    &Q^{(0)}_{\dw^{+-}\gamma} + 2G^{(0)}_{2\dw^{+-}\gamma} - \frac{4}{n}\left(1+\sqrt{1-\frac{n}{2}}\right) \left( \widetilde{G}^{(0)}_{\dw^{+-}\gamma} + 2G^{(0)}_{2\dw^{+-}\gamma}\right) \\
    & = -\frac{4}{n}\frac{1}{\gamma-\dw^{+-}} \Bigg\{\frac{\omega}{4}\left(2+\sqrt{4-2n}\right)\Bigg[G_{2\omega\gamma}\left(\gamma-r_1^{+-}\right)\left(\gamma-r_1^{--}\right) \notag\\
    &\hspace{6cm}- G^{(0)}_{2\omega\gamma}\left(\left(\gamma-\dw^{+-}\right)\left(\gamma-\dw^{--}\right) + 2 + \frac{1}{2}\sqrt{4-2n}\right)\Bigg] \notag\\
    &\hspace{2.4cm} +\frac{n}{4} \, \omega \, \Bigg[ \widetilde{Q}_{\omega\gamma}\left(\gamma-r_2^{+-}\right)\left(\gamma-r_2^{--}\right) - \widetilde{Q}^{(0)}_{\omega\gamma}\left(\left(\gamma-\dw^{+-}\right)\left(\gamma-\dw^{--}\right) + \frac{3}{2}\right) \notag\Bigg]  \notag \\
    &\hspace{2.4cm}-\left(\gamma\rightarrow\dw^{+-}\right) \Bigg\}\notag ,
\end{align}
\end{subequations}
where we have defined the roots of the second-order polynomials in $\gamma$ which multiply $G_{2\omega\gamma}$ and $\widetilde{Q}_{\omega\gamma}$ as 
\begin{subequations}\label{ch6rs}
\begin{align}
    \label{ch6r1alphabeta}
    &r_1^{\alpha\beta} = \frac{1}{2}\left[\omega + \alpha \, \sqrt{\omega^2 - 8 \, \left(1- \beta \, \sqrt{1-  \frac{n}{2}}\right) \, \left( 1 - \frac{2}{\omega \, \delta^{+,\beta}_\omega} \right)  
    }     \,\,\right], \\
    \label{ch6r2alphabeta}
    &r_2^{\alpha\beta} = \frac{1}{2}\left[\omega +\alpha \, \sqrt{\omega^2 - 2 - 4 \, \left(1- \beta \, \sqrt{1-  \frac{n}{2}}\right) \, \left( 1 - \frac{2}{\omega \, \delta^{+,\beta}_\omega} \right)  
    }    \,\,\right] \,. 
\end{align}
\end{subequations}
Note that by properly accounting for the poles at $\gamma = \dw^{++}$ and $\gamma = \dw^{+-}$ which lie to the right of the $\gamma$-contour (that is, by picking up these poles when we inverted the inverse Laplace transform in \eq{ch6conditions_b12}), we have ensured that there are no explicit poles at $\gamma = \dw^{++}$ and $\gamma = \dw^{+-}$ in Eqs.~\eqref{ch6ppcombo6} and \eqref{ch6pmcombo6}, respectively. This is consistent with the requirement that our inverse Laplace transform expressions \eqref{ch6doubleLaplacestart} remain well-defined with all the singularities of the integrand located to the left of the integration contours: for instance, requiring that $G_{2\omega\gamma}$ and $\widetilde{Q}_{\omega\gamma}$ have no singularities at $\gamma = \dw^{++}$ and $\gamma = \dw^{+-}$ appears to not lead to any contradictions in Eqs.~\eqref{ch6ppandpmcombo6}.

We now have two equations \eqref{ch6ppandpmcombo6} which we would like to solve for the double-Laplace images $G_{2\omega\gamma}$ and $\widetilde{Q}_{\omega\gamma}$: however, these functions appear in our equations with multiple different arguments, which complicates our task. The substitutions $\gamma\rightarrow \dw^{++}$ and $\gamma\rightarrow \dw^{+-}$ at the end of each of Eqs.~\eqref{ch6ppandpmcombo6} apply to the arguments of the double-Laplace images $G_{2\omega\gamma}$ and $Q_{\omega\gamma}$ as well, so that Eqs.~\eqref{ch6ppandpmcombo6} contain terms proportional to $G_{2\omega\dw^{++}}$, $\widetilde{Q}_{\omega\dw^{++}}$, $G_{2\omega\dw^{+-}}$, and $\widetilde{Q}_{\omega\dw^{+-}}$. At first glance, in addition to $G_{2\omega\gamma}$ and $\widetilde{Q}_{\omega\gamma}$ we have four other functions to solve for: $G_{2\omega\dw^{++}}$, $\widetilde{Q}_{\omega\dw^{++}}$, $G_{2\omega\dw^{+-}}$, and $\widetilde{Q}_{\omega\dw^{+-}}$. However, only two specific linear combinations of these four unknowns enter Eqs.~\eqref{ch6ppandpmcombo6}. They are 
\begin{subequations}
\begin{align}\label{ch6Zpp}
    &Z^{(++)}(\omega) \equiv - \frac{\omega}{4}\left(-2+\sqrt{4-2n}\right) 
    \, G_{2\omega\dw^{++}}\left(\dw^{++} - r_1^{++}\right)\left(\dw^{++} - r_1^{-+}\right) \\
    &\hspace{3cm}+ \frac{n}{4}\omega \widetilde{Q}_{\omega\dw^{++}} \left(\dw^{++} - r_2^{++}\right)\left(\dw^{++} - r_2^{-+}\right) , \notag \\
    \label{ch6Zpm}
    &Z^{(+-)}(\omega) \equiv - \frac{\omega}{4}\left(2+\sqrt{4-2n}\right) \, G_{2\omega\dw^{+-}}\left(\dw^{+-} - r_1^{+-}\right)\left(\dw^{+-} - r_1^{--}\right) \\
    &\hspace{3cm}- \frac{n}{4}\omega \widetilde{Q}_{\omega\dw^{+-}} \left(\dw^{+-} - r_2^{+-}\right)\left(\dw^{+-} - r_2^{--}\right) . \notag
\end{align}
\end{subequations}
In terms of these new objects, we re-write Eqs.~\eqref{ch6ppandpmcombo6} as 
\begin{subequations}\label{ch6ppandpmcombo6prime}
\begin{align}
    \label{ch6ppcombo6prime}
    &Q^{(0)}_{\dw^{++}\gamma} + 2G^{(0)}_{2\dw^{++}\gamma} - \frac{4}{n}\left(1-\sqrt{1-\frac{n}{2}}\right) \left( \widetilde{G}^{(0)}_{\dw^{++}\gamma} + 2G^{(0)}_{2\dw^{++}\gamma}\right) \\
    & = \frac{4}{n}\frac{1}{\left(\gamma-\dw^{++}\right)} \Bigg\{\frac{\omega}{4}\left(-2+\sqrt{4-2n}\right)\Bigg[G_{2\omega\gamma}\left(\gamma-r_1^{++}\right)\left(\gamma-r_1^{-+}\right) \notag \\
    &\hspace{6.4cm}- G^{(0)}_{2\omega\gamma}\left(\left(\gamma-\dw^{++}\right)\left(\gamma-\dw^{-+}\right) + 2 - \frac{1}{2}\sqrt{4-2n}\right) \notag \\
    &\hspace{6.4cm}+ G^{(0)}_{2\omega\dw^{++}}\left(2-\frac{1}{2}\sqrt{4-2n} \right) \Bigg] \notag\\
    &\hspace{2.4cm} -\frac{n}{4}\omega\Bigg[ \widetilde{Q}_{\omega\gamma}\left(\gamma-r_2^{++}\right)\left(\gamma-r_2^{-+}\right) \notag \\
    &\hspace{3.4cm}- \widetilde{Q}^{(0)}_{\omega\gamma}\left(\left(\gamma-\dw^{++}\right)\left(\gamma-\dw^{-+}\right) + \frac{3}{2}\right) + \frac{3}{2}\widetilde{Q}^{(0)}_{\omega\dw^{++}}\Bigg] + Z^{(++)}(\omega) \Bigg\} ,\notag \\  
    \label{ch6pmcombo6prime}
    &Q^{(0)}_{\dw^{+-}\gamma} + 2G^{(0)}_{2\dw^{+-}\gamma} - \frac{4}{n}\left(1+\sqrt{1-\frac{n}{2}}\right) \left( \widetilde{G}^{(0)}_{\dw^{+-}\gamma} + 2G^{(0)}_{2\dw^{+-}\gamma}\right) \\
    & = -\frac{4}{n}\frac{1}{\left(\gamma-\dw^{+-}\right)} \Bigg\{\frac{\omega}{4}\left(2+\sqrt{4-2n}\right)\Bigg[G_{2\omega\gamma}\left(\gamma-r_1^{+-}\right)\left(\gamma-r_1^{--}\right) \notag \\
    &\hspace{6.4cm}- G^{(0)}_{2\omega\gamma}\left(\left(\gamma-\dw^{+-}\right)\left(\gamma-\dw^{--}\right) + 2 + \frac{1}{2}\sqrt{4-2n}\right) \notag \\
    &\hspace{6.4cm}+ G^{(0)}_{2\omega\dw^{+-}}\left(2+\frac{1}{2}\sqrt{4-2n}\right)  \Bigg] \notag\\
    &\hspace{2.4cm} +\frac{n}{4}\omega\Bigg[ \widetilde{Q}_{\omega\gamma}\left(\gamma-r_2^{+-}\right)\left(\gamma-r_2^{--}\right) \notag \\
    &\hspace{3.4cm}- \widetilde{Q}^{(0)}_{\omega\gamma}\left(\left(\gamma-\dw^{+-}\right)\left(\gamma-\dw^{--}\right) + \frac{3}{2}\right) + \frac{3}{2}\widetilde{Q}^{(0)}_{\omega\dw^{+-}}\Bigg] + Z^{(+-)}(\omega) \Bigg\} .\notag 
\end{align}
\end{subequations}
Solving \eq{ch6ppcombo6prime} algebraically for $G_{2\omega\gamma}$ (still in terms of the unknowns $\widetilde{Q}_{\omega\gamma}$ and $Z^{(++)}(\omega)$), and substituting the result into \eq{ch6pmcombo6prime}, we obtain
\begin{align}\label{ch6pmcombo7}
    &\frac{2+\sqrt{4-2n}}{-2+\sqrt{4-2n}}\frac{\left(\gamma-r_1^{+-}\right)\left(\gamma-r_1^{--}\right)}{\left(\gamma-r_1^{++}\right)\left(\gamma-r_1^{-+}\right)}Z^{(++)}(\omega) - Z^{(+-)}(\omega) \\
    & = \frac{\omega}{4}\left(2+\sqrt{4-2n}\right)\notag \\
    &\hspace{1cm}\times\Bigg\{ \frac{\left(\gamma-r_1^{+-}\right)\left(\gamma-r_1^{--}\right)}{\left(\gamma-r_1^{++}\right)\left(\gamma-r_1^{-+}\right)} \bigg[G^{(0)}_{2\omega\gamma}\left(\left(\gamma-\dw^{++}\right)\left(\gamma-\dw^{-+}\right) + 2 - \frac{1}{2}\sqrt{4-2n}  \right) \notag \\
    &\hspace{5.8cm}- G^{(0)}_{2\omega\dw^{++}}\left(2-\frac{1}{2}\sqrt{4-2n}\right) \bigg] \notag \\
    &\hspace{.8cm}- G^{(0)}_{2\omega\gamma}\left(\left(\gamma-\dw^{+-}\right)\left(\gamma-\dw^{--}\right) + 2 + \frac{1}{2}\sqrt{4-2n}  \right) + G^{(0)}_{2\omega\dw^{+-}}\left(2+\frac{1}{2}\sqrt{4-2n}\right)\Bigg\} \notag \\
    & + \frac{n}{4}\omega\Bigg\{ \frac{2+\sqrt{4-2n}}{-2+\sqrt{4-2n}}\frac{\left(\gamma-r_1^{+-}\right)\left(\gamma-r_1^{--}\right)}{\left(\gamma-r_1^{++}\right)\left(\gamma-r_1^{-+}\right)} \notag \\
    &\hspace{1.5cm}\times\bigg[ -\widetilde{Q}^{(0)}_{\omega\gamma}\left(\left(\gamma-\dw^{++}\right)\left(\gamma-\dw^{-+}\right) + \frac{3}{2}\right) + \widetilde{Q}^{(0)}_{\omega\dw^{++}}\left(\frac{3}{2}\right) \notag \\
    &\hspace{1.4cm} + \frac{\left(\gamma-\dw^{++}\right)}{\omega} \left(Q^{(0)}_{\dw^{++}\gamma} + 2G^{(0)}_{2\dw^{++}\gamma} - \frac{4}{n}\left(1-\sqrt{1-\frac{n}{4}}\right)\left(\widetilde{G}^{(0)}_{\dw^{++}\gamma} + 2G^{(0)}_{2\dw^{++}\gamma}\right) \right) \bigg] \notag \\
    &\hspace{.5cm} - \widetilde{Q}^{(0)}_{\omega\gamma}\left(\left(\gamma-\dw^{+-}\right)\left(\gamma-\dw^{--}\right) + \frac{3}{2}  \right) + \widetilde{Q}^{(0)}_{\omega\dw^{+-}}\left(\frac{3}{2}\right) \notag \\
    &\hspace{.5cm}+  \frac{\left(\gamma-\dw^{+-}\right)}{\omega} \left(Q^{(0)}_{\dw^{+-}\gamma} + 2G^{(0)}_{2\dw^{+-}\gamma} - \frac{4}{n}\left(1+\sqrt{1-\frac{n}{4}}\right)\left(\widetilde{G}^{(0)}_{\dw^{+-}\gamma} + 2G^{(0)}_{2\dw^{+-}\gamma}\right) \right)
    \Bigg\} \notag \\
    & + \frac{n}{4}\omega \widetilde{Q}_{\omega\gamma} \left\{ \frac{2\sqrt{4-2n}}{\left(-2+\sqrt{4-2n}\right)} \frac{\left(\gamma-\gw^{++}\right)\left(\gamma-\gw^{+-}\right)\left(\gamma-\gw^{-+}\right)\left(\gamma-\gw^{--}\right)}{\left(\gamma-r_1^{++}\right)\left(\gamma-r_1^{-+}\right)} \right\} \notag.
\end{align}
In writing the last line of \eq{ch6pmcombo7} we have defined the following:
\begin{align}
    & 2\sqrt{4-2n}\left(\gamma - \gw^{++}\right)\left(\gamma - \gw^{+-}\right)\left(\gamma - \gw^{-+}\right)\left(\gamma - \gw^{--}\right) 
    \\
    &\hspace{1cm}\equiv \left(2+\sqrt{4-2n}\right)\left(\gamma-r_1^{+-}\right)\left(\gamma-r_1^{--}\right)\left(\gamma-r_2^{++}\right)\left(\gamma-r_2^{-+}\right) 
    \notag \\
    &\hspace{1.2cm}
    - \left(2-\sqrt{4-2n}\right)\left(\gamma-r_1^{++}\right)\left(\gamma-r_1^{-+}\right)\left(\gamma-r_2^{+-}\right)\left(\gamma-r_2^{--}\right), \notag
\end{align}
where the functions $\gw^{\pm\pm}$ are given by 

\begin{align}\label{ch6gammapmpmfull}
    &\gamma_\omega^{\pm\pm} = \frac{1}{2}\left[\omega \pm \sqrt{\omega^2 + s_1(\omega) \pm \sqrt{s_2(\omega)}}  \right] 
\end{align}
with
\begin{subequations}\label{ch6gammapmpmfull12}
\begin{align} 
    \label{ch6news1}
    &s_1(\omega) = -9 + \frac{2\left(8-3n \right)\left(\dw^{--} + \dw^{-+}\right) + 8\sqrt{4-2n}\left(\dw^{--} - \dw^{-+}\right)}{\omega \left(2-n\right)} , \\
    \label{ch6news2}
    &s_2(\omega) = \frac{1}{\left(2-n\right)^2}\frac{1}{\omega^2} \Bigg\{\omega^2\left(2-n\right)^2\left(49-16n\right) - 64\left(2-n\right)\left(8-3n\right) \\
    &\hspace{2.5cm}+ 8\omega\left(\dw^{--}+\dw^{-+} \right)\left(8 + 11n - 7n^2 \right) \notag \\
    &\hspace{2.5cm} + 16\omega\left(\dw^{--}-\dw^{-+}\right) \sqrt{4 - 2n}\left(2 + 5n -2n^2\right) + 8n\dw^{--}\dw^{-+}\left(16 - 7n\right)\Bigg\} .\notag
\end{align}
\end{subequations}
Note that in obtaining Eqs.~\eqref{ch6gammapmpmfull} and \eqref{ch6gammapmpmfull12} we extensively used the identities involving the functions $\dw^{\alpha\beta}$ in Eqs.~\eqref{ch6deltasumrule} and \eqref{ch6deltaproductrule}.

Equation~\eqref{ch6pmcombo7} contains three unknowns: $\widetilde{Q}_{\omega\gamma}$, $Z^{(++)}(\omega)$, and $Z^{(+-)}(\omega)$. However, by evaluating \eq{ch6pmcombo7} at particular values of $\gamma$, we could completely eliminate the term containing $\widetilde{Q}_{\omega\gamma}$ (the last line), as long as $\widetilde{Q}_{\omega\gamma}$ has no pole at our choice of $\gamma$.  As can be seen from Eqs.~\eqref{ch6gammapmpmfull}, we have the scalings
\begin{align}\label{ch6gammascaling}
    \gw^{++} \sim \gw^{+-} \sim \omega \qquad \text{when} \qquad \omega \rightarrow \infty ,
\end{align}
and so we conclude that $\widetilde{Q}_{\omega\gamma}$ cannot have a pole at either $\gamma = \gw^{++}$ or $\gamma = \gw^{+-}$, for the same reasoning discussed in the text after \eq{ch6polescalingslargew}, that is, to avoid having singularities to the right of the $\gamma$ integration contour. Thus, if we evaluate \eq{ch6pmcombo7} first for $\gamma = \gw^{++}$ then again for $\gamma = \gw^{+-}$, while requiring that $\widetilde{Q}_{\omega\gamma}$ is finite at these values of $\gamma$, we will obtain two equations for the two unknowns $Z^{(++)}(\omega)$ and $Z^{(+-)}(\omega)$, which we can solve. Then, having explicit expressions for the functions $Z^{(++)}(\omega)$ and $Z^{(+-)}(\omega)$, we can construct $G_{2\omega\gamma}$ and $\widetilde{Q}_{\omega\gamma}$ by using Eqs.~\eqref{ch6ppandpmcombo6prime}.
The system of two equations we can solve for $Z^{(++)}(\omega)$ and $Z^{(+-)}(\omega)$ is then
\begin{align}\label{ch6ZppZmmsystem}
    &\Bigg[-\frac{2+\sqrt{4-2n}}{-2+\sqrt{4-2n}}\frac{\left(\gamma-r_1^{+-}\right)\left(\gamma-r_1^{--}\right)}{\left(\gamma-r_1^{++}\right)\left(\gamma-r_1^{-+}\right)}Z^{(++)}(\omega) - Z^{(+-)}(\omega) \\
    & + \frac{\omega}{4}\left(2+\sqrt{4-2n}\right)\notag \\
    &\hspace{1cm}\times\Bigg\{ \frac{\left(\gamma-r_1^{+-}\right)\left(\gamma-r_1^{--}\right)}{\left(\gamma-r_1^{++}\right)\left(\gamma-r_1^{-+}\right)} \bigg[G^{(0)}_{2\omega\gamma}\left(\left(\gamma-\dw^{++}\right)\left(\gamma-\dw^{-+}\right) + 2 - \frac{1}{2}\sqrt{4-2n}  \right) \notag \\
    & \hspace{6cm}- G^{(0)}_{2\omega\dw^{++}}\left(2-\frac{1}{2}\sqrt{4-2n}\right) \bigg] \notag \\
    &\hspace{2cm}- G^{(0)}_{2\omega\gamma}\left(\left(\gamma-\dw^{+-}\right)\left(\gamma-\dw^{--}\right) + 2 + \frac{1}{2}\sqrt{4-2n}  \right) \notag \\
    &\hspace{2cm}+ G^{(0)}_{2\omega\dw^{+-}}\left(2+\frac{1}{2}\sqrt{4-2n}\right)\Bigg\} \notag \\
    & + \frac{n}{4}\omega\Bigg\{ \frac{2+\sqrt{4-2n}}{-2+\sqrt{4-2n}}\frac{\left(\gamma-r_1^{+-}\right)\left(\gamma-r_1^{--}\right)}{\left(\gamma-r_1^{++}\right)\left(\gamma-r_1^{-+}\right)} \notag \\
    &\hspace{1cm}\times\bigg[ -\widetilde{Q}^{(0)}_{\omega\gamma}\left(\left(\gamma-\dw^{++}\right)\left(\gamma-\dw^{-+}\right) + \frac{3}{2}\right) + \widetilde{Q}^{(0)}_{\omega\dw^{++}}\left(\frac{3}{2}\right) \notag \\
    &\hspace{1.3cm} + \frac{\left(\gamma-\dw^{++}\right)}{\omega} \left(Q^{(0)}_{\dw^{++}\gamma} + 2G^{(0)}_{2\dw^{++}\gamma} - \frac{4}{n}\left(1-\sqrt{1-\frac{n}{4}}\right)\left(\widetilde{G}^{(0)}_{\dw^{++}\gamma} + 2G^{(0)}_{2\dw^{++}\gamma}\right) \right) \bigg] \notag \\
    &\hspace{.7cm} - \widetilde{Q}^{(0)}_{\omega\gamma}\left(\left(\gamma-\dw^{+-}\right)\left(\gamma-\dw^{--}\right) + \frac{3}{2}  \right) + \widetilde{Q}^{(0)}_{\omega\dw^{+-}}\left(\frac{3}{2}\right) \notag \\
    &\hspace{.7cm}+  \frac{\left(\gamma-\dw^{+-}\right)}{\omega} \bigg(Q^{(0)}_{\dw^{+-}\gamma} + 2G^{(0)}_{2\dw^{+-}\gamma} \notag \\
    &\hspace{4cm}- \frac{4}{n}\left(1+\sqrt{1-\frac{n}{4}}\right)\left(\widetilde{G}^{(0)}_{\dw^{+-}\gamma} + 2G^{(0)}_{2\dw^{+-}\gamma}\right) \bigg)
    \Bigg\}\Bigg]_{\gamma=\gw^{++}, \gamma=\gw^{+-}} = 0\notag,
\end{align}
where the subscripts after the square bracket in the last line denote the fact that \eq{ch6ZppZmmsystem} contains two separate equations, one with the left-hand side evaluated at $\gamma = \gw^{++}$ and another one with the left-hand side evaluated at $\gamma = \gw^{+-}$.

Solving this system of equations and  substituting the results for $Z^{(++)}(\omega)$ and $Z^{(+-)}(\omega)$ back into Eqs.~\eqref{ch6ppandpmcombo6prime}, we explicitly construct $G_{2\omega\gamma}$ and $\widetilde{Q}_{\omega\gamma}$. After some significant algebra, the results can be written as
\begin{subequations}\label{ch6QtildeandG2full}
\begin{align}
    \label{ch6newQomegagamma}
    &\widetilde{Q}_{\omega\gamma} = \frac{1}{2\sqrt{4-2n}\left(\gamma-\gw^{++}\right)\left(\gamma-\gw^{+-}\right)\left(\gamma-\gw^{-+}\right)\left(\gamma-\gw^{--}\right) } \\
    &\hspace{2cm}\times\bigg[f^{(\widetilde{Q})}(\omega,\gamma) - \tfrac{2\left(\gamma-\gw^{+-}\right)\left(\gamma-\gw^{--}\right)}{\sqrt{s_2(\omega)}}f^{(\widetilde{Q})}(\omega,\gamma=\gw^{++}) \notag \\
    &\hspace{5cm}+ \tfrac{2\left(\gamma-\gw^{++}\right)\left(\gamma-\gw^{-+}\right)}{\sqrt{s_2(\omega)}}f^{(\widetilde{Q})}(\omega,\gamma=\gw^{+-})\bigg]  \, , \notag\\
    \label{ch6newG2omegagamma}
    &G_{2\omega\gamma} = \frac{-1}{{2\sqrt{4-2n}\left(\gamma-\gw^{++}\right)\left(\gamma-\gw^{+-}\right)\left(\gamma-\gw^{-+}\right)\left(\gamma-\gw^{--}\right) }} \\
    &\hspace{2cm}\times\bigg[f^{(G_2)}(\omega,\gamma) - \tfrac{2\left(\gamma-\gw^{+-}\right)\left(\gamma-\gw^{--}\right)}{\sqrt{s_2(\omega)}}f^{(G_2)}(\omega,\gamma=\gw^{++}) \notag \\
    &\hspace{5cm}+ \tfrac{2\left(\gamma-\gw^{++}\right)\left(\gamma-\gw^{-+}\right)}{\sqrt{s_2(\omega)}}f^{(G_2)}(\omega,\gamma=\gw^{+-}) \bigg]  \notag
\end{align}
\end{subequations}
with $\gw^{\alpha\beta}$ and $s_2(\omega)$ as defined in Eqs.~\eqref{ch6gammapmpmfull} and \eqref{ch6gammapmpmfull12}. We have defined two new quantities,
\begin{subequations}\label{ch6fQtandfG2}
\begin{align}
    \label{ch6fQt}
    &f^{(\widetilde{Q})}(\omega,\gamma) = \left(\gamma-r_1^{+-}\right)\left(\gamma-r_1^{--}\right)f^{(+)}(\omega,\gamma) - \left(\gamma-r_1^{++}\right)\left(\gamma-r_1^{-+}\right)f^{(-)}(\omega,\gamma) \, , \\
    \label{ch6fG2}
    &f^{(G_2)}(\omega,\gamma) = n \frac{\left(\gamma-r_2^{+-}\right)\left(\gamma-r_2^{--}\right)  }{2+\sqrt{4-2n}} f^{(+)}(\omega,\gamma) - n \frac{\left(\gamma-r_2^{++}\right)\left(\gamma-r_2^{-+}\right)  }{2-\sqrt{4-2n}} f^{(-)}(\omega,\gamma) ,
\end{align}
\end{subequations}
where $r_1^{\alpha\beta}$ and $r_2^{\alpha\beta}$ are defined in Eqs.~\eqref{ch6rs}, while
$f^{(+)}(\omega,\gamma)$ and $f^{(-)}(\omega,\gamma)$ are given by
\begin{subequations}\allowdisplaybreaks
\begin{align}
    \label{ch6fplus}
    &f^{(+)}(\omega,\gamma) = \bigg\{ 2G_{2\omega\gamma}^{(0)}\left[\left(\gamma-\dw^{++}\right)\left(\gamma-\dw^{-+}\right) + 2-\tfrac{1}{2}\sqrt{4-2n}  \,\right] \\
    &\hspace{2cm}+ \widetilde{Q}^{(0)}_{\omega\gamma} \left(2+\sqrt{4-2n}\,\right)\left[\left(\gamma-\dw^{++}\right)\left(\gamma-\dw^{-+}\right) + \frac{3}{2}\right] \notag\\
    &\hspace{2cm} - \left(2+\sqrt{4-2n}\right)\frac{\gamma-\dw^{++}}{\omega}\bigg[Q^{(0)}_{\dw^{++}\gamma} + 2G^{(0)}_{2\dw^{++}\gamma} \notag \\
    &\hspace{4.5cm}- \tfrac{4}{n}\left(1-\sqrt{1-\tfrac{n}{2}}\right)\left(\widetilde{G}^{(0)}_{\dw^{++}\gamma} + 2G^{(0)}_{2\dw^{++}\gamma}  \right)    \bigg]   \bigg\} - \left\{\gamma \rightarrow \dw^{++} \right\} \notag\\
    \label{ch6fminus}
    &f^{(-)}(\omega,\gamma) = \bigg\{ 2G_{2\omega\gamma}^{(0)}\left[\left(\gamma-\dw^{+-}\right)\left(\gamma-\dw^{--}\right) + 2+\tfrac{1}{2}\sqrt{4-2n}  \right] \\
    &\hspace{2cm}+ \widetilde{Q}^{(0)}_{\omega\gamma} \left(2-\sqrt{4-2n}\right)\left[\left(\gamma-\dw^{+-}\right)\left(\gamma-\dw^{--}\right) + \frac{3}{2}\right] \notag \\
    &\hspace{2cm} - \left(2-\sqrt{4-2n}\right)\frac{\gamma-\dw^{+-}}{\omega}\bigg[Q^{(0)}_{\dw^{+-}\gamma} + 2G^{(0)}_{2\dw^{+-}\gamma} \notag \\
    &\hspace{4.5cm}- \tfrac{4}{n}\left(1+\sqrt{1-\tfrac{n}{2}}\right)\left(\widetilde{G}^{(0)}_{\dw^{+-}\gamma} + 2G^{(0)}_{2\dw^{+-}\gamma}  \right)    \bigg]   \bigg\} - \left\{\gamma \rightarrow \dw^{+-} \right\} .\notag
\end{align}
\end{subequations}
All four of these new objects, $f^{(\widetilde{Q})}(\omega,\gamma)$, $f^{(G_2)}(\omega,\gamma)$, $f^{(+)}(\omega,\gamma)$, and $f^{(-)}(\omega,\gamma)$ are determined by the initial conditions/inhomogeneous terms in our evolution equations \eqref{ch6evoleqs_IR_1}. 

One can easily show using \eq{ch6gammapmpmfull} that 
\begin{subequations}
\begin{align}
    &\frac{2\left(\gamma - \gw^{+-}\right)\left(\gamma - \gw^{--}\right)}{\sqrt{s_2(\omega)}}\bigg|_{\gamma = \gw^{++}} = 1 \qquad \text{and} \\
    &\frac{2\left(\gamma - \gw^{++}\right)\left(\gamma - \gw^{-+}\right)}{\sqrt{s_2(\omega)}}\bigg|_{\gamma = \gw^{+-}} = -1.
\end{align}
\end{subequations}
This makes it clear that both $\widetilde{Q}_{\omega\gamma}$ and $G_{2\omega\gamma}$ as expressed in Eqs.~\eqref{ch6QtildeandG2full} have vanishing residues at both $\gamma = \gw^{++}$ and $\gamma = \gw^{+-}$, as indeed they must in order for the inverse Laplace transforms in Eqs.~\eqref{ch6doubleLaplaceG2} and \eqref{ch6doubleLaplaceQtilde} to be well-defined, with the integrands having no singularities to the right of the integration contours. 

Having explicitly constructed $\widetilde{Q}_{\omega\gamma}$ and $G_{2\omega\gamma}$, the full solution of the evolution Eqs.~\eqref{ch6evoleqs_IR_1} is now formally complete. In the next Section, we summarize the results of our calculation.


\section{Summary of the Solution}\label{ch6sec:solnsummary}

Returning to the original variables $zs$, $z's$, $\xoz^2$, $\xto^2$ (see \eq{originalvars}) and employing the definition 
\begin{align}\label{ch6alphabar}
    \bas = \frac{\as N_c}{2\pi}\,,
\end{align}
we collect all the pieces here which form the complete analytic solution to the modified large-$N_c\&N_f$ evolution Eqs.~\eqref{eq_LargeNcNf}. In addition, we undo the rescaling introduced in defining the variables in \eq{originalvars} by replacing
\begin{align}\label{ch6resc}
 & \omega \to \frac{\omega}{\sqrt{\bas}}, \ \gamma \to \frac{\gamma}{\sqrt{\bas}}, \ \widetilde{Q}_{\omega\gamma} \to \bas \, \widetilde{Q}_{\omega\gamma}, \ G_{2\omega\gamma} \to \bas \, G_{2\omega\gamma}, \\
 &G^{(0)}_{2\omega\gamma} \to \bas \,G^{(0)}_{2\omega\gamma}, \ \widetilde{G}^{(0)}_{\omega\gamma} \to \bas \, \widetilde{G}^{(0)}_{\omega\gamma}, \ Q^{(0)}_{\omega\gamma} \to \bas \, Q^{(0)}_{\omega\gamma}, \ \widetilde{Q}^{(0)}_{\omega\gamma} \to \bas \, \widetilde{Q}^{(0)}_{\omega\gamma}, \notag\\ &  \ r_1^{\alpha\beta} \to \frac{r_1^{\alpha\beta}}{\sqrt{\bas}}, \ r_2^{\alpha\beta} \to \frac{r_2^{\alpha\beta}}{\sqrt{\bas}}, \ f^{(\widetilde{Q})}(\omega,\gamma) \to \frac{f^{(\widetilde{Q})}(\omega,\gamma)}{\bas}, \  f^{(G_2)}(\omega,\gamma) \to \frac{f^{G_2}(\omega,\gamma)}{\bas}, \notag \\ & A_{\omega\gamma} \to \bas \, A_{\omega\gamma}, \  {\overline A}_{\omega\gamma} \to \bas \, {\overline A}_{\omega\gamma}, \ \widetilde{\Gamma}_{\omega}^{(\alpha\beta)}(\xoz^2) \to \sqrt{\bas} \, \widetilde{\Gamma}_{\omega}^{(\alpha\beta)}(\xoz^2), \notag \\
 &\delta_{\omega}^{\pm\pm} \to \frac{\delta_{\omega}^{\pm\pm}}{\sqrt{\bas}}, \ \gamma_{\omega}^{\pm\pm} \to \frac{\gamma_{\omega}^{\pm\pm}}{\sqrt{\bas}} ,\notag
\end{align}
with the arguments $\omega, \gamma$ of the double inverse Laplace transform images not reflecting the rescaling of $\omega, \gamma$. 
Recalling that $n = N_f/N_c$ we write
\begin{subequations}\label{ch6fullsoln}\allowdisplaybreaks
\begin{align}
    \label{ch6fullsolnQtilde}
    &\widetilde{Q}(\xoz^2,zs) = \wint \gint e^{\omega \, \ln(zs\xoz^2) + \gamma \, \ln\left(\tfrac{1}{\xoz^2\Lambda^2} \right)}\widetilde{Q}_{\omega\gamma}\,,\\
    \label{ch6fullsolnG2}
    &G_2(\xoz^2,zs) = \wint \gint e^{\omega\, \ln(zs\xoz^2) + \gamma\, \ln\left(\tfrac{1}{\xoz^2\Lambda^2} \right)}G_{2\omega\gamma}\,, \\
    \label{ch6fullsolnGtilde}
    &\widetilde{G}(\xoz^2,zs) = \wint \gint e^{\omega\, \ln(zs\xoz^2) + \gamma\, \ln\left(\tfrac{1}{\xoz^2\Lambda^2} \right)}\left[\frac{\omega\gamma}{2 \, \bas}\left(G_{2\omega\gamma} - G^{(0)}_{2\omega\gamma} \right) - 2 \, G_{2\omega\gamma}\right]\,, \\
    \label{ch6fullsolnQ}
    &Q(\xoz^2,zs) = \wint \gint e^{\omega\, \ln(zs\xoz^2) + \gamma\, \ln\left(\tfrac{1}{\xoz^2\Lambda^2} \right)}\left[-\frac{\omega\gamma}{\bas} \, \left(\widetilde{Q}_{\omega\gamma} - \widetilde{Q}^{(0)}_{\omega\gamma}\right) - 2 \, G_{2\omega\gamma}\right]\,, \\
    \label{ch6fullsolnGamma2}
    &\Gamma_2(\xoz^2,\xto^2,z's) = \wint \gint \bigg[e^{\omega\, \ln(z's\xto^2)+\gamma\, \ln\left(\tfrac{1}{\xoz^2\Lambda^2} \right)}\left(G_{2\omega\gamma} - G^{(0)}_{2\omega\gamma}  \right)  \\
    &\hspace{4.8cm}+ e^{\omega\, \ln(z's\xoz^2)+\gamma\, \ln\left(\tfrac{1}{\xoz^2\Lambda^2}\right)} G^{(0)}_{2\omega\gamma}  \bigg] \,, \notag\\
    \label{ch6fullsolnGammatilde}
    &\widetilde{\Gamma}(\xoz^2,\xto^2,z's) =  \wint \, e^{\omega\, \ln(z's\xto^2)}\sum_{\alpha,\beta = +,-} e^{\dw^{\alpha\beta}\, \ln\left(\tfrac{1}{\xoz^2\Lambda^2}\right)} \, \widetilde{\Gamma}_\omega^{(\alpha\beta)}(\xoz^2) \\
    &\hspace{1cm} + \wint\gint \bigg[e^{\omega\, \ln(z's\xto^2)+\gamma\, \ln\left(\tfrac{1}{\xto^2\Lambda^2}\right)}A_{\omega\gamma} \notag \\
    &\hspace{1.6cm}-2e^{\omega\, \ln(z's\xto^2)+\gamma\, \ln\left(\tfrac{1}{\xoz^2\Lambda^2}\right)}\left(G_{2\omega\gamma} - G^{(0)}_{2\omega\gamma}\right) -2e^{\omega\, \ln(z's\xoz^2)+\gamma\, \ln\left(\tfrac{1}{\xoz^2\Lambda^2}\right)}G^{(0)}_{2\omega\gamma} \bigg] , \notag \\
    \label{ch6fullsolnGammabar}
    &\overline{\Gamma}(\xoz^2,\xto^2,z's) = \wint \, e^{\omega\, \ln(z's\xto^2)}\sum_{\alpha,\beta = +,-} e^{\dw^{\alpha\beta}\, \ln\left(\tfrac{1}{\xto^2\Lambda^2}\right)} \frac{1 + \beta \, \sqrt{ 1 - \frac{N_f}{2 \, N_c} }}{N_f / (4 N_c) }  \,  \widetilde{\Gamma}_\omega^{(\alpha\beta)}(\xoz^2) \\
    &\hspace{1cm} + \wint\gint \bigg[ e^{\omega\, \ln(z's\xto^2)+\gamma\, \ln\left(\tfrac{1}{\xto^2\Lambda^2}\right)} \, {\overline A}_{\omega\gamma} \notag \\
    &\hspace{1.6cm}- 2e^{\omega\, \ln(z's\xto^2)+\gamma\, \ln\left(\tfrac{1}{\xoz^2\Lambda^2}\right)}\left(G_{2\omega\gamma} - G^{(0)}_{2\omega\gamma}\right) - 2e^{\omega\, \ln(z's\xoz^2)+\gamma\, \ln\left(\tfrac{1}{\xoz^2\Lambda^2}\right)}G^{(0)}_{2\omega\gamma} \bigg] , \notag
    \end{align}
\end{subequations}
\vspace{.5cm}
with 
\vspace{.5cm}
\begin{subequations}\label{ch6fullsoln1}\allowdisplaybreaks
\begin{align}
    \label{ch6fullsoln1Qomegagamma}
    &\widetilde{Q}_{\omega\gamma} = \frac{1}{2\sqrt{4-\frac{2N_f}{N_c}}\left(\gamma-\gw^{++}\right)\left(\gamma-\gw^{+-}\right)\left(\gamma-\gw^{-+}\right)\left(\gamma-\gw^{--}\right) }\\
    &\hspace{1.5cm}\times \bigg[f^{(\widetilde{Q})}(\omega,\gamma) - \tfrac{2\left(\gamma-\gw^{+-}\right)\left(\gamma-\gw^{--}\right)}{\bas \, \sqrt{s_2(\omega)}}f^{(\widetilde{Q})}(\omega,\gamma=\gw^{++}) \notag \\
    &\hspace{5cm}+ \tfrac{2\left(\gamma-\gw^{++}\right)\left(\gamma-\gw^{-+}\right)}{\bas \, \sqrt{s_2(\omega)}}f^{(\widetilde{Q})}(\omega,\gamma=\gw^{+-})\bigg] \,, \notag \\
    \label{ch6fullsoln1G2omegagamma}
    &G_{2\omega\gamma} = \frac{-1}{2\sqrt{4-\frac{2N_f}{N_c}}\left(\gamma-\gw^{++}\right)\left(\gamma-\gw^{+-}\right)\left(\gamma-\gw^{-+}\right)\left(\gamma-\gw^{--}\right) } \\
    &\hspace{1.5cm}\times \bigg[f^{(G_2)}(\omega,\gamma) - \frac{2\left(\gamma-\gw^{+-}\right)\left(\gamma-\gw^{--}\right)}{\bas \, \sqrt{s_2(\omega)}}f^{(G_2)}(\omega,\gamma=\gw^{++}) \notag \\
    &\hspace{5cm}+ \frac{2\left(\gamma-\gw^{++}\right)\left(\gamma-\gw^{-+}\right)}{\bas \, \sqrt{s_2(\omega)}}f^{(G_2)}(\omega,\gamma=\gw^{+-})\bigg] \,, \notag \\
    \label{ch6fullsoln1fQt}
    &f^{(\widetilde{Q})}(\omega,\gamma) = \left(\gamma-r_1^{+-}\right)\left(\gamma-r_1^{--}\right)f^{(+)}(\omega,\gamma) - \left(\gamma-r_1^{++}\right)\left(\gamma-r_1^{-+}\right)f^{(-)}(\omega,\gamma) \,,\\
    \label{ch6fullsoln1fG2}
    &f^{(G_2)}(\omega,\gamma) = \tfrac{N_f}{N_c} \tfrac{\left(\gamma-r_2^{+-}\right)\left(\gamma-r_2^{--}\right)  }{2+\sqrt{4-\frac{2N_f}{N_c}}} f^{(+)}(\omega,\gamma) - \tfrac{N_f}{N_c} \tfrac{\left(\gamma-r_2^{++}\right)\left(\gamma-r_2^{-+}\right)  }{2-\sqrt{4-\frac{2N_f}{N_c}}} f^{(-)}(\omega,\gamma) \,, \\
    \label{ch6fullsolnfplus}
    &f^{(+)}(\omega,\gamma) = \bigg\{ 2G_{2\omega\gamma}^{(0)}\left[\left(\gamma-\dw^{++}\right)\left(\gamma-\dw^{-+}\right) + \bas \, \left( 2-\tfrac{1}{2}\sqrt{4-\tfrac{2N_f}{N_c}} \right)  \right] \\
    &\hspace{1cm}+ \widetilde{Q}^{(0)}_{\omega\gamma} \left(2+\sqrt{4-\tfrac{2N_f}{N_c}}\right)\left[\left(\gamma-\dw^{++}\right)\left(\gamma-\dw^{-+}\right) + \frac{3}{2} \, \bas \right] \notag\\
    &\hspace{1cm} - \bas \left(2+\sqrt{4-\tfrac{2N_f}{N_c}}\right)\frac{\gamma-\dw^{++}}{\omega}\bigg[Q^{(0)}_{\dw^{++}\gamma} + 2G^{(0)}_{2\dw^{++}\gamma} \notag \\
    &\hspace{3.6cm}- \tfrac{4N_c}{N_f}\left(1-\sqrt{1-\tfrac{N_f}{2N_c}}\right)\left(\widetilde{G}^{(0)}_{\dw^{++}\gamma} + 2G^{(0)}_{2\dw^{++}\gamma}  \right)    \bigg]   \bigg\} - \left\{\gamma \rightarrow \dw^{++} \right\}\,, \notag\\
    \label{ch6fullsolnfminus}
    &f^{(-)}(\omega,\gamma) = \bigg\{ 2G_{2\omega\gamma}^{(0)}\left[\left(\gamma-\dw^{+-}\right)\left(\gamma-\dw^{--}\right) + \bas \, \left( 2+\tfrac{1}{2}\sqrt{4-\tfrac{2N_f}{N_c}} \right)  \right] \\
    &\hspace{1cm}+ \widetilde{Q}^{(0)}_{\omega\gamma} \left(2-\sqrt{4-\tfrac{2N_f}{N_c}}\right)\left[\left(\gamma-\dw^{+-}\right)\left(\gamma-\dw^{--}\right) + \frac{3}{2} \, \bas \right] \notag \\
    &\hspace{1cm} - \bas \left(2-\sqrt{4-\tfrac{2N_f}{N_c}}\right)\frac{\gamma-\dw^{+-}}{\omega}\bigg[Q^{(0)}_{\dw^{+-}\gamma} + 2G^{(0)}_{2\dw^{+-}\gamma} \notag \\
    &\hspace{3.6cm}- \tfrac{4N_c}{N_f}\left(1+\sqrt{1-\tfrac{N_f}{2N_c}}\right)\left(\widetilde{G}^{(0)}_{\dw^{+-}\gamma} + 2G^{(0)}_{2\dw^{+-}\gamma}  \right)    \bigg]   \bigg\} - \left\{\gamma \rightarrow \dw^{+-} \right\} \,,\notag \\
    \label{ch6fullsoln1r1alphabeta}
    &r_1^{\alpha\beta} = \frac{\omega}{2} \left[1 + \alpha \, \sqrt{1 - \frac{8 \, \bas}{\omega^2} \, \left(1- \beta \, \sqrt{1-  \frac{N_f}{2 \, N_c}}\right) \, \left( 1 - \frac{2 \, \bas}{\omega \, \delta^{+,\beta}_\omega} \right)  
    }     \,\,\right] \, ,  \\
    \label{ch6fullsoln1r2alphabeta}
    &r_2^{\alpha\beta} = \frac{\omega}{2} \left[1 + \alpha \, \sqrt{1 - \frac{2 \, \bas}{\omega^2} - \frac{4 \, \bas}{\omega^2} \, \left(1- \beta \, \sqrt{1-  \frac{N_f}{2 \, N_c}}\right) \, \left( 1 - \frac{2 \, \bas}{\omega \, \delta^{+,\beta}_\omega} \right)  
    }     \,\,\right] \, , \\
    \label{ch6fullsoln1doubleLaplaceG20}
    &G_2^{(0)}(\xoz^2,zs) = \wint \gint e^{\omega\, \ln(zs\xoz^2)+\gamma\, \ln\left(\tfrac{1}{\xoz^2\Lambda^2}\right)}G^{(0)}_{2\omega\gamma}\,, \\
    \label{ch6fullsoln1doubleLaplaceGtilde0}
    &\widetilde{G}^{(0)}(\xoz^2,zs) = \wint \gint e^{\omega\, \ln(zs\xoz^2)+\gamma\, \ln\left(\tfrac{1}{\xoz^2\Lambda^2}\right)}\widetilde{G}^{(0)}_{\omega\gamma}\,, \\
    \label{ch6fullsoln1doubleLaplaceQ0}
    &Q^{(0)}(\xoz^2,zs) = \wint \gint e^{\omega\, \ln(zs\xoz^2)+\gamma\, \ln\left(\tfrac{1}{\xoz^2\Lambda^2}\right)}Q^{(0)}_{\omega\gamma}\,, \\
    \label{ch6fullsoln1doubleLaplaceQtilde0}
    &\widetilde{Q}^{(0)}(\xoz^2,zs) = \wint \gint e^{\omega\, \ln(zs\xoz^2)+\gamma\, \ln\left(\tfrac{1}{\xoz^2\Lambda^2}\right)}\widetilde{Q}^{(0)}_{\omega\gamma}\,,\\
    \label{ch6fullsolnGammatildes}
    &  \widetilde{\Gamma}_{\omega}^{(\alpha\beta)}(\xoz^2) = e^{-\dw^{\alpha\beta}\, \ln\left(\tfrac{1}{\xoz^2\Lambda^2} \right) }\gint \, e^{\gamma \, \ln\left(\tfrac{1}{\xoz^2\Lambda^2} \right) } \frac{\beta}{8 (\dw^{\alpha\beta} - \dw^{-\alpha, \beta}) \, \sqrt{4 - \tfrac{2N_f}{N_c}}} \\ 
    & \times \, \Big\{ \left[ \, {\overline A}_{\omega\gamma} \tfrac{2N_f}{N_c} - 4 A_{\omega\gamma} \left(2- \beta \sqrt{4 - \tfrac{2N_f}{N_c}})\right) \right] (\dw^{-\alpha, \beta} - \gamma) \notag \\
    &\hspace{1cm}+ \tfrac{N_f}{N_c} \, \omega  \left( \widetilde{Q}_{\omega\gamma} - \widetilde{Q}^{(0)}_{\omega\gamma} \right) \left(3 - \beta \sqrt{4 - \tfrac{2N_f}{N_c}} - \frac{2}{\bas} \, \gamma \, \dw^{\alpha, \beta}  \right) \notag \\ 
    &\hspace{1cm} + 2 \, \omega \left(G_{2\omega\gamma} - G^{(0)}_{2\omega\gamma}\right) \left[ \left(2- \beta \sqrt{4 - \tfrac{2N_f}{N_c}}\right) \left(4 - \frac{1}{\bas} \, \gamma \, \dw^{\alpha, \beta}  \right) - \tfrac{2N_f}{N_c}   \right] \Big\} \notag \,,\\
    \label{ch6fullsoln1Aeq}
    & A_{\omega\gamma} =  \frac{1}{\left(\gamma-\dw^{++} \right)\left(\gamma-\dw^{+-} \right)\left(\gamma-\dw^{-+} \right)\left(\gamma-\dw^{--} \right)} \\
    &\times\Bigg\{  \frac{\omega\gamma}{2} \, \left[ \bas \, \left( 3 - \tfrac{N_f}{2 N_c} \right) - 3 \,  \gamma \, (\gamma - \omega) \right] \, \left(G_{2\omega\gamma} - G^{(0)}_{2\omega\gamma} \right)
    + \bas \, \tfrac{3 N_f}{8 N_c} \, \omega \, \gamma \, \left( \widetilde{Q}_{\omega\gamma} - \widetilde{Q}^{(0)}_{\omega\gamma} \right) \notag \\
    &\hspace{.8cm}
    + \bas \left[ 4 \, \gamma \, (\gamma - \omega) - \bas \, \left( 4 - \tfrac{N_f}{N_c} \right) \right] \, G_{2\omega\gamma} + \bas \, \tfrac{N_f}{2 N_c} \, \left[ \gamma \, (\gamma - \omega) - \bas \right] \, \widetilde{Q}_{\omega\gamma} \Bigg\}, \notag \\
    \label{ch6fullsoln1Abar_eq}
    & \overline{A}_{\omega\gamma} =  \frac{1}{\left(\gamma-\dw^{++} \right)\left(\gamma-\dw^{+-} \right)\left(\gamma-\dw^{-+} \right)\left(\gamma-\dw^{--} \right)} \\
    &\times\Bigg\{ \omega\gamma \, [2 \, \bas - \gamma \, (\gamma - \omega)] \, \left(G_{2\omega\gamma} - G^{(0)}_{2\omega\gamma} \right) + 4 \, \bas \, [\gamma \, (\gamma - \omega) - \bas ] \, G_{2\omega\gamma} \notag \\
    &\hspace{.8cm}+ \tfrac{3}{2} \, \omega\gamma \, [\gamma \, (\gamma - \omega) + \bas] \, \left( \widetilde{Q}_{\omega\gamma} - \widetilde{Q}^{(0)}_{\omega\gamma} \right) - \bas^2 \, \tfrac{N_f}{N_c} \, \widetilde{Q}_{\omega\gamma}  \Bigg\} \notag  \,,\\
    \label{ch6fullsoln1deltapms}
    &\delta_{\omega}^{\pm\pm} \equiv \frac{\omega}{2}\left[1 \pm \sqrt{1 \pm \frac{4 \, \bas}{\omega^2} \, \sqrt{1-\frac{N_f}{2N_c} } } \right] \,, \\
    \label{ch6fullsoln1neweigs}
    &\gamma_\omega^{\pm\pm} = \frac{\omega}{2}\left[1 \pm \sqrt{1 + \frac{\bas}{\omega^2} \, \left[ s_1(\omega) \pm \sqrt{s_2(\omega)}\right] }  \, \right] \,,\\
    \label{ch6fullsoln1news1}
    &s_1(\omega) = -9 + \frac{2\left( 8-3 \, \frac{N_f}{N_c} \right)\left(\dw^{--} + \dw^{-+}\right) + 8\sqrt{4-2 \, \frac{N_f}{N_c}}\left(\dw^{--} - \dw^{-+}\right)}{\omega \left(2-\frac{N_f}{N_c}\right)} \,,\\
    \label{ch6fullsoln1news2}
    & s_2(\omega) = \tfrac{1}{\left(2-\frac{N_f}{N_c}\right)^2} \, \tfrac{1}{\omega^2} \, \Bigg\{\omega^2\left(2-\tfrac{N_f}{N_c}\right)^2\left(49-16 \, \tfrac{N_f}{N_c}\right) - 64 \, \bas \, \left(2-\tfrac{N_f}{N_c} \right)\left(8-3 \, \tfrac{N_f}{N_c} \right) \\
    &\hspace{3.5cm} + 8 \, \omega \, \left(\dw^{--}+\dw^{-+} \right)\left[ 8 + 11 \, \tfrac{N_f}{N_c} - 7 \, \left( \tfrac{N_f}{N_c}\right)^2 \right] \notag \\
    &\hspace{3.5cm}
    + 16 \, \omega \, \left(\dw^{--}-\dw^{-+}\right) \sqrt{4 - 2\, \tfrac{N_f}{N_c}} \left[ 2 + 5\, \tfrac{N_f}{N_c} -2\, \left( \tfrac{N_f}{N_c} \right)^2\right] \notag \\ 
    &\hspace{3.5cm} + 8 \, \tfrac{N_f}{N_c} \, \dw^{--} \, \dw^{-+} \, \left(16 - 7 \, \tfrac{N_f}{N_c} \right)\Bigg\} .\notag
\end{align}
\end{subequations}
The solution contained in Eqs.~\eqref{ch6fullsoln} and \eqref{ch6fullsoln1} represents a fully analytic solution to the small-$x$, large-$N_c\&N_f$ evolution equations~\eqref{eq_LargeNcNf}. It is valid for any initial conditions $G^{(0)}_{2}(\xoz^2,zs)$, $\widetilde{G}^{(0)}(\xoz^2,zs)$, $Q^{(0)}(\xoz^2,zs)$, $\widetilde{Q}^{(0)}(\xoz^2,zs)$. This solution is the main result of this Chapter.

With the solution we have constructed here, we can straightforwardly express the gluon and flavor-singlet quark helicity PDFs within the DLA using Eqs.~\eqref{ch6pdfsfromdipoles}. We immediately obtain
\begin{subequations}\label{ch6fullsolnpdfs}
\begin{align}
    \label{ch6fullsolnDeltaG}
    &\Delta G(x,Q^2) = \frac{2N_c}{\as \pi^2}\wint\gint e^{\omega\, \ln(1/x) + \gamma\, \ln(Q^2/\Lambda^2)}G_{2\omega\gamma}\,, \\
    \label{ch6fullsolnDeltaSigma}
    &\Delta\Sigma(x,Q^2) = \frac{N_f}{\as \pi^2}\wint\gint e^{\omega \, \ln(1/x) + \gamma \, \ln(Q^2/\Lambda^2)}\widetilde{Q}_{\omega\gamma}\,.
\end{align}
\end{subequations}
Next, we employ \eq{ch6g1_DLA} with our double-Laplace solution and carry out the integrals over $\xoz^2$ and $z$ to obtain
\begin{align}\label{ch6fullsolng1}
    g_1(x,Q^2) = \sum_{f} \frac{Z_f^2}{\as 2\pi^2}\wint\gint \frac{\omega}{\omega-\gamma} \left(\widetilde{Q}_{\omega\gamma} - \widetilde{Q}^{(0)}_{\omega\gamma} \right)e^{\omega\, \ln(1/x) + \gamma\, \ln(Q^2/\Lambda^2)},
\end{align}
where we specify that $\text{Re}\,\omega > \text{Re}\,\gamma$ along their contours.


\section{Connecting to DGLAP}\label{ch6sec:DGLAP}

As the small-$x$ helicity evolution studied here contains the resummation parameter $\as\ln(1/x)\ln(Q^2/\Lambda^2)$ (along with the other double-log resummation parameter $\as \, \ln^2 (1/x)$), the solution constructed herein should contain the solution to the small-$x$ limit of the polarized DGLAP evolution equations \cite{Altarelli:1977zs,Dokshitzer:1977sg,Gribov:1972ri}. We can compare the DGLAP part of our results both to the predictions of BER \cite{Bartels:1996wc} and to the small-$x$, large-$N_c\&N_f$ limit of the existing finite order calculations \cite{Altarelli:1977zs,Dokshitzer:1977sg,Zijlstra:1993sh,Mertig:1995ny,Moch:1999eb,vanNeerven:2000uj,Vermaseren:2005qc,Moch:2014sna,Blumlein:2021ryt,Blumlein:2021lmf,Davies:2022ofz,Blumlein:2022gpp}. 

The analytic solution to the large-$N_c$ version of the small-$x$ helicity evolution constructed in Ch.~\ref{largeNcsoln.ch} allowed us to extract an analytic expression for the (small-$x$, large-$N_c$) $GG$ polarized anomalous dimension $\Delta \gamma_{GG} (\omega)$. This expression, when expanded in powers of $\alpha_s$, completely agreed with all three existing loops of the (small-$x$ and large-$N_c$ limit of the) finite-order calculations \cite{Altarelli:1977zs,Dokshitzer:1977sg,Zijlstra:1993sh,Mertig:1995ny,Moch:1999eb,vanNeerven:2000uj,Vermaseren:2005qc,Moch:2014sna,Blumlein:2021ryt,Blumlein:2021lmf,Davies:2022ofz,Blumlein:2022gpp} and agreed with the first three loops of the perturbative expansion of the BER $GG$ anomalous dimension,  derived in \cite{Borden:2023ugd} and Ch.~\ref{largeNcsoln.ch} using the BER IREE technique. However, $\Delta \gamma_{GG} (\omega)$ from Ch.~\ref{largeNcsoln.ch} disagreed in its overall functional shape with that of BER, which led to a numerically minor disagreement at four loops and beyond in the perturbative expansion of the anomalous dimensions. Here, in the large-$N_c\&N_f$ limit of the evolution, we now have access to all four polarized anomalous dimensions: $\Delta\gamma_{GG}(\omega)$, $\Delta\gamma_{qq}(\omega)$, $\Delta\gamma_{qG}(\omega)$, and $\Delta\gamma_{Gq}(\omega)$. The goal of this Section is to extract an analytic expression (at small-$x$, in the large-$N_c\&N_f$ limit) for each of these polarized anomalous dimensions. Since the large-$N_c$ limit can be taken as a $N_f/N_c \to 0$ sub-limit of the large-$N_c \& N_f$ calculation at hand, we see right away that the disagreement between BER and our $\Delta \gamma_{GG} (\omega)$ anomalous dimensions established in Ch.~\ref{largeNcsoln.ch} should also be contained in the present calculation.

The solution to the spin-dependent DGLAP equations at fixed coupling (e.g., in the DLA) can be written as
\begin{align}\label{ch6DGLAP}
    &\begin{pmatrix}
        \Delta \Sigma(x,Q^2) \\
        \Delta G(x,Q^2)
    \end{pmatrix}
    = \wint e^{\omega\ln(1/x)}
    \begin{pmatrix}
        \Delta\Sigma_{\omega}(Q^2) \\
        \Delta G_{\omega}(Q^2)
    \end{pmatrix}\\
    &\hspace{1.5cm}= \wint e^{\omega\ln(1/x)} \text{exp}\left\{\begin{pmatrix}
        \Delta\gamma_{qq}(\omega) & \Delta\gamma{qG}(\omega) \\
        \Delta\gamma_{Gq}(\omega) & \Delta\gamma{GG}(\omega)
    \end{pmatrix} \ln\frac{Q^2}{\Lambda^2}  \right\}
    \begin{pmatrix}
        \Delta\Sigma_{\omega}(\Lambda^2) \\
        \Delta G_{\omega}(\Lambda^2)
    \end{pmatrix} \,,\notag
\end{align}
with $\Delta\Sigma_{\omega}(\Lambda^2)$ and $\Delta G_{\omega}(\Lambda^2)$ specifying the initial conditions of the helicity PDFs for the evolution at the input scale $\Lambda^2$. As in \cite{Adamiak:2023okq}, let us define the eigenvalues of the anomalous dimension matrix (multiplied by $\ln (Q^2/\Lambda^2)$), 
\begin{subequations}\label{ch6anomdim_eigenvalues}
\begin{align}
    \label{ch6lambda1}
    &\lambda_1 = \frac{1}{2}\left[\Delta\gamma_{qq} + \Delta\gamma_{GG} + \sqrt{\left(\Delta\gamma_{qq} - \Delta\gamma_{GG} \right)^2 + 4 \, \Delta\gamma_{qG} \, \Delta\gamma_{Gq} } \, \right] \, \ln \frac{Q^2}{\Lambda^2} \,,\\
    \label{ch6lambda2}
    &\lambda_2 = \frac{1}{2}\left[\Delta\gamma_{qq} + \Delta\gamma_{GG} -\sqrt{\left(\Delta\gamma_{qq} - \Delta\gamma_{GG} \right)^2 + 4 \, \Delta\gamma_{qG} \, \Delta\gamma_{Gq} } \, \right] \, \ln \frac{Q^2}{\Lambda^2}, 
\end{align}
\end{subequations}
which we can use to exponentiate the matrix of anomalous dimensions in \eq{ch6DGLAP}. Employing these eigenvalues, we write the DGLAP equation's solution as
\begin{align}\label{ch6DGLAPineigs}
    &\begin{pmatrix}
        \Delta\Sigma(x,Q^2) \\
        \Delta G(x,Q^2)
    \end{pmatrix}
    = \wint e^{\omega\ln\tfrac{1}{x}} \begin{pmatrix}
        M_{11}& M_{12} \\
        M_{21} & M_{22}
    \end{pmatrix}
    \begin{pmatrix}
        \Delta\Sigma_{\omega}(\Lambda^2) \\
        \Delta G_{\omega}(\Lambda^2)
    \end{pmatrix} \,,
\end{align}
where the entries of the $2\times 2$ matrix are
\begin{subequations}\label{ch6ch6_dglapmatrix}
\begin{align}
    &M_{11} = \frac{e^{\lambda_1}+e^{\lambda_2}}{2} + \left(\Delta\gamma_{qq}(\omega) - \Delta\gamma_{GG}(\omega)\right)\frac{e^{\lambda_1}-e^{\lambda_2}}{2\left(\lambda_1-\lambda_2 \right)}\ln\tfrac{Q^2}{\Lambda^2} ,\\
    &M_{12} = \Delta\gamma_{qG}(\omega) \frac{e^{\lambda_1}-e^{\lambda_2}}{\lambda_1-\lambda_2}\ln\tfrac{Q^2}{\Lambda^2} ,\\
    &M_{21} = \Delta\gamma_{Gq}(\omega) \frac{e^{\lambda_1}-e^{\lambda_2}}{\lambda_1-\lambda_2}\ln\tfrac{Q^2}{\Lambda^2} ,\\
    &M_{22} = \frac{e^{\lambda_1}+e^{\lambda_2}}{2} - \left(\Delta\gamma_{qq}(\omega) - \Delta\gamma_{GG}(\omega) \right) \frac{e^{\lambda_1}-e^{\lambda_2}}{2\left(\lambda_1-\lambda_2 \right)}\ln\tfrac{Q^2}{\Lambda^2}.
\end{align}
\end{subequations}

Let us choose the simple initial conditions 
\begin{align}\label{ch6DGLAPics}
    \Delta G(x,\Lambda^2) = \frac{2N_c}{\as \pi^2} \quad \text{and} \quad \Delta \Sigma(x,\Lambda^2) = 0\, ,
\end{align}
which correspond in Mellin space to
\begin{align}\label{ch6DGLAPicsmellin}
    \Delta G_{\omega}(\Lambda^2) = \frac{2N_c}{\as\pi^2}\frac{1}{\omega} \quad \text{and}\quad \Delta\Sigma_{\omega}(\Lambda^2) = 0\,,
\end{align}
with which we obtain from \eq{ch6DGLAPineigs} the following expressions for $\Delta\Sigma(x,Q^2)$ and $\Delta G(x,Q^2)$:
\begin{subequations}\label{ch6pdfsfromdglap}
\begin{align}
    \label{ch6DeltaSigmaics1}
    &\Delta\Sigma(x,Q^2) = \frac{2N_c}{\as \pi^2}\wint e^{\omega\ln\tfrac{1}{x}} \frac{1}{\omega}\Delta\gamma_{qG}(\omega)\frac{e^{\lambda_1}-e^{\lambda_2}}{\lambda_1-\lambda_2}\ln\tfrac{Q^2}{\Lambda^2}\,,\\
    \label{ch6DeltaGics1}
    &\Delta G(x,Q^2) = \frac{2N_c}{\as \pi^2}\wint e^{\omega\ln\tfrac{1}{x}} \frac{1}{\omega}\\
    &\hspace{3.5cm}\times\left[\frac{e^{\lambda_1} + e^{\lambda_2}}{2} - \left(\Delta\gamma_{qq}(\omega)-\Delta\gamma_{GG}(\omega)\right)\frac{e^{\lambda_1}-e^{\lambda_2}}{2\left(\lambda_1-\lambda_2\right)}\ln\tfrac{Q^2}{\Lambda^2} \right] \,. \notag
\end{align}
\end{subequations}
In order to extract the anomalous dimensions, we would like to compare Eqs.~\eqref{ch6pdfsfromdglap} to the predictions for $\Delta\Sigma$ and $\Delta G$ obtained from the solution to the small-$x$ helicity evolution we constructed in the previous Sections. To match the initial conditions \eqref{ch6DGLAPics} we take the inhomogeneous terms for our helicity evolution to be (cf. Eqs.~\eqref{ch6pdfsfromdipoles})
\begin{align}\label{ch6DGLAPicsfordipoles}
    G^{(0)}_{2}(\xoz^2, zs) = 1\,,\quad\quad Q^{(0)}(\xoz^2,zs) = \widetilde{Q}^{(0)}(\xoz^2,zs) = \widetilde{G}^{(0)}(\xoz^2,zs) = 0\,,
\end{align}
which straightforwardly give (see Eqs.~\eqref{ch6fullsoln1doubleLaplaceG20}, \eqref{ch6fullsoln1doubleLaplaceGtilde0}, \eqref{ch6fullsoln1doubleLaplaceQ0}, and \eqref{ch6fullsoln1doubleLaplaceQtilde0})
\begin{align}\label{ch6DGLAPicsdoublelaplace}
    G_{2\omega\gamma}^{(0)} = \frac{1}{\omega\gamma}\,,\quad\quad Q^{(0)}_{\omega\gamma} = \widetilde{Q}^{(0)}_{\omega\gamma} = \widetilde{G}^{(0)}_{\omega\gamma} = 0\,.
\end{align}
Using Eqs.~\eqref{ch6DGLAPicsdoublelaplace}, we can construct expressions for the double-Laplace images $G_{2\omega\gamma}$ and $\widetilde{Q}_{\omega\gamma}$ by using Eqs.~\eqref{ch6fullsoln1}. The results are
\begin{subequations}\label{ch6laplaceimagesfromics}
\begin{align}
    \label{ch6G2omegagammafromics}
    &G_{2\omega\gamma} = \frac{-1}{2\sqrt{4-\tfrac{2N_f}{N_c}} \left(\gamma-\gw^{++}\right)\left(\gamma-\gw^{+-}\right)\left(\gamma-\gw^{-+}\right)\left(\gamma-\gw^{--}\right)} \\
    &\times\bigg\{ \frac{2N_f}{N_c}\frac{1}{\omega}\frac{\left(\gamma-r_2^{+-}\right)\left(\gamma-r_2^{--}\right)}{2+\sqrt{4-\tfrac{2N_f}{N_c}}}\left(\gamma-\dw^{++}\right) - \frac{2N_f}{N_c}\frac{1}{\omega}\frac{\left(\gamma-r_2^{++}\right)\left(\gamma-r_2^{-+}\right)}{2-\sqrt{4-\tfrac{2N_f}{N_c}}}\left(\gamma-\dw^{+-}\right) \notag\\
    &- \frac{2\left(\gamma-\gw^{+-}\right)\left(\gamma-\gw^{--}\right)}{\bas\sqrt{s_2(\omega)}}\bigg[\frac{2N_f}{N_c}\frac{1}{\omega}\frac{\left(\gw^{++}-r_2^{+-}\right)\left(\gw^{++}-r_2^{--}\right)}{2+\sqrt{4-\tfrac{2N_f}{N_c}}}\left(\gw^{++}-\dw^{++}\right) \notag\\
    &\hspace{5cm}- \frac{2N_f}{N_c}\frac{1}{\omega}\frac{\left(\gw^{++}-r_2^{++}\right)\left(\gw^{++}-r_2^{-+}\right)}{2-\sqrt{4-\tfrac{2N_f}{N_c}}}\left(\gw^{++}-\dw^{+-}\right) \bigg] \notag \\
    &+ \frac{2\left(\gamma-\gw^{++}\right)\left(\gamma-\gw^{-+}\right)}{\bas\sqrt{s_2(\omega)}}\bigg[\frac{2N_f}{N_c}\frac{1}{\omega}\frac{\left(\gw^{+-}-r_2^{+-}\right)\left(\gw^{+-}-r_2^{--}\right)}{2+\sqrt{4-\tfrac{2N_f}{N_c}}}\left(\gw^{+-}-\dw^{++}\right) \notag\\
    &\hspace{5cm}- \frac{2N_f}{N_c}\frac{1}{\omega}\frac{\left(\gw^{+-}-r_2^{++}\right)\left(\gw^{+-}-r_2^{-+}\right)}{2-\sqrt{4-\tfrac{2N_f}{N_c}}}\left(\gw^{+-}-\dw^{+-}\right) \bigg]     \bigg\} \, ,  \notag \\
    \label{ch6Qomegagammafromics}
    &\widetilde{Q}_{\omega\gamma} = \frac{1}{2\sqrt{4-\tfrac{2N_f}{N_c}} \left(\gamma-\gw^{++}\right)\left(\gamma-\gw^{+-}\right)\left(\gamma-\gw^{-+}\right)\left(\gamma-\gw^{--}\right)} \\
    &\times\bigg\{ \frac{2}{\omega}\left(\gamma-r_1^{+-}\right)\left(\gamma-r_1^{--}\right)\left(\gamma-\dw^{++}\right) - \frac{2}{\omega}\left(\gamma-r_1^{++}\right)\left(\gamma-r_1^{-+}\right)\left(\gamma-\dw^{+-}\right) \notag\\
    &- \frac{2\left(\gamma-\gw^{+-}\right)\left(\gamma-\gw^{--}\right)}{\bas\sqrt{s_2(\omega)}}\bigg[\frac{2}{\omega}\left(\gw^{++}-r_1^{+-}\right)\left(\gw^{++}-r_1^{--}\right)\left(\gw^{++}-\dw^{++}\right) \notag \\
    &\hspace{6cm}- \frac{2}{\omega}\left(\gw^{++}-r_1^{++}\right)\left(\gw^{++}-r_1^{-+}\right)\left(\gw^{++}-\dw^{+-}\right) \bigg] \notag \\
    &+ \frac{2\left(\gamma-\gw^{++}\right)\left(\gamma-\gw^{-+}\right)}{\bas\sqrt{s_2(\omega)}}\bigg[\frac{2}{\omega}\left(\gw^{+-}-r_1^{+-}\right)\left(\gw^{+-}-r_1^{--}\right)\left(\gw^{+-}-\dw^{++}\right) \notag \\
    &\hspace{6cm} - \frac{2}{\omega}\left(\gw^{+-}-r_1^{++}\right)\left(\gw^{+-}-r_1^{-+}\right)\left(\gw^{+-}-\dw^{+-}\right) \bigg] \bigg\} \notag \,.
\end{align}
\end{subequations}

Now with $\Delta G(x,Q^2)$ and $\Delta\Sigma(x,Q^2)$ given in terms of $G_{2\omega\gamma}$ and $\widetilde{Q}_{\omega\gamma}$ in Eqs.~\eqref{ch6fullsolnpdfs}, we would first like to carry out the integrals over $\gamma$. The only non-vanishing poles in the $\gamma$-plane contained in Eqs.~\eqref{ch6laplaceimagesfromics} are those at $\gamma = \gw^{--}$ and $\gamma = \gw^{-+}$. So we carry out the $\gamma$-integrals in Eqs.~\eqref{ch6fullsolnpdfs} by closing the contour to the left and picking up these two simple poles. Schematically,
\begin{subequations}\label{ch6DGLAPgammaintegrals}
\begin{align}
    \label{ch6DeltaGgammaintegral}
    &\Delta G(x,Q^2) = \frac{2N_c}{\as \pi^2}\wint e^{\omega\ln(1/x)} \bigg[e^{\gw^{--}\ln(Q^2/\Lambda^2)} \lim_{\gamma\rightarrow\gw^{--}}\left(\gamma-\gw^{--}\right)G_{2\omega\gamma} \\
    &\hspace{5.1cm}+ e^{\gw^{-+}\ln(Q^2/\Lambda^2)} \lim_{\gamma\rightarrow\gw^{-+}}\left(\gamma-\gw^{-+}\right)G_{2\omega\gamma} \bigg] \,,\notag \\
    \label{ch6DeltaSigmagammaintegral}
    &\Delta\Sigma(x,Q^2) = \frac{N_f}{\as \pi^2}\wint e^{\omega\ln(1/x)} \bigg[e^{\gw^{--}\ln(Q^2/\Lambda^2)} \lim_{\gamma\rightarrow\gw^{--}}\left(\gamma-\gw^{--}\right)\widetilde{Q}_{\omega\gamma} \\
    &\hspace{5.1cm}+ e^{\gw^{-+}\ln(Q^2/\Lambda^2)} \lim_{\gamma\rightarrow\gw^{-+}}\left(\gamma-\gw^{-+}\right)\widetilde{Q}_{\omega\gamma}\bigg] \,,\notag
\end{align}
\end{subequations}
with $G_{2\omega\gamma}$ and $\widetilde{Q}_{\omega\gamma}$ as written in Eqs.~ \eqref{ch6laplaceimagesfromics}. We can rewrite each of Eqs.~\eqref{ch6DGLAPgammaintegrals} in terms of the sum and difference of the two exponential structures.
\begin{subequations}\label{ch6DGLAPgammaintegrals1}
\begin{align}
    \label{ch6DeltaGgammaintegral1}
    &\Delta G(x,Q^2) = \frac{2N_c}{\as \pi^2}\wint e^{\omega\ln(1/x)} \\
    &\hspace{.2cm}\times\bigg[\left(e^{\gw^{--}\ln(Q^2/\Lambda^2)} + e^{\gw^{-+}\ln(Q^2/\Lambda^2)}\right)\notag\\
    &\hspace{3cm}\times\frac{1}{2}\left(\lim_{\gamma\rightarrow\gw^{--}}\left(\gamma-\gw^{--}\right)G_{2\omega\gamma} + \lim_{\gamma\rightarrow\gw^{-+}}\left(\gamma-\gw^{-+}\right)G_{2\omega\gamma} \right) \notag\\
    &\hspace{.4cm} + \left(e^{\gw^{--}\ln(Q^2/\Lambda^2)} - e^{\gw^{-+}\ln(Q^2/\Lambda^2)}\right)\notag \\
    &\hspace{3cm}\times\frac{1}{2}\left(\lim_{\gamma\rightarrow\gw^{--}}\left(\gamma-\gw^{--}\right)G_{2\omega\gamma} - \lim_{\gamma\rightarrow\gw^{-+}}\left(\gamma-\gw^{-+}\right)G_{2\omega\gamma} \right)\bigg] \,,\notag  \\
    \label{ch6DeltaSigmagammaintegral1}
    &\Delta\Sigma(x,Q^2) = \frac{N_f}{\as \pi^2}\wint e^{\omega\ln(1/x)} \\
    &\hspace{.4cm}\times\bigg[\left(e^{\gw^{--}\ln(Q^2/\Lambda^2)} + e^{\gw^{-+}\ln(Q^2/\Lambda^2)}\right)\notag \\
    &\hspace{3cm}\times\frac{1}{2}\left(\lim_{\gamma\rightarrow\gw^{--}}\left(\gamma-\gw^{--}\right)\widetilde{Q}_{\omega\gamma} + \lim_{\gamma\rightarrow\gw^{-+}}\left(\gamma-\gw^{-+}\right)\widetilde{Q}_{\omega\gamma} \right) \notag\\
    &\hspace{.4cm} + \left(e^{\gw^{--}\ln(Q^2/\Lambda^2)} - e^{\gw^{-+}\ln(Q^2/\Lambda^2)}\right)\notag\\
    &\hspace{3cm}\times\frac{1}{2}\left(\lim_{\gamma\rightarrow\gw^{--}}\left(\gamma-\gw^{--}\right)\widetilde{Q}_{\omega\gamma} - \lim_{\gamma\rightarrow\gw^{-+}}\left(\gamma-\gw^{-+}\right)\widetilde{Q}_{\omega\gamma} \right)\bigg] \,.\notag
\end{align}
\end{subequations}

Comparing Eqs.~\eqref{ch6DGLAPgammaintegrals1} to Eqs.~\eqref{ch6pdfsfromdglap}, we make several identifications. First, from the exponentials themselves, we conclude that
\begin{subequations}\label{ch6gammasareeigs}
\begin{align}
    &\frac{\lambda_1(\omega)}{\ln\tfrac{Q^2}{\Lambda^2}} = \gw^{--} \,,\\
    &\frac{\lambda_2(\omega)}{\ln\tfrac{Q^2}{\Lambda^2}} = \gw^{-+} \,.
\end{align}
\end{subequations}
That is, the functions $\gw^{--}$ and $\gw^{-+}$ (which correspond to the two pole structures that survive in the double-Laplace images $G_{2\omega\gamma}$ and $\widetilde{Q}_{\omega\gamma}$) are the eigenvalues of the anomalous dimension matrix. To cross-check this result against the finite-order calculations \cite{Altarelli:1977zs,Dokshitzer:1977sg,Zijlstra:1993sh,Mertig:1995ny,Moch:1999eb,vanNeerven:2000uj,Vermaseren:2005qc,Moch:2014sna,Blumlein:2021ryt,Blumlein:2021lmf,Davies:2022ofz,Blumlein:2022gpp} we can expand the quantities in Eqs.~\eqref{ch6gammasareeigs} (or equivalently in Eqs.~\eqref{ch6fullsoln1news1}) in powers of $\as$. For our functions $\gw^{--}$ and $\gw^{-+}$, we find ($\beta = \pm$)
\begin{align}\label{ch6eigenvalueexpansions}
    &\gamma^{-,\beta}_{\omega} =
    \left(\frac{\as N_c}{4\pi} \right)\frac{1}{2}\left[9 + \sqrt{49 - 16\tfrac{N_f}{N_c}}  \right] \frac{1}{\omega} \\
    &\hspace{0.9cm}+ \left(\frac{\as N_c}{4\pi} \right)^2 \frac{1}{2}\frac{1}{\left(49 - 16\tfrac{N_f}{N_c}\right)} \bigg[\left(49-16\tfrac{N_f}{N_c}\right)\left(33-8\tfrac{N_f}{N_c}\right) \notag \\
    &\hspace{8cm}-\beta \sqrt{49 - 16\tfrac{N_f}{N_c}}\left(217 - 80\tfrac{N_f}{N_c} \right) \bigg] \frac{1}{\omega^3} \notag \\ 
    &\hspace{0.9cm} + \left(\frac{\as N_c}{4\pi} \right)^3 \frac{1}{\left(49 - 16\tfrac{N_f}{N_c}\right)^2} \bigg[\left(49 - 16\tfrac{N_f}{N_c}\right)^2 \left(225 - 64\tfrac{N_f}{N_c} \right) \notag \\
    &\hspace{5.4cm} -\beta \sqrt{49 - 16\tfrac{N_f}{N_c}} \bigg(76489 - 60712\tfrac{N_f}{N_c} + 14784\left(\tfrac{N_f}{N_c}\right)^2 \notag \\
    &\hspace{8.5cm}-1024\left(\tfrac{N_f}{N_c}\right)^3 \bigg) \bigg] \frac{1}{\omega^5} \notag \\
    &\hspace{1.8cm} + \mathcal{O}\left(\as^4\right) \notag .
\end{align}
Meanwhile, the small-$x$ large-$N_c\&N_f$ limit of the polarized splitting functions calculated to three loops are \cite{Altarelli:1977zs,Dokshitzer:1977sg,Mertig:1995ny,Moch:2014sna} (with the bar over each splitting function to denote that it was calculated in the $\overline{\text{MS}}$ scheme)
\begin{subequations}\label{ch6msbarsplittingfuncs}
    \begin{align}
    \label{ch6Pqq}
    &\Delta \overline{P}_{qq}(x) = \left(\frac{\alpha_sN_c}{4\pi}\right) + \left(\frac{\alpha_s N_c}{4\pi}\right)^2 \left( \frac{1}{2}-2\frac{N_f}{N_c} \right)\ln^2\frac{1}{x} \\
    &\hspace{1.6cm}+ \left(\frac{\alpha_sN_c}{4\pi}\right)^3\frac{1}{12} \left( 1-20\frac{N_f}{N_c} \right) \ln^4\frac{1}{x} + {\cal O} (\alpha_s^4) \,, \notag \\
    \label{ch6eq:PqG}
    &\Delta \overline{P}_{qG}(x) =  - \left(\frac{\alpha_sN_c}{4\pi}\right)\frac{2N_f}{N_c} - \left(\frac{\alpha_sN_c}{4\pi}\right)^2 5\frac{N_f}{N_c}\ln^2\frac{1}{x} \\
    &\hspace{2.1cm}- \left(\frac{\alpha_sN_c}{4\pi}\right)^3\frac{1}{6}\frac{N_f}{N_c} \left( 34-4\frac{N_f}{N_c} \right) \ln^4\frac{1}{x} + {\cal O} (\alpha_s^4) \,, \notag \\
    \label{ch6eq:PGq}
    &\Delta \overline{P}_{Gq}(x) =   2\left(\frac{\alpha_sN_c}{4\pi}\right) + 5\left(\frac{\alpha_s N_c}{4\pi}\right)^2 \ln^2\frac{1}{x} \\
    &\hspace{2cm}+ \left(\frac{\alpha_s N_c}{4\pi}\right)^3\frac{1}{6} \left( 36-4\frac{N_f}{N_c} \right) \ln^4\frac{1}{x} + {\cal O} (\alpha_s^4)  \,,  \notag  \\
    \label{ch6PGG} 
    &\Delta \overline{P}_{GG}(x) =   8 \left(\frac{\alpha_sN_c}{4\pi}\right) + \left(\frac{\alpha_s N_c}{4\pi}\right)^2 \left( 16-2\frac{N_f}{N_c} \right) \ln^2\frac{1}{x} \\
    &\hspace{2.1cm}+ \left(\frac{\alpha_s N_c}{4\pi}\right)^3\frac{1}{3} \left( 56-11\frac{N_f}{N_c} \right) \ln^4\frac{1}{x} + {\cal O} (\alpha_s^4) \,.\notag 
\end{align}
\end{subequations}
We have explicitly checked that if one converts these finite-order polarized splitting functions $\Delta\overline{P}_{ij}(x)$ to the corresponding polarized anomalous dimensions $\Delta\overline{\gamma}_{ij}(\omega)$ (with the bar denoting the $\overline{\text{MS}}$ scheme again) defined by
\begin{align}\label{ch6anomdimdef}
    \Delta\overline{\gamma}_{ij}(\omega) = \int\limits_0^1\mathrm{d}x \,x^{\omega-1} \Delta\overline{P}_{ij}(x) 
\end{align}
and subsequently constructs the eigenvalues of the anomalous dimension matrix $\frac{\lambda_1}{\ln(Q^2/\Lambda^2)}$ and $\frac{\lambda_2}{\ln(Q^2/\Lambda^2)}$ via Eqs.~\eqref{ch6anomdim_eigenvalues}, one finds exactly the same perturbative expansion as that in \eq{ch6eigenvalueexpansions}.\footnote{Note that we also have the freedom to swap $\lambda_1 \leftrightarrow \lambda_2$ in Eqs.~\eqref{ch6gammasareeigs}. On the basis of comparing Eqs.~\eqref{ch6DGLAPgammaintegrals1} to Eqs.~\eqref{ch6pdfsfromdglap} without doing further calculations, we cannot uniquely identify the eigenvalues: the fact that the perturbative expansion \eqref{ch6eigenvalueexpansions} matches the finite-order calculations confirms that our choice in Eqs.~\eqref{ch6gammasareeigs} is correct.} 

Note that in Ch.~\ref{transitionops.ch}, the polarized splitting functions were extracted from the most recent version of the small-$x$ evolution equations (the version including the transition operators) order by order in $\as$, up to four loops. There, it was observed that the splitting functions predicted by the small-$x$ helicity evolution agree exactly with the full 3 loops of finite order calculations for $\Delta P_{qq}$ and $\Delta P_{GG}$, but disagree at the third loop for $\Delta P_{qG}$ and $\Delta P_{Gq}$ (though this is ultimately attributable to a scheme dependence \cite{Moch:2014sna}). However at fixed coupling, a scheme transformation of the anomalous dimension matrix reduces only to a rotation. Since rotations do not affect the eigenvalues of a matrix, the eigenvalues of the anomalous dimensions matrix are scheme invariant (in the approximation where one can neglect the running of the coupling, such as our DLA). 
Indeed, an anomalous dimension matrix in any scheme can then be rotated to a scheme where the new matrix is diagonal (the `eigenscheme'), the diagonal entries being the eigenvalues. Thus the anomalous dimension matrix eigenvalues calculated (at fixed coupling) in any scheme ought to agree, and so indeed our \eq{ch6eigenvalueexpansions} agrees with the equivalent expansion calculated in $\overline{\text{MS}}$, despite the fact that our splitting functions $\Delta P_{qG}(x)$ and $\Delta P_{Gq}(x)$ disagree at $\mathcal{O}(\as^3)$.

Returning to Eqs.~\eqref{ch6DGLAPgammaintegrals1} and their comparison to Eqs.~\eqref{ch6pdfsfromdglap}, we can make several additional identifications by matching the coefficients of the exponential structures:
\begin{subequations}\label{ch6PDFids}
\begin{align}
    \label{ch6PDFid1}
    &\frac{1}{\omega} = \left[ \lim_{\gamma\rightarrow\gw^{--}}\left(\gamma-\gw^{--}\right)G_{2\omega\gamma} + \lim_{\gamma\rightarrow\gw^{-+}}\left(\gamma-\gw^{-+}\right)G_{2\omega\gamma} \right] \,,\\
    \label{ch6PDFid2}
    &\Delta\gamma_{qq}(\omega) - \Delta\gamma_{GG}(\omega) \\
    &\hspace{1cm}= -\omega\left(\gw^{--} - \gw^{-+} \right)\left[ \lim_{\gamma\rightarrow\gw^{--}}\left(\gamma-\gw^{--}\right)G_{2\omega\gamma} - \lim_{\gamma\rightarrow\gw^{-+}}\left(\gamma-\gw^{-+}\right)G_{2\omega\gamma} \right] \,, \notag \\
    \label{ch6PDFid3}
    & 0 = \frac{1}{2}\left[ \lim_{\gamma\rightarrow\gw^{--}}\left(\gamma-\gw^{--}\right)\widetilde{Q}_{\omega\gamma} + \lim_{\gamma\rightarrow\gw^{-+}}\left(\gamma-\gw^{-+}\right)\widetilde{Q}_{\omega\gamma} \right] \,,\\
    \label{ch6PDFid4}
    &\Delta\gamma_{qG}(\omega) = \frac{N_f}{4N_c}\omega \left(\gw^{--} - \gw^{-+}\right)  \left[ \lim_{\gamma\rightarrow\gw^{--}}\left(\gamma-\gw^{--}\right)\widetilde{Q}_{\omega\gamma} - \lim_{\gamma\rightarrow\gw^{-+}}\left(\gamma-\gw^{-+}\right)\widetilde{Q}_{\omega\gamma} \right] \,,
\end{align}
\end{subequations}
where in obtaining Eqs.\eqref{ch6PDFid2} and \eqref{ch6PDFid4} we made use of the identities in Eqs.~\eqref{ch6gammasareeigs}.

Eqs.~\eqref{ch6PDFid1} and \eqref{ch6PDFid3} can be verified to be true by explicit calculation using $G_{2\omega\gamma}$ and $\widetilde{Q}_{\omega\gamma}$ as written (for our particular choice of initial conditions) in Eqs.~\eqref{ch6laplaceimagesfromics}. Meanwhile, we can supplement Eqs.~\eqref{ch6PDFid2} and \eqref{ch6PDFid4} with two additional equations. Using Eqs.~\eqref{ch6anomdim_eigenvalues} and \eqref{ch6gammasareeigs} we can write 
\begin{subequations}\label{ch6sumanddiffofeigs}
\begin{align}
    \label{ch6sumofeigs}
    &\Delta\gamma_{qq}(\omega) + \Delta\gamma_{GG}(\omega) = \gw^{--} + \gw^{-+} \quad\quad \text{and} \\
    \label{ch6diffofeigs}
    & \Delta\gamma_{Gq}(\omega) = \frac{1}{4\Delta\gamma_{qG}(\omega)} \left[\left(\gw^{--} - \gw^{-+}\right)^2 - \left(\Delta\gamma_{qq}(\omega) -\Delta\gamma_{GG}(\omega)\right)^2   \right]\,.
\end{align}
\end{subequations}
Together, Eqs.~\eqref{ch6PDFid2}, \eqref{ch6PDFid4}, \eqref{ch6sumofeigs}, \eqref{ch6diffofeigs} form a system of four equations that we can solve for each of the four polarized anomalous dimensions. After considerable algebra we obtain
\begin{subequations}\label{ch6alladims}
\begin{align}
    \label{ch6alladimsqq}
    &\Delta\gamma_{qq}(\omega) = \frac{1}{2}\bigg\{\gamma_{\omega}^{-+}+\gamma_{\omega}^{--} \\
    &\hspace{2.4cm}- \frac{\gamma_{\omega}^{-+}-\gamma_{\omega}^{--} }{\omega\left(2-\tfrac{N_f}{N_c}\right)\sqrt{s_2(\omega)} } \bigg[3\omega\left(6+\tfrac{N_f}{N_c}\right) - 2\left(8-\tfrac{N_f}{N_c}\right)\left(\delta_{\omega}^{++}+\delta_{\omega}^{+-}\right) \notag\\
    &\hspace{9.2cm}+ 8\sqrt{4-\tfrac{2N_f}{N_c}}\left(\delta_{\omega}^{++}-\delta_{\omega}^{+-}\right) \bigg] \bigg\} \,, \notag \\
    \label{ch6alladimsGG}
    &\Delta\gamma_{GG}(\omega) = \frac{1}{2}\bigg\{\gamma_{\omega}^{-+}+\gamma_{\omega}^{--} \\
    &\hspace{2.4cm}+ \frac{\gamma_{\omega}^{-+}-\gamma_{\omega}^{--} }{\omega\left(2-\tfrac{N_f}{N_c}\right)\sqrt{s_2(\omega)} } \bigg[3\omega\left(6+\tfrac{N_f}{N_c}\right) - 2\left(8-\tfrac{N_f}{N_c}\right)\left(\delta_{\omega}^{++}+\delta_{\omega}^{+-}\right)\notag \\
    &\hspace{9.2cm}+ 8\sqrt{4-\tfrac{2N_f}{N_c}}\left(\delta_{\omega}^{++}-\delta_{\omega}^{+-}\right)   \bigg]     \bigg\}\,, \notag \\
    \label{ch6alladimsqG}
    &\Delta\gamma_{qG}(\omega) = -\frac{N_f}{4N_c}\bigg\{\frac{\gamma_{\omega}^{-+}-\gamma_{\omega}^{--} }{\omega\left(2-\tfrac{N_f}{N_c}\right)\sqrt{s_2(\omega)} } \bigg[8\omega\left(2+\tfrac{N_f}{N_c}\right) - 16\left(\delta_{\omega}^{++}+\delta_{\omega}^{+-}\right) \\
    &\hspace{9.2cm}+ 8\sqrt{4-\tfrac{2N_f}{N_c}}\left(\delta_{\omega}^{++}-\delta_{\omega}^{+-}\right)   \bigg]     \bigg\}\,, \notag\\
    \label{ch6alladimsGq}
    &\Delta\gamma_{Gq}(\omega) = \frac{1}{4}\bigg\{ \frac{\gamma_{\omega}^{-+}-\gamma_{\omega}^{--} }{\omega\left(2-\tfrac{N_f}{N_c}\right)\sqrt{s_2(\omega)} } \bigg[8\omega\left(2+\tfrac{N_f}{N_c}\right) - 16\left(\delta_{\omega}^{++}+\delta_{\omega}^{+-}\right) \\
    &\hspace{9.2cm}+ 8\sqrt{4-\tfrac{2N_f}{N_c}}\left(\delta_{\omega}^{++}-\delta_{\omega}^{+-}\right)   \bigg]     \bigg\}. \notag
\end{align}
\end{subequations}
Thus in Eqs.~\eqref{ch6alladims} we have another important result of this Chapter --- fully resummed, all-order in $\as$ expressions for the (small-$x$ and large-$N_c\&N_f$) polarized DGLAP anomalous dimensions. 

Subsequently, we can expand each of these anomalous dimensions in powers of $\as$ to obtain
\begin{subequations}\label{ch6alladimexpansions}
    \begin{align}
    \label{ch6ourdeltagammaqqexpanded}
    &\Delta\gamma_{qq}(\omega) = \left(\frac{\as N_c}{4\pi}\right)\frac{1}{\omega} + \left(\frac{\as N_c}{4\pi}\right)^2 \left(1-4\tfrac{N_f}{N_c}\right) \frac{1}{\omega^3} \\
    &\hspace{1.55cm}+ \left(\frac{\as N_c}{4\pi}\right)^3\,2\left(1-20\tfrac{N_f}{N_c}\right)\frac{1}{\omega^5} + \left(\frac{\as N_c}{4\pi}\right)^4 \left(5-748\tfrac{N_f}{N_c} + 80\tfrac{N_f^2}{N_c^2}\right)\frac{1}{\omega^5} \notag \\
    &\hspace{1.5cm}+ \mathcal{O}\left(\as^5\right) \notag\,,\\
    \label{ch6ourdeltagammaggexpanded}
    &\Delta\gamma_{GG}(\omega) = \left(\frac{\as N_c}{4\pi}\right)\,8\,\frac{1}{\omega} + \left(\frac{\as N_c}{4\pi}\right)^2 \, 4\left(8-\tfrac{N_f}{N_c}\right)\frac{1}{\omega^3} \\
    &\hspace{1.7cm}+ \left(\frac{\as N_c}{4\pi}\right)^3\,8\left(56-11\tfrac{N_f}{N_c}\right) \frac{1}{\omega^5} +  \left(\frac{\as N_c}{4\pi}\right)^4\, 4\left(1984-549\tfrac{N_f}{N_c}+20\tfrac{N_f^2}{N_c^2}\right)\frac{1}{\omega^7} \notag \\
    &\hspace{1.7cm}+ \mathcal{O}\left(\as^5\right) \notag \,, \\
    \label{ch6ourdeltagammaqGexpanded}
    &\Delta\gamma_{qG}(\omega) = -\left(\frac{\as N_c}{4\pi}\right)\,2\tfrac{N_f}{N_c}\frac{1}{\omega} -\left(\frac{\as N_c}{4\pi}\right)^2\,10\tfrac{N_f}{N_c}\frac{1}{\omega^3} \\
    &\hspace{2cm}-\left(\frac{\as N_c}{4\pi}\right)^3\,4\tfrac{N_f}{N_c}\left(35-4\tfrac{N_f}{N_c}\right)\frac{1}{\omega^5} -\left(\frac{\as N_c}{4\pi}\right)^4\,2\tfrac{N_f}{N_c}\left(1213-224\tfrac{N_f}{N_c}\right)\frac{1}{\omega^7} \notag \\
    &\hspace{2cm}+ \mathcal{O}\left(\as^5\right) \notag \,, \\
    \label{ch6ourdeltagammaGqexpanded}
    &\Delta\gamma_{Gq}(\omega) = \left(\frac{\as N_c}{2\pi}\right)\frac{2}{\omega} + \left(\frac{\as N_c}{2\pi}\right)^2 \frac{10}{\omega^3} \\
    &\hspace{1.65cm}+ \left(\frac{\as N_c}{2\pi}\right)^3 \, 4\left(35-4\tfrac{N_f}{N_c}\right) \frac{1}{\omega^5} + \left(\frac{\as N_c}{2\pi}\right)^4 \, 2\left(1213 - 224\tfrac{N_f}{N_c}\right) \frac{1}{\omega^7} \notag \\
    &\hspace{1.65cm}+ \mathcal{O}\left(\as^5\right)\,. \notag
\end{align}
\end{subequations}
All the expansions in Eqs.~\eqref{ch6alladimexpansions} agree completely with those obtained by solving the small-$x$ evolution equations iteratively in Ch.~\ref{transitionops.ch}. The comparisons with BER and finite-order calculations remain the same as discussed in that Chapter. Equations~\eqref{ch6alladimexpansions} again exhibit full agreement with BER to three loops, with small disagreements beginning at four loops in all the polarized anomalous dimensions \cite{Bartels:1996wc,Blumlein:1996hb} just as in the large-$N_c$ case of Ch.~\ref{largeNcsoln.ch}. Equations~\eqref{ch6alladimexpansions} are in full agreement with all 3 loops of finite-order calculation for $\Delta\gamma_{qq}$ and $\Delta\gamma_{GG}$, but disagree with finite-order starting at three loops for $\Delta\gamma_{qG}$ and $\Delta\gamma_{Gq}$. This last disagreement (between the IREE results \cite{Bartels:1996wc,Blumlein:1996hb} and the finite-order calculations) was already known in \cite{Moch:2014sna} and was attributed to a scheme dependence there. The scheme transformation at small $x$ was explicitly constructed in Appendix B of \cite{Borden:2024bxa}.

Also of note is our prediction that, as shown by Eqs.~\eqref{ch6alladimsqG} and \eqref{ch6alladimsGq},
\begin{align}\label{ch6relationbtwqGandGqanomdims}
    \Delta\gamma_{Gq}(\omega) = -\frac{N_c}{N_f}\Delta\gamma_{qG}(\omega) \,,
\end{align}
which of course also implies
\begin{align}\label{ch6relationbtwqGandGqsplittingfuncs}
    \Delta P_{Gq}(x) = -\frac{N_c}{N_f}\Delta P_{qG}.
\end{align}
This relationship between the $qG$ and $Gq$ polarized splitting functions was observed in Ch.~\ref{transitionops.ch} where the splitting functions were constructed order by order in $\as$ up to four loops. However, here we have demonstrated that this prediction persists to all orders in the coupling (at small $x$ and large $N_c \& N_f$). The splitting functions obtained within the BER framework \cite{Blumlein:1996hb} obey the same property \eqref{ch6relationbtwqGandGqsplittingfuncs} up to and including the four-loop level (which is the order to which the BER splitting functions in the existing literature are known at large $N_c \& N_f$), even though the splitting functions we obtain at four loops disagree with those from \cite{Blumlein:1996hb}.


\section{\texorpdfstring{Small-$x$ Asymptotics}{Small x Asymptotics}}\label{ch6sec:asymptotics}

\subsection{The Intercept}
The leading small-$x$ asymptotic behavior of our polarized dipole amplitudes and thus also of the helicity distributions is governed by the singularity in the complex-$\omega$ plane with the largest real part. By studying our solution in Eqs.~\eqref{ch6fullsoln} and \eqref{ch6fullsoln1}, and assuming that the initial conditions contain no singularities in the complex-$\omega$ plane with a large real part of $\omega$, one can show that this leading singularity comes from a branch point of the large square root in the function $\gw^{--}$.\footnote{The right-most branch point in $\gw^{-+}$ does not have such a large real part as the right-most branch point of $\gw^{--}$; the branch points in $r_1^{\alpha\beta}$ and $r_2^{\alpha\beta}$ are not the branch points of ${\widetilde Q}_{\omega\gamma}$ and $G_{2\omega\gamma}$ because $r_1^{\alpha\beta}$ and $r_2^{\alpha\beta}$ enter the expressions for ${\widetilde Q}_{\omega\gamma}$ and $G_{2\omega\gamma}$ as $(\gamma - r_i^{\alpha\beta}) (\gamma - r_i^{-\alpha, \beta}) = \gamma^2 - \omega \gamma + r_i^{\alpha\beta} \, r_i^{-\alpha,\beta}$ in the numerator, for $i=1,2$. Since the large square roots in Eqs.~\eqref{ch6fullsoln1r1alphabeta} and \eqref{ch6fullsoln1r2alphabeta} disappear in the product $r_i^{\alpha\beta} \, r_i^{- \alpha, \beta}$, such terms do not generate new branch cuts. Being in the numerator, they do not generate new poles either.} Note that in this Section, we will work in the notation prior to the rescaling of \eq{ch6resc}, so that $\gw^{--}$, $s_1 (\omega)$, and $s_2 (\omega)$ are given by Eqs.~\eqref{ch6gammapmpmfull} and \eqref{ch6gammapmpmfull12}. Equating the argument of the large (outer) square root of $\gw^{--}$ to zero (cf. \eq{ch6gammapmpmfull}), we readily see that the intercept $\omega_b$ of the helicity distributions satisfies the algebraic equation
\begin{align}\label{ch6intercepteq}
    \omega_b^2 + s_1(\omega_b) - \sqrt{s_2(\omega_b)} = 0 ,
\end{align}
again with $s_1(\omega)$ and $s_2(\omega)$ defined in Eqs.~\eqref{ch6gammapmpmfull12}. Due to the complicated functions involved, this is a challenging equation to solve analytically. However, we can solve it numerically for choices of $N_f$ and $N_c$. In the following table we show the numerical values of the branch point $\omega_b$ for several values of $N_f$ with $N_c = 3$ obtained by a numerical solution of \eq{ch6intercepteq} and denoted $\omega_b^{\text{(us)}}$. 

For comparison, we also show, for each $N_f$, the corresponding prediction for this branch point we obtained from the calculation by BER IREE \cite{Bartels:1996wc}, in the large-$N_c\&N_f$ limit. The BER intercepts were calculated numerically by following the work of \cite{Bartels:1996wc} while taking the limit of large $N_c$ and $N_f$ in all the relevant formulas, after which we substituted $N_c = 3$ and the $N_f$ values indicated in Table~\ref{ch6tab:intercepts}. For $N_f = 2, 3, 4$ the BER intercepts  were previously presented in \cite{Adamiak:2023okq}. The $N_f =6$ case has to be taken as a limit $N_f \to 6$. 
\begin{table}[h!]
\centering
\begin{tabular}{|c|c|c|c|}
    \hline
    $N_f$ & $\omega_b^{(\text{us})}$ & $\omega_b^{(\text{BER})}$ &  $\omega_b^{(\text{BER})} - \omega_b^{(\text{us})}$ \\
    \hline
    2 & 3.54523 &  3.54816 & 0.00293  \\
    3 & 3.47910 & 3.48182 & 0.00272 \\
    4 & 3.40514 & 3.40757 & 0.00243 \\
    5 & 3.32036 & 3.32237 & 0.00201 \\
    6 & $3.21930^{(*)}$ & 3.22062 & 0.00132 \\
    7 & 3.08946 & 3.08943 & -0.00003\\
    8 & 2.88228 & 2.87704 & -0.00524\\
    \hline
\end{tabular}
\caption{The leading branch point $\omega_b$ for several values of $N_f$ with $N_c = 3$. $\omega_b^{(\text{us})}$ corresponds to our prediction based on the solution of the small-$x$ evolution in Eqs.~\eqref{eq_LargeNcNf} (that is, on numerically solving \eq{ch6intercepteq}), while $\omega_b^{(\text{BER})}$ corresponds to the predictions of  the IREE formalism by Bartels, Ermolaev, and Ryskin \cite{Bartels:1996wc} obtained here and in \cite{Adamiak:2023okq} while employing the large-$N_c\&N_f$ limit. Also shown in the last column are the differences between the predicted branch points, which are quite small numerically in comparison to the values of the branch points themselves. The asterisk for the $N_f = 6$ line denotes the case where an exact analytic expression for our intercept is available, given in \eq{ch6interceptNfequals2Nc}. }
\label{ch6tab:intercepts}
\end{table}

Comparing BER and our intercepts in Table~\ref{ch6tab:intercepts}, we conclude that the differences between the leading branch points are numerically very minor. Those of BER tend to be larger than ours for $N_f \leq 6$, while for $N_f \geq 7$ it appears that ours begin to grow larger with increasing $N_f$ (indeed, the predictions for $N_f \geq 7$ should probably be taken as a purely theoretical exercise).

It is also worth noting that the branch points we present here, which are based on the most recent version of the small-$x$ evolution as derived in Ch.~\ref{transitionops.ch}, are slightly different than those obtained from the previous version of the large-$N_c \& N_f$ helicity evolution \cite{Cougoulic:2022gbk}. That is, the quark-to-gluon and gluon-to-quark transition operators which were incorporated into the small-$x$ evolution in Ch.~\ref{transitionops.ch} modified the branch points slightly, tending to make them a bit larger than their values obtained from the evolution without the transition operators. For example, our (unpublished) exact analytic solution of the previous version of the large-$N_c \& N_f$ helicity evolution from \cite{Cougoulic:2022gbk}, derived before the work of Ch.~\ref{transitionops.ch}, yielded a leading branch point of 3.31621 for $N_f=4$ and $N_c=3$ (also obtained in a numerical solution of the same equations in \cite{Adamiak:2023okq}), as compared to the updated value from Table \ref{ch6tab:intercepts} of 3.40514. A similar trend is observed when comparing `our' intercepts from Table~\ref{ch6tab:intercepts} to those found in \cite{Adamiak:2023okq}, which were obtained by numerically solving the large-$N_c \& N_f$ helicity evolution equations from \cite{Cougoulic:2022gbk}.

The asterisk (*) in Table~\ref{ch6tab:intercepts} denotes the fact that when $N_f = 2N_c$, \eq{ch6intercepteq} becomes simple enough to solve analytically. The resulting right-most branch point is
\begin{align}\label{ch6interceptNfequals2Nc}
    \omega_b^{(N_f = 2N_c)} = \sqrt{\frac{1}{57}\left(266 + 38\times2^{2/3}\,\text{Re}\left[\left(137 + 9\,i\,\sqrt{107} \right)^{1/3} \right]\right)},
\end{align}
which has a qualitatively similar structure to the analytic intercept found from the large-$N_c$ evolution (\eq{intercept}). Note also that taking $N_f = 0$ in \eq{ch6intercepteq} and solving the equation, one obtains exactly the leading branch point from the large-$N_c$ helicity evolution (\eq{intercept}), as expected.

We conclude that all the helicity dependent quantities considered here grow with the same power of $1/x$ at small-$x$, driven by the leading branch point $\omega_b$ whose numerical values can be found by solving \eq{ch6intercepteq} for any choice of $N_c$ and $N_f$. The power law for the asymptotic behavior is thus (see Eqs.~\eqref{ch6fullsolnpdfs} and \eqref{ch6fullsolng1})
\begin{align}
    \Delta\Sigma(x,Q^2) \sim \Delta G(x,Q^2) \sim g_1(x,Q^2) \sim \left(\frac{1}{x}\right)^{\alpha_h} \,,
\end{align}
where the intercept is explicitly given by
\begin{align}\label{ch6alphah}
    \alpha_h \equiv \sqrt{\bas}\omega_b .
\end{align}


\subsection{Asymptotic Behavior: Integral Around the Leading Branch Cut}

Next we take a more detailed look at the asymptotic behavior of the helicity distributions. By considering the structure of our distributions in the complex-$\omega$ plane near the leading singularity (and not just the leading singularity itself as in the previous Subsection), we will be able to approximate the $\omega$ integrals in Eqs.~\eqref{ch6fullsolnpdfs} and obtain a more detailed description of the behavior of the helicity distributions in their small-$x$ asymptotics. A similar analysis was done in Appendix B of \cite{Kovchegov:2023yzd} (and presented here in Ch.~\ref{sec:largencasymptotics}) for the large-$N_c$ version of the small-$x$ helicity evolution. In this Section we follow very closely to that procedure.

Based on the correspondence with polarized DGLAP established in Sec. \ref{ch6sec:DGLAP} --- in particular using Eqs.~\eqref{ch6pdfsfromdglap} and \eqref{ch6alladims} --- we can write the helicity PDFs for the same initial conditions chosen in \eq{ch6DGLAPicsfordipoles} as
\begin{subequations}\label{ch6preasy_pdfs}
\begin{align}
    &\Delta\Sigma(y,t) = \frac{N_f}{\as 2\pi^2}\wint e^{\omega y}\frac{1}{\omega} \left(e^{t\gw^{--}} - e^{t\gw^{-+}} \right)\\
    &\hspace{3cm}\times\tfrac{8\omega\left(2 + \tfrac{N_f}{N_c}\right) - 16\left(\dw^{++}+\dw^{+-}\right) + 8\sqrt{4-\tfrac{2N_f}{N_c}}\left(\dw^{++}-\dw^{+-} \right)}{\omega\left(2-\tfrac{N_f}{N_c}\right)\sqrt{s_2(\omega)}} \,, \notag \\
    &\Delta G(y,t) = \frac{N_c}{\as\pi^2}\wint e^{\omega y}\frac{1}{\omega}\Bigg[\left(e^{t\gw^{--}} + e^{t\gw^{-+}} \right) \\
    &\hspace{2cm} - \left(e^{t\gw^{--}} - e^{t\gw^{-+}} \right) \tfrac{3\omega\left(6+\tfrac{N_f}{N_c}\right) - 2\left(8-\tfrac{N_f}{N_c}\right)\left(\dw^{++}+\dw^{+-}\right) + 8\sqrt{4-\tfrac{2N_f}{N_c}}\left(\dw^{++} - \dw^{+-}\right)}{\omega\left(2-\tfrac{N_f}{N_c}\right)\sqrt{s_2(\omega)}} \Bigg]\,. \notag
\end{align}
\end{subequations}
Note that here, as in the previous Subsection, we opt to work prior to the rescaling done in \eq{ch6resc} in order to avoid factors of $\sqrt{\bas}$. We have defined $y \equiv \sqrt{\bas}\ln(1/x)$ and $t \equiv \sqrt{\bas}\ln(Q^2/\Lambda^2)$.

\begin{figure}[ht]
\centering
\begin{subfigure}[b]{0.45\textwidth}
    \centering
    \includegraphics[width=0.95\linewidth]{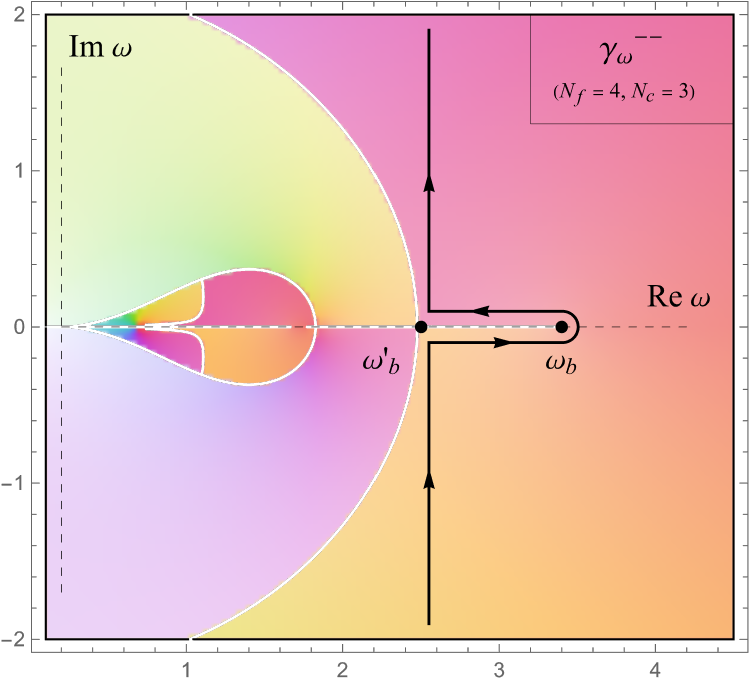}
    \label{ch6fig:eigplotsmm}
    \caption{$\gw^{--}$}
\end{subfigure}
\begin{subfigure}[b]{.45\textwidth}
    \centering
    \includegraphics[width=0.95\linewidth]{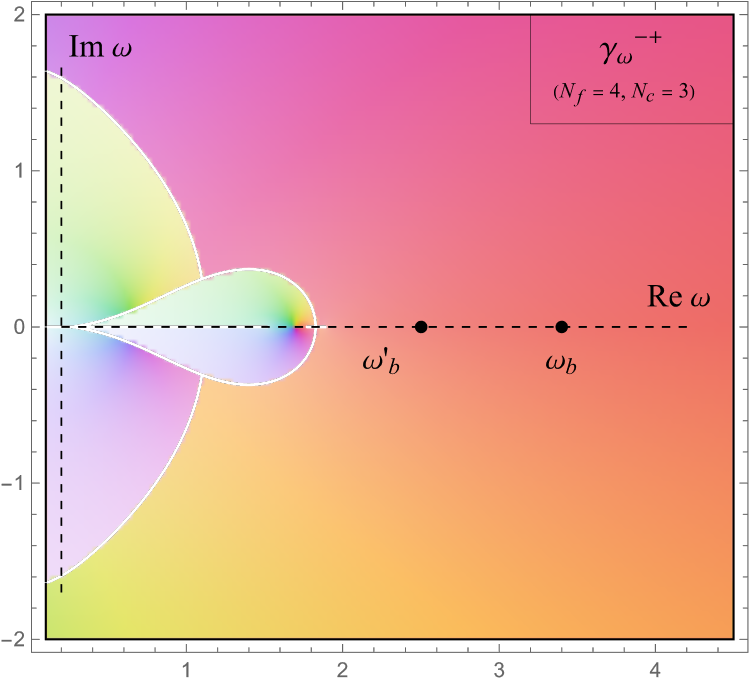}
    \label{ch6fig:eigplotsmp}
    \caption{$\gw^{-+}$}
\end{subfigure}
    \caption{Complex $\omega$-plane structure of the eigenvalues of the anomalous dimension matrix for $N_f = 4$ and $N_c=3$. The left plot shows $\gw^{--}$ with the distorted inverse-Laplace integration contour overlaid (black solid line), while the right shows $\gw^{-+}$ without the integration contour. White lines indicate the branch cuts, while black dashed lines denote the axes. The two rightmost branch points of $\gw^{--}$ are denoted by $\omega_b$ and $\omega_b'$. For comparison, they are also shown on the right panel as well: we observe that $\gw^{-+}$ has no discontinuity in the region between $\omega_b$ and $\omega_b'$. Note that in these plots, colors correspond to the Arg of the plotted function, while the intensity of color corresponds to magnitude (paler color corresponds to larger magnitude).}
    \label{ch6fig:eigplots}
\end{figure}

Now to obtain a more detailed description of the asymptotic behavior of these helicity distributions, we need to approximate the $\omega$ integrals in Eqs.~\eqref{ch6preasy_pdfs} in the vicinity of the rightmost branch point $\omega_b$, which itself comes from the function $\gw^{--}$. In Fig.~\ref{ch6fig:eigplots} we show the graphs illustrating the structure of $\gw^{--}$ and $\gw^{-+}$ in the complex-$\omega$ plane, concentrating on the region around the right-most singularity $\omega_b$ of $\gw^{--}$. Branch cuts are denoted by white lines, while black dashed lines denote the axes and the solid black line in the left panel of Fig.~\ref{ch6fig:eigplots} denotes the integration contour, with $\omega_b'$ the sub leading branch point of $\gw^{--}$.

As shown in Fig.~ \ref{ch6fig:eigplots}, we can wrap the integration contour around the leading branch point $\omega_b$. Then the vertical segments of the integration contour are sub-leading (since they have smaller real parts), and the $\omega$ integral can be approximated as the discontinuity across the leading branch cut on the real axis between $\omega_b'$ and $\omega_b$. Denoting the integrands, including pre-factors, of Eqs.~\eqref{ch6preasy_pdfs} as $\Delta\Sigma_{\omega}$ and $\Delta G_{\omega}$, we can again employ Eqs.~\eqref{ch4_asy_disc} (also see Eqs.~(B6) in \cite{Kovchegov:2023yzd} with the overall sign corrected)
\begin{subequations}\label{ch6preasy_disc}
\begin{align}
    &\Delta\Sigma(y,t) \approx - \lim_{\epsilon\rightarrow 0^+} \int\limits_0^{\infty}\frac{\mathrm{d}\xi}{2\pi i}\left(\Delta\Sigma_{\omega_b-\xi+i\epsilon} - \Delta\Sigma_{\omega_b-\xi-i\epsilon}\right) \,,\\
    &\Delta G(y,t) \approx - \lim_{\epsilon\rightarrow 0^+} \int\limits_0^{\infty}\frac{\mathrm{d}\xi}{2\pi i}\left(\Delta G_{\omega_b-\xi+i\epsilon} - \Delta G_{\omega_b-\xi-i\epsilon}\right) \,,
\end{align}
\end{subequations}
where we have defined $\omega = \omega_b - \xi$. Formally, we would integrate along the contour from $\omega = \omega_b$ to $\omega = \omega_b'$ or vice versa (equivalently, from $\xi = 0$ to $\xi = \omega_b - \omega_b'$). However the factor $e^{\omega y}$ in Eqs.~\eqref{ch6preasy_pdfs} ensures that the dominant contributions to the integrals come from larger values of $\omega$, so we can safely send $\omega_b'\rightarrow -\infty$ (equivalently $\xi\rightarrow \infty$).

Since none of the other functions of $\omega$ involved in Eqs.~\eqref{ch6preasy_pdfs} contain discontinuities in the immediate vicinity of the rightmost branch point/cut, the relevant discontinuity here is only that in $\gw^{--}$. Using Eqs.~\eqref{ch6preasy_pdfs} we can write the discontinuity in each helicity distribution across this branch cut as
\begin{subequations}\label{ch6preasy_disc1}
\begin{align}
    &\Delta\Sigma_{\omega+i\epsilon} - \Delta\Sigma_{\omega-i\epsilon} = \frac{N_f}{\as 2\pi^2} e^{\omega y }\frac{1}{\omega} \left(e^{t\gamma^{--}_{\omega+i\epsilon}} - e^{t\gamma^{--}_{\omega-i\epsilon}} \right)\\
    &\hspace{5.5cm}\times\tfrac{8\omega\left(2 + \tfrac{N_f}{N_c}\right) - 16\left(\dw^{++}+\dw^{+-}\right) + 8\sqrt{4-\tfrac{2N_f}{N_c}}\left(\dw^{++}-\dw^{+-} \right)}{\omega\left(2-\tfrac{N_f}{N_c}\right)\sqrt{s_2(\omega)}} \,, \notag \\
    &\Delta G_{\omega+i\epsilon} - \Delta G_{\omega-i\epsilon} = \frac{N_c}{\as\pi^2} e^{\omega y }\frac{1}{\omega} \left(e^{t\gamma^{--}_{\omega+i\epsilon}} - e^{t\gamma^{--}_{\omega-i\epsilon}} \right) \\
    &\hspace{4.5cm}\times\left(1 - \tfrac{3\omega\left(6+\tfrac{N_f}{N_c}\right) - 2\left(8-\tfrac{N_f}{N_c}\right)\left(\dw^{++}+\dw^{+-}\right) + 8\sqrt{4-\tfrac{2N_f}{N_c}}\left(\dw^{++} - \dw^{+-}\right)}{\omega\left(2-\tfrac{N_f}{N_c}\right)\sqrt{s_2(\omega)}} \right) \,. \notag
\end{align}
\end{subequations}
Then employing Eqs.~\eqref{ch6preasy_disc}, we can write
\begin{subequations}\label{ch6preasy_disc2}
\begin{align}
& \Delta\Sigma(y,t) \approx - \frac{N_f}{\as 2\pi^2}\lim_{\epsilon\rightarrow 0^+} \int\limits_0^\infty\frac{\mathrm{d}\xi}{2\pi i} \frac{e^{(\omega_b-\xi)y}}{\omega_b-\xi}\left(e^{t\gamma^{--}_{\omega_b-\xi+i\epsilon}} - e^{t\gamma^{--}_{\omega_b-\xi-i\epsilon}} \right) \\
    &\hspace{4cm}\times\tfrac{8\left(\omega_b-\xi\right)\left(2 + \tfrac{N_f}{N_c}\right) - 16\left(\delta_{\omega_b-\xi}^{++}+\delta_{\omega_b-\xi}^{+-}\right) + 8\sqrt{4-\tfrac{2N_f}{N_c}}\left(\delta_{\omega_b-\xi}^{++}-\delta_{\omega_b-\xi}^{+-} \right)}{\left(\omega_b-\xi\right)\left(2-\tfrac{N_f}{N_c}\right)\sqrt{s_2(\omega_b-\xi)}} \notag \,,\\
& \Delta G(y,t) \approx - \frac{N_c}{\as \pi^2}\lim_{\epsilon\rightarrow 0^+} \int\limits_0^\infty\frac{\mathrm{d}\xi}{2\pi i} \frac{e^{(\omega_b-\xi)y}}{\omega_b-\xi}\left(e^{t\gamma^{--}_{\omega_b-\xi+i\epsilon}} - e^{t\gamma^{--}_{\omega_b-\xi-i\epsilon}} \right) \\
    &\hspace{2.7cm}\times\left(1 - \tfrac{3\left(\omega_b-\xi\right)\left(6+\tfrac{N_f}{N_c}\right) - 2\left(8-\tfrac{N_f}{N_c}\right)\left(\delta_{\omega_b-\xi}^{++}+\delta_{\omega_b-\xi}^{+-}\right) + 8\sqrt{4-\tfrac{2N_f}{N_c}}\left(\delta_{\omega_b-\xi}^{++} - \delta_{\omega_b-\xi}^{+-}\right)}{\left(\omega_b-\xi\right)\left(2-\tfrac{N-f}{N_c}\right)\sqrt{s_2(\omega_b-\xi)}} \right) \,. \notag
\end{align}
\end{subequations}
Because $y\sim \ln(1/x)$ is very large, the integrals in Eqs.~\eqref{ch6preasy_disc2} are dominated by small values of $\xi$. Hence we can first expand the integrands in powers of $\xi$, then integrate the expansion term by term over $\xi$. We employ the following expansion of $\gamma^{--}_{\omega}$ around its branch point $\omega_b$:
\begin{align}\label{ch6preasy_gammaexpansion}
    \gamma^{--}_{\omega_b-\xi\pm i\epsilon} = \frac{\omega_b}{2} \mp i\frac{\omega_b}{2}\sqrt{C^{(1)}(\omega_b)} \xi^{1/2} - \frac{\xi}{2} \pm i\frac{\omega_b}{8}\frac{C^{(2)}(\omega_b)}{\sqrt{C^{(1)}(\omega_b)}} \xi^{3/2} + \mathcal{O}\left(\xi^{5/2}\right) \,,
\end{align}
where we define 
\begin{subequations}\label{ch6C1andC2}
\begin{align}
    \label{ch6C1}
    &C^{(1)}(\omega_b) = \frac{1}{\omega_b^2}\left[s_1'(\omega_b) - \frac{s_2'(\omega_b)}{2\left[s_1(\omega_b)+\omega_b^2\right]} \right] + \frac{2}{\omega_b} \,,\\
    \label{ch6C2}
    &C^{(2)}(\omega_b) = \frac{1}{\omega_b^2}\left[s_1''(\omega_b) - \frac{s_2''(\omega_b)}{2\left[s_1(\omega_b)+\omega_b^2\right]} + \frac{\left[s_2'(\omega_b)\right]^2}{4\left[s_1(\omega_b) + \omega_b^2\right]^3} \right] + \frac{2}{\omega_b^2}\,.
    \end{align}
\end{subequations}
In Eqs.~\eqref{ch6C1andC2} primes denote differentiation, the functions $s_1(\omega)$ and $s_2(\omega)$ are those defined in Eqs.~\eqref{ch6gammapmpmfull12} (prior to rescaling), and we have used \eq{ch6intercepteq} in several places to replace $\sqrt{s_2(\omega_b)}$ with $s_1(\omega_b) + \omega_b^2$ since it is somewhat easier to evaluate the latter.

Using the expansion in \eq{ch6preasy_gammaexpansion}, and also expanding the rest of the integrands (the parts multiplying $e^{(\omega_b-\xi)y}$) in $\xi$, we obtain for Eqs.~\eqref{ch6preasy_disc2}
\begin{subequations}\label{ch6preasy_disc3}
\begin{align}
    &\Delta \Sigma(y,t) \approx - \frac{N_f}{\as 2\pi^2} t e^{\omega_b\tfrac{t}{2}}\int\limits_0^\infty \frac{\mathrm{d}\xi}{2\pi}e^{\left(\omega_b-\xi\right)y} \frac{\omega_b\sqrt{C^{(1)}(\omega_b)}}{s_1(\omega_b) + \omega_b^2} \Bigg\{-F^{(\Delta\Sigma)}_{\omega_b}\xi^{1/2} \\
    &\hspace{.7cm}+ \bigg[F_{\omega_b}^{(\Delta\Sigma)\prime} -\frac{s_2'(\omega_b)}{2\left[s_1(\omega_b)+\omega_b^2 \right]^2}F^{(\Delta\Sigma)}_{\omega_b} \notag \\
    &\hspace{1.2cm}+ \left(\frac{C^{(2)}(\omega_b)}{4C^{(1)}(\omega_b)} + \frac{t}{2} + \frac{\omega_b^2}{24}t^2C^{(1)}(\omega_b)\right) F^{(\Delta\Sigma)}_{\omega_b}\bigg] \xi^{3/2} + \mathcal{O}\left(\xi^{5/2} \right) \Bigg\} \,,\notag \\
    &\Delta G(y,t) \approx - \frac{N_c}{\as \pi^2} t e^{\omega_b\tfrac{t}{2}}\int\limits_0^\infty \frac{\mathrm{d}\xi}{2\pi}e^{\left(\omega_b-\xi\right)y} \frac{\omega_b\sqrt{C^{(1)}(\omega_b)}}{s_1(\omega_b) + \omega_b^2} \Bigg\{-\frac{s_1(\omega_b)+\omega_b^2}{\omega_b} +F^{(\Delta G)}_{\omega_b}\xi^{1/2} \\
    &\hspace{.7cm}+ \bigg[-\frac{s_1(\omega_b)+\omega_b^2}{\omega_b^2} -F_{\omega_b}^{(\Delta G)\prime} +\frac{s_2'(\omega_b)}{2\left[s_1(\omega_b)+\omega_b^2 \right]^2}F^{(\Delta G)}_{\omega_b} \notag \\
    &\hspace{1cm}+ \left(\frac{C^{(2)}(\omega_b)}{4C^{(1)}(\omega_b)} + \frac{t}{2} + \frac{\omega_b^2}{24}t^2C^{(1)}(\omega_b)\right)\left(\frac{s_1(\omega_b)+\omega_b^2}{\omega_b}  - F^{(\Delta G)}_{\omega_b} \right)\bigg] \xi^{3/2} \notag\\
    &\hspace{1cm}+ \mathcal{O}\left(\xi^{5/2} \right) \Bigg\} \,.\notag
\end{align}
\end{subequations}
In Eqs.~\eqref{ch6preasy_disc3}, primes again denote derivatives and we have defined for brevity
\begin{subequations}\label{ch6FDeltaSigmaandFDeltaG}
\begin{align}
    \label{ch6FDeltaSigma}
    &F^{(\Delta\Sigma)}_{\omega} = \frac{8\omega\left(2+\tfrac{N_f}{N_c}\right) - 16\left(\dw^{++} + \dw^{+-}\right) + 8\sqrt{4-\frac{2N_f}{N_c}}\left(\dw^{++}-\dw^{+-}\right)}{\omega^2\left(2-\tfrac{N_f}{N_c}\right)} \,,\\
    &F^{(\Delta G)}_{\omega} = \frac{3\omega\left(6+\tfrac{N_f}{N_c}\right) - 2\left(8-\tfrac{N_f}{N_c}\right)\left(\dw^{++} + \dw^{+-}\right) + 8\sqrt{4-\frac{2N_f}{N_c}}\left(\dw^{++}-\dw^{+-}\right)}{\omega^2\left(2-\tfrac{N_f}{N_c}\right)}\,.
\end{align}
\end{subequations}
Next we simply carry out the straightforward integrals over $\xi$ in Eqs.~\eqref{ch6preasy_disc3} to obtain the full approximations for the asymptotic behavior of $\Delta\Sigma(y,t)$ and $\Delta G(y,t)$. The results can be written
\begin{subequations}\label{ch6preasymptoticexpansion}
\begin{align}
    \label{ch6deltasigmapreasym}
    &\Delta \Sigma(y,t) \approx \left[\frac{d_{1,q}(t)}{y^{3/2}} + \frac{d_{2,q}(t)}{y^{5/2}} + \mathcal{O}\left(\frac{1}{y^{7/2}} \right) \right]e^{\omega_b y} \,,\\
    \label{ch6deltagpreasym}
    &\Delta G(y,t) \approx \left[\frac{d_{1,G}(t)}{y^{3/2}} + \frac{d_{2,G}(t)}{y^{5/2}} + \mathcal{O}\left(\frac{1}{y^{7/2}} \right) \right]e^{\omega_b y} \,,
\end{align}
\end{subequations}
with the expansion coefficients given by
\begin{subequations}\label{ch6expansioncoefsds}
\begin{align}
    \label{ch6d1q}
    &d_{1,q}(t) = \frac{N_f}{\as 8 \pi^{5/2}}te^{\omega_b \tfrac{t}{2}} \,\frac{\omega_b\sqrt{C^{(1)}(\omega_b)}}{s_{1}(\omega_b) + \omega_b^2} F^{(\Delta\Sigma)}_{\omega_b} \,,\\
    \label{ch6d2q}
    &d_{2,q}(t) = - \frac{3}{16}\frac{N_f}{\as \pi^{5/2}}te^{\omega_b\tfrac{t}{2}}\,\frac{\omega_b\sqrt{C^{(1)}(\omega_b)}}{s_{1}(\omega_b) + \omega_b^2} \bigg\{ F^{(\Delta\Sigma)\prime}_{\omega_b} - \frac{F^{(\Delta\Sigma)}_{\omega_b}}{2\left[s_1(\omega_b)+\omega_b^2\right]^2}s_2'(\omega_b) \\
    &\hspace{4.9cm} + F^{(\Delta\Sigma)}_{\omega_b}\left[\frac{C^{(2)}(\omega_b)}{4C^{(1)}(\omega_b)} + \frac{t}{2} + \frac{\omega_b^2}{24}t^2 C^{(1)}(\omega_b)\right]  \bigg\} \notag \,,\\
    \label{ch6d1G}
    &d_{1,G}(t) = \frac{N_c}{\as 4 \pi^{5/2}}te^{\omega_b \tfrac{t}{2}} \,\frac{\omega_b\sqrt{C^{(1)}(\omega_b)}}{s_{1}(\omega_b) + \omega_b^2}\left[\frac{s_1(\omega_b)+\omega_b^2}{\omega_b} - F^{(\Delta G)}_{\omega_b}\right] \,,\\
    \label{ch6d2G}
    &d_{2,G}(t) = - \frac{3}{8}\frac{N_c}{\as \pi^{5/2}}te^{\omega_b\tfrac{t}{2}}\,\frac{\omega_b\sqrt{C^{(1)}(\omega_b)}}{s_{1}(\omega_b) + \omega_b^2} \bigg\{ -\frac{s_1(\omega_b)+\omega_b^2}{\omega_b^2} - F^{(\Delta G)\prime}_{\omega_b} \\
    &\hspace{1cm}+ \frac{F^{(\Delta G)}_{\omega_b}}{2\left[s_1(\omega_b)+\omega_b^2\right]^2}s_2'(\omega_b) + \left[\frac{s_1(\omega_b)+\omega_b^2}{\omega_b} - F^{(\Delta G)}_{\omega_b} \right] \notag\\
    &\hspace{6cm}\times
    \left[\frac{C^{(2)}(\omega_b)}{4C^{(1)}(\omega_b)} + \frac{t}{2} + \frac{\omega_b^2}{24}t^2 C^{(1)}(\omega_b)\right]  \bigg\} \,, \notag
\end{align}
\end{subequations}
where, as a reminder, we have defined $y = \sqrt{\bas}\ln(1/x)$ and $t = \sqrt{\bas}\ln(Q^2/\Lambda^2)$. Thus in Eqs.~\eqref{ch6preasymptoticexpansion} and \eqref{ch6expansioncoefsds} we have fully analytic expressions for the functional forms of $\Delta \Sigma$ and $\Delta G$ in the high energy asymptotic limit, now at large-$N_c\&N_f$.

Note that Eqs.~\eqref{ch6preasymptoticexpansion} and \eqref{ch6expansioncoefsds} are valid for the rightmost branch point $\omega_b$ which can be found for any $N_f$ and $N_c$. We can use the values of $\omega_b$ we obtained in Table~\ref{ch6tab:intercepts} to numerically compute the expansion coefficients (as functions of $t\sim\ln(Q^2/\Lambda^2)$) in Eqs.~\eqref{ch6expansioncoefsds}. We can also consider the asymptotic ratio of $\Delta G$ to $\Delta \Sigma$ by computing the ratio $d_{1,G}(t)/d_{1,q}(t)$ (this ratio was previously considered in \cite{Hatta:2018itc, Hatta:2016aoc, Boussarie:2019icw, Kovchegov:2023yzd}). Analytically, this can be written using Eqs.~\eqref{ch6d1q} and \eqref{ch6d1G} in a relatively simple form:
\begin{align}\label{ch6asymptratioanalytic}
    &\left(\frac{\Delta G}{\Delta \Sigma}\right)^{(\text{asympt})} \equiv \frac{d_{1,G}(t)}{d_{1,q}(t)} \\
    &\hspace{2.7cm}= \frac{N_c}{4N_f}\,\frac{\tfrac{-4N_f}{N_c}\left(\delta_{\omega_b}^{+-}+\delta_{\omega_b}^{++}\right) + 2\omega_b\left(2+\tfrac{3N_f}{N_c}\right) -\omega_b^3\left(2-\tfrac{N_f}{N_c}\right)}{2\left(\delta_{\omega_b}^{+-} +\delta_{\omega_b}^{++}\right) + \left(\delta_{\omega_b}^{+-}-\delta_{\omega_b}^{++}\right)\sqrt{4-\tfrac{2N_f}{N_c}} - \omega_b\left(2+\tfrac{N_f}{N_c}\right)}\,. \notag
\end{align}
In Table \ref{ch6tab:preasymptotics} we show, for various choices of $N_f$ with $N_c = 3$, the leading branch point $\omega_b$ (reproduced from Table \ref{ch6tab:intercepts}) along with the asymptotic ratio from \eq{ch6asymptratioanalytic}, evaluated numerically for each choice of $N_f$ and $N_c$.
\begin{table}[h!]
\centering
\begin{tabular}{|c|c|c|}
\hline
$N_f$ & $\omega_b$ & $(\Delta G/\Delta \Sigma)^{\text{(asympt)}}$  \\
\hline
2 & 3.5452 & -4.7871 \\
3 & 3.4791 & -3.0731 \\
4 & 3.4051 & -2.2075 \\
5 & 3.3204 & -1.6786 \\
6 & 3.2193 & -1.3143 \\
7 & 3.0895 & -1.0364 \\
8 & 2.8823 & -0.7872 \\
\hline
\end{tabular}
\caption{Table of small-$x$ intercepts and asymptotic ratios of $\Delta G$ to $\Delta \Sigma$ for values of $N_f$ with $N_c = 3$.}
\label{ch6tab:preasymptotics}
\end{table}
For $N_f = 4$ and $N_c = 3$, it was predicted in \cite{Kovchegov:2023yzd} from the large-$N_c$ version of the small-$x$ helicity evolution that the asymptotic relation between the hPDFs is $\Delta G(y,t) \approx - 3\Delta\Sigma(y,t)$. In \cite{Boussarie:2019icw} it was found, generalizing the formalism of BER (and working for any $N_c$ and $N_f$) that asymptotically $\Delta G(y,t) \approx - 2.29\Delta\Sigma(y,t)$ when $N_f = 4$, $N_c=3$. As can be seen in Table \ref{ch6tab:preasymptotics}, our prediction here for the asymptotic ratio, based on the most recent version of the large-$N_c\&N_f$ small-$x$ helicity evolution, is $\Delta G(y,t) \approx - 2.21\Delta\Sigma(y,t)$. We can see that considering the large-$N_c\&N_f$ version of the small-$x$ evolution has brought us closer to the predictions based on the BER formalism, although as with the intercepts and the polarized DGLAP anomalous dimensions (at four-loops and beyond), small disagreements still persist and probably cannot be entirely attributed to us working in the large-$N_c\&N_f$ approximation (while BER do not employ this limit).


\subsection{Asymptotic Behavior: The Saddle Point Method}

As a complimentary cross check for the results of the previous Section, we re-derive here the leading terms of asymptotic expansions of the hPDFs in Eqs.~\eqref{ch6preasymptoticexpansion}. In the previous Section we integrated the discontinuities of the integrands across the leading branch cut, but in this Section we will alternatively employ the saddle point method. The saddle point technique was recently used in \cite{Ermolaev:2025isl} to determine the small-$x$ asymptotics of hPDFs in the BER framework: that work inspired us to apply it here as well. Taking the same initial conditions from \eq{ch6DGLAPicsfordipoles}, we begin with the hPDFs in Eqs.~\eqref{ch6preasy_pdfs}. Making use of the notation in Eqs.~\eqref{ch6FDeltaSigmaandFDeltaG}, we can write more compactly
\begin{subequations}\label{ch6hpdfssaddlepoint}
\begin{align}
    \label{ch6deltasigmacompact}
    &\Delta\Sigma(y,t) = \frac{N_f}{\as 2\pi^2}\wint e^{\omega y}\left(e^{\gamma^{--}_{\omega}\,t} - e^{\gamma^{-+}_{\omega}\,t} \right)\frac{F^{(\Delta\Sigma)}_\omega}{\sqrt{s_2(\omega)}}\,, \\
    \label{ch6deltaGcompact}
    &\Delta G(y,t) = \frac{N_c}{\as\pi^2}\wint e^{\omega y}\frac{1}{\omega}\left[e^{\gamma^{--}_\omega\,t}\left(1 - \frac{\omega F^{(\Delta G)}_{\omega}}{\sqrt{s_2(\omega)}}   \right) + e^{\gamma^{-+}_\omega\,t}\left(1 + \frac{\omega F^{(\Delta G)}_{\omega}}{\sqrt{s_2(\omega)}}   \right)  \right]\,.
\end{align}
\end{subequations}
The saddle point of the terms containing the exponential $e^{\omega y + \gw^{--}t}$ for large $y$ and $t$ is determined by 
\begin{align}\label{ch6saddlepointdiffeq}
    \frac{\mathrm{d}}{\mathrm{d}\omega}\left(\omega y + \gamma^{--}_\omega t\right) = 0\,.
\end{align}
We will denote the saddle point (that is, the solution of \eq{ch6saddlepointdiffeq}) by $\omega = \omega_{sp}$. One can show that a saddle-point evaluation of the terms containing the exponential $e^{\omega y + \gw^{-+}t}$ leads to a sub-leading contribution at large $y$ in each of Eqs.~\eqref{ch6hpdfssaddlepoint}, as compared to the terms containing $e^{\omega y + \gw^{--}t}$: we can, therefore, discard the terms containing $e^{\omega y + \gw^{-+}t}$ in our evaluation below.

One can further show that $\omega_{sp}$ defined by \eq{ch6saddlepointdiffeq} lies in the vicinity (and to the right) of the leading branch point $\omega_b$ considered in the previous Section (as the ratio $y/t$ approaches infinity, $\omega_{sp}$ can be seen to approach $\omega_b$). Therefore, to employ the saddle method in order to approximate the integrals in Eqs.~\eqref{ch6hpdfssaddlepoint} we can again employ \eq{ch6preasy_gammaexpansion}, which contains our expansion of $\gamma^{--}_{\omega}$ near the leading branch point $\omega_b$. Employing \eq{ch6preasy_gammaexpansion}, now using $\omega = \omega_b + \xi$, we write
\begin{align}\label{ch6gammammexpansionsaddle}
    \gamma^{--}_{\omega_b+\xi} = \frac{\omega_b}{2} - \frac{\omega_b}{2}\sqrt{C^{(1)}(\omega_b)}\xi^{1/2} + \frac{\xi}{2} + \mathcal{O}\left(\xi^{3/2}\right)\,.
\end{align}
Using \eq{ch6gammammexpansionsaddle} in \eq{ch6saddlepointdiffeq}, and defining $\omega_{sp} = \omega_b+\xi_{sp}$, we find
\begin{align}\label{ch6xisp}
    \xi_{sp} = \frac{\omega_b^2C^{(1)}(\omega_b)}{4\left(   1+\frac{2y}{t}\right)^2} \approx \frac{\omega_b^2 C^{(1)}(\omega_b) t^2}{16 y^2}   \,.
\end{align}
In the last step we have assumed that $y\gg t$, as is proper for high-energy asymptotics: in this limit, indeed, $\xi_{sp}$ is small and the saddle point $\omega_{sp} = \omega_b + \xi_{sp}$ is close to the leading branch point $\omega_b$.

Neglecting the terms proportional to $e^{\gamma_\omega^{-+}t}$, we evaluate the hPDFs in Eqs.~\eqref{ch6hpdfssaddlepoint} around the saddle point $\omega_{sp}$:
\begin{subequations}\label{ch6hpdfssaddlepoint1}
\begin{align}
    &\Delta\Sigma(y,t) \approx \frac{N_f}{\as 2\pi^2} \wint e^{\omega_{sp}y + \gamma^{--}_{\omega_{sp}}\,t + \tfrac{1}{2}t\left(\omega-\omega_{sp}\right)^2 \left(\gamma^{--}_{\omega_{sp}}\right)''} \frac{F^{(\Delta\Sigma)}_{\omega_{sp}}}{\sqrt{s_2(\omega_{sp})}} \,,\\
    &\Delta G(y,t) \approx \frac{N_c}{\as \pi^2} \wint  e^{\omega_{sp}y + \gamma^{--}_{\omega_{sp}}\,t + \tfrac{1}{2}t\left(\omega-\omega_{sp}\right)^2 \left(\gamma^{--}_{\omega_{sp}}\right)''} \frac{1}{\omega_{sp}} \left(1 - \frac{\omega_{sp} F^{(\Delta G)}_{\omega_{sp}}}{\sqrt{s_2(\omega_{sp})}}\right)\,.
\end{align}
\end{subequations}
Now we evaluate the integrands around the saddle point. Employing the expansion in \eq{ch6gammammexpansionsaddle} along with the saddle point $\omega_{sp} = \omega_b + \xi_{sp}$, with $\xi_{sp}$ in \eq{ch6xisp},
we write
\begin{subequations}\label{ch6hpdfssaddlepoint2}
\begin{align}
    &\Delta\Sigma(y,t) \approx \frac{N_f}{\as2\pi^2} \frac{F^{(\Delta \Sigma)}_{\omega_b}}{\sqrt{s_2(\omega_b)}} e^{\omega_b y +\tfrac{\omega_b}{2}t - \tfrac{\omega_b^2 C^{(1)}(\omega_b)}{16}\tfrac{t^2}{y} }
    \int\limits_{-\infty}^{\infty}\frac{\mathrm{d}\nu}{2\pi} \text{exp}\left\{-\frac{t}{2}\frac{\left[1+(2y/t)\right]^3 }{\omega_b^2C^{(1)}(\omega_b)} \nu^2\right\} \,,\\
    &\Delta G(y,t) \approx \frac{N_c}{\as\pi^2}\frac{1}{\omega_b}\left(1 - \frac{\omega_b F^{(\Delta G)}_{\omega_b}}{\sqrt{s_2(\omega_b)}}\right) e^{\omega_b y +\tfrac{\omega_b}{2}t - \tfrac{\omega_b^2 C^{(1)}(\omega_b)}{16}\tfrac{t^2}{y} }\\
    &\hspace{8cm}\times
    \int\limits_{-\infty}^{\infty}\frac{\mathrm{d}\nu}{2\pi} \text{exp}\left\{-\frac{t}{2}\frac{\left[1+(2y/t)\right]^3 }{\omega_b^2C^{(1)}(\omega_b)} \nu^2\right\}\,,\notag
\end{align}
\end{subequations}
where we integrate along the vertical contour $\omega = \omega_b + i\nu$. Since we are interested in reproducing the leading high-$y$ asymptotics here, we have neglected higher powers of $\xi_{sp}$ in the parts of the integrands multiplying the exponentials: these pre-factors are now outside the $\nu$-integrals.

Carrying out the Gaussian integrals, we arrive at the small-$x$ asymptotics of the hPDFs,
\begin{subequations}\label{ch6hpdfssaddlepoint3}
\begin{align}
    &\Delta \Sigma(y,t) = \frac{e^{\omega_b y }}{y^{3/2}} \frac{N_f}{\as 8 \pi^{5/2}}\,t\,\,e^{\tfrac{\omega_b}{2}t - \tfrac{\omega_b^2 C^{(1)}(\omega_b)}{16}\tfrac{t^2}{y}}\,\, \frac{\omega_b\sqrt{C^{(1)}(\omega_b)}}{s_1(\omega_b) + \omega_b^2 } F^{(\Delta\Sigma)}_{\omega_b} \,,\\
    &\Delta G(y,t) = \frac{e^{\omega_b y}}{y^{3/2}} \frac{N_c}{\as 4\pi^{5/2}}\,t\,\,e^{\tfrac{\omega_b}{2}t - \tfrac{\omega_b^2 C^{(1)}(\omega_b)}{16}\tfrac{t^2}{y}}\,\,\frac{\omega_b\sqrt{C^{(1)}(\omega_b)}}{s_1(\omega_b) + \omega_b^2}\left(\frac{s_1(\omega_b) + \omega_b^2}{\omega_b} - F^{(\Delta G)}_{\omega_b}\right)\,,
\end{align}
\end{subequations}
where we employed the same substitution $\sqrt{s_2(\omega_b)} \rightarrow s_1(\omega_b) + \omega_b^2$ used in the previous Section, since the two quantities are equal at the branch point $\omega_b$, per \eq{ch6intercepteq}.
Note that in addition to the leading term in $y$ we have also obtained a diffusion term in the exponents of both of Eqs.~\eqref{ch6hpdfssaddlepoint3}:
\begin{align}\label{ch6diffusionterm}
    \text{exp}\left(-\frac{\omega_b^2C^{(1)}(\omega_b)}{16} \frac{t^2}{y}\right)\,.
\end{align}
This term is completely analogous to the similar diffusion term in the solution of the unpolarized BFKL \cite{Kuraev:1977fs,Balitsky:1978ic} evolution equation. If we neglect this diffusion term, putting the exponential in \eq{ch6diffusionterm} equal to 1, Eqs.~\eqref{ch6hpdfssaddlepoint3} would then exactly reproduce the first terms of the expansion in Eqs.~\eqref{ch6preasymptoticexpansion} (with the relevant coefficients found in Eqs.~\eqref{ch6d1q} and \eqref{ch6d1G}). Furthermore, expanding the exponential in \eq{ch6diffusionterm} to linear order in $t^2$ and using the result in Eqs.~\eqref{ch6hpdfssaddlepoint3} yields exactly the last (order-$t^3$ in the pre-factor) term of each of $d_{2,q}(t)$ and $d_{2,G}(t)$ in Eqs.~\eqref{ch6d2q} and \eqref{ch6d2G}, showing that our method of integration across the leading branch cut in the previous Section also captured (parts of) this diffusion term. Conversely, the diffusion term appears to capture all the leading-power of $t$ terms in the coefficients of the $1/y$ expansion of the pre-factors in Eqs.~\eqref{ch6preasymptoticexpansion}.  

The fact that, unlike the BFKL case, the helicity evolution allowed us to obtain the explicit expressions \eqref{ch6alladims} for the corresponding anomalous dimensions (cf.~\cite{Borden:2023ugd}) enabled us here to perform the asymptotic analyses both in the branch-cut integral and saddle-point methods.


\section{Chapter Summary}

In this Chapter we have analytically solved the most up-to-date version of the small-$x$ helicity evolution in the large-$N_c\&N_f$ limit, based on the quark-to-gluon and gluon-to-quark transition operators included in Ch.~\ref{transitionops.ch}. Our solution is based on a double-inverse-Laplace transform method and yields analytic expressions for all of the polarized dipole amplitudes. 

We have explicitly constructed the corresponding analytic expressions for the gluon and flavor singlet quark helicity PDFs, along with the $g_1$ structure function. We have successfully cross-checked our solution against the known solution to the spin-dependent DGLAP equations, and in doing so we have extracted analytic predictions for the (fully-resummed) eigenvalues of the matrix of DGLAP polarized anomalous dimensions. Further we extracted analytic predictions for all four individual polarized anomalous dimensions themselves, exact to all orders in $\as$. The expansions of these anomalous dimensions are fully consistent with the known finite order results, to three loops. We have also obtained numerical values for the intercept of the helicity distributions in Table \ref{ch6tab:intercepts}, along with a more detailed description of the asymptotic behavior of these distributions in Eqs.~\eqref{ch6preasymptoticexpansion}. Related to that, we also obtained explicit analytic and numerical predictions (at large-$N_C\&N_f$) for the asymptotic ratio of the gluon helicity PDF to the flavor singlet quark helicity PDF at small-$x$.

Just as with the analytic solution constructed in Ch.~\ref{largeNcsoln.ch} for the large-$N_c$ version of the small-$x$ helicity evolution, we find here the same general trend of full agreement with finite-order calculations and very minor disagreements with the predictions of the BER formalism. In particular, our polarized DGLAP anomalous dimensions are fully consistent with finite order calculations to the existing three loops \cite{Altarelli:1977zs,Dokshitzer:1977sg,Mertig:1995ny,Moch:2014sna} and agree completely with those predicted in the BER formalism to three loops \cite{Bartels:1996wc}. However, our anomalous dimensions and those of BER disagree beginning at the four-loop level \cite{Bartels:1996wc,Blumlein:1996hb}. We find a similarly small disagreement in the numerical values of the intercepts predicted by our solution and those predicted by BER, along with a small disagreement in the asymptotic ratio of $\Delta G$ to $\Delta \Sigma$ \cite{Bartels:1996wc, Boussarie:2019icw}.

Future calculations in the finite order framework for the four-loop polarized splitting functions could resolve the existing discrepancy between the predictions in this work and those of BER. Nevertheless, with the most general limit of the small-$x$ helicity evolution now completely solved, demonstrating solid agreement with existing calculations in several other paradigms, we have established the most complete and accurate description of the high energy spin structure of the proton yet achieved in the small-$x$ helicity evolution framework.

\chapter{Conclusions}
\label{conclusions.ch}

The spin of the proton is distributed among the angular momenta of its quarks and gluons and further spread across a rich phase space in Bjorken-$x$. The question of how all these contributions add up to a spin of $1/2$ --- the proton spin puzzle --- remains a fundamental test in our understanding of QCD and an outstanding question in the landscape of hadronic structure. The region of small-$x$, where the quarks and gluons carry only a small fraction of the proton's momentum, is an essential piece of the puzzle. 

In this dissertation, building on an extensive framework of small-$x$ helicity evolution, we have developed analytic solutions to helicity evolution equations which yield corresponding analytic descriptions of the quark and gluon helicity distributions at small-$x$. 

First we solved the large-$N_c$ evolution constructed originally in \cite{Cougoulic:2022gbk}. Our solution yielded analytic predictions for the small-$x$ intercept $\alpha_h$ and the polarized $GG$ DGLAP anomalous dimension $\Delta\gamma_{GG}(\omega)$. We also constructed explicit expressions for the hPDFs in the asymptotic limit. 

Next we demonstrated that the evolution equations in the more general large-$N_c\&N_f$ limit from \cite{Cougoulic:2022gbk} were incomplete, as they disagreed with robust finite order predictions for the polarized DGLAP splitting functions (and for the eigenvalues of the matrix of polarized DGLAP anomalous dimensions). We subsequently constructed the contributions of the quark-to-gluon and gluon-to-quark shock wave transition operators and incorporated these contributions into the large-$N_c\&N_f$ evolution. The result was a new set of evolution equations which, as we demonstrated order-by-order in $\as$, restored full consistency with finite order predictions for the DGLAP splitting functions.

Finally we applied the techniques developed in the large-$N_c$ solution to solve the more complicated set of modified large-$N_c\&N_f$ evolution equations. The result was a fully analytic description of the hPDFs and $g_1$ structure function (at small-$x$) in the more realistic large-$N_c\&N_f$ limit. We again extracted the small-$x$ intercept and constructed the explicit expressions for the hPDFs in the asymptotic limit, and we extracted analytic expressions exact to all orders in $\as$ for all four polarized DGLAP anomalous dimensions. 

From our predictions (both at large-$N_c$ and large-$N_c\&N_f$), we found complete consistency with the full existing extent of finite order calculations for the polarized anomalous dimensions. However for the small-$x$ intercept, for the anomalous dimensions beyond the precision of finite order calculations (beginning at four loops), and for the ratio of the hPDFs $\Delta G/\Delta \Sigma$ in the asymptotic limit, we found very minor disagreements with the predictions made in the framework of Bartels, Ermolaev, and Ryskin.

The analytic solutions we have constructed here form the most detailed and explicit description of small-$x$ quark and gluon spin yet derived in the small-$x$ helicity paradigm. The nature of our resummation results in predictions exact to all orders in $\as$ which can thus be compared, in principle, to the appropriate limit (small-$x$ and large-$N_c$ or large-$N_c\&N_f$) of any order of finite-order calculations --- in particular, the four-loop calculation of the polarized anomalous dimensions may illuminate the discrepancy between our predictions and those of BER at this order.

In addition, the explicit functional forms we have obtained for the hPDFs in the asymptotic limit can be essential ingredients for phenomenological studies (see \cite{Adamiak:2021ppq, Adamiak:2023yhz,Adamiak:2025dpw}). These predictions can be implemented as parametrizations of the hPDFs to be fit to data and extrapolated to small-$x$ (beyond the minimum $x$ accessible in a given experiment). Furthermore, recent efforts on the orbital angular momentum side of the small-$x$ formalism have resulted in a detailed description of quark and gluon OAM at small-$x$. In particular \cite{Manley:2024pcl} also employed double-inverse-Laplace techniques to solve small-$x$ evolution equations for the OAM distributions and to derive the functional forms of the OAM distributions in the asymptotic limit. These latter calculations can also serve as important inputs for phenomenological studies of orbital angular momentum at small-$x$, another key piece of the spin puzzle.

In conclusion, the tools and solutions we have developed in this dissertation constitute an important step in pushing the proton spin puzzle toward its eventual resolution and solidifying our understanding of the complicated picture of hadronic structure that emerges from Quantum Chromodynamics.



\providecommand{\href}[2]{#2}\begingroup\raggedright\endgroup


\appendix
\chapter{Comparison of some diagrams in BER IREE and shock wave approaches}
\label{berdisagreement.app}

The aim of this Appendix is to speculate on the possible origin of the minor disagreement between the result of BER \cite{Bartels:1996wc} and the analytic solution to the small-$x$, large-$N_c$ helicity evolution found in Ch.~\ref{largeNcsoln.ch}, as manifested in the difference between the intercepts (Eqs.~\eqref{intercept} and \eqref{BER_intercept2}) and the anomalous dimensions (Eqs.~\eqref{anomalous_dim} and \eqref{BER_anom_dim}). Admittedly, the author of this dissertation is not expert enough in the infrared evolution equations (IREE) framework to make any definitive statements, and the discussion below should be understood only as a possible origin of the discrepancy.  
 
The IREE \cite{Gorshkov:1966ht,Kirschner:1983di,Kirschner:1994rq,Kirschner:1994vc,Bartels:1995iu,Bartels:1996wc,Griffiths:1999dj} are based on evolving in the infrared cutoff on the transverse momenta of the quarks and gluons in a $2 \to 2$ forward scattering amplitude. In the original QCD version \cite{Kirschner:1983di}, the IREE for the Reggeon evolution were based on the following observation: the softest loop momentum integral can be driven either by one or two softest partons in the amplitude. Otherwise the amplitude is not double-logarithmic. Note that in this Appendix, soft and hard refer to the transverse momentum of the partons. If there is one softest parton driving the loop integral, then it must be a gluon and Gribov's bremsstrahlung theorem \cite{Gribov:1966hs,Gorshkov:1969yy} (also known as the soft-gluon theorem) applies, allowing one to keep only the diagrams where the soft gluon connects to the external legs. Since, by definition, the loop integral involving the bremsstrahlung gluon is the softest, the dependence on the IR cutoff $\Lambda$ enters the expression for the amplitude only through the transverse momentum part of the integral,
\begin{align}\label{IRlog}
\int\limits_{\Lambda^2} \frac{d k_T^2}{k_T^2}.
\end{align}
Differentiating the amplitude with respect to $\ln \Lambda^2$ would remove the contribution in \eq{IRlog}, thus truncating (removing) the soft gluon. 

If the softest loop involves two softest partons, in the Reggeon evolution of \cite{Kirschner:1983di} they must be quarks (to transfer the flavor between the projectile and the target) contributing two opposite ``rails" of the ladder: these softest quarks also contribute the logarithm of the IR cutoff in \eq{IRlog}. Truncating these soft quarks allows one to split the single forward $2 \to 2$ scattering diagram into two sub-diagrams, each of them containing a $2 \to 2$ forward scattering sub-process \cite{Kirschner:1983di}. These two observations led to the IREE for the $2 \to 2$ Reggeon scattering amplitude constructed in \cite{Kirschner:1983di}. Similar logic was applied in \cite{Bartels:1995iu} to construct double-logarithmic evolution equations for the flavor non-singlet helicity-dependent amplitude. The latter evolution was confirmed at large $N_c$ based on the $s$-channel shock wave approach in \cite{Kovchegov:2016zex}.  

In the flavor-singlet helicity evolution case \cite{Bartels:1996wc}, an additional category of diagrams was added: the two soft partons could be gluons, also comprising two opposite ``rails" of the ladder. Hence, for the flavor-singlet helicity evolution in \cite{Bartels:1996wc} one may have one softest parton dominating the softest loop integral in an amplitude (which has to be a gluon) or two partons, which could either be two quarks or two gluons, forming opposite ``rails" of the ladder. A consequence of this statement appears to be that for the evolution in \cite{Bartels:1996wc} to work, there should be no non-ladder hard gluons and no hard-gluon vertex corrections: only soft ``bremsstrahlung" non-ladder gluons are allowed, for which the soft-gluon theorem \cite{Gribov:1966hs,Gorshkov:1969yy} applies. In \cite{Bartels:1996wc}, starting after Eq.~(2.32) and until the end of Sec.~2, an argument is presented which appears to make the case that no such hard non-ladder gluons and vertex corrections exist in the flavor-singlet helicity evolution.

\begin{figure}[ht]
\begin{center}
\includegraphics[width= \textwidth]{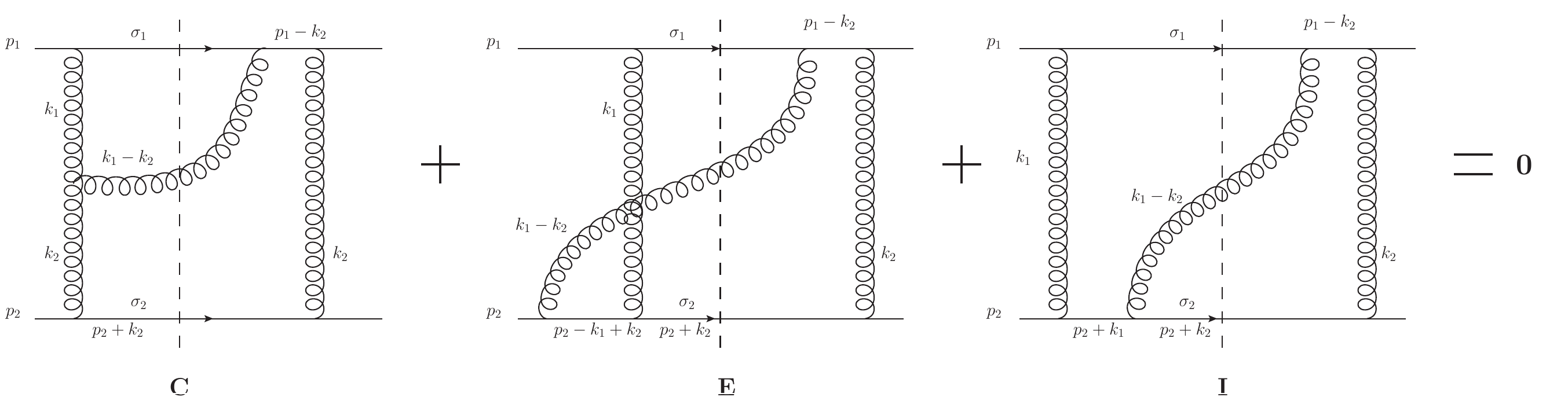} 
\caption{Diagram cancellations in the $k_{2 T} \gg k_{1 T}$ regime with the DLA accuracy, as outlined in \cite{Kirschner:1983di}. We use the diagram labeling from Appendix~B of \cite{Kovchegov:2016zex}. The diagrams are different by different connections of the soft gluon $k_1$ in the lower left corner. Vertical dashed line denotes the cut.}
\label{Feyn_gauge_canc1}
\end{center}
\end{figure}

To better understand this ``no hard non-ladder gluons" assertion, in Appendix~B of \cite{Kovchegov:2016zex} the types of non-ladder gluons were studied by an explicit calculation of several diagrams contributing to the helicity-dependent part of $qq \to qq$ forward scattering at the order $\as^3$. One such diagram, diagram~C in the nomenclature of \cite{Kovchegov:2016zex}, is shown here, in the left panel of \fig{Feyn_gauge_canc1}. The diagram is double-logarithmic: in the $p_1^+, p_2^- \gg k_1^+, k_2^-, k_{1 T}, k_{2 T} \gg k_1^-, k_2^+$ regime it gives (with the center-of-mass energy squared $s \approx 2 p_1^+ p_2^-$) \cite{Kovchegov:2016zex} 
\begin{align}\label{Ceq}
\int\limits_{\Lambda^2}^s d k_{1 T}^2  \int\limits_{\Lambda^2}^s \, d k_{2 T}^2 \ C \sim \int\limits_{\Lambda^2}^s \frac{d k_{1 T}^2}{k_{1 T}^2}  \int\limits_{\Lambda^2}^s \, \frac{d k_{2 T}^2}{k_{2 T}^2}
\end{align}
along with a logarithm of energy resulting from the longitudinal momentum integration (for simplicity we imagine working in a frame with $p_{1 T} = p_{2 T} =0$. The contribution C in \cite{Kovchegov:2016zex} does not explicitly include any of the integrals).

It seems that in the IREE framework, diagram C from the left panel of \fig{Feyn_gauge_canc1} should be separately considered in two different kinematic regions, $k_{1 T} \ll k_{2 T}$ and $k_{1 T} \gg k_{2 T}$. In either kinematic region, the non-ladder gluon $k_1 - k_2$ is hard, thus contributing a hard (cut) vertex correction in an apparent violation of the absence of such gluons in the IREE argued in \cite{Bartels:1996wc}. However, before reaching any conclusions, we should analyze diagram C in more detail. 

In the $k_{1 T} \ll k_{2 T}$ region the $k_1$ gluon is the softest in the diagram and the bremsstrahlung theorem applies. Following Sec.~3.3 of \cite{Kirschner:1983di}, we see that diagram~C for $k_{1 T} \ll k_{2 T}$ falls under the category of Fig.~7 in that reference, with the cut through the quark ($p_2 + k_2$) and gluon ($k_1 - k_2$) lines connecting an external leg to the rest of the diagram and with the soft uncut gluon ($k_1$) attaching in all possible ways to the three lines involved ($p_2$, $p_2 + k_2$, and $k_1 - k_2$) (the cut is mentioned in the text, but not shown explicitly in Fig.~7 of \cite{Kirschner:1983di}). These connections of the soft gluon $k_1$ are shown here in the diagrams C, E and I in \fig{Feyn_gauge_canc1}, using the diagram labeling from \cite{Kovchegov:2016zex}. Employing Eqs.~(B2) from \cite{Kovchegov:2016zex} we readily obtain (division by 4 and 2 is required to single out one diagram in the class of diagrams C, E and I, with the diagrams in each class related to each other by up-down and left-right symmetries)
\begin{align}\label{cancel1}
\left[ \frac{C}{4} + \frac{E}{4} + \frac{I}{2} \right]_{k_{2 T} \gg k_{1 T}} = g^6 \, C_F \, \sigma_1 \, \sigma_2 \, \frac{s}{k_{1 T}^2 \, k_{2 T}^2} \, \left[ -2 + 2 \, \frac{N_c^2 -2}{N_c^2} + \frac{4}{N_c^2} \right] =0. 
\end{align}
Here, as in \cite{Kovchegov:2016zex}, we keep only the part of the amplitude dependent on the polarizations $\sigma_1$ and $\sigma_2$ of the two colliding quarks. Also, $C_F = (N_c^2 -1)/(2 N_c)$ is the fundamental Casimir operator and $g$ is the QCD coupling.

\begin{figure}[ht]
\begin{center}
\includegraphics[width= 0.75 \textwidth]{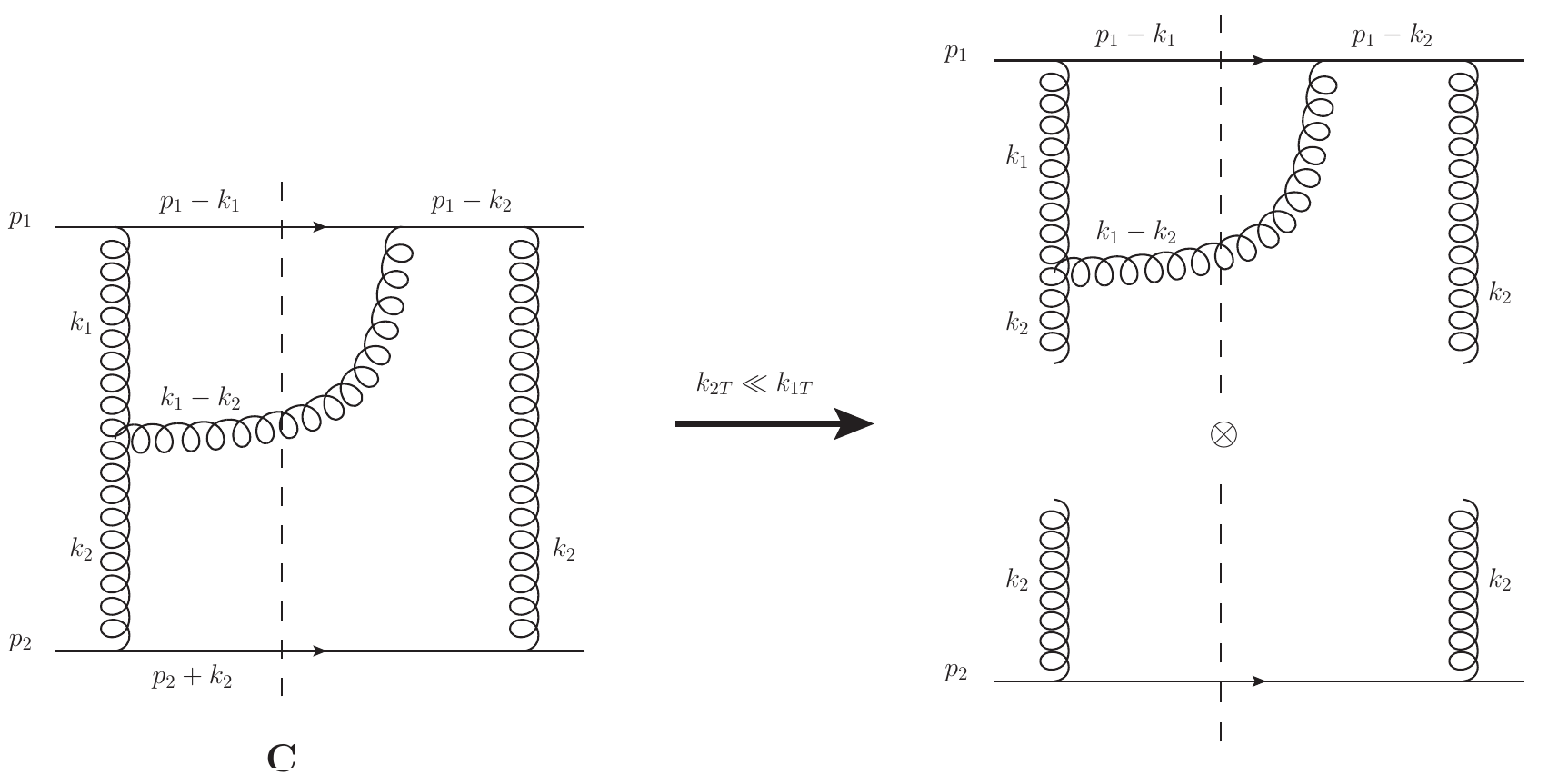} 
\caption{Possible decomposition of the diagram C from Appendix~B of \cite{Kovchegov:2016zex} in the $k_{2 T}  \ll k_{1 T}$ regime under the IREE from \cite{Bartels:1996wc}. The vertical dashed line denotes the cut.}
\label{fig:diagramC}
\end{center}
\end{figure}



In the $k_{1 T} \gg k_{2 T}$ region, both $k_2$ gluons become the softest internal partons in the diagram. According to the IREE framework one should then truncate these two gluons, splitting diagram C into two as shown in \fig{fig:diagramC}. The diagram on the top right of that figure, resulting from this splitting, appears to still be a non-ladder diagram with a hard gluon $k_1 - k_2$ and thus still appears to violate the IREE rules. However, imagine a situation where we attach a bremsstrahlung gluon to a single-rung ladder (an ``H"-shape diagram). By IREE rules, the transverse momentum of the external legs is negligibly small. Therefore, the transverse momentum in the loop formed by attaching a bremsstrahlung gluon to the single-rung ladder is going to be larger than that in the external legs, and by momentum conservation should be the same in the propagators of all the partons forming the loop. For the top right diagram in \fig{fig:diagramC} this implies that, in the $k_{1 T} \gg k_{2 T}$ regime, the transverse momenta of the gluon lines $k_1$ and $k_1 - k_2$ are comparable to each other and to the transverse momentum in the quark line $p_1 - k_1$. Therefore, at this low order in $\as$ (order-$\as^2$), there appears to be no difference between a diagram with a bremsstrahlung gluon and a diagram with a hard non-ladder gluon: the top right diagram in \fig{fig:diagramC} can be viewed as a bremsstrahlung gluon diagram. The remaining question is to identify which gluon is the bremsstrahlung one in the top right diagram of \fig{fig:diagramC}: is it the gluon $k_1$ or $k_1 - k_2$? According to \cite{Kirschner:1983di}, the bremsstrahlung gluon should carry longitudinal (``nonsense") polarization. Since the gluon $k_1 - k_2$ is cut, it can only be polarized transversely, and, hence, cannot be the bremsstrahlung gluon, leaving $k_1$ to be the bremsstrahlung gluon. Therefore, we can view the top right diagram in \fig{fig:diagramC} as a ladder made out of the gluons $k_1 - k_2$ and $k_2$ and the quark line, with the rung of the ladder given by the $p_1 - k_2$ quark line, and with the $k_1$ bremsstrahlung soft gluon attached to the ladder. Therefore, while initially appearing to violate the ``no hard non-ladder gluons"  argument, diagram C (along with the diagram B from Appendix~B of \cite{Kovchegov:2016zex}) can nevertheless be incorporated into the IREE derived by BER. Identifying the forward $2 \to 2$ quark and gluon scattering amplitudes with the anomalous dimensions at the same order in $\as$, per \cite{Kirschner:1983di}, we see that the above discussion appears to explain why the calculation of \cite{Bartels:1996wc} agrees with the polarized DGLAP anomalous dimensions to three loops \cite{Blumlein:1996hb}, that is, to order $\as^3$. 

\begin{figure}[ht]
\begin{center}
\includegraphics[width= \textwidth]{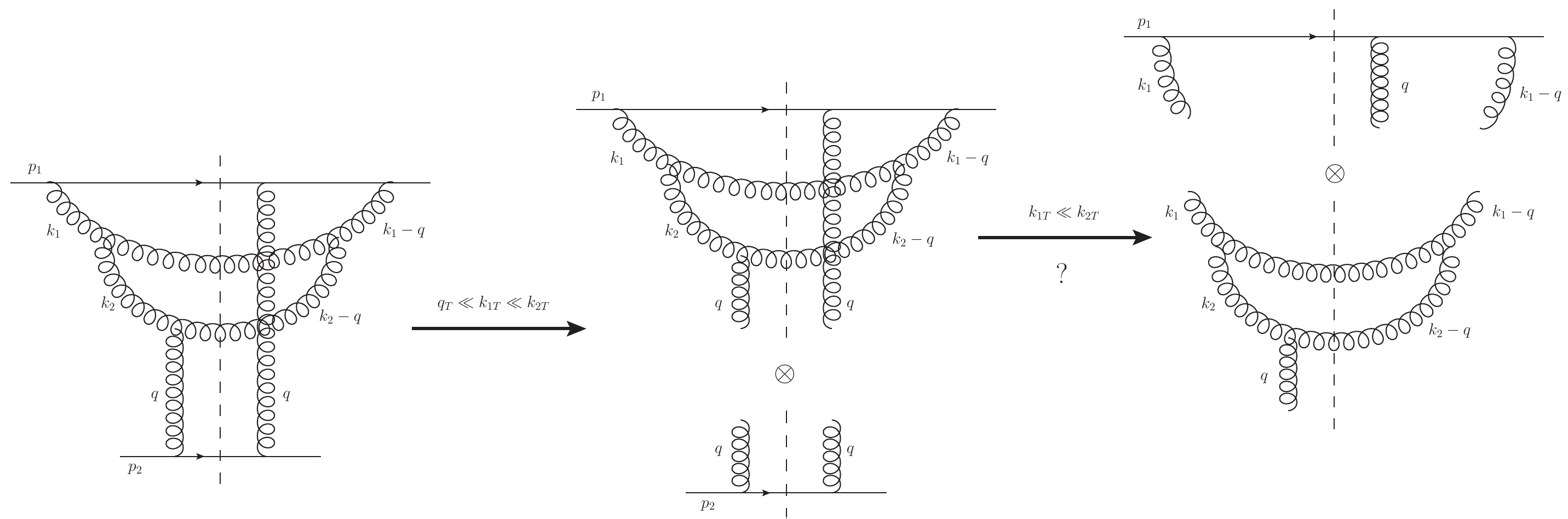} 
\caption{Possible decomposition of a diagram at the order $\as^4$ in the $q_T \ll k_{1 T} \ll k_{2 T}$ regime under the IREE from \cite{Bartels:1996wc}.}
\label{fig:4loops}
\end{center}
\end{figure}

We now want to investigate the hard non-ladder diagrams at higher orders in $\as$. Anomalous dimensions at four loops correspond to $2 \to 2$ forward scattering amplitudes at the order $\as^4$. While a systematic analysis of all order-$\as^4$ diagrams would be rather lengthy, we will consider one relevant diagram to illustrate a possible concern arising at this order. This order-$\as^4$ diagram is given in the left panel of \fig{fig:4loops}. This diagram is known to be double-logarithmic, at least in the small-$x$ helicity framework of \cite{Kovchegov:2015pbl, Kovchegov:2016zex, Kovchegov:2017lsr, Kovchegov:2018znm,Cougoulic:2022gbk}. Note that in \cite{Kovchegov:2015pbl, Kovchegov:2016zex, Kovchegov:2017lsr, Kovchegov:2018znm,Cougoulic:2022gbk} the calculations were performed in the light-cone gauge of the projectile (the upper quark line in \fig{fig:4loops}): it is possible that the diagram in \fig{fig:4loops} is not double-logarithmic in Feynman gauge employed in \cite{Bartels:1996wc}. However, the IREE technique developed in \cite{Kirschner:1983di} is stated to be gauge-invariant by the authors. Hence we proceed by assuming that the diagram in \fig{fig:4loops} is double-logarithmic in Feynman gauge as well. The authors of \cite{Cougoulic:2022gbk} also verified their calculations in background Feynman gauge.


We will concentrate on the kinematic region where $q_T \ll k_{1 T} \ll k_{2 T}$. In this region the diagram in the left panel of \fig{fig:4loops} is still double-logarithmic \cite{Cougoulic:2022gbk}. The diagram appears to be a one-loop vertex correction to diagram B from Appendix~B of \cite{Kovchegov:2016zex}, which in turn is quite similar to diagram C considered above. In the $q_T \ll k_{1 T} \ll k_{2 T}$ kinematic region, there are two softest gluons, the ones carrying momenta $q$ in \fig{fig:4loops} (these are the Glauber gluons in the formalism of \cite{Kovchegov:2015pbl, Kovchegov:2016zex, Kovchegov:2017lsr, Kovchegov:2018znm,Cougoulic:2022gbk}). Truncating these gluons splits the left diagram into the two diagrams in the middle panel of \fig{fig:4loops}. Just like in \fig{fig:diagramC}, the resulting top diagram appears to be non-ladder. One may wonder whether it is also included in the BER evolution, but this appears to be less clear. In the same $q_T \ll k_{1 T} \ll k_{2 T}$ kinematic region, the softest two gluons in the top diagram in the middle panel of \fig{fig:4loops} are $k_1$ and $k_1 -q$. The next question is whether these gluons are (i) ``rails" of some ladder or whether (ii) one of them is a bremsstrahlung gluon. In the case (i), by the IREE rules, truncating those gluons leads to the diagrams on the right of \fig{fig:4loops}. However, these diagrams form 3- and 5-point Green functions, whereas the IREE of \cite{Bartels:1996wc} only contain 4-point Green functions and not diagrams with an odd number of external legs. Therefore, if the ladder ``rails" interpretation from (i) is correct, it appears impossible to obtain the contribution of the diagram on the left of \fig{fig:4loops} in the $q_T \ll k_{1 T} \ll k_{2 T}$ kinematic region using the IREE framework. 

Option (ii), involving a bremsstrahlung gluon, appears to be in-line with our above interpretation of the top right diagram in \fig{fig:diagramC}. However, the top middle diagram in \fig{fig:4loops} has a significant difference from the top right diagram in \fig{fig:diagramC}: both gluons $k_1$ and $k_1 -q$ are not cut. Either of them may carry longitudinal polarization. Therefore it appears unclear which of these two gluons would be the bremsstrahlung one. Moreover, the position of gluons $k_1$ and $k_1 -q$ in the diagram appears to be rather ladder-like, making the applicability of Gribov's bremsstrahlung theorem \cite{Gribov:1966hs,Gorshkov:1969yy} questionable; after all, the bremsstrahlung theorem does not apply to two equally soft gluons forming the ``rails" of a ladder. It therefore appears unlikely that the diagram on the left of \fig{fig:4loops} in the $q_T \ll k_{1 T} \ll k_{2 T}$ kinematic region can be obtained from the IREE developed in \cite{Bartels:1996wc}.

However, with the author's very limited understanding of the IREE framework, we cannot reach a firm conclusion here. Indeed it is also possible that the diagram on the left of \fig{fig:4loops} is not double-logarithmic in Feynman gauge. Alternatively, it may also be possible to interpret this diagram in the BER IREE using some additional observation beyond what was presented in this Appendix. Yet again, in \cite{Bartels:1996wc}, BER do express concern about hard non-ladder diagrams and appear to argue that those are not double-logarithmic: the apparent violation of that argument found in \cite{Kovchegov:2016zex} should manifest itself at some order in $\as$. Moreover, if the concern expressed in this Appendix is correct, the fact that it applies only to a fairly high-order in $\as$ diagram may explain the numerically minor difference of the intercepts in Eqs.~\eqref{intercept} and \eqref{BER_intercept2} and the fact that the expansions \eqref{anomalous_dim_exp} and \eqref{BER_exp} for our and BER anomalous dimensions disagree only starting at order $\as^4$.

\chapter{Scheme Transformation}
\label{scheme.app}

The differences between the polarized small-$x$ large-$N_c \& N_f$ splitting functions in our/BER scheme \eqref{eq:DeltaP_BER} and in the $\overline{\text{MS}}$ scheme  \eqref{eq:DeltaP_MSbar} can be accounted for by a scheme transformation. In the following, we explicitly derive the scheme transformation matrix $Z$ that transforms $\Delta \mathbf{P}$ in our/BER scheme to $\Delta \overline{\mathbf{P}}$ in the $\overline{\text{MS}}$ scheme. In our notation, over-lined quantities are understood to be in the $\overline{\text{MS}}$ scheme. The derivation is performed in Mellin space, where the matrix of anomalous dimensions is
\begin{equation}
\Delta \gamma (\omega) = \int_0^1 dx\, x^{\omega - 1} \Delta \mathbf{P} (x).
\end{equation}
We follow the method in \cite{Moch:2014sna}. 

Let $\Delta f(\omega, \mu^2) \equiv (\Delta \Sigma(\omega, \mu^2), \Delta G(\omega, \mu^2)^T$ denote the quark and gluon helicity PDFs in Mellin space.
The hPDFs in the two schemes are related by
\begin{equation}
\Delta \overline{f}(\omega, \mu^2) = Z(\omega, \mu^2) \, \Delta f(\omega, \mu^2).
\end{equation}
The polarized DGLAP equations \eqref{DGLAP_diff} in Mellin space can be formally expressed as a matrix equation
\begin{equation}
\frac{\pd \Delta \overline{f}(\omega, \mu^2)}{ \pd \ln \mu^2} = \Delta \overline{\gamma}(\omega) \, \Delta \overline{f}(\omega, \mu^2),
\end{equation} 
which can be further written as 
\begin{subequations}
\begin{align}
&\frac{\pd Z(\omega, \mu^2)}{\pd \ln\mu^2} \Delta f(\omega, \mu^2) + Z(\omega, \mu^2)\frac{\pd \Delta f(\omega, \mu^2)}{\pd \ln\mu^2} = \Delta \overline{\gamma}(\omega) Z(\omega,\mu^2) \Delta f(\omega, \mu^2),\\
\Longrightarrow &\,\,\, \frac{\pd Z(\omega, \mu^2)}{\pd \ln\mu^2} \Delta f(\omega, \mu^2) + Z(\omega, \mu^2) \Delta \gamma(\omega) \Delta f(\omega, \mu^2)= \Delta \overline{\gamma}(\omega) Z(\omega, \mu^2) \Delta f(\omega, \mu^2).
\end{align}
\end{subequations}
One therefore obtains
\begin{equation}\label{eq:scheme_tranformation_eq}
\Delta \overline{\gamma}(\omega) = Z(\omega, \mu^2)\Delta \gamma(\omega) Z^{-1}(\omega, \mu^2)  + \frac{\pd Z(\omega, \mu^2)}{\pd \ln\mu^2} \, Z^{-1}(\omega, \mu^2).
\end{equation}
This is the equation to determine the scheme transformation matrix $Z$, given $\Delta \gamma$ and $\Delta \overline{\gamma}$.  In the following, we solve this equation perturbatively order by order in $a_s = \alpha_s(\mu^2)/4\pi$.  The anomalous dimensions have the expansions
\begin{subequations}
\begin{align}
&\Delta \overline{\gamma}(\omega) = a_s \Delta \overline{\gamma}^{(0)}_{\omega} + a_s^2 \Delta \overline{\gamma}^{(1)}_{\omega} + a_s^3 \Delta \overline{\gamma}^{(2)}_{\omega} + \ldots,\\
&\Delta \gamma(\omega) = a_s \Delta \gamma^{(0)}_{\omega} + a_s^2 \Delta \gamma^{(1)}_{\omega} + a_s^3 \Delta \gamma^{(2)}_{\omega} + \ldots.
\end{align}
\end{subequations}
The scheme transformation matrix $Z$ can also be expanded as 
\begin{subequations}
\begin{align}
&Z(\omega, \mu^2) = 1 + a_s Z_{\omega}^{(1)} + a_s^2 Z_{\omega}^{(2)} + \ldots,\\
&Z^{-1}(\omega, \mu^2) = 1- a_s Z_{\omega}^{(1)} - a_s^2\left(Z_{\omega}^{(2)}- \left[Z_{\omega}^{(1)}\right]^2\right) + \mathcal{O}(a_s^3).
\end{align}
\end{subequations}
The dependence on $\mu^2$ comes solely from the running coupling constant $a_s$. 
We start with the first term on the right-hand side of Eq.~\eqref{eq:scheme_tranformation_eq}.
\begin{align}
&Z(\omega, \mu^2)\Delta \gamma(\omega)Z^{-1}(\omega, \mu^2)
=  a_s \Delta \gamma^{(0)}_{\omega}+a_s^2\left(\Delta \gamma^{(1)}_{\omega}  + \left[Z^{(1)}_{\omega}, \Delta \gamma^{(0)}_{\omega}\right]\right) \\
&\hspace*{2cm}+a_s^3 \left(\Delta \gamma^{(2)}_{\omega} +\left[Z^{(1)}_{\omega}, \Delta \gamma^{(1)}_{\omega}\right] +  \left[Z^{(2)}_{\omega}, \Delta \gamma^{(0)}_{\omega}\right] - \left[Z^{(1)}_{\omega}, \Delta \gamma^{(0)}_{\omega}\right] Z^{(1)}_{\omega}\right) +\mathcal{O}(a_s^4). \notag
\end{align}
We then compute the second term on the right-hand side of Eq.~\eqref{eq:scheme_tranformation_eq} by first noting that 
\begin{equation}
\frac{ \pd Z(\omega, \mu^2)}{ \pd \ln\mu^2} = \frac{da_s}{d\ln\mu^2} Z^{(1)}_{\omega} + 2a_s \frac{da_s}{d\ln \mu^2} Z^{(2)}_{\omega} + \ldots .
\end{equation}
Here the running coupling constant satisfies the standard renormalization-group equation
\begin{equation}
\frac{da_s}{d\ln\mu^2} = \beta(a_s) = -\sum_{l=0}^{\infty} a_s^{l+2} \beta_l
\end{equation}
with the leading coefficients in the large $N_c\& N_f$ limit being 
\begin{subequations}
\begin{align}
&\beta_0 = \frac{11}{3}N_c-\frac{2}{3}N_f, \\
&\beta_1 = \frac{34}{3}N_c^2 - \frac{13}{3}N_cN_f.
\end{align}
\end{subequations}
One then obtains
\begin{equation}
\frac{\pd Z(\omega, \mu^2)}{\pd \ln\mu^2} Z^{-1}(\omega, \mu^2) = -a_s^2 \beta_0 Z^{(1)}_{\omega} -a_s^3 \left(\beta_1 Z^{(1)}_{\omega} + 2\beta_0 Z^{(2)}_{\omega} - \beta_0 \left[Z^{(1)}_{\omega}\right]^2\right) + \mathcal{O}(a_s^4).
\end{equation}
Using the above expressions, Eq.~\eqref{eq:scheme_tranformation_eq} becomes
\begin{align}
&a_s \Delta \overline{\gamma}^{(0)}_{\omega} + a_s^2 \Delta \overline{\gamma}^{(1)}_{\omega}+ a_s^3 \Delta \overline{\gamma}^{(2)}_{\omega} + \mathcal{O}(a_s^4)\\
&=  a_s \Delta \gamma^{(0)}_{\omega} +a_s^2\left(\Delta \gamma^{(1)}_{\omega}  + \left[Z^{(1)}_{\omega}, \Delta \gamma^{(0)}_{\omega}\right]- \beta_0 Z^{(1)}_{\omega}\right) \notag \\
&\hspace*{.4cm}+a_s^3 \Big(\Delta \gamma^{(2)}_{\omega} + \left[Z^{(1)}_{\omega}, \Delta \gamma^{(1)}_{\omega}\right] +  \left[Z^{(2)}_{\omega}, \Delta \gamma^{(0)}_{\omega}\right] \notag\\
 &\hspace*{2cm}- \left[Z^{(1)}_{\omega}, \Delta \gamma^{(0)}_{\omega}\right] Z^{(1)}_{\omega}  - \left(\beta_1 Z^{(1)}_{\omega} + 2\beta_0 Z^{(2)}_{\omega} - \beta_0 [Z^{(1)}_{\omega}]^2\right)\Big) +\mathcal{O}(a_s^4). \notag
\end{align}
Comparing the left-hand side and the right-hand side of the equation, one arrives at 
\begin{subequations}
\begin{align}
&\Delta \overline{\gamma}^{(0)}_{\omega} = \Delta \gamma^{(0)}_{\omega}, \\
& \Delta \overline{\gamma}^{(1)}_{\omega}  = \Delta \gamma_{\omega}^{(1)}  + \left[Z^{(1)}_{\omega}, \Delta \gamma_{\omega}^{(0)}\right]-\beta_0 Z^{(1)}_{\omega},\label{eq:ObyO_relation_(1)}\\
\begin{split}
& \Delta \overline{\gamma}_{\omega}^{(2)} =  \Delta \gamma_{\omega}^{(2)} + \left[Z^{(1)}_{\omega}, \Delta \gamma_{\omega}^{(1)}\right] +  \left[Z^{(2)}_{\omega}, \Delta \gamma_{\omega}^{(0)}\right] - \left[Z_{\omega}^{(1)}, \Delta \gamma_{\omega}^{(0)}\right] Z_{\omega}^{(1)} \\
&\qquad \qquad- \left(\beta_1 Z_{\omega}^{(1)} + 2\beta_0 Z_{\omega}^{(2)} - \beta_0 [Z_{\omega}^{(1)}]^2\right).  \label{eq:ObyO_relation_(2)}
\end{split}
\end{align}
\end{subequations}
From the explicit expressions for the splitting functions in Eqs.~\eqref{eq:DeltaP_BER}
and \eqref{eq:DeltaP_MSbar}, we have $\Delta \overline{\gamma}_{\omega}^{(0)} = \Delta \gamma_{\omega}^{(0)}$ and $\Delta \overline{\gamma}_{\omega}^{(1)} = \Delta \gamma_{\omega}^{(1)}$.  Eq.~\eqref{eq:ObyO_relation_(1)} becomes 
\begin{equation}
\left[Z_{\omega}^{(1)}, \Delta \gamma_{\omega}^{(0)}\right]-\beta_0 Z_{\omega}^{(1)}=0.
\end{equation}
One can readily see that the only solution is $Z_{\omega}^{(1)} =0$. As a result, Eq.~\eqref{eq:ObyO_relation_(2)} reduces to
\begin{equation}\label{eq:DeltaP(2)_equation}
\Delta \overline{\gamma}_{\omega}^{(2)} - \Delta \gamma_{\omega}^{(2)} =  \left[Z_{\omega}^{(2)}, \Delta \gamma_{\omega}^{(0)}\right]  - 2\beta_0 Z_{\omega}^{(2)}.
\end{equation}
Denoting the elements of the scheme transformation matrix at order $a_s^2$ as
\begin{equation}
Z^{(2)}_{\omega} = \begin{pmatrix}
Z^{(2)}_{qq}(\omega)\, & \,Z^{(2)}_{qG}(\omega)\\[10pt]
Z^{(2)}_{Gq}(\omega)\, & \,Z^{(2)}_{GG}(\omega)\\
\end{pmatrix} ,
\end{equation}
the four component equations from the matrix equation in Eq.~\eqref{eq:DeltaP(2)_equation} are
\begin{subequations}\label{eq:4components}
\begin{align}
&\Delta \overline{\gamma}^{(2)}_{qq}(\omega) - \Delta \gamma^{(2)}_{qq}(\omega) = Z^{(2)}_{qG}(\omega)\Delta \gamma^{(0)}_{Gq}(\omega) - \Delta \gamma^{(0)}_{qG}(\omega)Z^{(2)}_{Gq}(\omega) - 2\beta_0 Z^{(2)}_{qq}(\omega) =0,\\
&\Delta \overline{\gamma}^{(2)}_{GG}(\omega) - \Delta \gamma^{(2)}_{GG}(\omega) = Z^{(2)}_{Gq}(\omega)\Delta \gamma^{(0)}_{qG}(\omega) - \Delta \gamma^{(0)}_{Gq}(\omega) Z^{(2)}_{qG}(\omega) - 2\beta_0 Z^{(2)}_{GG}(\omega) =0, \\
&\Delta \overline{\gamma}^{(2)}_{qG}(\omega) - \Delta \gamma^{(2)}_{qG}(\omega) = \Delta \gamma^{(0)}_{qG}(\omega)\left(Z^{(2)}_{qq}(\omega)-Z^{(2)}_{GG}(\omega)\right) \\
&\hspace*{5cm}+ Z^{(2)}_{qG}(\omega)\left(\Delta \gamma^{(0)}_{GG}(\omega) - \Delta \gamma^{(0)}_{qq}(\omega)\right) , \notag\\
&\Delta \overline{\gamma}^{(2)}_{Gq}(\omega) - \Delta \gamma^{(2)}_{Gq}(\omega) = \Delta \gamma^{(0)}_{Gq}(\omega)\left(Z^{(2)}_{GG}(\omega)- Z^{(2)}_{qq}(\omega)\right) \\
&\hspace*{5cm}+ Z^{(2)}_{Gq}(\omega)\left(\Delta \gamma^{(0)}_{qq}(\omega) - \Delta \gamma^{(0)}_{GG}(\omega)\right). \notag
\end{align}
\end{subequations}
The solutions are 
\begin{subequations}\label{eq:formal_solutions_Z(2)}
    \begin{align}
        &Z^{(2)}_{GG}(\omega)=-Z^{(2)}_{qq}(\omega),\\
        \begin{split}
&Z^{(2)}_{qq}(\omega) = \frac{1}{2\beta_0} \left[\frac{\Delta \gamma^{(0)}_{Gq}(\omega)}{\Delta \gamma^{(0)}_{GG}(\omega) - \Delta \gamma^{(0)}_{qq}(\omega)} (\Delta \overline{\gamma}^{(2)}_{qG}(\omega) - \Delta \gamma^{(2)}_{qG}(\omega)) \right. \\
& \hspace*{3cm} \left. + \frac{\Delta \gamma^{(0)}_{qG}(\omega)}{\Delta \gamma^{(0)}_{GG}(\omega) - \Delta \gamma^{(0)}_{qq}(\omega)} (\Delta \overline{\gamma}^{(2)}_{Gq}(\omega) - \Delta \gamma^{(2)}_{Gq}(\omega)) \right],\\
\end{split}\\
&Z^{(2)}_{qG}(\omega) = \frac{\Delta \overline{\gamma}^{(2)}_{qG}(\omega) - \Delta \gamma^{(2)}_{qG}(\omega)}{\Delta \gamma^{(0)}_{GG}(\omega) - \Delta \gamma^{(0)}_{qq}(\omega)} - \frac{2\Delta \gamma^{(0)}_{qG}(\omega)}{\Delta \gamma^{(0)}_{GG}(\omega) - \Delta \gamma^{(0)}_{qq}(\omega)} Z^{(2)}_{qq}(\omega),\\
&Z^{(2)}_{Gq}(\omega) = - \frac{\Delta \overline{\gamma}^{(2)}_{Gq}(\omega) - \Delta \gamma^{(2)}_{Gq}(\omega)}{\Delta \gamma^{(0)}_{GG}(\omega) - \Delta \gamma^{(0)}_{qq}(\omega)}- \frac{2\Delta \gamma^{(0)}_{Gq}(\omega)}{\Delta \gamma^{(0)}_{GG}(\omega) - \Delta \gamma^{(0)}_{qq}(\omega)} Z^{(2)}_{qq}(\omega).
    \end{align}
\end{subequations}
The relevant anomalous dimensions at fixed orders are 
\begin{equation}
\Delta \gamma_{qq}^{(0)}(\omega) = N_c\frac{1}{\omega}, \quad \Delta \gamma_{GG}^{(0)}(\omega) = 8N_c \frac{1}{\omega}, \quad \Delta \gamma^{(0)}_{qG}(\omega) = -2N_f\frac{1}{\omega}, \quad \Delta \gamma^{(0)}_{Gq}(\omega) = 2N_c\frac{1}{\omega}
\end{equation}
and 
\begin{subequations}
\begin{align}
&\Delta \overline{\gamma}^{(2)}_{qG}(\omega) - \Delta \gamma^{(2)}_{qG}(\omega) = 4N_c^2N_f \frac{1}{\omega^5},\\
&\Delta \overline{\gamma}^{(2)}_{Gq}(\omega)- \Delta \gamma^{(2)}_{Gq}(\omega) = 4N_c^3 \frac{1}{\omega}.
\end{align}
\end{subequations}
Substituting these expressions into Eq.~\eqref{eq:formal_solutions_Z(2)}, one obtains 
\begin{subequations}
\begin{align}
&Z^{(2)}_{qq}(\omega)=-Z^{(2)}_{GG}(\omega)=0,\\
&Z^{(2)}_{qG}(\omega) = \frac{4}{7}N_cN_f \frac{1}{\omega^4},\\
&Z^{(2)}_{Gq}(\omega) = -\frac{4}{7} N_c^2 \frac{1}{\omega^4}.
\end{align}
\end{subequations}
In matrix form this can be written as
\begin{equation}
Z^{(2)}_{\omega} = \begin{pmatrix}
0 & \frac{4}{7} N_cN_f \frac{1}{\omega^4} \\[10pt]
-\frac{4}{7}N_c^2 \frac{1}{\omega^4} & 0 \\
\end{pmatrix}.
\end{equation}
Transforming back to $x$-space, we obtain
\begin{equation}
Z^{(2)}(x) = \begin{pmatrix}
0 & \frac{2}{21}N_cN_f \ln^3\frac{1}{x} \\[10pt]
-\frac{2}{21}N_c^2 \ln^3\frac{1}{x} & 0 \\
\end{pmatrix}.
\end{equation}

We conclude that the scheme transformation relating the $\overline{\text{MS}}$ splitting functions in Eqs.~\eqref{eq:DeltaP_MSbar} to our/BER splitting functions in   Eqs.~\eqref{eq:DeltaP_BER} up to three loops is given by the following matrix (where again $a_s = \alpha_s(\mu^2)/4\pi$): 
\begin{equation}
Z (x) = \begin{pmatrix}
x \, \delta (1-x) & \frac{2}{21} \, a_s^2 \, N_cN_f \ln^3\frac{1}{x} \\[10pt]
-\frac{2}{21} \, a_s^2 \, N_c^2 \ln^3\frac{1}{x} & x \, \delta (1-x) \\
\end{pmatrix}.
\end{equation}

\end{document}